# Scale-Free Identity: The Emergence of Social Network Science

Haiko Lietz

Para Oc 3

# Contents

# List of Figures

# List of Tables

# Acknowledgements

This work was sparked in 2006/2007 in Harrison White's seminars at Columbia University. In *Identity and Control* I have found a theory to embed into my longtime interest in fractal spaces and times. Three persons are central for how things have turned out. Klaus Liepelt at Mittweida University has been most helpful and instrumental finding into science and Lothar Krempel at MPIfG finding through science. Markus Strohmaier at GESIS supported this work all the way.

Three guys have become friends throughout many years of discussing research. Thank you Diego Rybski, Felix Scholkmann, and Arnim Bleier for valuable discussions and for making things fly.

I have benefited much from the technical help I have received. Thank you Thomas Gurney and Ulf Sandström for assistance with the author name disambiguation, Jim Moody for running the structural-cohesion algorithm, and Vladimir Batagelj and Andrej Mrvar for providing the wonderful and bug-free *Pajek* software and helping out within a minute.

Use of the database of the German Competence Centre for Bibliometrics is acknowledged.

# Summary

## Problem

Science is full of punctuating events that terminate periods during which styles of doing research are more or less reproduced. During the Constructivist Turn (about 1976), the sociology of science left the institutionalist program behind and turned towards the Sociology of Scientific Knowledge with its focus on context effects. Social Network Analysis is a way of studying agents embedded in contexts. During the Cultural Turn (about 1992), a fraction of this domain initiated Relational Sociology, strongly associated with the theory of Harrison C. White. Proponents advocate modeling social networks not purely structurally but as intertwined with cultural meaning. Then, in about 1998, physicists discovered social networks as representations of complex systems. Small-world and scale-free networks are the paradigmatic models of this Network Science, the emergence of which marks the Complexity Turn in Social Network Analysis. This work addresses the structure/culture, micro/macro, and stability/change problems. How useful is Relational Sociology's concept of identity to model scientific communities? What is the importance of emergence in modeling identity? What mechanism can explain stability as well as change? What is the contribution of Network Science modeling?

## Approach

Relying on various models and mechanisms of socio-cultural processes from Relational Sociology and Complexity Science, an identity model is developed and calibrated in a case study of Social Network Science. This research domain results from the union of Social Network Analysis and Network Science. A unique dataset of 25,760 scholarly articles from one century of research (1916–2012) is created. Clustering this set of publications, five subdomains are detected that are labeled Social Psychology, Economic Sociology, Social Network Analysis, Complexity Science, and Web Science. These identities are then analyzed in terms of authorship, citation, and word usage structures and dynamics. For this purpose, a graph theoretical data model is developed that allows comparisons across these three scholarly practices. In this model, authors, cited references, and words are treated as Durkheimian social facts. The scaling hypothesis of percolation theory is formulated for socio-cultural systems, namely that power-law size distributions like Lotka's, Bradford's, and Zipf's Law mean that the described identity resides at the phase transition between the stability and change of meaning. In this case, it can be diagnosed using bivariate scaling laws and Abbott's heuristic of fractal distinctions.

### Results

Identities are not dichotomies but dualities of social network and cultural domain, micro and macro phenomena, as well as stability and change. First, story sets that give direction to research fluctuate less, are less distinctive, and more inert than the individuals doing the research. Words have longer average lifetimes than authors, and word co-usage meaning structures are more centralized than co-authorship networks. Second, identities are scale-free. Not only are persons, groups, organizations, etc. manifestations of an idealized identity at different levels of socio-cultural complexity. Concrete identities also extend over multiple such levels. Third, six senses are diagnostic of different aspects of identity, and when they come together as process, a complex socio-cultural system comes into existence. The scaling hypothesis needs not be rejected. The convergence of identities to linear preferential attachment indicates that stability and change co-exist at a fractal phase transition. As expected from percolation theory, this state can be described by a number of scaling laws. Self-organization to criticality is expressed through hierarchically modular small-world social structures and self-similar meaning structures and dynamics. The evolution of the domain is convergent, i.e., it does not progress in a division of labor but in distinctions that repeat in themselves. Social Psychology is an exception because its story set is too different. All other subdomains continuously change through mating with other styles. The Complexity Turn of 1998 was not a scientific revolution because Social Network Science was not normal science until 2002. It was a scientific breakthrough that caused all subdomains but Social Psychology to markedly innovate.

### Contribution

A scale-free identity model with a corresponding data model is built that allows for studying the structure and dynamics of complex socio-cultural systems. The model is calibrated by operationalizing concepts from the toolbox of Relational Sociology (identity, control, autocatalysis, discipline, institution, style, switching, Bayesian fork, innovation, invention, ambage, and ambiguity) and studying the identity Social Network Science using a new data set. A mutual benefit that results from mating Relational Sociology and Network Science is identified. The latter can learn from the former that social systems are dualities of transactions and meaning and that studying multiple dimensions of the same system creates valuable insights about their identity. For the social sciences, the importance of Paretian thinking (scale invariance) is pointed out. The meaning of small-world and scale-free network models is that a mechanism of fractal optimization keeps identities adaptable through balancing stability and change. First steps are taken towards predicting change through a calculus of social and cultural uncertainty, to be developed in a framework of Computational Social Science.

# Introduction

In the mid 60s, the thing to do in Northern American sociology was Structural Functionalism, the grand theory of Talcott Parsons (1951) who aimed at a broadband systems explanation of all social phenomena. Parsons was the guiding spirit and chairman of the Department of Social Relations at Harvard University, but despite the name, the relational approach was pursued by Parsons strongest opponent Harrison C. White. Together with a group of graduate students, the physicist-turned-sociologist was developing a research program according to which social systems are not bunches of attributes and attitudes but composed of, and influenced by, social networks (H. C. White, 2008 [1965]; Schwartz, 2008).

From White's relational program, Mark Granovetter's paper "The strength of weak ties" (1973) emerged which shows that social networks can have both enabling and constraining effects. It is today the most cited paper in the social sciences.[1] But when it was submitted in 1969, framed as a contribution to the literature on alienation, reviewers were not convinced by the scholarship brought forward. An anonymous reviewer commented that "[t]he question of why some form of cohesion exists in large social contexts is indeed a good one, as Durkheim showed in *The Division of Labor*, but it must be answered in terms of '*what* ties together' not in terms of a definition of the ties themselves."[2] This critique is in line with the Parsonian view of the person as the source of ties, but this is not at all what the relational program was aiming at. The latter is well laid out by another breakthrough paper of White et al., François Lorrain and White's (1971) definition of structural equivalence. According to this idea, persons are a result of their social relations, not the other way around. In hindsight, White's program won over Parsons' as it not only created a renaissance at Harvard but triggered the establishment of Social Network Analysis (Freeman, 2004).

Granovetter succeeded getting his paper published by removing the distracting alienation frame and focusing on his ideational precursors. This tradition includes models of random networks that are biased towards closing triads in order to explain the clustering observed in social networks, developed by psychologist Anatol Rapoport and co-workers (Solomonoff & Rapoport, 1951; Rapoport, 1957). Another precursor was the experimental discovery of social psychologist Stanley Milgram that an arbitrary pair of persons is, on average, connected at a very short social distance of six or through five intermediaries (Milgram, 1967; Travers & Milgram, 1969).

Another lineage of network modeling was initiated by mathematicians Paul Erdős and Alfréd Rényi in 1959. Their fundamental finding about random graphs is that, when the creation of a connection or edge is independent of all edges already created and the nodes have more than one edge on average (an average degree larger than one), then nodes will form a giant connected component. The theory of random graphs became the standard tool for physicists because it provided a model for how strongly parts of a system are coupled. However, in 1996 physicist Duncan J. Watts and mathematician Steven H. Strogatz were hooked by the idea that social networks, not random graphs, could deliver valuable insights into the puzzle how biological oscillators synchronize. Two years later, Watts and Strogatz (1998) were able to show that Milgram's small world can be explained by a network that interpolates between complete randomness and complete order. Between these extremes, nodes are both connected at short average distance and they cluster into groups. More than that, such complex networks also exist in nature and technology. Small-world networks are complex because the name-giving property emerges from collective dynamics.

[1] en.wikipedia.org/wiki/Mark_Granovetter, visited July 19th, 2015

[2] scatter.wordpress.com/2014/10/13/granovetter-rejection/, visited July 19th, 2015

Only a year later, another property of complex networks was discovered. The group of physicist Albert-László Barabási had just crawled a part of the World Wide Web and was curious if a giant component had emerged. Departing from the randomness of the Erdős/Rényi model, the natural thing to do was to study the degree distribution. The result was "a shocking departure from everything I learned during my ... journey into networks," Barabási (2015, ch. 0, p. 10) recalls. Nodes did not have a characteristic number of links. The degree distribution was instead a power law, a mathematical function first observed in the social sciences in the context of wealth by the economist Vilfredo Pareto (1896–97). The important property of power laws is that they are typically not subject to the Central Limit Theorem of probability theory. Barabási: "The power law observed by us predicted that the Web has hubs, nodes with a huge number of links, outliers forbidden in a random universe. None of the existing models could account for them." The simple evolutionary model of Barabási and Réka Albert (1999) then constituted the breakthrough for scale-free networks that have no intrinsic scale. All it took was adding preferential attachment to the Erdős/Rényi model, that the probability for a node to acquire a new edge is proportional to its present degree. Besides the Web, two of Watts and Strogatz's small-world networks, the actor collaboration graph and the power grid, are also scale-free.

The Watts/Strogatz and Barabási/Albert models have been roaringly successful. Within ten years after publication, the papers were among the ten most cited ones in physics (Barabási, 2012b). Barabási (2002) and Watts (2003) wrote bestsellers about their work. Urry (2005) sees these books as part of a Complexity Turn that has been transforming the natural and social sciences since about the mid 90s. Sociologist Linton C. Freeman, a pioneer of Social Network Analysis and chronicler of the research domain, calls the attention social networks have received by physicists a "revolution" (Freeman, 2011). The discovery had not come easy. In a revealing statement in his book *Six Degrees: The Science of a Connected Age*, Watts (2003) tells why he and Strogatz had *not* discovered scale-free networks. "We were so convinced that nonnormal degree distributions weren't relevant that we never thought to look at which networks actually *had* normal degree distributions and which ones did not. We had the data sitting there staring at us for almost two years, and it would have taken all of half an hour to check it, but we never did." (p. 103)

The Complexity Turn involved some "resentment" among social network analysts that physicists had somehow "invaded" their turf (Bonacich, 2004). Sentiments were still being voiced many years after the fact, e.g., by Freeman (2008, 2011). The reason was that many physicists that entered this research domain after 1998 did not acknowledge earlier works on social networks and instead proclaimed a "new science," like Barabási in his book *Linked: The New Science of Networks* (2002). Barabási and Albert were not aware of precursors to their work (cf. Barabási, 2015, ch. 0, p. 12). The fact that degree distributions of empirical social networks were more skewed than normal had been observed by the founder of Social Network Analysis, psychiatrist Jacob L. Moreno (with psychologist Helen Jennings) as early as 1938. Preferential attachment had been popularized as the Matthew Effect by sociologist Robert Merton (1973 [1968]) and had indeed been formulated as a mechanism of social network formation by physicist Derek J. de Solla Price (1986 [1965], 1976). Other than Barabási, Watts (1999) actively approached the social sciences and published an article in the *American Journal of Sociology* as White hired him as a sociology professor at Columbia University. In a review article, Watts (2004) defended the label of a "new" science because it "does capture the sense of excitement surrounding what is unquestionably a fast developing field – new papers are appearing almost daily – and also the unprecedented degree of synthesis that this excitement has generated across the many disciplines in which network related problems arise." (p. 243)

### Research Problem and Contribution

These introductory episodes are accounts of friction at the boundaries of scientific communities. Change is resisted. Granovetter wanted to change reductionist sociology but got to learn its tenacity. Watts deviated from typical physicists' routines but, eventually, the habit to think in terms of plain randomness kept him from testing for power-law behavior. Barabási made the test, but his path kept him from delving into the Social Network Analysis literature. His path led towards Network Science, the self-chosen name of the multidisciplinary community that studies networks and their models (Brandes, Robins, McCranie, & Wasserman, 2013). How useful is the concept of identity to model communities and the friction at their boundaries? What is the importance of emergence in modeling identity? What mechanism can explain stability and change, the maintenance and revision of identity boundaries? What is the contribution of Network Science modeling?

In this work, we study identities as scientific communities in Social Network Science, the research domain that results from the union of what we have introduced as Social Network Analysis and Network Science. We construct a unique dataset of 25,760 scholarly articles from one century of research (1916–2012). Clustering this set of publications, five subdomains are detected that are labeled Social Psychology, Economic Sociology, Social Network Analysis, Complexity Science, and Web Science (ordered by age). These identities are then analyzed in terms of co-authorship, reference co-citation, and word co-usage structures and dynamics. For this purpose, a graph theoretical data model is developed that allows comparisons across these three scholarly practices. In this model, authors, cited references, and words are treated as Durkheimian social facts.

We will show that studying the practices in parallel carries the benefit of being able to show that identities as collectives are network domains, dualities of social networks and cultural domains. Subdomains continue to exist as story sets while their parts (authors) are exchanged. Identities are more than the sum of their parts. Just studying "social networks" misses the emergent meaning structures and vice versa. In this duality lies the explanation for the outer and inner resistance that Granovetter, the Social Network Analysis community, and the Network Science community encountered. The reviewers of Granovetter's paper and the physicists like Barabási were not ignorant. Meaning structures of social facts, besides emerging *from* transactions in social networks, also formulate expectations *for* behavior in networks. They provide the stories that can only be communicated in a given network domain – here: research domain. The stories offered by Granovetter and by Social Network Analysis were just too unexpected. Allowing them would have meant for the reviewers and the physicists to change.

Based on a model of identity as a feedback loop of emergence and constraint, we will show that, during the Complexity Turn of 1998, indeed a new science was born, "new" because the mainstream of Social Network Science acquired a radically new story set. Before the turn, stories were related to social scientific concepts like positions, roles, hierarchical structures, and centrality (Freeman, 2004, 2011), but after the turn, the first two concepts were forgotten and the last two reinvented by physicists. Despite this almost complete change of face, the Complexity Turn was not a Kuhnian scientific revolution for the simple reason that the domain had not been normal science before. Social Network Science only became normal in 2002, i.e., the turn was a breakthrough that gave a boost to an emerging science. After the turn, research in Complexity Science was itself subject to changes. And other than commonly portrayed (Freeman, 2004; Scott, 2012), the anthropological research in Social Psychology is somewhat decoupled from the whole domain.

We are theoretically at home in Relational Sociology. The central assumption is that identities – persons, groups, organizations, etc. – are not origins but emergent outcomes of processes that are simultaneously social and cultural. We take concepts and mechanisms from *Identity and Control*, the toolbox of H. C. White (2008), to model six senses of identity. The first two senses correspond to the parts and the whole, the third sense to paths in time, and the fourth sense to retrospective consciousness. The fifth sense of identity called style is most directly related to the Matthew Effect as the prime mechanism of identity construction. But style is much more, it is a sensibility of processing punctuations and allowing some change while maintaining a degree of recognizability or stability in shifting contexts. We propose a sixth sense of identity: what identity could be. The relational ontology and much of our terminology is inspired by, and derived from, the work of H. C. White. Padgett and Powell (2012a) push this tradition of network theory in the direction of large-scale historical transformations and contribute concrete modeling of reproduction and socio-cultural life. Abbott (2001a, 2001b) sharpens the approach of using stories as units of analysis and provides the heuristic of fractal distinctions according to which decisions regarding stability and change historically repeat in themselves. S. Fuchs (2001) has worked out the philosophical foundation.

To aid the explanation of the simultaneity of stability and change through style and the Matthew Effect, we offer an interpretation of percolation theory in socio-cultural contexts. Percolation theory is mostly used by physicists to predict the behavior of substances like water or a polymer at their respective critical point. For example, at 0°C and normal pressure, the percolation threshold called the liquid-to-solid phase transition, water co-exists in clusters of ice and clusters of liquid water. The regime of the phase transition has a fractal structure, i.e., cluster size is distributed by a power law, and bivariate power laws called scaling laws describe the system (Bunde & Havlin, 1996). This scaling hypothesis also applies to random graphs which builds the bridge to identities and network domains. Here, it means that power laws point at a fractal phase transition where disorder and order or stability and change co-exist at all length scales. We show that size distributions like Lotka's Law (1926), Bradford's Law (1985 [1934]), and Zipf's Law (2012 [1949]) and numerous scaling laws are diagnostic of identity emergence. They signal that a coherent story set has emerged which provides long-term stability but is changed in rare extreme events. It is in this context that Abbott's heuristic of fractal distinctions best unfolds its meaning. It provides a sociological understanding of an idealized scale-free identity that the five subdomains, several research fronts, as well as individual scholars are manifestations of.

Besides Relational Sociology, we contribute to Complexity Science, Computational Social Science, historical sociology, and bibliometrics. We hope that social scientists will learn about the opportunities that lie in Paretian thinking and modeling complexity, and that natural and computer scientists involved with studying "social systems" will learn about the aspect of meaning which makes society different from nature. In this sense, our work is well in line with that called for by the Gulbenkian Commission on the Restructuring of the Social Sciences (Wallerstein, 1996) and practiced in the "Information Society as a Complex System" project (Lane, Pumain, van der Leeuw, & West, 2009). We practice a historical and analytical sociology of complexity and process (Hedström & Bearman, 2009a).

The reader can expect answers to the following questions, among others: How can identity be modeled? How does the Matthew Effect work? How do small worlds emerge? How can the six senses of identity as well as other central concepts from White's toolbox like discipline, institution, switching, Bayesian fork, ambage, or ambiguity be operationalized? What is the difference between social facts, events, stories, and narratives? What are transaction structures and meaning structures, how are they operationalized, and how do they relate to co-authorship, citation, and "co-word" networks? Is autocatalysis a useful concept? Do boundaries exist and how can identities be delineated? Questions related to scaling are: How does stability relate to change? How do fractal distinctions add to the explanation? What do power-law distributions mean? How can scaling laws be defined and interpreted? Regarding large-scale socio-cultural formations, we answer questions like: How can emerging science, normal science, paradigms, and scientific revolutions be operationalized? What is innovation and invention and how are they related? How can research fronts and invisible colleges be understood as meaning structures? There are also epistemological issues: Why is it important to account for different levels of observation? What can Relational Sociology and Complexity Science learn from each other? What are the chances for prediction?

## Content

There are three chapters. The first one is theoretical, it delivers the concepts and tools for our empirical analysis. In the first of three section, we lay out our *paradigms*, the story sets from which we depart. Historical narratives and turning points, namely the Constructivist Turn in the sociology of science, the Harvard renaissance in sociology, and the Cultural Turn in Social Network Analysis are discussed and used to introduce the idea of fractal distinctions. The feedback mechanism central to our identity model is traced back to Durkheim. Kuhn's phase model of scientific progress is shortly reviewed. In the second section, we present the *rules* of method by which we abide. Allowing for variation means rejecting persons as society's atoms and modeling identities as outcomes of network process. The first two senses of identity, the basic duality of transactions and meaning, and the scaling hypothesis are introduced. Taking a narrative approach means explaining identities as processes in network domains. Concepts like story, event, social fact, and narrative are disentangled, and inspiration is drawn from modeling identities as living systems. Accounting for observation means distinguishing between transactions in social networks, selections of social facts, and observations of identities. Doing this improves epistemological precision. In the third section, we collect and anneal various network *models* to get to the core of the identity mechanism. First, we discuss the boundary problem, how permeable boundaries emerge from fractal distinctions, and what properties static meaning structures have. Second, disciplines are introduced as blueprints for orderly reproduction from which predictability arises. Identities in the third and fourth sense are developed, and preferential attachment is identified as the mechanism how identities get to the phase transition. Third, the phase transition is identified as an optimal state for balancing stability and change, the ultimate form of control performed by identities in the fifth sense. Meaning structures function as a memory that resists change but gets reset by punctuations which weaken stability. The state of research about our case, Social Network Science, is reviewed throughout this chapter.

The second chapter is about operationalization and comes in two sections. In the first section, we discuss our *data*. Bibliographic data perfectly enables the narrative approach. Our method is bibliometrics which is very relational and provides approaches to confront the boundary problem. Identities are the objects of observation and stories the units of analysis. Our data model is a bipartite graph of publications and social facts connected by selections. The scientific practices of authorship, citation, and word usage are treated equally to enable comparisons. In the second section, we lay out our *methodology*. To exemplify our methods and expectations, an idealtypical knowledge production process based on the concept of disciplines is discussed. The procedure for detecting the subdomains of Social Network Science and for constructing overlapping meaning structures and levels of complexity is explained. Social Network Science is diagnosed from the three perspectives of events, collaboration, and criticality. Each is focused on a different sense of identity and is guided by a specific research question that formulates expectations for calibrating the model.

The third chapter reports the results of our case study. The first of three sections is about the *delineation* of Social Network Science. A publication set is retrieved in a procedure that mimics the identity model of social facts selected by actors – here: publications. The solution is evaluated through manual coding of a sample. The second section describes the meaning structures and size/lifetime distributions at three *levels* of observation. Power laws signal that the scaling hypothesis needs not be rejected, i.e., Social Network Science is at a phase transition and can be diagnosed through scaling laws. The third section portrays the *emergence* of the domain and its five subdomains. The different perspectives reveal that turning points demarcate nested periods of stability, a structural hole separates the most forcefully densifying subdomains in collaboration networks, and identities permanently recontextualize their paradigms from different directions. To ease reading, each section ends with a summary.

Finally, we discuss how the case study evaluates the identity model, how we intend to test the selection mechanism in an agent-based simulation, how our analysis has extended and enriched the story of Social Network Analysis, what methodological issues we have faced, and what practical applications our work may turn out to have, especially regarding the predictability of change. A glossary lists definitions and operationalizations of central concepts. It may be consulted to further learn about the content of this work or while reading to keep track of arguments that often involve combinations of multiple concepts.

# 1. Identity and Control

## 1.1. Paradigms

### 1.1.1. Fractal Distinctions and the New Sociology of Science

The notion of friction in defining identity – I am because I am distinct – is a central idea in sociology. In *Chaos of Disciplines*, sociologist Andrew Abbott (2001a) discusses the identity of sociology itself. His perspective is that sociology and social science are structured in a fractal distinction in space and time. The idea is that dichotomies like positive/interpretive, realist/constructivist, or causal/narrative not only texture sociology at some 'macro' level, but these distinctions repeat within themselves. As a consequence, even though there are poles, there is also everything in between – sociology is intersticial, or in other words: it has strongly permeable boundaries. To lay out what this means for us, we turn towards paradigms and a "Paradigm for the sociology of science" (Gieryn, 2010).

Two persons are credited for having helped the paradigm term get to fame: Historian Thomas S. Kuhn has introduced it in *The Structure of Scientific Revolutions* and sociologist Robert K. Merton has used it occasionally over six decades in his works about science and social structure (Wray, 2011, pp. 50). While Kuhn used the paradigm concept in an observational way, as a philosopher of science, Merton used it in a normative way how science should be. Merton's essay "The normative structure of science" (1973 [1942]) is a foundational text in the sociology of science. It posits that a quadruple of norms ensures the orderly functioning of science: (universalism) Scientists should be objective, (communism) science is collective action, i.e., scientists must communicate their findings to their community, (disinterestedness) scientists should do research not for personal gain, and (organized scepticism) even far-fetched knowledge claims should be critically examined.

This part of Merton's work is best interpreted in light of the fact that he was as assistant professor of Talcott Parsons at Harvard University. Both drew inspiration from the sociology of Émile Durkheim. According to Durkheim (1893), societies like those of hunter-gatherers are characterized by "mechanical solidarity," i.e., order is maintained through tradition and collectively shared norms. In modern societies like industrial society, however, persons are strongly dependent on each other because of a division of labor which is mirrored in a division of norms. Such societies cohere through "organic solidarity." Parsons aimed at a grand explanation of individual behavior through norms and expectations. In Parsonian *structural functionalism*, society is hierarchically divided into systems that control themselves to fulfill a social function, e.g., to produce scientific knowledge. Systems are real and autonomous, i.e., theoretically relevant action does not occur outside the system. Interaction is symbolic between roles, where roles are insti-tutionalized behaviors meeting functional expectations. This is Parsons' interpretation of organic solidarity. Despite their reality, systems are only found in the representations of persons which are intrinsically determined by roles (Parsons, 1968 [1937]). Merton (1934) agreed in principle.

Besides the functionalist Merton, there is also a structuralist Merton who was seeking middle-range theory, i.e., focusing on partial explanation of phenomena observed in different social domains through identification of the central causal mechanisms (Hedström & Udéhn, 2009). Early on, he had pointed out that there are also unintended consequences and dysfunctions in social life (Merton, 1968 [1949], ch. 3). For example, according to the scientific norm of disinterestedness, the reward for scientific achievement is recognition, not influence, authority, or personal gain. According to structural functionalist modeling, recognition is attributed in the "reward system." When studying interviews of Nobel laureates, Merton found that scientists are awarded the more recognition, the more they have already been awarded. He called this the *Matthew Effect* in science, following the Gospel of Matthew (Merton, 1973 [1968]). But the effect is dysfunctional because, according to the norm of universalism, a Nobel laureate and a graduate student should receive the same reward for a knowledge claim. Such dysfunctions remain cardinal problems of functionalist modeling. The Matthew Effect is a sociological precursor of Barabási and Albert's (1999) preferential-attachment mechanism, mentioned in the introduction, that generates scale-free networks.

In any case, laying out norms for science, Merton set the stage for the *institutionalist program* of how to study knowledge production sociologically: to not study it in its sociocultural context. This is a realist position because knowledge production is considered to be an autonomously operating, bounded system that can be isolated from external effects. This program was the main paradigm of the sociology of science until the end of the 60s, and its demise was partly initiated by a work on paradigms themselves.

### The Constructivist Turn

Kuhn (2012 [1962]) has proposed that modern natural science is divided into alternating periods of normal science and scientific revolutions. In *normal science*, there is a relatively firm consensus among practicing researchers not only what constitutes a problem but also how this or these problems out to be solved. Knowledge production proceeds cumulatively and as planned. After some time, typically after generations, a crisis emerges as the governing consensus is not a good guide to scientific puzzle solving anymore and anomalies pile up. The crisis is resolved in a scientific revolution during which a new consensus is established that subsequently guides in a new period of normal science. In Hoyningen-Huene's (1993, pp. 135) reading, a Kuhnian *paradigm* is a set of rules that is exemplary for how puzzles are solved in normal science. This consensus not only has a locally normative aspect in the sense of a blueprint for solving concrete puzzles. It also consists of a global "mandate to generalize the locally normative aspect, projecting it on further research in the relevant field" (p. 135). Paradigms are not necessarily stated explicitly:

> [C]onsensus isn't primarily, let alone exclusively, established ... by explicit necessary and sufficient criteria for concept application; ... by laws or theo-

> ries conceived in abstraction from their consummated application to concrete individual cases ...; [or] by any kind of explicit, unequivocal methodological precepts such as recipies for problem choice, the evaluation of problem solutions, crisis identification, theory improvement, theory evaluation, theory comparison, theory rejection, and so forth. Kuhn calls all such explicit guides to action “rules,” in slight departure from common usage. Concrete problem solutions consensually accepted by the community, by contract, play their globally normative role in an implicit way, serving as exemplary models for scientific practice, as a source of analogies. (p. 136)

Kuhn’s phase model, in short, consists of periods of normal science in which paradigms serve as exemplars. Such phases of stability eventually enter crisis and are terminated by scientific revolutions during which paradigms shift. Both normal science and scientific revolutions are collective phenomena. This gets most evident in the concept of *incommensurability.* Paradigms are constellations of positions taken by scientific communities, proponents of competing paradigms practice science in domains that are distinct to some degree, and no one group wants to grant the other the non-empirical assumptions they need to defend their turf. They may not even agree on what actually constitutes a puzzle to be solved. This is also related to language. New paradigms are born from old ones and hence use some of the old words. But interpretations differ. In the new paradigm, old concepts, terms, and experiments acquire new meaning which causes misunderstandings among the competing communities. Conversations among proponents of different paradigms are only possible to some degree (Kuhn, 2012 [1962], pp. 148). The most basic aspect of incommensurability, the one related to distinct domains of practice, is one of Kuhn’s most sociological ones:

> [T]he proponents of competing paradigms practice their trades in different worlds. ... Practicing in different worlds, ... scientists see different things when they look from the same point in the same direction. Again, that is not to say that they can see anything they please. Both are looking at the world, and what they look at has not changed. But in some areas they see different things, and they see them in different relations one to the other. That is why a law that cannot even be demonstrated to one group of scientists may occasionally seem intuitively obvious to another. Equally, it is why, before they can hope to communicate fully, one group or the other must experience the conversion that we have been calling a paradigm shift. (p. 150)

A paradigm shift – used synonymously with scientific revolution – is, therefore, a transition from one social network with a certain meaning to another one with a different meaning. Because of the dependence of knowledge production on socio-cultural systems, Kuhn takes a constructivist position – even if he never thought so himself. The constructivism also gets very clear in comparison to Fleck (1979 [1935]), whom Kuhn (2012 [1962], pp. vi) acknowledged to have anticipated many of his ideas. Hacking writes in the introductory essay of the 50th anniversary edition of *Structure* that, in 1974, Kuhn had reemphasized that ”paradigm entered the book hand in hand with scientific community" and that "Kuhn's scientific community matches Fleck's notion of a 'thought-collective,' characterized by a 'thought-style,' which many readers now see as analogous to a paradigm" (Kuhn, 2012 [1962], p. xxi). Fleck is most clear in stating the constructivist inseparability of social action and cultural meaning, a notion most central to our own work.

In the early 70s, the Mertonian program came under fire from a group of constructivists, central figures of which were Nigel Gilbert, Steve Woolgar, Bruno Latour, Harry Collins, Michael Mulkay, and David Bloor, among others. If one were to sum their position up under one headline, it would be that context matters. They criticized the institutional program for epistemological realism, i.e., for reducing the production of scientific knowledge to inputs (norms) and outputs (knowledge) and hiding the scientific communities' knowledge production processes in a black box. They called for an empirical program that takes the duality of action and meaning seriously and deprives the science system as theorized of its decontextualization. Their style came to be known as the Sociology of Scientific Knowledge and, in hindsight, their victory constituded the *Constructivist Turn*. Certainly, Kuhn's theory was an important achievement in paving the way for this new sociology of science. Bloor (1991 [1976]) from the Edinburgh school wrote one of the manifestos, the "strong programme:" The Sociology of Scientific Knowledge ought to study the socio-cultural conditions of knowledge production, study both successful and unsuccessful knowledge claims, treat those alike, and be potentially self-referential, i.e., applicable to itself.

"By 1981, however, there were sharp philosophical discontinuities within the group", writes Abbott (2001a, pp. 78) in presenting his idea of fractal distinctions. By that year, the group had split into four camps. The first was closest in position to the original strong programme of the Edinburgh school. They focused on the contextual factors that influence how scientists produce knowledge, "on the role of interests in scientific change" (B. Barnes & MacKenzie, 1979), and came to be known as the "interests" school. Schools two and three dealt with the interpretive flexibility in observing nature. The Bath school argued for an "empirical program of relativism" (EPOR). They tended to focus on unsuccessful knowledge claims, recognized "the potential local interpretive flexibility of science which prevents experimentation, by itself, from being decisive" and set out to study "mechanisms which limit interpretive flexibility and thus allow controversies to come to an end" (Collins, 1981, p. 4). The third camp homed in on the inner workings of the institutionalist black box by employing techniques of discourse analysis (Gilbert & Mulkay, 1984). The fourth camp is called the "constructivist" school because they most strongly cast the realism of knowledge into doubt. Not only are observations of nature embedded in socio-cultural contexts, but knowledge is "manufactured." Not only is there flexibility in interpreting observations, but even the observations are constructed (Knorr-Cetina, 1981). This last group is unified by the ethnographic method applied in research laboratories (Latour & Woolgar, 1979) which is why they are also called the "laboratory studies" school. The difference between these schools is apparent and, in fact, open attacks began on the Edinburgh school.

In 1981, Woolgar of the constructivist school criticized the originators of the new sociology of science for being too realist in their focus on interests.

> This was a classic fractination attack; the Edinburgh school was attacked for not going far enough. Moreover, ... attack came also from the realist side ... Speaking for the earlier mainstream version of the sociology of science – Merton's "institutions" position – Gieryn claimed that the strong programme's philosophical justifications had little connection with its empirical work. ... Replies to Gieryn locate precisely the successive splits of fractionation. For EPOR, Collins replied that choice of problem was a matter of taste and that EPOR had decided to view science as a prototype of knowledge, precisely because if it could be shown to be socially grounded, then obviously any form of knowledge was so grounded. Discourse analysts Mulkay and Gilbert hastily remarked that while everything Gieryn said was probably true about EPOR and the constructivists, it did not apply to discourse analysis, which concerned "how ... scientists construct their versions of what is going on in science." For the constructivists, Knorr-Cetina replied by distinguishing between "epistemic relativism" (which assumes that interpretations of nature are grounded in historical place and time) and "judgmental relativism" (which believes that there is no reality). ... Thus, in responding to a (relative) conservative, each sociology of science went to some lengths to prove that it took less for granted than the others. (Abbott, 2001a, pp. 79)

We need not detail that, in 1985, the constructivists shot back at the scientific mainstream for being realist, because the ground should be set: The Sociology of Scientific Knowledge is intersticial. This is a result of a *fractal distinction* in time.[1] In space, the fractal distinction means that a culture is split into two or more subcultures, but each subculture is again split according to the original split. When these splits play out in time, an identity evolves along a fractally distinct path. In the early 70s, the sociology of natural science split into a constructivist and a realist position. But in the 80s, the constructivist position is again split into a constructivist and a realist subposition. As a result, the constructivist constructivists and the realist realists are the poles, but the realist constructivists and the constructivist realists are culturally quite similar (Abbott, 2001a).

Figure 1.1a is an attempt to visualize the paths along which the sociology of science changed. The main message is that distinctions only exist to some degree (S. Fuchs, 2001, pp. 50). If we zoom into the structuring dichotomy, we see the same dichotomy again. As a result, it becomes difficult to locate the various schools. This is not so much the case with the laboratory-studies school because they attacked both the realist constructivists and later the realists from the outmost constructivist pole. It is also not too difficult for the interests school and the relativists because they are certainly not as radically constructivist as Knorr-Cetina et al. But where should Merton be located? Where might Gieryn be placed who defended Merton in the 80s?

Merton is not obviously located on the realist extreme pole. He was always aware that prior expectations shape perception and that "[c]harisma becomes institutionalized,

[1] Natural scientists will object to our use of the term fractal because, for the distinction to be fractal, it would need to be split infinitely. What we really mean is self-similarity and these two concepts will be covered shortly.

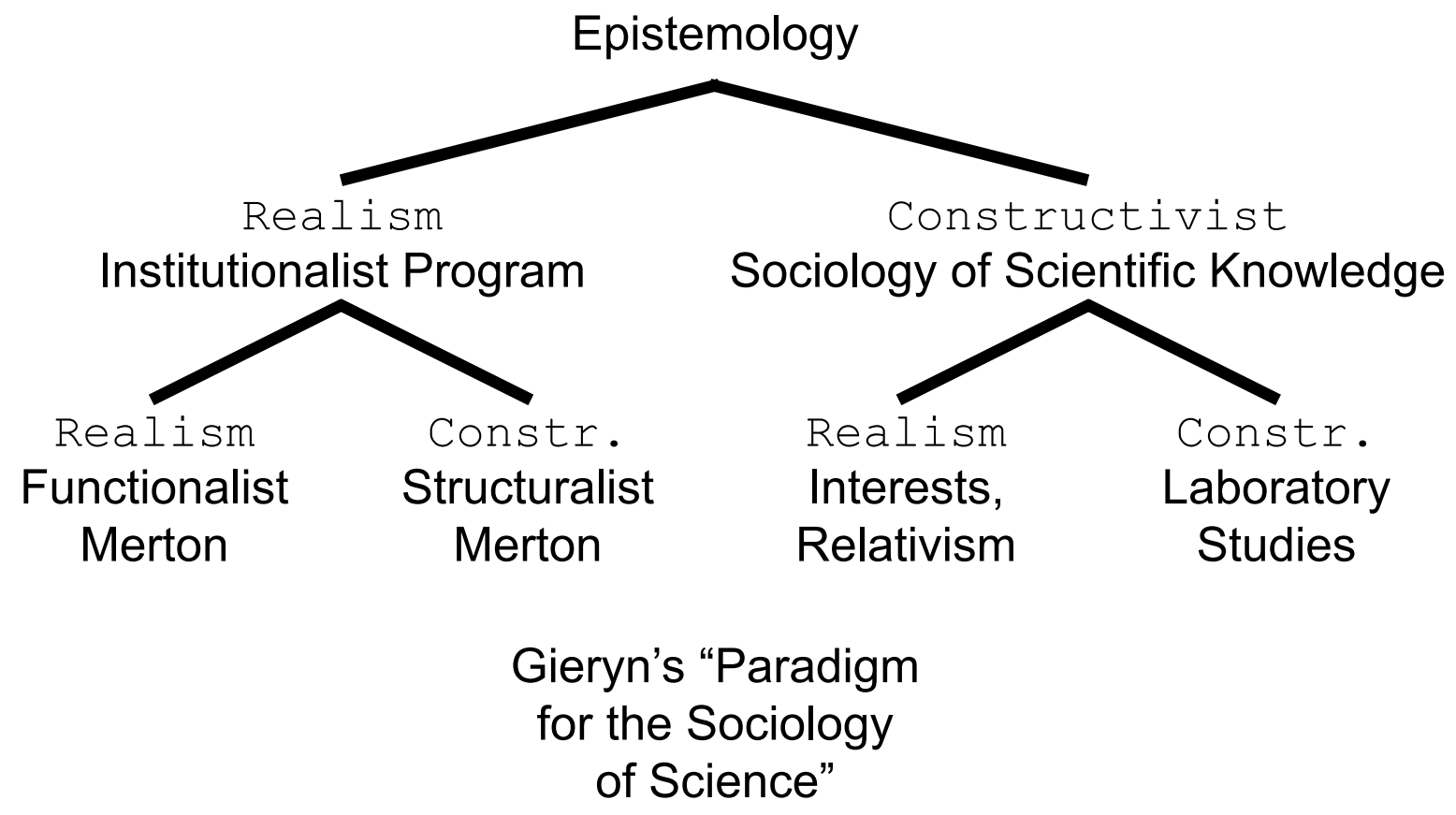

(a) Epistemology

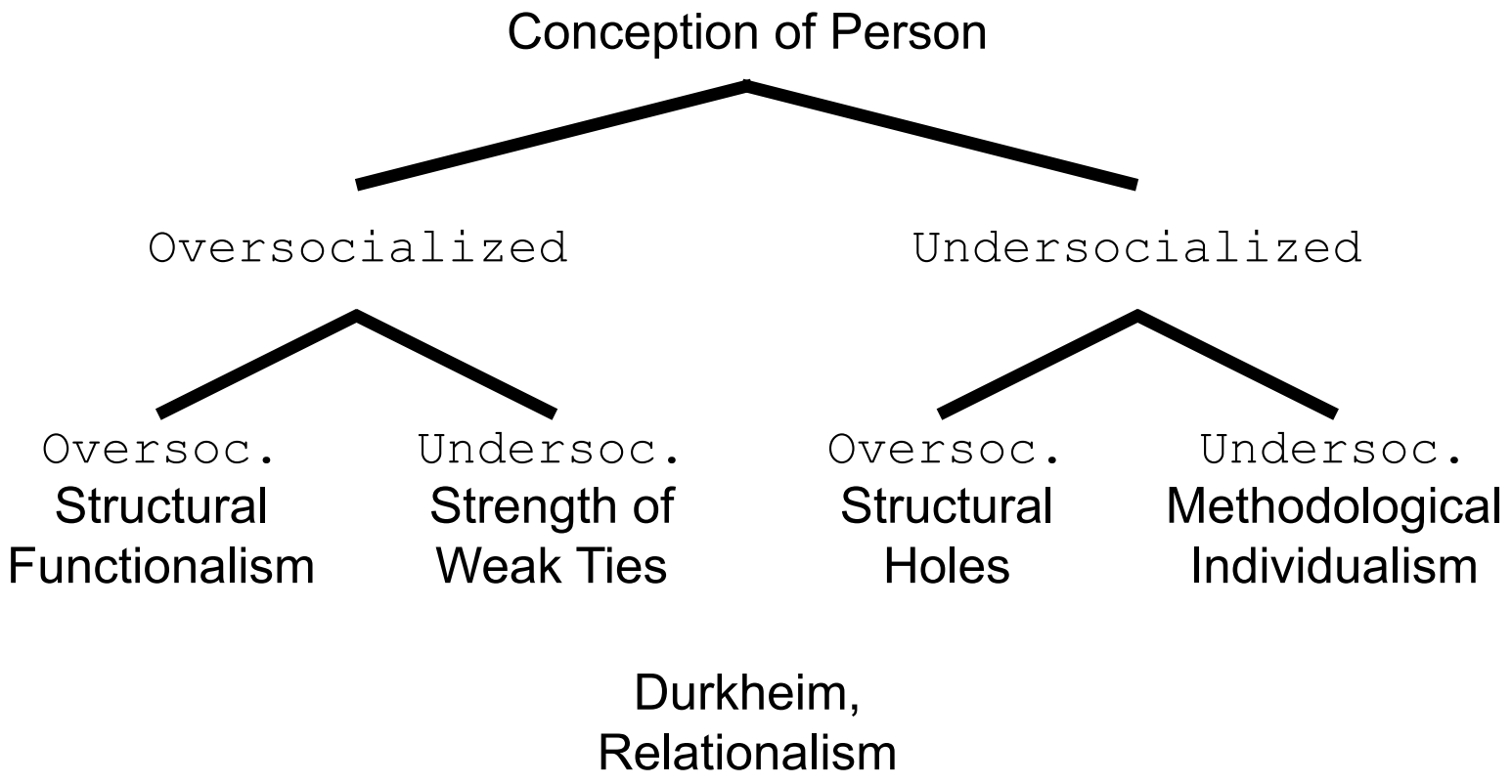

(b) Conception of Person

**Figure 1.1.: Fractal distinctions**

A fractal distinction is a cultural distinction that repeats within itself. (a) Until the 70s, the sociology of science progressed along the intellectual lines of Merton's institutionalist program which is based on a realist epistemology. In the 80s, a constructivist school matured but also split into camps (the interests and relativist schools) that took up the concerns of the defeated realist paradigm, thereby creating a self-similar meaning structure. Permeable boundaries result from fractal distinctions: constructivist realists and realist constructivists may be hard to distinguish or may collaborate. Gieryn's paradigm, e.g., cannot be located at any of the branches. (b) In the 70s, over- and undersocialized conceptions of persons coexisted in structural functionalism and methodological individualism, respectively, and were challenged from proponents of relationalism. Network arguments as similar as the "strength of weak ties" and the benefits of "structural holes" can be placed on different branches, constituting a fractal distinction. Durkheim's circular causality and modern relationalism are too balanced to be located at any of the branches.

in the form of schools of thought and research establishments" (Merton, 1973 [1968], p. 453). To order the perceived chaos, he formulated a "paradigm" ("scheme of analysis") that should guide sociologists in studying knowledge production (Merton, 1973 [1945], pp. 11). He did so, as with the rules for orderly scientific conduct, in a manner that is rather realist realist than constructivist realist.

42 years after Merton, Gieryn (2010) updated Merton's "paradigm:" "It turns out that my paradigm for the sociology of science will look different in substance and intent than Merton's, and these differences will politely subvert – in a self-negating, ironic way – the very idea of paradigms as Merton saw them" (p. 114). His paradigm consists of contradictions of principle: Science is social *and* cognitive, cooperative *and* competitive, institutionalized *and* emergent, and so on. In other words, the research program of today's Gieryn is best placed right in the middle of figure 1.1a. The sociology of science ought to be as it is: interstitial.

#### Summary

Kuhn's *Structure of Scientific Revolutions* consists of periods of normal science that are disrupted by scientific revolutions during which paradigms shift. In the 40s, Merton had formulated the realist institutionalist program for the sociology of science according to which science is driven by norms. In the 70s, this paradigm was contested by a school of inquiry relying on a constructivist paradigm. In the 80s, this New Sociology of Science itself split into multiple camps that more or less leaned back to the realist paradigm. Abbott calls such splits fractal distinctions because they repeat in themselves. In today's sociology of science, the view prevails that scientists are both influenced by norms and knowledge is still constructed socially.

### 1.1.2. Durkheim's Theory of Feedback

What kind of persons are Merton's scientists? If science is not dysfunctional, they are only driven by the inner force to follow the systemic norms of knowledge production. Eventually, they will even act irrationally if rational behavior is against the institutionalized ideals they have accepted as their own. Granovetter calls this an "oversocialized" conception of persons. They are "overwhelmingly sensitive to the opinions of others and hence obedient to the dictates of consensually developed systems of norms and values, internalized through socialization, so that obedience is not perceived as a burden" (Granovetter, 1985, p. 483). Structure has a constraining function. The antithesis, the "undersocialized" conception of persons, can be found in the *methodological individualism* of the utilitarian tradition and rational choice theory. There, the primary ontological phenomenon is the individual and all properties of larger-scale structure can be explained in terms of the individuals' properties in combination, like free will and purposive action. Individuals have a capacity to act independently. Behavior is rationally determined from within and may violate collectively or individually held values. Causation flows from individuals up to social structure. Structure has an enabling function (Granovetter, 1985, p. 483–7; Emirbayer, 1997, p. 283–5).

As different as these models of man are, they are similar in that persons are atomized essences that exist prior to experience, i.e., they give form to action. Both structural functionalism and methodological individualism rely on an essentialist or *reductionist ontology*. This is obvious for individualism where social structure may even be completely absent. But it is also true for functionalism because individual behavior is fully determined (Granovetter, 1985, p. 485; Emirbayer, 1997, pp. 282–91; Erikson, 2013, pp. 224). The puzzle that at least structural functionalism is trying to solve is closing the micro/macro gap between individuals and social structure, a problem that persists since the very beginning of sociology. To this day, Durkheim is not only credited for having kickstarted the science of social life, but for having opened the micro/macro gap as well (Erikson, 2013, p. 231). His œvre is often considered to pose the dilemma that either social structure is reducible to individuals or one proposes a dualistic ontology (Sawyer, 2005, pp. 100). In Merton's (1934) reading, Durkheim's theory is hardly reconcilable with empirical work. We think the dilemma is an illusion.

In distancing sociology from psychology, Durkheim (1982 [1895]) argued for a realism of social structure. In *The Rules of Sociological Method*, he claimed that, just like persons have mental states and individual behavior, social groups have collective consciousness. But collective behavior constitutes a reality sui generis, unique in its characteristics and distinct from the individual behavior which manifests it. He defined a *social fact* as a collective behavior "which is general over the whole of a given society whilst having an existence of its own, independent of its individual manifestations" (p. 59). At the same time, social facts "are only realised by men: they are the product of human activity" (p. 62). This is emergence, upward causation in its purest form, as defined in complexity theory: *Emergence* is a bottom-up process whereby the macrobehavior of a system results from the interactions of its parts and is "not only more than but very different from the sum of its parts." In other words, the system cannot be reduced to its parts without losing explanatory power because in the process of emergence a new quality comes into existence at a larger scale. This is why emergentism is counter to reductionism (Anderson, 1972).

But emergence is only half of Durkheim's story because a social fact is also "any way of acting, whether fixed or not, capable of exerting over the individual an external constraint" (1982 [1895], p. 59). In seeking an objective criterion to characterize the constraining potential of facts, he turned towards their degree of fixation or consolidation:

> [There] exists a whole range of gradations which, without any break in continuity, join the most clearly delineated structural facts to those free currents of social life which are not yet caught in any definite mould. This therefore signifies that the differences between them concern only the degree to which they have become consolidated. Both are forms of life at varying stages of crystallisation. (p. 58)

"Consolidated social life" like moral rules, popular sayings, or facts of social structure (p. 82) are what we call institutions. In the preface to the second edition, Durkheim acknowledged that sociology could be defined as the science of institutions, their genesis and their functioning. Institutions are causal from the top down. But since, in the first place, they have emerged from the interactions of the parts which they then influence, social facts are the central pieces of a feedback loop. *Feedback* is a process of circular causality.

"Durkheim's dilemma" is not what it seems to be. Because of emergence, the father of sociology is likely to be misunderstood if read with a reductionist mindset. Durkheim explicitly rejected

> a doctrinal meaning relating to the essence of social things – if, for instance, it is meant that they are reducible to the other cosmic forces. Sociology has no need to take sides between the grand hypotheses which divide the metaphysicians. Nor has it to affirm free will rather than determinism. All that it asks to be granted it, is that the principle of causality should be applicable to social phenomena. (p. 159)

Social life is a duality of emergence and downward causation. Merton and others had confused this with a duality of ontologies for individuals and collectives. In fact, this is what Parsons tried to avoid by rejecting emergence and localizing norms in persons. Instead, he constructed a micro/macro gap that Durkheim describes does not exist. The real dilemma consists in an incommensurability of the structural functionalists' and Durkheim's paradigms (cf. Sawyer, 2005, ch. 6).

#### Summary

We have traced the concept of feedback back to Durkheim who proposed a circularly causal mechanism of social facts as emergent social phenomena that constrain individuals in their behavior. His ontology is non-reductionist and – because of feedback – beyond emergentist. Facts exist in varying degrees of consolidation and attain the status of institutions if they crystallize or persist. What we need now is an exact mechanism how social formations and institutions emerge and constrain. It derives from the part of the definition of emergence not yet discussed: interactions of a system's parts – in other words: networks.

### 1.1.3. Social Network Analysis and the Harvard Breakthrough

Social network theory can be traced back to sociologist Georg Simmel who, in 1908, introduced the *triad*, a network consisting of three nodes, as the main social building block. But sustainable empirical network research only started in the 30s under the auspices of psychiatrist Jacob L. Moreno. His *Who Shall Survive?* (1934) remains the first book on social networks. "Sociometry," as the method was called, was focused on substructures and motives that give groups their form. Moreno and co-workers repeatedly observed that, in relationship questionnaires, few persons are typically chosen above chance, i.e., they uncovered early and robust evidence that the mechanism that generates size distributions in social networks is different from a purely random (Gaussian or Poisson) process (Moreno & Jennings, 1938).

Scott (2012) describes three cross-cutting lineages of Social Network Analysis: sociometry and two ethnographical paths. For each lineage, we review those works that seem most important to us and which we will refer to in the remainder of our work. Sociometry was part of a social psychological path based on Gestalt psychology and Kurt Lewin's field theory. Psychologist Fritz Heider (1946) proposed that persons always aim to avoid stress in social relations, i.e., they try to create balanced triads where 'my enemy's enemy is my friend' (and variants of that rule). Cartwright and Harary (1956) generalized this framework to larger graphs and proved that a group splits into up to two subgroups when all triads are balanced. Graph theoretical approaches then diverged. One group stayed on the Simmelian path and studied clustering through the triad census, a classification of triads, developed by sociologists James A. Davis, Paul W. Holland, and Samuel Leinhardt (J. A. Davis, 1967, 1970; P. W. Holland & Leinhardt, 1971). Another group aimed at larger populations. Rapoport and Horvath (1961) conducted what is maybe the first attempt to study a complete social network and avoid sampling effects. Studying a three-digit-sized friendship network, they confirmed that its degree distribution is more skewed than a purely random process can explain. The *degree* of a network node is the number of its edges. Rapoport and Horvath were among the first to explain highly skewed distributions by a version of the Matthew Effect, precisely a Pólya or Yule process. Like Merton, they are the narrative ancestors of modern preferential attachment.

A few years later, Travers and Milgram (1969) studied the small-world problem and showed empirically, though not convincingly, that a randomly chosen pair of persons has an average social network distance of only six.[2] As stated in the introduction, the small-world problem has been formalized by Watts and Strogatz in 1998. For a network to have the small-world property, it must not only have an average path length as short as a random graph of similar size but also a much higher probability of triads (clustering) than the random graph. Interestingly, Rapoport and Horvath (1961) had also presented early hints at the modern formulation of the small-world problem because the likelihood of being indirectly connected to the rest of the network through a friend is smaller the better the friend is. In other words, clustering interferes with distance. Independent from these graph theoretical approaches, Bavelas's (1950) group explored the efficiency of different communication network structures. They found that star- or Y-shaped networks with short distances are faster and less error-prone than chain- or circle-shaped networks with long distances. In the face of complexity, short distances proved to make the network more efficient. Unfortunately, they did not study any clustering effects. This work was foundational for later formalizations of centrality in Social Network Analysis (Freeman, 1977, 1978–1979).

The other two lineages originated in the ethnography of Alfred Radcliffe-Brown, a founder of structural functionalism. At Harvard University, where Parsons was running the Department of Social Relations, socio-anthropologists lead by Elton Mayo and W. Lloyd Warner used field methods to investigate industrial communities. Among other things, they revealed that informal networks differ from formal structures (1939). This finding is relevant in light of fractal distinctions because, whatever formal structures of differentiation there may be, real networks are largely overlapping. The informal networks were later reanalyzed by George C. Homans (1950), who generalized the modularity property of networks in non-mathematical terms. Along the other branch, British social anthropologists tried to offer an alternative to Parsons' abstract theory of normative action and observed integration processes in small-group and local settings. Among the most notable works, John A. Barnes (1954) studied the emergence of a class structure during industrialization not as a macro property of society but as a networked process in a local community. In 1957, Siegfried F. Nadel presented a non-normative structural role theory where relations are not symbolic but concrete social ties. In the same year, Elizabeth Bott developed the study of ego-centered networks. She found that highly clustered networks yield common patterns of behavior (Wolfe, 1978; Freeman, 2004, ch. 2–7; Scott, 2012, ch. 2).

[2]Their result stands on shaky statistical grounds but has been confirmed by now. The small-world problem was raised about a decade before Travers and Milgram by Pool and Kochen (1978 [about 1958]).

According to Freeman (2004), the chronicler of Social Network Analysis, these three branches made progress but never succeeded to establish the network paradigm in the wider social science community. Even though recent breakthroughs on small-world and scale-free networks have their roots in early Social Network Analysis, time was not quite ripe. This was because they never combined the four aspects that make up the research domain today: the use of empirical data, visualizations, and mathematical/computational modeling. Development did not take off because a convincing model was lacking for the Durkheimian duality of emergence and downward causation – until a "breakthrough" occurred and lead to a "renaissance" at Harvard (Freeman, 2004, ch. 8; Scott, 2012, p. 34–8).

### Renaissance at Harvard

In the 60s, also at Harvard University, the group of White was developing a social network research program that devoted equal space to the enabling and constraining functions of social networks (H. C. White, 2008 [1965]). This school of thought had grown out of opposition to the "attributes and attitudes perspective" most closely associated with Parsons (Schwartz, 2008). But like Parsons, White was influenced by cybernetics, an interdisciplinary research program about *Control and Communication in the Animal and the Machine*, to use the definition by Norbert Wiener. According to H. C. White (1973),

> [e]ach of us lives under erratic bombardment of all kinds of messages in a large and complex web which yet is different from, though tied to, the web of any neighbor. ... Not only is a person overloaded with respect to the total of all message flows he might attend to, but he must pare down his attention span through a priority scheme so that he responds to fewer streams of messages than he could handle in a deterministic utopia. (pp. 45–6)

Based on such an understanding of social life as an implicit, self-organized reduction of socio-cultural uncertainty, the group's work soon began to bear fruit. In 1973, White's student Granovetter published results from his dissertation that job hints were more likely to diffuse through social relations with low levels of trust, intimacy, and reciprocity ("weak ties") than through institutionalized relations ("strong ties"). His paper on "The strength of weak ties" is today the most-cited one in the social sciences. This is an undersocialized oversocialized conception of person because structure is considered to primarily constrain action. This interpretation is supported by Granovetter's second example where a lack of enabling social structure prevents a social movement from emerging. Almost 20 years later, Burt (1992) interpreted Granovetter's idea in a new light. The "structural holes" argument says that strong ties indeed constrain action but actors can strategically and actively reap benefits from spinning weak ties. Structurally, there is no difference, but conceptionally, this is a shift to oversocialized undersocialization because social structure is now primarily seen to emerge from rational decisions. This conceptional interstitiality of social theory is illustrated in figure 1.1b.

The extent to which over- and undersocialized conceptions of person are balanced in the new sociological perspective of White et al. is most evident in their search for aggregation principles. When persons are members of formal groups (where membership rosters are available), they embed in collectives based on shared interests or values. On the other hand, the particular patterning of a person's affiliations at least partly determines its identity (Breiger, 1974). This perspective of individuals defined by their ties in a social network is the motive of *structural equivalence*. Two persons are structurally equivalent when they are tied to the same others. Equivalence weakens as the two tie sets are decreasingly identical (Lorrain & White, 1971). From these patterns also arises a grouping of individuals with similar tie profiles into blocks. Another name for block is *position*. But while Parsons' roles were imposed top-down, White et al.'s positions emerge bottom-up. The corresponding framework to aggregate network nodes according to degrees of structural equivalence and depict the result in an image matrix is called *blockmodeling*. This procedure partitions network nodes into two types of blocks. In "1-blocks" nodes are structurally equivalent through the presence of ties, in "0-blocks" through the absence of ties. The unique modeling opportunity is that "0-blocks" can be on the matrix diagonal, i.e., persons on similar positions need not be connected to each other. It follows that a social network consists of multiple positions – such as core and periphery – which can be filled by concrete persons or, in general, social facts (H. C. White, Boorman, & Breiger, 1976).

As in the case of the Sociology of Scientific Knowledge (Abbott, 2001a, p. 80), it took White's group about one and a half decades until it attacked the enemy. In 1981, White engaged the methodological individualists in microeconomics. In the basic neoclassical market model, price is signaled by aggregated supply and demand in equilibrium. Product quality does not matter in these linear equations and producer interaction leads to market failure. White criticizes that this just assumes markets and does not explain them. He proposes a sociological model where price is not set by producers observing buyers but by producers observing what revenue other producers receive for what volume shipped, producing a market profile in the process. Quality is not observable but enters the equation in form of a non-linear producer/buyer trade-off over variation in producer's quality that is coupled to another trade-off over variation in product volume. This model does not assume that market failure results from the interaction of producers but predicts it when producers do not manage to sustain the market through interaction. Likewise, rational, profit-maximizing behavior is not a model assumption but a possible way of constructing a market (H. C. White, 2008, p. 138-141).

Shortly afterwards, Granovetter (1985) criticized the school of new institutionalism that takes neoclassical markets for granted but adds that frequent, risky, and costly economic transactions are disciplined within the governance structure of firms. Neither was there evidence in economic sociology that something like structureless markets existed outside textbooks, nor would the oversocialized view that employees internalize the interests of the firm and avoid conflict stand scrutiny. There are no reasons to not consider markets and firms as networks themselves, Granovetter concluded. All economic action is embedded in social networks.

Today, White is considered "a man who has started sociological revolutions, introduced new techniques, and trained one of the finest groups of students in the discipline" (Abbott, 1994, p. 895). According to Freeman (2004), he deserves credit for having established the field: "Once this generation started to produce, they published so much important theory and research focused on social networks that social scientists everywhere, regardless of their field, could no longer ignore the idea. By the end of the 1970s, then, social network analysis came to be universally recognized among social scientists" (p. 127). The breakthrough that facilitated the renaissance of Social Network Analysis was the combined use of empirical data, visualizations, and explicit modeling. The CONCOR algorithm of structural equivalence (Breiger, Boorman, & Arabie, 1975) provided the first network clustering method. In today's language, it was the first community detection algorithm, though with the distinguished possibility of identifying incohesive node sets. Through splitting blocks into subblocks one could probe the hierarchical "architecture of complexity" (Simon, 1962). For the researchers to come, these were the "structural ideas to make sense out of field work and ... data analyses", and CONCOR was the tool to solve a "methodological problem for which there was no method" (Levine and Carley, 1997, quoted in Freeman, 2004, p. 126). There is evidence for a sustained renaissance. A study of literature use in Social Network Analysis showed that, at least until 1990, the domain was accumulating knowledge systematically with a focus on positional analysis (Hummon & Carley, 1993). 20 years after the Harvard breakthrough, Social Network Analysis was institutionalized with its own software package (UciNet), society (INSNA), journal (*Social Networks*), and conference (Sunbelt) (Freeman, 2004, ch. 9).

The theoretical model underlying the algorithm has the deep meaning that the relations between persons, groups, and other social formations are the ontological phenomena of a mechanism that generates not only social structures but also persons. Groups emerge out of persons in interaction, and relations in groups partly determine a person's individuality or opportunity set. The way the mechanism works gets most clear in White's model of markets which are social facts. The market profile emerges from producer transactions and is repeatedly updated. As such, it gives all market participants orientation how to maximize profit:

> Market structure is used by producers for decision guidance, functioning as a much needed mechanism in an otherwise highly uncertain setting. The mechanism, however, is not a clever formula preferred by an academic, but one that is built from the behavior of market participants themselves, and

> reproduced through this behavior. The mechanism is not only built by participants, it is also simple enough for them to use. Markets as mechanisms are created, used, and hence reproduced by participating producers (Leifer, 1985, p. 443).

One consequence of this is that a general mechanism for emergence and downward causation will not be found in anything static. The making and remaking of transactions and meanings calls for a dynamic approach that unfolds the Durkheimian feedback dynamics through which emergent institutions can only impact behavior.

Interestingly, the Harvard perspective itself evolved along the branches of a fractal distinction. The early draft of the research program not only devoted equal space to emergence and constraint, it also claimed that every structure also harbors a meaning (H. C. White, 2008 [1965]). The cultural side of structure was sidestepped for a focus on their relational embodiment which was the practical way to engage the attributes-and-attitudes paradigm (Wellman, 1988). But the task was only postponed until White published a full-fledged structural theory of social action called *Identity and Control* in 1992. This elaboration of the duality of social network and cultural domain is a milestone in the *Cultural Turn* that saw a still small fraction of sociology embrace the meaning encoded in social networks (Emirbayer & Goodwin, 1994; Fuhse, 2009; Mützel, 2009). It is also a major orientation towards socio-cultural process, many concepts of which still await quantitative modeling and empirical treatment. Some are virtually unexplored. But most importantly, it is rigorously relational, i.e., the primary ontological phenomenon is process. This line of inquiry is called *Relational Sociology*, after an influential manifesto by Emirbayer (1997), and revolves around the completely revised edition of *Identity and Control* (H. C. White, 2008). This work is a toolbox for Mertonian middle-range theory. It provides the concepts to model the basic mechanisms of socio-cultural life. Recently, *The Emergence of Organizations and Markets* by political scientists John F. Padgett and educational scientist Walter W. Powell (2012a) has received a lot of attention. Granovetter praises it on the back cover as "the essential starting point for those seeking new and exciting theoretical departures."

## Summary

Network theory developed as a fractal distinction as the group of White et al. formed the synthesis of the oversocialization thesis and the undersocialization antithesis. The network approach complements Durkheim's more systems oriented feedback loop through the interaction of the system parts. The embedding of actors into networks leads towards a mechanism of emergence and constraint that is necessarily dynamic. Persons, groups, or organizations are socio-cultural processes that implicitly seek guidance and control over otherwise stochastic context.

## 1.2. Rules

### 1.2.1. Allowing for Variation

The sociology of science progressed by defeating the old institutionalist program during the constructivist turn. But subsequently it took up the "concerns of the defeated" (Abbott, 2001a, p. 22). In sociology, the functionalist and individualist interpretations of structure were challenged from within Social Network Analysis which can itself be distinguished into schools leaning towards constraining or enabling interpretations of structure. The Harvard school of social network theory defeated a part of its own program just to take it up at a later point. These observations exemplify the permeability of boundaries in social science, i.e., it evolves along the paths of a fractal distinction (Abbott, 2001a).

In two aspects, this model of scientific change contradicts what we have discussed so far. First, it is at odds with Kuhn's phase model which consists of periods of normal science interrupted by grand scientific revolutions. Even through there are revolutions of this type, there are also smaller changes that occur much more frequently and which divide normal science into sub-periods. Second, fractal distinctions are contrary to a straight division of labor because scientists that have taken different paths end up doing the same thing. The fractal closes the gap between the poles. But why do social formations change at all?

Relational Sociology's answer is quite simple. The reproduction of old patterns is important for maintaining an identity, but similarly important is a constant rate of renewal to get fresh action (H. C. White, 2008; Stark, 2011; Padgett & Powell, 2012a). This idea can be traced back to Pareto (1935 [1916]) who, in his book *The Mind and Society*, laid out his theory that social elites, to be stable, must combine conservative elements, represented by the violent character of lions, with elements of change, represented by cunning foxes. H. C. White (2008) puts it this way: "Sociology has to account for chaos and normality together" (p. 1). It shall be our goal to explain stability and change through one mechanism. To only explain stability, we could stick to long-held assumptions about atomistic, attribute-carrying actors. But to explain change, we must do away with atomized essences that exist prior to experience. Our *first rule of method* is contributed by sociologist Stephan S. Fuchs whose integration of network and systems theory is a passionate call *Against Essentialism* (2001):

> The most important rule of method is to allow for variation, and to turn essences or natural kinds into dependent variables that covary with other variables. One corollary of this rule is that there are no constants, only variables, or that constants are *held* constant within a network of variables chosen by an observer. A constant remains constant only until further notice, or until it is again allowed to vary and regains its degrees of freedom. (p. 50)

Allowing for variation means that modeling the former essences becomes the main task.

Relationalism is a bold departure from mainstream sociological thinking because it does not start from persons as fundamental building blocks. The things of social life are *identities* which can be persons, groups, organizations, research domains, markets, and even larger socio-cultural formations. Allowing for variation amounts to a *relational ontology*, i.e., identities are what they are through their relations. The relationalism of H. C. White searches out "self-similarity of social organization, according to which much the same dynamic processes apply over and over again across different levels and scopes" (p. 18). Abbott's fractal distinction is a prime example of such theorizing. In this regard, relationalism is similar to systems theory whose objects are also not localized at a specific scale.

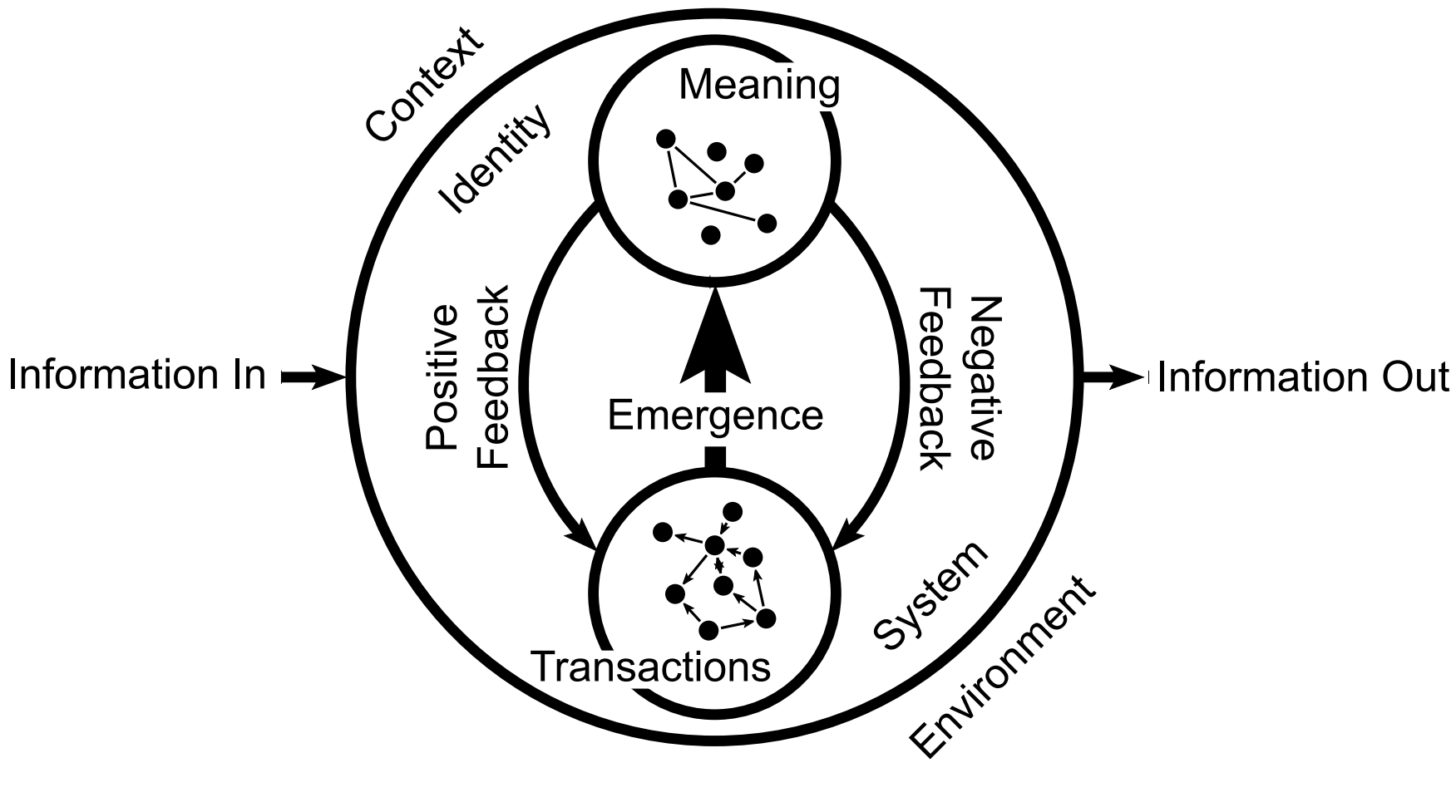

**Figure 1.2.: Identity modeled as a feedback loop**
An identity is an outcome of network process. It consists of parts in transaction, identities in the first sense, depicted in the lower structure. Each part selects, or is influenced by, a number of social facts. From these selections, a meaning structure of social facts in co-selection emerges, shown in the upper circle. These facts are stories, operationalized as words, cited references, and authors. Causality is circular, i.e., facts are both causes and effects of selections. When rich nodes, e.g., regarding the number of selections, get richer (poorer), feedback is positive (negative). When the meaning structure has an emergent pattern, identity exists in the second sense. Identities are always embedded in contexts or higher-level meaning structures (a system environment) from where new information in the form of social facts originates. Adaption to this influx as well as to endogenous events is control. "Social networks" best correspond to transaction structures and "cultural networks" or domains to meaning structures, but both do not exist separate from each other.

Identities seek *control*, i.e., they aim at finding footing in otherwise stochastic networks. Control is not something identities have, it is a way they *are*. It is a way of self-organizing to a state where identities can live and prosper. This strongly distinguishes them from the essences of structural functionalism and methodological individualism because those were externally controlled by norms or internally self-controlled. As nodes in a net-work, identities are both independent and dependent variables, both cause and effect. As causes, identities are social facts. Since causality can only play out over time, identities are necessarily processes. Relational Sociology is an analytical sociology of process (Hedström & Bearman, 2009b). The basic mechanism of our identity model is depicted in figure 1.2. It is based on Durkheimian feedback and synthesizes modeling approaches in Relational Sociology (D. R. White & Johansen, 2005, p. 9; Hedström & Bearman, 2009b, p. 11; Fuhse, 2009, p. 54; Godart & White, 2010, p. 580) and Complexity Science (J. H. Holland, 1995). Throughout this chapter, we will describe it in more detail.

Let us first define the basic terms we are going to use throughout our work. The large circle is an identity to be modeled. It consists of two structures. The one in the lower small circle is the *transaction structure*. Its nodes are identities in the *first sense*, identity as the fundamental building block or smallest unit of analysis (H. C. White, 2008, pp. 10, 17). We denote the sense $k$ of identity as a number leading the term, as $k$-identity. 1-identities are in *transaction*, i.e., they are talking or exchanging things. We use the *trans*action term of Emirbayer (1997) to distinguish transactions or relations among 1-identities from the *self* action of rational actors and the *inter*action of variables in linear modeling. Transaction structures are accounts of *talk* or what goes on in a network of identities.

1-identities in transaction select social facts. By *selection* we mean that they are influenced by facts which can be, e.g., norms, topics, but also persons one chooses to talk to. Selection implies activity. 1-identities that can select social facts have a capacity to act. These are building blocks with bodies and minds called *agents*, like persons, groups, etc. 1-identities influence each other through transactions regarding which selections they make. From the selections made, a *meaning structure* (Mohr, 1998; Fuhse, 2009) emerges, shown in the upper circle. Nodes in meaning structures are social facts that exert an influence on the 1-identities in transaction. These facts can be the norms or topics, but also agents. Facts give each other meaning or contextualize each other through *co-selection*. Two facts are co-selected when they are selected by a 1-identity. Selections are bidirected. An arc from a 1-identity to a fact represents emergence while an arc from a fact to a 1-identity represents constraint. When the feedback loop of emergence and constraint is operative, we speak of identity in the *second sense*, identity that has found footing through collective control, where "collective" subsumes 1-identities (H. C. White, 2008, p. 10, 17).

The model is most intuitive when the identity we are modeling is a social group. In this case, the nodes in the transaction network could be persons. To keep it simple, the facts persons select are other persons. The meaning structure then encodes which person is central or peripheral, which dyads are weak or strong, etc. It is the blueprint of the group's identity, it defines what the group is and what it is not. As this information influences transactions, the feedback loop is closed. The Matthew Effect is the prime example of positive feedback because the rich get richer and the poor get poorer. Persons in a group that preferentially reproduce strong ties contribute to a positive feedback loop that can potentially reinforce structural holes (Cartwright & Harary, 1956). Persons that bridge structural holes the more, the stronger those are, represent a negative feedback corrective in such groups. In negative feedback, the poor get richer and the rich get poorer.

Of course, the group also interacts with other groups or somehow processes external information such as from perturbing events. This potential source of change is depicted by the information in/out arrows. Other identities and events provide the context for the group we are modeling. Context is what an identity embeds into. It is shaped by, and it influences, identities (H. C. White, 2008, pp. xxi). In models of complex adaptive systems, context is usually called the environment (J. H. Holland, 1995). *Context* is the meaning structure of a higher-level identity that an identity is a part of. Consider some social form of organization. The group we have just modeled is now a node in both structures of figure 1.2, and the group's context is the upper meaning structure. *Levels* of embeddedness, layers of emergence, result from such modeling (H. C. White, 2008, pp. xviii). A group constitutes a higher level than a person because it emerges from it. Yet the group does not exist without the person and the person not without the group. Breiger's (1974) duality of persons and groups generalizes in both directions. It is easy to see how the model can be similarly applied to higher-level identities that provide context to organizations. It is a consequence of the systems approach that persons, too, consist of parts. A first understanding is that the parts of a person are the social facts that give meaning to the person's transactions at a higher level.

Meaning is the macrobehavior of a system that emerges from the micromotives of its parts (Schelling, 2006 [1978]). Two prominent macro meanings have been featured in the introduction: the small-world property and the scale-free property of complex networks. By complex we mean that networks are large enough for emergent properties to be identifiable (J. H. Holland, 1995). Even though Travers and Milgram's (1969) small-world experiment stood on shaky statistical ground at the time, there can be no doubt today that identities are connected through short average path lengths while being embedded in clusters of tight transaction (Leskovec & Horvitz, 2008; Ugander, Karrer, Backstrom, & Marlow, 2011). Contrary to speculations (H. C. White, 1970a), strongly connected nodes, so-called *hubs*, are not necessary to facilitate the small-world effect (Dodds, Muhamad, & Watts, 2003).

The scale-free model states that the degree in a network, the number of a node's edges, is distributed according to a power law which gives hubs a much higher probability than an exponential-type distribution. In sociometry, skewed degree distributions had been known at least since the small-group studies of Moreno and Jennings (1938), but actual power laws had first been observed in domains where large-scale data had been available early on. The beginning of this kind of analysis is attributed to Pareto (1896–97) who amassed large amounts of data from many countries to show that the distributions of income and wealth resemble power laws. A number of power laws are confirmed beyond doubt to describe the science system (to be discussed later in the chapter), namely Lotka's Law (1926) about the power-law distribution of the number of works written by scientists, Bradford's Law (1985 [1934]) about the distribution of attention given to scientific journals, and Zipf's Law (2012 [1949]) about word frequencies.

### The Scaling Hypothesis

Historically, physics has been successful spawning ideas in sociology (Leifer, 1992; Scott, 2011). Scholars past and present, from Pareto (1935 [1916], p. 160) and Durkheim (1982 [1895], p. 58) to H. C. White (2008, p. 67) and Padgett and Powell (2012a, p. 11), have referred to crystallization to describe the degree of inertia or the rate of consolidation of social systems. H. C. White (2008) advocates generally seeking guidance from mathematical formulations of the natural sciences (p. 367). He proposes that an identity embeds into contexts that can be "modeled on temperature" and refers to contexts "as if it were in the analogue of a gaseous or liquid state" (p. 144). Identities nest, concatenate, and combine into "turbulent social polymer gels" (p. 144), larger-scale orderings (p. 255) where "levels of social organization, such as cities and organizations and families, ... mix and blur" (p. 280). All these are hints at percolation theory.

The basic phenomenology of percolation theory can be observed in a simple random graph. An Erdős-Rényi graph is a random network with $N$ nodes where each edge has a probability $p$ of existing. When $p$ is increased from 0, the size of an average component (cluster of reachable nodes) will be negligibly small until the critical threshold $p_c = 1/N$ is reached. At this critical point of a one edge per node on average, a giant component emerges that contains a large fraction of the nodes (Erdős & Rényi, 1959). Above the critical point, most nodes are in the largest component, the emergence of which is called *percolation*. As we reach $p = 1$, the graph is fully connected. The interesting thing about the Erdős-Rényi graph is that, even though it is the standard example for exponential degree distributions, its cluster size distribution is scale-invariant at the percolation threshold. At the critical point, the probability to observe a component of size $s$ is $p(s) \propto s^{-2.5}$ (Barabási, 2015, ch. 3, p. 40). This is a power law, and its exponent $\alpha = 2.5$ indicates that the variance of the component sizes is infinite (cf. Appendix A). Scale invariance means that varying the scale of observing $s$ does not result in a different observation.

While random graphs provide a bridge to social network theory, references to gases, liquids, and gels unfold their meaning as we turn towards thermodynamics and statistical mechanics. In the latter, the branch of physics studying percolation, temperature takes the role of $p$. For example, at the critical point of water, at 374°C temperature and 218atm pressure, water coexists in the gaseous and liquid state. At the liquid-to-gas transition of water, liquid drops of all sizes exist just like clusters of all sizes in the random graph. In the most general form, a *phase transition* is a system's shift from one state to another. A phase transition always occurs at a critical point. Phase transitions are critical phenomena and at its critical point, a system is at criticality. In polymer chemistry, a sol is a collection of molecules or monomers. A gel is a macromolecule of bonded monomers, analogous to the giant component in a random graph or the large cluster of liquid water molecules. Gelation is the sol-to-gel phase transition, and it involves the emergence of a giant network that eventually and indirectly connects all monomers (Stauffer, Coniglio, & Adam, 1982; Bunde & Havlin, 1996, p. 51–5).

A "turbulent social polymer gel," in White's usage, is a social network that connects identities from many different domains. Like monomers in a polymer, identities are indi-rectly related, i.e., global communication is available and provides context for identities.

> Temperature is important, analogous to how the larger context induces level of activity in a social example, because temperature calibrates the level of order and disorder. But the existence of phases, here liquid and solid, is a statement that "order and disorder" as a congeries of local arrangements must be referred to a wholistic context. ... Transition from one wholistic context to another is a radical disjunction. Yet it must take some tangible form of intermediate behavior. ... The transition is a system state of its own, although indeed typically it is narrowly bounded in terms of relevant parameters. ... The central point from this analogy is to think of contract as the phase transition itself, with "status" referred to the phase regimes on either one "side" or the other of the transitional regime. (H. C. White, 2008, p. 175)

In other words, what keeps society together is not an order imposed top-down, e.g., by the state. Social formations reside at the critical point of simultaneous order and disorder where stability and change co-exist. This point is characterized by the onset of global communication, i.e., temperature is low enough to enable some context, but still high enough to prevent uniformity.

Our work extends the *scaling hypothesis* of percolation theory to the analysis of complex socio-cultural systems. The hypothesis states that the critical point of a system is described by a power law of component sizes and that the phase transition is described by a set of *scaling laws*, power laws that describe the narrow regime of the phase transition by putting system parameters into relationship (Wilson, 1979; Bunde & Havlin, 1996, p. 55–7). The onset of "global communication" in complex socio-cultural systems at the critical point is related to the scaling law that relates the so-called correlation length to how far away the system is from its critical point. Translated into sociological terms, the correlation length is the mean network distance of identities deprived of context, identities that do not participate in socio-cultural life as represented by the ongoings in the giant component. Above the phase transition, at supercritical temperature, no global context exists and all communication is local. These identities are in the analogue of a disordered gaseous state without a collective identity. At subcritical temperature, communication of all isolates is local but, essentially, communication in the giant cluster is only global. There is only order. At criticality, there is no typical correlation length, i.e., communication is both global and local, there is both order and disorder.

The scaling hypothesis connects the phenomenon of phase transitions, like the spontaneous freezing of water, with the description by power laws. Lotka's, Bradford's, and Zipf's Laws provide strong evidence that the hypothesis needs not be rejected for the science system, and, in sections 1.3.1 through 1.3.3, we will discuss more evidence in that direction, also regarding scaling laws. Our point is that percolation theory can also be applied to socio-cultural systems. The observation of a power law is indicative of an event of emergence. The scaling hypothesis is a possibility to explain simultaneous order and disorder.

How is this cross-sectional notion of order/disorder related to the longitudinal notions of stability/change and normality/chaos? Biologist Stuart A. Kauffman (1993, ch. 5) proposes to think of spatial connections in coupled systems as potential sources of change. He concludes that genetic networks self-organize to the power-law state because only the co-existence of stability and change ensures evolvability. If every 1-identity was influenced by many or all other 1-identities, social facts could very quickly or instantaneously propagate through the whole system which could tip in an instant. If that was realistic, no scientific community could develop a research program along some paradigm, no company could rely on suppliers to deliver components, and nobody could rely on a friend to get help in a moment of need. Every identity in every situation would immediately reorder its priorities and head off on a different path until, shortly after, another situation causes a rescheduling. Society would tumble into total chaos. From this perspective, *chaos*, the extreme sensitivity of a system to initial conditions, is synonymous with *change* and causally related to spatial order. Analogously, *normality*, the extreme insensitivity of a system to initial conditions, is synonymous with *stability* and causally related to spatial disorder. The critical point prevents total chaos through decoupling of components but allows some change through coupling. "[S]ocial formations emerge as insulation in between counteractions of identities seeking control, insulation that is too extreme to permit a temperature that pervades a whole system." (H. C. White, 2008, p. 144; cf. Simon, 1962). In less mathematical ways, the scaling hypothesis is stated as a trade-off among exploration and exploitation by Stark (2011, p. 175–83) and among self-reproduction and invention by Padgett and Powell (2012b, p. 10–1).

The scaling hypothesis allows physicists to organize very different systems into a few classes, each of which is described by the same set of parameters. This observation is called *universality*. Besides enabling the formulation of laws, universality is also of practical interest because, when studying a given problem, one can simply study one system of a universality class and still obtain valid results for all systems in that class (Stanley, 1999). In biology, the analysis of scaling relationships is called allometry. Using data on the physiological properties of, e.g., mammals it has been shown that an elephant is just a blown-up gorilla which is just a blown-up mouse (West, Brown, & Enquist, 1999). "This concept means that dynamically and organizationally, all mammals are, on the average, scaled manifestations of a single idealized mammal, whose properties are determined as a function of its size" (Bettencourt, Lobo, Helbing, Kühnert, & West, 2007, p. 7302).

Percolation theory is a very general mathematical formulation that has been used most extensively in the natural sciences, but applications in the social sciences can be found. Blockmodeling (H. C. White et al., 1976) is derived from the renormalization procedure of statistical mechanics whereby a system at criticality is coarse-grained. Market profiles (H. C. White, 1981) are scaling laws and the corresponding exponents are used to classify different kinds of markets. A recent research project found great benefit of applying the methods of allometric scaling to identify the universal power-law behavior of urban systems (Lane et al., 2009).

The idea behind social scaling is that persons, groups, and organizations are merely different manifestations of the same idealized type of identity. Power laws signal *scale*

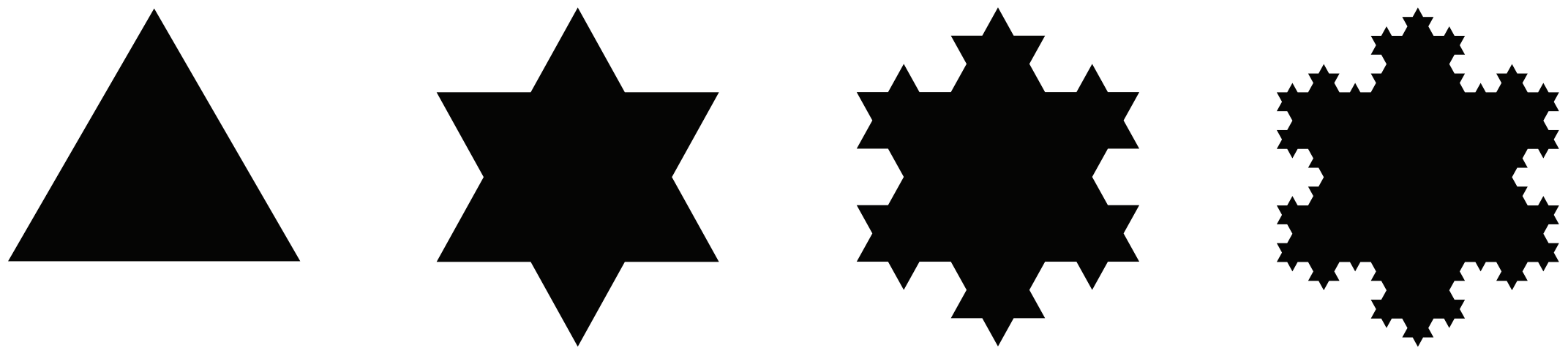

**Figure 1.3.: Construction and meaning of the Koch snowflake**
Starting from an equilateral triangular area, an edge is divided into three equal segments and, to the middle segment, an outward-pointing equilateral triangular area is attached with the middle segment forming the base. This geometrical transformation results, per iteration, in a multiplication by 4/3 of the total edge length of the resulting area. This has the effect that, as segment length decreases, the total edge length converges to infinity. Imagine that the initial triangle has an edge length of 27. In the first iteration, 12 segments of length 9 are needed to describe the form. In the second iteration, 48 segments of length 3 are needed, and in the third iteration 192 edges of length 1. The frequency distribution of segment lengths is described by a power law with exponent $D \approx 1.26$. $D$ is the fractal dimension of the Koch snowflake and quantifies the broken dimensionality of something closer to a line than to an area.

*invariance*, i.e., observations do not vary at different scales of observation. Scale invariance is the mathematical formulation of *self-similarity*, the property of a system that its parts are similar to the whole. Everybody who has ever looked at a magnified image of a snowflake knows how beautiful this self-organized regularity of nature can be. A snowflake is born from a liquid-to-solid phase transition in clouds when droplets of supercooled water crystallize around, e.g., a dust particle. As the seed of the snowflake grows and falls to the earth, more and more water molecules attach to it and, depending on atmospheric conditions, dendritic, plate-like, or hybrid ice crystals emerge. The regularity that strikes us as beautiful is the repetition of form as we zoom into a part of the crystal. Snowflakes belong to the most beautiful fractals nature has to offer. Fractals are self-similar. Like the Erdős-Rényi graph at criticality, the Koch snowflake is self-similar, but unlike the graph, its self-similarity is purely deterministic.

Fractals were introduced by mathematician Benoit B. Mandelbrot (1982). Mathematically, a *fractal* is an infinite set with non-integer dimensionality $D$ (p. 15). Real fractals only exist in mathematics because, even if they span multiple orders of magnitude, nature and society are finite. Mandelbrot showed that we can study our snowflake using a geometric model of it. Figure 1.3 explains how the model is constructed and that the form has a fractal dimension $D \approx 1.26$ which is the exponent of a power-law distribution of lengths describing the fractal. A line has $D = 1$, an area $D = 2$, and the Koch snowflake is a form closer to a line than an area. Self-similarity hides in non-integer dimensions. Mandelbrot proved that the fractal dimension can be derived from the geometric construction rules, in our case: $D = \log(4)/\log(3)$. The model is called a "Koch island" because Mandelbrot originally proved fractality for coastlines.

The introduction of fractals is an "epistemological breakthrough of considering seriously and quantitatively complex irregular structures" (Sornette, 2006a, p. 123). Conceptionally, they reflect the jump from translational invariance to scale invariance. Translational invariance means thinking in averages: A scientist has x works on average; a scientific journal gets y citations on average; a word is used z times on average. But Lotka's, Bradford's, and Zipf's Laws show that such averages are mathematically not even defined because the distributions typically are not in the domain of attraction of the normal distribution (for a mathematical discussion, see appendix A). Identities are self-similar like snowflakes. This is Paretian thinking. Abbott deserves credit for having provided a heuristic for the observation of scale invariance in social systems. When power laws are detected, it is worth searching for fractal distinctions.

#### Summary

Sociology must simultaneously explain stability and change, but to explain the latter, we must give up persons as essences of social life. Instead we must allow for variation, our first rule of method, and turn towards variable identities as outcomes as well as building blocks of social systems. Using a relational ontology, identity is modeled as an open system of a meaning structure that emerges from, and feeds back onto, transactions in a social network. Identities are self-similar. Persons, groups, organizations, and so on are manifestations of the same idealtypical feedback process. Levels emerge, i.e., organizations are combinations of, and contexts for, groups which are combinations of, and contexts for, persons and so on. We seek guidance in the scaling hypothesis of percolation theory which, applied to social formations, states that self-similar identities reside at the phase transition between disorder and order. The hypothesis needs not be rejected when power laws like Lotka's, Bradford's, and Zipf's Law are detected. Scaling laws describing the phase transition, and Abbotts fractal distinctions are diagnostic of simultaneous stability and change, normality and chaos.

### 1.2.2. Taking a Narrative Approach

According to Tilly (1995), "sociologists search almost instinctively for general, invariant models" of processes of change, i.e., they try to find general categories of recurrent events. "Sociologists suppose that if they had recognized the category when the process began they would have been able to predict its outcome" (p. 1594). Practically, one starts with a group of outcomes to be explained, e.g., political or scientific revolutions, and identifies a set or variables that, in interaction, explain the outcome, unveiling a category of emergence. This is a linear and variable-centered approach to study historical change. It is like Newtonian mechanics where the initial condition and the "law" determines the outcome (Wallerstein, 1996). In this model, stories are spun to describe insights about attributes measured as variables. In the variable-centered model of causality, big effects must certainly have had big causes. Big changes cannot result from endogenous dynamics, they are necessarily caused 'from outside' (Abbott, 2001b, chs. 1, 6).

For the linear model to be functional and valid, processes must to be in equilibrium.

> An equilibrium is a situation in which some motion or activity or adjustment or response has died away, leaving something stationary, at rest, "in balance," or in which several things that have been interacting, adjusting to each other and to each other's adjustment, are at least adjusted, in balance, at rest. ... An equilibrium is simply a result. (Schelling, 2006 [1978], p. 25–7)

In thermodynamics, *equilibrium* is the absence of a throughput of energy. In sociology, equilibrium means that an identity is isolated from context, when there is no throughput of information. But socio-cultural life is eventful and hinges on events. These influence which further sequence of events is initiated. Things do not fall into place the same way if one event in the chain is absent, replaced, or contextualized differently (Bearman, Moody, & Faris, 2002). The rest of Schelling's quote is: "The body of a hanged man is in equilibrium when it finally stops swinging, but nobody is going to insist that the man is all right."

The relational ontology calls for a narrative approach to studying change. For example, there are two ways of asking why people have certain kinds of careers. If we ask "*why* people have certain kinds of careers", we are interested in explaining an outcome and the underlying process falls into place. This is the linear model. If we ask "why people have *certain kinds of careers*", we are interested in the narratives that produced the outcome (Abbott, 2001b, p. 161). The latter aims at a generalization of process and mechanism, not of story and event. In other words, we try to identify the middle-range coherence of phenomena observed in different domains of social life. Such analytical sociology, in other words, aims at uncovering the universal behavior of similar socio-cultural systems (Hedström & Bearman, 2009b).

Our *second rule of method* is to reject the assumptions of the linear model and use a narrative approach "in the broad sense of processual, action-based approaches to social reality, approaches that are based on stories" (Abbott, 2001b, p. 185). We shall seek patterns and changes in stories and events. Stories are units of analysis, not words used to convey interpretations of causal interactions of variables. Theorizing narrative means getting a grip on story, story set, event, contingency, and path. The first understanding of *story* is that of talk connecting persons or combinations of persons in transaction structures. This is the interpretation offered in the previous section of what goes on in the lower circle of our identity model. But the identity-as-process approach only unfolds its full power if we get rid of the idea that nodes in "social networks" are necessarily persons or groups and so on. This gets easier to grasp when we consider social facts as stories and meaning structures as story sets.

Consider paradigms as a set of stories accepted by scientific communities (Mische & White, 1998, p. 711). Today's abstract interpretation of a Kuhnian paradigm is a set of rules accepted by the majority of a science that has acquired some degree of stability or institutionalization and is located in the core of a cultural network (Hoyningen-Huene, 1993; S. Fuchs, 2001). Cultural networks are meaning structures. Paradigms are just the agreed-upon facts, but there are may others, from specific knowledge claims to wannabe-paradigms. This is an understanding of social fact as a story. Essentially, identity is a unity of actor and story. It is an account of *what* goes on (transactions as stories) and *how* stories are spun that way (story set as meaning). This inseparability of transaction and meaning, of actor and story, is captured in the term *network domain*, where network stands for transaction structure and domain for the storied culture that gives meaning to the identity (H. C. White, 1995; Mische & White, 1998; Fuhse, 2009; Fontdevila, Opazo, & White, 2011; Padgett, 2012b).

*Contingency* is a possibility of path, "a way of speaking about the unpredictable nature of final outcomes, given some set of initial conditions" (Mahoney & Schensul, 2006, p. 461). *Events* are contingencies of meaning. An alteration of path may or may not result. They can be small encounters or big movements, endgenous or exogenous, of a socio-cultural, technological, or natural sort. Social facts are events, but events are only social facts when they have an effect. Events can be 1-identities. "Events strung together are a story. This is story as realized in time, just as ties realize storysets in social space" (H. C. White, 2008, p. 186). Events and social facts have sizes and durations. We have already encountered size in terms of distributions. The duration or lifetime of an event is marked by the period during which it is not perturbed by other events. They are contingencies that spark identities or other events but are themselves stable stories. This is a recursive definition. Events are not perturbed but they perturb other events. Consequently, events are enduring stories that are not perturbed. They are in quasi-equilibrium. When an event is perturbed, it can continue to go by the same name, but its story and meaning as a social fact will have changed. It will have acquired a new identity.

Control projects of identities aim at constraining possible events because each event that is allowed to be influential requires a modest or great adjustment of past control strategies (Padgett, 2012b). Beginnings, ends, and turning points are not assumed ex ante, as in the linear approach of classical comparative history. The meaning of events and social facts is conditional on their position in a sequence of events (Abbott, 2001b, p. 240–3). Narrative analysis seeks out patterns of stories. Story sets and narratives are such pattern. *Story sets* are patterns in space. *Narratives* are chains of events, story sets in time. Narratives are paths along which identities maneuver through socio-cultural spaces and times. Identity as path is identity in a *third sense* (3-identity) (H. C. White, 2008, pp. 10, 17; cf. Padgett, 2012a, p. 59). Each network domain calls one story set its own that provides a semantic envelope for emergence. Narratives are not devoid of change, they can contain turning points (H. C. White, 1995; Abbott, 2001b, ch. 8; Godart & White, 2010). "The only assumption that need be made is that an event can affect only events beginning after it" (Abbott, 2001b, p. 176). Another pattern is institutionalization. A story or social fact that is selected for a relatively long time is an *institution*, i.e., social facts are institutionalized to varying extents (Durkheim, 1982 [1895], ch. 1; H. C. White, 2008, ch. 5).

Lifetimes of social facts and events differ like their sizes. The variation of time horizon is consequential: "Since the narrative moves from event to event, there is no necessity even to assume a regularly spaced observation every year. A small, highly specifiable event can be seen as affecting large, diffuse events. The only necessary assumptions involve event orderability." (Abbott, 2001b, p. 176) Sociologist Peter S. Bearman has shown that narrative analysis need not be interpretive but can be quite positive. To identify patterns and paths, the repertoire of network analysis can be directed not only at networks of identities and facts but also at narrative networks of events in time (Bearman, 1993; Bearman & Stovel, 2000; Bearman et al., 2002). By allowing for variation, the narrative approach is able to study identity formation as a historical process that produces both stability and change (Tilly, 1995; H. C. White, 1995; Abbott, 2001b; H. C. White, 2008; Padgett, 2012b).

In the narrative approach, causality is not found in independent and identically distributed variables but in networks linking identities and events through stories. What makes the departure from essentialism so bold is that not just scientific communities or other social collectives are basically story sets and sequences of events. Persons too are identities in all three senses. They consist of basic building blocks, now necessarily stories and events, they have found footing in some way, and they have local and global narrative histories of control, of moving through socio-cultural space-time. A short way of saying this is: "In the short run, actors create relations; in the long run, relations create actors" (Padgett & Powell, 2012a, p. 2). From day to day, identities spin stories and span transaction networks in the process. From this talk collectively accepted story sets emerge that build up and flow into enduring and robust narratives – one way to see the meaning structure in our model – that lastingly constrain short-term transaction. Being biografies, identities tend to stay on track provided that they are not perturbed by events. If they are, they may or may not succeed to normalize the perturbation, depending how strongly they are identities in the third sense of path.

The narrative approach offers a refined understanding of fractal distinctions. In social science, a major narrative based on a functionalist conception of persons was initiated by the foundational work of Parsons in the 30s. At the same time, a network research lineage emerged. The sociometric path begun by Moreno split into several more or less graph theoretical story sets. In parallel, anthropologists devoted to network analysis split off from the functionalist narrative. In the 70s, the group of White tied together loose ends from sociometry and anthropology and developed the first algorithm for blocking networks. This breakthrough punctuated many other paths and brought strands together in Social Network Analysis. This arm subsequently focused on the story line of structural analysis but formed branches that again leaned towards and away from the oversocialized conception of persons. In the 90s, White departed from this path and refocused on the duality of transactions and meaning. Even though narratives permanently diverge, they also converge following contingent events like the innovation of blockmodeling. In the short run, scientists argue, but in the long run, it is all about networks.

### Complexity

Identities find footing through self-organization. To control contingencies and unpredictabilities, they maintain adaptability and adjust to the chaos of society. The order that is supposedly maintained at the phase transition emerges spontaneously (H. C. White, 1973; Godart & White, 2010; Fontdevila et al., 2011). Kauffman defines self-organization as the spontaneous emergence of order at the critical point: "order for free" (Kauffman, 1993). In our context, *self-organization* is the spontaneous emergence of control. None of the fundamental building blocks of an identity calls the shots, control is not even necessarily attained in collective consciousness. It is the result of a subtle mechanism to be identified (J. H. Holland, 1995).

Chemist Ilya Prigogine (1997) has contributed greatly to understanding the complexity of open systems – like identities. The throughput of information that is associated with embedding into contexts results in identities being out of, and often far from, equilibrium. Complexity means turning towards statistical mechanics and leaving Newtonian mechanics behind. The variable-based approach of identifying a set of variables that generally leads to social change amounts to modeling a Newtonian deterministic process where time is reversible. The narrative approach is more like the thermodynamics of statistical mechanics. Changes of temperature associated with embeddedness introduce an arrow of time because the second law of thermodynamics states that entropy, the disorderliness of a system, can only increase. Complex socio-cultural systems are irreversible and stochastic. Non-linear dynamics result. Emergence depends on accidents, contingencies, endogenous and exogenous events (Brunk, 2002). The information output is not directly proportional to the input. The same input does not necessarily produce the same output in another context. "Searching for simple input-cause and output-effect is a fantasy in densely connected feedback systems. Analyzing the percolation of perturbations through existing networks, no matter how data intensive, is essential for understanding system response to the perturbation." (Padgett & Powell, 2012b, pp. 26)

Throughput of energy is the most important criterion in the definition of biological life, and if we allow information to replace energy then socio-cultural systems are alive as well. Padgett (2012a) has introduced the chemical concept of autocatalysis to aid modeling of control in social life. In chemistry, catalysis is the acceleration of a chemical reaction by an additional substance. If the catalytic substance is produced by the reaction itself, the latter is *autocatalytic*, i.e., the chemical reaction reproduces itself. Because the production cycle is closed, autocatalysis is a positive feedback loop. We use autocatalysis synonymously with *reproduction*, the iteration of transactions over time. Autocatalysis requires a throughput of energy, and autocatalytic systems are the minimal forms of life.

In the Eigen/Schuster hypercycle, a model of autocatalysis at the levels of pre-genetic molecules, the emergence of life is almost inevitable. Ribonucleic acids (RNA) that have a role in expressing genes reproduce autocatalytically and produce, based on information stored in genes, the catalysts (enzymes) that other types of RNA need for their reproduction. The hypercycle model is credited for having demonstrated the negative consequences of self-reproduction alone. If RNA autocatalysis did not produce the enzymes that catalyze other RNA, then all types would reproduce "selfishly," the best-reproducing type would crowd out all others, and variety would disappear. Padgett (2012a) has translated this into a sociological context:

> It is as if in evolution the complexity of individuals can be increased through the social control of their reproduction by other types of individuals with whom they are in interaction. "Social control" has the connotation, and the reality, of decreasing variability at the micro level of the individual. Eigen's pro-

> found and counterintuitive observation, however, is that social-interactional control actually increases complexity and variability at the higher population level of the system itself when that control involves catalysis. (p. 50)

The hypercycle model can explain the emergence of identities as an "altruistic" process of collective control. However, it can only explain the emergence of up to four types of living chemical species from simple rules of production. Padgett, McMahan, and Zhong (2012) and Padgett (2012b) show that this "complexity barrier" can be transcended and that overlapping social networks as well as culture can emerge from simple interaction principles. We will return to this imagery in later sections.

#### Summary

In this section, we have set the stage for modeling the meaning aspect of identity. Our second rule of method is to take a narrative approach that allows the dissection of history from the bottom up. In this approach, stories are units of analysis that appear as relations in transaction structures and as nodes in meaning structures. They vary in terms of size and degree of institutionalization. Events are contingencies of meaning, they have the ability to punctuate identities and other events. Many contingent events form a dynamic and non-linear story set which is far from equilibrium, i.e., small perturbations can have large effects. This modeling leads to an understanding of "social networks" as network domains, dualities of actor and meaning. Inspiration is drawn from modeling identities as living systems. The hypercycle model shows how complexity as variability is maintained when transactions are altruistic.

### 1.2.3. Accounting for Observation

The scaling hypothesis of section 1.2.1 posits that self-similar identities are in a state of permanent emergence. Another way to understand the phase transition is as a tipping point. Tipping is the spontaneous switching of meaning when a "critical mass" is reached (Granovetter, 1978; Schelling, 2006 [1978], ch. 3; Kim & Bearman, 1997). During a scientific revolution a new paradigm crystallizes and establishes. The system tips at a macro scale. The old paradigm may survive as a building block in the cultural periphery of the community, but it dies as a collective contingency, as a 3-identity. As the community observes itself it will look to have reinvented its narratives and blueprints of path, it will realize that it is different than before. An identity that is conscious about its identity and path – about its self – is an identity in the *fourth sense* (4-identity) (H. C. White, 2008, pp. 11, 17). Self-consciousness can be identified using blockmodeling. Blocks from structural equivalence are identities consisting of parts in the first sense of identity. A 0-block identity is not conscious about itself because the parts do not transact. Self-consciousness emerges from homophilous action and shows empirically as a 1-block (Bearman, 1993; Bearman & Stovel, 2000). There are two reasons why we need to incorporate a notion of observation into our analysis. The first is the self-consciousness of identities just stated. The second reason is that the tools we employ in our analysis derive from the very same identity we study, Social Network Science.

Our point of departure for the concept of observation is sociologist Niklas Luhmann's (1995 [1984]) social systems theory. Like identities arise from stories in Relational Sociology, social systems arise from communications in Luhmann's theory. The concept of autopoiesis states that systems reproduce self-referentially, i.e., they create a distinction between themselves and the environment. Inside the constructed boundary, they reduce complexity by selecting certain parts to be connected, not unlike a complex adaptive system that self-organizes to the critical point between normality and chaos. Essentially, distinctions are iterated, i.e., communications are similar to themselves over time. Self-referential systems are self-similar because they are homeostatic, they look similar no matter if they are observed over short or long time scales (S. Fuchs, 2001, p. 272–5). While in the previous section social self-similarity refers to the percolation threshold, autopoiesis stresses a stable and trivial side of self-similarity. This is exactly the problem because Luhmannian systems always resist change. If they were matter, they would freeze in. This is because social systems are closed, walled off from the environment by hypothesis. Stochasticity does have its place in systems theory in the form of complexity to be reduced, but this complexity is always inside the boundary of the system and is determined in the initial distinction which constitutes the boundary (Luhmann, 1995 [1984], pp. 119). Luhmann defines social life as autocatalytic and enclosed, not as far from equilibrium and autocatalytic. This has caused Padgett (2012a) to say that Luhmann has taken the good idea of autopoiesis or autocatalysis and "run off with it in the wrong direction" (p. 56). Despite the need to retheorize inherent essentialist tendencies (S. Fuchs, 2001; H. C. White, Fuhse, Thiemann, & Buchholz, 2007; Fontdevila et al., 2011), systems theory has pioneered self-referentiality in the sense of social theory that is capable of explaining itself as laid out by S. Fuchs (2001, ch. 1) and practiced by H. C. White (2008, p. 135–141).

Our *third rule of method* is to account for observation. We encounter observation already in Durkheim (1982 [1895]) who used the systems term as "a specific reality which has its own characteristics" (p. 129), "things which exists in itself" (p. 143), but also "legal and moral rules" (p. 52), "ethics" (p. 66), and "precepts" (p. 66). This puts Durkheimian systems close in meaning to institutionalized social facts. These facts are not only emergent but also constrain identities in transactions. For facts to influence identities they need to be observed. The practical corollary of this rule is that observation turns identities into social facts. Social facts are identities, but identities are only social facts when observed.

To observe means to make a distinction, e.g., self/other, inside/outside, or true/false in science. In Abbott's (2001a) case these are fractal and create permeable boundaries. To observe does not mean absence of story because one cannot not communicate. There are different levels of observation. Note that these are not the same levels are not the layers of emergence defined before. On the first level of observation, an identity observes the social, technological, or natural world. This observation takes place in the transaction structures of our identity model. For example, an identity may observe what happens when an apple falls from a tree or when persons transact in a group. The observation may serve the goal to decide if it is true that the apple accelerates or it may be directed at a group outside identity's own group. First-level observation is about *what* goes on in the world. This level is realist because the world to observe certainly is there.

On the second level, an identity observes observers of the world. This is the observation of meaning structures by identities in transaction structures. If the identity watching the apple is Isaac Newton, his observers could be scientists that observe how he thought about acceleration because they may want to learn that way of observation. Or an identity may observe how identities in another group behave in contrast to those in his own group. In both examples, observation of observations is about control, about learning norms such as scientific paradigms or social conventions. Social facts only arise in second-level observation because they require an observer to observe how the facts influence members of the collective from whose transactions they have emerged. Social facts are constructed because they are observed, and therefore interpreted, observations. That something is constructed by an observer does not mean that it is not real. Second-level observation is about *how* social facts constrain transactions inside/outside a collective and are central to the emergence of identity (S. Fuchs, 2001, p. 24–9).

Further levels exist. In our work, we observe on up to three levels. On the first level, we analyze social networks to gain new insights about their structure and dynamics. This is us as social network scientists. On the second level, we observe how other scientists analyze social networks to get an idea what the standard practices are and to be able to address our peers. We learn, e.g., by reading other scientists reports and tracing back their citations to related work. This is our role as social network scientists trying to find footing in Social Network Science. On the third level, we observe how social network scientists collectively embed in Social Network Science. We do so by analyzing the narrative flows and meaning structures emerging from, e.g., citations made in the domain. This is our role as sociologists of Social Network Science. If one of our own papers was part of these networks, we would even observe the products of our own second-level observations. Sociology is "constructivism about science and culture" (p. 67) because the cultural thing of knowledge is produced socially (Fleck, 1979 [1935]; Kuhn, 2012 [1962]).

We treat social facts as if they were things. This was Durkheim's (1982 [1895], ch. 2) argument when establishing sociology versus psychology and is supposed to prevent subjective interpretations of constructions. This does not mean that we "treat persons as nonproblematic and fixed" to positivistically explain social network structures (J. L. Martin, 2009, p. 14). Representing identities, which we have constructed as stochastic narrative flows, as simple nodes in a network means that we pragmatically accept to lose information. It is our way as third-level observers to actively transform identities and events into data points that "work" for theorizing (Leifer, 1992).

## Observing Complexity

Social facts carry meaning which is a qualitative category. Nothing is better suited than the snowflake to show that emergence has a qualitative aspect. After all, the whole is not only more than but also very different from the sum of a system's parts. What emerges is qualitatively different. Quality emerges from quantity. Fractals like ice crystals are even beautiful. Durkheim argued that persons obey the laws of psychology but sociology is not just applied psychology. This argument can be scaled down to the bottom of the hierarchy of the sciences: molecules obey the laws of many-body physics but chemistry is not just applied many-body physics; atoms obey the laws of elementary particle physics but many-body physics is not just applied elementary particle physics (Anderson, 1972). As we move from the most realist to the most constructivist science, general laws give room to particular laws of pattern and predictability within universality classes (Watts, 2014). Science does not get "soft" at the natural-to-social science transition but once natural and social species are observed in contingent co-evolution (Bak, 1996, p. 7–9). The universality classes of evolutionary theory are the branches of the tree of life. The social sciences have universality classes too. They are just fuzzier than in the natural sciences.

Social systems are different from all systems beneath them. The impossibility to isolate systems from the environment is most pronounced. It is Luhmann's (1995 [1984]) fallacy to have encapsulated social systems which is already an observer's decision (Padgett, 2012a, pp. 36). Not even organisms can be isolated, they die when energy throughput is cut off. Complexity is not reduced inside a system but uncertainty is reduced through complexity by forming networked systems, i.e., control is achieved collectively. This shows as higher clustering coefficients in social networks than in technological or biological networks (Newman & Park, 2003). People do not communicate through pheromones like ants, but identities communicate through talk and stories (H. C. White, 1995). Social systems are the only ones capable of symbolic communication. If the rules of communication are changed, new emergent structures result (Sawyer, 2005, ch. 10; Padgett, 2012b).

Identities process information, but information is not bound by conservation laws. Aggregation of knowledge from many sources can result in synergy or redundancy which endows groups with "problem-solving strategies that are superior to those possible among noninteracting individuals and, in turn, may provide a selection drive toward collective cooperation and coordination" (Bettencourt, 2009). This "selection drive," the absence of general laws in social systems, is what allows greater complexity to emerge.

> Evolutionary progress to more complex hypercycles will never be smooth, no matter what the buffers. System collapses of some magnitude are both inevitable and useful for clearing the way to subsequent advances. But ragged and stochastic drifts upward in technological complexity are increasingly likely as more sophistication in communication is enabled. Even though bacteria and humans both evolve, humans evolve faster because of their communication and cooperation through product and identity symbols. (Padgett, 2012b, p. 113)

To analyze this complexity, we need the tools of mathematics like graph theory and percolation theory (H. C. White, 2000). Any measure of complexity must describe the amount of regularity in a system (Gell-Mann, 1995). Power laws are a possibility because they are measures of fractal order. But since emergence has a qualitative aspect, power laws do to. Mathematics needs to be supplemented by interpretation of meaning to be social. Modeling narratives is sociology's contribution to Complexity Science (H. C. White, 1997).

#### Summary

Our third rule of method is to account for observation. This is a necessary step because social facts can only have an effect on transactions if they are observed and because we are observers of the very same identity we take our tools from. Treating our observables as things is our pragmatic way of "denying the data" to escape the constructivist trap. Social systems are different from non-social systems. Language is a new level of the complexity of life. The fact that information is not bound by conservation laws propels society into ever higher levels of complexity. Yet, we can expect to find laws of pattern and process within narrow universality classes. Because of emergence, scaling laws as metrics of complexity have a qualitative dimension. The modeling of meaning is what sociology contributes to Complexity Science.

## 1.3. Models

### 1.3.1. Identities Emerge from Distinctions

It is impossible to draw clear-cut boundaries around identities. Consider the long-standing question of what demarcates science from non-science. From the emergentist perspective (S. Fuchs, 2001, pp. 84), pinning down science logically with essential delineation criteria – its empirical basis, objective truth, the use of method, or others – fail because it will always be possible to find violations of these criteria. There is no "one science", just a bunch of research domains that are typically combined into disciplines and finally subsumed under the labels "science" and "arts." Each domain has a cultural *core* in its meaning structure that contains the facts that matter for the specialty as a whole at a given moment in time and which were selected because they are either true, up-to-date, ideationally close, or reputationally rich. Each specialty is unique and it would be impossible to boil science down to a core that matters equally for all research domains. The boundary of science that separates it from non-science is constructed from the boundaries of its specialties and defining it means taking emergence and change seriously: "'demarcation' is a social, not logical, activity" (p. 91).

Abbott (2001b, ch. 9) has presented a formal model of spatial boundary emergence of social organization from distinctions. His example is the domain of social work which did not exist in 1870 but was established by 1920. Social work "emerged when actors began to hook up the women from psychiatric work with the scientifically trained workers from the kindergartens with the non-church group in friendly visiting and the child workers in probation. All those people were placed 'within' social work, and the others ruled outside it" (p. 271). Figure 1.4 is our attempt of visualization. Think of letter $a$ as skills of workers in psychiatry, $b$ in kindergartens, $c$ in visiting, and $d$ in probation. A *distinction* is a network position that workers can take. It consists of observations of which skills are present (inside) and which are absent (outside). These are second-level observations how work is accomplished in this and other positions. All workers taking a specific position are hence structurally equivalent in the sense that they select the same social facts. Blockmodels generalized that way reveal boundaries (H. C. White et al., 1976, p. 737; Doreian, Batagelj, & Ferligoj, 2004; Padgett, 2012b, p. 113). The top-left distinction in figure 1.4a is a position with the skill to work in psychiatry $a$, another skill $f$ needed or helpful there, and the observation that the skills needed to work in kindergartens, visiting, and probation $b$–$d$ are absent. The skill $e$ is the one to work in social work. In a pre-identity state, there are only proto-boundaries, distinctions that can be coupled to result in actual boundaries (cf. S. Fuchs, 2001, p. 20). Figure 1.4b is the situation in 1870. Distinctions have been recognized to be compatible and are coupled to bound the four identities of psychiatric work, kindergarten work, friendly visiting, and child work in probation. Proto-boundaries are now boundaries and $a$–$d$ are primary skills. Figure 1.4c is the situation in 1920. Skill $e$ has been recognized by four positions to be compatible, and social work emerges in the moment that it is made primary.

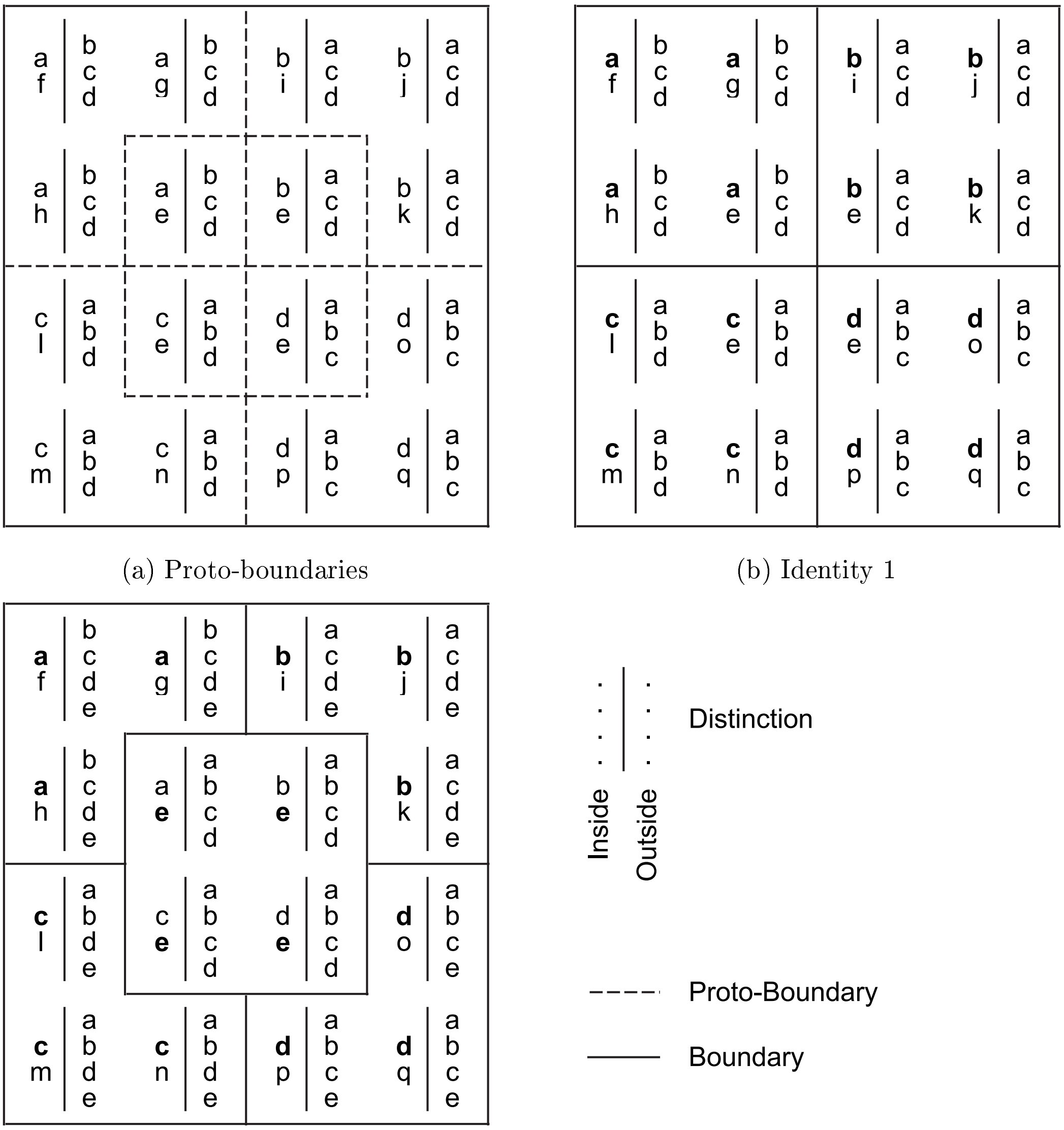

(a) Proto-boundaries

(b) Identity 1

(c) Identity 2

**Figure 1.4.: Boundary emergence from coupling distinctions**
Distinctions are positions that observe skills (letters) to be inside or outside. (a) Proto-boundaries exist where skills are potentially incompatible. When skills are recognized to be compatible and are coupled, proto-boundaries turn into boundaries. Yet boundaries remain permeable when skills overlap and it is just a matter which skill is primary, marked in bold. From (b) to (c) the boundary shifts because skill *e* switches from secondary to primary.

Abbott's (2001b, ch. 9) argument is that there is nothing wrong with distinctions, which can be mechanisms to wall off context, but that it would be wrong to just make them on one scale. The boundaries in figures 1.4b and 1.4c are actually permeable. Fractal distinctions can be modeled if we think of figure 1.4b as another situation in 1920. In that figure, a person with skills $a$ and $e$ works in psychiatry with the skill to be in social work. In figure 1.4c, another person with the same position is in social work with the psychiatric skill. That is Abbott's very idea of socio-cultural self-similarity. You come along different paths but still work on similar things. Boundaries never lose their character of being just proto-boundaries. We can define *boundary* now as a fuzzy region in the periphery of a core that results from inside/outside distinctions at multiple levels.

The emergence of identity can also be described using the autocatalysis framework of Padgett and Powell (2012a). To do so, let us translate Abbott's case of social work into our science context. Social work and the four communities it emerged from are network domains (research domains). Skills are rules of method, and primary skills are paradigmatic rules that are superordinate to the other skill. The Eigen/Schuster hypercycle model we have introduced in section 1.2.2 can explain how non-overlapping network domains like in figure 1.4c can emerge from the altruistic reproduction of domain-specific rules. The easy translation is mutual "learning by doing." Reproduction is altruistic because each of the four rules produces an output that catalyzes the reproduction of another rule which needs that output as input. We can think of altruistic autocatalysis as team work. Altruism is a plausible assumption because it mimics teams where scientists with different skills are brought together for knowledge production (Guimerà, Uzzi, Spiro, & Amaral, 2005). Recall that this kind of collective control, while decreasing variability of the positions, maintains diversity of the network, and that there is a complexity barrier that disallows the autocatalysis of more than four roles.

The original hypercycle model is non-spatial, i.e., rules can, in principle, altruistically catalyze all other rules. Technically, this is achieved by implementing the model on a fully connected graph. This is as if a rule (a scientists with some skill) searches for compatible rules (for scientists which can use what one has to offer) in all of science, without narrowing down search to scientists in, e.g., the same research domain. Padgett et al. (2012) have shown through computer simulations of agent-based models that such random, unlimited, and actually unrealistic catalysis is the reason for the complexity barrier that networks can harbor no more than four rules (or positions for scientists). When social space is introduced, when rules can only catalyze neighboring rules, larger autocatalytic networks can be sustained. Introducing an interaction grid has the effect of introducing cultural similarity – scientists do not search in far-away domains. Thus, to explain the four bounded networks shown in figure 1.4c, four original hypercycle models would suffice. But if each network harbored more than four rules, physical proximity would need to be introduced into the model. A research domain could not emerge if groups did not create local ecologies of progress that lower the barrier for more complex organization to emerge.

To explain overlapping networks with permeable boundaries or proto-boundaries, as shown in figure 1.4a, Padgett et al. have introduced another fix. Once physical proximity has been introduced, there are two ways of learning by doing. A lesson is learned when a rule leads to a transaction between two positions. Source reproduction is when the sending position gets to reproduce, target reproduction when the receiving position gets to reproduce. Ultimately, target reproduction strengthens free riders that are literally free to also couple into other autocatalytic cycles. In the long run, "altruistic" target reproduction enables the emergence of overlapping network domains and vice versa. Domains are sustained if they are themselves related to other networks (Padgett et al., 2012, p. 85–8). The principle of target reproduction is at the heart of the science system because reward is attributed to targets through citations (Merton, 1988).

These multiple overlapping network domains are the different worlds of society. The women from psychiatric work populate a different network than the scientifically trained workers from the kindergartens, the friendly visitors, or the child workers in probation, yet social work emerged in the intersection. The institutionalist program is another network than the Sociology of Scientific Knowledge, yet they overlap and house Gieryn's "Paradigm for the Sociology of Science." The oversocialized conception of person structures another domain than the undersocialized conception, yet Social Network Analysis is somewhere in the social and cultural intersection. On a higher level, all merge into science or art as opposed to the economy, law, religion, etc. On lower levels, scientists leave work and switch into the home domain. Or sports or wherever. The multitude and difference of network domains has caused sociologists like Durkheim to recognize the division-of-labor and division-of norms society whose parts are mutually dependent. *Switching* among network domains is the source of meaning. Switching is like zapping between TV channels (Mische & White, 1998, p. 704). During such cross-sectional transitions, identity decouples from one network domain and couples into another one. Stories and narratives are altered in the process. During switching, identities become conscious about themselves (Godart & White, 2010; Fontdevila et al., 2011). "Everything is what it is in relation to what it is not, not yet, or not anymore" (S. Fuchs, 2001, p. 339).

## Meaning Structures I

The meaning that emerges from switching is empirically accessible through the traces left by talk and communication (Carley, 1991, 1994; Mohr, 1998; Fuhse, 2009). In our case study, we restrict ourselves to the analysis of bibliographic data. Therefore, our review is mostly focused on, though not restricted to, networks studied in bibliometrics, the statistical analysis of such data. An early figure in bibliometrics is physicist Derek J. de Solla Price (1986 [1963]). He understood that the various power-law size distributions are results of network dynamics coupled to logistic growth and that they convey meaning. Lotka's Law of the number of works written by scientists, e.g., does not allow to set an arbitrary threshold above which a scientist can be regarded to be excellent.[3] Price proposed that this self-similarity of productivity is a result of growth and scientific collaboration. At the time of writing, he saw science in the middle of a phase transition from "little science," where every scientist could read all the relevant works in his domain, to a "big science" of overlapping subdomains. A *subdomain* is a part or substructure of a network domain. When a domain exceeds 100 scientists, so his rule of thumb, an *invisible college* emerges that acts as a representative and recognizable body of the domain and whose members "claim to be reasonably in touch with everyone else" (p. 119). The emergence of an invisible college often goes hand in hand with the founding of a new scientific journal tailored to the need of the specialty. As the network grows, a lower class of "fractional authors" develops, scientists which serve the elite as co-authors (p. 90). Scale invariance of productivity is a broadly confirmed phenomenon (Pao, 1986; Nicholls, 1989; Newman, 2001b). A concept closely related to that of the invisible college is the *research front*, the core of active and current research in a domain (Price & Beaver, 1986 [1966]).

Price (1986 [1963], ch. 3) noticed overlaps of invisible colleges and "multiple discoveries," innovations independently made by numerous scientists, but did not develop this systematically in the sense of fractal distinctions. A step in this direction was made by Crane (1972) who continued Price's focus on social structure and identified a network hierarchy. On a basic level, social networks consist of "groups of collaborators" which are coordinated by their most productive members who also set the local standards. On a higher level, the group leaders connect in "communication networks," close in meaning to invisible colleges, where the domain is coordinated as a whole. Distinguishing these types earned criticism because both are essentially network (collaboration) and domain (communication) but meaning is not modeled. Lievrouw (1989) made this criticism explicit, pointed out the need to conceptualize invisible colleges as processes, and added "specialty" to the definition to account for the meaning in scientific community elites. Today's definition is in line with our understanding of network domains: "An invisible college is a set of interacting scholars or scientists who share similar research interests concerning a subject specialty, who often produce publications relevant to this subject and who communicate both formally and informally with one another to work towards important goals in the subject, even though they may belong to geographically distant

[3] Price was aware that the number of works is not a measure of scientific excellence, but he used the example to explain scale invariance.

research affiliates." (Zuccala, 2006, p. 155)

Despite this progress, the various social and cultural scientific networks are mostly studied separately, exceptions withstanding. *Co-authorship* networks link (aggregates of) authors if they jointly author publications (Newman, 2001b). They are understood to be imperfect proxies for social collaboration (Katz & Martin, 1997). Most co-authorship networks, like those of publications in the *arXiv* and *Medline* databases (Newman, 2001b), social science (Moody, 2004), social psychology, economics, ecology, astronomy (Guimerà et al., 2005), or physics (Sun, Kaur, Milojević, Flammini, & Menczer, 2013), and other social networks, like those of email exchange (Ebel, Mielsch, & Bornholdt, 2002), instant messaging (Leskovec & Horvitz, 2008), jazz musicians playing in bands (Gleiser & Danon, 2003), or the Broadway musical industry (Guimerà et al., 2005), are not scale-free or only have power-law regimes, i.e., they are not scale-free in the upper tail of the distribution. This is plausible because persons cannot maintain an extremely large number of contacts for reasons of maintenance costs (Amaral, Scala, Barthelemy, & Stanley, 2000). In other words, one cannot "enact the role of 'friend' with large numbers of other people" (Simon, 1962, p. 477). Only the co-authorship network of high-energy physicists is scale-free because they have papers with 1,000 and more authors.[4] The network of inter-organizational collaborations in the biotech industry is scale-free because organizations can handle more contacts than persons (Powell, White, Koput, & Owen-Smith, 2005). The small-world property (Watts & Strogatz, 1998) has been demonstrated for co-authorship (Newman, 2001b, 2001c; Liu, Bollen, Nelson, & Van de Sompel, 2005) and other social networks (Kogut & Walker, 2001; G. F. Davis, Yoo, & Baker, 2003; Uzzi & Spiro, 2005).

What does it mean when networks are small worlds? Let us call them Watts/Strogatz networks. According to a rather consensual sociological definition, cohesion is a mechanism that bridges different levels of social organization and affects the attitudes and behaviors of persons in a group (Friedkin, 2004). This corresponds to the Durkheimian mechanism according to which meaning emerges from transactions and feeds back onto actors. Structurally, cohesion is measured as network density, the fraction of possible ties actually realized. Cohesion stresses that networks are locally dense and clustered. Local cohesion is the first property of Watts/Strogatz networks. They consist of many overlapping network domains with local clustering and meaning. Because networks carry meaning, the average path length as a measure of the typical distance among node pairs can be interpreted as the typical difference of local cultures. In the completely ordered regime, the typical distance is very large and an innovation can only spill over to very similar cultures. At the other end of the spectrum, in the completely disordered regime of the random Erdős-Rényi graph, cultures cannot even form because there are no clustered transactions from which meaning could emerge. Small worlds are both locally cohesive and globally connected through short average path lengths. Such behavior is realized by 'melting' the frozen state where a network domain only overlaps with very similar ones. Rewiring even just a few links randomly will drastically enable that potentially very different cultures influence each other. At some point, the network will be both ordered and disordered at the same time, allowing both local and global cascades or search. In fact, the rewiring probability has the role of temperature in critical phenomena and a scaling law describes the typical distance across which meaning can flow as a function of that probability (Newman & Watts, 1999).

[4]Note that the degree distribution of co-authorship networks is not Lotka's Law, which is the distribution of how many publications an author has written.

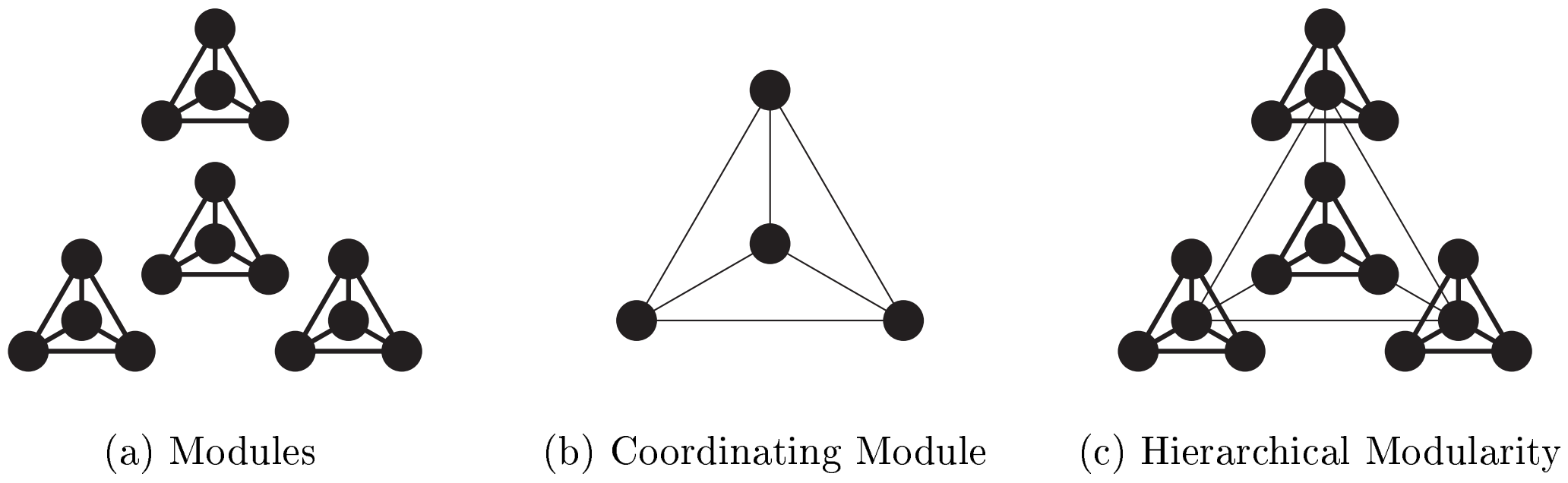

**Figure 1.5.: Idealtype of hierarchically modular identity**
(a) At a lower level, multiple modules exist in isolation. These 1-identities are distinctions, second-level observations of how other modules operate differently. Boundaries are sharp. (b) A coordinating module is anchored in the low-level modules. (c) Taken together, the domain is a 2-identity of 1-identities that have collectively found footing. It is a whole, yet with cultural differences in the parts. The coordinating module forms the core of the network and all nodes not in the core are the periphery. As one imagines higher levels of modules embedded into modules, the difference between core and periphery becomes increasingly fine-structured. As distinctions become fractal, the boundary between identity and context becomes permeable. The width of the ties is meant to be proportional to the "amount of time, the emotional intensity, the intimacy (mutual confiding), and the reciprocal services" (Granovetter, 1973). The strength of the weak ties is coordination.

Some networks, e.g. an email exchange network (Ebel et al., 2002), are both scale-free in a regime of the degree distribution and a small world. More social networks are small worlds than scale-free. The Microsoft Messenger instant-messaging network, e.g., a network of planetary scale (180 million nodes), is locally cohesive and globally connected but degrees are not power-law distributed (Leskovec & Horvitz, 2008). This network has a particular type of what Durkheim (1982 [1895]) calls social morphology:

> We know that societies are made up of a number of parts added on to each other. ... It is known in fact that the constituent parts of every society are themselves societies of a simpler kind. A people is produced by the combination of two or more peoples that have preceded it. If therefore we knew the simplest society that ever existed, in order to make our classification we should only have to follow the way in which these simple societies joined together and how these new composites also combined. (p. 111–2)

Figure 1.5c shows such a structure. It consists of four modules (figure 1.5a) that are themselves connected in a similar but higher-level module (1.5b). This structure is scale-invariant because the parts are similar to the whole. Ravasz and Barabási (2003) have shown that such hierarchically modular networks are described by a scaling relationship of node degree and clustering coefficient. The higher the degree, the lower the clustering coefficients, as can be seen in figure 1.5c. The instant-messaging network is such a formation (Leskovec & Horvitz, 2008). There is evidence that hierarchical modularity is due to its ability to optimize flows. Arenas, Danon, Díaz-Guilera, Gleiser, and Guimerà (2004) find that modules in an email exchange network, the jazz network, and several co-authorship networks derived from the *arXiv* branch out like a river basin that optimizes water transport. Think of information instead of water. As expected, self-similarity leaves its mark in the form of power laws.

We can think of Crane's (1972) finding in this light. The low-level modules are her "groups of collaborators" and their most productive members form a higher-level "communication network." Both are collaboration networks, but with different meaning. The higher-level nodes have the role to coordinate the whole and the nodes that are only on the lower level are Price's (1986 [1963]) "fractional authors," scientists with a different role of "sorcerer's apprentices." (pp. 79). As the network gets bigger than shown in figure 1.5c, many more levels will exist and the system will gain the character of a kinship system (D. R. White & Johansen, 2005) as the former servants start their own research group and have their own fractional authors. Therefore, even though author productivity is scale-invariant, it is not meaning-invariant. The authors in the fat tail of the power law are the sorcerers and those at the other end are the apprentices.

There is further evidence for hierarchical modularity. An individual's core support group typically consists of 3–5 persons. Sympathy groups consist of 12–20 persons that are typically contacted once a month. Studies of hunter-gatherer societies have further revealed that groups of 30–50 persons called bands gather in overnight camps which are drawn from clans or regional groups of typically 150 persons. Even higher, ethnographers have identified megabands of about 500 persons and tribes of 1,000–2,000 persons. Statistical analyses have detected discrete scale invariance with factors of 3–4, i.e., a whole is about three to four times as big as its parts (Zhou, Sornette, Hill, & Dunbar, 2005; Hamilton, Milne, Walker, Burger, & Brown, 2007). Similar morphologies have been found for administrative organizations (Johnson, 1982), military brigades (Zhou et al., 2005), and groups of avatars in a virtual world (B. Fuchs, Sornette, & Thurner, 2014).

Modules, hierarchical or not, are meaningful because they are candidates for network domains and, therefore, offer switching opportunities for meaning to arise. There is a plethora of methods to detect modules. We distinguish between role models (Reichardt & White, 2007) and cohesion models (D. R. White & Harary, 2001). Blockmodeling based on structural equivalence (Breiger et al., 1975; H. C. White et al., 1976) is a unique case of role modeling because, besides identifying blocks in the invisible-college sense, it is capable of detecting 0-blocks which are not aware of themselves. Modularity maximization is another special case of role modeling which only produces 1-blocks on the diagonal (Reichardt & White, 2007). Community detection, as it is most often referred to, is usually not described in a positional framework, but positions are exactly what communities are. Nodes in a community are structurally equivalent in the sense that most ties are to nodes inside the module and few are to outside nodes.

Modularity, developed by physicists Mark E. J. Newman and Michelle Girvan (2004), quantifies the extent to which network ties or observations are internalized in modules or communities. Inside/outside observations (boundaries) are distinctions that give meaning to a community. In the autocatalysis framework, transactions inside the boundary of an identity are *constitutive ties*. They are devoted to reproduction and teaching the skills needed to reproduce the identity over time. These "selfish" ties are necessary to keep the identity alive. Observations of outside identities are *relational ties*. These are cross-sectional stories devoted to information and product exchange. If the identity to be modeled is a group, constitutive ties are the transactions among the persons constituting the group, and relational ties are transactions with members of other groups. "Altruistic" relational ties are sufficient to keep the identity alive because they maintain diversity and switching opportunities and enable control (Padgett, 2012b). Like identity, modularity is a question of scale. Therefore, a practical way to maximize modularity and still be able to identify modules at different length scales is to cluster nodes bottom-up (Blondel, Guillaume, Lambiotte, & Lefebvre, 2008). Modularity maximization maximizes the number of constitutive ties and minimizes the number of relational ties. Blockmodeling allows proto-identities without constitutive ties.

Cohesion models do not detect communities but how smaller, denser structures are embedded or nested in larger, less dense contexts. A simple model divides a network into a core and a periphery (Borgatti & Everett, 2000). The sophisticated model goes by the name of *structural cohesion*. A network is a $k$-component if no more than $k$ nodes need to be removed to split it into more than one component (D. R. White & Harary, 2001; Moody & White, 2003). The network in figure 1.5c is a 1-component because it falls apart if just one of the coordinating nodes is removed. Each of the four low-level modules as well as the coordinating module is a 3-component. Structural cohesion is especially suited to detect results of Abbottian fractal distinctions. If distinctions are fractal, in a research context, scientists with different cultural backgrounds will work on similar things and likely collaborate. The nested levels of structural cohesion represent the permeability of boundary resulting from fractal distinctions. Such networks are not expected to show multiple peaks, like islands in a sea, but more or less a single island with some steepness.

Moody (2004) has shown that the co-authorship network of sociology as represented by scholarly journal articles from 1963–1999 is not a multi-peaked landscape, as would be expected if research proceeded in a division of labor. Social science, in other words, is largely intersticial, as Abbott (2001a) has theoretically argued. Interestingly, the co-authorship network exhibits neither a scale-free degree distribution nor an average distance comparable to a random graph. While we do not expect the former to be the case in transaction structures like this, the latter means that sociology is intersticial even though its different subdomains are relatively distant in terms of meaning. A modularity maximization algorithm like the one of Blondel et al. (2008) would have likely detected subdomains which, however, presumably do not match commonsense categories of social science disciplines.

Another example of structural cohesion is the biotech industry of 1988–1999, as ana-lyzed by Powell et al. (2005). This case is a very significant example for the importance of relational ties and the impracticality of the neo-classical model of markets (Granovetter, 1985). In the 80s, hundreds of small Dedicated Biotechnology Firms were founded. This was a decade of divided labor and mutual dependence. The firms, e.g., had the skills to develop a medicine, but not to bring it to the market. Large pharmaceutical companies, on the other hand, had the marketing skills but not the licensing skills that the universities had. Those were not skilled in finance, but venture capital firms were. Organizaions cooperated in Durkheimian organic solidarity, dealing with skills that others did not have. By the 90s, some firms had grown up to sizable and multifunctional companies, but despite the fact that they now combined many skills under one roof, the market did not turn into a gas. In the period of study, the biotech industry had a clear single-peaked identity. The market acted as a whole with a core that housed short-lived collaborations of organizations of even the same type. Powell et al. argue that this dense network kept the players alive exactly because each one could potentially outreproduce its competitors. If a winner took most, it would destroy the market mechanism that gives all players orientation, lose information sources to mitigate uncertainty, scare of venture capitalists that cannot afford to lose their money when their only investment goes down, and so on. In other words, Dedicated Biotechnology Firms and other organizational species altruistically catalyzed each other because a part cannot exist without the whole that it is a part of. The degree to which something is a whole with a multifunctional core called an *open elite* and a permeable boundary (periphery) is measurable by structural cohesion (Powell & Owen-Smith, 2012; D. R. White, Owen-Smith, Moody, & Powell, 2004). In general, co-authorship and direct citation networks are found to stronger resemble core/periphery structures than networks resulting from the exchange of emails or messages in an online community (Holme, 2005).

### Meaning Structures II

Up until here, we have discussed how the meaning of social networks can be measured. We now turn to networks related to word usages and citations where meaning is apparent and the social aspect is hidden (Yan & Ding, 2012). Both words and cited references are social facts or concepts that stand for meanings expressed in symbolic communication (Small, 1978; Carley & Kaufer, 1993; Padgett, 2012b). What Lotka's Law is for author productivity, Zipf's Law is for word usage. Zipf (2012 [1949], ch. 2) found that the rank-frequency distribution of word usage is a power law and attributed it to an optimization principle he called the "principle of least effort." *Optimization* is the improvement or maintenance of control. He modeled the latter as a trade-off between the opposing forces of unification and diversification. The former acts towards small vocabularies of ambiguous words, the latter towards large vocabularies of unambiguous words. He found that also the average number of meanings for bins of frequency ranks is scale-invariant, i.e., the less frequently a word is used, the fewer meanings it has.[5] Put differently, a word acquires its meaning by embedding into context, a network of words. Analyses of word networks have revealed that these contexts are both scale-free and have the small-world property (Solé, Ferrer i Cancho, Montoya, & Valverde, 2002, 2010). Unlike in these studies, we are interested in contexts that are not devoid of social relations. Language contexts are constantly negotiated in every situation of social life, especially when identities zap from one network domain to another (Fontdevila et al., 2011). Zipf, Complexity Science, and Relational Sociology, then, place word usage at the transition where not just poetry or instruction manuals are possible, but where language can be used flexibly.

[5] Obviously, the most frequent words are function words like "the" and "of." Zipf demonstrated self-similarity of meaning for ranks lower than 500.

According to Carley (1994), the analysis of language and text is a direct window on the mind and on culture. Text is "an encapsulation of a portion of the author's mental model" and "individual mental models provide insight into the relationship between individual cognition and culture as both cognition and culture evolve through, and concurrent with, the evolution of language" (pp. 291). Semantic network analysis is like content analysis without discarding the relations among concepts. Concepts as nodes in meaning structures are automatically or manually classified by a thesaurus into parts of speech or different classes like agent, knowledge, resources, etc. The resulting networks depict how concepts or stories relate in sentences or larger semantic units and can be analyzed using the tools of information science or network analysis. The meaning of a concept depends on the way it is embedded in the network. If concepts are persons, groups, or organizations, social networks described in document corpora can be revealed. According to this framework, symbols are concepts shared by a social network under observation (Carley & Kaufer, 1993; Diesner & Carley, 2005).

Another approach called *co-word* analysis emerged from the Latourian laboratory-studies school in the Sociology of Scientific Knowledge. Callon and Law (1986) embed text analysis in Actor-Network Theory. According to this theory, the frequent co-usage of words signals that an "actor world" has been successfully "translated," i.e., an idea has been able to mobilize a critical mass of actors. In Relational Sociology, this would be expressed as an identity that has consciously formed around a story set (Mützel, 2009). Proponents have tried hard to translate Actor-Network Theory (Rip, 1988) but without much success of rooting the theory in scientometrics, the bibliometric analysis of science (Braam, Moed, & Van Raan, 1991a). Co-word analysis can be done with either theoretical background and, as a method, is a standard tool and typically involves the interpretation of maps (Courtial, 1994; Börner & Scharnhorst, 2009).

References cited in texts are the other main class of concepts in scientometrics. Other than the use of words, the debate what a citation actually means has a long history (Small, 1978; Cozzens, 1989; Van Raan, 1998). From our perspective as third-level observers of science it is important that citations have changed their meaning over time. In the times of little science, a citation was a reference to a scientist or his œvre, i.e., a first-level observation. In today's big science, it is a second-level observation, i.e., scientists are influenced by practices in scientific communities (Leydesdorff, 1998). Second, to study knowledge production as a narrative, we utilize the property that citations have a citing and a cited side and can be chained together in a *direct citation* network. Therefore, we use the historical bibliography approach of the pioneers or bibliometrics according to which a citation signals that a reference is symbolic of the idea or concept in the citing publication. The text around the citation can be analyzed to specify the particular idea (Small, 1978).

Bradford's Law about the skewed distribution of attention received by scientific journals was the first diagnosis of citation scale invariance. The latter is today one of the most certain macro results about knowledge production (Price, 1986 [1965]; Seglen, 1992; Redner, 1998; Milojević, 2010). Power-law citation distributions have been found for 17 out of 22 disciplines (Albarrán & Ruiz-Castillo, 2011) and 140 out of 219 fields (Albarrán, Crespo, Ortuño, & Ruiz-Castillo, 2011). This has caused an ongoing debate in the bibliometrics community which has been relying on invalid linear methods for very long. Like with Lotka's distribution of productivity, there is no intrinsic threshold above which a cited reference is excellent. But certainly the highest cited concepts are what we call paradigms.

Direct citation networks can be transformed into two other types of networks. In the *bibliographic coupling* network, citing publications are the nodes and edges give similarities according to the shared citation of references (Kessler, 1963). These networks are best used to identify research domains and fronts on coherent knowledge discovery paths (Boyack & Klavans, 2010). Role models are the primary tools to detect subdomains. Brandes and Pich (2011) have shown for our case of Social Network Science that, until the Complexity Turn, the domain had consisted of basically a structuralist and a social psychologist network domain. Seven years after the turn, a very coherent and focused cluster of Complexity Science publications had attached to the structuralist domain with hardly any coupling to the social psychologists.

In the *co-citation* network, on the other hand, cited references are the nodes and edges tell how often a pair of references is cited together. It is commonly understood that co-citation networks are the primary meaning structures of scientific communities and that highly cohesive cores shelter paradigms (Small, 1973, 1980). Lazer, Mergel, and Friedman (2009) showed through co-citation analysis of two leading American sociology journals from 1990–2005 that works of physicists had increasingly been picked up in sociology and brought classics like small-world research to new life.

These networks are ideally probed using cohesion models to identify the core "institutions, lifeworld, common sense, tacit background, or practices" (S. Fuchs, 2001, p. 218). A consistent method to identify the nested meaning structure of science is to identify how co-citation clusters embed into higher level formations (Small, 1993). Van Raan (1991) has shown that, in 1984, this structure was self-similar. In other words, knowledge is hierarchically modular like the toy network in figure 1.5c. "[S]cience is structured in terms of (small) research specialties, whereas at higher levels the co-citation clusters become increasingly extensive in size and therefore represent more and more higher hierarchical structures like subfields, fields and disciplines" (pp. 441). The importance of the scale invariance discovered by Van Raan is that, quantitatively, concept symbols are similar on all levels of sociological observation. But like invisible colleges are not quantitatively but qualitatively different from the many groups at lower levels of meaning, the most cited concept symbols increasingly acquire the character of paradigms as we observe higher levels of meaning. The same is true for words as concept symbols. The fractality of co-citation clusters adds a spatial dimension to the power-law distributions derived from direct citation networks and builds a bridge to the hierarchical modularity of social structure (Crane, 1972; Guimerà, Danon, Díaz-Guilera, Giralt, & Arenas, 2003; Ravasz & Barabási, 2003).

### Multivariate Analyses

The self-similarity of the science system is demonstrated by scaling laws of citation impact. Katz (1999) pioneered these studies by showing that, on average, research domains acquire the more citations per paper the larger they are. Increasing returns to scale are demonstrated by a superlinear scaling relationship of publications and citations. In percolation theory, such scaling laws describe how a system behaves close to its critical point. Here, it demonstrates the Matthew Effect which states in its most general form that there is an advantage in large size (Merton, 1973 [1968]). Such an effect also describes recognition in lower strata of science, namely specialties (Katz, 2000), universities (Van Raan, 2008), and research groups (Van Raan, 2006).

Small to medium-scale analyses of social and cognitive meaning structures have revealed that these are neither independent nor identical. Strong support for fractal distinctions comes from a bibliometric analysis of Science and Technology Studies. Social interaction among authors cognitively belonging to different subdomains is much stronger than expected based on cognitive similarity. Scientists in the field even disagree on where the cleavages are (Van Den Besselaar, 2001). The observation of a group of 21 leading authors in information policy suggests a large overlap of collaborative and intellectual structure (Rowlands, 1999). Yet in information science, social networks based on co-authorship do not map neatly to the cultural networks based on word usage and citation. In a multidimensional-scaling space where the two dimensions can be freely interpreted, co-authorship networks feature in the social/non-citation regime while co-word, co-citation, direct citation, and bibliographic coupling networks cluster in the cognitive/citation regime (Yan & Ding, 2012). Social scientists belonging to invisible colleges are found to consolidate mainstream research while peripheral authors called “social climbers” are most likely to introduce fresh ideas. Altruistic collaboration among members of different strata is found to be a sufficient condition for domain evolution (Mutschke & Quan Haase, 2001). Different behavior of different levels is confirmed for chemical research. Multiple specialties in chemistry identified by co-citation clusters are embedded in larger topics of coherent word usage. Language, in other words, is more general or fuzzy than citation (Braam et al., 1991a). An analysis of an elite circle of interdisciplinary researchers sponsored to study human development finds that citations do not covary with social ties like collegiality or advice seeking. However, the group increasingly became aware of its identity – it gained a 4-identity – as shown by an increase of intra-group citations (H. D. White, Wellman, & Nazer, 2004). Large-scale timeseries analyses, discussed later, are hoped to draw a clearer picture.

#### Summary

Boundaries of identities emerge from the coupling of roles or inside/outside distinctions. As these are made at multiple scales, permeable boundaries arise in which multiple network domains overlap. Altruistic behavior, the property of identities to actively catalyze each other, is modeled as the source of multiple network domains which, in turn, enable distinctions and the emergence of meaning. In uncertain contexts, identities are expected to actively seek relational ties to other identities to maintain diversity and collectively gain control. The ubiquity of power-law size distributions of authorships, citations, and word usages is strong evidence in favor of the scaling hypothesis. These distributions are scale-invariant but not meaning-invariant, i.e., facts in different regimes of power laws have different meanings. Structural analyses and modeling point to a self-similar hierarchally modular structure of network domains. Practically, boundaries are best detected on the level of meanings, not transactions. Role models are suited to detect network domains, cohesion models to probe their nested structure.

### 1.3.2. Chaos is Disciplined

Network domains emerge from collective distinctions and attain some degree of duration – institutionalization or crystallization, in Durkheim's words – when the emergent story set causes distinctions to be iterated. As we saw, autocatalysis can give rise to multiple such domains which overlap in identities that share stories or social facts. Boundaries are permeable because stories come in sets. Not all stories in such sets are equally ranked. Only few take the lead while most will be subordinate. The major story of Granovetter's strength-of-weak-ties argument is oversocialization, the minor one is undersocialization. In Burt's structural-holes argument, the priorities are changed. In the end, both stories are very similar. Only few social facts, like paradigms, are major stories and get selected into the fat tail of size distributions. In the body of the distribution, we find a medium number of stories of medium guidance, and in the head, there are the many stories with little influence. The more we leave the tail and enter the head, narratives become small and narrow. Yet, those small facts whose sharing hurts no paradigm create the permeable boundaries of small-world networks where meaning is locally cohesive and yet globally connected (Zipf, 2012 [1949]; Carley, 1991; Solé et al., 2010). This generalizes to all social facts, to words, cited references, and authors. Our hierarchically modular working model (figure 1.5c) is not meant to represent just co-authorship, but network domains in general. Cumulated over some duration, these structures have power-law signatures in many ways, often in terms of node degree. To get to the structural mechanism that generates these distributions, we turn towards biased randomness, the Matthew Effect, but will shortly see that it does not make sense to decouple preferential attachment from homophily, an optimization mechanism that integrates structure and meaning.

In its dynamic version of preferential attachment, the Matthew Effect states that the advantage in large size accumulates over time. It is a positive-feedback mechanism. Merton (1973 [1968]) chose the name according to the Gospel of Matthew which says: "For unto every one that hath shall be given, and he shall have abundance: but from him that hath not shall be taken even that which he hath" (cited from Merton, 1973 [1968]). The first part until the colon says that the rich get richer, the second part says that the poor get poorer. It is the first part that most modeling efforts have focused on. Yule (1925) explained the distribution of the number of species in a biological family by speciation, the splitting of one species into two. It is clear that large species have a growth advantage. If speciation rates are constant, a family with $x$ species will grow proportionally to $x$. Rich families with many species get richer. The Yule process is Markovian because present success explains future success. Simon (1955) and Price (1976) showed that mathematical sophistications of the Yule process can explain the Zipf and Lotka distributions and the scale invariance of cited references in science, respectively.

Barabási and Albert (1999) have translated the Yule process of cumulative advantage into a network framework and given it the name *preferential attachment*. The Barabási/Albert model says that a power-law degree distribution with exponent $\alpha \approx 3$ emerges when the probability that a new node $j$ attaches to an existing node $i$ is linearly proportional to $i$'s degree. If preferential attachment is sublinear, the degree distribution will have a stretched exponential[6] form, and when it is superlinear, few nodes will be dominantly selected and the network will become star-like, connected by superhubs (Krapivsky, Redner, & Leyvraz, 2000). Practically, the attachment parameter is determined by predicting present degree from past degree through a scaling law (Jeong, Néda, & Barabási, 2003; Newman, 2001a; Golosovsky & Solomon, 2013). Barabási and Albert (1999) deserve credit for having provided a simple network mechanism capable of explaining the emergent phenomenon of scale invariance. Preferential attachment is *how* identities self-organize to the critical point. From social transactions emerges a probability distribution which gives further transactions meaning. As identities collectively learn from past transactions, frequently selected social facts stand a greater chance of getting selected in the future. Major stories are learned and provide paths through otherwise chaotic context (Reali & Griffiths, 2010).

It is well possible that the networks studied by Moreno and Jennings (1938) and Rapoport and Horvath (1961) had emerged from a sublinear rich-get-richer effect. Today, the Matthew Effect has been established as a system dynamics property both inside (Merton, 1988) and outside (Perc, 2014) the sociology of science. Preferential attachment has been shown to govern the growth of scientific collaboration and citation networks. Co-authorship networks as manifested in the *arXiv* (60,000 authors) and *Medline* (1.6 million authors) publication databases are not scale-free but are driven by linear preferential attachment (Newman, 2001a). Networks of mathematics and neuro-science, however, seem to be scale-free but governed by a sublinear Matthew Effect (Barabási et al., 2002; Jeong et al., 2003). These results contradict the model, possibly because different methods to measure the Matthew exponent were used.[7] Certainly, these results show that scale invariance in co-authorship networks only stands a chance, if at all, if networks are integrated over many years.

[6]We can think of the stretched exponential as a probability distribution that allows not as large values as a power law but larger ones than an exponential.

[7]Barabási et al. used logarithmic binning to determine scale invariance which has been shown to be error-prone (Clauset, Shalizi, & Newman, 2009).

A degree distribution is a distribution of meaning, and the Matthew exponent puts a number in the underlying generative process. Further insights await if we turn towards analyses of the actual network structures. Guimerà et al. (2005) want to know how team assembly mechanisms determine collaboration network structure. Let us consider a research domain that initially only consists of a number of unconnected wannabe-scientists. At time 1, four of them bring together their skills and form a team, e.g., in a speciation process described by the hypercycle model. Our research domain is an open system, i.e., we need not worry about funding. Each following time step, a new team of size $m = 4$ is instantiated and for each of the four positions probability $p$ determines if it is filled by a scientist already in the network, who is now a sorcerer, or by an apprentice standing in line to show how he can change the world. If $p = 1$, the four sorcerers will never teach any apprentice as the future team will always be the present team. The network will look like in figure 1.5b and it will lock in to a path of eternal autocatalytic self-reproduction. If $p = 0$, apprentices will never find their sorcerers because, each time step, the old team dies and a new team of four is born. The cumulative network will look like in figure 1.5a. Due to the lack of teaching, each new team will reinvent science and head off on a completely new epistemological path. $p = 1$ leads to one module with infinite lifetime while $p = 0$ leads to an infinite number of modules with unit lifetime.

If $0 < p < 1$, a second criterion is needed to determine how positions in new teams are filled. If a position is filled by a sorcerer and another is to be filled by a second sorcerer, probability $q$ determines if the second sorcerer is drawn from the first's previous collaborators. In other words, $q$ represents the inclination for sorcerers to reproduce their old team rather than learn from another sorcerer. Knowledge production prospers between stability and change. Simulations show that there is a phase transition in $p, q$ parameter space where a giant component of collaboration emerges. In this network, sorcerers repeat old constitutive ties, transact with new sorcerers, teach apprentices, and apprentices interact with other apprentices. Figure 1.5c idealizes this state. At criticality, we expect the research domain to be fractal, i.e., we expect that there is no characteristic size of co-authorship clusters. We suppose that, at the critical point – which actually has the form of a line in parameter space –, a research domain finds the optimum conditions for maintaining a research program along a path defined by a paradigm while being open for creative chaos.

Guimerà et al. (2005) have modeled the collaborative process that likely produced the empirical parameters of four real disciplines (1955–2004). They find that economics is right at criticality, social psychology and ecology are close to it. Only astronomy is far from criticality. All disciplines have quite similar probabilities to reproduce the composition of old teams ($q = 0.77 \pm 0.03$). But astronomy has a much higher probability $p = 0.76$ to not teach apprentices (for the other three disciplines, $p = 0.57 \pm 0.01$). The first three disciplines reproduce their ties less than 25% of the times. Astronomy is much more stable. Its tie reproduction rate is 39%. These percentages can be interpreted as the extent to which the disciplines self-reproduce their communications in Luhmannian autopoiesis.

Palla, Barabási, and Vicsek (2007) have progressed understanding by showing that co-authorship communities of different size also have different lifetimes. More precisely, the time that a module lives is a function of its size and the degree to which it reproduces itself: the most long-lived small groups heavily reproduce their parts and transactions while the most long-lived large groups exchange their members to a much larger extent. The optimal reproduction rate of a group depends on its size. This is shown for the social structure captured in a part of the *arXiv* (30,000 authors over 142 months). Similar results are obtained for a dynamic network of phone calls (4 million users over 52 weeks).

Sun et al. (2013) have observed that, in the bibliographic history of the American Physical Society since 1910, collaborative modularity bursts in times of, or shortly after, the foundation of major new journals. They propose that, just like Price had observed, this formation of collaboration modules is paralleled by the emergence of invisible colleges, and that scientific disciplines are fundamentally reflected in collaboration. Besides preferential attachment and the reproduction of past ties, their model of structural evolution incorporates two other attachment principles: triadic closure and *homophily*. The latter is the principle that similarity breeds connectivity – identities prefer to interact with culturally similar ones (Kossinets & Watts, 2009). From these dynamics, clusters emerge that are either split or merged if the result maximizes modularity. This way, the authors can well explain degree and productivity distributions but not so much disciplinary sizes and mixtures.

### Identity as Path

Four mechanisms of attachment have been introduced: preferential attachment proportional to degree, reproduction of past ties, triadic closure, and homophily. In the model by Sun et al. (2013), all mechanisms are active in parallel. In contrast, Moody (2004) has hypothesized that different structural mechanisms are independent of each other. What we are trying to explain by these mechanisms is 3-identity, identity as a narrative. Figure 1.6 is an attempt to visualize an idealtype. Each time step, parts of the network are reproduced. The low-level modules of figure 1.5 have a higher frequency of occurrence than the coordinating module. Hence, the latter have a longer lifetime. Clarity of modeling is gained at the expense of losing growth. Tie reproduction proceeds in linear preferential attachment as future degree is proportional to present degree. Triadic closure and homophily go hand in hand at the lower level to form cohesive, internally similar, but externally different modules. Identity of the whole is maintained while preserving diversity of the parts. The network on the right is the historical, cumulative, and retrospective outcome of identity's path as well as a blueprint that gives identity a future direction. It corresponds to figure 1.5c. When the identity is conscious about its path, it acquires a 4-identity.

The mixture of mechanisms is demonstrated in further studies. We already saw that the collaboration network of the biotech industry is scale-free and has a multifunctional core. The coordinating module in our idealtype is such an open elite that organizes diversity. Powell et al. (2005) find that none of the structural mechanisms they studied is solely or independently responsible for this structure. Homophily, the selection of collaboration partners on the basis of their similarity to previous organizations, is most strongly driven by physical proximity but is always attenuated by a preference to maintain the diversity of ties. If collaboration partners similar to previous ones are selected, organizations make sure that they are still different in some way like age or size. Homophily is systematically traded off against diversity. On the one hand, diversity is sought to generate novelty and change. On the other hand, trends cascading through the network are controlled by reproducing previous ties. Preferential attachment is instrumental in forming the cohesive core that coordinates the whole network, but it is hard to distinguish from multiconnectivity, the formation of structurally cohesive networks with large $k$-component values, because the core is also where the hubs are. Most importantly, the topology of the biotech industry is the result of the industry being on a path contingent on events generated in a positive feedback loop of finance and R&D. Due to autocatalysis, core is synonymous to path. Financial support fuels expensive research, progress attracts venture capital which fuels further research. This trajectory is formed by various mechanisms acting in concert. In the process, the collaboration network becomes increasingly cohesive but maintains a multifunctional core, just as if it was the result of fractal distinctions (cf. D. R. White et al., 2004).

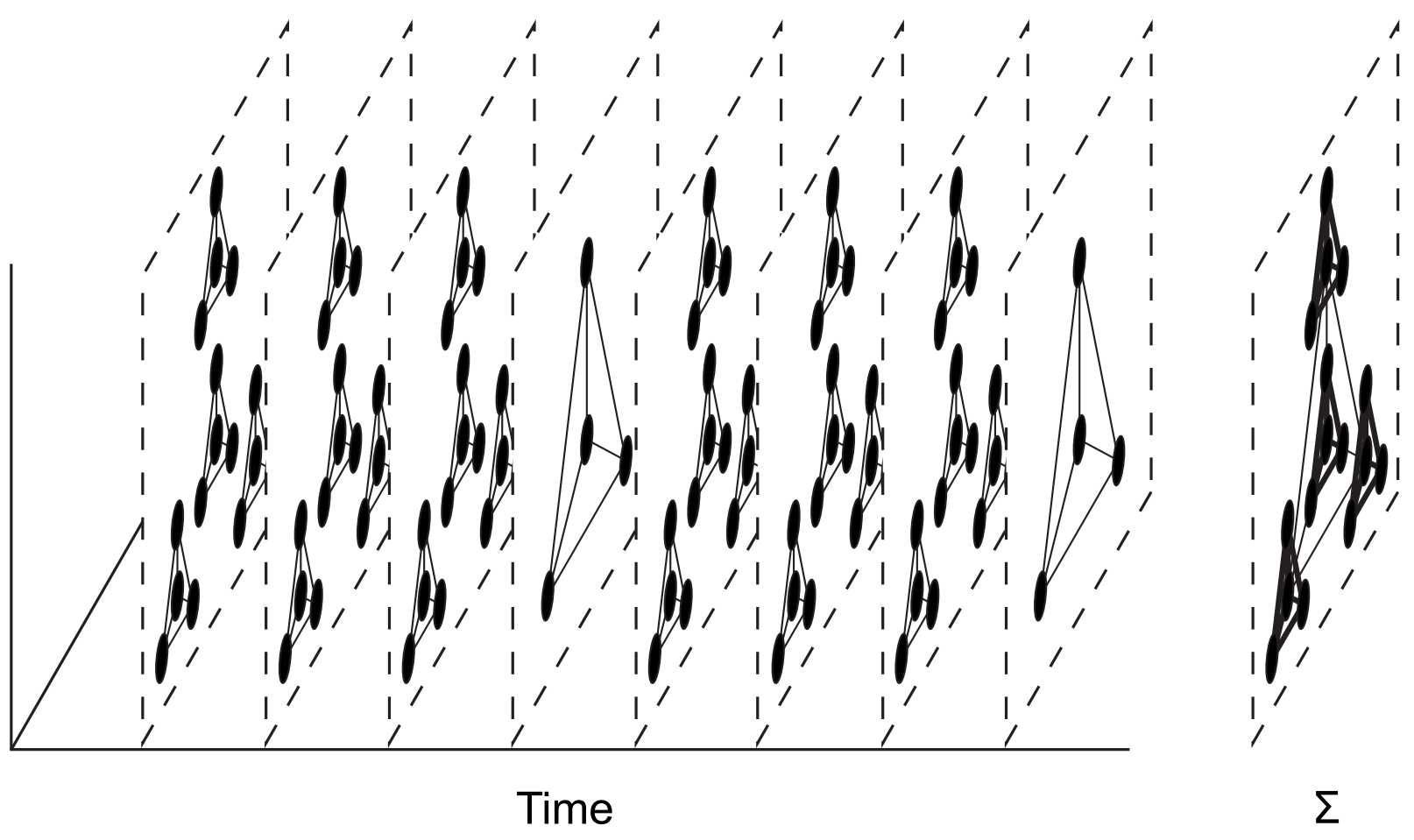

**Figure 1.6.: Idealtype of council discipline**
Council disciplines are structural meta mechanisms that give direction to identities and organize diversity in hierarchically modular reproduction. The network on the right, identical to figure 1.5c, is the historical outcome of, and blueprint for, reproduction. Network nodes are positions and are coupled according to prestige. The four lower-level modules (figure 1.5a) are culturally distinct and are coordinated by a higher-level module (figure 1.5b) that resembles a multifunctional core that consists of positions with high prestige (degree). Low levels of organization are associated with high frequency (short lifetime), i.e., peripheral positions have higher rates of fluctuation than low-frequency (long-lifetime) core positions. The process depicts an identity in the third sense as path. When the identity becomes conscious about itself, it observes the meaning structure on the right. An easy example of council discipline is an organization that persists in homeostasis as positions are differently filled. Other examples are co-citation and co-word processes.

Kossinets and Watts (2009) discuss evidence that homophily is itself a cumulative advantage mechanism. They analyze the email exchange of 30,000 university students, faculty, and staff, enriched by attribute data, and demonstrate that proximate persons are culturally more similar than persons at large network distance. But if transactions are determined by who is available and available persons tend to be similar, then the question comes up if homophily can be chosen or is structurally induced. Persons do hook up with distant similars, but nevertheless similarity and proximity are strongly correlated. Kossinets and Watts propose that the preferred establishment of ties to close alters, the 'closest' example of which is triadic closure, and homophily form a self-enforcing feedback loop that causes network domains to increasingly cohere both structurally and ideationally. Such co-evolutionary densification of network and domain is confirmed for a biologist community of 15,000 authors studied over a period of eight years. Roth and Cointet (2010) operationalize the research domain as a hybrid network of co-authorship and word usage ties. Collaborative cohesion is shown to be a main driver but the resulting triads are also semantically cohesive. "Scientists favor interaction with scientists working on similar topics, but not too much, granting a bit of diversity" (p. 26). As soon as co-authorships have been formed, authors also become more similar semantically. In autocatalysis, similarity breeds connectivity and proximity creates similarity (Carley, 1986).

Both studies demonstrate that the small-world model (Watts & Strogatz, 1998), even though it is purely structural, does harbor meaning. The more proximal network nodes are, the more similar they are culturally. Random edges connect subdomains potentially distant in terms of transactions and meaning. If a 4-identity is one that is conscious about itself being a story set on a trajectory, then cohesion is the phenomenological side of collective consciousness (Bearman, 1993; Bearman & Stovel, 2000). Leskovec, Kleinberg, and Faloutsos (2005) have introduced a metric that quantifies the extent to which a 3-identity is bound to become a cohesive network. It is demonstrated for growing co-authorship and direct citation networks (*arXiv* dataset) that, when the number of edges per year is plotted against the number of nodes per year, the exponent $\beta$ (equation A.5) of the scaling relationship for all years is indicative of densification dynamics. From percolation theory we know that a giant component emerges at the critical point of one edge per node on average. If $\beta = 1$, the average degree is stationary, but if $\beta > 1$, the average degree increases with time and the network becomes denser as it grows in terms of nodes. If a network is not in the ordered regime yet, it will inevitably percolate sooner or later and enter the frozen regime. Leskovec et al. also show that densifying networks become more compact. Besides the increase of the average clustering coefficient the effective network diameter decreases.

Bettencourt, Kaiser, and Kaur (2009) have applied this framework to Kuhn's phase model of scientific change. Research domains are expected to undergo a topological phase transition as they mature from a gas-like collection of isolated teams to a cohesive collaboration network coordinated by an invisible college. The authors study eight fields differing in size and temporal patterns of development and show that all but one exhibit superlinear densification dynamics. The field that does not densify is the only one considered by the scientific mainstream to be "pathological." In another work, Bettencourt, Kaiser, Kaur, Castillo-Chávez, and Wojick (2008) demonstrate that the dynamics that produce Lotka's Law can also be described by an exponent. In a scaling relationship of the number of new publications vs. the number of new authors, the exponent tells if there are decreasing or increasing returns of author productivity as the domain grows. Superlinear scaling indicates synergies from collective knowledge production.

Groups or organizations will increasingly find their own identity if they strengthen collaboration. If densification is superlinear, the socio-cultural condensation of network domains exhibits a cumulative advantage (Leskovec et al., 2005; Kossinets & Watts, 2009; Bettencourt et al., 2009; Roth & Cointet, 2010). As discussed earlier, scaling laws of impact reveal that publications in higher volumes of knowledge production are more highly cited on average if the works find footing in the community. As they do, even the fraction of uncited publications will drop (Katz, 1999, 2000; Van Raan, 2006, 2008). Similarly, as market schedules, scaling laws offer guidance what revenue to expect for which production volume (H. C. White, 1981; Leifer, 1985).

What does it mean that co-authorship networks constructed from publications in the *arXiv* are not scale-free (Newman, 2001b), preferential attachment is sublinear (Newman, 2001a), but densification is superlinear (Leskovec et al., 2005)? We have already stated the possibility that the contradiction between the form of the degree distribution and the attachment dynamics may be due to measurement error. The Matthew Effect is always at work, feedback is always positive, when $\beta_{\text{attachment}} > 0$. For sublinear preferential attachment with $\beta_{\text{attachment}} = 0.8$, an author with two co-authors is predicted to also have two in the future, but one with 20 is expected to have only eleven co-authors in the future, not 20 as in linear preferential attachment.[8] Sublinear preferential attachment still equals a cumulative advantage of size. Regarding densification, superlinear scaling indicates that there is an increasing return of the average degree to size. If the densification exponent was $\beta_{\text{densification}} = 0.8$ and returns were decreasing, the Matthew Effect would still be at work because the total number of edges would still increase with the total number of nodes. But the average number of edges per node would decrease and the network would disintegrate. In other words, network feedback is only positive when densification is superlinear. Then it resembles an autocatalytic process as in biochemistry: collaboration increases creativity, creativity increases the ability to reproduce, more collaboration results, and so on (cf. Padgett, 2012a, p. 39).

### Citation Analyses

In networks of direct citation, preferential attachment has dominantly been found to be linear (Jeong et al., 2003; Redner, 2005; Milojević, 2010; Eom & Fortunato, 2011). This is in good agreement with the fact that citation distributions are one of the most plausible power laws to be found in the sociology of science (Price, 1986 [1965]; Seglen, 1992; Redner, 1998; Milojević, 2010; Albarrán & Ruiz-Castillo, 2011; Albarrán et al.,

[8] Assuming a unit normalization constant.

2011). Only Golosovsky and Solomon (2013) find a Matthew exponent increasing from 1 to 1.28 over time in the physics discipline.

The simplicity of the Barabási/Albert model is both a virtue and handicap. Two deficiencies are the first-mover advantage that “older ... vertices increase their connectivity at the expense of the younger” (Barabási & Albert, 1999, p. 511) and that the model only knows ever-rising degrees. This is in contrast to empirical observations especially of citation networks. Price (1986 [1963], pp. 70) argued that specialties coordinated by invisible colleges form because knowledge grows exponentially and selection and specialization is the only option to control the chaos resulting from information overflow. In the period Price was studying, cited references in chemistry had a half-life of 15 years, i.e., a reference 15 years older than another one is half as likely to get cited. Half-life is a citing-side measure of scientific memory. Recent references are more likely to be remembered than old ones. In 1998, memory was even shorter with a half-life of about seven years for the post-war period (Van Raan, 2000). Milojević (2012) has demonstrated for five disciplines, including one from the social sciences, that productive researchers and especially the most collaborative ones, both senior and junior, have the shortest reference memories and are, therefore, the most progressive ones. Contrary to the Barabási/Albert model, forgetting old publications, then, is a way of avoiding lock-in to a frozen path (cf. Roth & Cointet, 2010).

The invalidity of the first mover advantage has also been demonstrated for the other side of the citation interface, the cited side. Some publications are known to be “sleeping beauties,” i.e., they acquire most of their citations only very late in their life (van Raan, 2004). A similar phenomenon is known for the World Wide Web where new pages are able to outreproduce and even displace old ones. Huberman and Adamic (1999) have proposed to explain scale-free networks not through preferential attachment but through multiplicative growth combined with different growth rates. Bianconi and Barabási (2001) instead generalize preferential attachment and introduce a fitness term that allows nodes to have different growth rates and, therefore, new nodes to outreproduce old ones. The way *fitness* plays out corresponds to Darwin’s idea of natural selection in the sense that fitter nodes with higher growth rates can attract a share of ties that increases over time. The Bianconi/Barabási model represents a fit-get-richer mechanism.

Another fix to the Barabási/Albert model became necessary because cumulative advantage only sets in after a critical mass of links has been reached. As Price (1976) had proposed ad hoc, Dorogovtsev, Mendes, and Samukhin (2000) introduce an initial attractiveness $k_0$ that governs the probability for young nodes to get new links. In the co-authorship network of nanoscience, $k_0 = 20$ (Milojević, 2010), and in the direct citation network of physics papers, $k_0 = 7$ (Eom & Fortunato, 2011). Initial attractiveness is related to the fact that recognition is generally higher in the time following publication and decays after a peak number of years which differs for disciplines (J. Wang, 2013). In physics, the peak is about three years after publication for comparable works (D. Wang, Song, & Barabási, 2013). Only few large bursts of recognition – the sleeping beauties – occur later than ten years after publication (Eom & Fortunato, 2011).

D. Wang et al. (2013) have integrated all these growth mechanisms in a model of citation dynamics and show that all publications in physics have a universal citation history. The trick is to model the continuously growing cumulative number of citations as a logistic function of time. Sleeping beauties or rapidly recognized papers differ according to immediacy, bursts are explained by short longevity, and the rise to prominence with different growth rates is governed by preferential attachment and fitness. Using five citation years for training the model, it is capable of predicting the correct citation range for 93.5% of all papers published in 1960 in the *Physical Review* family of journals 25 years into the future. J. Wang (2013) has shown that the number of years needed to predict the citation ranking of papers 31 years after their publication is different according to discipline. To achieve a correlation of 0.8, three years are sufficient for physics, but five are needed for the social sciences.

Recall that a superlinear cumulative advantage leads to the emergence of superhubs. Golosovsky and Solomon (2013) find superlinear attachment in a physics citation network and show that, when publications become superhubs, they become immortal. Mathematically, if a publication exceeds a critical number of citations – the authors have empirically determined a threshold of about 600 citations –, its lifetime becomes infinite. A discipline-specific tipping point above which publications and the concepts they symbolize are not forgotten anymore and are collectively archived is reasonable (Van Raan, 2000). Merton (1968 [1949], pp. 35) has described such a phenomenon as "obliteration by incorporation." Social facts that are considered to be consensual by a scientific community are not necessarily getting cited anymore. They are forgotten (obliterated) in the sense that they are commonsense (incorporated into collective memory). Einstein's paper "Die Feldgleichungen der Gravitation" is a prime example. The general-relativity paradigm of astronomy is cited only 611 times.[9] Like D. Wang et al.'s (2013) discovery of universal logistic citation histories, obliteration by incorporation confirms Price's Law that eventually all growth must come to an end.

### Modeling Disciplines

A 2-identity is a coupling of distinctions, an outcome of collective control. 3-identities are paths through socio-cultural space-time and 4-identities are such paths made conscious, self-observations of careers and story sets. As we have seen, many attachment mechanisms govern how 3-identities embed into context, namely preferential attachment in all its variants, reproduction of past ties, cohesion creation like triadic closure, and homophily. We adopt the term *discipline* to mean the process that brings together some of these mechanisms for the purpose of ordering chaos (H. C. White, 2008, ch. 3). Disciplines are meta mechanisms of collective control that bring together attachment mechanisms to result in 3-identities. They are not identities but blueprints for 2-identities to evolve along some path. Densification is the sure sign that chaos is getting disciplined (p. 63–9).

We have already encountered the three species of discipline White proposes. Each one relies on a different valuation to induce order. In a 3-identity, one discipline likely dominates but in principle all three species are more or less active. The idealtype in figure 1.5 is a *council* discipline (H. C. White, 2008, p. 86–95). Councils organize diversity, like open elites (D. R. White et al., 2004), also known as "heterarchies" (Stark, 2011, p. 23–7). They couple 1-identities according to prestige, in the figure measurable by degree. Peripheral nodes, the apprentices, have lower prestige than the sorcerers in the coordinating module. Homophily varies at scale, it differs at levels, such that diversity of the parts is maintained while the network is a unit as a whole. But we also think of co-citation or co-word networks as councils. Paradigmatic concepts are located in the network core and give meaning via prestige to peripheral social facts (S. Fuchs, 2001, p. 287–92). Modules as structurally equivalent sets are best found by role models (Reichardt & White, 2007).

[9] Checked using *Google Scholar* on March 31st, 2015. We are not implying that this supports Golosovsky and Solomon threshold because there are papers with many more citations.

*Arena* discipline (H. C. White, 2008, p. 95–104) creates boundaries through inside/outside distinctions, like in Abbott's model (figure 1.4). It works towards creating modules (1-blocks in the diagonal of a blockmodel and 0-blocks otherwise), such as scientific specialties, subfields, fields, and disciplines, identifiable through role models. Each arena can then be screened using cohesion models. The valuation order is purity. Homophily and cohesion feature strong as mechanisms but specialization goes along with dependence, as in organic solidarity. The permeability of boundary can be described by cohesion models (D. R. White & Harary, 2001).

*Interface* discipline (H. C. White, 2008, p. 80–6) is associated with directed ties and channels flows like in a direct citation network or a production market. Interfaces operate around the value of quality – of knowledge or industrial products. Networks are often tree-like so cohesion mechanisms are least expressed. It is strongly associated with creating hierarchy, i.e, with preferential attachment (Nakano & White, 2006).

The reproduction of past ties is actually not an attachment mechanism. It just means that relations persist. Disciplines make networks durable through feedback. The blueprint (meaning structure) that formats transactions emerges continuously from transactions. It is recognized ex post and gives direction ex ante. The Matthew Effect is the most apparent feedback mechanism. It is plain biased randomness (Barabási & Albert, 1999). Triadic closure as the ultimate cohesion mechanism requires second-level observations of node transactions up to a distance of two. The observed meaning structure then is the basis for attachment strategies. Larger cycles require farther observations (Burt, 1992). Homophily requires observations of not only what goes on in a network but how stories are spun. Connecting to similars requires observing collective distinctions which then influence transactions (Powell et al., 2005).

S. Fuchs (2001, ch. 5) associates low levels of society with high frequency, short lifetime, inclusion, and concreteness. Conversely, high levels are characterized by low frequency, long lifetime, exclusion, and abstractness. Short-lived encounters end but groups persist. When groups end, organizations are not questioned (p. 199–203). Organizations are also momentary like encounters, just on larger time scales. "Cellular autocatalysis of cells and biographies usually crystallizes on a slower time frame than does production autocatalysis of products and skills" (Padgett & Powell, 2012a, p. 11). Distinctions are made at all scales, from the top split in the trees of figure 1.1 down the hierarchy. But at higher levels, communication becomes increasingly and massively parallel (S. Fuchs, 2001, p. 201). Therefore, distinctions are fractal and permeable boundaries emerge.

Fuchs' structural model is captured in figure 1.6 and shows that disciplines couple social facts with varying lifetimes. This corresponds to Elias's (1991 [1939], p. 47–54) observation that social formations are simultaneously very solid and liquid. On a day-by-day basis, individuals are free to roam on paths they chose, but at the end of the day, most will have chosen their path along the cultural mainstreams provided by society or the rules of the organization. Similarly in Kuhn's epistemology – other than in Popper's –, scientists are free to solve puzzles differently, but contingent on common rules of practice. This is nothing else but saying that in the short run, actors create relations, but in the long run, relations create actors. It is all a matter of socio-cultural scale (Padgett & Powell, 2012a).

The next modeling step is to incorporate *homeostasis*, the phenomenon that a meaning structure is stable while 1-identities are exchanged. This is necessary because parts of a system can die, fail, or leave. Disciplines organize the reproduction of positions, not concrete social facts. A node as position in figure 1.6 can be taken by multiple social facts at the same time or different facts over time. For example, if the blueprint network of figure 1.6 represents the ideational structure of the Sociology of Scientific Knowledge, the nodes in the network core symbolize the constructivist epistemology. Then, the four lower-level modules correspond to the social facts that govern the four camps described in section 1.1.1. Camps and sympathy groups come and go all the time without altering the paradigm. Only occasionally, the identity of the whole clan changes, as was the case when the sociology of science left the institutionalist program behind. In general, exchangeable core facts are contextualized by short-lived peripheral facts. Because disciplines are structural blueprints, pure network, they endure all change. Councils will always be hierarchically modular, arenas will always build permeable boundaries, and interfaces will channel production flows.

### Summary

Disciplines are structural blueprints of relations among positions that put order into chaotic socio-cultural life and give identities direction. There are three basic types for the purposes of the organization of diversity (council), boundary creation (arena), and production flow (interface). Disciplines are meta mechanisms that bring together simultaneously operating tie formation rules like preferential attachment, homophily, and cohesion. Preferential attachment is how social systems attain criticality. The Barabási/Albert model operationalizes the rich-get-richer aspect of the Matthew Effect and has been extended by Darwinian fitness. Phase transitions acquire meaning as points of reference for the degree of longevity and homeostasis.

Multiple levels are disciplined simultaneously. In hierarchically modular structures reproduced by council discipline, low levels are associated with high frequency, short lifetime, inclusion, and concreteness, high levels with low frequency, long lifetime, exclusion, and abstractness. Multifunctional network cores maintain the identity as a whole while preserving diversity of the parts. Disciplines give rise to universality of process and predictability of outcome. A scaling law of social network densification quantifies the extent to which an identity is becoming conscious about itself.

### 1.3.3. Styles Mate to Change

Disciplines can explain stability but not change. They are inert and merely perpetuate an existing structural order. Disciplines create "the impression of moving 'in a given direction' (for example, toward the moon or along an academic career), even if the outcome of the projected trajectory is still unresolved in absolute terms" (Mische & White, 1998, p. 712). Disciplines do not care if they are punctuated because they are just positional protocols. Perturbations only affect identities. The extent to which the core of an identity's story set is affected determines if the identity gains a new meaning. A *punctuation* is an interruption of a reproduction of a story set that necessitates a reaction because the event is felt to matter. Punctuations "shake up configurations of acceptability and availability, perhaps opening up new spaces (drawn from crosscutting [network domains]) as 'running grounds' for conversations, while others are closed off, and transitional possibilities are rearranged" (Mische & White, 1998, p. 711). Punctuations introduce new stories or reshuffle an existing story set. H. C. White (1995) calls such punctuations *Bayesian forks* because the punctuating events are contingencies that open up the future for a new direction, potentially terminating a narrative.

According to Abbott (2001b), what makes a punctuation a punctuation "is the passage of sufficient time 'on the new course' such that it becomes clear that direction has indeed been changed" (p. 245). Change of direction is change of identity and can only be decided in retrospective, or in other words, the meaning of an event is conditional on its position in a sequence of events. The stages along chains of events are "transitions" which can be, but need not be, "radical shifts" (p. 243). Abbott sees an "implicit image of the social sciences as comprising a number of programmed inertial trajectories, with strong constraints on their number and desirability" (p. 248). A path (trajectory) results as a meaning structure unfolds in time. In this trace of discipline, concrete facts can change positions. These are cross-sectional maneuvers and the better a fact reproduces, the likelier it gets to be in the core. Punctuations act longitudinally and can take the form of new facts entering the positional system or of events exerting some kind of influence. Figure 1.5 shows one subdomain where the nodes of the coordinating module, the sorcerers, constitute the core and the peripheral nodes, the apprentices, are in their basin of attraction. The space that contains all the social facts that can potentially be selected is called a *public*. It is a highly sparse meaning structure of overlapping domains in which switchings occur and from which punctuations originate.

#### Structure and Memory

Dynamics of change in network domains have been studied in a number of computer simulations. The duality of network and domain requires the simultaneous modeling of social agents and social facts. These facts are concept symbols, like stories, words, or cited references. Our point of departure is the social-autocatalysis model of Padgett et al. (2012), the only one that is capable of explaining the birth of multiple network domains needed for meaning to emerge. Recall that multiple overlapping domains are the result of free riding enabled by altruistic agent behavior and that the complexity barrier of just four positions per domain can be broken if agents are allowed to only interact with agents neighboring on a grid, a so-called Moore neighborhood. The grid mimics social networks and has the effect of memory in the sense that agents have "learned" to only search their immediate environment for successful transactions (Padgett, 2012b, p. 96). *Memory* is the function of a meaning structure to recall successful transactions. It is the condition for autocatalysis/reproduction.

Powerful explanations have been made using static grids as social networks. Axelrod (1997) equips grid positions with a set of cultural features, each of which can have one from a set of traits. A random pair of neighboring nodes is chosen and, with a probability equal to their cultural similarity, they transact. If they do, they align a random feature in which they differ. Despite these local dynamics of cultural convergence, the grid polarizes into global cultures with perfect boundaries. Inside a cultural arena, all agents are similar, but all arenas have completely different sets of traits. The problem with such grids is not only that the average distance is long. The bigger problem is that social networks are kept constant and cannot even evolve short paths or a fractal dimension over time because meaning structures are only allowed to alter dynamics on networks, not the networks themselves.

Instead of letting positions interact on a grid, Padgett (2012b, p. 93–100) introduces memory and learning into the autocatalysis model. This allows agents to enter a transaction with any other agent based solely on memorizing which agent is structurally equivalent in terms of selections, no matter what the social distance is. Learning occurs through communications which involve receipts to originators of a transaction that a relational tie was successfully initiated.

This mechanism is implemented in Carley (1991) "constructural model." Initially, each agent knows a subset of facts known in the public. Agents know the fact overlap with other agents – they know who they can transact with – and, based on this knowledge, a randomly drawn agent learns a fact from the agent he is structurally most equivalent with. All proto-boundaries that may have existed vanish over time, provided that the capacity to learn is unlimited and all agents indirectly share at least one fact. Two groups in an organization with hardly any knowledge overlap will eventually merge and form a fully connected clique where each tie has the same probability of occurrence. All facts diffuse like a piece of ice in warm water. This system tends towards equilibrium because it is kept constant as a whole, there is no throughput of information or personnel.

Agents continuously learn and what they learn affects future behavior. Social structure emerges from cultural similarity memorized in meaning structure. The "past is encoded into the present in patterns of connection that we call structure" (Abbott, 2001b, p. 257). This is nothing else but the attachment mechanism of homophily. But it has now gained a temporal dimension. In arena discipline, not only similarity breeds connectivity, but homophily breeds further homophily. Homophily is a positive feedback mechanism like Kossinets and Watts (2009) suggest.

When a new identity embeds into an existing network domain, the new identity must be taught how to talk and who to talk to. The teaching material is saved in the meaning structure. While learning is cross-sectional, teaching is longitudinal or genealogical: "Cell inheritance in the social context means passing production skills and relational protocols down from experienced cells to less experienced cells" (Padgett & Powell, 2012a, p. 10). Like target reproduction, teaching (biographical autocatalysis) is altruistic and indirectly selfish, it is the long-term solution to the short-term threat of cell death or personnel turnover. Identities learn from the past but teach for the future (Padgett, 2012b, p. 100–8). Like cross-sectional learning, longitudinal teaching does not require complex mental cognition. Because teaching imprints domain knowledge into network structure, complex overlapping formations (families) can emerge. This is the condition for group splits (Palla et al., 2007) and the source of hierarchical modularity (Durkheim, 1982 [1895], p. 111–2).

When social formations are modeled as open, equilibrium makes room for growth, homeostasis, and adaption. In the constructural model, when just one person is added to one of two overlapping groups, the time to equilibrium (until every agent has learned all facts and group boundaries have dissolved) is prolonged, i.e., the organization lives longer. The life of simpler cultures is less prolonged because there is less to teach to the newcomer. Carley's simulations lend support to Padgett's model that short-term costs have long-term benefits. When a new piece of information finds into a group, discovered inside or let in from outside, the group must adapt its protocols but increases its cultural distinctiveness and enhances its lifetime. Simultaneously, new facts increase intra-group diversity because, in reality, knowledge is not immediately shared and some will never be. The simpler a group's culture and the less bounded it is, the more punctuations are disruptive. The model suggests that things that serve to disrupt the natural tendency to stability "are critical to the perpetual endurance of groups simply *because* they disrupt this tendency to stability" (Carley, 1991, p. 351).

Open systems can self-organize in a way that they are not segregated into regions, like in the Axelrod model, but globally connected. In the Bak/Sneppen model of evolution (Bak & Sneppen, 1993), species arranged on a grid initially have random fitness. The least fit species mutates and with it all its neighbors. As this process continues, adaptions of all spatial and temporal length scales take place. This has also been demonstrated for dynamic network substrates (Garlaschelli, Capocci, & Caldarelli, 2007). Cascades create not only rapid and mass forgetting (extinctions) as the basis for subsequent learning, but also rapid and mass teaching (speciation). Forgetting is the past-side of a Bayesian fork. The flushing of memory resets reproductive flows and creates possibilities to "get action." New cores can now emerge and give a new direction (H. C. White, 2008, ch. 7, especially p. 304). But the tendency to normalize punctuations and resist change is always present (S. Fuchs, 2001, p. 266–9).

Evolutionary models like this have been proven successful to also model the history of science. They show the difficulty of new ideas to rise to prominence in systems with strong inertia. Bornholdt, Jensen, and Sneppen (2011) model opinion formation on a grid and show that the forgetting of old ideas is crucial for the emergence of new scientific paradigms. But paradigms are just the most dominant states and stay in memory longest. Smaller symbols or rules of practice exist at all sizes and lifetimes. Sterman and Wittenberg (1999) discuss a very elaborate model of scientific revolutions, especially regarding social dynamics and path dependence. A multitude of positive feedback loops operates to sustain normal science:

> Rising confidence and successful puzzle-solving boost practitioner confidence, leading to more focused and successful effort, articulation and improvement of theory and technique, and still greater success in puzzle solving, further boosting confidence and attracting still more members. Rising confidence, skill, and familiarity with the paradigm increasingly condition practitioner perceptions and expectations, suppressing the recognition of anomalies; a low level of anomalies further increases practitioners' confidence in and commitment to the theory. (p. 330)

In the model, these cumulative advantages cause competing paradigms and their associated network domains to go extinct. In other words, only one core is allowed. But the same feedback dynamics that create path dependence also accelerate crisis. As anomalies rise, confidence in the paradigm begins to fall, scientists turn their back to puzzle solving and discover new anomalies. It follows that ideas can be ahead of their time, like Pareto's or Moreno's. To achieve initial attractiveness, they need to be articulated at a time when the old core, paradigm, or attractor shows signs of destabilization. But once a new idea achieves critical mass, it will tip into existence and pull in related work "with citation avalanches on all scales – from small cascades reflecting quasi-continuous scientific progress all the way up to scientific revolutions" (Mazloumian, Eom, Helbing, Lozano, & Fortunato, 2011, pp. 4).

### Randomness and Optimization

Learning naturally involves a cumulative advantage because, once an agent has been registered as compatible, it is more likely to be a transaction partner next time. Reali and Griffiths (2010) have shown that Zipf's Law emerges when learning is iterated. Again, no mental cognition is required to explain the emergence of social life, just explicit memory that reduces search efforts, like Zipf (2012 [1949]) had proposed. But how does this go together with the homophily mechanism?

A recent model has revived the debate of Simon and Mandelbrot of whether power laws stem from biased randomness or an optimization process (Barabási, 2012a). Papadopoulos, Kitsak, Serrano, Boguñá, and Krioukov (2012) have presented a version of preferential attachment that incorporates homophily. Nodes attach not simply to highly connected nodes but they trade connectivity off against similarity. Cultural similarity is interpreted geometrically as the angular distance on a circle centered in the network core. The model thus mimics the contextualization of the core from different directions. Empirically, it turns out that angular coordinates predict country-level attachment in a email network well. Because networks have historically emerged inseparable from meaning and identities are addressed that have been addressed before, homophily complements preferential attachment as *how* social facts constrain transactions. Clearly, the Matthew Effect was never meant to be purely structural, so Papadopoulos et al. have actually restored the meaning dimension intended by Merton (1973 [1968]).

Before Travers and Milgram's (1969) empirical treatment of the small-world problem, Pool and Kochen (1978 [about 1958]) had treated it theoretically. If social life was purely random and clustering was totally absent, they figured, two randomly picked individuals in a society would very probably not know each other but still be connected by only two intermediaries.[10] They were aware of the fact of clustering but did not manage to come up with a satisfactory model because, at the time, not much was known about large-scale structure. Pool and Kochen did set the stage for research to come by considering that search in social networks is influenced by cohesion (triads and other $k$-components), different network domains (which they call "circles of acquaintances"), and skewed degree distributions (cf. Moreno & Jennings, 1938; Rapoport & Horvath, 1961; H. C. White, 1970a). So what do we know about search today and how does it relate to randomness and optimization?

Search is both path dependent and contingent, i.e., events have both a past and a future. Researchers, e.g., re*search* for solutions of scientific puzzles. They do so along attractive narratives called paradigms. Stark (2011) formulates the risk when past success breeds further success and grounds future course:

> Each evening during their hunting season, the Naskapi Indians of the Labrador peninsula determined whether they would look for game on the next day's hunt by holding a caribou shoulder bone over the fire. Examining the smoke deposits on the caribou bone, a shaman would read out, for the hunting party, the points of orientation of the next day's search. In this way, the Naskapi introduced a randomizing element to confound a short-term rationality that would have concluded that the one best way to find game would be to look again tomorrow where they had found game today. By following the divergent daily maps of smoke on the caribou bone, they avoided locking in to early successes that, while taking them to game in the short run, would have depleted in the long run the caribou stock in that quadrant and reduced the likelihood of successful hunting. (pp. 175)

Randomness not only prevents exploitation. Exploration can well be a source of control. In fact, a group identity emerges simply when the critical point of one edge per node on average is exceeded (Erdős & Rényi, 1959). When simple memory is possible, talk can emerge from pure randomness. In human problem solving, reports or clues from trial and error consolidate into conventions that are memorized. As rules of thumb, they supply story sets as guidance (H. C. White, 2008, pp. 28).

Simon (1962) provides the famous parable of the watchmakers Tempus and Hora to lay out how further complexity can arise once a first take on randomness has been achieved. Tempus builds a watch from scratch, he assembles 1,000 parts to make a whole. Hora builds 100 modules of ten parts each, assembles ten of these modules into meta modules, and finally makes the whole out of ten meta modules. Tempus assembles just one thing in the process while Hora assembles 111 things, but when they get interrupted and return to work, Tempus will have to start from scratch again while Hora can continue assembling modules or even meta modules. In the face of punctuations, it will take Tempus much longer to finish a watch than Hora. Simon proposes that all emergence, natural or social, proceeds step by step through "stable intermediate forms."

[10] They assumed that the individual knows 800 others and society consists of 160 million people.

Solé and Valverde (2004) show that hierarchically modular networks are optimal because they self-organize to the critical point. Proteomic, ecological, technological, and also language networks occupy only a small region in a parameter space. They all favor self-similarity in terms of degree as well as some amount of randomness and modularity.[11] Solé and Valverde reproduce these structures and find that they strike a balance between global connectivity (correlation) and diversity. Once a seed of order has been achieved, higher levels of complexity are inevitable.

An important piece of the solution to the small-world puzzle of global searchability has been provided by Kleinberg (2000). Like in the Watts/Strogatz model, search is modeled on a grid. But differently, search distance is scale-invariant, i.e., the distance bridged by a random edge decays with a fat tail. If search is local and jumps are only minimal like in a random walk on the grid, it will take many steps to leave the neighborhood and find a culturally distant target. On the other hand, if short and long jumps on the grid are just as likely, like in the Watts/Strogatz model, the searcher will not be able to home in on the target because a jump that might have decreased the distance to the target might be followed by another long jump that increases the target distance again. Kleinberg shows that, if the power-law exponent governing the likelihood of a jump of a certain distance is equal to the dimension of the grid, search is optimized to the critical point where few long-distance jumps can lead to the target domain where the target can then be found at short distance.

Scale-free networks enable efficient search for the same reason. Not search distance is self-similar but the networks themselves are (Adamic, Lukose, Puniyani, & Huberman, 2001). Scale-free networks can be thought of as grids with a fractal dimension (Roberson & ben-Avraham, 2006). Hubs that reach out to many nodes in close neighborhoods and some nodes in distant network domains provide much of global connectivity (H. C. White, 1970a). In their problem-solving experiments, Bavelas (1950) and coworkers have observed that networks with short average path lengths like the star are more efficient than stringy networks like the circle. Ferrer i Cancho and Solé (2003) argue that complex systems avoid both architectures. Star networks easily fail when the central nodes are defunct and stringy networks have long average path lengths. Short distance can be achieved by fully connecting a graph which, however, is maximally costly.[12] Simulations show that stringy and star-like networks are the poles in a continuum that results from a trade-off among short distance and cost efficiency (link density). Scale-free networks strike a balance and are, therefore, preferred in network evolution. At small scale, scale-free networks resemble Bavelas's stars.

Ferrer i Cancho and Solé's Mandelbrotian optimization argument seems counter to Simon's (1955) biased-randomness explanation of power laws. But the positions are reconcilable. Optimization is not an alternative mechanism behind scale-free networks but a reason *why* the mechanism of preferential attachment – in concert with homophily – operates to create skewed degree distributions. Sublinear or no preferential attach-ment, under efficiency constraints, generates long average paths. Superlinear attachment creates short paths but the network is not robust. Both extremes are inefficient and hierarchically modular structures prevail. Power laws improve control (Solé & Valverde, 2004). Further reconciliation derives from search. Kleinberg's fractal search is an example of a biased random walk where the search distance slowly decays with a power law tail, called a *Lévy flight* by Mandelbrot (1982, ch. 32). Lévy flights are optimal designs for transport in networks. Biased randomness is an optimization principle (Li et al., 2010). From this perspective, the Matthew Effect enables the self-organization of (socio-cultural) systems to the fractal stability-to-change transition where evolution is further sped up.

[11] Unfortunately, the authors do not include social networks into their analysis, but as we have seen, they are also hierarchical even though more condensed (Newman & Park, 2003).

[12] This is the same logic as Kauffman's (1993, ch. 5): stars are frozen but cliques are chaotic.

### Style

Searching means making a distinction. Local search maps to staying in the home arena, long-distance search maps to transcending the arena, and fractal search maps to occasional selection of culturally distant facts to get action. Small-world networks from Lévy flights are cohesive but fractally distinct. The generative model for *feedback networks* by D. R. White, Kejžar, Tsallis, and White (2006) unifies this insight with preferential attachment. The model generates single-peaked networks from stochastic search and growth. It is based on the intuition that Lévy flights are desired to build alliances (Powell et al., 2005) or avoid incest (D. R. White & Johansen, 2005). Naturally, when a node randomly drawn with a probability proportional to its degree (first parameter) attracts another node at a distance governed by a fractal random walk (second parameter) and with a tendency to be in the core (third parameter), then short distances are more likely to succeed because of a network's limited diameter. Successful short-distance searches cause the network to densify. A core, necessary for reproductive stability, emerges and the average distance decreases. Lévy flights counter this autocatalytic tendency to stability. Search at long distance increases the likelihood of not finding a node. Failed searches cause new nodes to enter the network and the average distance to increase. The new nodes contextualize the core. They are the potential free riders that create overlaps with other network domains and maintain diversity and switching opportunities inside the public (Padgett et al., 2012). They are the necessary sources of innovation and new meaning that balance the path-dependent trajectory caused by autocatalysis. When the network is scale-free, tunable by the first parameter, then learning and teaching is balanced at all length scales.

Cores reduce uncertainty. The stronger the core, the more uncertainty is sensed to be reduced. But there is a continuum. Degrees of uncertainty can be perceived to be constraining or enabling. So far we have treated all social facts equally, like authors, cited references, and words. Now we make a distinction. Social uncertainty or *ambage* is the uncertainty in a meaning structure of agentic social facts. This is "fuzz in the concrete embodiment" of identity (H. C. White, 2008, p. 58). Cultural uncertainty or *ambiguity* is the uncertainty in a meaning structure of facts that cannot make selections. It is high, e.g., when no paradigm gives direction to a research domain. This is "fuzz in the rules of perception and interpretation" (H. C. White, 2008, p. 58). H. C. White (2008,

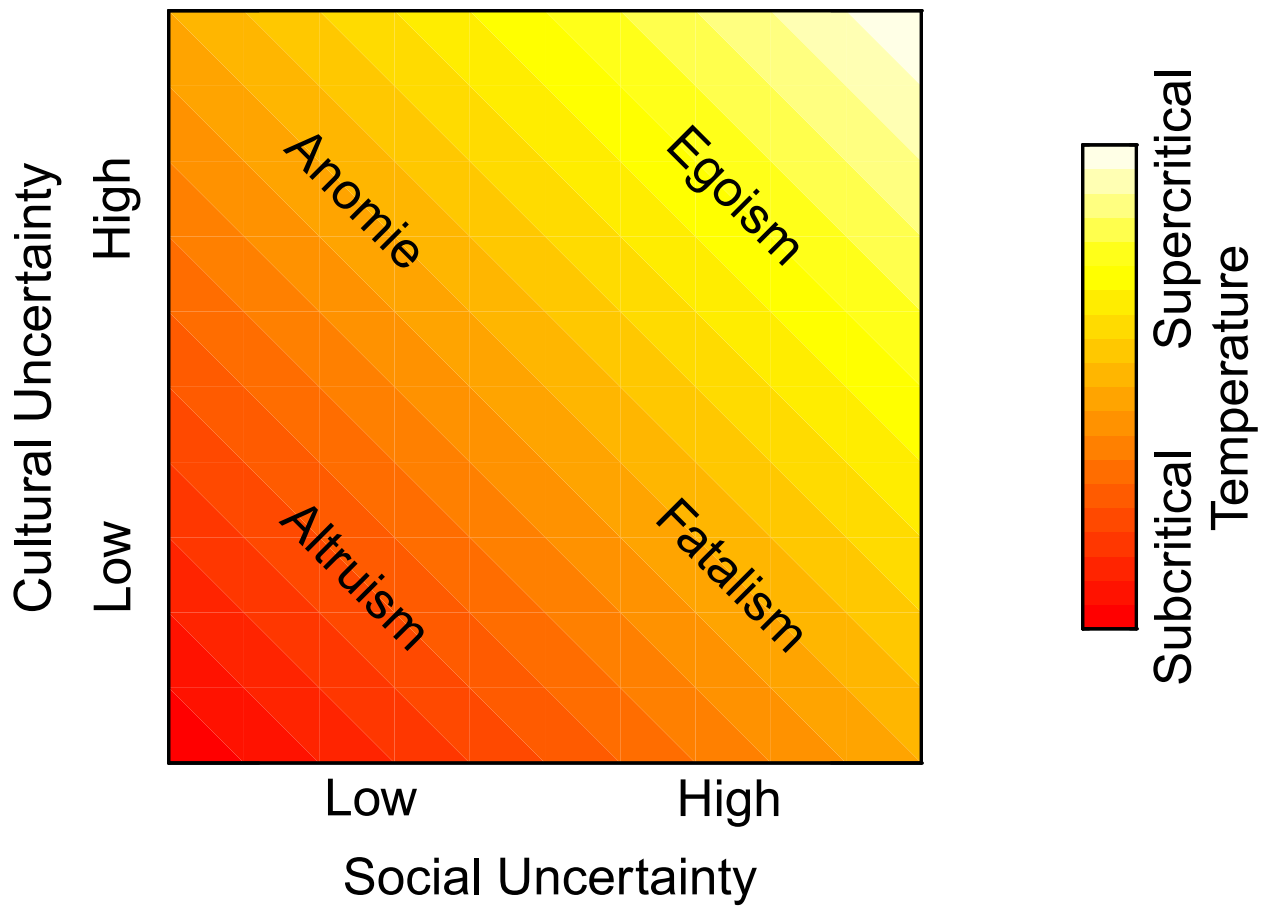

**Figure 1.7.: Trade-off among social and cultural uncertainty**
Socio-cultural life harbors two kinds of uncertainty. Social uncertainty is a measure how much the live social network an identity embeds into is perceived to be constraining or enabling. Cultural uncertainty is a measure how much the social facts (stories and norms) an identity is influenced by are perceived to be constraining or enabling. A certain degree of contingency is modeled on, and proportional to, socio-cultural temperature. Identities in contexts with low social and cultural uncertainty lock into a path – they "freeze." This "altruistic" state of slow motion (low temperature and contingency) is not only the cheapest in terms of social cost but also the most predictable one. Context with high social and cultural uncertainty bear the risk of "melting," i.e., of not finding an observable path through socio-cultural space-time. This "egoistic" state (high temperature and contingency) is most costly and paths are least predictable. The state of critical temperature is optimal in terms of control. In culturally uncertain (certain) contexts, identities decrease (increase) social uncertainty through forming (leaving) cohesive clusters. The critical state is characterized by a self-similar story set. Quadrants in the plane are labeled following Bearman (1991).

p. 70–2, 144–5, 294–6) proposes that control is a systematic trade-off among ambage, ambiguity, and contingency. Socio-cultural complexity requires identities to continuously renew themselves and find "fresh ways to gain control" (p. 296). The work of Guimerà et al. (2005), reported in section 1.3.2, is well suited to lay out what such a trade-off can look like when ambiguity is neglected. In combination, parameters $p$ and $q$ define the uncertainty in social networks. When $p$ and $q$ are high, ambage is low because teams are strongly autocatalytic or reproduced. The social future is relatively determined, paths are strongly Markovian, and there is a strong immunity to punctuations, i.e., contingency is weak. When $p$ and $q$ are low, ambage is high because networks strongly renew themselves. The future is relatively open and events can be strongly contingent. The two parameters act like the search (second) parameter in the feedback network model.

Figure 1.7 is an attempt at modeling the trade-off. The two types of uncertainty form the axes. Contingency is modeled on socio-cultural *temperature*, as the extent to which the state of percolation allows punctuating facts to take a core position. We discuss this model in regard to a *fifth sense* of identity (5-identity) called style (H. C. White, 2008, p. 18, ch. 4). A *style* is an identity's coherent way of embedding into updated context. Styles are more than paths and reproduction. Commonsense is always present, but styles are sensibilities that add an "interpretive tone" (p. 112) to autocatalysis. They adapt directions to unforeseen events. Styles are activated by punctuations, i.e., they are involved in deciding the path taken at a Bayesian fork. They are patterns of making distinctions which social facts from the public are inside or outside. Adaptions to contingencies by way of fractal distinctions are the reason why the path of an identity is predictable only to some extent. To observers, styles are narratives – networks of events that are "attempts to describe causality" (Godart & White, 2010, p. 579). A career, e.g., is a succession of transitions among multiple network domains. Overall, a certain direction is maintained, but various events at various distances have an impact and alter the path. When identities are conscious about where they are heading, they can purposely employ narratives to mobilize for social action (Kim & Bearman, 1997). The longer story sets have gone unquestioned, the more they are institutionalized and spelled out in rhetorics. An identity then acquires "a coherent institutional habitus" (S. Fuchs, 2001, p. 247; see how meanings travel from switchings to rhetorics in Godart & White, 2010, fig. 5; cf. Fontdevila et al., 2011).

Styles tune ambage and ambiguity. Analogies of these uncertainties are discussed in Durkheim's analysis of religion and suicide. In *Suicide* (1951 [1897]), Durkheim considers the fact that, more than a century ago, suicide was more common among protestants than catholics or jews. His four types of suicide map to different values of social and cultural uncertainty:

Altruistic suicide (social overintegration) → Low structural uncertainty

Egoistic suicide (social underintegration) → High structural uncertainty

Fatalistic suicide (cultural overregulation) → Low cultural uncertainty

Anomic suicide (cultural underregulation) → High cultural uncertainty

Protestantism, Durkheim concluded, provides not enough social integration and, therefore, causes higher levels of egoistical suicide.

A century later, however, suicide rates could not be explained by church affiliation anymore. While catholicism still seemed to provide some "protection," various denominations of protestantism now also were associated with low rates. Pescosolido and Georgianna (1989) show empirically that "participation in networks ..., rather than church membership per se, represent a more subtle but significant influence than is often realized" (p. 43). Survey results indicate that network proxies – the fraction of persons weekly attending services and of persons having friends in the same religion – explain suicide rates much better. The authors hypothesize that the likelihood to commit suicide is a U-shaped function of network density, with socially underintegrated non-affiliated atheists and socially overintegrated "greedy group" cults forming the poles with high suicide rates. Among the religions, catholicism is supposed to mark the point where suicide is least probable. According to our model, social underintegration is problematic when associated with cultural underregulation because an identity that does not embed into a discipline has no source of meaning. Similarly, social overintegration is problematic when associated with cultural overregulation. Bearman (1991) has interpreted Durkheim relationally and provided a network model where these problematic combinations correspond to egoism (high ambiguity and high ambage) and altruism (low ambiguity and low ambage), respectively. In this model, compatible with ours, anomie (high ambiguity and low ambage) and fatalism (low ambiguity and high ambage) are not pathological forms but rather desired states that enable control through an uncertainty trade-off. These labels are used in figure 1.7. The poles in Pescosolido and Georgianna's model conform to altruism and egoism, so their independent density variable maps to temperature whose gradient is parallel to the bottom-left to top-right diagonal in figure 1.7.

Uzzi and Spiro (2005) have found such a U-curve empirically for the success of the Broadway musical industry. The small-world coefficient $sw$ is used as a measure of the connectivity of creative teams (producers, composer, lyricists, etc.). Seasons with too low or too high $sw$ are less successful commercially and in terms of critics' reviews. $sw$ is a compound indicator of the normalized clustering coefficient and normalized average distance, i.e., it subsumes the uncertainty factors density and distance. Then $sw$ resembles socio-cultural temperature and success is maximized near the critical bottom-left to top-right diagonal. Similarly, the citation impact of scientific publications is associated with flexible assembly mechanisms of the teams that produce the publications. Impact is positively associated with the probability $p$ to search co-authors already in the domain, suggesting that expertise is important for success. Impact is, however, negatively associated with the probability $q$ to reproduce old teams, which indicates that diversity is fundamental. This result holds for social psychology, economics, and ecology, but not for astronomy. For the only discipline found to reside far from criticality, correlation of structure and impact is insignificant (Guimerà et al., 2005). Theoretically, such behavior is expected for an identity that operates in isolation, absent external expectations, and solely reproduces itself (H. C. White, 1970b). Clearly, astronomy is the discipline with the least societal relevance of the four. Its information output is finding limited attention.

Astronomy's social structure is like that of a cult in the sense of a high-density group whose "culture is relatively settled, institutionalized, and unwilling to change much" (S. Fuchs, 2001, p. 215). Fleck (1979 [1935]) has attributed identities with such behavior a "tenacity." What we call a paradigm, he calls a thought style, "directed perception, with corresponding mental and objective assimilation of what has been so perceived" (p. 99). Fleck discusses tactics to immunize styles, from ridiculing criticism over concealing contradictions to praising own achievements, and concludes that these are only reasonable for the thought collectives voicing them (p. 40–53). Styles are conscious when punctuated but unconscious when closer to equilibrium. Interpretive inertia can be explained in terms of Kuhnian incommensurability. The disability of research domains to speak the same language is proportional to their distinctiveness. When similarity is low or absent, they cannot understand each other even if they wanted. The sometimes religious character of science is a consequence of the way identities discipline chaos. The systems theoretical perspective is that cores are eminently realist, "they cannot imagine things to be very different from what they are" (S. Fuchs, 2001, p. 34).

Tenacity arises necessarily when both ambage and ambiguity are low. Low ambage means superlinear densification and, for the most part, subcritical socio-cultural temperature. Superlinear densification leads to the emergence and consolidation of a core (Leskovec et al., 2005; Bettencourt et al., 2009) and the stronger the core, the fewer degrees of freedom a social fact in the respective story set has (S. Fuchs, 2001, p. 217). Any superlinear scaling, if maintained too long, puts a complex system at risk of unsustainability (Bettencourt et al., 2007). Low degrees of ambage are only sustainable if coupled to high levels of ambiguity. Low/low cannot be a control strategy because it locks a style into an equilibrium that – on the long run – deprives it of learning and switching opportunities to renew its meaning and eventually freezes it to death, like in Carley's constructural model. But high ambage and high ambiguity is no solution either. We call this the "melting" scenario because high levels of ambage impede coherence which is even weakened by the tendency of the modules or system parts to renew themselves and change direction. Such identities have a hard time maintaining their constitutive ties, there tend to be apprentices without any teachers.

The reason that most facts are quickly forgotten is that styles optimize selection in second-level observation. Identities do not and cannot observe all social facts that float around in the public. "The rate of advance in a science depends on its ignorance" (S. Fuchs, 2001, p. 85). Selection criteria for social facts in science are codes (a social fact is deemed to be true or false), timeliness (with varying thresholds for different levels of meaning), closure (a social facts is culturally close or distant, inside or outside), and prestige (of knowledge producers or products) (p. 84–92). Making a selection is making a distinction. Once making distinctions – an autocatalytic production rule – has been learned and complexity is syncopated through teaching, it begets further complexity in a positive-feedback loop. Evolution speeds up and future searches or selections are increasingly successful (Adamic et al., 2001; Watts, Dodds, & Newman, 2002; Reali & Griffiths, 2010). A power-law size distribution – Lotka's, Bradford's, and Zipf's Law – is the outcome of a style self-tuning to criticality, a "by-product of counting and sheer accumulation" (H. C. White, 2008, p. 148). It functions as a priority scheme that allows identities to pare down attention to few commonsense social facts in meaning structures. The scaling exponents of these distributions have the role of context variables (H. C. White, 1973). When optimization is why the Matthew Effect operates, styles are the identities *who* optimize. The size distribution signals the self-similarity of selections and provides selection rules, at which meaning scale which story set best enables communication or is "allowed" (Abbott, 2001a).

From the perspective of style, subcritical lock-in is tempting because it is cheap to have – exploitation is always easier than exploration. Hot regimes, on the other hand, are quite costly. They involve a lot of teaching to not lose organizational experience. Style, we conclude, is a trade-off among ambage, ambiguity, and contingency. *When nobody knows the path, you better gather in groups; when everybody is on the same path, you better stray.* Styles trade stability, reproduction, experience, long lifetime, and teaching off against change, renewal, diversity, short lifetime, and learning. When they fail to do so, death and suicide are not the wrong metaphors (Padgett & Powell, 2012a).

A vivid example of self-similar style is the identity of a multi-generational Turkish nomad clan, studied by D. R. White and Johansen (2005, p. 272–91). Kinship systems necessarily have to avoid incest to not diminish their reproductive fitness. Freezing (inbreeding) is a real threat to reproduction. Genealogical data for the clan comprises 1,100 persons, five generations, and allows 234 types of blood kinship.[13] It turns out that 192 of these types occur only once or twice but offspring with a father's brother's daughter occurs 32 times. The distribution is a power law. Such self-similarity emerges generally in societies with a high proportion of marriages related by blood (D. R. White & Houseman, 2002, pp. 78). D. R. White and Johansen also test if the nomad clan displays a searchable small-world network and find that the probability to marry a cousin indeed scales with a cousin's genealogical distance. Marriage in the clan, the result proposes, is a Lévy flight and resembles strategic control (optimization) to get unique access to valued resources (p. 79).

Similarly, in the biotech industry of the 90s, optimization was guaranteed by a scale-free collaboration style that allowed control in a culturally uncertain fast-changing field. In the 80s, the situation had been very different. Dedicated Biotechnology Firms, pharmaceutical companies, universities, and venture capital firms were highly specialized and mutually dependent on each others skills, making this a socially uncertain field. Culturally, however, the field was relatively certain. Drug development was in equilibrium, with proven rules of thumb for development duration and time to market. This had changed by the 90s, as a new market structure had emerged, based on medicines from genetic engineering, which required organizations to be tremendously innovative. As the field turned culturally uncertain, the organizations responded by reducing ambage. The cooperation of competitors in an open-elite partnership system was the only possible style in the face of high ambiguity. As a whole, the field ventured from the fatalism quadrant to the anomie quadrant in figure 1.7. But it remained at the critical temperature all the time (Powell et al., 2005; Powell & Owen-Smith, 2012).

### Control Through Tinkering

A general mechanism of change is the mating of styles (H. C. White, 2008, p. 160–5), fleshed out by Padgett and Powell (2012b, p. 12–5, chs. 6, 13, 15) as transposition and refunctionality. The emergence of the biotech industry as discussed can be traced back to a punctuating event, the innovation of genetic engineering in the 70s. What proved disruptive was the discovery that parts of different genetic molecules can be recombined to produce something, first and foremost medicines, that evolution has not come up with.[14] In the 80s, many scientists sensed that a window of opportunity had opened. They were not projecting an industry but left the university labs and found support by venture capitalists. The new style – the Dedicated Biotechnology Firm situated at the intersection of science and finance – emerged because two old styles were mated:

[13] Two generations (e.g., father's brother's daughter), three generations (e.g., father's father's sister's son's daughter) up to five.

[14] Note that genetic engineering is itself a 'mating' principle to produce novelty.

the newborn entrepreneurs introduced scientific practices into the start-up domain and commercial practices into science (Powell & Sandholtz, 2012). An early form of venture capital had emerged in the same way. Now the setting is Renaissance Florence. All we need to do is replace university labs by domestic bankers, venture capital by the city council, and Dedicated Biotechnology Firms by international merchants. The event that triggered the partnership style of early venture capitalism from the transposition of economic and political styles was the Ciompi revolt of 1378 which caused the city council to charge domestic bankers with rebuilding international economic relations shattered by war (Padgett, 2012c).

Burt's (1992) theory of structural holes is a direct example of the principle that styles mate to change.[15] Burtian brokerage is situated in social networks, where identities can reap benefits from spanning otherwise unconnected or weakly coupled modules. Managers in a large electronic company, e.g., have increased probabilities to be creative when they are positioned near structural holes (Burt, 2005, p. 28–46). In their desire to innovate, scientists maintain contacts to colleagues in other research domains (Crane, 1972). The hypothesis of creativity from collaborative brokerage finds support, but such a position can also hamper the diffusion of novelty (Fleming, Mingo, & Chen, 2007). Effects are domain-specific. While productivity is associated with brokerage in nanoscience, it is associated with embeddedness into dense networks in astronomy. The latter is at the very periphery of the network of scientific disciplines (Jansen, von Görtz, & Heidler, 2010). Together with earlier results that its co-authorship structure is not at a critical point, this suggests that the periphery may offer niches where identities can prosper in local optima. From the perspective of the system of scientific disciplines, astronomy is highly specialized.

Chen et al. (2009) have translated brokerage to meaning structures and presented a model based on the premise that "bridging structural holes in a knowledge space is a valuable and viable mechanism for understanding and arriving at transformative scientific discoveries" (p. 196). Predictions of future impact are based on a paper's co-citation betweenness and on burstiness. *Betweenness* is a brokerage measure how much a node in a network is between other nodes in terms of shortest paths (Freeman, 1977). A paper with high co-citation betweenness is potentially a transformative discovery because it belongs to multiple story sets or styles. A paper bursts if it ends a period where its number of citations relative to other papers is small and starts a period in which its relative number of citations is reasonably large (Kleinberg, 2003). While many citations show that a paper has diffused in a community, potentially slowly, high burstiness indicates that it has diffused rapidly. Chen et al. show for two of three specialties that foundational publications had high betweenness scores and experienced strong bursts of impact. A correlation of betweenness and impact is also confirmed for direct citation networks of two different specialties (Shibata, Kajikawa, & Matsushima, 2007). In all of science, while most papers cite references from conventional (statistically expected) combinations of journals, papers that cite atypical combinations are higher cited on average (Uzzi,

[15] Burt builds on Granovetter's (1973) model of the strength of weak ties. Weak ties are strong because they provide shortcuts to distant knowledge pools.

Mukherjee, Stringer, & Jones, 2013).

Our model rests on the principle that control is maintained through tinkering. Tinkering or mating can be applied to the way Hora builds watches: the stable intermediate modules need not be combined the same way each and every time. New combinations make new watches. Padgett and Powell (2012b) call this transposition and refunctionality, "a new purpose for an old tool" (p. 12). Sources of change can be any kind of event from inside (encounters, turnover, and avalanches at all levels of an organizational hierarchy) or outside (public) of identity (S. Fuchs, 2001, p. 211, 229; Sornette, 2006b; Padgett & Powell, 2012b, p. 26–8). Punctuations from outside can only be sensed by styles if the extent of insulation is not too big, if network domains overlap, like science and economy in case of the biotech industry. When overlap is given, small causes can have big effects. In Padgett and Powell's (2012a, p. 7–11) terminology, *organizational innovation* is the punctuation of style, the transposition of practices and relational ties across network domains. *Organizational invention* is the second stage, resisted by autocatalysis, when the innovation is absorbed and causes an adaption of teaching protocols for reproduction. The stronger invention is, the more punctuating social facts cascade through a network domain and tip the style into a new direction. *Systemic invention* is when a punctuation permanently changes styles and is not restricted to a single network domain. In science, when events affect the network core, a paradigm shift occurs. In the terminology of Mische and White (1998, p. 709–14), weak and strong organizational invention and systemic invention map to passages mediated by a public, cascades, and a phase transition of the whole public. The magnitude of a burst is an indicator, to what extent a style has been tipped by a punctuation (Kleinberg, 2003).

As identities are continuously bombarded by events, they adapt their paths according to their very own style. Criticality is the point of reference for specialization or innovation, depending on social and cultural uncertainty. Lock-in, we hypothesize, will sooner or later cause groups to split when all necessary reproduction skills have been duplicated through teaching. Similarly, before identities die heat death, they will likely form clusters and, thereby, return to the critical point. This is a probable explanation for the merger and split dynamics of collaboration observed by Palla et al. (2007) and Sun et al. (2013). Strong evidence also comes from analyses of cultural dynamics. The theory of fractal distinctions predicts that narratives diverge into several narratives or styles only to also converge at all spatial and temporal length scales – winners of a paradigm shift take up the concerns of the defeated. When the citation network of the American Physical Society is restricted to network science, it can be observed that topics last for about 20 years until their identity has changed so much that one cannot speak of the same story set anymore. According to that study (Lancichinetti & Fortunato, 2012), network analysts in physics crossed the threshold of ten papers per year in 1998, the year Watts and Strogatz presented the small-world model. As expected, citation paths of Social Network Analysis (1978–1990) (Hummon & Carley, 1993) and Complex Network Analysis (1967–2002) (Garfield, 2004) show no sign of only diverging.

Even more clearly, the rise and fall of research domains as depicted by co-word networks exibits a "phylomemetic" pattern that is related to network density. Chavalarias and Cointet (2013) study story sets as proxies of research specialties, which is a fair as-sumption, and uncover that, over a 20-year period in embryology, decreasing conceptual density is a sure sign of a specialty's demise. Identities persist when they are conceptually stabilized by a core which constantly reinvents itself. Adaption comes through convergence (mergers of story sets) and divergence (splits). Mergers are preceded by the emergence of new words and are events of word recombination. Splits have slightly different meaning in the sequence of events. They are events of word emergence and are followed by the recombination of words. Fundamentally, density significantly decreases before one of these events and increases subsequently. Bayesian forks turn on the heat to melt frozen structures and erase memory to allow new narratives to emerge. But while mergers seem to be recombinatory reactions to events by identities already reactive, splits may primarily be the events themselves, generated endogenously from forceful densification. This is certainly food for thought but further substantiates our hypothesis that control involves an uncertainty trade-off.

Chavalarias and Cointet (2013) show that their results are valid for various levels of organizational complexity, small and large cluster sizes, i.e., the typical lifecycle they have uncovered is size-independent. Van Raan (2000) has proven mathematically that, if an exponential aging dynamic is coupled to exponential growth, a research domain will differentiate into a hierarchically modular structure with a power-law size distribution of specialties, as detected empirically (1991).

It is the careers of social facts and events and the accompanying statistics that we want to study to potentially identify a middle-range order and enrich sociological theory by a perspective of social scaling. The way we have modeled identity in this chapter is applicable not only to research domains, organizations, and groups. Persons, too, are complex outcomes and origins of narratives, styles that teach, learn, and adapt in the daily selection of social facts. Studying life at the critical point requires self-similar theory (Abbott, 2001a; H. C. White, 2008; Lane et al., 2009).

## Summary

Meaning structures have the function of memory from which identities in the first sense can reproduce transactions. Punctuations are potentially large perturbations of autocatalytic flows that effectively turn on the heat, melt or erase structural memory, and create an open future. Styles, a fifth sense of identity, are sensibilities of making distinctions and adapting meaning structures to such Bayesian forks. The percolation threshold allows styles to strike a balance of reproductive stability and renewing change. It provides optimal evolvability in terms of global connectivity, modular diversity, relational economy, and structural memory.

Scale invariance is the phenomenological outcome of *what* goes on in transaction structures. Preferential attachment in concert with homophily is *how* meaning structures become self-similar. Optimization is the reason *why* preferential attachment operates. And identities as styles are *who* optimize through preferential attachment. Power laws are outcomes of the adaptive walks of styles in meaning space. The importance of small-world and scale-free networks, integrated in feedback networks, is that they balance the stability and change of paths through fractal search.

Styles alter their paths, modeled on socio-cultural temperature, by trading social integration off against cultural regulation. Fractal distinctions gain importance as a heuristic for modeling system dynamics as the divergence and convergence of styles at all length scales. As multiple cores innovate due to the same punctuation, the whole dynamic resembles systemic invention or, in science, a scientific revolution.

# 2. Operationalization of Theory

## 2.1. Data

### 2.1.1. Bibliographic Data and Bibliometrics

Citation, word usage, and authorship analysis is a powerful method of historical research into science (Hummon & Carley, 1993; Small, 1980; Barabási et al., 2002; Garfield, 2004; Lancichinetti & Fortunato, 2012; Chavalarias & Cointet, 2013). For the analysis of historical narratives and the evolution of a multidisciplinary research domain, two datasets can be used: the *Web of Science* (WoS) and *Google Scholar*. We use WoS because it has a history of 60 years of quality assurance and can be easily queried, while the latter is a black box and data retrieval is more complicated.[1]

WoS is a child of linguist Eugene Garfield who had started drafting it in the 50s as an early version of a "world brain," plead for by H. G. Wells about a decade before. Besides being a search tool, Garfield (1955) envisioned it to be "particularly useful in historical research when one is trying to evaluate the significance of a particular work and its impact on the literature and thinking of the period" (p. 109). The problem with searching a lexical index, Garfield reasoned, is that one often does not know the right terms to search for. To find literature on a desired topic, it is much easier to pinpoint a set of relevant references and then retrieve publications that cite these references. WoS is effectively designed around the way identities find footing through Durkheimian feedback. To find publications, identify the core of what you are looking for and the periphery will follow (Garfield, 1964).

Not all scientific communication is covered by WoS. In the early days, a rule of thumb could be derived from Bradford's Law that 500–1,000 journals cover 95% of the significant literature in a given discipline. But journals largely overlap: due to Garfield's Law of concentration, which states that "the tail of the literature of one discipline consists, in a large part, of the cores of the literature of other disciplines" (Garfield, 1979a, p. 23), 95% of the cited literature in natural science at the end of the 80s was covered by just 3,000 journals in WoS. This is prime evidence for the permeability of boundaries in science. The database producers select only core journals based on citation impact but all references of covered papers are captured (p. 21–3). Today, WoS covers roughly 12,000 journals plus 150,000 individual conference proceedings in about 250 disciplines.[2]

Much effort has been put into building WoS but deficiencies and discontinuities are clearly present. Essentially, the database excludes books and works that do not provide at least an English abstract.[3] This significantly reduces the coverage in disciplines with regional importance, like the social sciences. 1990 constituted a break in database development. Conference proceedings were added and author keywords were included. But in practice, even abstracts are not available before 1991. Many data fields are preprocessed. Most importantly, strings of cited references are standardized, e.g., *American Journal of Sociology* is coded as `AM J SOCIOL`. Despite pre-processing, data cleaning and disambiguation is necessary, discussed in appendix B.

[1] The *Scopus* database does not allow historical analyses because publications before 1996 do not have references.

[2] thomsonreuters.com/en/products-services/scholarly-scientific-research/scholarly-search-and-discovery/web-of-science-core-collection.html, visited April 30th, 2015.

Traditionally, bibliographic data belongs to information science which is classically split into scientometrics and information retrieval (Van Den Besselaar & Heimeriks, 2006). Scientometrics is the quantitative analysis of science and becomes *bibliometrics*, when bibliographic data such as from the WoS is used (Broadus, 1987). Bibliographic data is deeply relational and, therefore, network analysis has a long history in bibliometrics. Methodological developments have largely benefited from the application of bibliometrics as a science management tool because indicators have political consequences and many stakeholders critically observe evaluation practices. Normalization of matrices and metrics is important if domains with different publication characteristics are to be compared (Garfield, 1979b; Waltman et al., 2012).

In impact analysis, normalization is necessary because domains have different citation potentials. A citation from a publication in a domain where 10 references per paper are cited on average has greater value than one from where 30 references are cited. In the former domain each citation is three times more valuable because the citation potential is just a third of the latter domain. Classically, normalization is achieved by dividing observed citation rates by an expected citation rate which is the average number of citations a reference gets in a subject domain (Schubert & Braun, 1986). Because citation distributions typically do not have finite variance (Albarrán & Ruiz-Castillo, 2011; Albarrán et al., 2011), the bibliometrics community is constantly discussing alternative indicators. Leydesdorff and Opthof (2010) have proposed to weight each citation by the inverse reference list length to alternatively account for citation potentials. This method is known as fractional counting and is also used in productivity analysis. Here, the idea is that an author only gets credited for a quarter publication if he has three co-authors (Gauffriau, Larsen, Maye, Roulin-Perriard, & von Ins, 2007). This allows to compare, e.g., a domain where research is typically done in teams with one where authors write alone. Batagelj and Cerinšek (2013) have generalized normalization using fractional counting in a matrix framework.

A central topic in information retrieval is the boundary problem how to delineate a set of publications on a specific topic. Methodologically, this problem brings us to the point where we necessarily violate our first rule of method to allow for variation because, to collect data, we will have to draw a sharp boundary that does not exist. This is most consequential because, by selecting the 1-identities that constitute the 2-identity, we determine all future analysis. The classical approach is to identify terms and search them in the title, abstract, and author keywords of publications. Because keywords selected by authors are very subjective, the WoS producers scan the words used in the cited references and offer the most important ones as *KeyWords Plus* (Garfield & Sher, 1993). Such an approach requires extensive pre-search knowledge about the domain to be delineated (Maghrebi, Abbasi, Amiri, Monsefi, & Harati, 2011). To aid searching, WoS experts tag journals with one or more of about 250 disciplines. The assumption is that, e.g., all papers published in journals tagged "Physics, Condensed Matter" actually belong to that subdomain of physics. It is easy to show that research domains whose boundaries overlap below the level of journals cannot be delineated with a journal classification system because one will never be able to reduce false positives and false negatives at the same time. Eventually, every boundary definition amounts to keeping something constant, but with coarse classification systems, boundaries are kept constant at high levels (S. Fuchs, 2001, p. 199).

[3] Books were included in 2011 but coverage is meager.

Suitable approaches to the boundary problem involve retrieval of individual publications through hybrid lexical/citation-based search. These are bibliometric retrieval strategies based on sociological models of research domains in context (Glänzel, 2015; Zitt, 2015). Authorship or co-authorship is not used for retrieval because authors often work in multiple domains. Both words and cited references have strengths and weaknesses for delineation. Language use crystallizes on slower time frames than citation (Braam, Moed, & Van Raan, 1991b; Padgett, 2012a, p. 60). Words are much more ambiguous and less precise than references, and word distributions are much more skewed. Therefore, co-word networks are much denser, and their boundaries much fuzzier, than co-citation networks. The latter maps are less evocative. On the other hand, language maps and lexical retrieval methods are easy to use. Boundaries from cited references are less permeable but certainty may be deceptive because of uncertainty in the reason for citations. Taken together, lexical delineation tends to introduce false positives (low precision but high recall) while citation-based approaches tend to introduce false negatives (low recall but high precision). Therefore, hybrid approaches utilizing both meaning dimensions aim at increasing both precision and recall (Braam et al., 1991a; Glänzel & Thijs, 2011; Glänzel, 2015; Zitt, 2015).

Two sophisticated delineation methods have been proposed. The one of Zitt and Bassecoulard (2006) is deeply rooted in our identity model according to which an identity is defined through a core of social facts the story set of which is shared by all 1-identities. One starts with a highly precise, lexically retrieved *seed* of publications. This seed $A$ cites a set of references $B$. From $B$, all references that receive less than $Y$ citations from $A$ are removed, resulting in set $C$. $Y$ is the first parameter that makes sure that the references in $C$ are generic in terms of retrieval. From $C$, all references are removed that receive less than $U$ percent of all their citations from $A$, resulting in the cited core $D$. $U$ is the second parameter that makes sure that references in $C$ are specific to the domain to be delineated. It requires knowledge about numbers of citations from all of science $S$. Finally, from $S$, all publications are retrieved that cite at least $X$ core references, resulting in the retrieval solution $E$. $X$ is the third parameter that makes sure that publications in $E$ are relevant. It is a drawback that, to compute a paper's specificity, direct access to the full WoS database is required, which only a few research centers worldwide have. Domain knowledge is only required to create the precise seed. Retrieval scores high in terms of precision and recall (Laurens, Zitt, & Bassecoulard, 2010).

Mogoutov and Kahane (2007) start with a low-precision publication seed $A$. Their case is nanoscience and the seed basically consists of publications using the string `*NANO*`. Title words of all seed publications are pre-processed (stemmed, stop words removed) and statistically significant word sequences ($n$-grams) are extracted. For simplicity, we will call $n$-grams words too. Because the seed $A$ is imprecise and harbors culturally different subdomains, it is partitioned into communities $A_j$ based on journal citation density. For each $A_j$, all words are removed that are less specific than a threshold $I$, resulting in a set of lexical descriptors per subdomain.[4] These descriptors are enriched by specific author keywords. Finally, subdomains are delineated by querying all of science $S$ for publications using the descriptors. As in the first method, publications are retrieved that select a set of core facts (now words instead of references), but differently, cores for multiple subdomains are identified. This method requires domain knowledge not in creating the seed but, later, in filtering the extracted search terms. Both methods are capable of retrieving publications of complex and emerging research domains.

#### Summary

The *Web of Science* is our database of choice because it perfectly enables the narrative approach. It contains the core papers of science, selected according to citations of scholarly journals. Books and works without English abstracts are not covered and are not part of our analysis as citing items. Our method is bibliometrics, the statistical analysis of bibliographical data, which is devoted to studying, delineating, and evaluating the science system. Hybrid methods of delineating network domains involve the identification of core references and words and the retrieval of publications that cite or use those social facts.

### 2.1.2. Data Model of Selection

There is a confusion of names for the various bibliometric networks discussed so far. Co-authorship and co-citation are used homogeneously. They are networks of social facts – authors and cited references, respectively – and they are labeled by the corresponding *practices*, authorship and citation. Co-word networks, though similarly constructed, are labeled by the facts with a coupling prefix. Here, we unify names and practices and translate the identity model into a data model of selection. Publications are the fundamental building blocks of the identities we observe. They take the role of actors, i.e., they select (are authored by, cite, use) social facts (authors, references, words). Our *objects of observation*, the things for which results are reported in tables and through graphs, are identities. These can be large like the domain of Social Network Science or its subdomains but also individual authors in the domain. Our *units of analysis* are stories, foremost in the form of the nodes (social facts) and edges (co-selections of facts) in meaning structures, but also as transactions like direct citation. Results about objects are derived from the statistical analysis of relations from publications – in other words: from bibliometrics.

[4] Mogoutov and Kahane compute specificity differently than Zitt and Bassecoulard (2006).

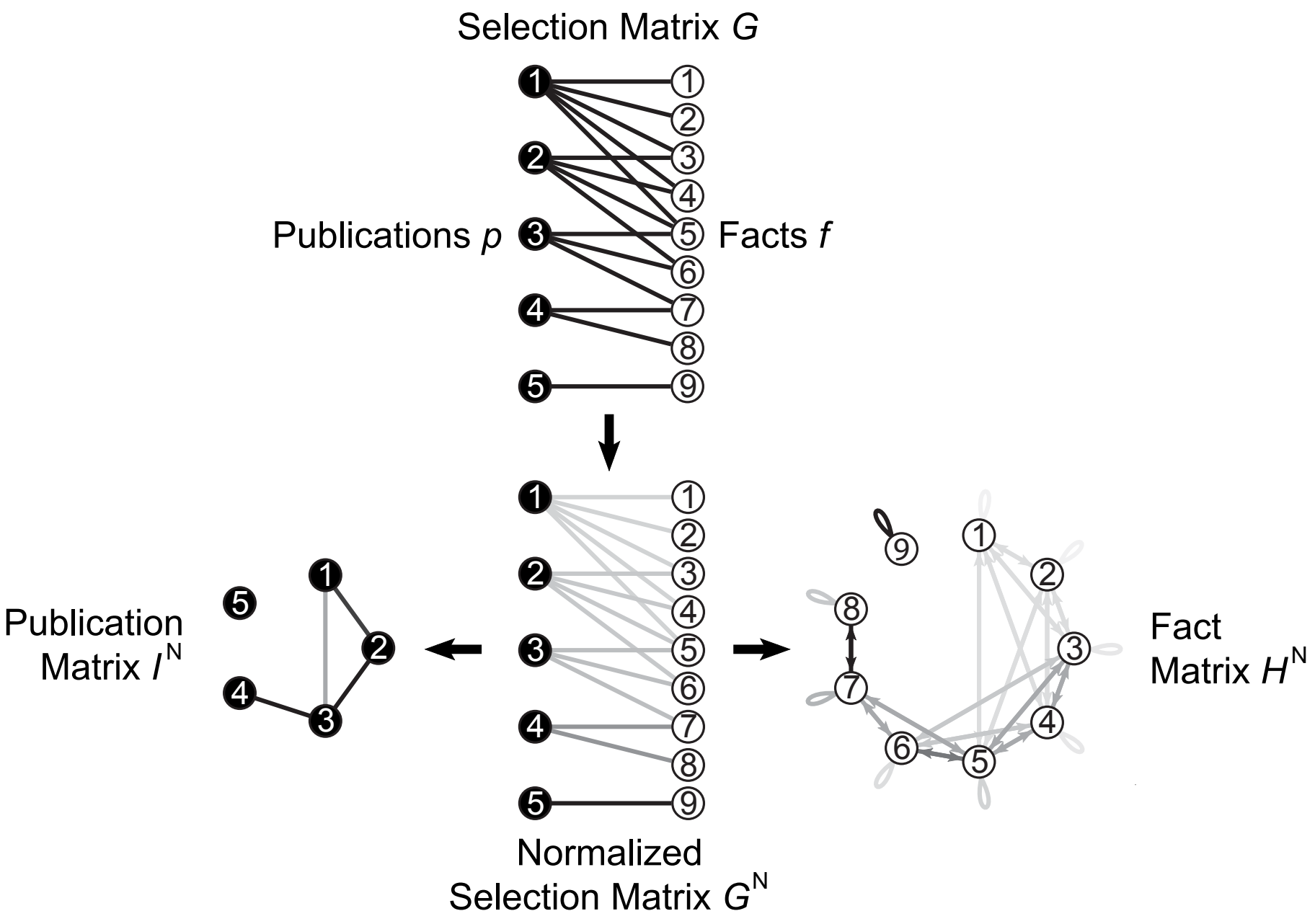

**Figure 2.1.: Data model of selection**
Publications $p$ (matrix rows), the fundamental building blocks of identities, select social facts $f$ (columns). The bipartite selection matrix $G$ is normalized by dividing selection edges by row sums, i.e., all publication's selections sum to 1. The normalized selection matrix $G^{\mathrm{N}}$ is transformed into the publication matrix $I^{\mathrm{N}}$ and the fact matrix $H^{\mathrm{N}}$. $I^{\mathrm{N}}$ serves the analytical purpose of identifying subdomains represented by papers. $H^{\mathrm{N}}$ is a meaning structure of facts coupled by additive co-selections. Arrows indicate that facts are catalytic. Ties are symmetric, i.e., facts are co-selected. Facts 7 and 8 are most autocatalytic because they are the sole influences of publication 4. Loops have the important meaning of self-reproduction. Fact 9 is most "selfish" because it does not influence any publication in companion with another fact.

Figure 2.1 depicts the data model of selection, a generalization of bibliographic networks discussed in the literature (Kessler, 1963; Small, 1973; Callon & Law, 1986; Newman, 2001b). It treats all social facts equally and, thereby, allows the comparison of different scientific practices. The basic intuition is to use publications as 1-identities, the fundamental building blocks. They are agentic because they have the capacity to select social facts. Publications are coupled through selected facts to form analytical networks. Those have the sole purpose of identifying subdomains for the analysis of facts. When publications are coupled through the citation practice, this is the so-called bibliographic coupling network. Our general name for this analytical network is 'publication co-selection' network. The other side of bipartite-matrix multiplication is called the 'fact co-selection' network, like 'author co-authorship,' 'reference co-citation,' and 'word

co-usage.'

Formally, all data is organized as a bipartite *selection graph* $\mathcal{G} = (\mathcal{F}, \mathcal{P}, \mathcal{S}, w)$ which contains two sets $\mathcal{F}$ and $\mathcal{P}$ of vertices, a set $\mathcal{S}$ of edges, and $w$ are edge weights. $\mathcal{P} = \{p_i : i \in \mathbb{N}, \max(i) = m\}$ is a set of $m$ publications. $\mathcal{F} = \{f_j : j \in \mathbb{N}, \max(j) = n\}$ is a set of $n$ social facts. Social facts are authoring authors $\mathcal{F}_{\text{aut}}$, cited references $\mathcal{F}_{\text{ref}}$, or used words $\mathcal{F}_{\text{wrd}}$. $\mathcal{S} = \{s_{ij} : i \in \mathcal{P}, j \in \mathcal{F}\}$ is a set of edges between $\mathcal{P}$ and $\mathcal{F}$, representing the selections of facts by publications. Selection means that publications attach to facts. Meaning structures with cores and fuzzy boundaries emerge from these practices and simultaneously are part of the public from which publications can select.

$\mathcal{G}$ is simple, i.e., there are no parallel edges. Publications select authors through authorships $\mathcal{S}_{\text{aut}}$, references through citations $\mathcal{S}_{\text{cit}}$, or words through usages $\mathcal{S}_{\text{use}}$. To each edge $s_{ij} \in \mathcal{S}$, a weight $w_{ij} \in [0, 1]$ is mapped. $G = [w_{ij}]$, with $w_{ij} = w(i, j)$ for $(i, j) \in \mathcal{S}$ and $w_{ij} = 0$ otherwise, is the bipartite $m \times n$ *selection matrix* assigned to graph $\mathcal{G}$. $G^{\text{T}} = [w_{ij}]^{\text{T}} = [w_{ji}]$ is the *transposed selection matrix* with dimensionality $n \times m$.

To account for the possibility that publications in different subdomains select different numbers of facts, network normalization is necessary. In a network with rows $i$ and columns $j$, a node $i$'s outdegree is $k_i^{\text{out}} = \sum_j w_{ij}$ and a node $j$'s indegree is $k_j^{\text{in}} = \sum_i w_{ij}$. In symmetric networks, $k^{\text{out}} = k^{\text{in}}$. Batagelj and Cerinšek (2013) have generalized normalization using fractional counting in a matrix framework. Normalization requires a binary selection matrix $G$ where each existing edge has a weight of 1. Then, $G^{\text{N}} = \text{diag}(1/\max(1, k_i))G$ is the *normalized selection matrix* where $k_i$ is the number of facts publication $i$ selects. If, for example, a publication has three authors, each authorship gets a weight of 1/3.

Statistics based on degrees of facts in selection matrices have very natural interpretations. For binary matrices $G$,

> the degree $k_j$ is the number of publications that select (are authored by/cite/use) fact (author/reference/word) $f_j$ and
>
> $K_j = k_j/m$ is the fraction of all publications that select (are authored by/cite/use) fact (author/reference/word) $f_j$.

For normalized matrices $G^{\text{N}}$,

> the weighted degree $k_j^{\text{N}}$ is the normalized number of publications that select (are authored by/cite/use) fact (author/reference/word) $f_j$ and
>
> $K^{\text{N}}(f_j) = k_j^{\text{N}}/m$ is the fraction of all selections of (authorships by/citations of/usages of) fact (author/reference/word) $f_j$.

Through matrix multiplication, the unipartite weighted fact co-selection matrix $H$ and publication co-selection matrix $I$ are obtained. In the unnormalized case:

> $H = G^{\text{T}}G$ is an undirected fact matrix where edge weights $w_{jr} \in \mathbb{N}$ give the number of publications that co-select (are co-authored by/co-cite/co-use) facts (authors/references/words) $f_j$ and $f_r$ (Batagelj & Cerinšek, 2013, sec. 3.2).

$I = GG^{\mathrm{T}}$ is an undirected publication matrix where edge weights $w_{iq} \in \mathbb{N}$ give the number of facts (authors/references/words) co-selected by (co-authoring/co-cited by/co-used by) publications $p_i$ and $p_q$.

In the normalized case:

$H^{\mathrm{N}} = G^{\mathrm{T}}G^{\mathrm{N}}$ is a symmetric directed fact matrix where arc weights $w_{jr} \in [0, \infty)$ give the normalized number of publications that co-select (are co-authored by/co-cite/co-use) facts (authors/references/words) $f_j$ and $f_r$ (Batagelj & Cerinšek, 2013, sec. 3.3). The network's total weight $w_\Sigma = \sum_{j,r} w_{jr}$ including $j = r$ equals the number of selections in the graph because weights are additive. The weighted degree $k_j$ equals the number of publications that have selected (been authored by/cited/used) a fact, i.e., Lotka's, Bradford's, and Zipf's Law can be studied using distributions of weighted degrees. The weight of an arc is interpreted as the strength with which a source fact catalyzes a target fact. For example, for a publication that select three facts, an autocatalytic clique is generated where each arc, including self-loops, has a weight of 1/3. The diagonal tells that each fact in the triad autocatalyses itself with a strength of 1/3. This is a general translation of the hypercycle model into a bipartite network framework (Padgett, 2012a). In the case of authors as facts, it is also possible to interpret arc weights including loops as fractional contributions to authorship. The more authors a paper has, the more the former are fractional authors. As diagonals are summed over multiple papers, authors with small scores are likely apprentices Price, 1986 [1963], pp. 79.

$I^{\mathrm{N}} = G^{\mathrm{N}}(G^{\mathrm{N}})^{\mathrm{T}}$ is an undirected publication matrix where edge weights $w_{iq} \in [0, 1]$ are the products of normalized selections of (authorships by/citations of/usages of) facts (authors of/references by/words by) publications $p_i$ and $p_q$. Weights can be interpreted as publication similarities. The total weight $w_\Sigma = \frac{1}{2}\sum_{i,q} w_{iq}$ or the weighted degrees have no particular meaning because weights are not additive. This network is the complementary transformation of the one described by Batagelj and Cerinšek (2013, sec. 3.4).

### Summary

Identities are our objects of observation, and stories are our units of analysis. Selection is modeled as a bipartite graph of publications representing 1-identities or agents, social facts, and edges from publications to facts representing selections. All facts are treated equally which allows the comparison of different scientific practices within a network domain. Matrix normalization through fractional counting of selections by publications further enables comparison across different network domains. Normalized meaning structures have a natural interpretation of autocatalytic structures where matrix diagonals quantify the strength of self-reproduction. This data model can be a general framework for empirical analyses. As the special case of the citation practice shows, where selecting publications and selected facts are of the same type, all social structures, even unipartite ones, are amenable to this framework.

## 2.2. Methodology

### 2.2.1. Idealtype of Knowledge Production

In this section, we present an idealtypical process of knowledge production to explain our metrics and be able to put results into perspective later on. The test of a network model is if it generates structural and dynamical properties observed in reality, like scale-free degree distributions, clustering, hierarchy, core/periphery structure, short average path lengths, overlap, and densification. Most network models are probabilistic. The discussion in chapter 1 culminated in the feedback network model (D. R. White et al., 2006) which combines many of these properties in a sound sociological framework. In contrast, our idealtype is a deterministic graph process. It also has all properties we want to demonstrate, and it is largely amenable to exact mathematical solutions. But importantly, it is based on a hierarchical construction principle of deterministic fractals. Like the Koch snowflake (figure 1.3), its self-similarity can be visually grasped, which is our intention. The model is influenced by, and closely resembles, the deterministic network model presented by Ravasz and Barabási (2003) but is sociologicalized by distinguishing between the three scientific practices authorship, citation, and word usage, as well as the disciplines arena, council, and interface.

The idealtypical emergence of a research domain as a knowledge production process is shown and fully described in figure 2.2. The process resembles the self-similar growth of disciplines (the positional blueprints) coupled through publications. Direct citation is governed by interface discipline which channels the flow of knowledge. All fact co-selection graphs are hierarchically modular council disciplines with multifunctional cores. Domain evolution is modeled in a genealogical way. Each time step, families teach a new generation of scholars which, in the process and subsequently, produce new knowledge that contextualizes the core. Short-term teaching pays off in terms of long-term learning from new generations.

Overlap of different network subdomains is of greatest importance because switching is the source of meaning and fresh action. The way overlap is built into the idealtype is closely related to the method of detecting multifunctionality. Most community detection algorithms perform hard clustering, i.e., overlap cannot be detected (Breiger et al., 1975; Reichardt & White, 2007; Blondel et al., 2008). Here, we exploit the mathematical property that a hard partitioning of the publication matrix $I$ of 1-identities translates to link communities (co-selection communities) in the fact matrix $H$ (cf. figure 2.1). Therefore, social facts can be embedded into multiple network domains represented by one story set each (Ahn, Bagrow, & Lehmann, 2010). Each story set is represented by a *type of tie*. For hard-clustering publications we use the Louvain algorithm of Blondel et al. (2008) because it is methodologically grounded in a hierarchical model of modules with different lifetimes, works with weighted ties, and is very fast. The Louvain method clusters publications or nodes in an undirected network from the bottom up. Initially, all nodes

**Figure 2.2.: Idealtype of knowledge production**

The sequence of events depicted on the following three pages describes the self-similar emergence of a research domain as a knowledge production process. Publications are knowledge products and knowledge is described through social facts selected by publications. At $t = 1$, an initial publication $p_1$ exists, authored by author $f_1^{\text{aut}}$ and using word $f_1^{\text{wrd}}$. This paper is the founding work of the emerging domain described by the first word. At $t = 2$, the founder of the domain starts a group of four – the sorcerer teaches three apprentices – and together they publish three new papers. Each paper cites the initial paper and uses the initial word, but also introduces a new word that contextualizes the first word. The direct citation network (a) shows publications as ellipses. They are differently colored to show different subdomains that grow around them. Citations are colored according to the subdomain of the cited reference. Publications are the only nodes that are hard-clustered, i.e., they belong to just one subdomain. Because of matrix multiplication, blocking of publications maps to blocking of co-selections in fact networks. The co-usage network (b) shows words as diamonds. Social fact nodes are never colored because they can be selected by multiple subdomains. Words $f_1^{\text{wrd}}$ and $f_2^{\text{wrd}}$, e.g., are co-used once in publication $p_2$ which belongs to the green subdomain. Accordingly, $f_1^{\text{wrd}}$ and $f_3^{\text{wrd}}$ are coupled through a blue relation because $p_3$ uses both words, and the relation between $f_1^{\text{wrd}}$ and $f_4^{\text{wrd}}$ is purple. The four subdomains overlap lexically in the first word. The co-citation network is not shown because it always looks like the co-usage network at $t - 1$. The co-authorship network (c) shows authors as boxes. It depicts the 4-clique representing the initial group, a fully connected network with four nodes. All co-authorships have a weight of three because a sorcerer hires three apprentices with which three new papers are produced. Because each of the new works belongs to a different subdomain, the three relations per author pair are actually green, blue, and purple, but since they are superimposed, the relations appear thin and grayish. Whenever this happens, a node pair is shared by different network subdomains. $t = 2$ is the only time where different types of co-authorship relation overlap. At $t = 3$, the training of the three apprentices is finished, and the now four sorcerers start their own groups. $f_2^{\text{aut}}$ proceeds along the lines of $f_2^{\text{wrd}}$, $f_3^{\text{aut}}$ focuses on $f_3^{\text{wrd}}$, and $f_4^{\text{aut}}$ on $f_4^{\text{wrd}}$. Each group writes three papers that all cite their sorcerer's paper in addition to the founding work, and each introduces a new word, co-using them with the parent and parent's parent word (if possible). Word $f_1^{\text{wrd}}$ is now co-used four times with $f_2^{\text{wrd}}$, $f_3^{\text{wrd}}$, and $f_4^{\text{wrd}}$. A multifunctional but uncohesive lexical core is emerging. Hardly visible, within the red subdomain, word $f_1^{\text{wrd}}$ is co-used once with $f_5^{\text{wrd}}$, $f_6^{\text{wrd}}$, and $f_7^{\text{wrd}}$. Inside the core, the founding word has increased its multifunctional importance as the connector of different subdomains. From this time on, new co-authorship 4-cliques are fully rooted in subdomains, i.e., overlap is facilitated only by the first four authors. At $t = 4$ each node 'splits' into four again. All networks now consist of 64 facts and the core exists as a hierarchy of hubs in all networks. The co-citation network looks like the co-usage network at $t = 3$. The matrices of these networks are given in appendix C.

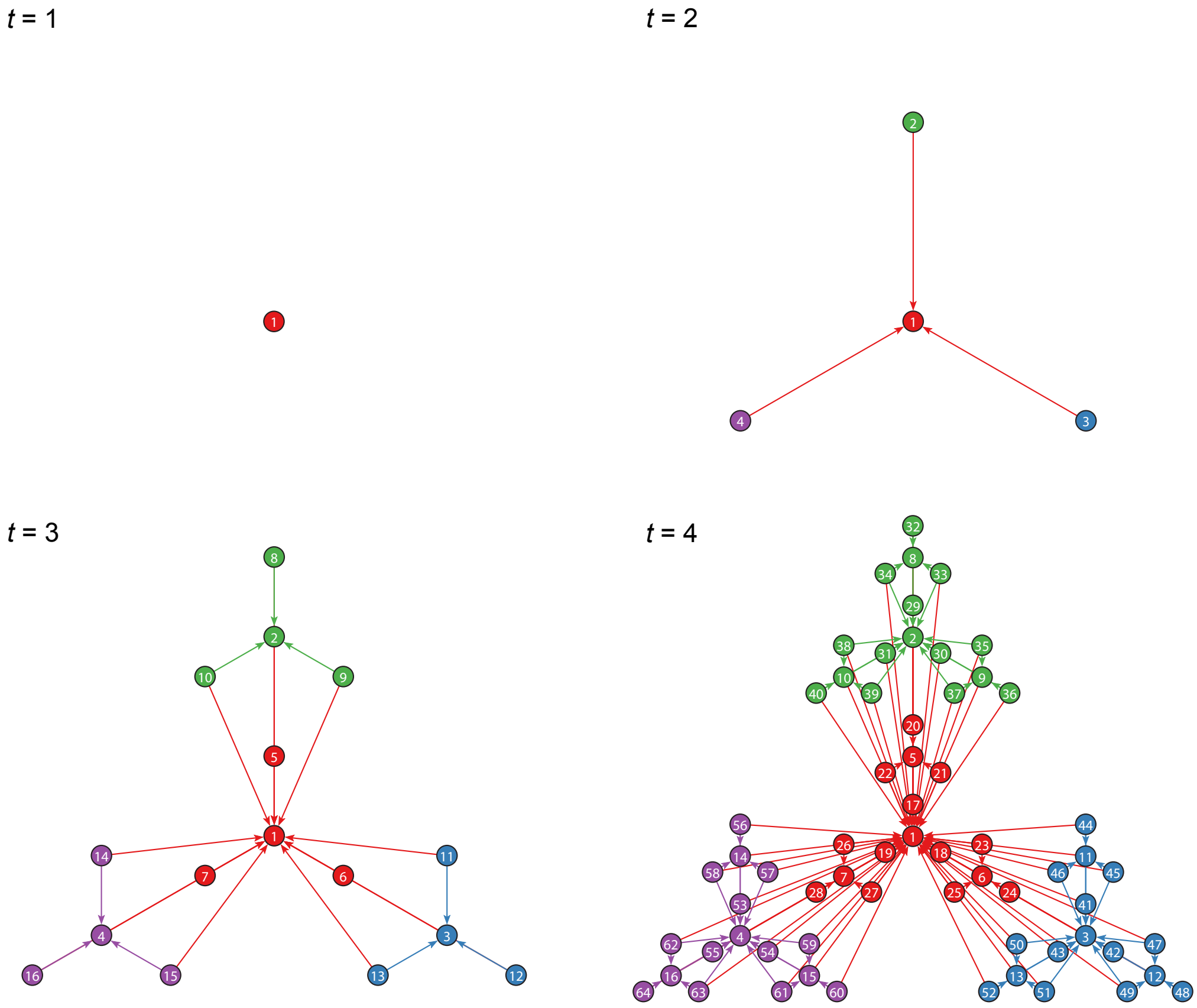

(a) Direct citation

**Figure 2.2.: Idealtype of knowledge production**

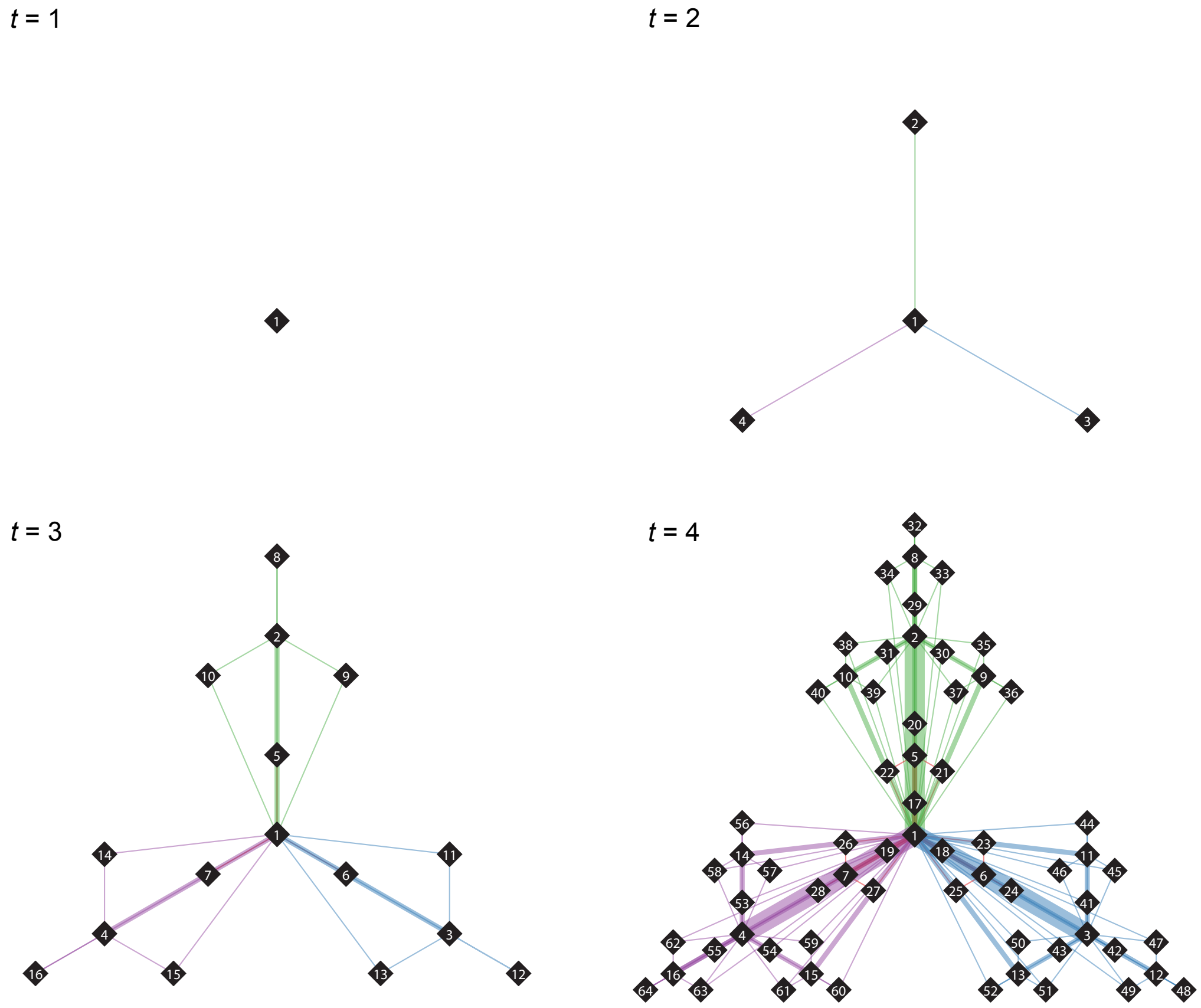

(b) Word co-usage

**Figure 2.2.: Idealtype of knowledge production**

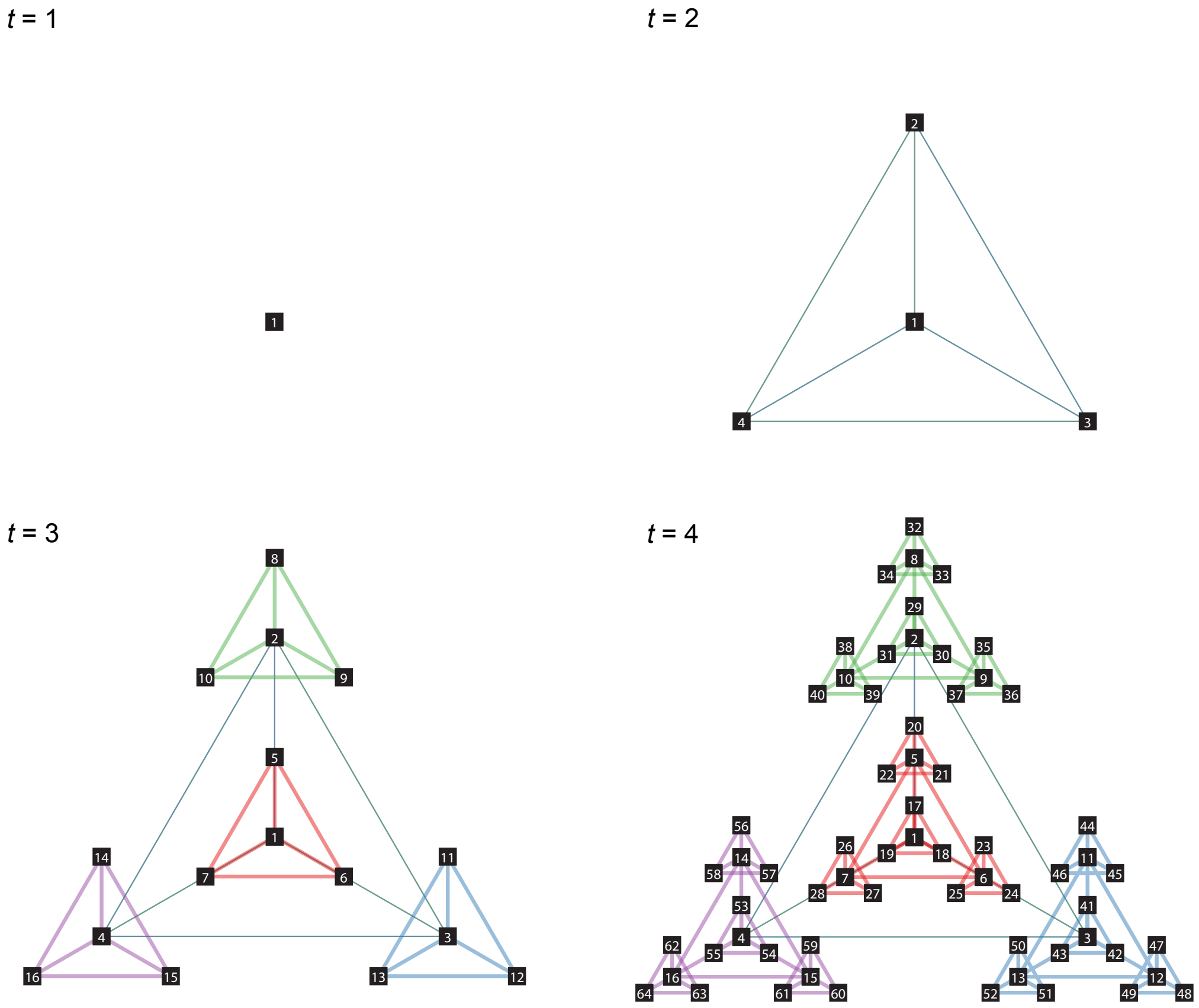

(c) Author co-authorship

**Figure 2.2.: Idealtype of knowledge production**

are their own community. Then, the algorithm iterates through the nodes and moves nodes $i$, randomly chosen, into the community of the neighbor $q$ for which modularity is maximized. Modularity, the fraction of edge weights internalized in communities minus the expected fraction, is quantified by

$$Q = \frac{1}{2w_\Sigma} \sum_{i,q} \left[ w_{iq} - \gamma \frac{k_i k_q}{2w_\Sigma} \right] \tag{2.1}$$

where $\gamma > 0$ is a resolution parameter. If modularity cannot be increased by moving $i$, it is not moved. When the local optimum is found, the nodes in a community are merged such that nodes are now the communities found on the previous level. For coarser and coarser versions of the network, modularity is maximized until $Q$ cannot be improved anymore. The communities at the highest level reached are the solution of the algorithm. In the standard method, $\gamma = 1$. When $\gamma$ is increased (decreased), the expected fraction of edge weights internalized in communities is increased (decreased) and smaller (larger) communities are detected.[5]

Louvain community detection (Blondel et al., 2008) harmonizes with our model of hierarchically modular identities. The resolution parameter $\gamma$ can be interpreted as a time scale. Small (large) communities at low (high) levels of social organizations have short (long) lifetimes (Delvenne, Yaliraki, & Barahona, 2010). In our analyses, we always use a unit resolution parameter. Communities in publication matrices are blocks of nodes that are structurally equivalent in the sense that they have similar selection profiles. Nodes in a community share the same story set. Practically, we do not detect blocks in the bipartite selection matrix (Doreian et al., 2004), but we detect communities in the unipartite publication matrix. The speed we gain comes a the cost of not being able to model 0-blocks. But this is only a theoretical cost because, when subdomains are partitioned through publication co-citation and/or co-usage, the emergence of a conscious 4-identity can be probed by studying densification dynamics in author co-authorship networks.

Overlap in a fact co-selection network is obtained through a sequence of steps. First, the domain in form of a publication matrix is hard-clustered, resulting in a set of subdomains. Second, for each subdomain, the selection matrix is transformed into a fact matrix. Third, the fact matrices for all subdomains are 'stacked' on each other. As a result, facts can have as many types of tie as there are subdomains. In the co-authorship network of our idealtype (figure 2.2c), the authors $f_1^{\text{aut}}$ through $f_4^{\text{aut}}$ forming the topmost multifunctional invisible college have three types of tie: a green/blue/purple one from co-authoring $p_2/p_3/p_4$. This is co-switching of identities.

As events, knowledge products acquire their meaning through their position in a chain of events. Typically, networks for which time series are available are divided into a succession of snapshots of length $\Delta t$ and communities are detected for the slices which are then matched for continuities. The smaller $\Delta t$ is, the shorter the periods through which events acquire their meaning. The small-world publication by Travers and Milgram (1969), e.g., had a different meaning in the 70s compared to the time following the Complexity Turn when it was revived by physicists (Garfield, 2004; Lazer, Mergel, & Friedman, 2009). When subdomains are allowed to change, merging and splitting events can be identified (Palla et al., 2007; Lancichinetti & Fortunato, 2012; Chavalarias & Cointet, 2013; Sun et al., 2013).

[5]We use a multilevel refinement method where entire communities can be moved to improve modularity (Rotta & Noack, 2011). Modularity was introduced by Newman and Girvan (2004) and the resolution parameter by Reichardt and Bornholdt (2004).

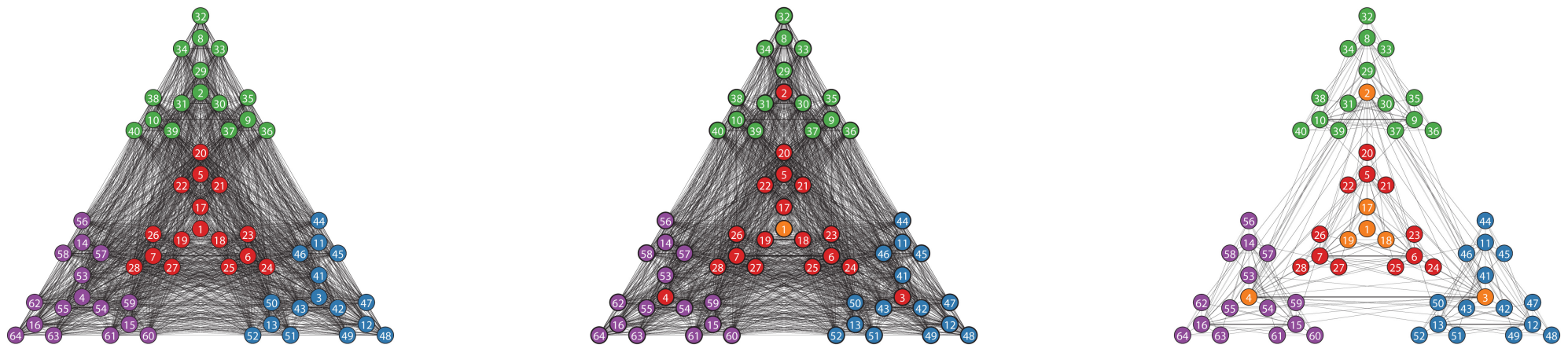

(a) Word coupling (b) Reference coupling (c) Author coupling

**Figure 2.3.: Subdomain detection in idealtypical publication networks**
Communities of publications are detected using the Louvain method. The network obtained from word coupling (a) is maximally dense ($D = 1.00$) and modularity is low ($Q = 0.10$), but the partitioning corresponds to the one used to construct the idealtype. The reference coupling network (b) is not fully connected ($D = 0.97$) because the initial publication is not citing and the solution is slightly worse both quantitatively ($Q = 0.08$) and qualitatively than the one obtained from word coupling. The author-coupled network (c) is least dense ($D = 0.14$) and, therefore, most modular ($Q = 0.63$). Another solution that occurs about once in three runs identifies seven subdomains.

Our approach in constructing the idealtype and studying the case is different. We chose $\Delta t = t_{\max} - t_{\min} + 1$, i.e., our subdomains are time-invariant. Figure 2.3 shows community detection results when publications are coupled through different social facts. Word coupling gives the solution used for constructing the idealtype. Here and later in the case study, subdomains are detected in normalized publication matrices $I^{\mathrm{N}}$. The number of shared facts being constant, similarity increases with a decreasing number of selections. This accounts for different subdomain practices. Integrating over generations (time-invariant clustering) is consequential because it determines the way we gain insights about our case. Essentially, we take a third-level observer position at $t_{\max}$. Whatever boundaries and labels of subdomains result, they are complete histories as wholes. This, however, does not mean that the meaning of subdomains is kept constant because the social facts that are locally selected over time do change. In other words, we delineate the subdomains from an ex post perspective and study how those publications configure their cores over time.

### Constructing Levels of Analysis

Two central concepts of our selection calculus are those of levels of analysis and conditional lifetimes. These concepts are related to fractionally counted selections. We

**Table 2.1.: Usages and conditional lifetimes of words in the idealtype**

| | | $k$ | | | | | $k^{\mathrm{N}}$ | | | | | $K$ | | | | | $K^{\mathrm{N}}$ | | | | | | $\lambda$ | | |
|---|---|---|---|---|---|---|---|---|---|---|---|---|---|---|---|---|---|---|---|---|---|---|---|---|---|
| $\rho$ | $j$ | $t_1$ | $t_2$ | $t_3$ | $t_4$ | $t_\Sigma$ | $t_1$ | $t_2$ | $t_3$ | $t_4$ | $t_\Sigma$ | $t_1$ | $t_2$ | $t_3$ | $t_4$ | $t_\Sigma$ | $t_1$ | $t_2$ | $t_3$ | $t_4$ | $t_\Sigma$ | $\Sigma_\rho$ | **100%** | **75%** | **50%** |
| 1 | 1 | 1 | 3 | 12 | 48 | 64 | 1.0 | 1.5 | 4.5 | 14 | 21 | 100% | 100% | 100% | 100% | 100% | 100% | 50% | 38% | 30% | 33% | 33.2% | 4 | 4 | 3 |
| 2 | 2 | | 1 | 3 | 12 | 16 | | 0.5 | 1.0 | 3.3 | 4.8 | | 33% | 25% | 25% | 25% | | 17% | 8.3% | 6.8% | 7.4% | 40.6% | 3 | 3 | 0 |
| 3 | 3 | | 1 | 3 | 12 | 16 | | 0.5 | 1.0 | 3.3 | 4.8 | | 33% | 25% | 25% | 25% | | 17% | 8.3% | 6.8% | 7.4% | 48.0% | 3 | 3 | 0 |
| 4 | 4 | | 1 | 3 | 12 | 16 | | 0.5 | 1.0 | 3.3 | 4.8 | | 33% | 25% | 25% | 25% | | 17% | 8.3% | 6.8% | 7.4% | 55.5% | 3 | 3 | 0 |
| 5 | 5 | | | 1 | 3 | 4 | | | 0.5 | 1.0 | 1.5 | | | 8.3% | 6.3% | 6.3% | | | 4.2% | 2.1% | 2.3% | 57.8% | 2 | 1 | 0 |
| 6 | 6 | | | 1 | 3 | 4 | | | 0.5 | 1.0 | 1.5 | | | 8.3% | 6.3% | 6.3% | | | 4.2% | 2.1% | 2.3% | 60.2% | 2 | 1 | 0 |
| 7 | 7 | | | 1 | 3 | 4 | | | 0.5 | 1.0 | 1.5 | | | 8.3% | 6.3% | 6.3% | | | 4.2% | 2.1% | 2.3% | 62.5% | 2 | 1 | 0 |
| 8 | 8 | | | 1 | 3 | 4 | | | 0.3 | 0.8 | 1.1 | | | 8.3% | 6.3% | 6.3% | | | 2.8% | 1.6% | 1.7% | 64.2% | 2 | 0 | 0 |
| 9 | 9 | | | 1 | 3 | 4 | | | 0.3 | 0.8 | 1.1 | | | 8.3% | 6.3% | 6.3% | | | 2.8% | 1.6% | 1.7% | 65.9% | 2 | 0 | 0 |
| 10 | 10 | | | 1 | 3 | 4 | | | 0.3 | 0.8 | 1.1 | | | 8.3% | 6.3% | 6.3% | | | 2.8% | 1.6% | 1.7% | 67.6% | 2 | 0 | 0 |
| 11 | 11 | | | 1 | 3 | 4 | | | 0.3 | 0.8 | 1.1 | | | 8.3% | 6.3% | 6.3% | | | 2.8% | 1.6% | 1.7% | 69.3% | 2 | 0 | 0 |
| 12 | 12 | | | 1 | 3 | 4 | | | 0.3 | 0.8 | 1.1 | | | 8.3% | 6.3% | 6.3% | | | 2.8% | 1.6% | 1.7% | 71.0% | 2 | 0 | 0 |
| 13 | 13 | | | 1 | 3 | 4 | | | 0.3 | 0.8 | 1.1 | | | 8.3% | 6.3% | 6.3% | | | 2.8% | 1.6% | 1.7% | 72.7% | 2 | 0 | 0 |
| 14 | 14 | | | 1 | 3 | 4 | | | 0.3 | 0.8 | 1.1 | | | 8.3% | 6.3% | 6.3% | | | 2.8% | 1.6% | 1.7% | 74.3% | 2 | 0 | 0 |
| 15 | 15 | | | 1 | 3 | 4 | | | 0.3 | 0.8 | 1.1 | | | 8.3% | 6.3% | 6.3% | | | 2.8% | 1.6% | 1.7% | 76.0% | 2 | 0 | 0 |
| 16 | 16 | | | 1 | 3 | 4 | | | 0.3 | 0.8 | 1.1 | | | 8.3% | 6.3% | 6.3% | | | 2.8% | 1.6% | 1.7% | 77.7% | 2 | 0 | 0 |
| 17 | 17 | | | | 1 | 1 | | | | 0.5 | 0.5 | | | | 2.1% | 1.6% | | | | 1.0% | 0.8% | 78.5% | 1 | 0 | 0 |
| 18 | 18 | | | | 1 | 1 | | | | 0.5 | 0.5 | | | | 2.1% | 1.6% | | | | 1.0% | 0.8% | 79.3% | 1 | 0 | 0 |
| 19 | 19 | | | | 1 | 1 | | | | 0.5 | 0.5 | | | | 2.1% | 1.6% | | | | 1.0% | 0.8% | 80.1% | 1 | 0 | 0 |
| 20 | 20 | | | | 1 | 1 | | | | 0.3 | 0.3 | | | | 2.1% | 1.6% | | | | 0.7% | 0.5% | 80.6% | 1 | 0 | 0 |
| 21 | 21 | | | | 1 | 1 | | | | 0.3 | 0.3 | | | | 2.1% | 1.6% | | | | 0.7% | 0.5% | 81.1% | 1 | 0 | 0 |
| 22 | 22 | | | | 1 | 1 | | | | 0.3 | 0.3 | | | | 2.1% | 1.6% | | | | 0.7% | 0.5% | 81.6% | 1 | 0 | 0 |
| 23 | 23 | | | | 1 | 1 | | | | 0.3 | 0.3 | | | | 2.1% | 1.6% | | | | 0.7% | 0.5% | 82.2% | 1 | 0 | 0 |
| 24 | 24 | | | | 1 | 1 | | | | 0.3 | 0.3 | | | | 2.1% | 1.6% | | | | 0.7% | 0.5% | 82.7% | 1 | 0 | 0 |
| 25 | 25 | | | | 1 | 1 | | | | 0.3 | 0.3 | | | | 2.1% | 1.6% | | | | 0.7% | 0.5% | 83.2% | 1 | 0 | 0 |
| 26 | 26 | | | | 1 | 1 | | | | 0.3 | 0.3 | | | | 2.1% | 1.6% | | | | 0.7% | 0.5% | 83.7% | 1 | 0 | 0 |
| 27 | 27 | | | | 1 | 1 | | | | 0.3 | 0.3 | | | | 2.1% | 1.6% | | | | 0.7% | 0.5% | 84.2% | 1 | 0 | 0 |
| 28 | 28 | | | | 1 | 1 | | | | 0.3 | 0.3 | | | | 2.1% | 1.6% | | | | 0.7% | 0.5% | 84.8% | 1 | 0 | 0 |
| 29 | 29 | | | | 1 | 1 | | | | 0.3 | 0.3 | | | | 2.1% | 1.6% | | | | 0.7% | 0.5% | 85.3% | 1 | 0 | 0 |
| 30 | 30 | | | | 1 | 1 | | | | 0.3 | 0.3 | | | | 2.1% | 1.6% | | | | 0.7% | 0.5% | 85.8% | 1 | 0 | 0 |
| 31 | 31 | | | | 1 | 1 | | | | 0.3 | 0.3 | | | | 2.1% | 1.6% | | | | 0.7% | 0.5% | 86.3% | 1 | 0 | 0 |
| 32 | 41 | | | | 1 | 1 | | | | 0.3 | 0.3 | | | | 2.1% | 1.6% | | | | 0.7% | 0.5% | 86.8% | 1 | 0 | 0 |
| 33 | 42 | | | | 1 | 1 | | | | 0.3 | 0.3 | | | | 2.1% | 1.6% | | | | 0.7% | 0.5% | 87.4% | 1 | 0 | 0 |
| 34 | 43 | | | | 1 | 1 | | | | 0.3 | 0.3 | | | | 2.1% | 1.6% | | | | 0.7% | 0.5% | 87.9% | 1 | 0 | 0 |
| 35 | 53 | | | | 1 | 1 | | | | 0.3 | 0.3 | | | | 2.1% | 1.6% | | | | 0.7% | 0.5% | 88.4% | 1 | 0 | 0 |
| 36 | 54 | | | | 1 | 1 | | | | 0.3 | 0.3 | | | | 2.1% | 1.6% | | | | 0.7% | 0.5% | 88.9% | 1 | 0 | 0 |
| 37 | 55 | | | | 1 | 1 | | | | 0.3 | 0.3 | | | | 2.1% | 1.6% | | | | 0.7% | 0.5% | 89.5% | 1 | 0 | 0 |
| 38 | 32 | | | | 1 | 1 | | | | 0.3 | 0.3 | | | | 2.1% | 1.6% | | | | 0.5% | 0.4% | 89.8% | 1 | 0 | 0 |
| 39 | 33 | | | | 1 | 1 | | | | 0.3 | 0.3 | | | | 2.1% | 1.6% | | | | 0.5% | 0.4% | 90.2% | 1 | 0 | 0 |
| 40 | 34 | | | | 1 | 1 | | | | 0.3 | 0.3 | | | | 2.1% | 1.6% | | | | 0.5% | 0.4% | 90.6% | 1 | 0 | 0 |
| 41 | 35 | | | | 1 | 1 | | | | 0.3 | 0.3 | | | | 2.1% | 1.6% | | | | 0.5% | 0.4% | 91.0% | 1 | 0 | 0 |
| 42 | 36 | | | | 1 | 1 | | | | 0.3 | 0.3 | | | | 2.1% | 1.6% | | | | 0.5% | 0.4% | 91.4% | 1 | 0 | 0 |
| 43 | 37 | | | | 1 | 1 | | | | 0.3 | 0.3 | | | | 2.1% | 1.6% | | | | 0.5% | 0.4% | 91.8% | 1 | 0 | 0 |
| 44 | 38 | | | | 1 | 1 | | | | 0.3 | 0.3 | | | | 2.1% | 1.6% | | | | 0.5% | 0.4% | 92.2% | 1 | 0 | 0 |
| 45 | 39 | | | | 1 | 1 | | | | 0.3 | 0.3 | | | | 2.1% | 1.6% | | | | 0.5% | 0.4% | 92.6% | 1 | 0 | 0 |
| 46 | 40 | | | | 1 | 1 | | | | 0.3 | 0.3 | | | | 2.1% | 1.6% | | | | 0.5% | 0.4% | 93.0% | 1 | 0 | 0 |
| 47 | 44 | | | | 1 | 1 | | | | 0.3 | 0.3 | | | | 2.1% | 1.6% | | | | 0.5% | 0.4% | 93.4% | 1 | 0 | 0 |
| 48 | 45 | | | | 1 | 1 | | | | 0.3 | 0.3 | | | | 2.1% | 1.6% | | | | 0.5% | 0.4% | 93.8% | 1 | 0 | 0 |
| 49 | 46 | | | | 1 | 1 | | | | 0.3 | 0.3 | | | | 2.1% | 1.6% | | | | 0.5% | 0.4% | 94.1% | 1 | 0 | 0 |
| 50 | 47 | | | | 1 | 1 | | | | 0.3 | 0.3 | | | | 2.1% | 1.6% | | | | 0.5% | 0.4% | 94.5% | 1 | 0 | 0 |
| 51 | 48 | | | | 1 | 1 | | | | 0.3 | 0.3 | | | | 2.1% | 1.6% | | | | 0.5% | 0.4% | 94.9% | 1 | 0 | 0 |
| 52 | 49 | | | | 1 | 1 | | | | 0.3 | 0.3 | | | | 2.1% | 1.6% | | | | 0.5% | 0.4% | 95.3% | 1 | 0 | 0 |
| 53 | 50 | | | | 1 | 1 | | | | 0.3 | 0.3 | | | | 2.1% | 1.6% | | | | 0.5% | 0.4% | 95.7% | 1 | 0 | 0 |
| 54 | 51 | | | | 1 | 1 | | | | 0.3 | 0.3 | | | | 2.1% | 1.6% | | | | 0.5% | 0.4% | 96.1% | 1 | 0 | 0 |
| 55 | 52 | | | | 1 | 1 | | | | 0.3 | 0.3 | | | | 2.1% | 1.6% | | | | 0.5% | 0.4% | 96.5% | 1 | 0 | 0 |
| 56 | 56 | | | | 1 | 1 | | | | 0.3 | 0.3 | | | | 2.1% | 1.6% | | | | 0.5% | 0.4% | 96.9% | 1 | 0 | 0 |
| 57 | 57 | | | | 1 | 1 | | | | 0.3 | 0.3 | | | | 2.1% | 1.6% | | | | 0.5% | 0.4% | 97.3% | 1 | 0 | 0 |
| 58 | 58 | | | | 1 | 1 | | | | 0.3 | 0.3 | | | | 2.1% | 1.6% | | | | 0.5% | 0.4% | 97.7% | 1 | 0 | 0 |
| 59 | 59 | | | | 1 | 1 | | | | 0.3 | 0.3 | | | | 2.1% | 1.6% | | | | 0.5% | 0.4% | 98.0% | 1 | 0 | 0 |
| 60 | 60 | | | | 1 | 1 | | | | 0.3 | 0.3 | | | | 2.1% | 1.6% | | | | 0.5% | 0.4% | 98.4% | 1 | 0 | 0 |
| 61 | 61 | | | | 1 | 1 | | | | 0.3 | 0.3 | | | | 2.1% | 1.6% | | | | 0.5% | 0.4% | 98.8% | 1 | 0 | 0 |
| 62 | 62 | | | | 1 | 1 | | | | 0.3 | 0.3 | | | | 2.1% | 1.6% | | | | 0.5% | 0.4% | 99.2% | 1 | 0 | 0 |
| 63 | 63 | | | | 1 | 1 | | | | 0.3 | 0.3 | | | | 2.1% | 1.6% | | | | 0.5% | 0.4% | 99.6% | 1 | 0 | 0 |
| 64 | 64 | | | | 1 | 1 | | | | 0.3 | 0.3 | | | | 2.1% | 1.6% | | | | 0.5% | 0.4% | 100% | 1 | 0 | 0 |

describe the calculus for the usage of 64 words by 64 publications in the idealtype. Data is given in table 2.1. As defined in the previous section, $k_j$ is the raw number of fact $j$'s selections, $k_j^{\mathrm{N}}$ the fractionally counted number of selections, $K_j$ the fraction of all publications, and $K_j^{\mathrm{N}}$ the fraction of all selections. These counts are determined for time points $t_1$ through $t_4$ and the integrated (historical) time period $t_\Sigma$. For example, word $f_1^{\mathrm{wrd}}$ is used by all 48 publications authored at $t_4$, hence $k_1(t_4) = 48$ and $K_1(t_4) = 48/48 = 100\%$. But the 48 publications that select the word also use other words. Three papers use $f_1^{\mathrm{wrd}}$ as one out of two words, 18 as one of of three, and 27 as one out of four, and therefore $k_1^{\mathrm{N}}(t_4) = 3/2 + 18/3 + 27/4 \approx 14$ and $K_1^{\mathrm{N}}(t_4) \approx 14/48 \approx 30\%$.

We are interested in studying the depth of meaning. The idea of levels is to start from the macro level of all (or many) social facts and home in on the micro level by increasingly filtering out less influential facts. The micro level then holds the core facts of the meaning structure. To construct *levels of analysis* $\theta$, facts are ranked descendingly by their historical fraction of all selections in the full time period, $K^{\mathrm{N}}(t_\Sigma)$. Ranks are indexed $\rho$. By definition, $K^{\mathrm{N}}$ sums to unity over all ranks, shown in the column labeled $\Sigma_\rho$ which fully reads $\Sigma_\rho K^{\mathrm{N}}(t_\Sigma)$. Facts belong to a level $\theta$ if this cumulative fraction $\Sigma_\rho K^{\mathrm{N}}(t_\Sigma) \leq \theta$. In the example, let the macro level be complete, $\theta^{\mathrm{macro}} = 100\%$. When $\Sigma_\rho K^{\mathrm{N}}(t_\Sigma) \leq 100\%$, then $\mathcal{F}^{\mathrm{wrd,macro}} = \{f_j^{\mathrm{wrd}} : j = 1, ..., 64\}$.

Setting thresholds to construct levels is subjective and depends on how large sets of facts are desired to be. We want the meso level to be at $\theta^{\mathrm{meso}} = 75\%$ of all selections. This threshold would set the cutoff after $f_{14}^{\mathrm{wrd}}$, but since $K_j^{\mathrm{N}}(t_\Sigma) \approx 1.7\%$ for $j = 8, ..., 18$, we cut off at a cumulative fraction of 62.5% which results in words $j = 1, ..., 7$. Though of the same generation, facts $j = 5, 6, 7$ are different from facts $j = 8, ..., 16$ because only the formers' parent is the founding word. Finally, core facts on the micro level should account for half of all selections, $\theta^{\mathrm{micro}} = 50\%$, but because of a tie of ranks, $\mathcal{F}^{\mathrm{wrd,micro}} = \{f_1^{\mathrm{wrd}}\}$.

Institutions are operationalized through the lifetimes of facts. Lifetimes differ according to levels of analysis. The *conditional lifetime* $\lambda_j|\theta$ is the number of time points at which fact $j$ has a momentary selection fraction not smaller than the historical selection fraction of the least influential fact at level $\theta$. For core facts to be institutions, they must be selected many times per time point. Mathematically,

$$\lambda_j|\theta = \sum_t \tau(K_j^{\mathrm{N}}(t), K_{\mathrm{min}}^{\mathrm{N}}(t_\Sigma)|\theta), \tag{2.2}$$

where the function $\tau(x, y) = 1$ if $x \geq y$ and 0 otherwise. Because $K_{\mathrm{min}}^{\mathrm{N}}(t_\Sigma)$ is conditional on $\theta$, $\lambda_j$ is too. For example, for the macro level, $K_{\mathrm{min}}^{\mathrm{N}}(t_\Sigma)|\theta^{\mathrm{macro}} \approx 0.4\%$. Since $K^{\mathrm{N}}(t) \geq 0.4\%$ for all facts, macro level lifetimes $\lambda|\theta^{\mathrm{macro}}$, shown in the third column from the right, correspond to the number of time points in which a fact is selected at all. For the meso level, $K_{\mathrm{min}}^{\mathrm{N}}(t_\Sigma)|\theta^{\mathrm{meso}} \approx 2.3\%$. Facts $j = 1, ..., 4$ are always above that fraction, but $j = 5, 6, 7$ only at the first of two time points. Finally, the founding word is only above its historical selection fraction of 33% at the first three time points and, therefore, has a conditional lifetime of three, not four.

Scientific paradigms are social facts that influence the majority of agents in a community during periods of normal science. But these periods are relatively long compared to the short moment when paradigms shift. Paradigms are not only strongly selected but also strongly selected for a relatively long time. They are institutionalized. The final step in constructing levels of analysis is limiting meso and micro levels to institutionalized facts whose lifetimes are not smaller than the average conditional lifetime $\bar{\lambda}|\theta$. Therefore, facts at the micro level are the paradigms of research subdomains. The average lifetime of facts at the meso level constructed so far is $\bar{\lambda}|\theta^{\text{meso}} \approx 2.3$. Since only the first four words have lifetimes larger than that, the final meso level fact set is $\mathcal{F}^{\text{wrd,meso}} = \{f_j^{\text{wrd}} : j = 1, ..., 4\}$. $\mathcal{F}^{\text{wrd,micro}}$ is not changed. The first word is the paradigm of our idealtype.

### Summary

An idealtypical knowledge production process has been presented to exemplify our methods. Direct citation as interface discipline and networks of social facts co-selected by publications as council/arena disciplines are modeled as self-similar growth processes. Subdomains are sets of publications that are structurally equivalent in terms of which social facts they select. Overlap in fact networks is obtained because subdomains of publications translate into link communities of facts with one type of tie per subdomain. Subdomains are time-invariant, i.e., they are wholes from the perspective of the last observational point in time. Selection years are counted to determine the lifetimes of social facts. Lifetimes are conditional on levels of meaning. The more micro the level of analysis, the higher selection fractions must be to count. Levels serve to screen the depth of meaning structures. At the micro level, facts resemble scientific paradigms which are highly selected and institutionalized.

## 2.2.2. Research Questions and Design

The scaling hypothesis states that power laws are signatures of a system at or near a critical phase transition and, in the context of social systems, it provides a framework to model identities as a process between stability and change. The idealtypical knowledge production process of the previous section was introduced for three reasons. First, it allows to demonstrate various scaling laws that can be used to diagnose identities. Second, it visualizes the different types of networks and meaning structures often subsumed under the label of "social networks." Third, it serves as a benchmark for later empirical results for Social Network Science. As we study our case, we will encounter operationalizations of, or quantitative approaches to, five senses of identity. Abbott's (2001a) fractal distinctions serve as a heuristic throughout large parts of our analysis.

Size distributions like Bradford's, Zipf's, and Lotka's Law are the most-often diagnosed signs of self-similarity. In our case, they are the litmus test for the scaling hypothesis. In the idealtype, selection of all three types of facts grows exponentially (figure 2.4a) but, since only the practices of citation and word usage are governed by (linear) preferential attachment (2.4h), only Bradford's and Zipf's Law emerge (2.4b). The first-mover advantage fully plays out culturally in this deterministic network process. The third chapter has five sections. The first is concerned with delineating the set of publications as our em-

**Figure 2.4.: Statistics of idealtype**

Statistics on the following page are shown for authorship (boxes), citation (ellipses), and word usage (diamonds). The first author, publication, and word are often not perfectly fitted because of their foundational role for the domain, but fits are perfect if cumulative distributions are studied. All distributions and plots shown are non-cumulative. (a) The number of social facts grows exponentially. (b) Citation and word usage is scale-invariant with $\alpha = 1$ but authorship decays exponentially (dashed line). (c) The co-authorship network is not scale-free because degrees also decay exponentially (dashed line). (d) But it is hierarchically modular as indicated by an inverse scaling relationship of author degree and clustering coefficient in the $t \to \infty$ limit (details in the text). Gray nodes result from increasing $t$ and black nodes are averages from logarithmic binning. While (d) indicates the existence of clustering, (e) shows that network diameter, and, consequently, average path length, grows logarithmically (dashed line) with network size, so co-authorship occurs in a small world. (f) Over time, the number of co-authorship edges scales with the number of author nodes. The edge-per-node ratio is constantly 6/4 so a giant component can emerge, but modules are only connected in the co-authorship network integrated over time. (g) The autocatalytic reproduction of co-authorship ties is not more likely in the core where degrees are large. (h) Linear preferential attachment is present for the selection of words and references but is absent for authors. (i) The lifetime of facts decays exponentially (dashed line) which is a direct consequence of exponential growth (a) and the not-forgetting of facts.

pirical basis. The second describes the three levels of analysis. In the last three sections, stability and change in Social Network Science is diagnosed from the three perspectives of events, collaboration, and criticality, each guided by a research question.

#### Events

The first perspective is most directly narrative and engages Social Network Science as a 3-identity, path, or sequence of events. It is empirically associated with the concepts of interface discipline, quasi-equilibrium, and punctuations. The goal is to identity fractal distinctions if they exist. *Does Social Network Science change along diverging and converging paths?* Network subdomains split when intellectual schools proceed along different paths, and they merge when schools with different pasts find common ground. A fractal distinction requires an identity's distinction to be revoked in time. A selection of a story set that had been cast outside before is an event of convergence. Kuhnian normal science terminates in an extreme punctuation. As Bayesian forks, the latter effectively erase structural memory and open up new opportunities of action and selection (H. C. White, 1995; Mische & White, 1998; Chavalarias & Cointet, 2013). We want to know if punctuations can be identified in Social Network Science that resemble major shifts, especially the Complexity Turn.

To operationalize 3-identity and uncover the narratives of Social Network Science, we resort to historiographic bibliometrics (Garfield, 1955; Hummon & Carley, 1993; Garfield,

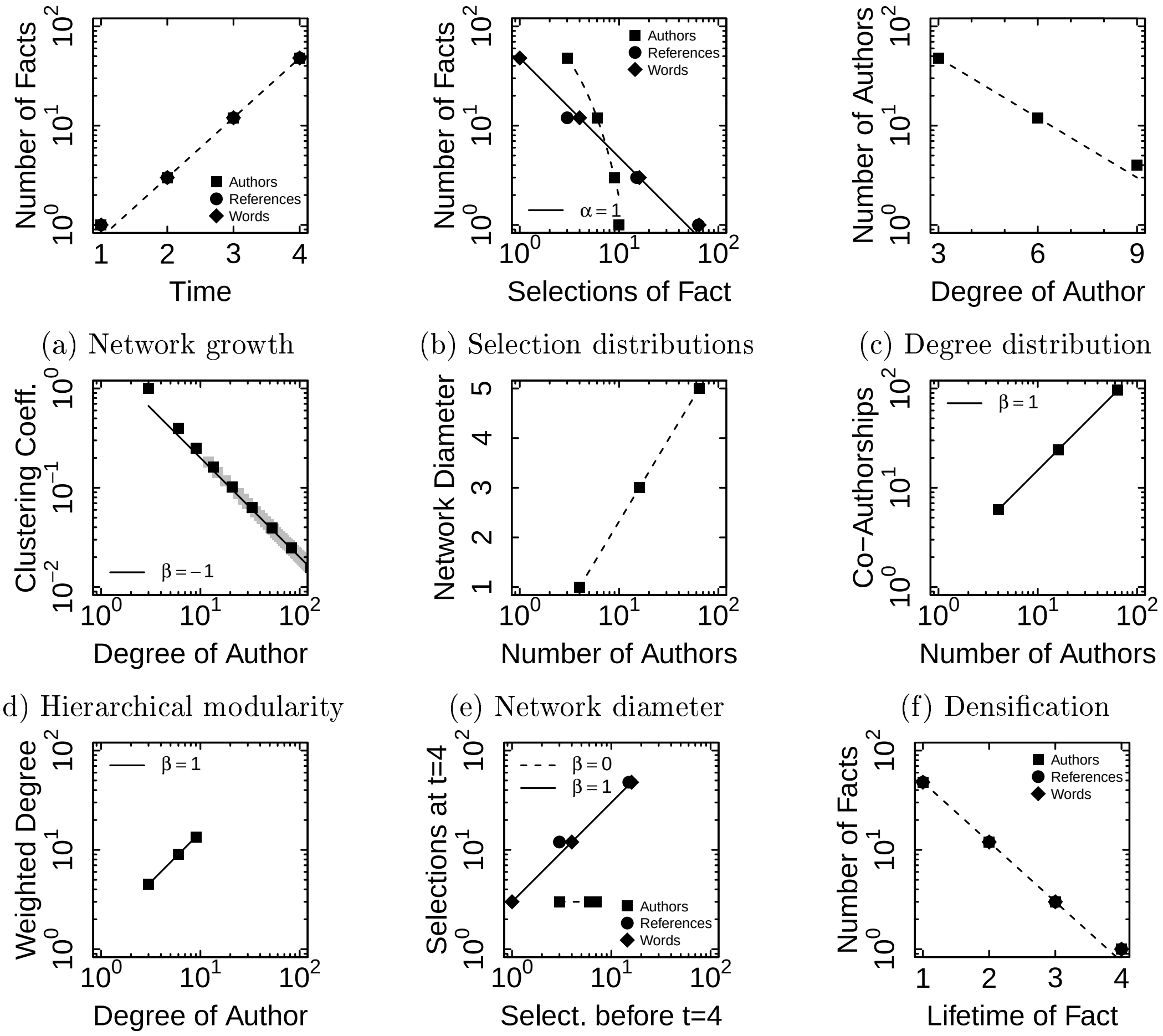

(a) Network growth (b) Selection distributions (c) Degree distribution
(d) Hierarchical modularity (e) Network diameter (f) Densification
(g) Autocatalysis (h) Preferential attachment (i) Lifetime distribution

**Figure 2.4.: Statistics of idealtype**

2004), precisely, the analysis of citation paths as introduced by Hummon and Doreian (1989) and developed into the search-path-count method by Batagelj (2003). Direct citation networks of publications/references are interface disciplines that, in retrospective, depict the historical search efforts in a domain. Given such a network, the algorithm counts the different search paths that traverse a particular citation. Details are given in the caption of figure 2.5. Like the way we have partitioned papers into subdomains, search path counting implies that we take a third-level observer position at $t_{\max}$ and treat history as a whole.

The evolution of the idealtype is clearly not convergent because no paths merge over time. The way knowledge production is modeled resembles a Durkheimian division of labor with scientific communities specializing in parallel. In fact, all that unifies the four subdomains is citing the foundational publication and using the initial word (cf. figure 2.2). Boundaries become decreasingly permeable with time. Contrary to the idealtype, fractal distinctions exhibit Bayesian forks. Paradigm shifts and weaker punctuations show as moments in time where past citation paths converge and future paths diverge. search-path-count analysis is an appropriate method because a citation relation's weight is contingent on its position in a chain of citations.

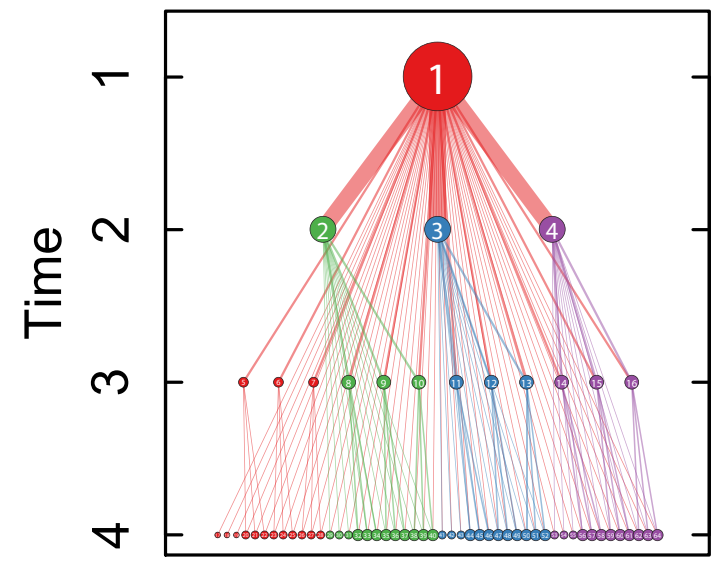

**Figure 2.5.: Narrative tree of idealtype**
Historiographic citation analysis uncovers the evolutionary lineages of thought. In the genealogical tree of the idealtype, citations flow from the bottom to the top publications. Node colors show subdomain affiliation and citations are colored by subdomain of target reference. Arrows are not shown for clarity. Technically, an artificial source $p_{\text{src}}$ is added that cites all references not getting cited, and a sink $p_{\text{snk}}$ is added that is getting cited by all publications not citing a reference. The citation weight $w^{\text{search}}$ is the number of different search paths from $p_{\text{src}}$ to $p_{\text{snk}}$ that traverse a particular citation of $p_j$ by $p_i$. For example, $w^{\text{search}}(5, 1) = 3$ because three paths come in from $p_{20}$, $p_{21}$, and $p_{22}$. The publication size is the number of different search paths from $p_{\text{src}}$ to $p_{\text{snk}}$ that traverse a particular reference $p_i$. The idealtype does not exhibit fractal distinctions because paths only diverge.

### Collaboration

The second perspective tackles social structure operationalized through co-authorship. It engages the spatial side of fractal distinctions and pursues an operationalization of self-conscious 4-identity. Main concepts are council and arena discipline, structural cohesion, and autocatalysis. A network that results from fractal distinctions would be multifunctional in a single core – it would have an open elite (Moody & White, 2003; D. R. White et al., 2004; Powell et al., 2005). Networks at the phase transition are also hierarchically modular or discretely scale-invariant small worlds because these architectures are optimal in terms of connectivity, cost efficiency, searchability, and diversity (Watts & Strogatz, 1998; Kleinberg, 2000; Ferrer i Cancho & Solé, 2003; Solé & Valverde, 2004; Zhou et al., 2005; D. R. White et al., 2006). *Does Social Network Science have an open-elite hierarchically modular small-world structure?*

The co-authorship idealtype is not scale-free (figure 2.4c) which is expected, as the literature shows. But it resembles a perfect hierarchically modular small-world council discipline, indicated by discrete scale invariance with a factor of four and an inverse scaling relationship of clustering coefficient versus degree (2.4d). We measure hierarchical modularity for co-authorship networks integrated over time. In the scaling relationship of equation A.5, $x$ is an author's degree and $y$ is its transitivity-based clustering coefficient (Ravasz & Barabási, 2003; Humphries & Gurney, 2008).

Diversity is organized hierarchically through short paths (figure 2.4e) with cores or invisible colleges at different levels. The whole network at $t = 4$ is coordinated by the most prestigious (highest degree) authors $f_1^{\text{aut}}$ through $f_4^{\text{aut}}$. This cluster is truly diverse, i.e., the core is an open elite. But if we zoomed into, e.g., the red subdomain and identify communities there, authors $f_1^{\text{aut}}$ and $f_5^{\text{aut}}$ through $f_7^{\text{aut}}$ would coordinate a similarly diverse structure. Homophily, the attachment mechanism of arena discipline, acquires a different meaning at different levels of organization. Because the idealtype does not emerge in fractal distinctions, its core/periphery structure is hierarchically flat and presents itself level by level. To probe Social Network Science for fractal distinctions in meaning space, we will study the co-authorship network using the structural-cohesion model described in section 1.3.1 (D. R. White & Harary, 2001; Moody & White, 2003; Moody, 2004).

The emergence of a network core presupposes densification. Density and structural cohesion are outcomes of the collective dynamics of control. In this process, structure acts as a memory. We operationalize Durkheim's collective consciousness and White's 4-identity through the *densification exponent*. The parameter is a measure for the percolation strength of 4-identity, it quantifies the extent to which an identity is collectively finding footing (Durkheim, 1982 [1895], ch. 1; Leskovec et al., 2005; H. C. White, 2008, ch. 1; Bettencourt et al., 2009; Bearman & Stovel, 2000). Practically, we estimate the exponent $\beta$ in the scaling relationship of equation A.5 where $x(t)$ is the number of authors active at $t$ and $y(t)$ is the number of undirected edges (excluding self-loops) in the author co-authorship network $H^{\text{N}}$ at $t$ (Leskovec et al., 2005). The idealtype's densification exponent is $\beta = 1$ (figure 2.4f). Knowledge production progresses by continuous growth and teaching. At each time step but the first, a set of unconnected quads is borne, but it is only in the cumulative network that global connectivity emerges, facilitated through levels.[6] The process is permanently critical, it is permanently at the percolation threshold, because no two authors co-author a paper at a second time step, co-authorship ties are not reproduced.

Autocatalysis is the (co-)reproduction of social facts (D. R. White et al., 2006; Padgett et al., 2012; Padgett, 2012b). This concept empirically addresses structure as a memory needed to reproduce ties. A cross-sectional scaling law (figure 2.4g) demonstrates that macro level autocatalysis in idealtype authorship is also critical. Now, $x$ in equation A.5 is an author's degree and $y$ is its weighted degree, self-reproductions excluded. $\beta = 1$ means that the autocatalytic strength of an average co-authorship tie is the same for authors with small and large degrees. In Granovetter's terms, the *autocatalysis exponent* tells to which extent strong ties exist in cores where degrees are large rather than in peripheries.

[6] In time, the cumulative number of co-authorship edges scales as a shifted power law with the number of author nodes: edges = 2(nodes − 1). If a non-shifted relationship is fitted, scaling converges from $\beta > 1$ to linearity for $t \to \infty$.

### Criticality

The third perspective sheds light onto the role of the Matthew Effect and involves an operationalization of style or 5-identity. According to our model, styles are learned sensibilities which fact to select from the public, and they result in power-law size distributions (Adamic et al., 2001; Reali & Griffiths, 2010). Identities self-organize to the percolation threshold where stability and change are balanced. We do not only want to know if power laws emerge, but we also want to uncover the dynamics of change that stochastic preferential attachment brings. *Does, and do the subdomains of, Social Network Science continuously reconfigure their network cores?* To answer this question, we borrow a concept from Darwin (2003 [1859]). Fit social facts acquire increasing shares of influence in a population by getting selected. A fact's fitness is the ability to outreproduce others. To not get outreproduced or even displaced by fitter ones, facts must continuously maintain their fitness. But facts are not agentic, they acquire their fitness through selections by agents, publications in our case. They are the 'traits' that do or do not prevail in a population of identities.

The classical bibliometric way to measure the fitness of scientific ideas is to count how often a publication is cited and divide that count by the number of citations an average publication in the same field receives (Van Raan, 2006). Since citation distributions are highly skewed and often scale-invariant, this method is flawed. Realistic approaches to measuring fitness must not be decoupled from the Matthew Effect as the prime mechanism how systems self-organize. Preferential attachment is autocatalytic because a fact's selection success breeds further success. In the original Barabási/Albert model of linear preferential attachment, all nodes are equally fit on average – they have identical growth rates (Barabási & Albert, 1999). This is given for references and words in our idealtype (figure 2.4h). Only when fitness modulates preferential attachment to include a fit-get-richer effect, identities can outreproduce others (Bianconi & Barabási, 2001). To predict the scientific impact of references, D. Wang et al. 2013 empirically determine one fitness score over the whole lifetime of a publication. Keeping fitness constant amounts to a retrospective third-level observation. While this allows to predict future impact with good precision, it is incompatible with the Darwinian meaning of fitness as a property that varies as species or facts adapt to changing environments or contexts.

Empirically, we proceed in two steps. First, we measure the extent to which the Matthew Effect governs social selection. Recall from section 1.3.2 that the attachment exponent is determined by predicting the number of present selections from the number of past selections through a scaling law. Prediction is possible because interface disciplines produce stability through orderly reproduction. Because only linear preferential attachment generates power laws and drives an identity to the critical point, the Matthew exponent $\beta$ quantifies the strength of the 5-identity. Only at $\beta = 1$ a style optimizes control. In a second step, we reconstruct the second-level observational adaptive walks story sets perform in meaning structures over time, i.e., we measure the fitness of social

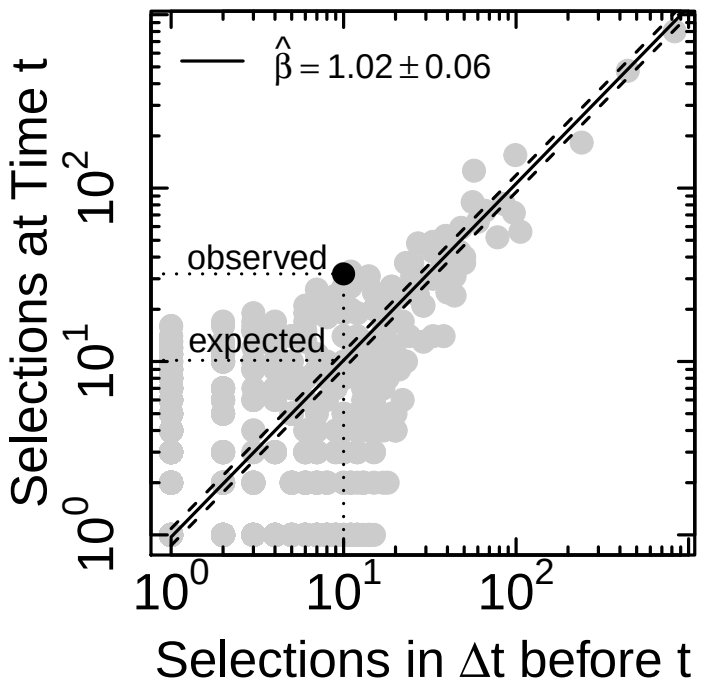

**Figure 2.6.: Measuring the fitness of social facts**

The fitness of a social fact is a function of past and present selection and quantifies the deviation from expectation according to preferential attachment. Gray data points show a scaling relationship synthesized from $y \propto x$ with noise added in both variables. Each data point corresponds to the number of selections of a social fact in the time period $\Delta t$ before $t$ (x) and at time $t$ (y). Linear preferential attachment is recovered by a fit of a scaling law with $\hat{\beta} = 1.02 \pm 0.06$. The black point is a fact whose fitness is to be determined. It was selected ten times before $t$ and 32 times at $t$, but due to linear preferential attachment, its expected number of selections at $t$ is also ten. Therefore, its fitness $\eta = 32/10$, i.e., the fact is 3.2 time fitter than an average fact with similar past selection success. Mathematical details are given in the text. Dashed lines depict uncertainty in the estimation of the normalization constant $D$.

facts selected by styles. Individual fitness scores are measured in the process of determining the Matthew exponent of collective preferential attachment. In words, fitness is the observed number of a fact's selections at present divided by the number of selections that can be expected based on past selections. Fitness is not predictable, it quantifies the error in predicting selection success. The procedure to measure the Matthew exponent and fitness is described in figure 2.6 for full counting of selections. In our case study, we will measure fitness using fractional counting to ensure comparability of subdomains. Mathematically, the fitness $\eta$ of a fact $j$ at time $t$ is

$$\eta_j(t) = \frac{k_j^{\mathrm{N}}(t)}{D(t-\Delta t) k_j^{\mathrm{N}}(t-\Delta t)^{\beta}}, \tag{2.3}$$

where $D$ is a normalization constant, $\beta$ is the preferential-attachment or Matthew exponent, and $\Delta t$ is the size of the memory before $t$ used for prediction. Fitness scores are in the $[0, \infty[$ interval. 0 means that a fact is not getting selected anymore, and 1 (2) means that it is getting selected (twice) as many times as is expected for an average fact with the same number of past selections. Uncertainty in fitting a scaling law to the data, shown as dashed lines in the figure, is used to construct uncertainty intervals for the fitness scores.

## Summary

In this final methodological section, our research design was laid out. In five empirical sections we will study five senses of identity. We follow three research questions. The first question is if Social Network Science changes along diverging and converging paths. 3-identities operationalized as citation paths will be studied for merging events that resemble fractal distinctions. The idealtype exhibits divergent evolution or a division of labor. The second question is if the author co-authorship network has an open-elite hierarchically modular small-world structure. The idealtype has these properties that are expected when a network is a result of fractal distinctions. The percolation strength of 4-identities will be measured through network densification. The third question is if Social Network Science and its subdomains continuously reconfigure their network cores. The strength of 5-identities is operationalized through the Matthew exponent of preferential attachment and the result of styles as changes in the fitness of social facts. Fitness is the error in predicting selection success.

# 3. Social Network Science

## 3.1. Delineation

Social Network Science or parts of it have been analyzed in a number of studies (Hummon & Carley, 1993; Freeman, 2004; Garfield, 2004; Shibata et al., 2007; Lazer, Mergel, & Friedman, 2009; Brandes & Pich, 2011; Freeman, 2011; Lancichinetti & Fortunato, 2012; Batagelj & Cerinšek, 2013). The best bibliographic dataset is the SN5 dataset (Batagelj & Cerinšek, 2013), retrieved from the *Web of Science* in 2007. It contains papers that use the search term `SOCIAL NETWORK*` in either title, abstract, author keywords, or *KeyWords Plus* or have been published in the journal *Social Networks* plus the "most frequently cited works" of those papers.[1] Because SN5 is outdated and also contains papers that talk of "social networks" metaphorically, we decided to delineate Social Network Science anew by combining the hybrid retrieval methods of Zitt and Bassecoulard (2006) and Mogoutov and Kahane (2007).

The difficulty is to draw a sharp retrieval boundary that is permeable in reality. If the boundary is allowed to be very uncertain, then we can allow many more publications into the domain than when the boundary is supposed to be very certain. We assume that relevant papers not necessarily use `SOCIAL NETWORK*` in title, abstract, or author keywords. On the other hand, not all papers using that 2-gram should automatically be inside the boundary. But all relevant papers should use the words `SOCIAL` and `NETWORK*`. Therefore, we created two datasets: the *public* contains 44,307 publications using `SOCIAL` and `NETWORK*`. These are papers that are at least remotely relevant for Social Network Science. The *seed* is a subset of the public and contains 23,568 papers using `SOCIAL NETWORK*`. These sets will be used to create the *solution* set. Publication years in the public range from 1953 to 2014.[2]

This data was then pre-processed to be analyzable. Author names were disambiguated semi-automatically for the solution. Each publication and reference was transformed into a key such that a cited reference can be matched to a citing publication. Granovetter's (1973) paper, e.g., has the key `GRANOVET_1973_A_1360`. All titles, abstracts, and author keywords were preprocessed and stemmed. All words (incl. $n$-grams or words that consist of $n$ words) used by at least one author in the seed as a keyword represent the vocabulary. A word of the vocabulary is selected by a publication if it is used in either title, abstract, or author keywords (for details see appendix B).

[1] The dataset can be downloaded at vlado.fmf.uni-lj.si/pub/networks/pajek/WoS2Pajek/WoS2Pajek.htm, visited May 1st, 2015.

[2] Sets were retrieved on November 6th, 2013, from the *Web of Science* online interface www.webofknowledge.com. Publications were delivered for the products *Science Citation Index Expanded* (publication years since 1900), *Social Sciences Citation Index* (since 1900), *Arts & Humanities Citation Index* (since 1975), *Conference Proceedings Citation Index – Science* (since 1990), and *Conference Proceedings Citation Index – Social Science & Humanities* (since 1990). The query for the public was `TS=(SOCIAL and (NETWORK or NETWORKS))` and that for the seed was `TS=(SOCIAL NETWORK or SOCIAL NETWORKS)`. That means, papers on "social networking" are excluded. Publications where the search terms only occur as *KeyWords Plus* were filtered out ex post because these keywords were found to be unreliable. Results include all document types (articles, reviews, letters, editorials, corrections, etc.).

Based on the description of Scott (2012), discussed in section 1.1.3, we expect a social psychological path with a strong graph-theoretical focus, a diverging ethnographical lineage, a structuralist narrative following the Harvard breakthrough, and a recent development driven by physics. These paths belong to different scientific disciplines with different styles of practice. Therefore, we do not delineate Social Network Science as if it had one core, but we delineate the domain's subdomains that revolve around different story sets. We use the procedure proposed by Zitt and Bassecoulard (2006) because it delineates an identity in a very sociological way by identifying its story set and retrieving the building blocks or 1-identities that select these facts. This method is enhanced through Mogoutov and Kahane's (2007) focus on subdomains. Building on the seed and public sets, the delineation steps are the following. First, publications in the seed $A$ are partitioned into subdomains $A_j$. Second, for each subdomain, the cited and used cores $C_j$ are identified. From $C_j$, unspecific facts are removed to arrive at the sets $D_j$. To compute specificity, the Zitt/Bassecoulard method requires knowledge about which facts in $C_j$ are selected how many times in all of science $S$. Since had not had access to the full *Web of Science* at the time the domain was delineated, the decision was made to let the public set of at least remotely relevant papers be a proxy for $S$. Then, publications $E_j$ in $S$ which select facts in specific cores $D_j$ are identified. The third step is originally not intended by the Zitt/Bassecoulard method. It consists of taking the union of $E_j$, partitioning this interim solution into subdomains $A'_j$, and adding the cores $C'_j$ cited by $A'_j$ to the interim solution. Adding the most cited references to the citing publications is supposed to aid a more complete citation path analysis.

## Partitioning the Seed

Publications in the seed are transformed into a time-invariant normalized publication matrix, i.e., similarity scores are constructed for all publications from all years through structural equivalence of selections. For the idealtype, figure 2.3 reveals that communities differ when the similarity of publications used for clustering is constructed using different social facts. Figure 3.1 shows that the size distributions of facts are power laws except for the word distribution which is still very skewed. Concentration increases from authorship over citation to word usage. Only the latter does not show scaling behavior, leading to the conjecture that the way natural language was processed did not produce a natural representation of the system. The power-law skewness is an important condition for the validity of our identity model. Regarding the delineation, it means that, per practice, a set of few social facts exists that many publications select. We will only have to select few social facts to bound the subdomains, i.e., the procedure will be efficient.

A smaller exponent for keyword usage than for citation means that the publications

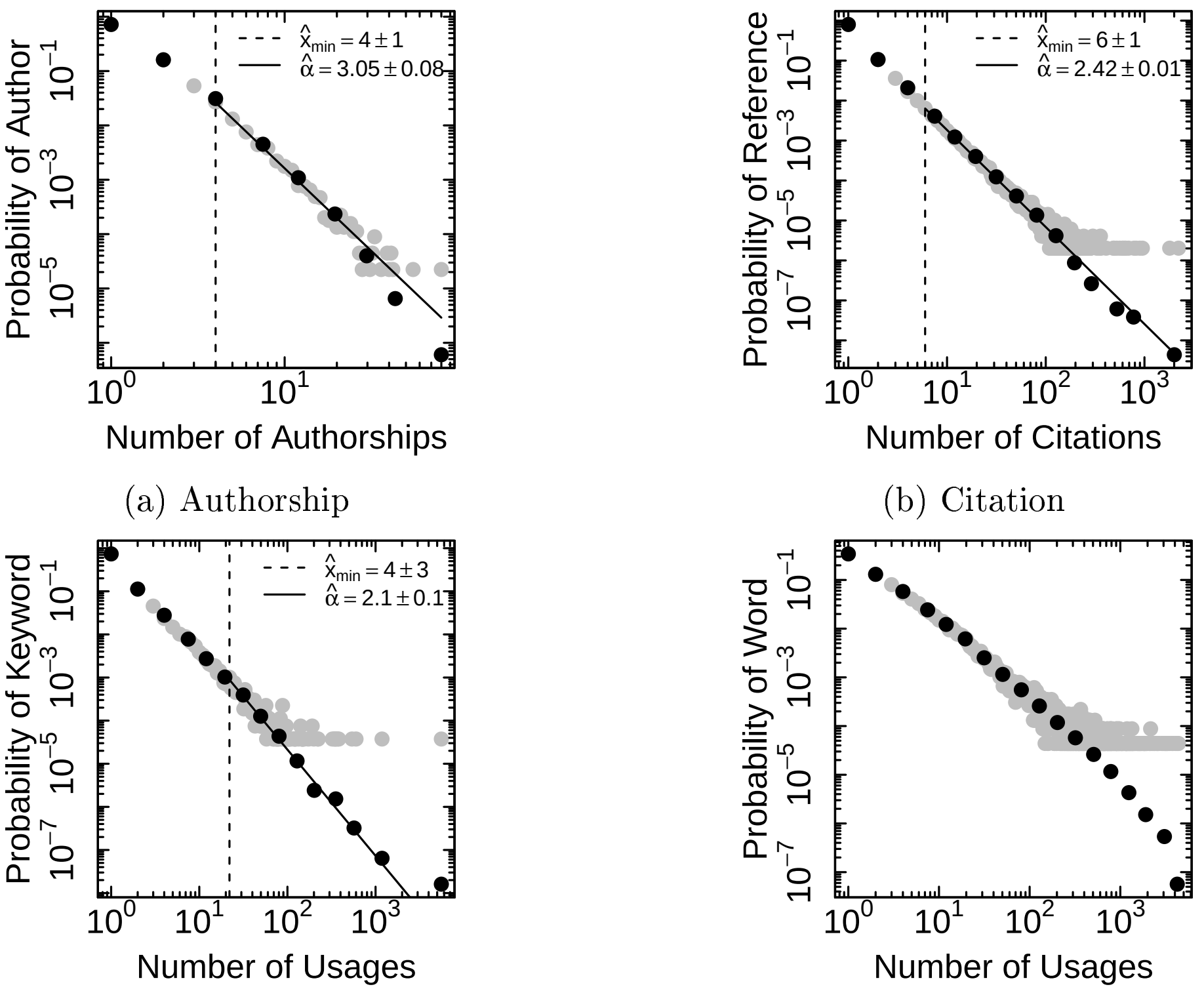

(a) Authorship

(b) Citation

(c) Author keyword usage

(d) Word usage

**Figure 3.1.: Seed size distributions**

For authorship ($p = 0.74$), citation ($p = 0.90$), and keyword usage ($p = 0.79$), power laws are reasonable fits ($p > 0.1$) above a lower cutoff $x_{min}$ marked with a vertical dashed line. Exponents for the power-law distributions differ and ranges are consistent with those reported in the literature. For authorship (a), $\alpha \leq 3$, i.e., infinite variance cannot be excluded. The exponent of citation (b) is larger than that of author keyword usage (c), i.e., concentration to few keywords is stronger than to few references. A hypothetical power-law fit for word usage (d) with $\alpha = 1.74 \pm 0.01$ must be rejected due to curvature ($p = 0.00$). But an average word usage $\bar{x} = 30 \pm 156$ is rather meaningless, as indicated by the giant standard deviation. Occasionally, as in (c), the parameters estimated from the data (cutoff depicted at $x = 22$) deviate from the ones estimated by bootstrapping ($\hat{x}_{min} = 4 \pm 3$).

which select the most-used word are more than those which select the most-cited reference. Accordingly, we see in table 3.1 that networks from author coupling are two orders of magnitude more sparse than from reference coupling and three orders more than when words are used. The three coupling methods are also differently distinctive. We reject authors for coupling publications because only 41.1% of all publications are at least indirectly connected through authors. A modularity of $Q_{\mathrm{R}} = 0.46$ for reference coupling means that these facts are less distinctive than authors ($Q_{\mathrm{A}} = 0.96$) but more than words ($Q_W = 0.14$). In any case, all further results are contingent on our choice of using references, words, or both for coupling papers. In table 3.2, the three solutions are compared. They are robust in that they all detect five non-trivial communities which can be labeled Social Psychology (SP), Economic Sociology (ES), Social Network Analysis (SNA), Complexity Science (CS), and Web Science (WS) by examining the top subject categories and social facts. Only reference coupling results in two subdomains for Social Network Analysis. Subdomains from different fact coupling also results in the same temporal ordering. Social Psychology is the oldest subdomain and Web Science is the newest.

Subdomain boundaries are fairly permeable because (a) consensus in table 3.1 are as low as 0.92 and (b) the choice of facts for coupling has a large effect on the size of the subdomains. Social Psychology is much larger when delineated lexically or the hybrid way. To quantify the consensus in community detection, we use the adjusted Rand index. It counts similarly partitioned pairs of publications and compares the result to a null hypothesis. 1 means that two solutions are identical. The index is highly dependent on the number of clusters. We arrive at means and standard deviations by comparing the solutions of ten runs. The consensus of reference coupling is $0.95 \pm 0.02$ and $0.97 \pm 0.01$ of word coupling (table 3.1), but the inter-coupling consensus of reference and word coupling is only $0.243 \pm 0.002$. Since we are lacking a gold standard, we do not have an objective criterion to evaluate the clustering solutions. We chose hybrid publication

**Table 3.1.: Seed subdomain statistics for different coupling methods**

Rows represent methods of coupling publications through authors (A), cited references (R), used words (W), and combinations thereof. Network density increases from top to bottom, the publication co-authorship network is three orders of magnitude sparser than when publications are coupled through word co-usage. In all networks that involve word coupling, every second edge is actually realized. Accordingly, the size of the largest (giant) component is only 41.1% for author coupling but 99.6% for word coupling. Modularity $Q$ decreases with density. Consensus (means and standard deviations) tells how much clustering solutions are reproducible. We use the adjusted Rand index, a standard measure for the similarity of two clustering solutions, to compute consensus. A score of 1 means that two solutions agree perfectly. For hybrid clustering, tie strengths are averaged.

| | **Network** | | **Clustering** | | **Comp.** |
|---|---|---|---|---|---|
| | **Density** | **Giant** | **Modularity** | **Consensus** | **Time [min]** |
| **A** | 0.0002 | 41.1% | 0.96 | $0.78 \pm 0.03$ | $\ll 1$ |
| **R** | 0.0332 | 93.0% | 0.46 | $0.95 \pm 0.02$ | 2 |
| **AR** | 0.0333 | 96.1% | 0.60 | $0.92 \pm 0.02$ | 2 |
| **W** | 0.5315 | 99.6% | 0.14 | $0.97 \pm 0.01$ | $37 \pm 2$ |
| **AW** | 0.5315 | 99.7% | 0.14 | $0.79 \pm 0.18$ | $35 \pm 2$ |
| **RW** | 0.5414 | 99.8% | 0.14 | $0.92 \pm 0.14$ | $43 \pm 4$ |
| **ARW** | 0.5414 | 99.9% | 0.15 | $0.81 \pm 0.17$ | $45 \pm 2$ |

**Table 3.2.: Seed subdomain clusters from different coupling methods**
Columns are publications coupled through references, words (lexical), or both references and words (hybrid coupling). Each subdomain solution per method is described by its size in publications, publication year quartiles, top 5 subject categories, and top 5 social facts. Numbers are selections $k$, and ranks are given in brackets. Rankings are based on $tf * idf$ scores where $tf = k$ is the number of selections in a subdomain and $idf = \log(1/K)$ is the logarithm of the inverse fraction of selecting publications in the whole domain. Subdomain clusters for different coupling methods are quite robust in terms of fact rankings. Interpretation and comparison of subject categories and social facts allows a coherent labeling of five subdomains (shown on this and the following four pages): (a) Social Psychology (SP), (b) Economic Sociology (ES), (c) Social Network Analysis (SNA), which is split into two for reference coupling, (d) Complexity Science (CS), and (e) Web Science (WS). Subdomains are listed by median publication year. Publications are classified into multiple subject categories in the *Web of Science* database.

(a) Social Psychology

| | **Citation-based** | **Lexical** | **Hybrid** |
|---|---|---|---|
| **Publications** | 5,088 | 6,885 | 6,850 |
| **Year Quartiles** | 1999/2007/2011 | 2001/2008/2011 | 2001/2008/2011 |
| **Subject Category** | | **Classifications** | |
| Public, Env. & Occup. Health | 1,041 (1) | 1,197 (1) | 1,216 (1) |
| Psychiatry | 732 (2) | 753 (2) | 761 (2) |
| Gerontology | 433 (3) | 536 (3) | 539 (3) |
| Psychology, Multidisciplinary | 380 (4) | 440 (4) | 445 (4) |
| Social Science, Biomedical | | | 361 (5) |
| Psychology, Developmental | | 345 (5) | |
| Psychology, Clinical | 304 (5) | | |
| **Author Keyword** | | **Usages** | |
| `SOCIAL_SUPPORT` | 527 (1) | 561 (1) | 567 (1) |
| `DEPRESSION` | 143 (2) | 145 (2) | 145 (2) |
| `HIV` | 117 (3) | 118 (5) | 121 (5) |
| `SOCIAL_CAPITAL` | | 190 (3) | 177 (3) |
| `GENDER` | | 129 (4) | 131 (4) |
| `QUALITY_OF_LIFE` | 99 (4) | | |
| `AGE` | 95 (5) | | |
| **Reference** | | **Citations** | |
| `BERKMAN_1979_A_186` | 334 (1) | 322 (1) | 331 (1) |
| `COHEN_1985_P_310` | 252 (2) | 257 (2) | 261 (2) |
| `RADLOFF_1977_A_385` | 234 (3) | 230 (3) | 233 (3) |
| `HOUSE_1988_S_540` | 219 (4) | 224 (4) | 227 (4) |
| `GRANOVET_1973_A_1360` | | 390 (5) | 346 (5) |
| `COBB_1976_P_300` | 173 (5) | | |
| **Author** | | **Authorships** | |
| `LATKIN,_CA` | 51 (1) | 47 (1) | 49 (1) |
| `CHRISTAKIS,_NA` | 31 (4) | 43 (2) | 40 (2) |
| `LITWIN,_H` | 32 (2) | 34 (3) | 34 (3) |
| `LATKIN,_C` | 28 (3) | 27 (4) | 28 (4) |
| `BERKMAN,_LF` | 26 (5) | 25 (5) | 25 (5) |

**Table 3.2.: Seed subdomain clusters from different coupling methods**

(b) Economic Sociology

| | **Citation-based** | **Lexical** | **Hybrid** |
|---|---|---|---|
| **Publications** | 4,983 | 3,963 | 3,780 |
| **Year Quartiles** | 2006/2009/2011 | 2005/2009/2011 | 2005/2009/2011 |
| **Subject Category** | | **Classifications** | |
| Management | 789 (2) | 590 (1) | 598 (1) |
| Sociology | 856 (1) | 511 (2) | 514 (2) |
| Business | 538 (3) | 426 (3) | 430 (3) |
| Economics | 448 (4) | 263 (5) | 263 (5) |
| Geography | 284 (5) | 237 (4) | 245 (4) |
| **Author Keyword** | | **Usages** | |
| SOCIAL_CAPITAL | 467 (1) | 230 (1) | 249 (1) |
| MIGRATION | 62 (5) | 50 (3) | 49 (2) |
| COMMUNITY | | 66 (2) | 56 (4) |
| INNOVATION | | 52 (4) | 53 (3) |
| ENTREPRENEURSHIP | | 42 (5) | 42 (5) |
| SOCIAL_NETWORK_ANALYSIS | 217 (2) | | |
| TRUST | 92 (3) | | |
| GENDER | 78 (4) | | |
| **Reference** | | **Citations** | |
| GRANOVET_1973_A_1360 | 1502 (1) | 597 (1) | 625 (1) |
| BURT_1992_STRUCTURAL | 762 (2) | 390 (2) | 415 (2) |
| PUTNAM_2000_BOWLING | 606 (3) | 241 (4) | 264 (4) |
| GRANOVET_1985_A_481 | 484 (4) | 302 (3) | 321 (3) |
| COLEMAN_1988_A_95 | | 222 (5) | 238 (5) |
| COLEMAN_1990_FDN | 453 (5) | | |
| **Author** | | **Authorships** | |
| KILDUFF,_M | 16 (4) | 10 (2) | 10 (2) |
| HOSSAIN,_L | 18 (3) | 9 (4) | |
| FOLKE,_C | | 11 (1) | 11 (1) |
| SORENSON,_O | | 8 (3) | 8 (3) |
| DUNBAR,_RIM | 19 (1) | | |
| JACKSON,_MO | 18 (2) | | |
| BRASS,_DJ | 14 (5) | | |
| BODIN,_O | | 7 (5) | |
| ERNSTSON,_H | | | 7 (4) |
| JONES,_N | | | 7 (5) |

**Table 3.2.: Seed subdomain clusters from different coupling methods**

(c) Social Network Analysis

| | Citation-based | | Lexical | Hybrid |
|---|---|---|---|---|
| **Publications** | 1,187 | 1,982 | 2,931 | 2,802 |
| **Year Quartiles** | 06/09/11 | 07/10/11 | 08/10/12 | 07/10/12 |
| **Subject Category** | **Classifications** | | | |
| Comp. Sci., Artificial Intelligence | 131 (3) | 220 (2) | 417 (2) | 360 (2) |
| Comp. Sci., Information Systems | | 357 (1) | 602 (1) | 567 (1) |
| Comp. Sci., Theory & Methods | | 254 (3) | 457 (3) | 400 (3) |
| Management | | 189 (4) | 272 (5) | 281 (4) |
| Sociology | 124 (4) | 169 (5) | | |
| Comp. Sci., Interdisciplinary Applications | | | 252 (4) | |
| Information Science & Library Science | | | | 252 (5) |
| Zoology | 99 (1) | | | |
| Anthropology | 96 (2) | | | |
| Behavioral Sciences | 67 (5) | | | |
| **Author Keyword** | **Usages** | | | |
| SOCIAL_NETWORK_ANALYSIS | 194 (1) | 420 (1) | 1,097 (1) | 1,105 (1) |
| CENTRALITY | 24 (2) | 36 (3) | 49 (4) | 53 (3) |
| NETWORK_ANALYSIS | | 53 (2) | 60 (2) | 68 (2) |
| DATA_MINE | | 23 (4) | 53 (3) | 54 (4) |
| CLUSTER | | 20 (5) | 47 (5) | |
| KNOWLEDGE_MANAGEMENT | | | | 37 (5) |
| VISUALIZATION | 14 (4) | | | |
| ASSOCIATION | 13 (3) | | | |
| SOCIAL_STRUCTURE | 13 (5) | | | |
| **Reference** | **Citations** | | | |
| FREEMAN_1979_S_215 | 195 (1) | 322 (3) | 394 (2) | 432 (2) |
| BORGATTI_2002_UCINET | 182 (2) | 307 (4) | 375 (3) | 405 (3) |
| HANNEMAN_2005_INTRO | 81 (4) | 128 (5) | 185 (5) | 195 (5) |
| WASSERMA_1994_SOCIAL | | 1,817 (1) | 931 (1) | 1,068 (1) |
| SCOTT_2000_SOCIAL | | 361 (2) | 299 (4) | 320 (4) |
| SCOTT_1991_SOCIAL | 93 (3) | | | |
| FREEMAN_1977_S_35 | 59 (5) | | | |
| **Author** | **Authorships** | | | |
| KAZIENKO,_P | | 24 (1) | 23 (1) | 22 (1) |
| LEYDESDORFF,_L | 13 (1) | | 19 (2) | 20 (2) |
| PARK,_HW | | | 17 (3) | 17 (3) |
| MUSIAL,_K | | 19 (2) | | |
| VALENTE,_TW | | 17 (3) | | |
| BUTTS,_CT | | 16 (4) | | |
| HOSSAIN,_L | | | | 16 (4) |
| SNIJDERS,_TAB | | 16 (5) | | |
| BRANDES,_U | | | 14 (4) | |
| CHEN,_HC | | | | 14 (5) |
| DOREIAN,_P | 12 (3) | | | |
| LEE,_PC | | | 12 (5) | |
| SUEUR,_C | 12 (2) | | | |
| REYNOLDS,_RG | 10 (4) | | | |
| VARGAS-QUESADA,_B | 9 (5) | | | |

**Table 3.2.: Seed subdomain clusters from different coupling methods**

(d) Complexity Science

| | Citation-based | Lexical | Hybrid |
|---|---|---|---|
| **Publications** | 2,893 | 3,508 | 4,066 |
| **Year Quartiles** | 2008/2010/2012 | 2007/2010/2012 | 2007/2010/2012 |
| **Subject Category** | | **Classifications** | |
| Comp. Sci., Information Systems | 655 (2) | 597 (1) | 747 (2) |
| Comp. Sci., Theory & Methods | 636 (1) | 520 (2) | 679 (1) |
| Comp. Sci., Artificial Intelligence | 455 (3) | 421 (3) | 568 (3) |
| Engineering, Electrical & Electronic | 435 (5) | 345 (4) | 435 (4) |
| Physics, Multidisc. | 257 (4) | | 237 (5) |
| Comp. Sci., Interdisc. Applications | | 265 (5) | |
| **Author Keyword** | | **Usages** | |
| COMPLEX_NETWORK | 145 (1) | 87 (1) | 127 (1) |
| SMALL_WORLD | 50 (4) | 44 (4) | 49 (5) |
| AGENT_BASE_MODEL | | 77 (2) | 82 (2) |
| CLUSTER | 55 (5) | | 55 (4) |
| SIMULATION | | 43 (5) | 50 (3) |
| SOCIAL_NETWORK_ANALYSIS | 172 (2) | | |
| COMMUNITY_DETECTION | 75 (3) | | |
| TRUST | | 58 (3) | |
| **Reference** | | **Citations** | |
| WATTS_1998_N_440 | 777 (1) | 441 (2) | 641 (1) |
| BARABASI_1999_S_509 | 737 (2) | 439 (1) | 602 (2) |
| NEWMAN_2003_S_167 | 513 (3) | 263 (4) | 405 (3) |
| ALBERT_2002_R_47 | 442 (4) | 278 (3) | 382 (4) |
| WASSERMA_1994_SOCIAL | | 390 (5) | 481 (5) |
| GIRVAN_2002_P_7821 | 325 (5) | | |
| **Author** | | **Authorships** | |
| NEWMAN,_MEJ | 25 (2) | 17 (5) | 24 (2) |
| CHRISTAKIS,_NA | 31 (1) | | 31 (1) |
| SNIJDERS,_TAB | | 22 (1) | 24 (3) |
| JACKSON,_MO | | 19 (2) | 23 (4) |
| WANG,_L | 19 (5) | | 19 (5) |
| WU,_B | 21 (4) | | |
| FRANKS,_DW | 20 (3) | | |
| SANCHEZ,_A | | 17 (3) | |
| SANTOS,_FC | | 16 (4) | |

**Table 3.2.: Seed subdomain clusters from different coupling methods**

(e) Web Science

| | **Citation-based** | **Lexical** | **Hybrid** |
|---|---|---|---|
| **Publications** | 5,591 | 6,152 | 5,954 |
| **Year Quartiles** | 2009/2011/2012 | 2009/2011/2012 | 2009/2011/2012 |
| **Subject Category** | | **Classifications** | |
| Comp. Sci., Information Systems | 1,538 (1) | 1,709 (1) | 1,645 (1) |
| Comp. Sci., Theory & Methods | 1,261 (2) | 1,467 (2) | 1,396 (2) |
| Eng., Electrical & Electronic | 914 (3) | 1,099 (3) | 1,051 (3) |
| Comp. Sci., Artificial Intelligence | 757 (4) | 793 (5) | 735 (5) |
| Telecommunications | 537 (5) | 673 (4) | 638 (4) |
| **Author Keyword** | | **Usages** | |
| FACEBOOK | 304 | 306 | 322 |
| SOCIAL_NETWORK_SITE | 307 | 277 | 295 |
| WEB_2.0 | 272 | 271 | 283 |
| SOCIAL_MEDIA | 232 | 250 | 255 |
| INTERNET | 244 | 229 | 239 |
| **Reference** | | **Citations** | |
| BOYD_2008_J_210 | 293 (1) | 243 (1) | 275 (1) |
| ELLISON_2007_J_1143 | 234 (2) | 187 (2) | 212 (2) |
| O'REILLY_2005_WHAT | 158 (3) | 130 (5) | 144 (3) |
| BOYD_2007_J | 135 (4) | | 122 (4) |
| ELLISON_2007_J | 112 (5) | | 113 (5) |
| WASSERMA_1994_SOCIAL | | 295 (3) | |
| GRANOVET_1973_A_1360 | | 252 (4) | |
| **Author** | | **Authorships** | |
| ZHANG,_J | 18 (4) | 17 (2) | 17 (2) |
| JUNG,_JJ | 17 (3) | 15 (3) | 15 (3) |
| TURBAN,_E | 15 (5) | 14 (5) | 14 (5) |
| ZHANG,_Y | | 17 (1) | 17 (1) |
| LEE,_S | | 15 (4) | 15 (4) |
| MORENO,_MA | 26 (1) | | |
| CHRISTAKIS,_DA | 19 (2) | | |

similarities for clustering because all results are fairly robust and hybrid methods balance the advantages and disadvantages of citation-based and lexical approaches (Braam et al., 1991a; Glänzel & Thijs, 2011; Zitt, 2015). Having excluded author coupling, this prevents either references or words from determining future results.

#### Hybrid Subdomain Retrieval

The next step is to identify the story set cores of the seed subdomains we have just identified. Other than in the Zitt/Bassecoulard method, we do not only retrieve the solution through citations of reference cores but also through usages of word cores. Each subdomain core must be both generic and specific, and each feature is controlled by a parameter. In the Zitt/Bassecoulard method, genericness is ensured through requiring core facts to each have at least $Y$ selections from the seed. If we chose the same absolute selection threshold for all subdomains, then small subdomains and those with a less skewed size distribution would be punished. To ensure that all subdomain cores are equally generic, we take advantage of the situation that few facts are selected by, or retrieve, many publications. $Y_{\max}$ is a cumulative sum of selection fractions $K^{\mathrm{N}}$. For example, when $Y_{\max} = 0.1$, then the facts that accumulate the top ten percent of all selections are chosen for retrieval. To determine $Y_{\max}$, facts are ranked descendingly by $tf * idf$ weights (described in the caption of table 3.2). This weighting decreases the ranks of facts when they are highly selected in multiple subdomains.

Specificity is originally ensured through requiring core facts to each get at least $U$ percent of their total selections from the seed. As specified above, we could not compute specificity the intended way but chose to use the public to act as a proxy for all of science. But the Zitt/Bassecoulard method also necessitates a precise seed because only then $U$ is meaningful. Our seed is not precise because it also contains papers that use the word `SOCIAL_NETWORK` metaphorically. To ensure specific cores, we took a sample of 1,000 papers from the public, 499 of which are in the seed, and manually distinguished if they should be inside or outside Social Network Science. This allows stating objective quality criteria like recall and precision, but objectivity is gained by subjectifying the uncertainty of boundary. Appendix D gives 20 examples for each category to follow the process. The goal was to define a science of social networks, not a sociological network science. Papers were ruled inside or relevant when they are truly relational and outside or irrelevant when the `NETWORK` concept is used metaphorically or in non-social contexts. Papers about engineering networked social systems were ruled inside when they were not purely about issues of implementation. A fact was then characterized as specific at a level of, e.g., 10% when at least that fraction of the sampled papers the fact retrieves was ruled inside the domain. $U_{\min}$ is our measure of specificity. We do not use the relevance parameter $E$ of the Zitt/Bassecoulard method because the sample coding already allows us to determine recall and precision.

Figure 3.2 demonstrates how the two parameters influence the recall and precision of publication sets in total and broken down to subdomains. Precision and specificity are directly related. While precision is a measure for the fraction of retrieved documents that are relevant, specificity is a measure for the precision that a social fact generates. The

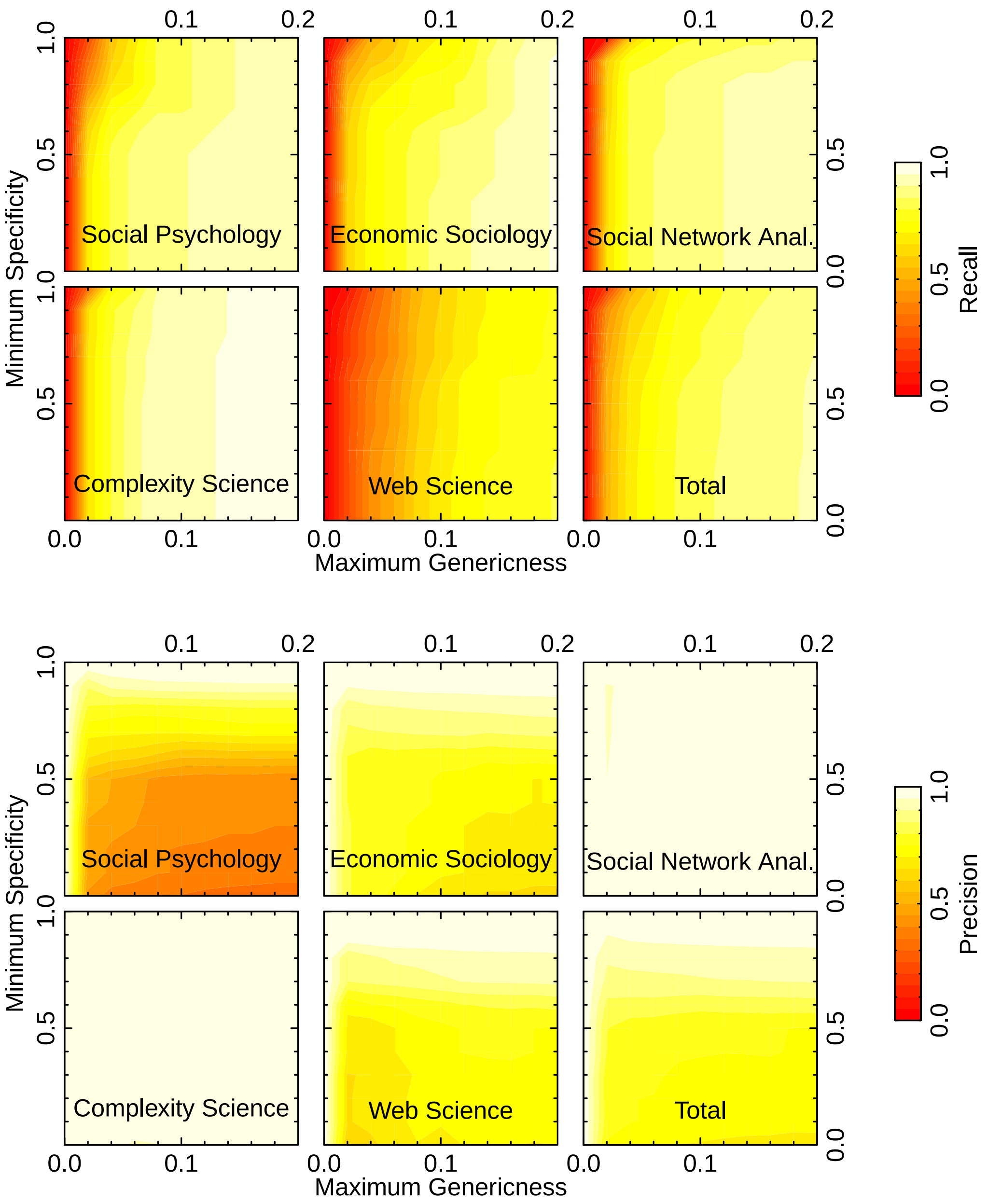

(a) Citation-based retrieval

**Figure 3.2.: Recall and precision of publication retrieval**
The retrieval procedure mimics identities, i.e., publications are retrieved when they select core facts. Maximum genericness ensures that retrieval is efficient. The larger it is, the more the core consists of larger numbers of decreasingly selected facts. Minimum specificity ensures that that retrieval is accurate. The larger it is, the larger is the fraction of sampled papers which are ruled inside Social Network Science. Publication sets from citation-based retrieval (a) and lexical retrieval (b) are evaluated using ...

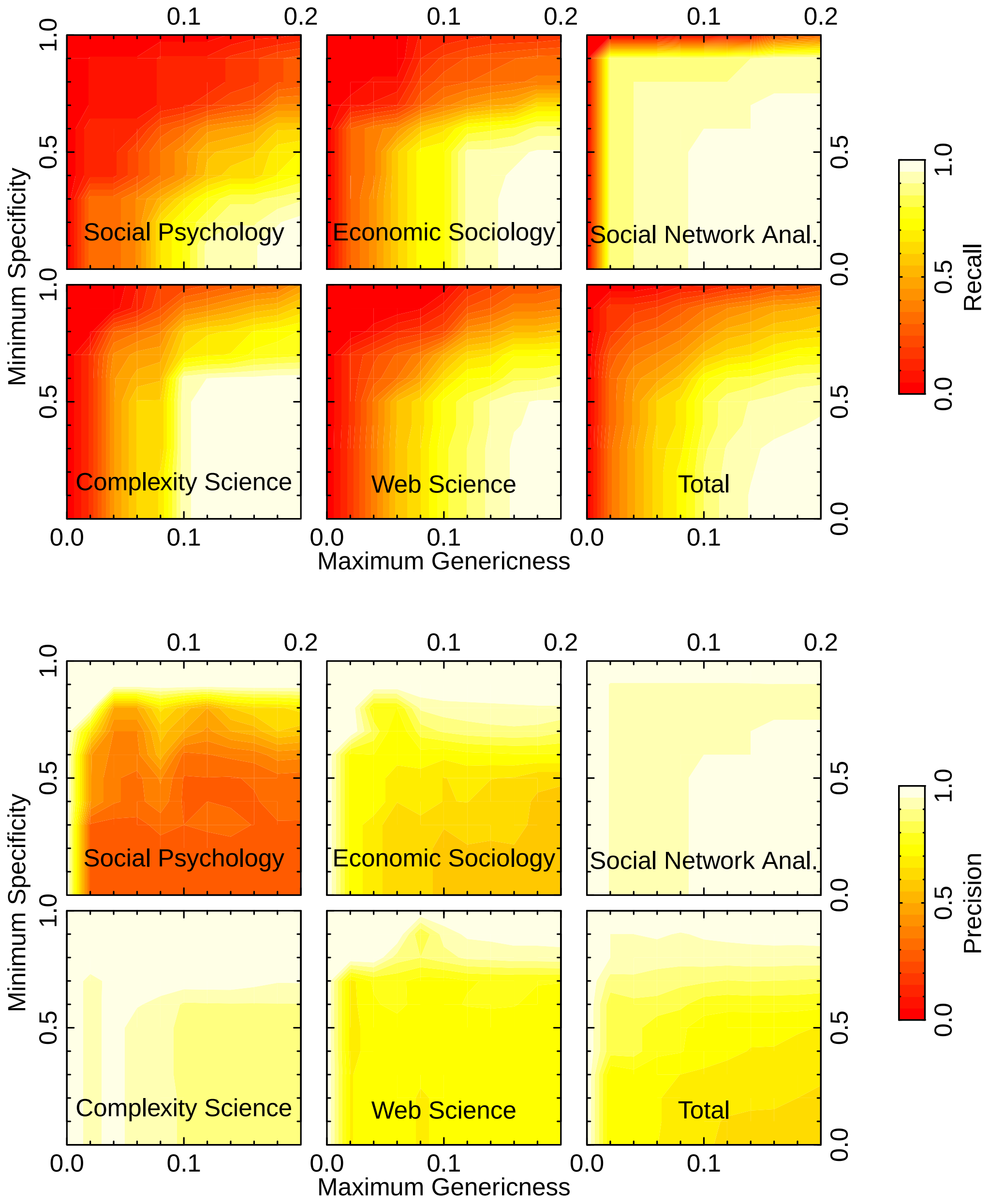

(b) Lexical retrieval

**Figure 3.2.: Recall and precision of publication retrieval**
... recall (fraction of relevant documents that are retrieved) and precision (fraction of retrieved documents that are relevant). For the example of Social Psychology (SP), the papers that cite one of the references in a core with a genericness of 0.1 and a specificity of 0.5 recall the SP subset of the seed to 89% and are 43% precise. For lexical retrieval, recall of the SP seed is 43% at a precision of 29%. In general, recall is higher for citation-based retrieval, i.e., reference cores are more generic ceteris paribus.

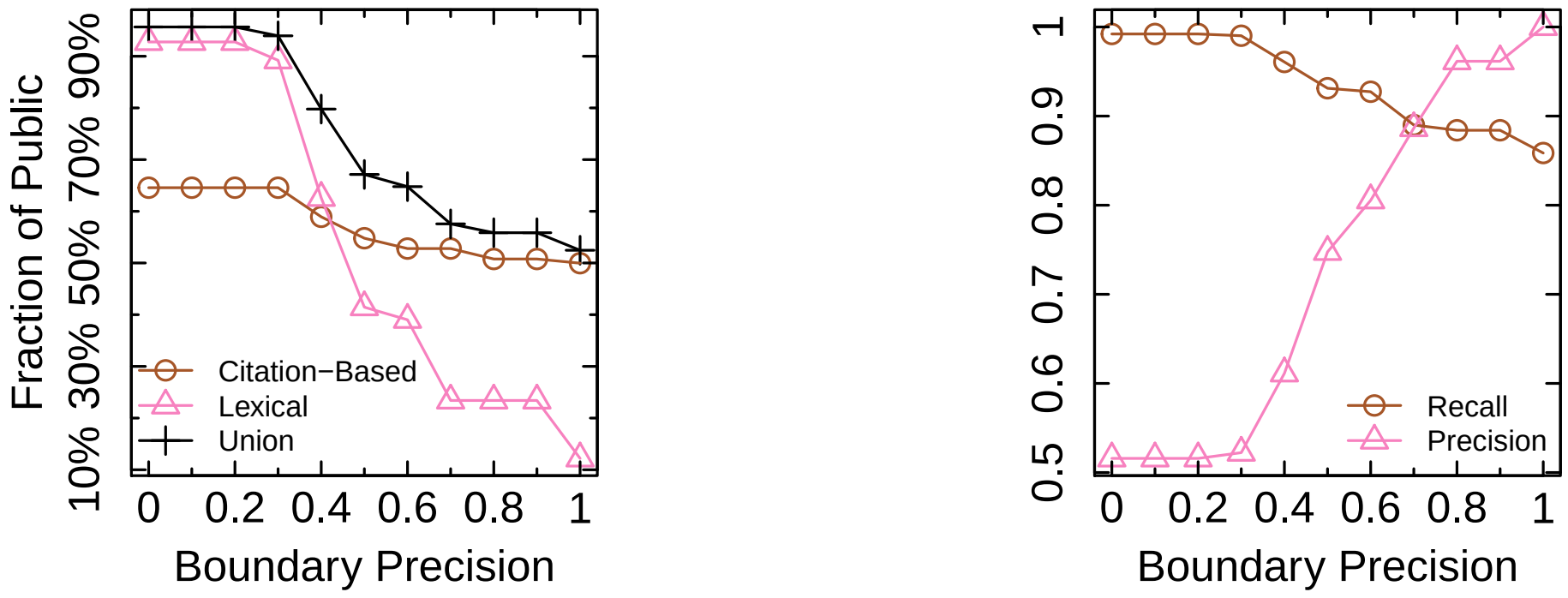

(a) Efficiency (b) Quality

**Figure 3.3.: Efficiency and quality of delineation**

Overall efficiency and quality of the retrieval procedure with respect to the public set of 44,307 publications. (a) The retrieved fraction of those documents decreases with increasing boundary precision levels defined in table 3.3, but less so for citation-based retrieval. The black trend gives the set union as the result of hybrid retrieval. (b) Opposing trends of recall and precision according to hybrid retrieval of an evaluated sample.

**Table 3.3.: Boundary precision of subdomains**

A boundary precision is introduced because facts that form the core of one subdomain can cause the precision of another subdomain to drop below the minimum precision set by $U_{\mathrm{min}}$. Boundary precision quantifies the minimum quality of all subdomain boundaries. For example, for citation-based retrieval and a desired boundary precision of at least 0.5, the best overall recall is obtained for $Y_{\mathrm{max}} = 0.20$ and $U_{\mathrm{min}} = 0.6$. For lexical retrieval and the same parameters, recall is lower. Recall and precision are means and standard deviations from the five subdomains.

| **Retrieval** | **Boundary precision** | $Y_{\mathrm{max}}$ | $U_{\mathrm{min}}$ | **Recall** | **Precision** |
|---|---|---|---|---|---|
| | 0.0, ..., 0.3 | 0.20 | 0.0 | $0.92 \pm 0.06$ | $0.71 \pm 0.24$ |
| | 0.4 | 0.20 | 0.3 | $0.92 \pm 0.07$ | $0.74 \pm 0.21$ |
| **Citation-based** | 0.5 | 0.20 | 0.6 | $0.92 \pm 0.06$ | $0.82 \pm 0.15$ |
| | 0.6, ..., 0.7 | 0.20 | 0.7 | $0.91 \pm 0.07$ | $0.89 \pm 0.09$ |
| | 0.8, ..., 0.9 | 0.20 | 0.9 | $0.90 \pm 0.07$ | $0.97 \pm 0.02$ |
| | 1.0 | 0.20 | 1.0 | $0.89 \pm 0.07$ | 1.00 |
| | 0.0, ..., 0.2 | 0.20 | 0.0 | $0.99 \pm 0.01$ | $0.68 \pm 0.24$ |
| | 0.3 | 0.20 | 0.4 | $0.94 \pm 0.09$ | $0.68 \pm 0.23$ |
| | 0.4 | 0.20 | 0.6 | $0.86 \pm 0.14$ | $0.75 \pm 0.19$ |
| **Lexical** | 0.5 | 0.20 | 0.7 | $0.71 \pm 0.19$ | $0.82 \pm 0.14$ |
| | 0.6 | 0.18 | 0.7 | $0.71 \pm 0.18$ | $0.83 \pm 0.14$ |
| | 0.7, ..., 0.9 | 0.20 | 0.9 | $0.51 \pm 0.24$ | $0.98 \pm 0.02$ |
| | 1.0 | 0.20 | 1.0 | $0.26 \pm 0.12$ | 1.00 |

figures reveal two things. First, recall is higher for citation-based retrieval ceteris paribus. Reference cores are more generic or, put differently, at similar genericness, lexical retrieval is associated with a lower recall because language use is relatively imprecise. Second, idiosyncrasies of subdomains point at differences of ideational closure or cultural coherence (S. Fuchs, 2001, p. 55). For Social Network Analysis, few references and words suffice to retrieve a large fraction of relevant publications. Accordingly, `SOCIAL_NETWORK_ANALYSIS` is used by 1,105 or 39% of 2,802 papers. 38% cite `WASSERMA_1994_SOCIAL` (table 3.2c). Social Psychology is at the other extreme. Lexical retrieval is either good regarding recall or precision, but not both. Only 8% of the papers use `SOCIAL_SUPPORT` and 5% cite `BERKMAN_1979_A_186` (table 3.2a).

Our goal is to delineate Social Network Science through one set of social facts chosen by one parameter setting, not one set and setting for each subdomain. This complicates the procedure because a core that retrieves a set of publications which is precise at some level in the total is not necessarily equally precise for all subdomains. For citation-based retrieval, setting the minimum specificity to $U_{\min} = 0.5$ ($Y_{\max} = 0.1$) results in an overall precision of 0.76 but, for Social Psychology, precision (0.43) is below the desired specificity. To circumvent this problem, we introduce a *boundary precision* which quantifies the minimum precision of all subdomain boundaries given a parameter setting. For each such precision level, that parameter setting is identified which maximizes overall recall. The result is shown in table 3.3. With only one exception, $Y_{\max} = 0.20$ is the value of choice.

Figure 3.3 reports the efficiency and quality of publication retrieval with respect to the public set conditioned on boundary precision. Figure 3.3a gives the fraction of papers retrieved from the public. Different upper bounds are visible. For citation-based retrieval, no more than 65% of 44,308 papers can be retrieved because $Y_{\max}$ does not get bigger than 0.2. Lexical cores can retrieve 93% at low precision but the fraction quickly drops when boundaries are required to be less fuzzy. Figure 3.3b reports the quality of hybrid retrieval, the set union from the citation-based and lexical approaches. Recall is at a satisfactorily high level for all precision levels. At this point, we decide that we want our confidence into the precision of the boundary to be 80%. The corresponding parameters

**Table 3.4.: Number of social facts used for retrieval**

To delineate Social Network Science, a genericness $Y_{\max} = 0.2$ is defined. The five subdomains contribute different numbers of facts to retrieval. The selection fraction also translates to different absolute numbers of selections. These are only minima due to the $tf * idf$ ranking of facts.

| | References | Citations | Words | Usages |
|---|---|---|---|---|
| **Social Psychology** | 4,685 | 5 | 42 | 33 |
| **Economic Sociology** | 3,443 | 4 | 90 | 8 |
| **Social Network Analysis** | 1,004 | 7 | 15 | 20 |
| **Complexity Science** | 1,162 | 8 | 76 | 12 |
| **Web Science** | 3,583 | 4 | 35 | 34 |

are $Y_{\max} = 0.2$ and $U_{\min} = 0.9$. Note that the minimum fact specificity had to be set higher than 0.8 to achieve the stated boundary precision. We see in table 3.4 that the subdomains contribute different numbers of facts to the overall retrieval core and that a genericness of 20% translates to different absolute selection thresholds. While Web Science's core references are at least cited four times, an absolute threshold for Complexity Science would have to be eight. In other words, the original Zitt/Bassecoulard method of not distinguishing subdomains is only applicable to research domains that do not consist of subdomains with idiosyncratic selection practices. The retrieval setting results in 24,748 papers, slightly more than in the seed. We call this set the *interim solution*.

### Creating the Solution

To create the solution set, the last retrieval step is to add the cited core of the interim solution to the latter. This step is supposed to reconstruct historically complete citation

**Table 3.5.: Core description and sourcing of interim solution**

The final retrieval step is adding the interim solution's cited cores with genericness $Y_{\max} = 0.2$ to the interim solution. Clustering the interim solution results in five communities which are labeled by their most used word. Labels map to the seed subdomains of table 3.2. The first data row gives the size of each community's cited core. The core of the `SOCIAL_SUPPORT` community which maps to Social Psychology contains 2,914 references. 73% of these references are articles and 7% are chapters according to the definition given in appendix B. The rest are books. Domain closure is the fraction of the cited articles that are interim solution publications. 30% of the `SOCIAL_SUPPORT` core references are themselves contained in the domain. Subdomain closure is the fraction of the cited articles that are publications belonging to the same community. 21% of the core references are themselves contained in the `SOCIAL_SUPPORT` community. Sourcing rates tell which fraction of a community's core could successfully be retrieved in the *Web of Science* and added to the interim solution. Rates are generally high for articles. Chapters include conference proceedings, some of which are included in the database.

| | SOCIAL_ SUPPORT | SOCIAL_ CAPITAL | SOCIAL_ NETWORK_ ANALYSIS | COMPLEX_ NETWORK | SOCIAL_ MEDIA |
|---|---|---|---|---|---|
| **Mapping** | SP | ES | SNA | CS | WS |
| | | | **Core statistics** | | |
| **References** | 2,914 | 3,338 | 981 | 1,143 | 2,768 |
| **Article fraction** | 73% | 55% | 71% | 79% | 46% |
| **Chapter fraction** | 7% | 6% | 5% | 3% | 19% |
| **Closure (domain)** | 30% | 23% | 38% | 29% | 23% |
| **Closure (subdomain)** | 21% | 15% | 15% | 21% | 9% |
| | | | **Sourcing rates** | | |
| **Articles** | 93% | 90% | 91% | 87% | 78% |
| **Chapters** | 3% | 11% | 14% | 22% | 15% |
| **Total** | 68% | 50% | 66% | 70% | 38% |

paths. Again, we proceed by identifying the cores of the interim subdomains. Louvain community detection ($Q = 0.12$) in the hybrid network of publications (coupled through references and words) expectedly results in five clusters which are mapped to seed subdomains through the most used word (table 3.5). Clustering consensus is 99% instead of 92% for the seed (cf. table 3.1). Cited cores are also smaller than for the seed (cf. table 3.4). Both results are clear evidence that the delineation procedure has created more compact subdomains and more certain subdomain boundaries.

By counting the fraction of cited references that are also contained in the interim solution, we can determine the degree of closure of a subdomain. Books are not covered by the *Web of Science* products we had accessed. Consequently, founding books like Moreno's *Who Shall Survive?* (1934) can only show up as cited references, not as citing publications. Social Network Analysis is most self-contained with respect to the whole domain: 38% of its core references are domain publications. When references are required to belong to the citing subdomain, Social Psychology and Complexity Science are most closed. Web Science, the youngest subdomain, least cites its own publications. As expected for a subdomain strongly rooted in computer science, it cites a large fraction of book chapters or, in this case, conference proceedings articles. Because those are covered in the *Web of Science* database to a lesser extent than articles and to avoid related artifacts, closure is computed for journal articles only.

Only 38% of Web Science's 2,768 core references could be identified in the database – could be sourced – and added to the interim solution. Its article sourcing rate is smallest, too. Complexity Science has the largest sourcing rate overall (70%) and for chapters (22%). 93% of Social Psychology's cited articles were successfully added to the interim solution (table 3.5).[3] In total, 4,965 core references were added to the interim solution. In the resulting set, some publications or references were removed to prevent meaningless results, artifacts, or the failure of algorithms.[4] Finally, all publications published after 2012 were removed because those years were not completely covered in the database.

### Describing the Solution

The desired solution of the delineation procedure contains 25,760 journal articles and conference proceedings articles. These cite 574,036 references and use 26,906 words. Following the disambiguation of author names, 45,580 author identities remain. Table 3.6 is a description of the five final subdomains of Social Network Science. Labels are derived from subject categories and social facts and match those of the seed. The consensus of

[3] 9,142 unique cited references were chosen for complementing the interim solution. Publication identifiers for references published not earlier than 1980 were queried using the bibliometric database of the Competence Centre for Bibliometrics (www.bibliometrie.info). Heavily cited references that could not be found as well as references published before 1980 were queried using the online interface www.webofknowledge.com. The primary search criterion was the doi, the secondary criteria were the tagged meta data.

[4] 61 publications were removed because they did not have unique matchkeys. Five articles with an `ANONYMOUS` author were removed. Furthermore, we removed citations from a publication to a reference with the identical matchkey, references with Chinese letters, and references without cited author or source names or publication years.

**Table 3.6.: Subdomains in Social Network Science**

Each subdomain is described by its size in publications, publication year quartiles, top 5 *Web of Science* subject categories, and top 10 social facts ranked by $tf * idf$ (as described in the caption of table 3.2). $k$ is the number of selections, $k^{\mathrm{N}}$ the normalized number of selections, $K$ the fraction of all publications that select the fact, and $K^{\mathrm{N}}$ the fraction of all selections in the subdomain. The lifetime $\lambda|100\%$ is the number of years in which a fact is selected at least once.

(a) Social Psychology

| **Publications** | 5,945 | | | | |
|---|---|---|---|---|---|
| **Year Quartiles** | 1997/2005/2010 | | | | |
| **Subject Category** | **Classifications** | | | | |
| Public, Env. & Occup. Health | 1,049 | | | | |
| Sociology | 757 | | | | |
| Psychology, Multidisc. | 497 | | | | |
| Psychology, Social | 497 | | | | |
| Psychology, Developmental | 478 | | | | |
| **Word** | $k$ | $k^{\mathrm{N}}$ | $K$ | $K^{\mathrm{N}}$ | $\lambda\|100\%$ |
| SOCIAL_SUPPORT | 1,397 | 291.55 | 23.50% | 4.90% | 37 |
| FRIEND | 978 | 126.89 | 16.45% | 2.13% | 34 |
| COMMUNITY | 636 | 77.66 | 10.70% | 1.31% | 29 |
| FRIENDSHIP | 358 | 48.91 | 6.02% | 0.82% | 32 |
| SOCIAL_SUPPORT_NETWORK | 277 | 51.28 | 4.66% | 0.86% | 31 |
| SOCIAL_RELATIONSHIP | 306 | 43.16 | 5.15% | 0.73% | 28 |
| LONGITUDINAL_STUDY | 240 | 31.53 | 4.04% | 0.53% | 26 |
| EMOTIONAL_SUPPORT | 215 | 26.53 | 3.62% | 0.45% | 23 |
| LONELINESS | 205 | 36.15 | 3.45% | 0.61% | 32 |
| SOCIAL_CAPITAL | 341 | 38.44 | 5.74% | 0.65% | 18 |
| **Reference** | $k$ | $k^{\mathrm{N}}$ | $K$ | $K^{\mathrm{N}}$ | $\lambda\|100\%$ |
| BERKMAN_1979_A_186 | 384 | 9.25 | 6.46% | 0.16% | 33 |
| COHEN_1985_P_310 | 361 | 7.46 | 6.07% | 0.13% | 27 |
| HOUSE_1988_S_540 | 289 | 6.57 | 4.86% | 0.11% | 24 |
| COBB_1976_P_300 | 272 | 6.34 | 4.58% | 0.11% | 34 |
| RADLOFF_1977_A_385 | 261 | 6.10 | 4.39% | 0.10% | 31 |
| FISCHER_1982_DWELL | 291 | 7.26 | 4.89% | 0.12% | 31 |
| GRANOVET_1973_A_1360 | 472 | 10.03 | 7.94% | 0.17% | 36 |
| ROOK_1984_J_1097 | 183 | 3.74 | 3.08% | 0.06% | 29 |
| PUTNAM_2000_BOWLING | 266 | 5.67 | 4.47% | 0.10% | 12 |
| HOUSE_1981_WORK | 183 | 4.17 | 3.08% | 0.07% | 31 |
| **Author** | $k$ | $k^{\mathrm{N}}$ | $K$ | $K^{\mathrm{N}}$ | $\lambda\|100\%$ |
| LATKIN,_CARL | 68 | 17.78 | 1.14% | 0.30% | 17 |
| BERKMAN,_LISA | 42 | 13.95 | 0.71% | 0.23% | 24 |
| KAWACHI,_ICHIRO | 31 | 7.71 | 0.52% | 0.13% | 14 |
| LITWIN,_HOWARD | 28 | 22.00 | 0.47% | 0.37% | 15 |
| ANTONUCCI,_TONI | 24 | 8.33 | 0.40% | 0.14% | 17 |
| DUNBAR,_ROBIN_I_M | 22 | 11.08 | 0.37% | 0.19% | 10 |
| VALENTE,_THOMAS | 22 | 7.13 | 0.37% | 0.12% | 11 |
| CELENTANO,_DAVID | 20 | 2.96 | 0.34% | 0.05% | 16 |
| COHEN,_SHELDON | 19 | 6.99 | 0.32% | 0.12% | 17 |
| KRAUSE,_NEAL | 19 | 12.23 | 0.32% | 0.21% | 15 |

**Table 3.6.: Subdomains in Social Network Science**

(b) Economic Sociology

| **Publications** | 7,554 | | | | |
|---|---|---|---|---|---|
| **Year Quartiles** | 2001/2007/2010 | | | | |
| **Subject Category** | **Classifications** | | | | |
| Management | 1,783 | | | | |
| Business | 1,293 | | | | |
| Sociology | 1,357 | | | | |
| Geography | 665 | | | | |
| Economics | 628 | | | | |
| **Word** | $k$ | $k^{\mathrm{N}}$ | $K$ | $K^{\mathrm{N}}$ | $\lambda\|100\%$ |
| SOCIAL_CAPITAL | 1,210 | 147.52 | 16.02% | 1.95% | 22 |
| ORGANIZATIONAL | 614 | 74.90 | 8.13% | 0.99% | 29 |
| INNOVATION | 582 | 68.71 | 7.70% | 0.91% | 30 |
| COMMUNITY | 896 | 94.89 | 11.86% | 1.26% | 23 |
| OPPORTUNITY | 555 | 58.94 | 7.35% | 0.78% | 24 |
| TRUST | 538 | 60.01 | 7.12% | 0.79% | 24 |
| AGENCY | 324 | 39.07 | 4.29% | 0.52% | 23 |
| GOVERNANCE | 291 | 31.21 | 3.85% | 0.41% | 22 |
| ENTREPRENEUR | 264 | 26.75 | 3.49% | 0.35% | 21 |
| INFORMAL | 284 | 29.83 | 3.76% | 0.39% | 28 |
| **Reference** | $k$ | $k^{\mathrm{N}}$ | $K$ | $K^{\mathrm{N}}$ | $\lambda\|100\%$ |
| GRANOVET_1973_A_1360 | 1,143 | 23.01 | 15.13% | 0.30% | 38 |
| GRANOVET_1985_A_481 | 841 | 16.26 | 11.13% | 0.22% | 27 |
| BURT_1992_STRUCTURAL | 884 | 17.45 | 11.70% | 0.23% | 21 |
| COLEMAN_1988_A_10.1086/228943 | 799 | 15.71 | 10.58% | 0.21% | 22 |
| COLEMAN_1990_FDN | 586 | 10.88 | 7.76% | 0.14% | 18 |
| PUTNAM_1993_MAKING | 505 | 10.27 | 6.69% | 0.14% | 18 |
| PUTNAM_2000_BOWLING | 521 | 11.43 | 6.90% | 0.15% | 13 |
| UZZI_1997_A_35 | 374 | 6.65 | 4.95% | 0.09% | 16 |
| NAHAPIET_1998_A_242 | 371 | 7.56 | 4.91% | 0.10% | 14 |
| PORTES_1998_A_1 | 370 | 6.83 | 4.90% | 0.09% | 14 |
| **Author** | $k$ | $k^{\mathrm{N}}$ | $K$ | $K^{\mathrm{N}}$ | $\lambda\|100\%$ |
| FOLKE,_CARL | 15 | 4.71 | 0.20% | 0.06% | 8 |
| SORENSON,_OLAV | 13 | 7.33 | 0.17% | 0.10% | 7 |
| GULATI,_RANJAY | 13 | 7.83 | 0.17% | 0.10% | 8 |
| CROSS,_ROB | 13 | 5.12 | 0.17% | 0.07% | 5 |
| EISENHARDT,_KATHLEEN | 13 | 8.00 | 0.17% | 0.11% | 11 |
| YEUNG,_H_W_C | 12 | 9.90 | 0.16% | 0.13% | 8 |
| STUART,_TOBY | 12 | 6.33 | 0.16% | 0.08% | 10 |
| BURT,_RONALD_S | 13 | 10.42 | 0.17% | 0.14% | 11 |
| MCEVILY,_BILL | 11 | 4.33 | 0.15% | 0.06% | 8 |
| PORTES,_A | 11 | 6.50 | 0.15% | 0.09% | 8 |

**Table 3.6.: Subdomains in Social Network Science**

(c) Social Network Analysis

| **Publications** | 2,459 | | | | |
|---|---|---|---|---|---|
| **Year Quartiles** | 2006/2009/2011 | | | | |
| **Subject Category** | **Classifications** | | | | |
| Computer Science, Information Systems | 500 | | | | |
| Computer Science, Theory & Methods | 350 | | | | |
| Computer Science, Artificial Intelligence | 305 | | | | |
| Information Science & Library Science | 279 | | | | |
| Computer Science, Interdisc. Applications | 241 | | | | |
| **Word** | $k$ | $k^{N}$ | $K$ | $K^{N}$ | $\lambda$\|100% |
| SOCIAL_NETWORK_ANALYSIS | 1,846 | 271.81 | 75.07% | 11.05% | 34 |
| NETWORK_ANALYSIS | 248 | 32.97 | 10.09% | 1.34% | 26 |
| SOCIAL_NETWORK_ANALYSIS_SNA | 190 | 19.89 | 7.73% | 0.81% | 12 |
| CENTRALITY | 205 | 24.90 | 8.34% | 1.01% | 23 |
| COMMUNITY | 304 | 31.40 | 12.36% | 1.28% | 18 |
| COLLABORATION | 179 | 16.94 | 7.28% | 0.69% | 14 |
| SOCIAL_STRUCTURE | 159 | 32.74 | 6.47% | 1.33% | 38 |
| NETWORK_STRUCTURE | 136 | 11.82 | 5.53% | 0.48% | 16 |
| WEB | 124 | 11.31 | 5.04% | 0.46% | 11 |
| DATA_MINE | 83 | 8.96 | 3.38% | 0.36% | 11 |
| **Reference** | $k$ | $k^{N}$ | $K$ | $K^{N}$ | $\lambda$\|100% |
| WASSERMA_1994_SOCIAL | 1,034 | 51.37 | 42.05% | 2.09% | 19 |
| FREEMAN_1979_S_215 | 407 | 14.37 | 16.55% | 0.58% | 28 |
| BORGATTI_2002_UCINET | 362 | 12.25 | 14.72% | 0.50% | 10 |
| SCOTT_2000_SOCIAL | 286 | 10.81 | 11.63% | 0.44% | 11 |
| HANNEMAN_2005_INTRO | 156 | 5.19 | 6.34% | 0.21% | 8 |
| SCOTT_1991_SOCIAL | 154 | 9.55 | 6.26% | 0.39% | 18 |
| BURT_1992_STRUCTURAL | 207 | 5.10 | 8.42% | 0.21% | 18 |
| GRANOVET_1973_A_1360 | 250 | 7.01 | 10.17% | 0.29% | 30 |
| FREEMAN_1977_S_35 | 112 | 3.72 | 4.55% | 0.15% | 19 |
| DE_2005_EXPLORATORY | 88 | 4.14 | 3.58% | 0.17% | 7 |
| **Author** | $k$ | $k^{N}$ | $K$ | $K^{N}$ | $\lambda$\|100% |
| LEYDESDORFF,_LOET | 22 | 14.17 | 0.89% | 0.58% | 11 |
| KAZIENKO,_PRZEMYSLAW | 20 | 7.50 | 0.81% | 0.31% | 6 |
| GLOOR,_PETER | 16 | 5.76 | 0.65% | 0.23% | 7 |
| BRANDES,_ULRIK | 14 | 5.61 | 0.57% | 0.23% | 10 |
| PARK,_HAN_WOO | 14 | 6.25 | 0.57% | 0.25% | 6 |
| DOREIAN,_PATRICK | 13 | 7.83 | 0.53% | 0.32% | 11 |
| CARLEY,_KATHLEEN | 13 | 5.39 | 0.53% | 0.22% | 9 |
| HOSSAIN,_LIAQUAT | 12 | 5.00 | 0.49% | 0.20% | 5 |
| CHEN,_CHAOMEI | 9 | 3.46 | 0.37% | 0.14% | 6 |
| SHNEIDERMAN,_BEN | 9 | 3.40 | 0.37% | 0.14% | 6 |

**Table 3.6.: Subdomains in Social Network Science**

(d) Complexity Science

| **Publications** | 6,031 | | | | |
|---|---|---|---|---|---|
| **Year Quartiles** | 2005/2009/2011 | | | | |
| **Subject Category** | **Classifications** | | | | |
| Computer Science, Theory & Methods | 996 | | | | |
| Computer Science, Information Systems | 1,073 | | | | |
| Computer Science, Artificial Intelligence | 881 | | | | |
| Engineering, Electrical & Electronic | 799 | | | | |
| Physics, Multidisc. | 531 | | | | |
| **Word** | $k$ | $k^{\mathrm{N}}$ | $K$ | $K^{\mathrm{N}}$ | $\lambda\|100\%$ |
| COMPLEX_NETWORK | 594 | 67.25 | 9.85% | 1.12% | 15 |
| COMMUNITY | 655 | 72.23 | 10.86% | 1.20% | 19 |
| NETWORK_STRUCTURE | 367 | 43.22 | 6.09% | 0.72% | 25 |
| CONNECTIVITY | 315 | 35.81 | 5.22% | 0.59% | 20 |
| SMALL_WORLD | 279 | 33.79 | 4.63% | 0.56% | 18 |
| AVERAGE | 332 | 39.50 | 5.50% | 0.65% | 20 |
| SCALE_FREE_NETWORK | 256 | 29.76 | 4.24% | 0.49% | 13 |
| EVOLVE | 319 | 33.97 | 5.29% | 0.56% | 17 |
| COMMUNITY_STRUCTURE | 258 | 26.59 | 4.28% | 0.44% | 14 |
| COOPERATION | 302 | 31.12 | 5.01% | 0.52% | 20 |
| **Reference** | $k$ | $k^{\mathrm{N}}$ | $K$ | $K^{\mathrm{N}}$ | $\lambda\|100\%$ |
| WATTS_1998_N_440 | 1,101 | 46.12 | 18.26% | 0.76% | 14 |
| BARABASI_1999_S_509 | 1,018 | 40.41 | 16.88% | 0.67% | 13 |
| ALBERT_2002_R_47 | 708 | 27.94 | 11.74% | 0.46% | 11 |
| NEWMAN_2003_S_167 | 675 | 26.54 | 11.19% | 0.44% | 10 |
| WASSERMA_1994_SOCIAL | 623 | 20.08 | 10.33% | 0.33% | 18 |
| GIRVAN_2002_P_7821 | 355 | 13.49 | 5.89% | 0.22% | 10 |
| PASTOR-S_2001_P_3200 | 275 | 9.32 | 4.56% | 0.15% | 12 |
| AMARAL_2000_P_11149 | 273 | 8.74 | 4.53% | 0.14% | 13 |
| WATTS_1999_SMALL | 290 | 12.03 | 4.81% | 0.20% | 14 |
| STROGATZ_2001_N_268 | 255 | 8.65 | 4.23% | 0.14% | 12 |
| **Author** | $k$ | $k^{\mathrm{N}}$ | $K$ | $K^{\mathrm{N}}$ | $\lambda\|100\%$ |
| NEWMAN,_M_E_J | 43 | 26.73 | 0.71% | 0.44% | 12 |
| BARABASI,_ALBERT | 41 | 14.14 | 0.68% | 0.23% | 12 |
| VESPIGNANI,_ALESSANDRO | 26 | 8.79 | 0.43% | 0.15% | 12 |
| LATORA,_VITO | 25 | 6.90 | 0.41% | 0.11% | 11 |
| NOWAK,_MARTIN | 23 | 7.82 | 0.38% | 0.13% | 12 |
| PACHECO,_JORGE | 22 | 7.32 | 0.36% | 0.12% | 8 |
| EGUILUZ,_VICTOR | 22 | 7.10 | 0.36% | 0.12% | 10 |
| SNIJDERS,_TOM_A_B | 23 | 11.48 | 0.38% | 0.19% | 15 |
| KLEINBERG,_JON | 22 | 8.90 | 0.36% | 0.15% | 11 |
| MORENO,_YAMIR | 22 | 5.72 | 0.36% | 0.09% | 9 |

**Table 3.6.: Subdomains in Social Network Science**

(e) Web Science

| **Publications** | 3,771 | | | | |
|---|---|---|---|---|---|
| **Year Quartiles** | 2008/2010/2011 | | | | |
| **Subject Category** | **Classifications** | | | | |
| Computer Science, Information Systems | 1,315 | | | | |
| Computer Science, Theory & Methods | 982 | | | | |
| Engineering, Electrical & Electronic | 733 | | | | |
| Computer Science, Artificial Intelligence | 621 | | | | |
| Computer Science, Software Engineering | 460 | | | | |
| **Word** | $k$ | $k^{\mathrm{N}}$ | $K$ | $K^{\mathrm{N}}$ | $\lambda$\|100% |
| USER | 1,989 | 187.69 | 52.74% | 4.98% | 23 |
| INTERNET | 687 | 63.25 | 18.22% | 1.68% | 18 |
| FACEBOOK | 560 | 51.14 | 14.85% | 1.36% | 7 |
| SOCIAL_NETWORK_SITE | 462 | 44.44 | 12.25% | 1.18% | 8 |
| WEB | 516 | 48.70 | 13.68% | 1.29% | 16 |
| TWITTER | 395 | 38.20 | 10.47% | 1.01% | 5 |
| WEB_2.0 | 371 | 32.84 | 9.84% | 0.87% | 7 |
| ONLINE_SOCIAL_NETWORK | 349 | 35.78 | 9.25% | 0.95% | 9 |
| BLOG | 330 | 27.52 | 8.75% | 0.73% | 9 |
| SOCIAL_MEDIA | 317 | 27.26 | 8.41% | 0.72% | 8 |
| **Reference** | $k$ | $k^{\mathrm{N}}$ | $K$ | $K^{\mathrm{N}}$ | $\lambda$\|100% |
| O'REILLY_2005_WHAT | 143 | 6.82 | 3.79% | 0.18% | 7 |
| STEINFIE_2007_J_1143 | 104 | 3.30 | 2.76% | 0.09% | 5 |
| BOYD_2007_J_210 | 88 | 3.34 | 2.33% | 0.09% | 4 |
| ELLISON_2007_J | 83 | 2.67 | 2.20% | 0.07% | 5 |
| BOYD_2007_J | 82 | 3.05 | 2.17% | 0.08% | 5 |
| GRANOVET_1973_A_1360 | 171 | 5.64 | 4.53% | 0.15% | 21 |
| ADOMAVIC_2005_I_734 | 68 | 3.66 | 1.80% | 0.10% | 7 |
| DONATH_2004_B_71 | 67 | 1.81 | 1.78% | 0.05% | 7 |
| HERLOCKE_2004_A_5 | 63 | 2.95 | 1.67% | 0.08% | 8 |
| GOLDER_2006_J_198 | 57 | 2.77 | 1.51% | 0.07% | 7 |
| **Author** | $k$ | $k^{\mathrm{N}}$ | $K$ | $K^{\mathrm{N}}$ | $\lambda$\|100% |
| JUNG,_JASON_J | 19 | 15.58 | 0.50% | 0.41% | 7 |
| THELWALL,_MIKE | 12 | 7.33 | 0.32% | 0.19% | 6 |
| CARMINATI,_BARBARA | 10 | 3.23 | 0.27% | 0.09% | 6 |
| FERRARI,_ELENA | 10 | 3.23 | 0.27% | 0.09% | 6 |
| SUNDARAM,_HARI | 10 | 3.20 | 0.27% | 0.08% | 4 |
| PASSARELLA,_ANDREA | 10 | 2.85 | 0.27% | 0.08% | 7 |
| DECKER,_STEFAN | 9 | 2.78 | 0.24% | 0.07% | 5 |
| GOLBECK,_JENNIFER | 9 | 6.17 | 0.24% | 0.16% | 6 |
| LIN,_YU_RU | 9 | 2.20 | 0.24% | 0.06% | 3 |
| ALMEIDA,_VIRGILIO | 8 | 1.66 | 0.21% | 0.04% | 3 |

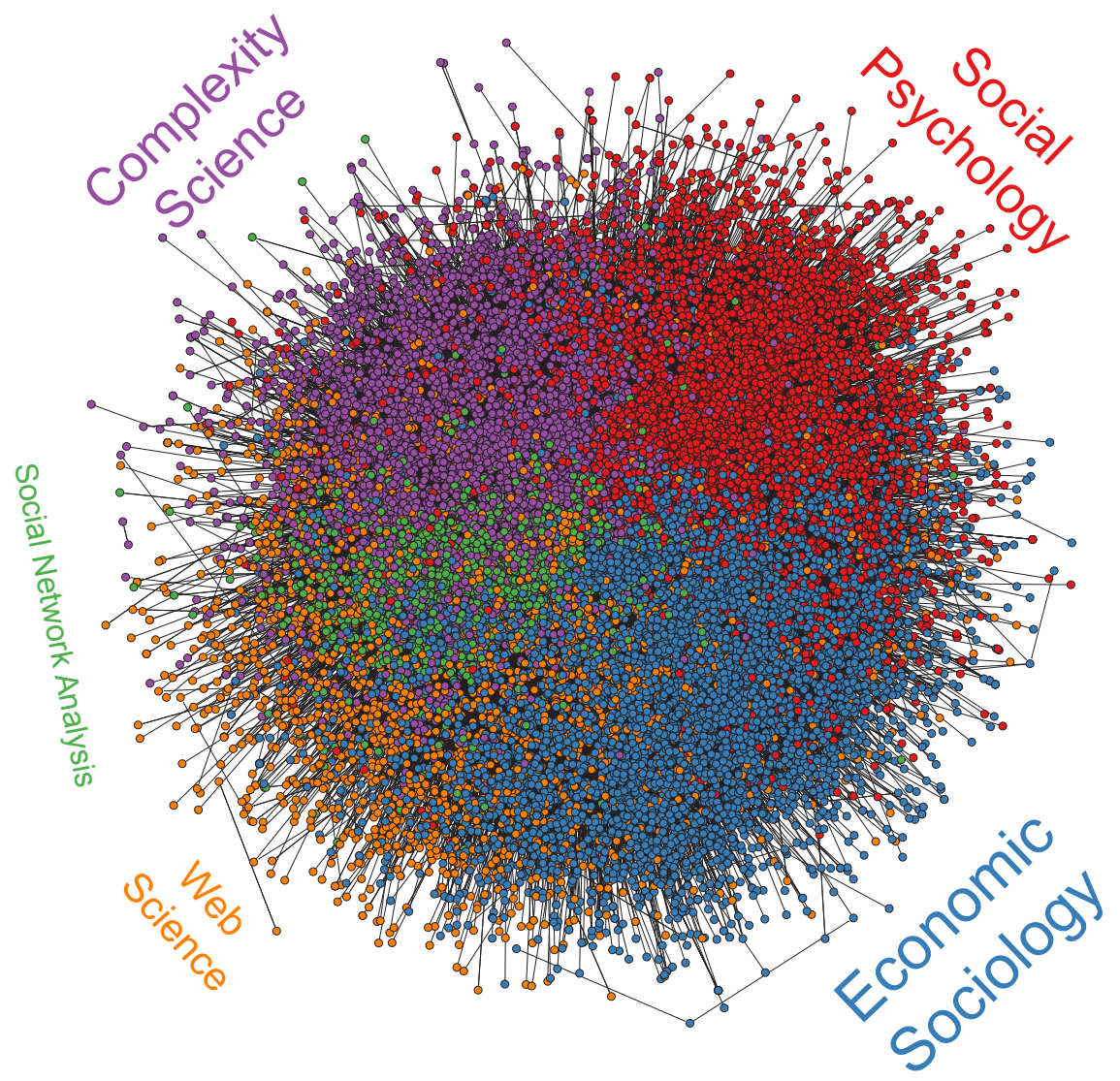

**Figure 3.4.: Subdomains of Social Network Science**
Largest bicomponent (23,958 publications, strongest edge per node) of the normalized hybrid publication network, where publications are coupled through the citation of references and the usage of words. Five subdomains are found for the full network (25,760 publications, density $D = 0.57$) with modularity $Q = 0.13$ and consensus $C = 0.91 \pm 0.07$ using the Louvain method: Social Psychology (red), Economic Sociology (blue), Social Network Analysis (green), Complexity Science (purple), and Web Science (orange). Subdomains are described in table 3.6. Label size corresponds to the number of publications. The network is layouted using the Fruchterman/Reingold algorithm in *Pajek* on the top 0.1% of the strongest similarities.

detecting these five identities is $0.91 \pm 0.07$, i.e., adding the core to the interim solution has reduced the consensus from the interim solution. Boundaries are fuzzier again. Figure 3.4 depicts the normalized publication network of the domain, with colors indicating subdomains. Again, publication similarities are averages of citation-based and lexical similarities. Modularity is low ($Q = 0.13$) because more than every second edge exists ($D = 0.57$).

Some classifications of publications are counterintuitive. A paper of Heider on balance theory (`HEIDER_1946_J_107`) is not in Social Psychology but in Economic Sociology, together with Cartwright and Harary's graph theoretical generalization (`CARTWRIG_1956_-P_277`) as well as foundational works of the Harvard school, like Granovetter's "The strength of weak ties" (`GRANOVET_1973_A_1360`) and H. C. White et al.'s paper on blockmodeling (`WHITE_1976_A_730`). This makes sense because these papers belong to the sociometry tradition initiated by Moreno.

The importance of fractional selection counting in the construction of similarity scores is demonstrated by the average number of references per publication which is a charac-

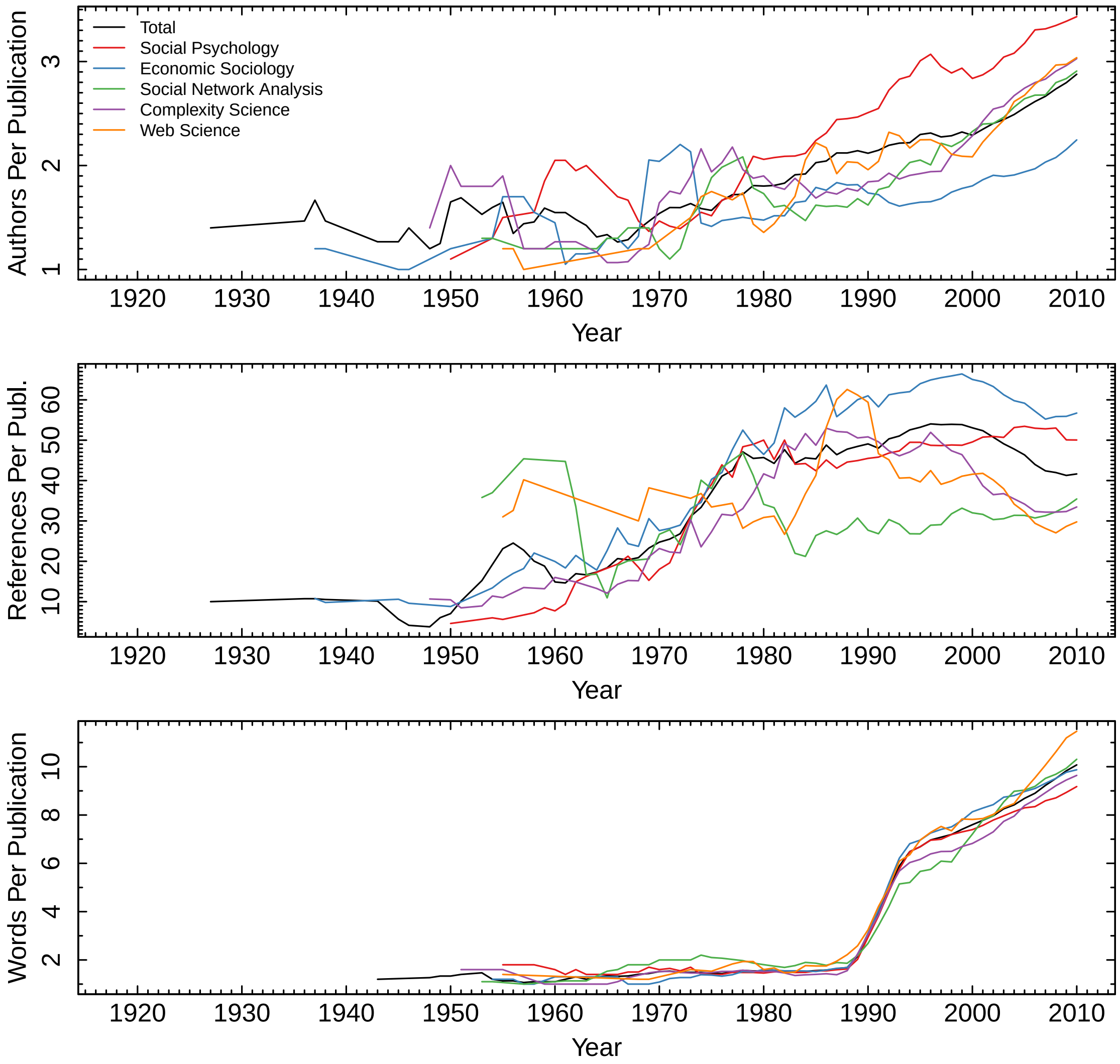

**Figure 3.5.: Average number of facts per publication**

The number of facts that an average paper selects is characteristic for subdomains. The average number of authors per publication is an indicator of how much a subdomain constitutes team science. The average length of reference lists is more or less steady since the late 70s and even exhibits a tendency to decline in three subdomains. The sudden increase of word usage is due to the inclusion of abstracts and author keywords into the *Web of Science* database starting in 1990. While this will have an effect on the applicability of our metrics, the bias is the same for all subdomains. Scores are 5-year averages.

teristic score for each subdomain, depicted in figure 3.5. The fact that an average paper in Economic Sociology cited almost twice as many references in 2010 than an average paper in Web Science means that a citation in the latter subdomain is twice as valuable. Citation counts $k^{\mathrm{N}}$ account for these differences but are still affected by the size of the respective subdomain. Publication fractions $K$ account for size differences but not for different citation practices. Only the citation fractions $K^{\mathrm{N}}$, which we summed to use as the genericness retrieval parameter, are comparable across subdomains. The reference `WASSERMA_1994_SOCIAL` and word `SOCIAL_NETWORK_ANALYSIS` are about ten times more common in Social Network Analysis than the top reference `O'REILLY_2005_WHAT` in Web Science or the top word `COMPLEX_NETWORK` in Complexity Science. For the authorship practice, the normalization effect is strongest, indicative of Priceian fractional authorship.

After the domain had been delineated, we gained access to the full *Web of Science* database for publications not published before 1980. This allowed further reducing the

**Table 3.7.: Generality of Words**

For each subdomain, the top five words for each generality class are given. In the 90% generality class, no more that 90% of the word usages are outside Social Network Science.

| | Generality | | |
|---|---|---|---|
| | **90%** | **99%** | **99.9%** |
| **SP** | SOCIAL_SUPPORT_NETWORK | SOCIAL_SUPPORT | SAMPLE |
| | SOCIAL_RELATIONSHIP | FRIEND | SUPPORT |
| | SOCIAL_CAPITAL | COMMUNITY | RELATIONSHIP |
| | SUPPORT_NETWORK | FRIENDSHIP | STUDY |
| | SOCIAL_TIE | LONGITUDINAL_STUDY | FAMILY |
| **ES** | SOCIAL_CAPITAL | ORGANIZATIONAL | PROCESS |
| | ACTOR_NETWORK_THEORY | INNOVATION | ROLE |
| | EMBEDDEDNESS | COMMUNITY | ORGANIZATION |
| | FOUNDATION | OPPORTUNITY | DEVELOPMENT |
| | SOCIAL_STRUCTURE | TRUST | ECONOMIC |
| **SNA** | SOCIAL_NETWORK_ANALYSIS | NETWORK_ANALYSIS | ANALYSIS |
| | SOCIAL_NETWORK_ANALYSIS_SNA | CENTRALITY | METHOD |
| | SOCIAL_STRUCTURE | COMMUNITY | USE |
| | BETWEENNESS | COLLABORATION | PAPER |
| | BETWEENNESS_CENTRALITY | NETWORK_STRUCTURE | DATA |
| **CS** | SCALE_FREE_NETWORK | COMPLEX_NETWORK | MODEL |
| | DEGREE_DISTRIBUTION | COMMUNITY | NODE |
| | SMALL_WORLD_NETWORK | NETWORK_STRUCTURE | ALGORITHM |
| | REAL_WORLD_NETWORK | CONNECTIVITY | SIMULATION |
| | CLUSTER_COEFFICIENT | SMALL_WORLD | PAPER |
| **WS** | FACEBOOK | USER | INFORMATION |
| | SOCIAL_NETWORK_SITE | INTERNET | PAPER |
| | TWITTER | WEB | SHARE |
| | WEB_2.0 | COMMUNITY | CONTENT |
| | ONLINE_SOCIAL_NETWORK | PRIVACY | APPLICATION |

vocabulary to those words that are specific in the original sense of the Zitt/Bassecoulard method. To distinguish this attribute from the specificity defined above, it is called generality. Table 3.7 lists the most used words per subdomain and generality class. In table 3.6, top words are listed that are not more general than 90%. This generality class of 23,027 words is also used throughout our work because it sufficiently removes general language. Using the most general top words, the subdomains can be distinguished through main methodologies. These are helpful putting subdomain labels into perspective. Web Science, e.g., dates back to 1916. This is a consequence of our decision to treat subdomains as complete histories. What we detect as Web Science is a set of structurally equivalent publications that cite similar references and use similar words. The label was chosen because the most-selected references and words revolve around the web which, in turn, is because 75 percent of the publications are not older than 2008. The methodological descriptors designate Web Science as applied computer science. Complexity Science contributes algorithmic modeling while Social Network Analysis is the methodological powerhouse of Social Network Science. Economic Sociology is largely concerned with the analysis of processes and roles. Work in Social Psychology, a dominant topic of which is public health, revolves around the sampling method.

### Summary

Delineating a research domain is a complicated problem because an identity is a core/periphery structure with a fuzzy boundary that can only be quantified with some precision. Because our goal was to delineate a multidisciplinary science of social networks, we chose to delineate the subdomains of Social Network Science and unify them. Our publication retrieval method mimics identities as dualities of emergence and downward causation, i.e., we have identified the cores of subdomain seeds and retrieved publications that select these cores. To asses the quality of our solution, we have sampled and evaluated 1,000 papers. Confidence into the precision of the domain boundary is always as low as of the least precise subdomain boundary because cores of subdomains overlap and social facts that are precise in one subdomain are not necessarily precise in another one.

The whole procedure results in a solution set of 25,760 articles that are partitioned into Social Psychology, Economic Sociology, Social Network Analysis, Complexity Science, and Web Science. These subdomains are self-contained or ideationally closed to varying degrees. Such tendencies are consolidated by the delineation method because it is essentially a feedback mechanism. To not let our method determine results too strongly, a hybrid delineation approach of citation-based and lexical retrieval was followed. For the same reason, communities were detected in a hybrid way. Language is generally less distinctive and counters the centrifugal forces of the citation practice. The authorship practice was not used for boundary creation or community detection so that it can be purely reactive to the path dependence that is still created by hybrid delineation.

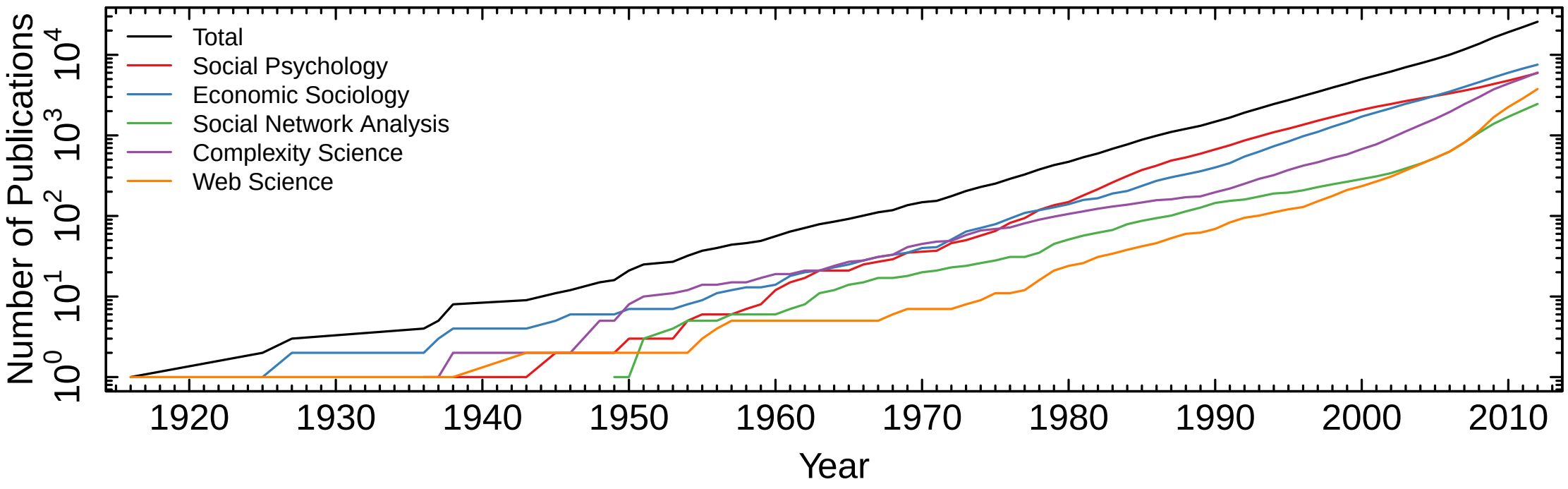

(a) Number of publications

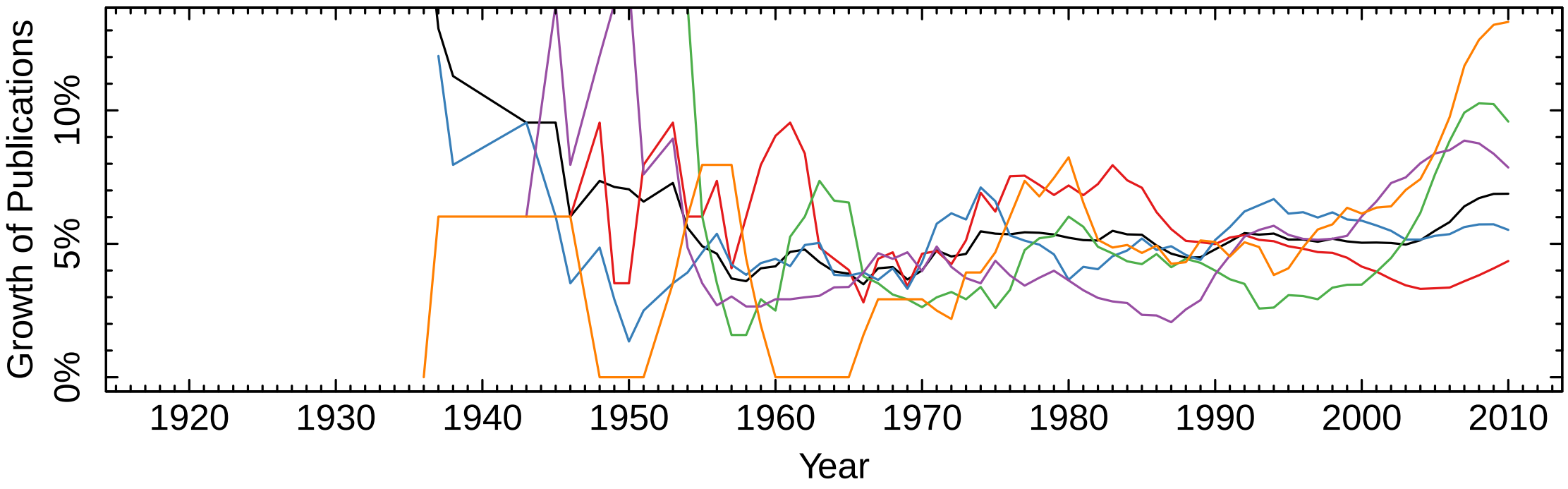

(b) Growth of publications

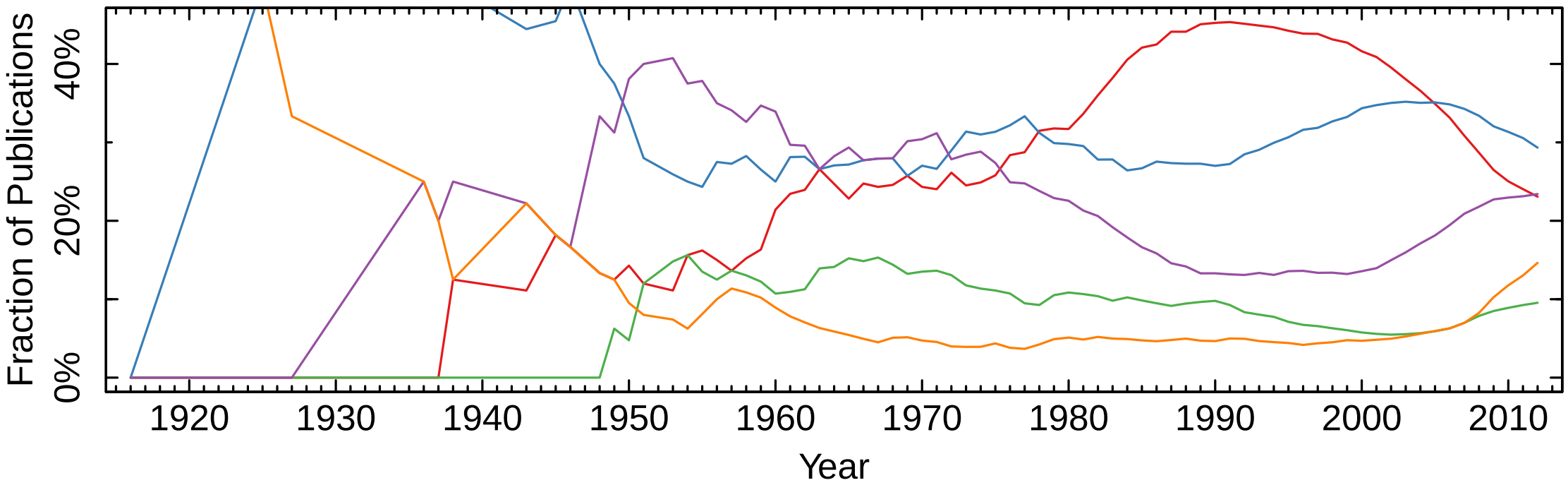

(c) Fraction of publications

**Figure 3.6.: Cumulative growth curves**

Exponential growth shows as a straight line on a logarithmic y-axis. (a) Cumulative number of publications. (b) Annual $\log(m(t+1)/m(t))$ growth rate of publications, smoothed by a 5-year average. (c) Non-cumulative fraction of publications by subdomain. (d) The cumulative number of publication is best mirrored by the cumulative number of authors. (e) Reference growth is least differentiated. (f) The sharp jump in word growth is due to the inclusion of abstracts and author keywords into the *Web of Science* database starting in 1990.

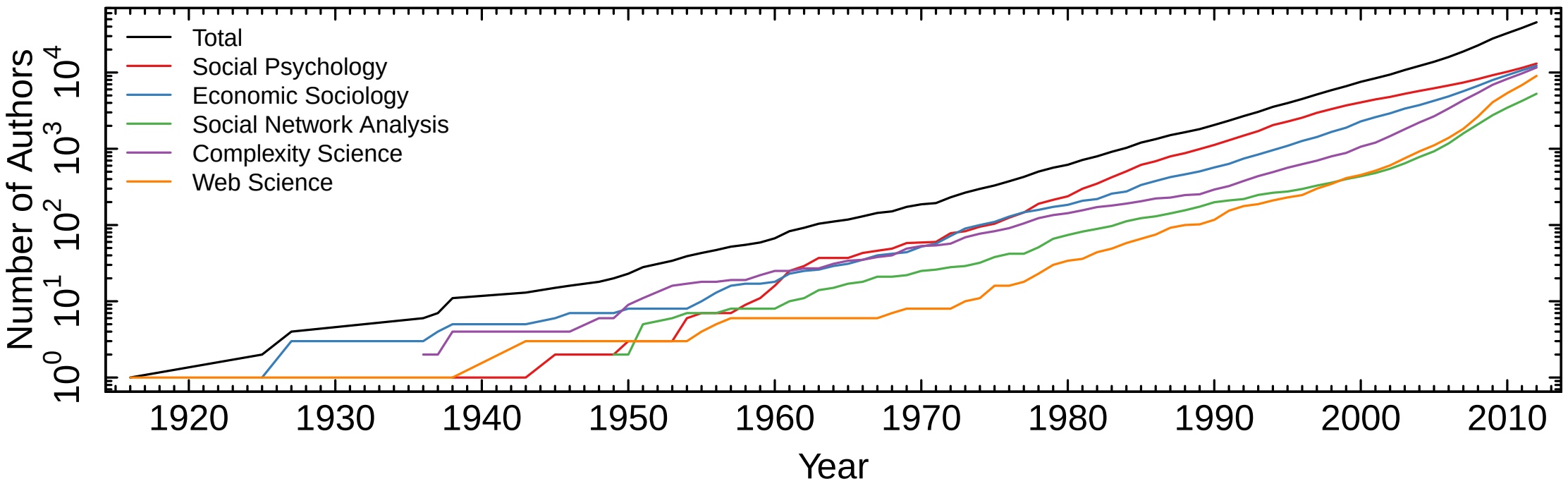

(d) Number of authors

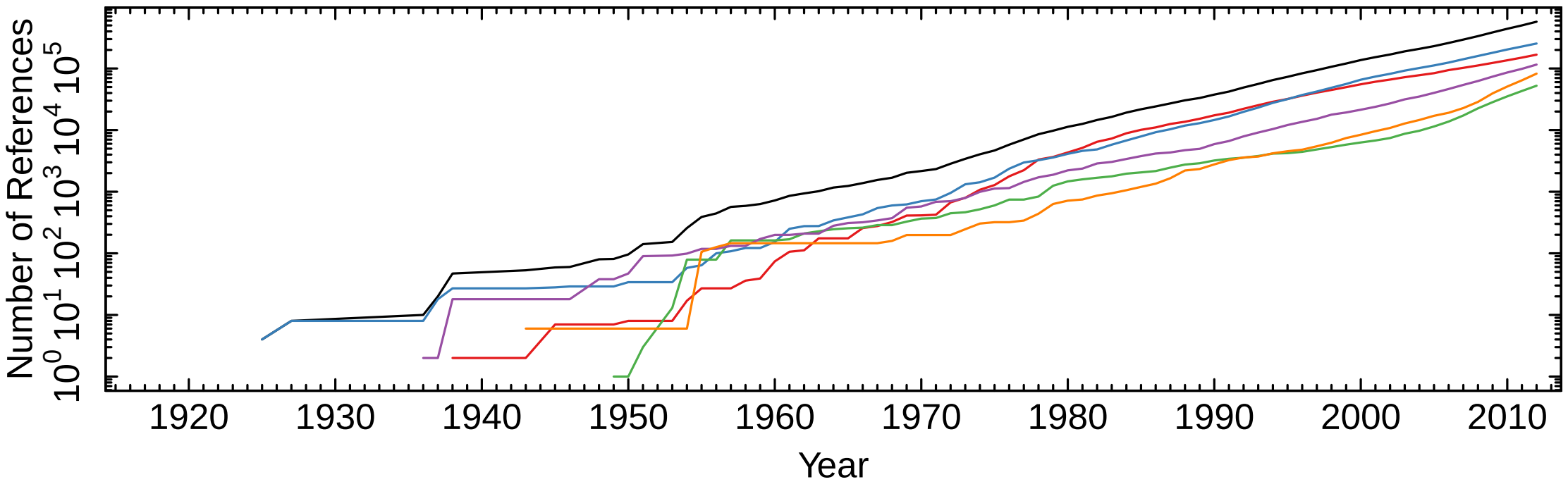

(e) Number of references

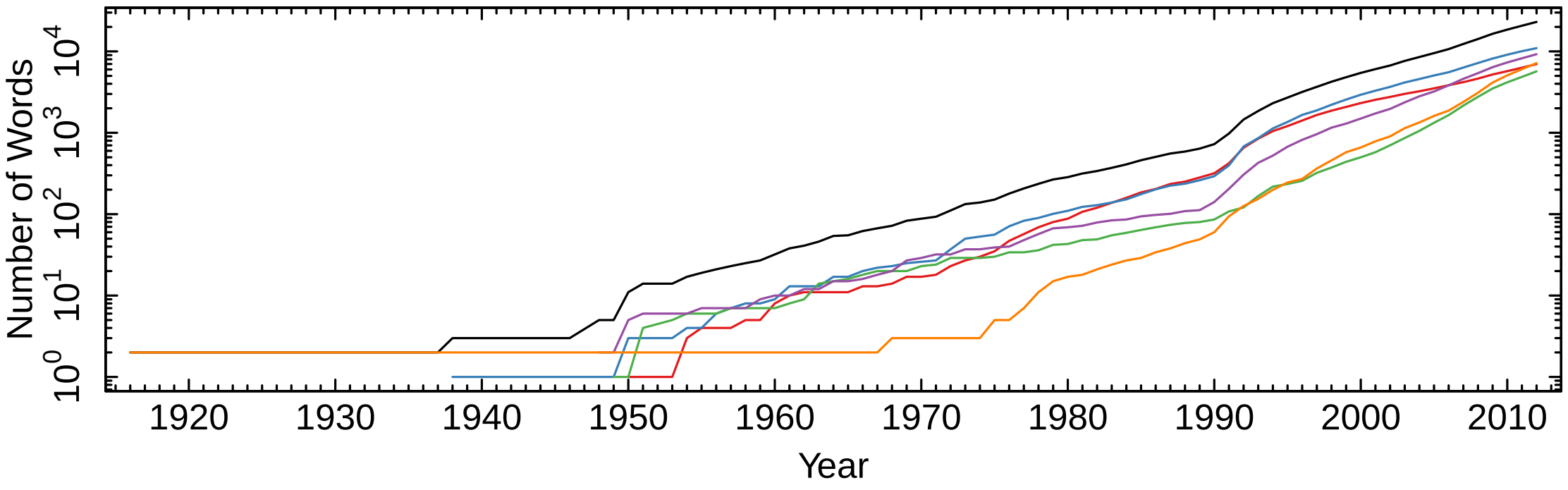

(f) Number of words

**Figure 3.6.: Cumulative growth curves**

## 3.2. Levels

Social Network Science was born in 1916. The founding paper, so to say, is HANIFAN_-1916_A_130 by Lyda J. Hanifan who provided the first definition of SOCIAL_CAPITAL as the constitutive and relational ties that an identity is formed by. It is artificially attributed to Web Science. And recall that a subdomain label tells what had become of the subdomain all throughout its history, not at any point in time. Figure 3.6 shows the cumulative growth curves for the domain and its subdomains. Publications and authors, which are strongly correlated, are the primary growth indicators. Essentially, the domain is emergent: 75% of the publications have been published since 2002. Domain growth can be distinguished into three periods: An initial period until the mid 70s during which subdomain growth fluctuated wildly. A maturation period during which domain growth stabilized at 5%. And an explosive period since the late 90s that, sooner or later, saw all subdomains accelerate their growth rates (figure 3.6b). Social Psychology is the oldest subdomain with a quarter of the papers published before 1997. During the mid 70s to mid 80s, it has the highest growth rate and increases its share of publications from 26% to 42% at the cost of Complexity Science. But it is not able to sustain this high growth rate and continuously decelerates until the mid 90s. Like the total, Economic Sociology has been growing rather constantly since maturation began. Social Network Analysis and Complexity Science exhibit accelerated growth beginning in the late 90s and lasting for a decade. The increasing importance of computer science can be seen in the growth dynamics of Web Science which started growing last of all subdomains and experienced the steepest growth since the mid 90s. Together with Complexity Science, it benefited most in the explosion period. From 2000 to 2012, they more than doubled their share of publications. In terms of authors, Web Science almost managed to reach par with Social Psychology, Economic Science, and Complexity Science. Unless noted, the following results in this section are for Social Network Science integrated over all years.

If the domain is at a critical point, then there are no typical length scales. Regarding the size (number of selections) of social facts, Lotka's and Bradford's Law are basically confirmed for the domain and subdomains, but Zipf's Law only for the subdomains. This may be a consequence of the delineation procedure which was done for subdomains. Cf. figure 3.7 and table 3.8 for parameters and test results. Fractal sizes are the litmus test for our theory. If we had not found power laws, we would have to reject the scaling hypothesis. They indicate hierarchies of meaning. The most productive author in Social Network Science (74 papers) is epidemiologist Carl Latkin (LATKIN,_CARL). The most cited reference is Wasserman & Faust's WASSERMA_1994_SOCIAL, a handbook on SOCIAL_NETWORK_ANALYSIS methods with 2,475 citations. The most frequently used word is COMMUNITY (3,091 usages). Since the domain was initiated in 1916, we say that such facts occur only once in a century. But facts with 40 authorships, 900 citations, or 1,000 word usages are expected once in a decade.[5]

[5]Social facts are ranked descendingly by the number of selections and frequencies are inverse ranks (Roberts & Turcotte, 1998).

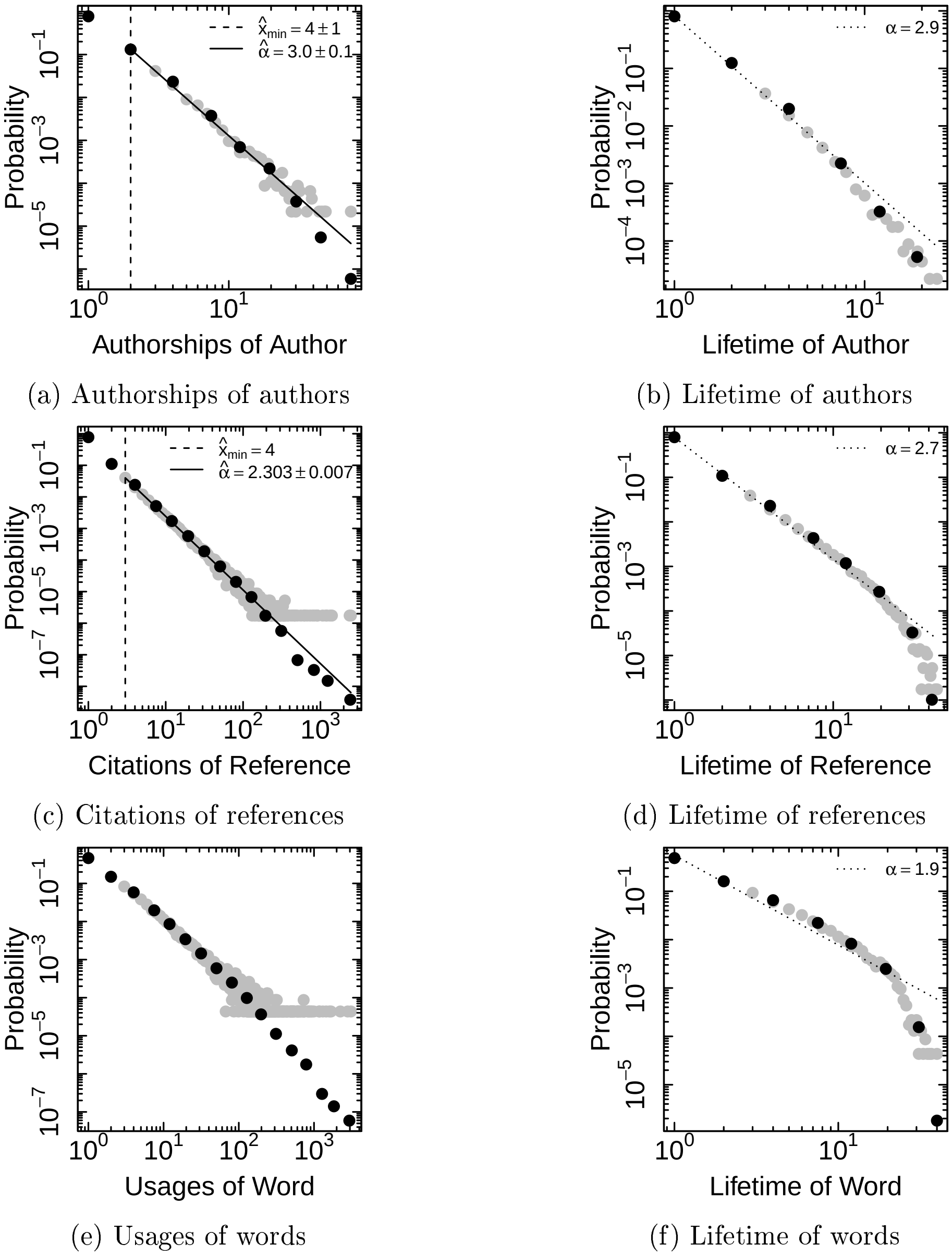

**Figure 3.7.: Size and lifetime distributions**

Distributions for the number of authorships by authors (a) and citations of references (c) are plausibly fit by power-law distributions. No characteristic size describes these distributions ($\alpha \leq 3$). A hypothetical power-law fit for word usage (e) with $\alpha = 2.0 \pm 0.1$ must be rejected. Nevertheless, the distribution is heavy-tailed ($\bar{x} = 9 \pm 51$). Parameters and goodness-of-fit test results for the domain and its five subdomains are given in table 3.8 and show that word usage is self-similar on the subdomain level. The distributions of unconditional lifetimes are not scale-invariant, i.e., there are typical lifetimes for all three facts: (b) $\bar{x} = 1 \pm 1$, (d) $\bar{x} = 2 \pm 2$, and (f) $\bar{x} = 3 \pm 4$ years. Still, given the skewness, it is questionable what these averages really mean. Scaling may prevail below an upper cutoff, and power-law hypotheses are shown as dotted lines.

### Levels of Analysis

To study hierarchies of meaning, we construct three levels of analysis. Figure 3.8 gives the details of construction, figure 3.9 explains what further steps were taken to visualize the levels and presents the networks. The motivation is to identify the paradigms of Social Network Science and, in case of the authorship practice, the invisible college. Regarding the identification of paradigmatic social facts, consult table 3.6a for an example: In Social Psychology, `LATKIN,_CARL` is responsible for 0.30% of all authorships. The first three ($tf * idf$-ranked) authors together account for no more than 1% of all selections. The cutoff is made at 5% and results in a set of 54 authors that are paradigmatic for Social Psychology. Two scholars stand out because they are highly productive in three subdomains: `CARLEY,_KATHLEEN` and `WELLMAN,_BARRY`, both students of Harrison C. White. Top authors for all subdomains but Social Psychology, namely `FOLKE,_CARL` in Economic Sociology, `LEYDESDORFF,_LOET` in Social Network Analysis, `NEWMAN,_M_E_J` in Complexity Science, and `JUNG,_JASON_J` in Web Science (table 3.6), are missing in the micro level network visualizations of figure 3.9c because they are not embedded in a bicomponent at that level.

**Table 3.8.: Size distribution parameters**

Power-law size distribution statistics are given for three practices on the levels of the whole domain and the subdomains. Power laws are plausible fits if $p > 0.1$. In general, word usage is most ($\alpha \approx 2$) and authorship is least ($\alpha \approx 3$) concentrated. Lower bounds are always very small, i.e., almost the whole distributions are self-similar. Word usage is scale-invariant only on the subdomain level which explains the large fluctuations in the total's lower bound.

| | | $\hat{x}_{\min}$ | $\hat{\alpha}$ | $p$ |
|---|---|---|---|---|
| **Authorship** | **Total** | $4 \pm 1$ | $2.9 \pm 0.1$ | 0.30 |
| | **Social Psychol.** | $3 \pm 1$ | $2.9 \pm 0.2$ | 0.07 |
| | **Economic Sociol.** | $4 \pm 1$ | $3.5 \pm 0.7$ | 0.46 |
| | **Soc. Netw. Anal.** | $4 \pm 1$ | $3.3 \pm 0.2$ | 0.74 |
| | **Complexity Sci.** | $4 \pm 1$ | $2.81 \pm 0.07$ | 0.24 |
| | **Web Science** | $4 \pm 1$ | $3.8 \pm 0.7$ | 0.19 |
| **Citation** | **Total** | 4 | $2.303 \pm 0.007$ | 0.39 |
| | **Social Psychol.** | $4 \pm 1$ | $2.43 \pm 0.03$ | 0.31 |
| | **Economic Sociol.** | 4 | $2.34 \pm 0.01$ | 0.77 |
| | **Soc. Netw. Anal.** | $4 \pm 1$ | $2.45 \pm 0.03$ | 0.97 |
| | **Complexity Sci.** | $5 \pm 2$ | $2.33 \pm 0.04$ | 0.83 |
| | **Web Science** | $4 \pm 1$ | $2.67 \pm 0.05$ | 0.89 |
| **Word usage** | **Total** | $10 \pm 14$ | $2.0 \pm 0.1$ | 0.00 |
| | **Social Psychol.** | $3 \pm 1$ | $1.96 \pm 0.03$ | 0.53 |
| | **Economic Sociol.** | $3 \pm 2$ | $1.95 \pm 0.09$ | 0.00 |
| | **Soc. Netw. Anal.** | $2 \pm 1$ | $2.05 \pm 0.04$ | 0.55 |
| | **Complexity Sci.** | $2 \pm 1$ | $1.92 \pm 0.04$ | 0.27 |
| | **Web Science** | $2 \pm 1$ | $1.91 \pm 0.03$ | 0.46 |

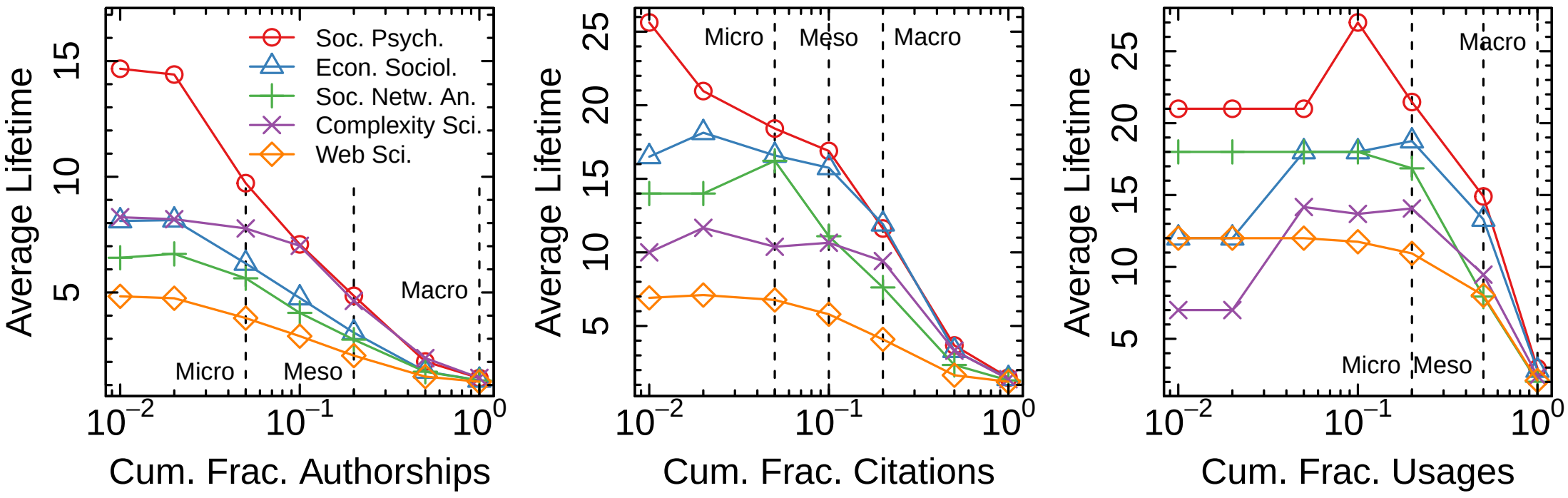

(a) Authorship (b) Citation (c) Word usage

**Figure 3.8.: Construction of levels for analysis and visualization**

Social facts cover five to six orders of magnitude. There are 45,580 authors, 574,036 cited references, and 23,027 used words in Social Network Science (cf. figure 3.6). Three levels are identified for analysis and visualization: a macro level that contains as much of the full sets as feasible and necessary, a meso level where the moderately selected facts are removed, and a micro level that consists only of core facts or paradigms of the meaning structures. The basic procedure to delineate levels of observation is described in section 2.2.1. Facts are ranked by $k \log(1/K)$ which gives more weight to facts weakly selected in other subdomains. Choices of cumulative fraction cutoffs for levels are shown as dashed lines and given in the table below. The macro level of citation is restricted to references that attract 20% of all selections to be comparable in size to the other practices.

| **Practice** | **Level** | **Cumulative selection fraction** | **Above average lifetime** |
|---|---|---|---|
| Authorship | Macro | 100% | No |
| | Meso | 20% | No |
| | Micro | 5% | No |
| Citation | Macro | 20% | No |
| | Meso | 10% | Yes |
| | Micro | 5% | Yes |
| Word usage | Macro | 100% | No |
| | Meso | 50% | Yes |
| | Micro | 20% | Yes |

For levels, average lifetimes are determined. Subfigures (a–c) show that highly selected facts have, by and large, longer average lifetimes, but there are differences for subdomains. Social Psychology and Economic Sociology memorize facts longer than, e.g., Web Science because they are older. In the final step of delineating meso and micro levels, we remove all facts that have shorter than average conditional lifetimes. This ensures that facts are not only highly selected but also strongly institutionalized. Authorship is exempted from this rule because it destroys network connectivity too much.

**Figure 3.9.: Fact networks at different levels**

Networks are fact co-selection graphs $H^{\mathrm{N}}$ shown for the three scientific practices and three different levels of observation. Node size is autocatalysis or self-reproduction in the whole population. For example, per publication, an author's self-reproduction is the smaller, the more co-authors he has. As an author writes multiple publications, these fractions accumulate. An author with a self-reproduction of 2 can, e.g., be the sole author of two papers or of four papers with each one co-author. Edge weights are fractionally counted co-selections. Subdomains are overlapping link communities of Social Psychology (red), Economic Sociology (blue), Social Network Analysis (green), Complexity Science (purple), and Web Science (orange). The construction of levels is described in figure 3.8. For the visualizations of co-citation and word co-usage shown here, edges are filtered to increase sparsity. Parameters are given in the table below. All filters are applied per subdomain. After filtering nodes and edges, the five subdomains are stacked on each other, pruned, and layouted. Stacking of subdomains is exemplified for co-authorship in figure 3.10. The stack is then reduced to the largest or all bicomponents (2-components), i.e., the periphery is cut off. In the case of co-authorship, the meso level represents the invisible college of Social Network Science. Force-directed layout algorithms are abbreviated FR (Fruchterman/Reingold) and KK (Kamada/Kawai). FR is fast but places structurally equivalent nodes on top of each other. This is desirable for large and dense networks because it naturally visualizes cores. For small and sparse networks, the slower KK is preferred. Networks are layouted in *Pajek*.

| Practice | Level | Cumulative edge fraction | Bicomponent | Number of nodes | Layout |
|---|---|---|---|---|---|
| Authorship | Macro | 100% | Largest | 4,790 | KK |
| | Meso | 100% | Largest | 243 | KK |
| | Micro | 100% | All | 29 | KK |
| Citation | Macro | 5% | Largest | 5,624 | FR |
| | Meso | 2% | Largest | 285 | FR |
| | Micro | 1% | All | 31 | KK |
| Word usage | Macro | 5% | Largest | 3,622 | FR |
| | Meso | 2% | Largest | 323 | FR |
| | Micro | 1% | All | 16 | KK |

In general, of the three practices, authorship is most fissured and fine-structured. Authors at the micro level are prominent group leaders and nodes can, therefore, be understood to represent research groups. Citation is multi-peaked with more pronounced subfield-boundaries. Social Psychology has two cores. The core farther from the network center revolves around `BERKMAN_1979_A_186`. Web Science's micro level co-citation core is badly visible and is a star with `GRANOVET_1973_A_1360` in the center. Word co-usage is a much more coherent single-peaked meaning structure. The one concept that all subdomains use at the micro level is `COMMUNITY`.

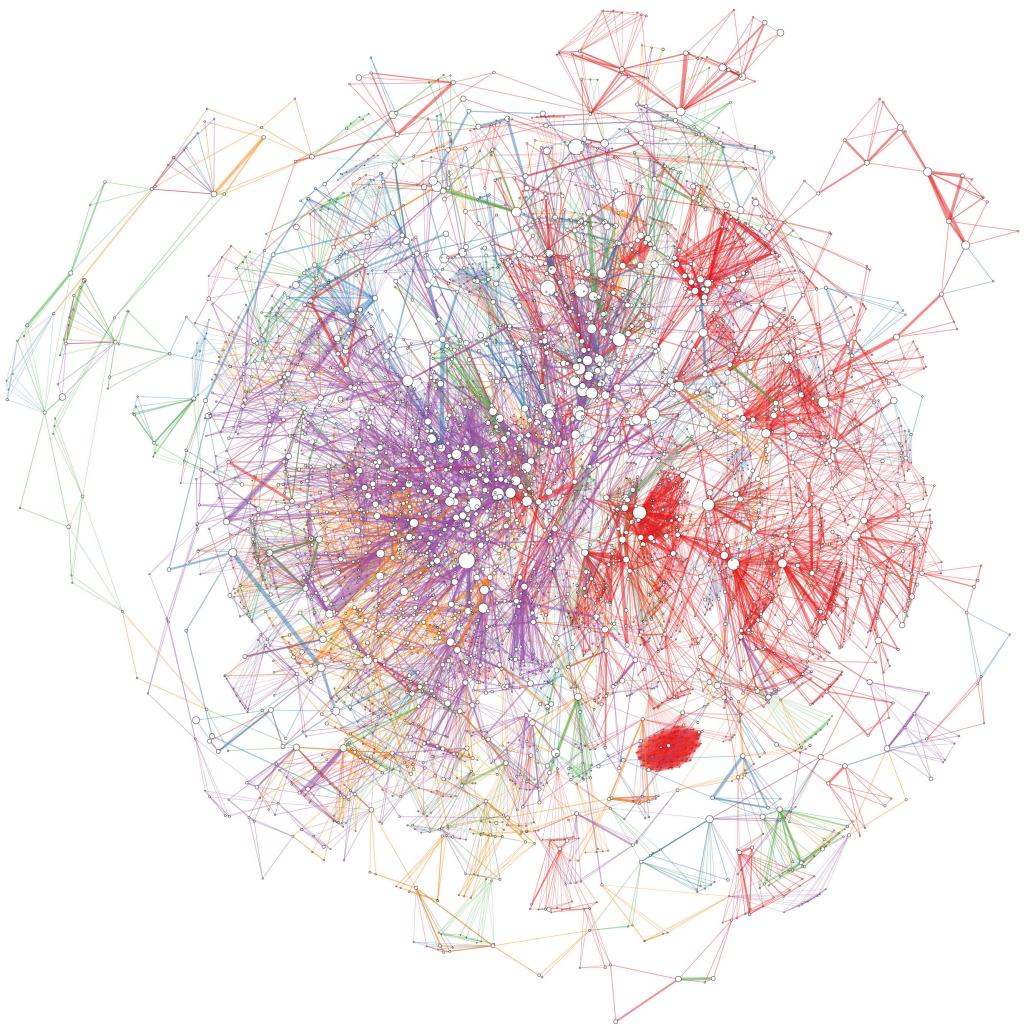

(a) Macro level authorship

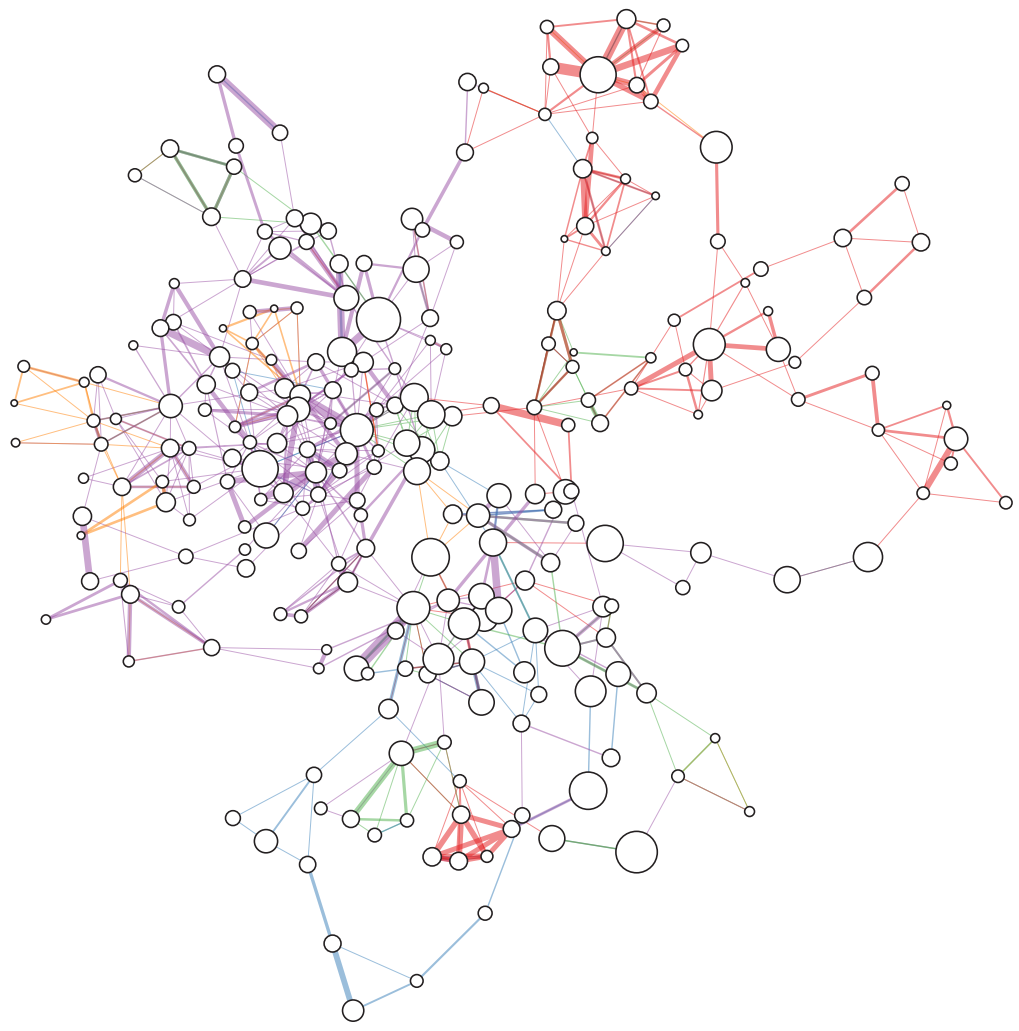

(b) Meso level authorship

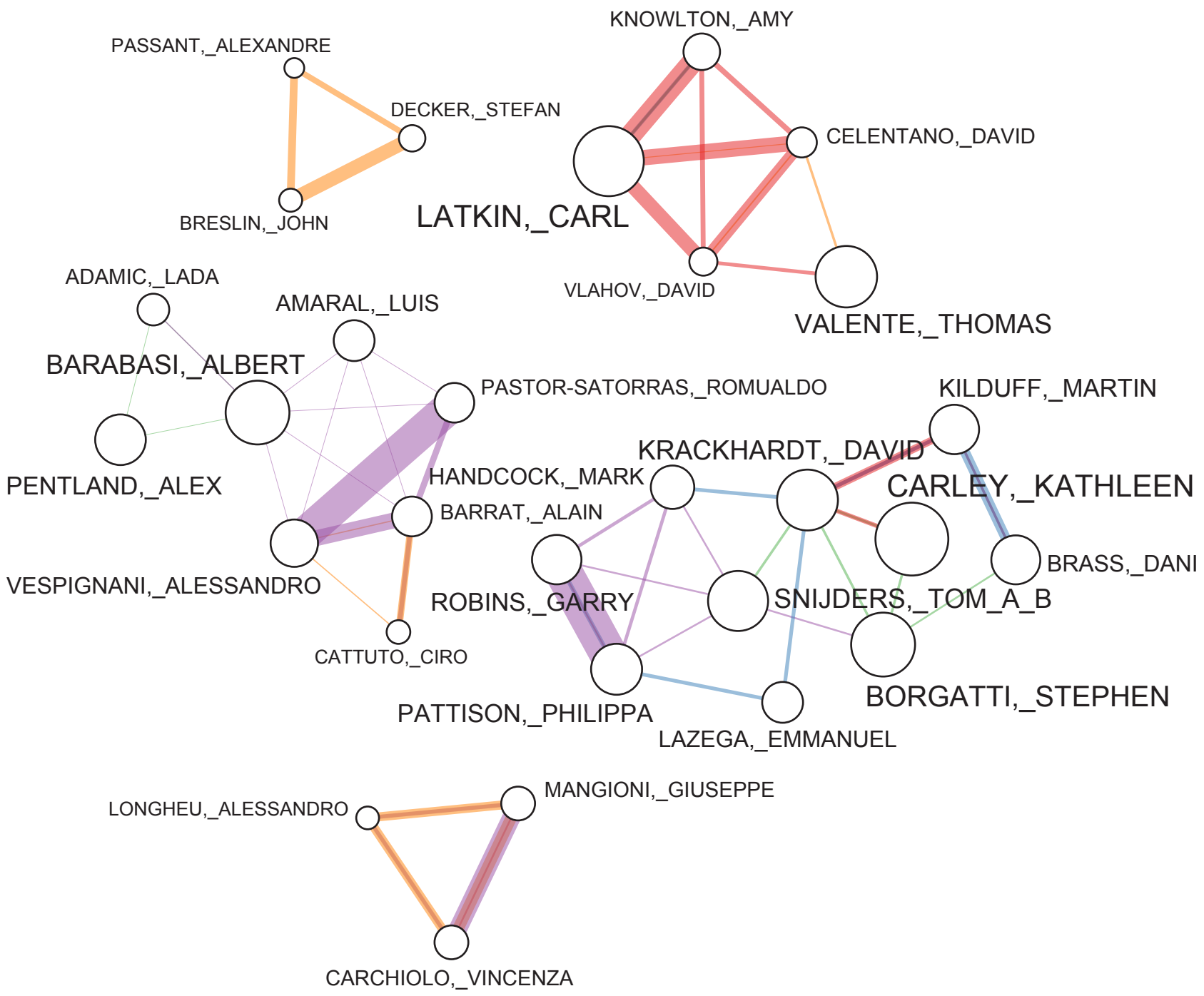

(c) Micro level authorship

**Figure 3.9.: Fact networks at different levels**

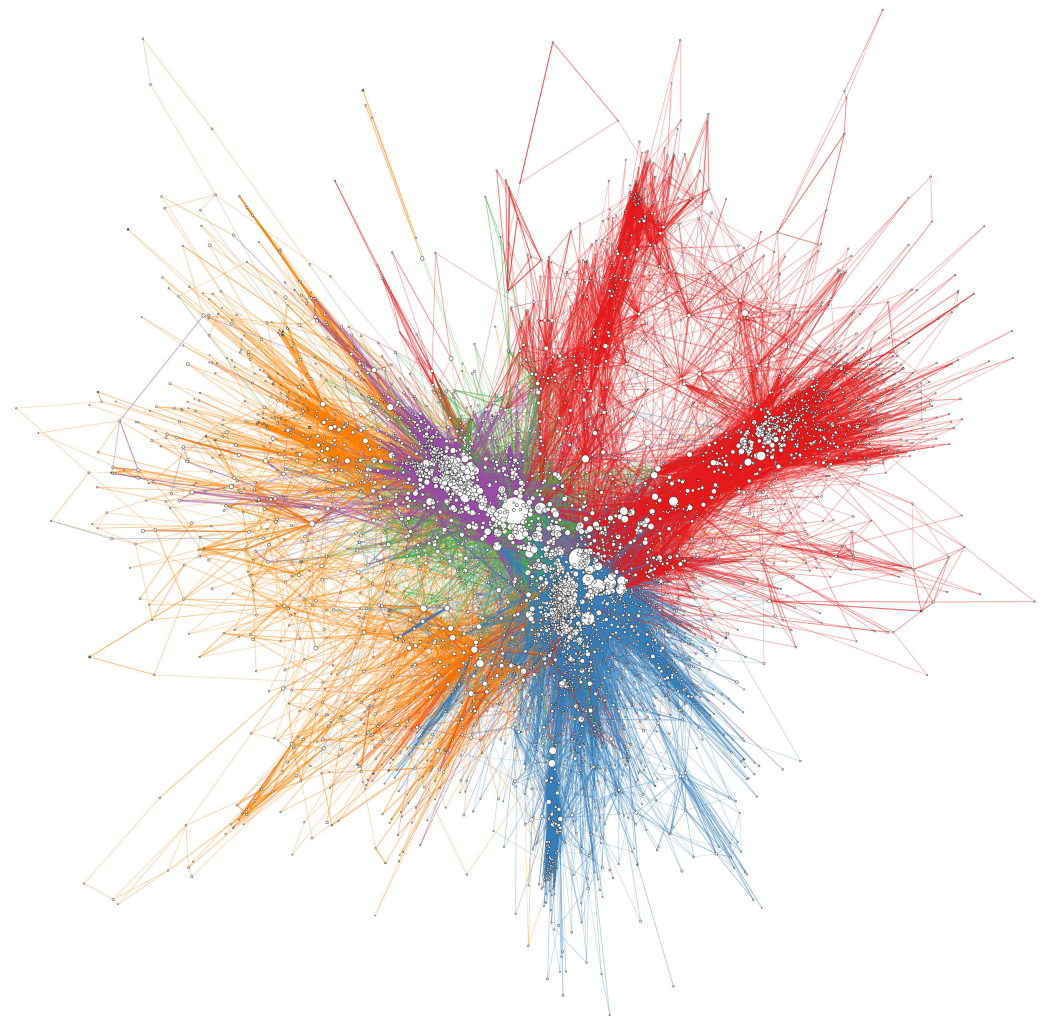

(d) Macro level citation

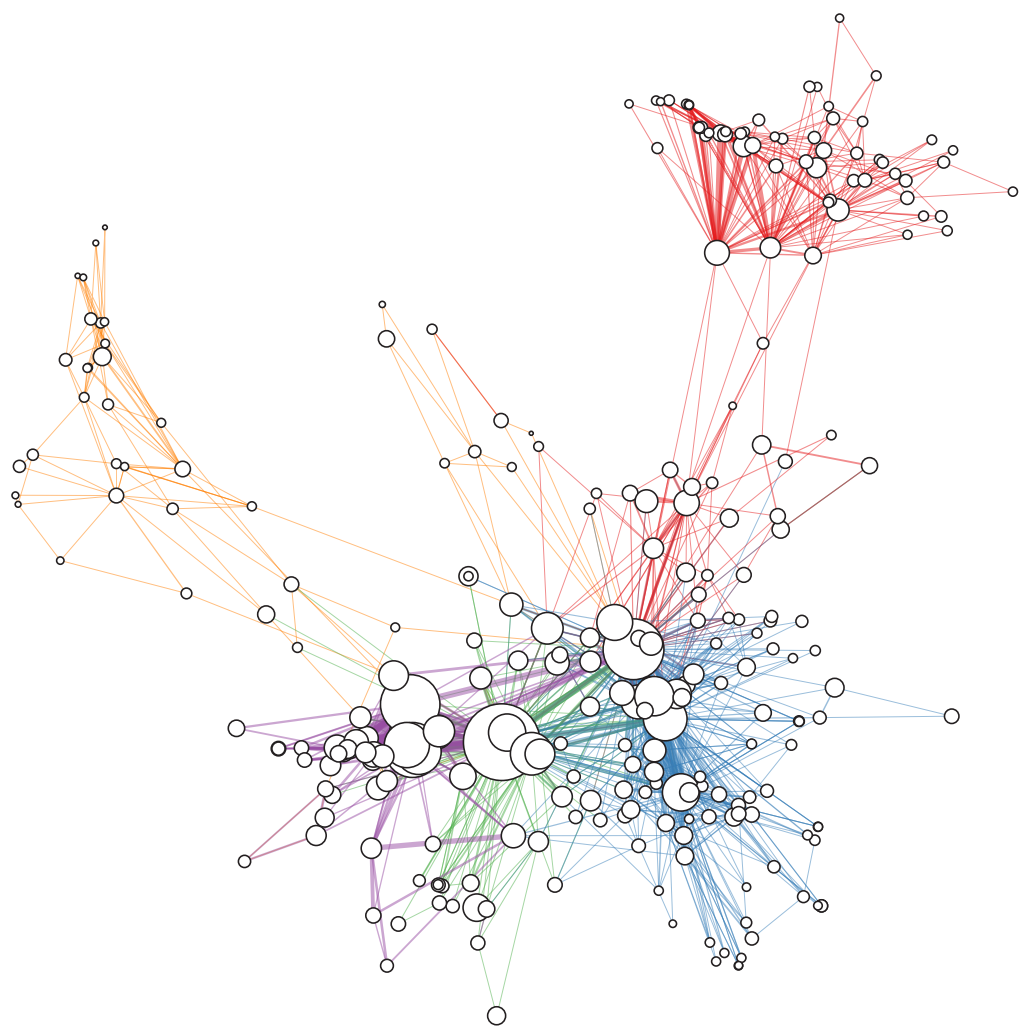

(e) Meso level citation

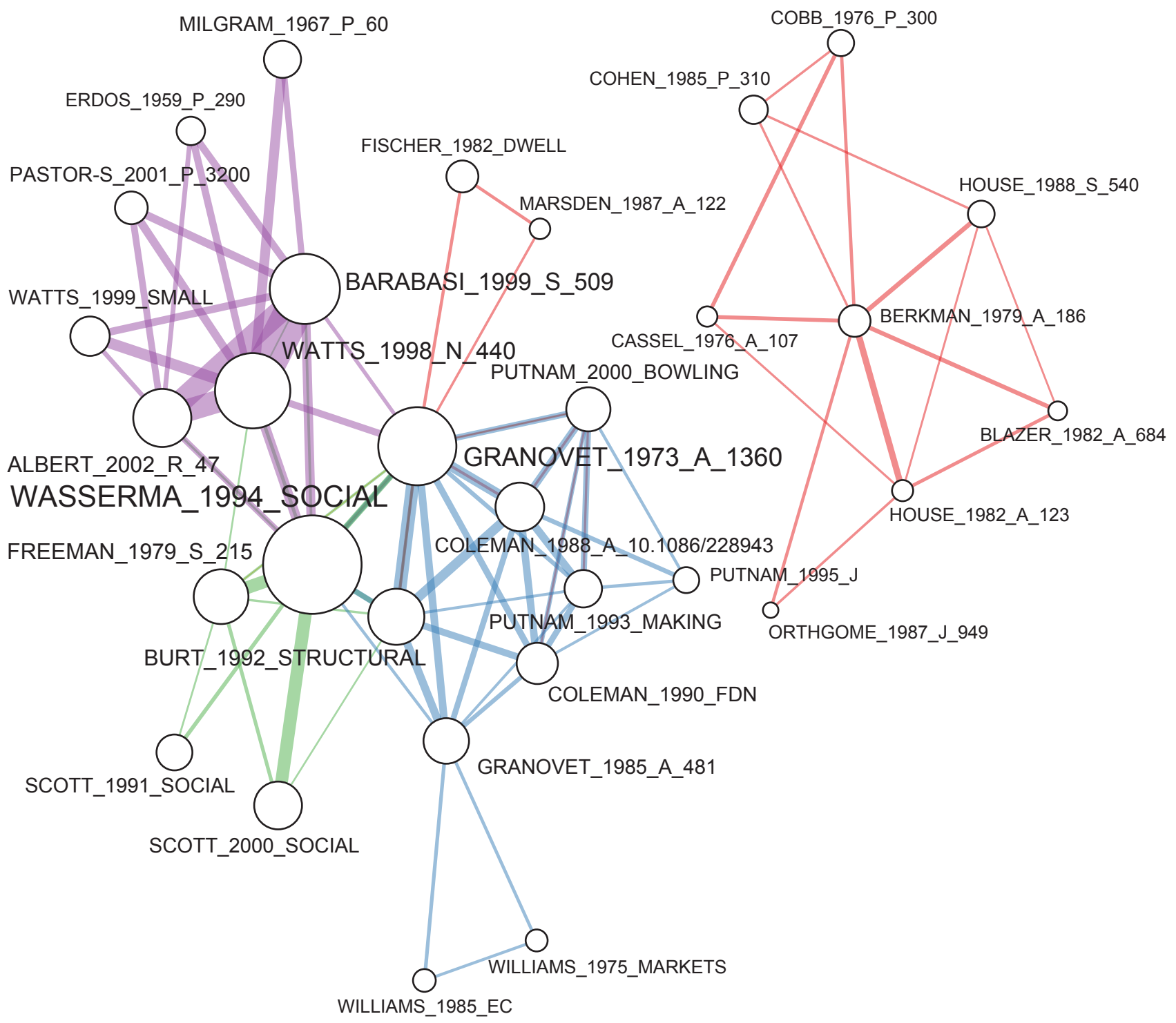

(f) Micro level citation

**Figure 3.9.: Fact networks at different levels**

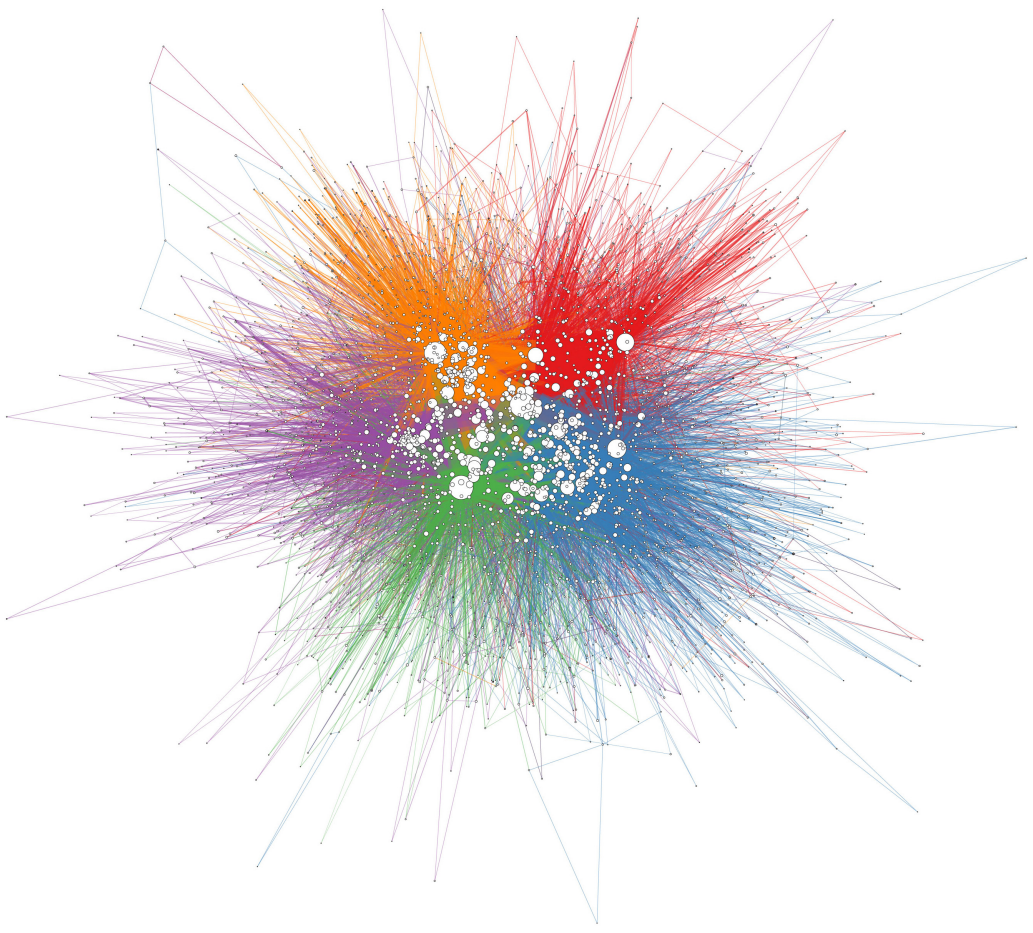

(g) Macro level word usage

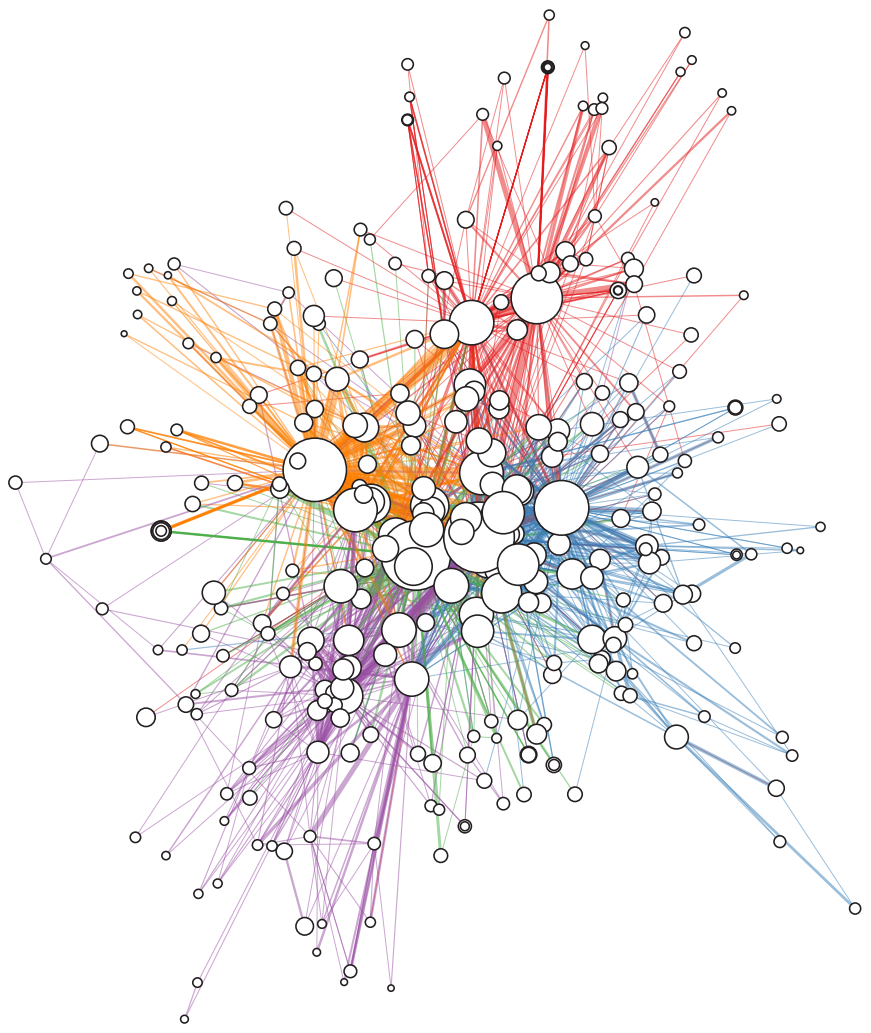

(h) Meso level word usage

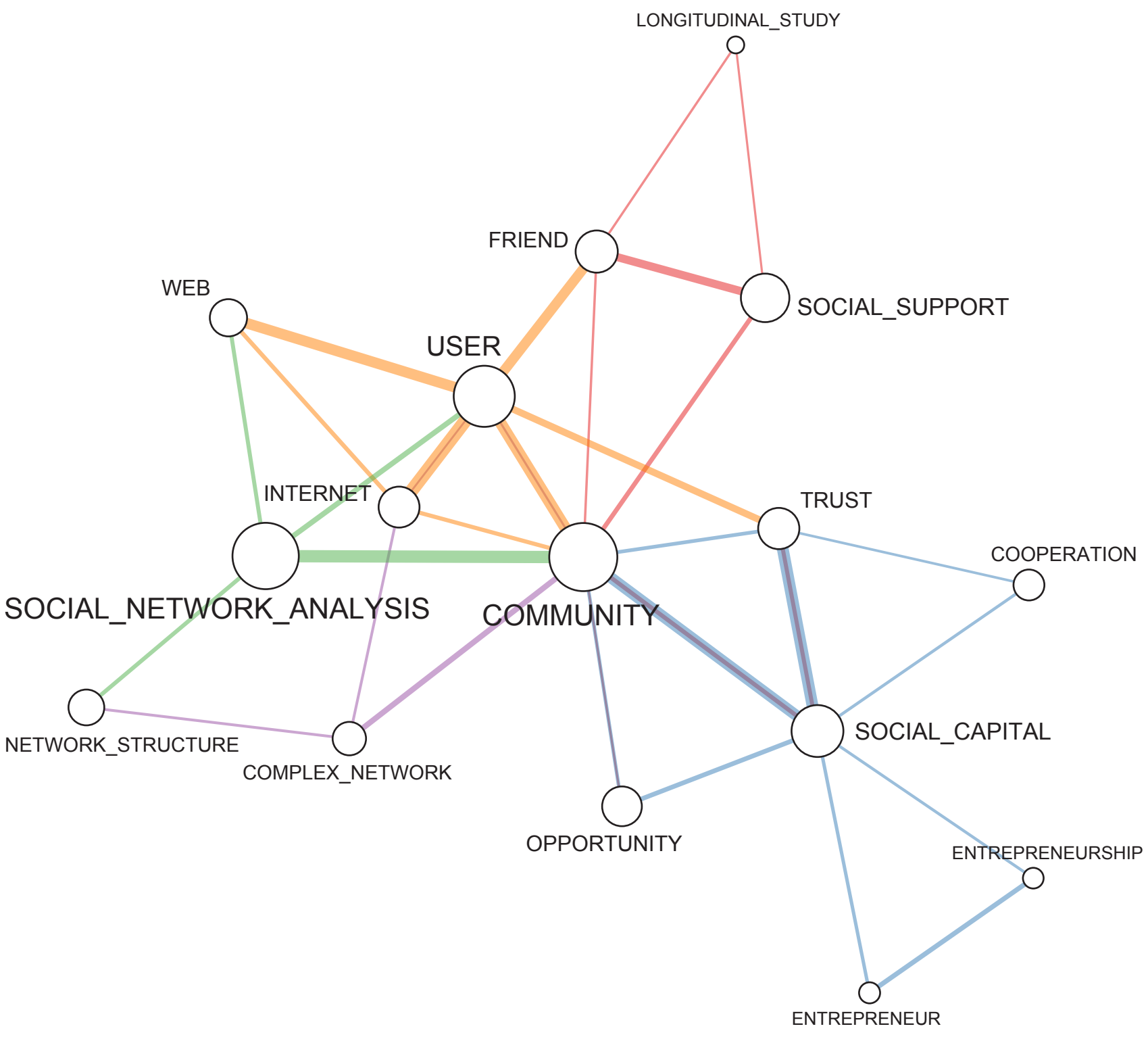

(i) Micro level word usage

**Figure 3.9.: Fact networks at different levels**

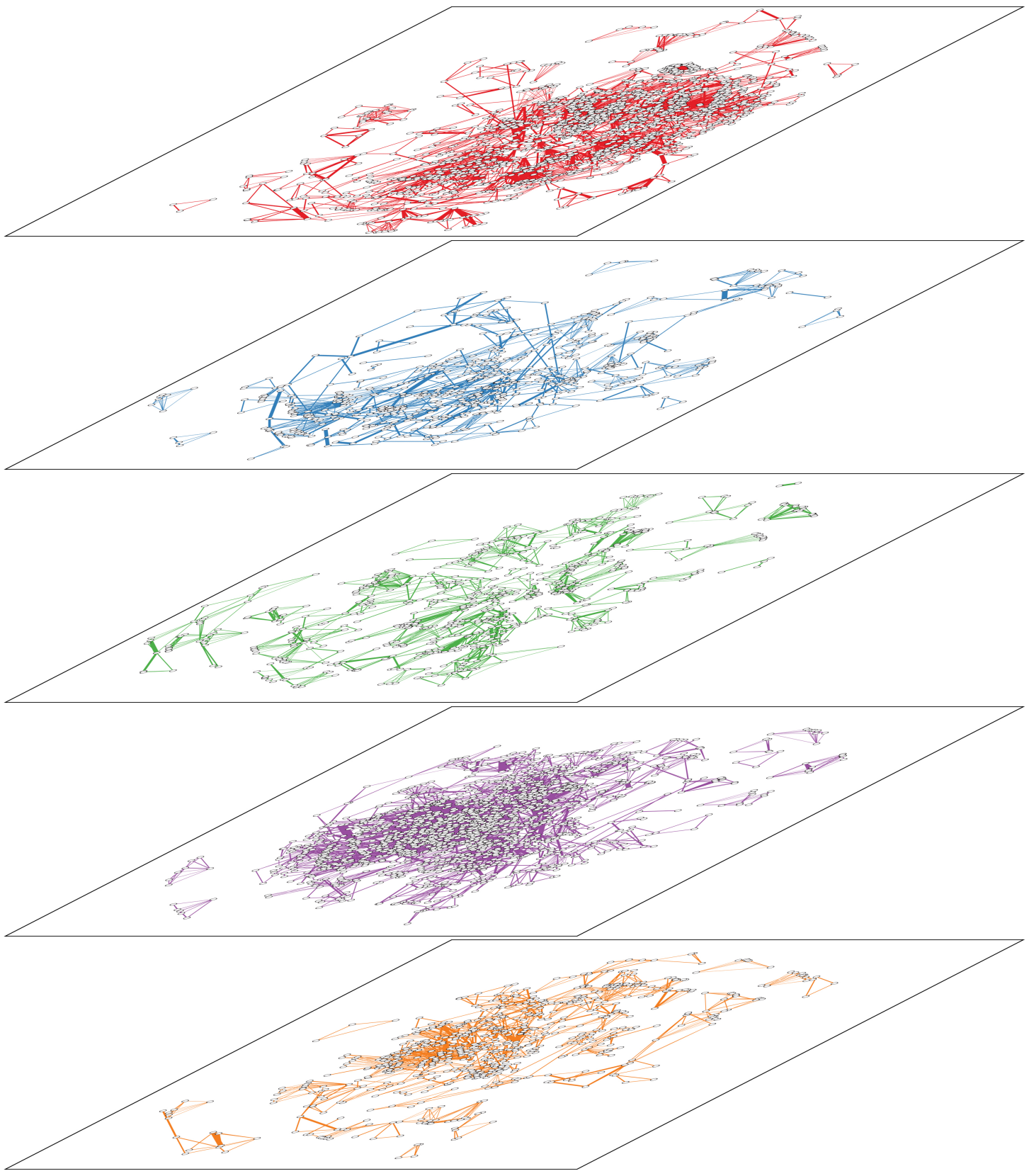

**Figure 3.10.: Authorship subdomains and switching**
Author co-authorship subdomains ($H^{\mathrm{N}}$ networks) are stacked on each other. There are five types of tie or story sets, one each for the five subdomains, from top to bottom: Social Psychology (SP, red), Economic Sociology (ES, blue), Social Network Analysis (SNA, green), Complexity Science (CS, purple), and Web Science (WS, orange). There are 45,583 authors in Social Network Science, 4,790 of which belong to the largest bicomponent. Overlap and meaning results from authors that switch among multiple subdomains. Overall, 90% of all authors are specialized to just one subdomain. 9% publish in two and 1% in three subdomains. Only 139 authors switch among four and 23 among all five subdomains. The diagonal of the table shows how many authors publish at least one paper in a subdomain. Off-diagonal cells give the cosine similarity of pairs of subdomains. The higher the similarity, the more their normalized authorship profiles are structurally equivalent and the less distinct they are in terms of meaning.

| | **SP** | **ES** | **SNA** | **CS** | **WS** |
|---|---|---|---|---|---|
| **SP** | 18,094 | 0.13 | 0.10 | 0.08 | 0.06 |
| **ES** | | 15,326 | 0.15 | 0.11 | 0.06 |
| **SNA** | | | 6,757 | 0.18 | 0.13 |
| **CS** | | | | 17,053 | 0.11 |
| **WS** | | | | | 10,997 |

The invisible college is identified using a combination of author and co-authorship filtering. Price (1986 [1963], p. 74) reasoned that a research domain of ten or hundred thousand scientists is coordinated by the 100 most prestigious of them. Therefore, the meso levels should consist of a three-digit number of facts. The author (node) and co-authorship (edge) thresholds are given in the captions of figures 3.8 and 3.9. Removing, for each subdomain, authors below a cumulative authorship fraction of 20% and stacking the five networks on each other as depicted in figure 3.10 results in a network of 2,553 authors. This set is still too large to function as an invisible college. The next step is to reduce this network to the largest bicomponent because members of the invisible college are "reasonably in touch with everyone else" (p. 119). The resulting invisible college (figure 3.9b) consists of 243 authors that coordinate Social Network Science.

In figure 3.9c, the collaboration network of all paradigmatic authors is reduced to all bicomponents. This reveals that the invisible college consists of islands of extreme productivity. Node size corresponds to self-reproduction, a measure that arises naturally from network normalization and resembles Price's notion of fractional productivity. The largest 2-component with `KRACKHARDT,_DAVID` as most connected and `CARLEY,_KATHLEEN` as most productive author is multifunctional, i.e., knowledge is produced at the interface of multiple subdomains. Other micro-level bicomponents are basically unifunctional, one belonging to Complexity Science, one to Social Psychology, and two to Web Science. Reproduced (strongly weighted) collaborations are mostly embedded in triads which supports our model that autocatalytic control is a collective or altruistic process.

At the meso level, there are no significant bicomponents besides the largest one. The fraction of top 20% authors that are embedded in the invisible college is a measure of the extent to which subdomains participate in coordinating the whole domain. Top Complexity Science authors are most strongly represented in the meso level network (26%), followed by Social Psychology and Social Network Analysis (both 13%), Economic Sociology (6%), and last Web Science (3%). These fractions can be explained by the average team size in subdomains (figure 3.5) in combination with their age. Web Science was in 2012 as much a team science as Complexity Science, but it is younger.

There is an important message in these fractions. On the one hand, it is trivial that subdomains whose practice is more team-oriented have a stronger presence in networks pruned to be cohesive. But on the other hand, only embeddedness translates into influence. In this sense, team science is advantaged. It takes more effort for subdomains like Economic Sociology, whose average number of authors per publication has long been below two, to compete for the network core than for Social Psychology whose authorship teams are today larger than three on average.

Eye inspection of the meso-level authorship network (the invisible college) proposes that there are three clusters (clockwise): a Social Psychology cluster, a multifunctional cluster, and a Complexity Science cluster with pockets of Web Science. The macro level, which contains all authors belonging to the largest bicomponent, shows a Social Psychology cluster somewhat detached from a multifunctional one dominated by Complexity Science in the core. Multifunctionality is expected when fractal distinctions are at play. In section 3.3.2, we will pursue this issue using a cohesion model. For macro level authorship, we demonstrate in figure 3.10 how the stacking of subdomains leads to opportunities of switching and the emergence of meaning. CARLEY,_KATHLEEN and WELLMAN,_BARRY belong to the 23 of all 45,583 authors that publish in all five subdomains.

Co-authorship networks are deeply cultural because every tie that co-constitutes a team or relates teams is symbolic communication. Turning to the levels and networks of the citation and word usage practice, the actual stories become apparent. The micro level structures we encounter here are network cores containing the institutionalized commonsense concepts that scientific paradigms are. This does not mean that authors cannot be paradigms too. LATKIN,_CARL and NEWMAN,_M_E_J are as much exemplars as their methods of studying the public risk associated with INJECTION_DRUG_USE and a COMPLEX_NETWORK, respectively.

While two subdomains dominate large areas of the co-authorship network, the word co-usage network unveils a rather coherent topic space where each subdomain sheds light on SOCIAL_NETWORKs from a different perspective. This meets our expectation that word usage fluctuates much less than authorship. Story sets overlap in the unifying COMMUNITY concept. But it means something else for each subdomain. Social Psychology has been treating it mainly as a source of SOCIAL_SUPPORT, Economic Sociology as an ingredient of SOCIAL_CAPITAL, Social Network Analysis as a target of SOCIAL_NETWORK_ANALYSIS methods, Complexity Science as a property of COMPLEX_NETWORKs, and Web Science as a group of USERs. As we zoom out to the meso and macro level, no structural features emerge other than more fine-structured neighborhoods. The clockwise positioning of subdomains in order of their actual age is a striking and robust – i.e., reproducibility of layout – property of the topic network. The observation that Social Psychology has least word overlap with Complexity Science mirrors a similar co-authorship divide.

As weakly coupled authors in Economic Sociology may be, its paradigm is most pronounced in terms of cited references. It is the only subdomain whose co-citation core or attractor includes a 6-clique, i.e., in the history of the subdomain, six references have been catalyzing each other in self-organization getting selected. The central reference is GRANOVET_1985_A_481, Granovetter's (1985) paper on embeddedness which has come to symbolize the ubiquity of networks in business and the economy. Other than to authors and words, a fixed year is attributed to references. The left component of the micro level co-citation network in figure 3.9f has an average publication year of 1988, and this is the average year of subsets for Economic Sociology, Complexity Science, and Web Science. Complexity Science also has a very pronounced co-citation core that consists of five 4-cliques, all including the WATTS_1998_N_440–BARABASI_1999_S_509 dyad. Web Science is least developed, without a clique and centered on GRANOVET_1973_A_1360. Social Network Analysis cites two 4-cliques, both containing the triad of FREEMAN_1979_S_215, BURT_1992_STRUCTURAL, and WASSERMA_1994_SOCIAL which represents the subdomain's focus on NETWORK_STRUCTURE. This subparadigm is most recent (1992).

Social Psychology's micro level network has eleven triads but two story sets. The newer core (1988) is centered on GRANOVET_1973_A_1360 and PUTNAM_2000_BOWLING, a book on the decline of social capital in US American society. The older core (1982) basically consists of eight triads including the central BERKMAN_1979_A_186. This is a classical study which shows that adults in California who lack social and community ties (consti-tutive ties) are more likely to die than persons embedded in `SOCIAL_SUPPORT` networks. The seven co-cited references all deal with the relationship of social embeddedness with stress, health, and mortality. Compared to the newer line of research, this path of inquiry tends to refer to social networks metaphorically, is more small-group oriented, and less structuralist. As we zoom out to the meso level, the older core forms a remote peak in the upper-right periphery. Another peak emerges in the upper-left periphery in form of a Web Science cluster focused on the development of collaborative-filtering methods for recommender systems. The more central Web Science cluster deals with issues related to `WEB_2.0`. On the macro level, co-citation presents itself as a fissured practice with multiple decoupled cores. `WASSERMA_1994_SOCIAL` is the one consensus reference that belongs to the paradigm of all five subdomains. `GRANOVET_1973_A_1360` is at the heart of all subdomains but Complexity Science.

### Lifetimes

The lifetimes of facts are not self-similar as the right column of figure 3.7 shows. Fluctuations are largest in the authorship practice and smallest in word usage. Authors live $1 \pm 1$ year on average, references $2 \pm 2$ years, and words $3 \pm 4$ years. Lifetime scaling is a valid hypothesis below an upper cutoff. The exponents of the power-law hypotheses, shown as dotted lines in figures 3.7b, 3.7d, and 3.7f, point in the same direction as discussed above, namely that language is more inert than authorship. The probability of a word to live for ten years is an order of magnitude larger than that of an author. References are closer to authors than to words. These results complement the visual findings that word networks are most and author networks least coherent. Together, they reveal that stories are more enduring than the concrete authors that work on the scientific problems the stories represent. Stories provide the network positions that are then filled by persons.

Figure 3.8 generally reveals that highly selected social facts (at low cumulative fractions) have long (unconditional) lifetimes, i.e., the facts in the upper tails of size distributions are also in the fat tails of lifetime distributions (cf. figure 3.7). Our definition of the meso and micro level includes that facts have lifetimes larger than the average conditional lifetime. Only authorship is exempted from this rule because it destroys network connectivity too much. Subdomains have different average fact lifetimes at similar cumulative selection fractions. For example, references at the subdomain meso levels are required to collectively attract up to 10% of all citations. Figure 3.8b shows that lifetime minima for references to belong to that level are 17 for Social Psychology, 16 for Economic Sociology, 12 for Social Network Analysis, 11 for Complexity Science, and 6 for Web Science.

This difference of lifetime finds explanation in the different ages of subdomains. But despite these differences, memory is universal, i.e., the dynamic selection of social facts is of the same form in different subdomains, just at different lifetime scales. In figure 3.11, we rescale distributions of conditional lifetimes. Memory is universal because all curves collapse onto a single curve. This holds for all practices and levels. Because memory is embodied in meaning structures, universality means that subdomains with markedly different contexts and ages have similarly institutionalized meanings down to paradigms.

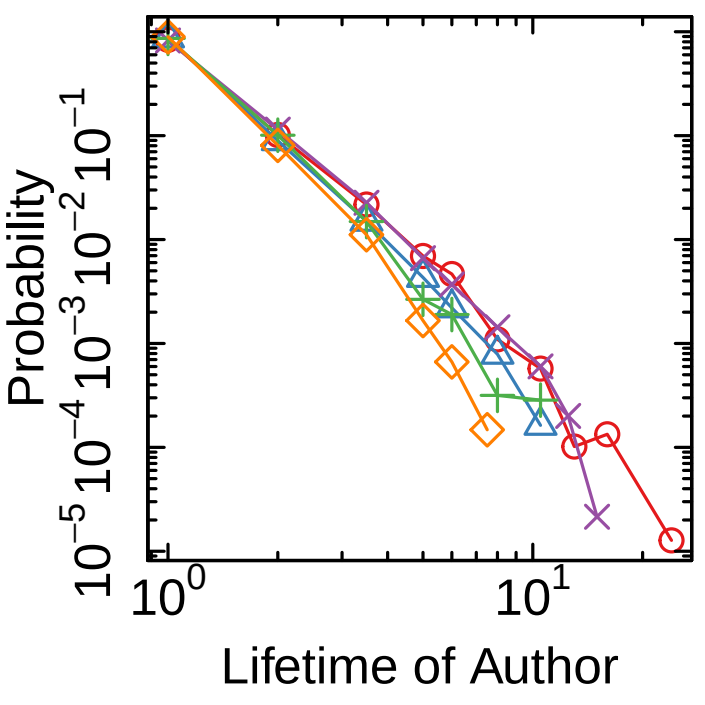

(a) Macro level authorship

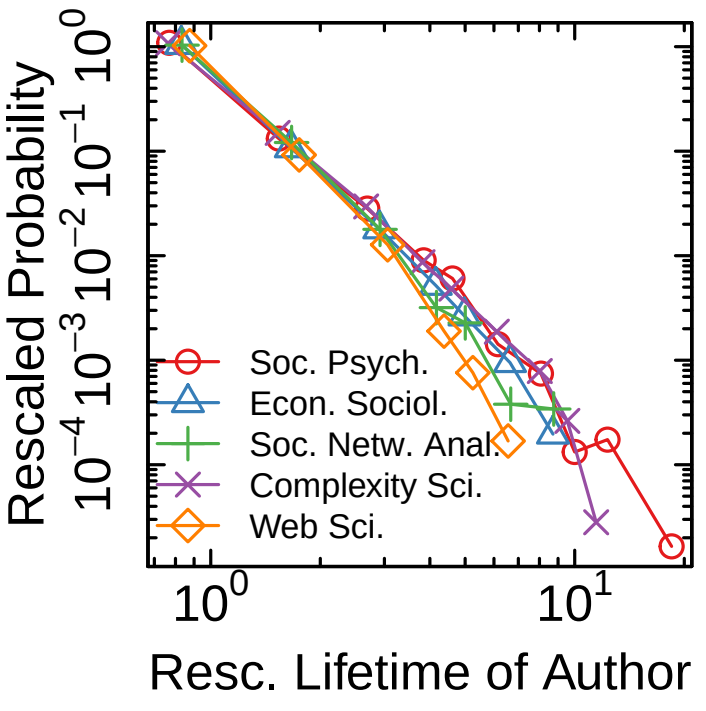

(b) Macro level authorship (rescaled)

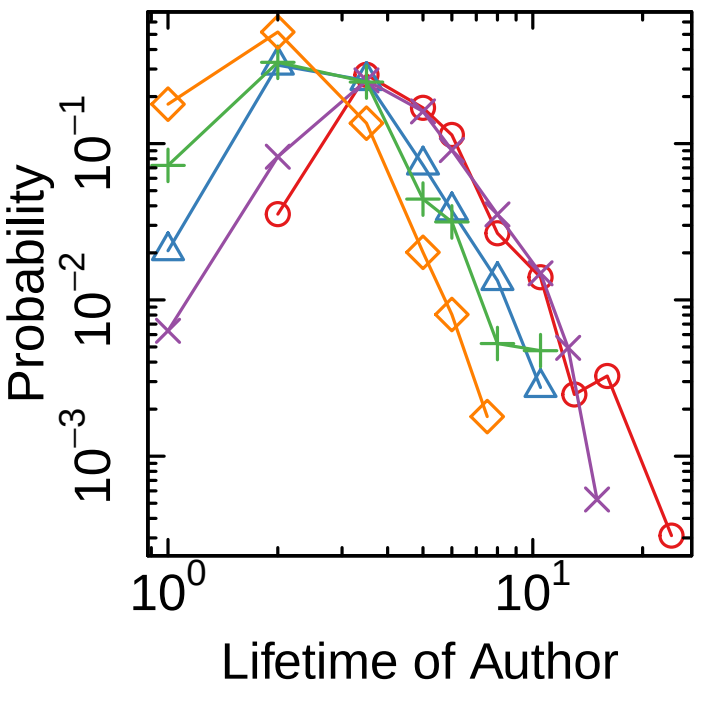

(c) Meso level authorship

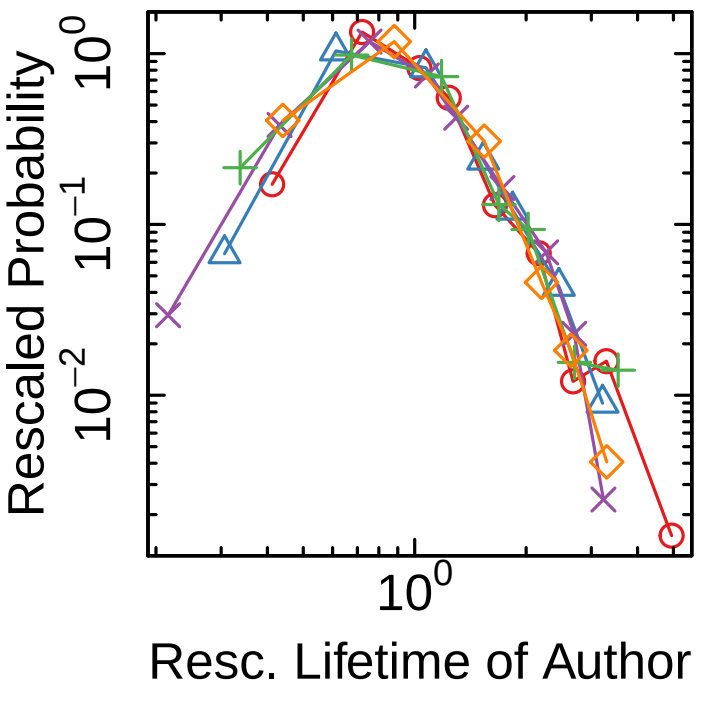

(d) Meso level authorship (rescaled)

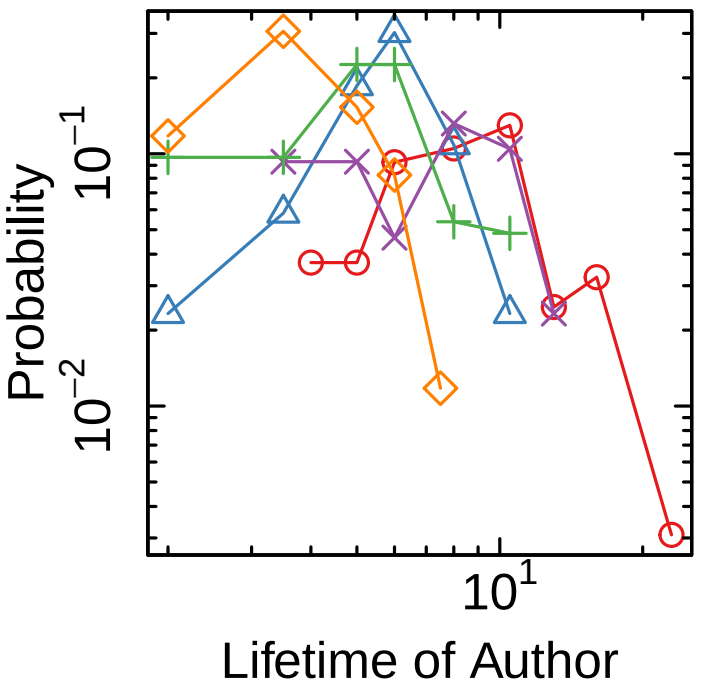

(e) Micro level authorship

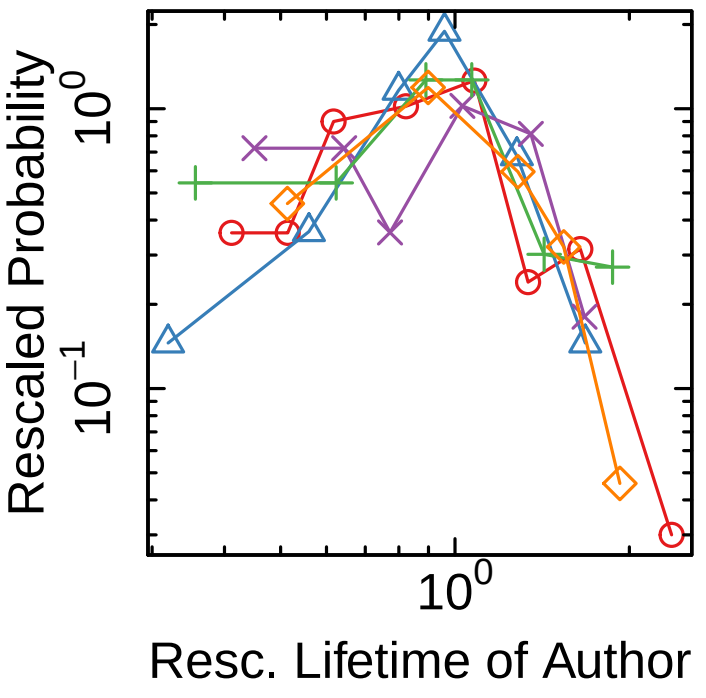

(f) Micro level authorship (rescaled)

**Figure 3.11.: Universality of memory**

Despite subfields being relatively old or young (cf. table 3.6), memory is universal. In other words, selection behavior is of the same form, just at a different temporal scale. Universality is demonstrated for all three practices and levels. The left columns show original lifetime distributions. In the right columns, distributions are collapsed by dividing lifetimes by the average conditional lifetime of the respective subdomain and multiplying probabilities with that average. Universality also holds for the noisy micro level of paradigms.

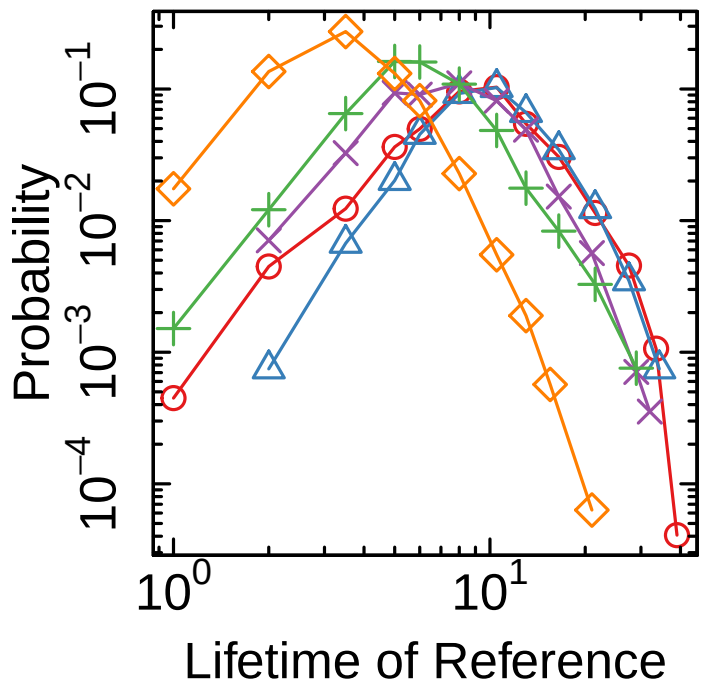

(g) Macro level citation

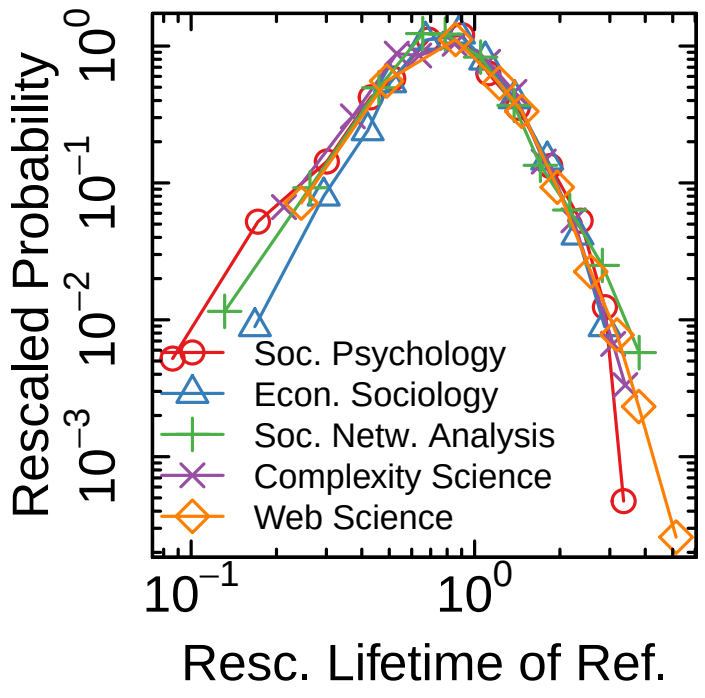

(h) Macro level citation (rescaled)

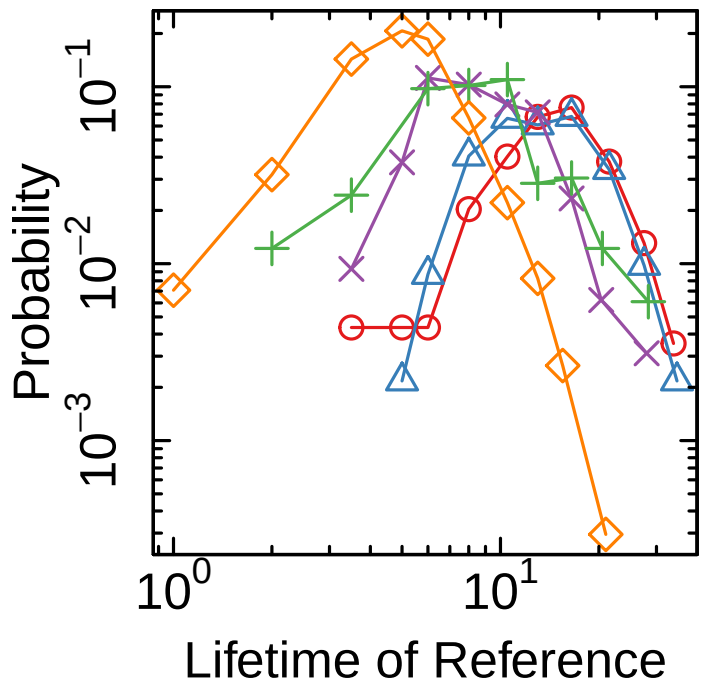

(i) Meso level citation

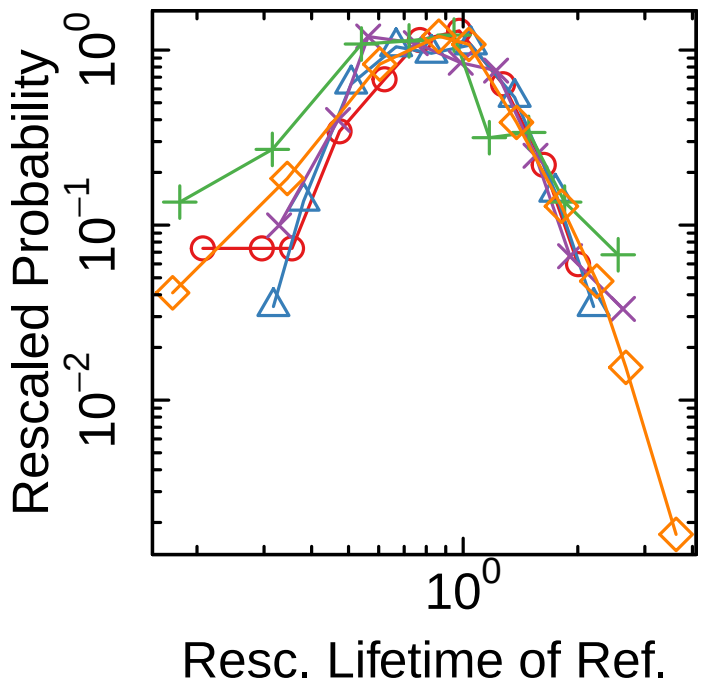

(j) Meso level citation (rescaled)

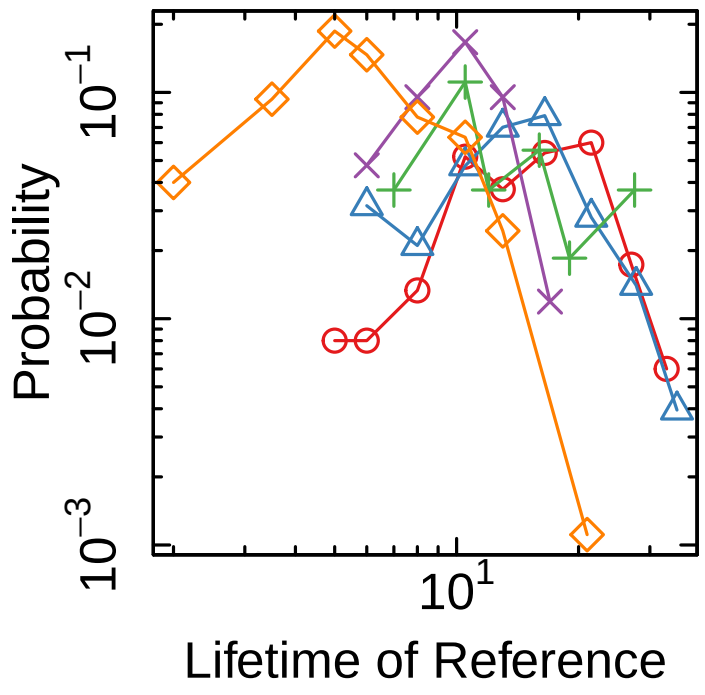

(k) Micro level citation

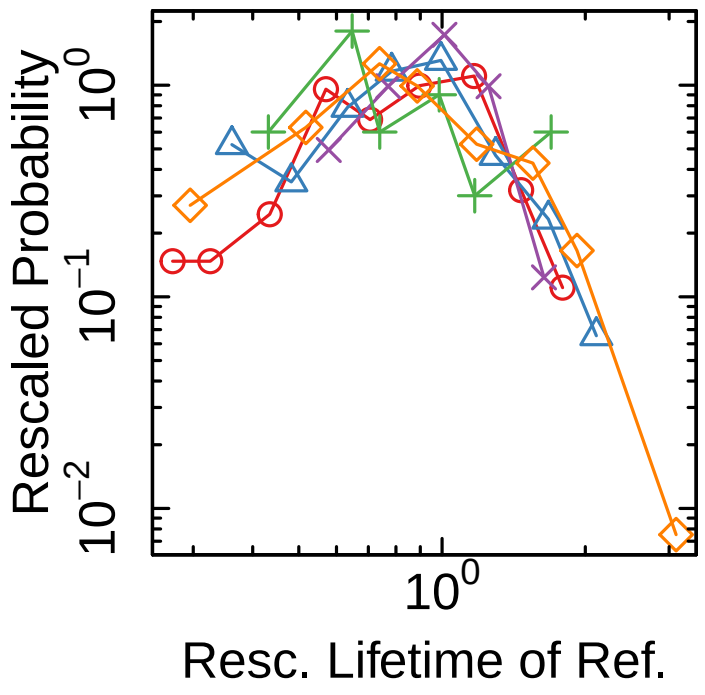

(l) Micro level citation (rescaled)

**Figure 3.11.: Universality of memory**

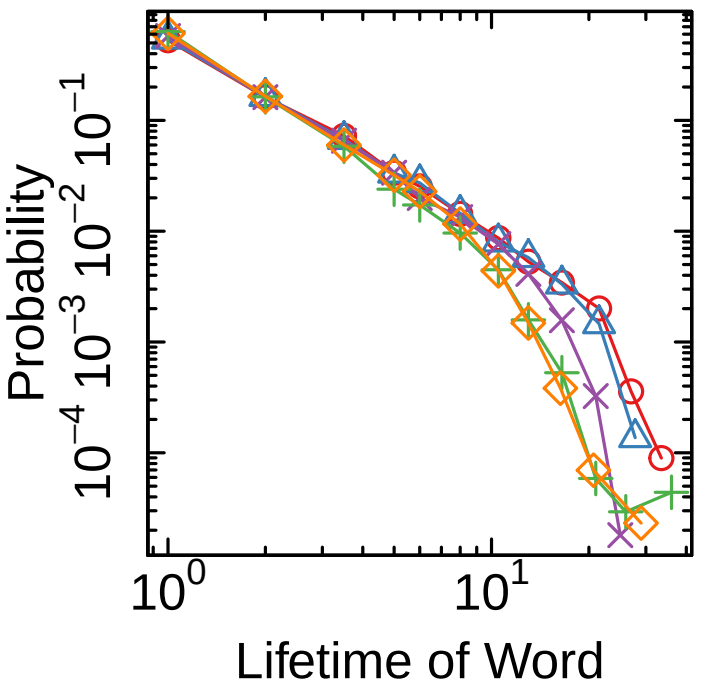

(m) Macro level word usage

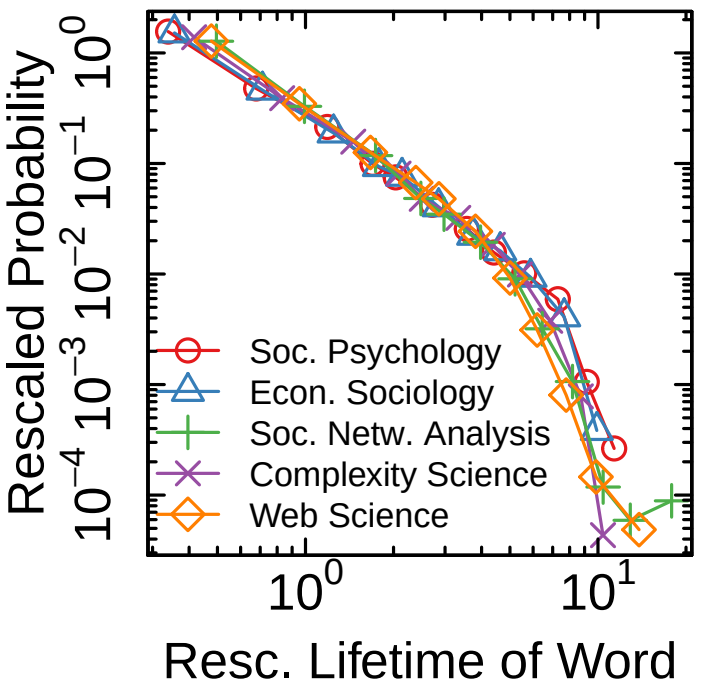

(n) Macro level word usage (rescaled)

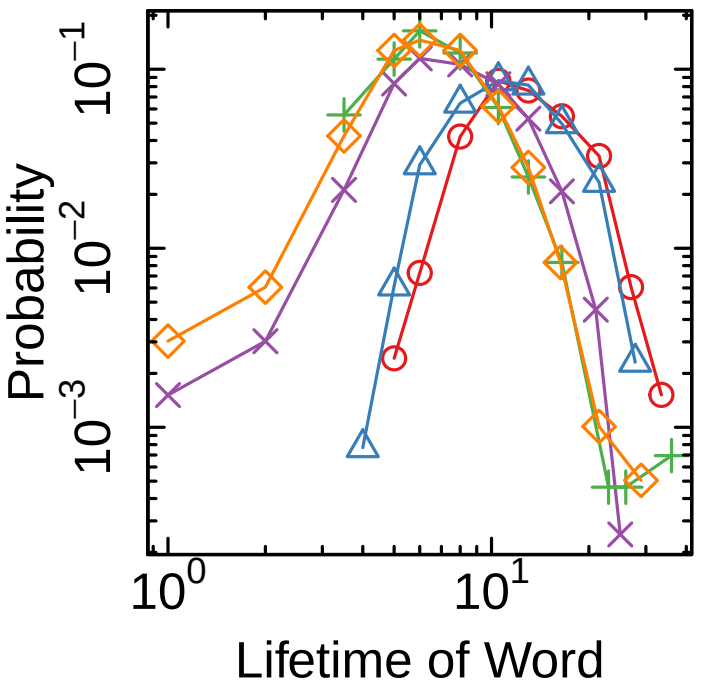

(o) Meso level word usage

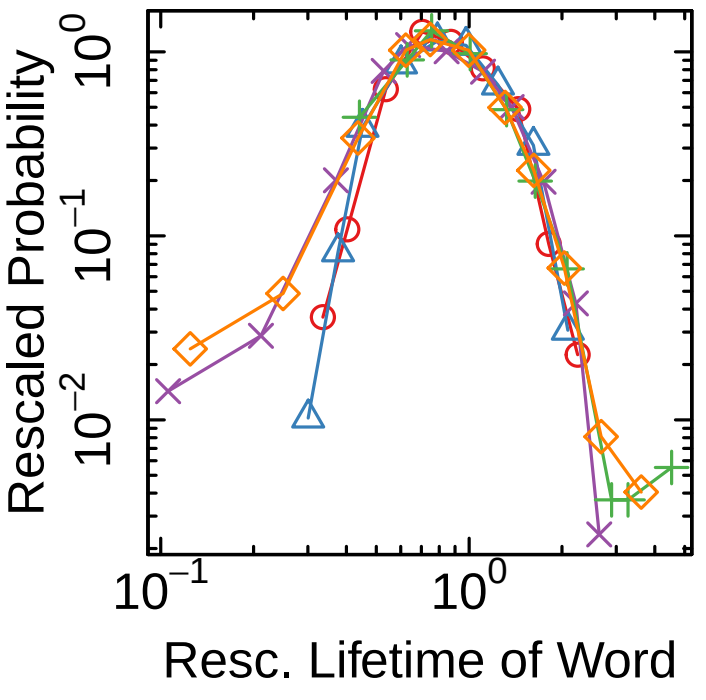

(p) Meso level word usage (rescaled)

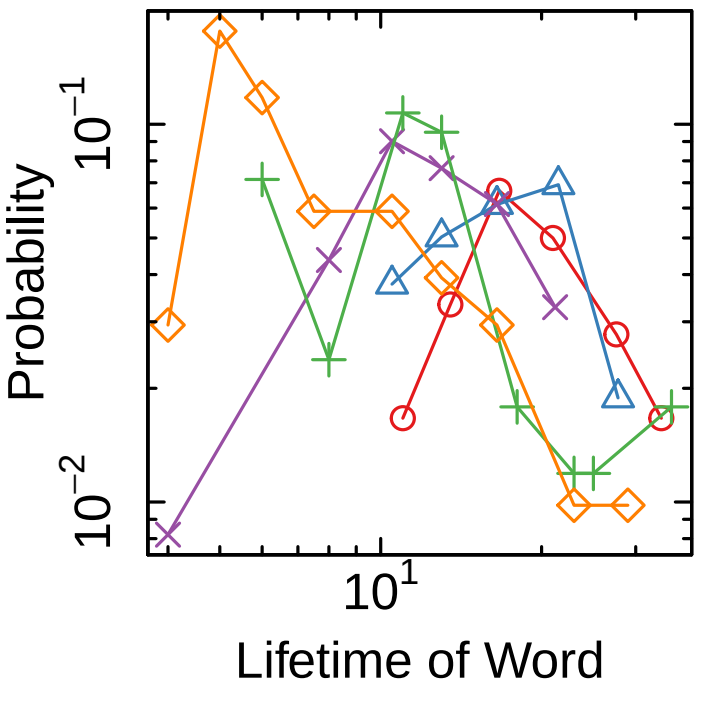

(q) Micro level word usage

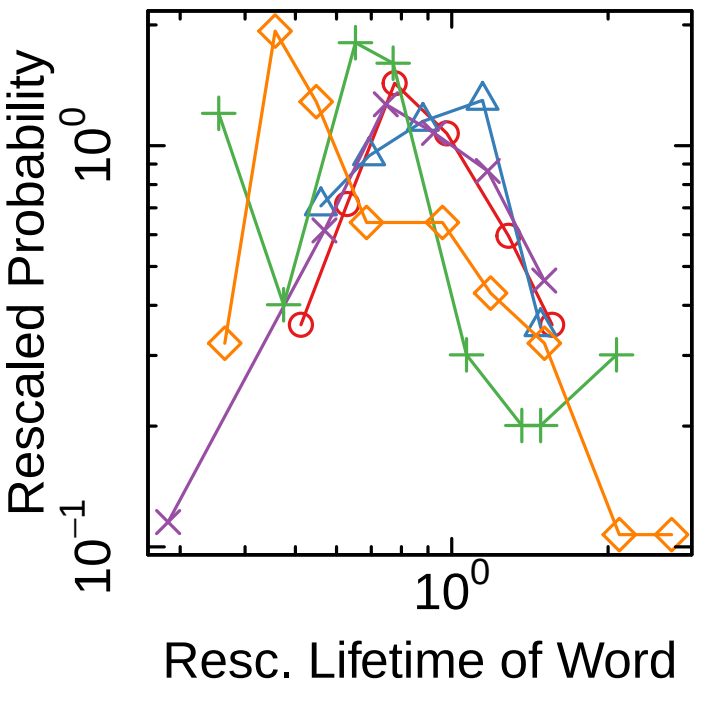

(r) Micro level word usage (rescaled)

**Figure 3.11.: Universality of memory**

It also shows that the normalization underlying the construction of levels is fair.

#### Summary

Social Network Science has experienced constant growth since the 70s and an accelerated growth period starting in the late 90s which lasted about a decade. The domain is fully emerging with 75% of all papers published since 2002. The five subdomains embed into a context that is self-similar regarding the authorship and citation practices. Word usage is self-similar at the level of the five identities. The scaling hypothesis needs not be rejected. Highly selected facts also live long. Lifetime distributions are not scale-invariant but universal, i.e., the meaning structures of subdomains are similarly institutionalized. Young subdomains are like old subdomains translated to another lifetime scale. This is true for three different levels of analysis: a macro level of all social facts, a meso level of medium-to-highly-selected facts, and a micro level of paradigmatic facts. Meso level co-authorship, when reduced to the largest bicomponent, resembles the invisible college of Social Network Science. Each subdomain has such a social circle of scholars that participates to a different degree in coordinating the whole. Subdomains like Social Psychology and Complexity Science with large author-per-paper ratios are embedded to a larger extent and, therefore, have an advantage coordinating the domain through social influence. The two subdomains take core positions in the collaboration structure, but there is a cleavage between the two.

Subdomains are homeostatic. Research topics represented by words are stable story sets and harbor positions to be filled by authors. References fluctuate less than authors but more than words. Citation provides different perspectives on topics. Low social coherence does not mean low conceptual coherence, like Economic Sociology demonstrates whose average team size is smallest but whose citation core is structurally most cohesive. The whole domain has a methodological consensus that the concept of `COMMUNITY` is analyzed using the methods developed by the subdomain of Social Network Analysis and described in `WASSERMA_1994_SOCIAL`. But meanings of these concepts differ as each subdomain contextualizes them differently. Social Psychology has aligned itself with the other subdomains by moving from an ego-centered to a `SOCIAL_CAPITAL` approach. Complexity Science revolves around the paradigm of the Watts/Strogatz and Barabási/Albert models introduced in the late 90s. Web Science is youngest and concerned with studying and designing web services.

## 3.3. Emergence

### 3.3.1. Bayesian Forks Demarcate Periods of Stability

To see if Social Network Science changes along paths of fractal distinctions, we search for periods of stability and moments of change in its genealogical tree. Constructing the 3-identity as a chain of events is possible using the interface network of direct citation. Because we do not know the citations made by cited references that cannot be sourced back to papers in our dataset, we restrict ourselves to the chains constructed by the journal papers in our dataset. Since the final step of delineating Social Network Science was adding the cited cores to the interim solution set, we are confident that the results we obtain are valid. But keep in mind that our method punishes subdomains with large fractions of book citations. Search path counts are computed for 22,664 papers from 1916 to 2012, and results are reported in figure 3.12.[6] During our analysis, we reduce the weighted citation network to the top 1% of the most traversed citations and to the main citation path (759 papers). This mercilessly removes references and citations that do not participate in the *mainstream*, the main narrative of the domain. In other words, we get a first glimpse into which research is most reproductive in the contest of ideas to give direction to Social Network Science.

The mainstream of knowledge production began in 1946 with Heider's balance theory (`HEIDER_1946_J_107`) and the graph theoretical sophistication by Cartwright and Harary (`CARTWRIG_1956_P_277`). It then flowed through the triad census papers of Davis, Holland, and Leinhardt (`DAVIS_1967_H_181`, `DAVIS_1970_A_843`, and `HOLLAND_1971_C_107`) before the Harvard school gave sociometry, as this line of research was known, direction. For 25 years, three papers by the group of White would be unmatched regarding the extent to which they channeled research in Social Network Science: Granovetter's "Strength of weak ties" (`GRANOVET_1973_A_1360`), Breiger et al.'s CONCOR algorithm (`BREIGER_1975_J_328`), and H. C. White et al.'s introduction of blockmodels (`WHITE_1976_A_730`) (cf. inset in figure 3.12b).

Independent of this narrative, a parallel chain of events was initiated in 1954 by J. A. Barnes' anthropological study (`BARNES_1954_H_39`), closely followed by a paper by Bott on observed roles in urban families (`BOTT_1955_H_345`). Both paths began to prosper in the 70s. During their heydays in the 80s, they each and annually produced about seven papers that historically turned out to channel the dominant search efforts. It is encouraging that our quantitative analysis reproduces Wolfe's (1978) qualitative observations that, in anthropology, "the beginning of serious theoretical thinking about networks must be credited to J. A. Barnes (1954), followed almost immediately by Elizabeth Bott (1957, 1959)" (p. 54).

The two streams are identified through the detection of clusters in the network of 759 references connected by the top 1% of the most weighted citations (figures 3.12a and 3.12c). We use the Girvan/Newman algorithm which partitions references by iteratively removing citation arcs with the strongest betweenness, i.e., it draws boundaries at bottlenecks of search flow. In figures 3.12e and 3.12f, the story sets are shown that are spun in each of the citation communities identified that way.[7] As can already be seen in figure 3.12a, the black cluster practically belongs to Social Psychology and is strongly focused on the study of `SOCIAL_SUPPORT`. Its central reference is Cobb's paper on "Social support as a moderator of life stress" (COBB_1976_P_300), i.e., it is attracted by the older of Social Psychology's co-citation cores (figure 3.9f). Based on central words and references, the black cluster is labeled Social Support Studies.

[6]The original direct citation network of 25,760 nodes is not acyclic because papers in an issue cite each other or because of matching errors. Such cycles are removed. The search-path-count algorithm described in figure 2.5 is applied to the 22,664 papers in the largest component. We have explored computing search flows for individual subdomains but given up on the idea because citation crosses such boundaries.

[7]Because abstracts and author keywords are not available for the bulk of papers in the early two clusters, we report results for word usage in titles only.

The pink event cluster is Structuralism. WHITE_1976_A_730 is the most central reference and forms the strongest co-citation triad – the BLOCKMODELing triad – with BREIGER_1975_J_328 and Lorrain and White's introduction of STRUCTURAL_EQUIVALENCE (WHITE_1971_J_49).[8] BLOCKMODEL is also among the most frequently used words but the strongest is SOCIAL_STRUCTURE. Other notable concepts are GROUP_STRUCTURE and SMALL_WORLD. Though the pink cluster is dominated by Economic Sociology, it combines papers from many subdomains (table in caption of figure 3.12).

From 1946 onwards, the main path of Social Network Science had been paved by Structuralism, but by 1981, research in Social Support Studies had become such a concerted effort that the latter attracted the main path. In that year, Hammer proposed a structural approach to the study of schizophrenia (HAMMER_1981_S_45) and cited Wellman's structural analysis of the intimate networks of East Yorkers (WELLMAN_1979_A_1201). 1981 also marks a period during which the identity of Social Support Studies began to change. Its emergence had begun in the mid 70s and coincided with the rise of more statistical approaches in psychology and health science which crowded out the observational approaches of anthropology/sociology and clinical medicine. From the late 70s/early 80s onward, while psychology maintained its influence and clinical medicine became irrelevant, the health-science approach began its decline and sociology rose to be on par with psychology.[9] Also from that time onward, the older co-citation core of Social Psychology centered on BERKMAN_1979_A_186 lost attractiveness, measured as the annual fraction of citing publications, while the newer core centered on GRANOVET_1973_A_1360 became more attractive. The return to sociological and structural methods that had become secondary in the 70s constitutes a fractal distinction. An early voice of sociological structuralism was Mueller who had recognized social networks as "a promising direction for research on the relationship of the social environment to psychiatric disorder" (MUELLER_1980_S_147). By 1990, the narrative had become very structural. Wellman's explanation of SOCIAL_SUPPORT (WELLMAN_1990_A_558) is Harvard-school relationalism.

But by 1990, Social Support Studies had already lost its ability to compete with other search paths in the domain. From then on, Social Psychology's publication share continuously decreased (figure 3.6c). SOCIAL_SUPPORT as a research topic continued to grow and reached 118 papers in 2012, but its time in the mainstream was over. We split the direct citation network four times, i.e., we identify that many turning points where common search paths converge or diverge. We have called these events Bayesian forks.

[8]Note that the paper that introduced structural equivalence (WHITE_1971_J_49) is falsely attributed to White as first author which is either because the many papers that have cited it have not read but simply copied it from other work's reference lists, or the concept is so strongly connected to the paradigm WHITE,_HARRISON_C that LORRAIN,_FRANCOIS was unintentionally denied his due role.

[9]This analysis is based on an abstract analysis of the papers on the main path and in the co-cited core as well as on the journal classifications of the 91 papers in Social Support Studies aggregated into disciplines using the OECD category scheme (ipscience-help.thomsonreuters.com/incitesLive/globalComparisonsGroup/globalComparisons/subjAreaSchemesGroup/oecd.html, visited June 11th, 2015).

The layouted clustered network in figure 3.12d helps visualize these events by highlighting and labeling references with high betweenness centrality. The first Bayesian fork occurs in 1993 and concludes Social Support Studies as an identity in its own as it merges with Structuralism. The last paper of Social Support Studies `PORTES_1993_A_1320` is already attributed to Economic Sociology. It is cited by `GULATI_1995_A_619`, a study how the structure of social networks affects the formation of business alliances. Here is where the main path flows back into Structuralism. One may regard the mid 70s to the mid 80s as a period of a division of labor, but certainly it is strong evidence for fractal distinctions that two paths with distinct roots merge.

### The Complexity Turn

In 1993, the last year of Social Support Studies, Structuralism still had eleven papers contributing to the citation mainstream, but in the late 90s, the reign of Structuralism itself ended. By that time, paths had converged to four papers that served as most recent connectors for the next big thing to come: Complexity Science. The transition from the merger of Structuralism and Social Support Studies to Complex Network Analysis, the *Complexity Turn* described by Urry (2005), is the most significant turning point in the life of Social Network Science. The Bayesian fork is identified by the first split of the citation network (left inset of figure 3.12c). We choose to precisely locate it in 1998 where the number of mainstream references is minimized (inset of figure 3.12a).

Four heavily traversed citations, three of them directed at diffusion studies, occurred at this major event (figure 3.12d). In their analysis of scaling in small-world networks, Newman and Watts (`NEWMAN_1999_P_7332`) cited Valente's threshold model of the diffusion of innovations (`VALENTE_1996_S_69`). Strogatz, Watts' co-author on the small-world paper, wrote an early review on `COMPLEX_NETWORK`s (`STROGATZ_2001_N_268`) and cited a paper in microbiology that puts the role of social networks in the spread of diseases into perspective (`WALLINGA_1999_T_372`). And Newman's guidance on scientific collaboration networks (`NEWMAN_2001_P_10.1103/PhysRevE.64.016131`) cited two papers by Davis et al. on the diffusion of innovation in, and the structure of, corporate networks (`DAVIS_1997_A_1` and `DAVIS_1999_A_215`). In terms of guiding the mainstream research path, Newman's paper and part II (`...PhysRevE.64.016132`) are the most important works in the 3-identity of Social Network Science since the introduction of `BLOCKMODEL`s in 1976. Note that roles of references are acquired in citation sequences. References need not be highly cited to have important roles. In narrative analysis, the right citation by the right reference can be sufficient to elevate a social fact's importance. On the other hand, highly cited references need not have top channeling positions in the flow network. This is why paradigmatic papers like `WATTS_1998_N_440`, `BARABASI_1999_S_509`, and `ALBERT_2002_R_47` do not show up. They were filtered out by the top 1% edge selection rule and are not on the main path.

In the previous section, we have identified the period from the Harvard breakthrough until the Complexity Turn as a maturation phase during which Social Network Science grew constantly. The turn is correlated with changes in subdomain growth rates – the explosions that followed maturation (figure 3.6) –, and in section 3.3.3 we will see that

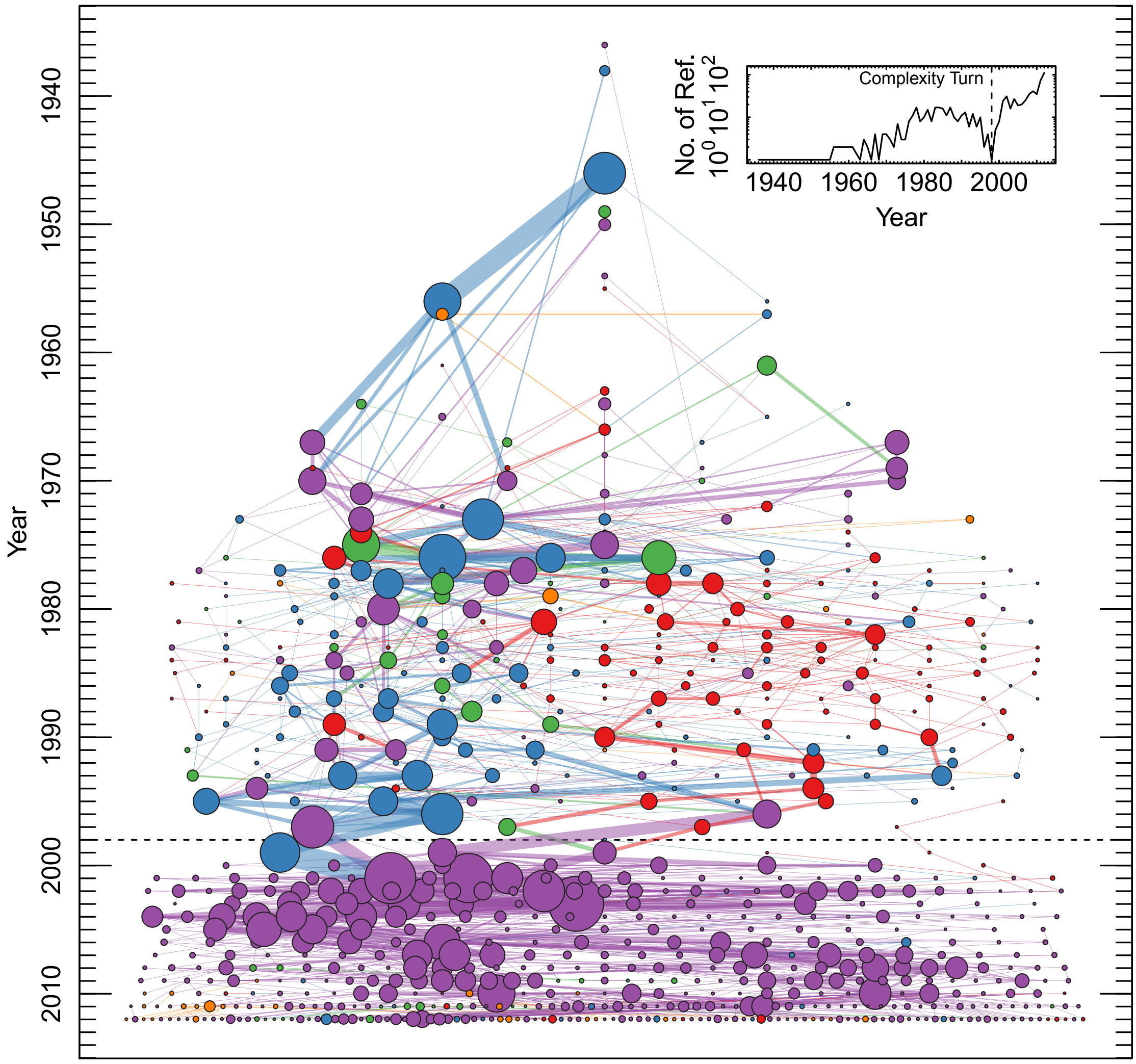

(a) Citation mainstream

**Figure 3.12.: Historical flow of citations**
Search path networks reveal along which intellectual paths knowledge has 'flown' over three quarters of a century. The size of arcs and nodes corresponds to the number of paths that traverse an arc and node. Citations are colored by target node. In (a–c), citations flow from younger to older publications (or of the same age). Arrows are omitted for clarity. In (a–b), colors correspond to subdomains. (a) The top 1% of the most heavily weighted citations, the mainstream, connects 759 references. Three historical periods are readily observable. Starting in 1936, the domain grows and focuses on merely a few references. After about the late 70s, references are many and highly-traversed references belong to all subdomains, though there seem to be two parallel paths in the 80s, the right one belonging to Social Psychology. After a turn in the late 90s, the majority of ...

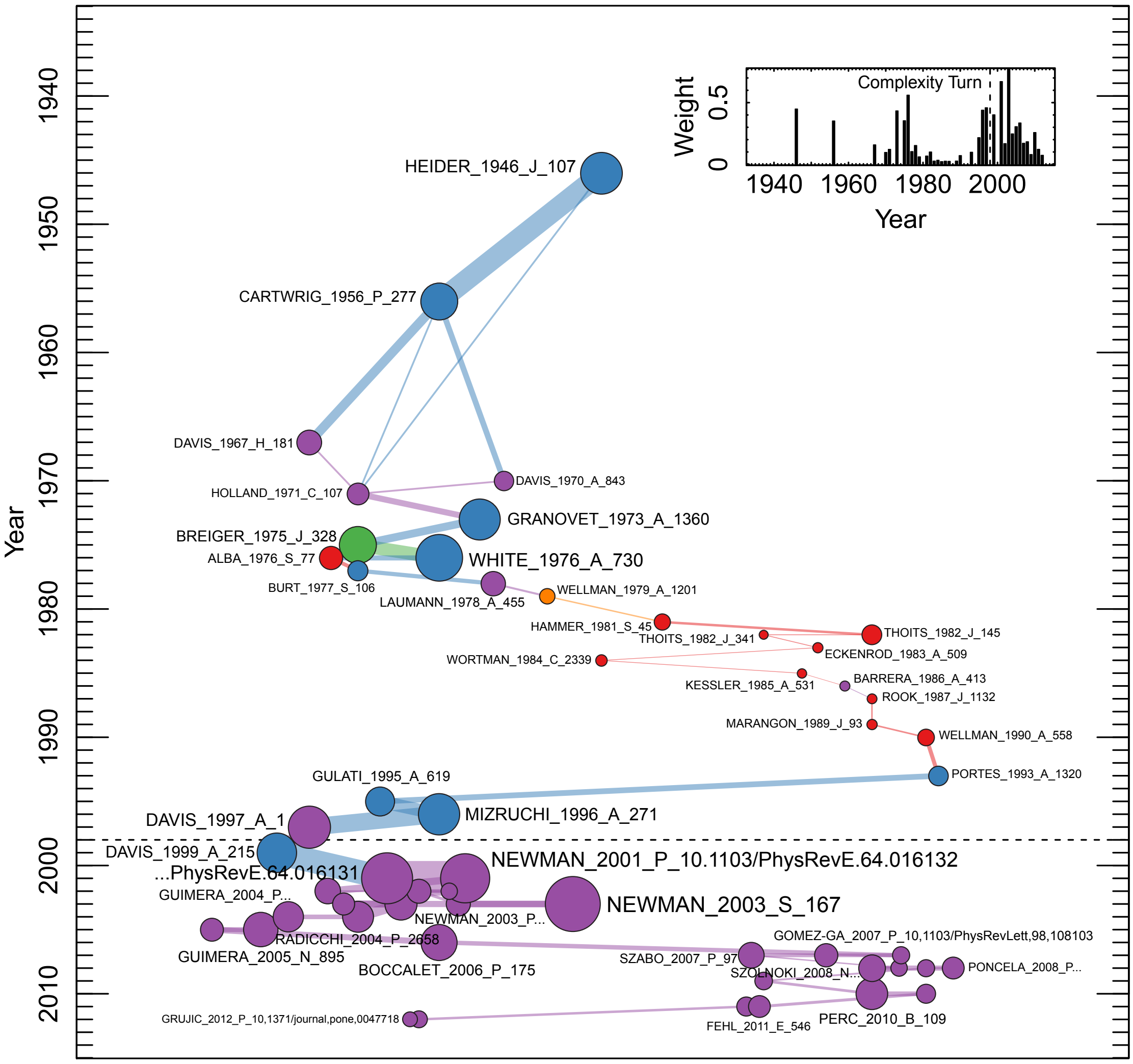

(b) Main citation path

**Figure 3.12.: Historical flow of citations**

... attention flows through papers belonging to Complexity Science. During 1996–1999 the number of references drops markedly, 1998 being the minimum (inset). Therefore, we label that year the “Complexity Turn.” (b) The main path is a reduction of the first network to a single path that contains the strongest weights. It reveals that Social Network Science is conceptually rooted in the balance theory of Heider and the triad census of Davis, Holland, and Leinhardt, was then coined by the Harvard school of White et al., subsequently dominated by Social Psychology in the 80s, and finally tipped into Complexity Science, a subdomain heavily influenced by the work of Newman, in around 1998. The subdomain(s) through which the main path flows strongly shape the core at that time. The inset shows the maximum normalized citation weight per ...

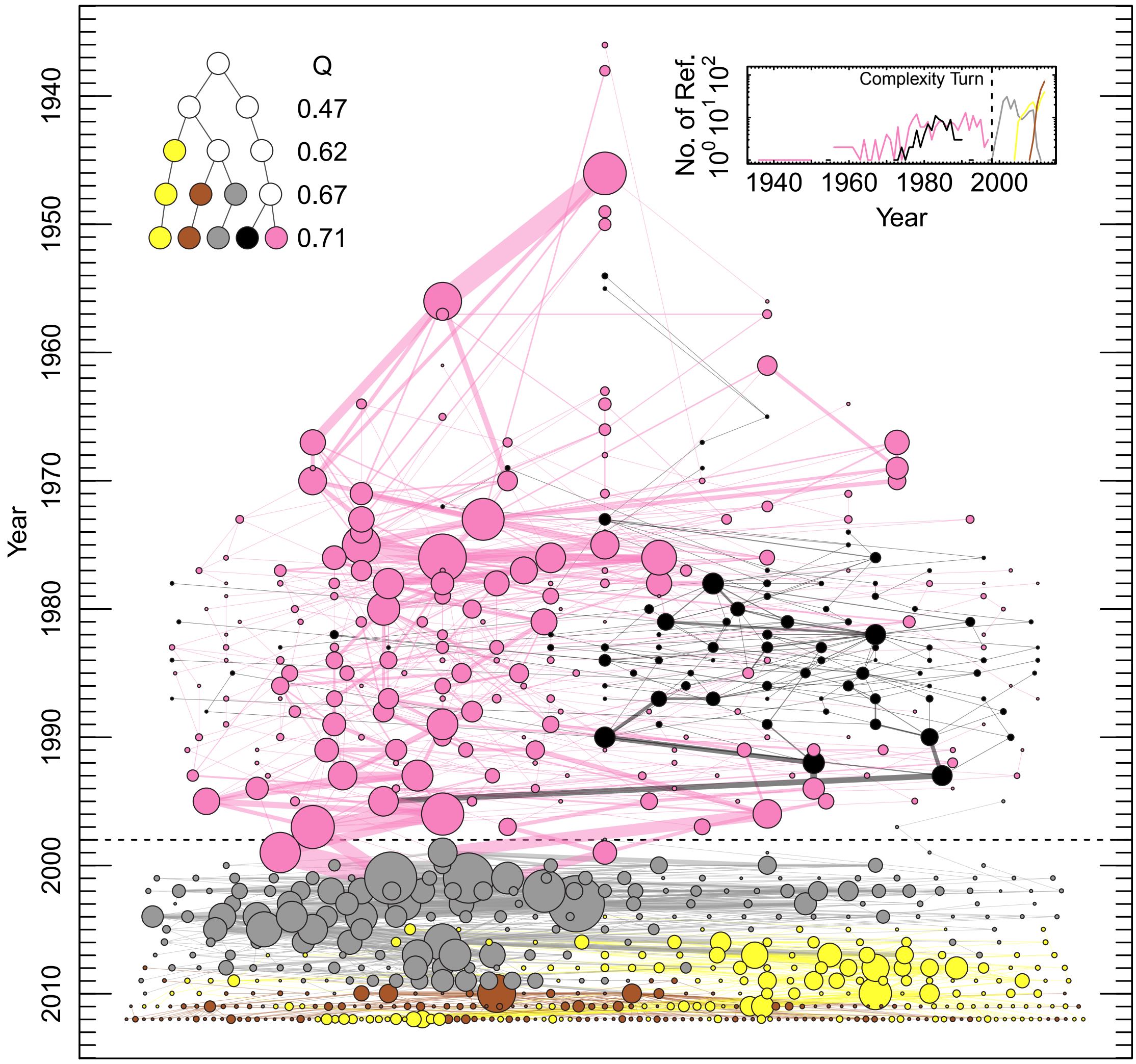

(c) Partitioned top 1% citation weights

**Figure 3.12.: Historical flow of citations**
... year. (c) Nested periods of normal science detected in the first network by iterative removal of arcs with the highest betweenness (Girvan & Newman, 2002). The dendogram in the left inset shows which periods and modularities result at which split. First, history is split into pre- and post-turn narratives. Then, post-turn periods of Complexity Science are distinguished. Last, the Social Psychology episode of the 80s is singled out as a coherent stream in the pre-turn era. The right inset gives annual frequencies (number of papers in the mainstream) for each of the five periods. (d) A spring-embedded layout reveals main events. Pink and black periods were born independent of each other with founding papers by Heider and Barnes, respectively. Arc width is citation weight, node size now mirrors betweenness. Large nodes indicate Bayesian forks where paths ...

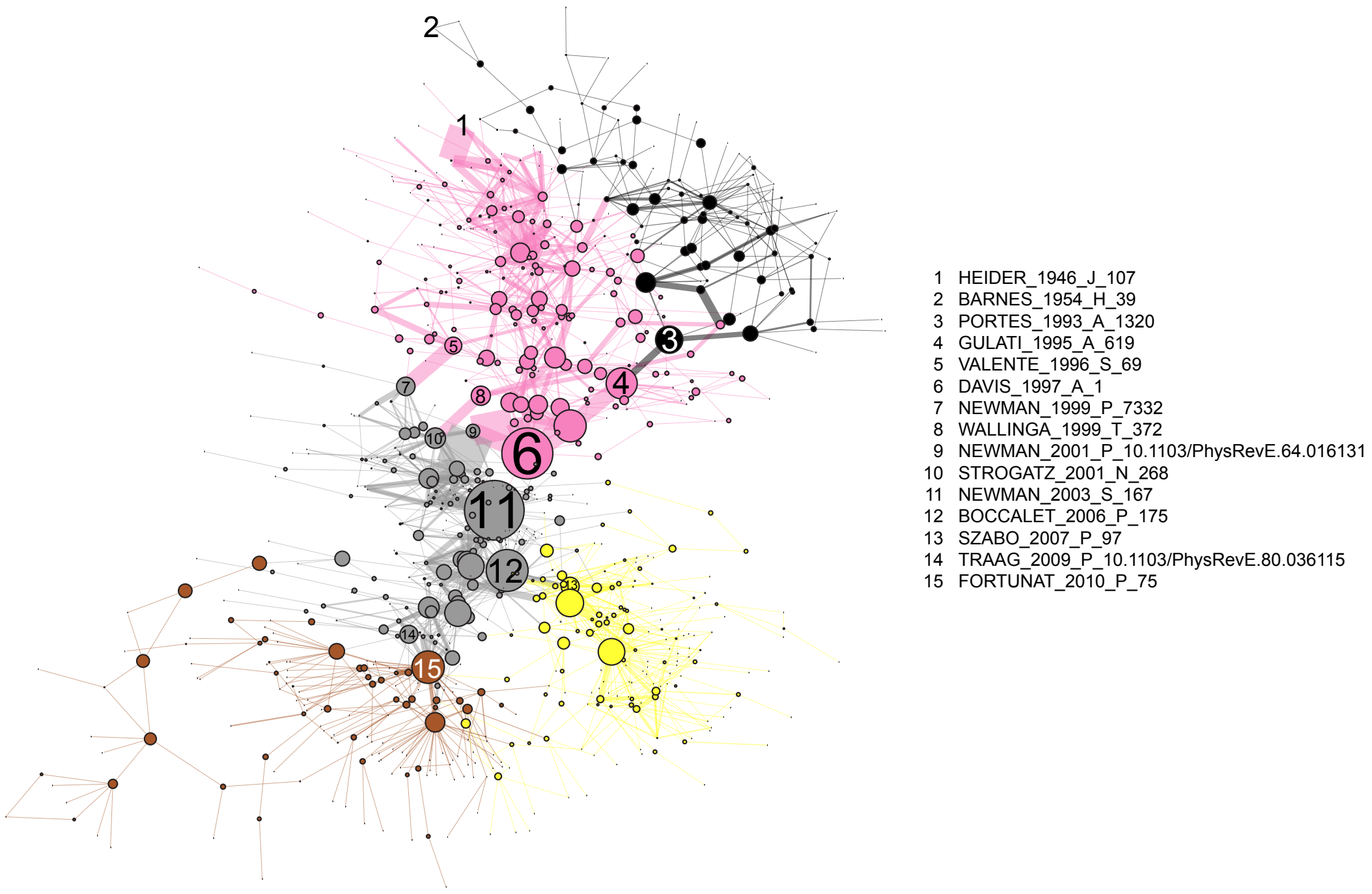

(d) Layouted top 1% citation weights

**Figure 3.12.: Historical flow of citations**

... converge or diverge. Towards the Complexity Turn, paths converge, but after the turn, they diverge, making the case for convergent evolution or fractal distinctions. For the five clusters of cited references identified in (c), reference co-citation (e) and word co-usage (f) networks are constructed and stacked on each other. Facts correspond to what we have defined as paradigms in figure 3.8 not filtered by lifetimes and edges. Node size is self-reproduction (autocatalysis) summed over all stacks. Story sets overlap most strongly for the three post-turn paths. Word usage is only shown for the post-turn clusters because abstracts are only available as data since 1990. Based on these story sets, clusters are given labels shown in the table below. Size is the number of references in each cluster, and fractions tell which subdomains dominate.

| | | | | Subdomain fraction | | | | |
|---|---|---|---|---|---|---|---|---|
| | **Period** | **Color** | **Size** | **SP** | **ES** | **SNA** | **CS** | **WS** |
| **Structuralism** | 1936–1998 | Pink | 197 | 14% | 44% | 13% | 26% | 3% |
| **Social Support Studies** | 1954–1993 | Black | 91 | 81% | 9% | 2% | 5% | 2% |
| **Complex Netw. Anal. I** | 1998–2010 | Gray | 174 | 3% | 2% | 2% | 91% | 1% |
| **Evolutionary Game Theory** | since 2006 | Yellow | 158 | 3% | 3% | 2% | 89% | 3% |
| **Complex Netw. Anal. II** | since 2010 | Brown | 139 | 5% | 6% | 11% | 64% | 14% |

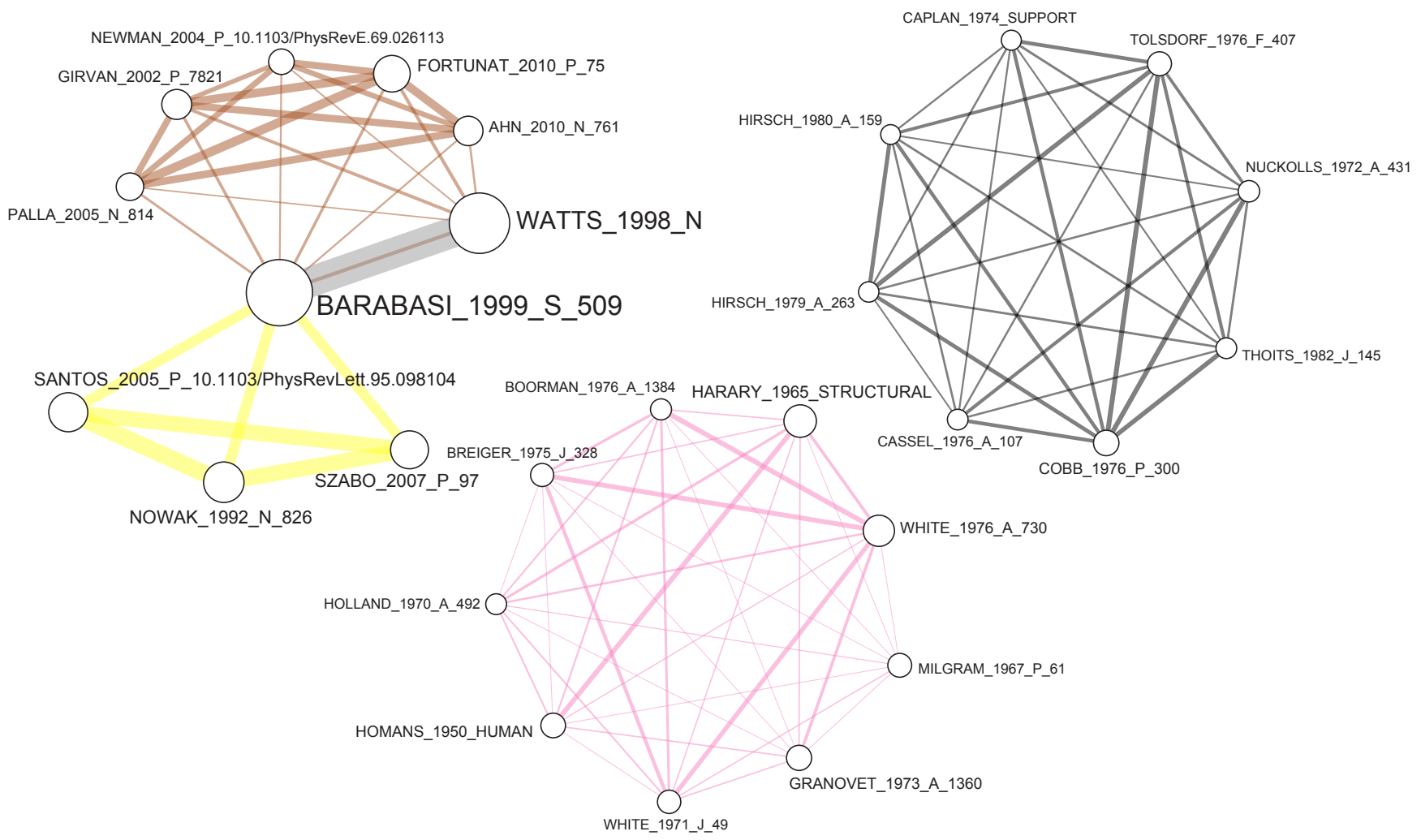

(e) Story sets (references)

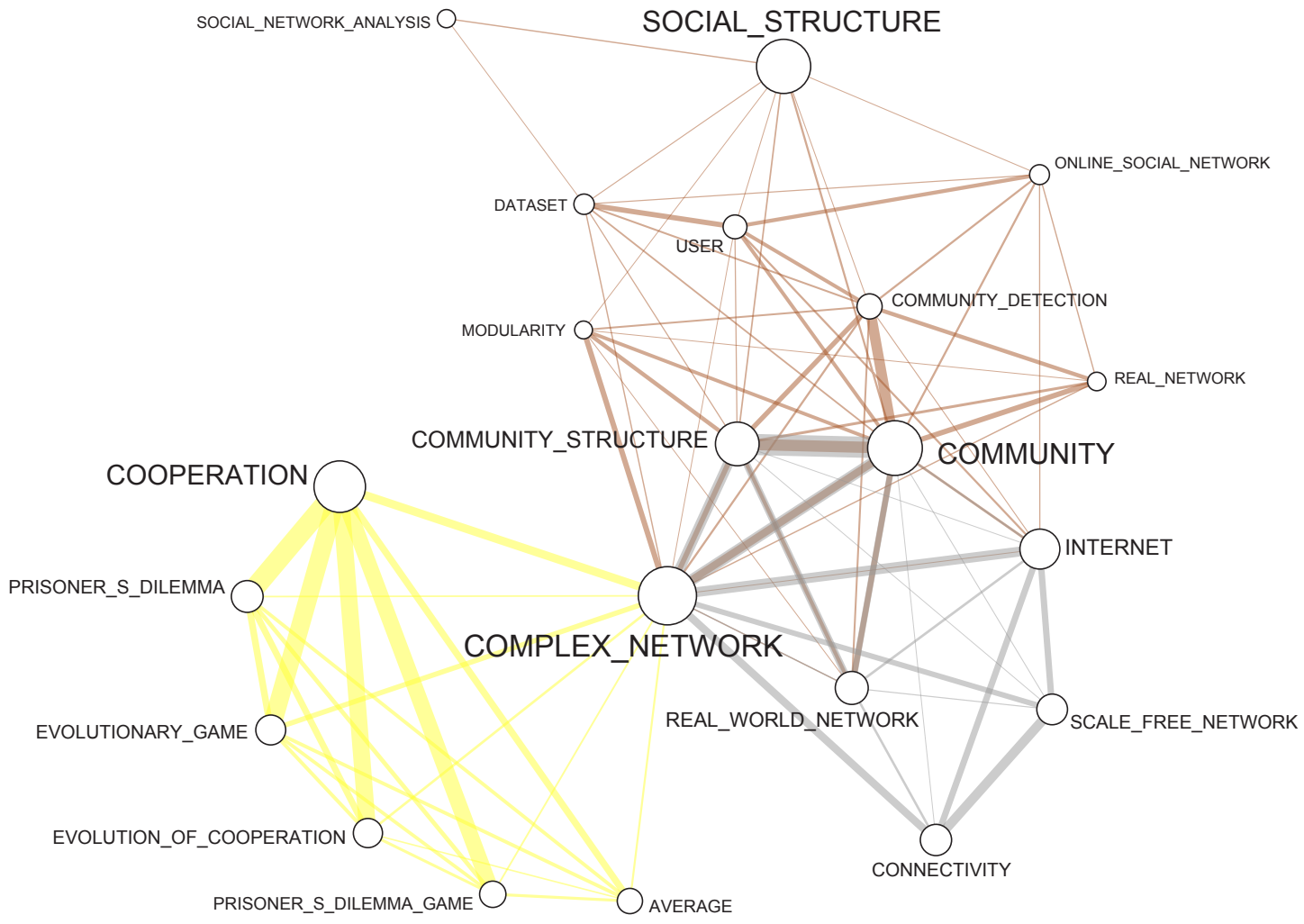

(f) Story sets (words)

**Figure 3.12.: Historical flow of citations**

the innovations in Complexity Science are actually causal. The rates of Complexity Science, Social Network Analysis, and Web Science start increasing drastically one, two, and four years after the transition. Social Network Analysis reaches a growth rate of 10% and Web Science a staggering 13%. Economic Sociology and Social Psychology, the two subdomains with the largest publication fraction in the pre-turn period, are not positively affected. The growth of the others caused Economic Sociology to lose publication shares. Social Psychology was already on the decline and lost shares at an increasing rate. 14 years after the turn, Social Psychology and Complexity Science met at 23%. The former had dramatically lost 20 percentage points and the latter had gained ten.

Up until here, Scott (2012) has reviewed the domain. In contrast to his three genealogical lineages, our empirical analysis only identifies one anthropological branch, the one that includes Barnes and Bott, besides what we call Structuralism. If Structuralism was, judged by the most frequent unspecific words of its main subdomains Economic Sociology, Social Network Analysis, and Complexity Science, mainly concerned with `ROLE ANALYSIS`, `METHOD`ology, and `MODEL`ing (table 3.7), attention is focused on `MODEL`s and `ALGORITHM`s since the Complexity Turn. But Complex Network Analysis also changed in the years to follow. Immediately after the turn, the mainstream of research heavily concentrated on `SMALL_WORLD_NETWORK`s (`WATTS_1998_N_440`), `SCALE_FREE_NETWORK`s (`BARABASI_1999_S_509`), and the analysis of `COMMUNITY_STRUCTURE`.

In the physicist-type research style, reviews have important roles as summaries of the state of the art from which further progress is made. Newman's *SIAM Review* (`NEWMAN_2003_S_167`) is the most important work in the history of Social Network Science – again, judged solely by how much it channels search –, matched only by his works on collaboration networks (ranks 2 and 3) and `WHITE_1976_A_730` (rank 4). Another important review is Boccaletti et al.'s "Complex networks: Structure and dynamics" (`BOCCALET_2006_P_175`). Eight years after the turn, it represents the first milestone after which a second mainstream 3-identity of Social Network Science emerged. Figure 3.12d shows that Evolutionary Game Theory split off from Complex Network Analysis. The main path has been in Evolutionary Game Theory since 2007 when Szabó & Fáth's review of "Evolutionary games on graphs" (`SZABO_2007_P_97`) cited `BOCCALET_2006_P_175`. Its cultural core is quite distinct from Complex Network Analysis, `COMPLEX_NETWORK`s and `BARABASI_1999_S_509` being the overlapping core facts. Its story set is centered on `COOPERATION`, a story weakly shared with Structuralism. But while the latter mainly treated `COOPERATION` as an `INTERORGANIZATIONAL` phenomenon, the former aims at an explanation through games like the `PRISONER_S_DILEMMA`.

Shortly after the emergence of Evolutionary Game Theory, the other 3-identity of Social Network Science underwent change through recontextualization. Technically, Complex Network Analysis transformed in 2010 when Fortunato's review "Community detection in graphs" (`FORTUNAT_2010_P_75`) cited a paper on clustering networks with positive and negative links (`TRAAG_2009_P_10.1103/PhysRevE.80.036115`). Because Complex Network Analysis I (dead) and II (alive) share a strong triad in their story set (`COMPLEX_NETWORK`, `COMMUNITY`, and `COMMUNITY_STRUCTURE`), the turning point marks a change of 2-identity or face. The emergence of Complex Network Analysis II is due to the increasing convergence of Complexity Science and Web Science. The latter's type of tie increased its share in the 3-identity from 1% to 13% and is mirrored in a context shift from concepts related to CONNECTIVITY and the INTERNET to the SOCIAL_STRUCTURE of, and USER behavior in, ONLINE_SOCIAL_NETWORKs.

How can we characterize the history of, and the events in, Social Network Science regarding Kuhn's phase model? Hummon and Carley (1993) find that GRANOVET_1973_A_-1360 and the papers of the BLOCKMODELing triad are the four references most frequently cited by papers in the journal *Social Networks* from 1978 to 1990. Because these works were followed by papers that built on this foundation, the authors conclude that the citation pattern in that journal and period is consistent with Kuhnian normal science of autocatalytic progress. In our much larger and more refined dataset, this episode corresponds to the period from the mid 70s to 1998, from the publication of the four references to the Complexity Turn. Indeed, this chain of events has the prime characteristic of normal science: unpunctuated reproduction in 'equilibrium.' On the main citation path, WHITE_1976_A_730 was unmatched for a quarter of a century, and in that time, Structuralism grew constantly and uninterruptedly.

If the period after the Harvard breakthrough is normal science, the due question is if the Complexity Turn constitutes a scientific revolution. The extent to which a Bayesian fork is a scientific revolution is the extent to which the paradigm shifts in the situation. We operationalize the magnitude of a paradigm shift as the degree of structural equivalence regarding the selection of paradigmatic references in two successive paths. Technically, we compute the cosine similarity of distributions of the top 20% references.[10] Since Complex Network Analysis merely changed face around 2010, similarity is largest (40%) for the Complex Network Analysis I → II transition. The similarity of the Complex Network Analysis I and Evolutionary Game Theory paths is second largest (22%). As expected, the strongest Bayesian fork is most disruptive. Not a single paradigmatic reference is kept during the Structuralism → Complex Network Analysis I transition. In other words, since the co-citation core of Social Network Science was completely renewed in the first detected turning point, the Complexity Turn exhibits a defining property of a Kuhnian paradigm shift. If it really is a revolution depends on whether or not the new main core also attracts the other subdomains of SNS, to be seen in section 3.3.3.

Surprisingly, Social Support Studies and Structuralism are almost as dissimilar (4%) as the pre- and post-turn periods. Does that mean that the corresponding event is almost comparable to the Complexity Turn? No, because both events are of a fundamentally different type. The event involving Social Support Studies and Structuralism, on the one hand, is a merger of two research paths with independent roots and very different topics to study – social networks as metaphors for, or substances of, SOCIAL_SUPPORT versus SOCIAL_STRUCTURE as objects or organizational embodiments – which eventu-ally converged methodologically. Overlap of Social Support Studies and Structuralism consists of only four paradigmatic references. While the social anthropological books BOTT_1957_FAMILY and MITCHELL_1969_SOCIAL were cited throughout Social Support Studies, only four of 20 citations to the later structuralist works GRANOVET_1973_A_1360 and POOL_1978_S_5 stem from before 1981, corroborating our finding that the main path crossed into Social Support Studies the moment the latter turned structural. The merger of Social Support Studies and Structuralism is not terminal as a whole, just for the former.

[10]In figure 3.8, the top 20% references constitute the macro level, not the micro level of paradigms. But since the citation clusters are smaller than the subdomains, 20% result in a reasonable set of references small enough to qualify as paradigms and large enough to allow reliable statistics. Cosine similarity is a standard geometrical measure in information science specified as the cross product of the citation vectors divided by the square roots of the individual vector cross products. In another work, we have shown that discontinuities of structural equivalence in conversational practices of politicians on *Twitter* point at punctuating events (Lietz, Wagner, Bleier, & Strohmaier, 2014).

The Complexity Turn, on the other hand, is an event during which the identity of Social Network Science was completely transformed. Before the turn, the two narratives converge and the paths are further narrowed down to four transition citations. After the turn, paths massively diverge. But the question what makes 3-identities different in type is worth staying with. We have split the citation network four times, and the first split identified the Complexity Turn which separates two research periods in 'equilibrium.' The second split reveals that the post-turn event is actually punctuated itself and consists of two events which are even more in equilibrium. The third split identifies the Complex Network Analysis I → II transformation, a Bayesian fork of a similar type as the Complexity Turn. Periods of normal science are embedded in periods that are again more normal or in equilibrium than the period they are themselves embedded in. Scientific revolutions do demarcate phases of normal science, as Kuhn suggested, but in a self-similar way. It is just that, as the number of splits increases, turning points soon lose the character of revolutions. In the Complex Network Analysis I → II transition, the shift does not have the characteristics of a revolution – the core is only recontextualized as the domain moves on.

Despite transformations there is some continuity in the Structuralism → Complex Network Analysis I → Complex Network Analysis II lineage. In the highly selective subset of the citation mainstream, the analysis of SOCIAL_STRUCTURE or SOCIAL_NETWORK_ANALYSIS is the unifying story. Regarding structure as a macro phenomenon, the SMALL_WORLD problem symbolized by the works of Milgram (MILGRAM_1967_P_61) and Pool and Kochen (POOL_1978_S_5) was carried over to Complex Network Analysis by WATTS_1998_N_440. The analysis of power-law distributions was popularized by BARABASI_1999_S_509. Methodologically, the means to the end of SOCIAL_NETWORK_ANALYSIS have changed from BLOCKMODELing to COMMUNITY_DETECTION, the two special cases of role modeling (Reichardt & White, 2007). From the two most successful breakthroughs of the Harvard group, the principle of STRUCTURAL_EQUIVALENCE was not institutionalized beyond Structuralism. It was not carried into Complex Network Analysis. "The strength of weak ties," however, has survived as the ultimate sociological insight on how structure constrains and enables diffusion. Following its transformation, Complex Network Analysis increasingly turned towards SOCIAL_NETWORK_ANALYSIS (figure 3.12f). GRANOVET_1973_A_1360 almost made it into Complex Network Analysis II's core. Without adopting another style of research or deviating from the new paradigm, Complex Network Analysis partially returned to a structuralist root of the domain – about a decade after it had abandoned it in the Complexity Turn. This is the second fractal distinction we have identified in the narrative history of Social Network Science.

### Summary

We have modeled Social Network Science as a 3-identity by reconstructing historical citation search paths. Structuralism emerged as the main narrative in 1936 and consolidated after the breakthroughs by White et al. (1973–1976) in which fundamental insights about structural equivalence in, and the diffusion properties of, social networks were discovered. The corresponding papers were unmatched, and the Harvard renaissance lasted, until, in 1998, Structuralism tipped into Complex Network Analysis with its paradigm of small-world and scale-free networks. This Complexity Turn has the characteristics of a Kuhnian scientific revolution because this largest Bayesian fork erased almost all memory. The small-world problem is the main continuity in the history of Social Network Science as revealed by the citation mainstream. Blockmodeling was not carried over to Complex Network Analysis and replaced by large-scale community-detection methods. Social Network Analysis, Complexity Science, and especially Web Science have been able to translate the event into increased publication shares at the cost of the old subdomains, Social Psychology and Economic Sociology, that show almost no signs of being impacted by the Complexity Turn in terms of growth.

Social Network Science changes along diverging and converging paths. There are two major fractal distinctions. Before the turn, Structuralism merged with the formerly anthropological Social Support Studies when the latter had returned to its relational roots. After the turn, Complex Network Analysis took up structuralist concerns in a partial change of face or 2-identity. Bayesian forks demarcate stable periods of reproduction, but this happens at multiple scales. Events are only nearly decomposable to an equilibrium. The post-turn transformation of Complex Network Analysis is an event of the Complexity-Turn type but does not constitute a scientific revolution because its core was not deeply affected. Though the dynamics are self-similar, only the major turning points qualify as paradigm shifts. The only major diverging event saw the emergence of Evolutionary Game Theory which was still the most forceful research path in 2012.

### 3.3.2. Research Fronts Are Separated by a Structural Hole

In the preceding section we have identified two fractal distinctions in time – episodes during which an identity "takes up the concerns of the defeated." In this section, we want to know if the social structure of Social Network Science is a hierarchically modular small-world network with a multifunctional core because that is the expected outcome of fractal distinctions in meaning space. If, e.g., a sociologist embraces the methods of Complexity Science, and a physicist turns towards the sociological roots of Social Network Science, cultural boundaries blur and structural holes in the collaboration landscape disappear. Our point of departure is the co-authorship bicomponent of figure 3.9a. The meso level reveals the invisible college that coordinates the research domain which is broadly differentiated into a cluster dominated by Complexity Science, one dominated by Social Psychology, and a multifunctional one.

The left contour plot of figure 3.13a shows that the macro collaboration network is a core/periphery structure because the density of $k$-components with high $k$-scores is

**Figure 3.13.: Structural cohesion of collaboration**

The structural-cohesion algorithm identifies components in which no more then $k$ nodes must be removed in order to disconnect the $k$-component. Contour plots in (a) are for the author co-authorship bicomponent shown in figure 3.9a. In the left plot, the third dimension (structural cohesion) corresponds to maximum $k$-scores of authors up to 5 when $k$-components are not allowed to be fully-connected cliques. All $k$-cliques are $(k-1)$-components, and by not allowing k-components to be cliques we create meaningful solutions. The map uncovers a clear core/periphery structure because the density of high-$k$ components is higher in the core than in the periphery. 5-components can easily be in the periphery simply because any two papers with more than 5 authors but with 5 authors in common form a 5-component. The following table shows that a giant 3-component is embedded in the largest bicomponent. Size is the number of authors and year is the mode of the publications generating the relations. Subdomain fractions correspond to the proportion of fractionally counted co-authorships in Social Psychology (SP), Economic Sociology (ES), Social Network Analysis (SNA), Complexity Science (CS), and Web Science (WS). SP and CS have the largest tie shares, but at $k > 3$ and no less than 20 authors, a cleavage emerges between the two, with either SP or CS having the lion's share. Only one 4-component is multifunctional. The position of the communities is seen in the contour plots. Each $k$-component is represented by the author with the highest self-reproduction (cf. figure 3.9). For this author and the whole $k$-component, degree and average degree, respectively, are given. Three 5-components are embedded in three parent 4-components, corresponding to two smaller SP communities and one large CS community. The right contour plot in (a) shows only those three 5-components and the lower-$k$ components they are embedded.

| | | | | Co-authorship share | | | | | | | Avg. |
|---|---|---|---|---|---|---|---|---|---|---|---|
| ***id*** | ***k*** | **Size** | **Year** | **SP** | **ES** | **SNA** | **CS** | **WS** | **Author** | **Degree** | **degree** |
| 1 | 2 | 4,790 | 2012 | 31% | 12% | 10% | 37% | 10% | NEWMAN,_M_E_J | 24 | 9 |
| 2 | 3 | 1,404 | 2012 | 26% | 7% | 8% | 51% | 8% | NEWMAN,_M_E_J | 18 | 10 |
| 3 | 4 | 224 | 2012 | 7% | 3% | 4% | 74% | 11% | DUNBAR,_ROBIN_I_M | 14 | 11 |
| 4 | 4 | 45 | 2007 | 77% | 6% | 3% | 5% | 9% | FRIEDMAN,_SAMUEL_R | 42 | 12 |
| 5 | 4 | 35 | 2007 | 93% | 2% | 2% | 1% | 3% | LATKIN,_CARL | 33 | 10 |
| 6 | 4 | 29 | 2011 | 24% | 2% | 35% | 40% | 0% | JAMES,_RICHARD | 18 | 12 |
| 7 | 4 | 28 | 2011 | 0% | 0% | 1% | 94% | 5% | MARATHE,_MADHAV | 27 | 9 |
| 8 | 4 | 24 | 2006 | 93% | 0% | 0% | 7% | 0% | KELLY,_JEFFREY_A | 23 | 12 |
| 9 | 4 | 22 | 2002 | 89% | 2% | 2% | 6% | 1% | FARMER,_THOMAS | 21 | 8 |
| 10 | 5 | 118 | 2012 | 7% | 3% | 3% | 78% | 8% | BARABASI,_ALBERT | 38 | 13 |
| 11 | 5 | 35 | 2007 | 89% | 2% | 2% | 6% | 1% | FRIEDMAN,_SAMUEL_R | 32 | 13 |
| 12 | 5 | 21 | 2011 | 93% | 3% | 1% | 2% | 2% | LATKIN,_CARL | 20 | 10 |

While, at $k = 3$, the collaboration network can be said to potentially be a result of fractal distinctions, multifunctionality is definitely absent at $k = 5$. The collaboration structure of communities 10, 11, and 12 is shown in (b). There is not only a structural hole between CS and SP, but also between the two SP communities 11 and 12. Author names are ranked descendingly by self-reproduction. Authors named in bold form the minimum set of nodes from which the whole $k$-component can be reached in one step.

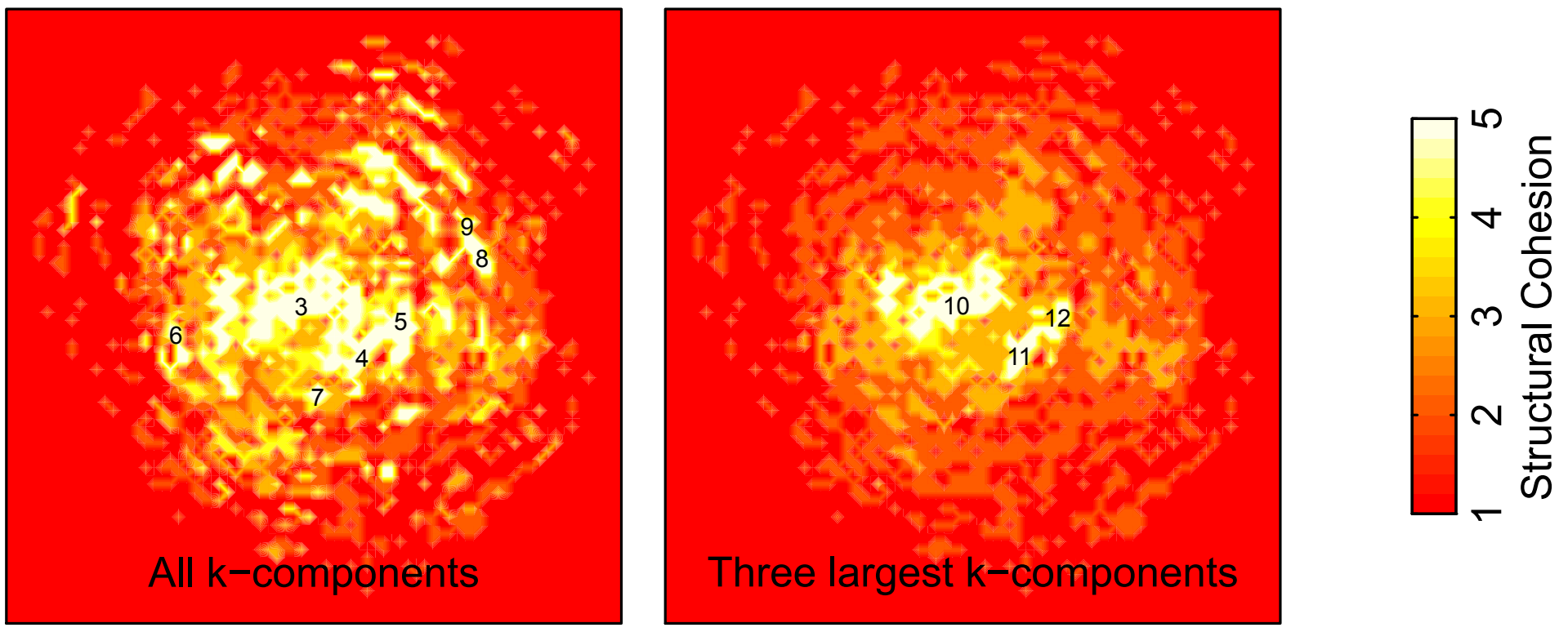

(a) Contour plot of collaboration landscape

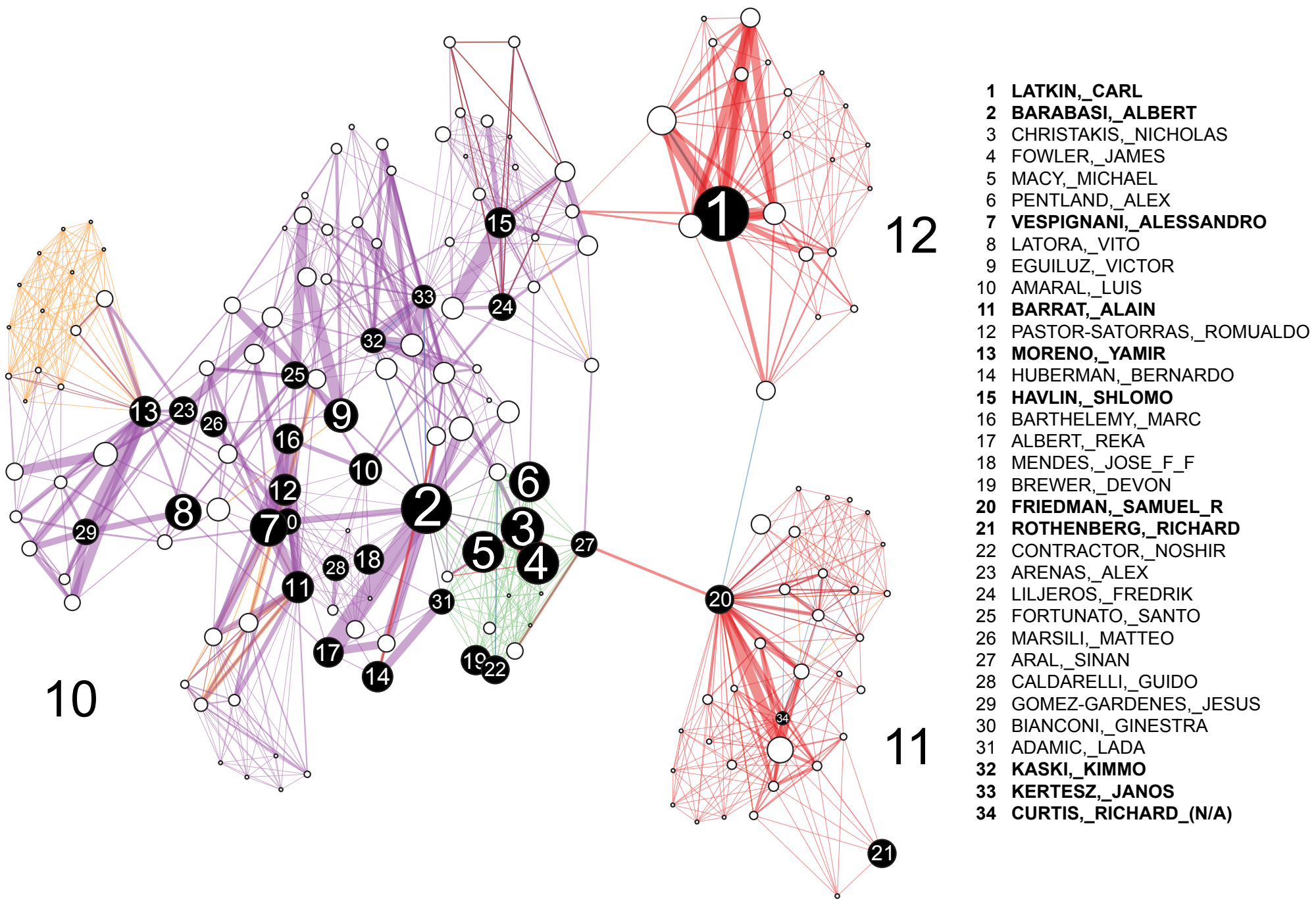

(b) Co-authorship network of three largest 5-components

**Figure 3.13.: Structural cohesion of collaboration**

larger towards the center. Again, methodological details are given in the caption of the figure. To get rid of the many 'almost-cliques' that still populate the scenery, we only study the substructures that consist of at least 20 authors and are not generated by a trivial number of cliques. A giant 3-component is embedded in the bicomponent, and NEWMAN,_M_E_J is the fractionally most productive or most self-reproductive author in both. This corresponds to his important role as the author of the most important papers on the citation main path. At $k = 3$, the core is still very multifunctional, though, compared to $k = 2$, Complexity Science has an increased share at the expense of Social Psychology and Economic Sociology.

The cohesion model is well suited to detect an open elite because it uncovers structural holes. In our case, global connectivity and multifunctionality break down at $k > 3$. There are seven 4-components, three of which harbor 5-components and are depicted in the right contour plot. A structural hole exists between community 10 and the communities 11 and 12. As we inspect the actual network of the nodes making up these 5-components (figure 3.13b), we see that the whole structure is only 1-connected and the hole also separates the Social Psychology clusters. Other than the multitopical invisible college of figure 3.9b, communities 6 to 12 are research fronts, cohesive conglomerates of groups doing research on singular topics. However, they differ in terms of age. Communities 6 and 7 are very recent research fronts on ANIMAL_BAHAVIOR and SOCIAL_CONTAGION, respectively. Community 8 on HIV_PREVENTION is a former front with a mode year of 2006. Even older is Social Psychology community 9 on SCHOOL_DROPOUT.[11]

Community 10 embedded in 3 is the top research front of the domain, regarding both size and age. It resembles the 3-identity Complex Network Analysis identified in the previous section. Compared to the face of Complex Network Analysis II, the story set is slightly more shifted towards the analysis of USER behavior in the WEB. The most central author is BARABASI,_ALBERT who has almost three times more co-authors than an average scholar in this community. Only six more authors are required as departure points to reach all nodes in the 5-component at unit distance. These authors are, descendingly ranked by Priceian productivity, VESPIGNANI,_ALESSANDRO, BARRAT,_ALAIN, MORENO,_YAMIR, HAVLIN,_SHLOMO, KASKI,_KIMMO, and KERTESZ,_JANOS. All these physicists are professors (all but one) and direct a group in a different country (USA, France, Spain, Israel, Finland, and Hungary). 78% of the collaborations occur in the Complexity Science type of tie.

Community 11 is concerned with epidemiology and is strongly associated with FRIEDMAN,_SAMUEL_R. Studies are focused on the TRANSMISSION_DYNAMICs of infections like HIV and GONORRHEA in RISK_NETWORKs due to DRUG_INJECTION or UNPROTECTED_SEX. Community 12 is LATKIN,_CARL's quasi-group in epidemiology on RISK_NETWORKs. Other than Friedman's wider group, it is more focused on SOCIAL_SUPPORT and the PERSONAL_NETWORK of the individual INJECTION_DRUG_USER. Latkin's group started a few years later and is a more recent research front. The share of Social Psychology in Complex Net-work Analysis is basically due to the collaboration dyad of CHRISTAKIS,_NICHOLAS and FOWLER,_JAMES who study health-related issues in the context of large-scale network structure, e.g., in the highly cited paper "The collective dynamics of smoking in a large social network" (CHRISTAK_2008_N_2249). This research may be exemplary for what results when Social Psychology and Complexity Science mate.

[11] These research fronts are not complete. LATORA,_VITO, e.g., is part of community 10 but also the most focal author in a fourth 5-component (of size 24). This community is not object of our analysis because it only consists of large cliques produced by three papers and is, therefore, trivial.

We propose the following relationship between invisible colleges and research fronts: The invisible college basically consists of the most prestigious group or lab leaders. They do not necessarily co-author papers but may coordinate the domain in committees or through editorials. The green clique in community 10 is an example. This editorial call for the establishment of Computational Social Science (LAZER_2009_S_721) brings together many prestigious hubs. In an invisible college, arena discipline is muted because coordination requires permeable boundaries. Research fronts, on the other hand, contain sorcerers of the invisible college but also the apprentices in the sorcerers' research groups. Arena discipline plays out much stronger because research fronts are dedicated to specific research problems. The invisible college of Web Science is most strongly represented in the research fronts, i.e., through collaboration, this subdomain has the strongest influence on the domain's path. 58 scholars coordinate the subdomain, 60% of which contribute to research fronts at the level of 4-components. Complexity Science takes the second rank with 47%. Fractions for Social Network Analysis and Social Psychology are 38% and 30%, respectively. Economic Sociology, at the lowest rank, participates least in shaping the domain's path. Only 29% of its 72 invisible-college members are embedded in 4-components. Why is it that, even though productive authors in Economic Sociology and Web Science are underrepresented in the invisible college of the whole domain (cf. section 3.2), these subdomains form the opposite ends of the spectrum in terms of influence on the research fronts?

We have already seen that, by 2012, Economic Sociology had the lowest authors-per-paper rate of all subdomains and Social Psychology had the highest (figure 3.5). The densification exponents in figure 3.14 quantify the extent to which the domain or its subdomains as 4-identities become constitutively cohesive or conscious about their selves. They are parameters of percolation. In agreement with authors-per-paper rates, the exponent for all years is largest for Social Psychology (1.31) and smallest for Economic Sociology (1.18). The other subdomains are all close to the score of the total (1.24). These scaling laws mean that, all throughout history, 4-identities are longitudinally self-similar. Annual slices of co-authorship structures are scaled versions of each other. The goodness of these fits is always $r^2 \geq 0.97$. While densification is self-similar over a whole century at this goodness of fit, a more careful analysis unveils that the domain as a whole densifies to a much lesser extent (1.13) and with a better fit ($r^2 = 0.99$) when we restrict scaling to the years after the Complexity Turn. In that case, Social Psychology is still among the two most densifying subdomains (1.18) but Social Network Analysis's parameter is barely distinguishable from unity. Economic Sociology dramatically increases densification from 1.18 to 1.41.

Densification exponents, rather than being invariant over time, can indicate differences of emergence (Lietz & Riechert, 2013; cf. Bettencourt et al., 2008). To deepen our understanding why Economic Sociology and Web Science participate so differently

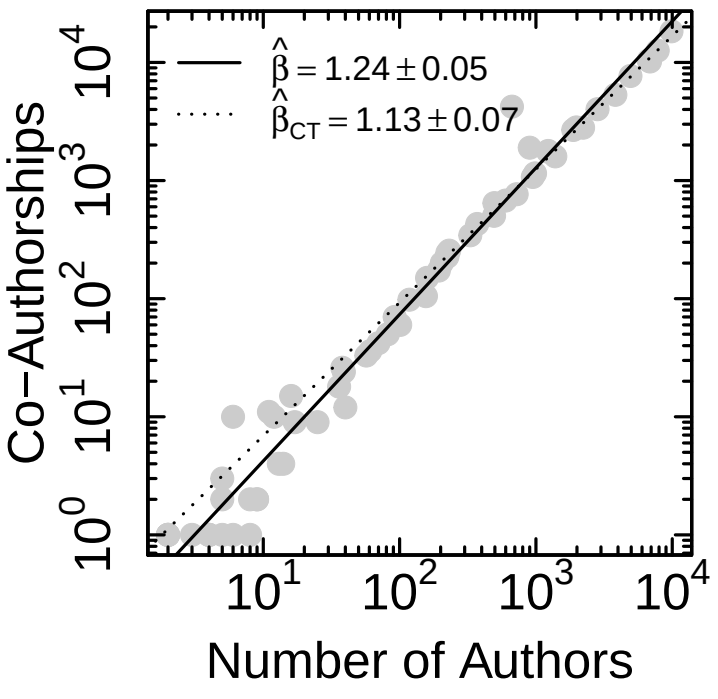

(a) Total

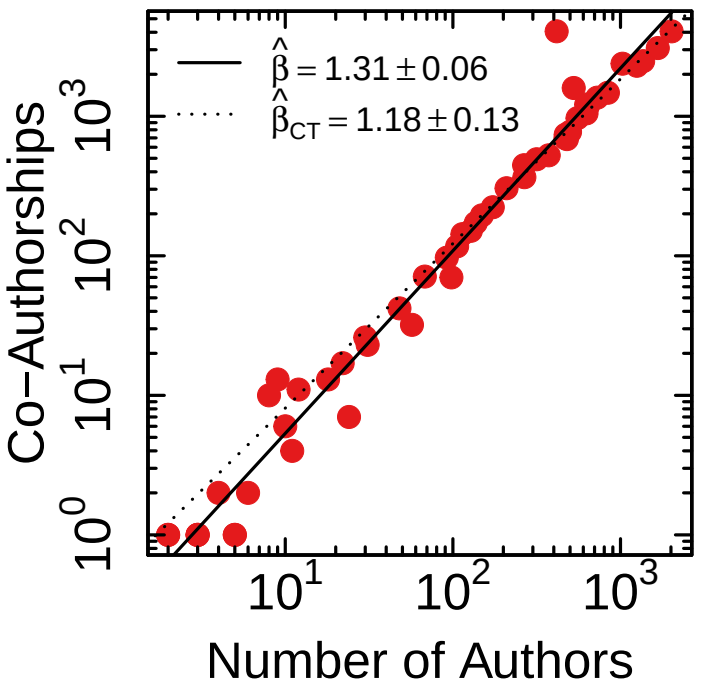

(b) Social Psychology

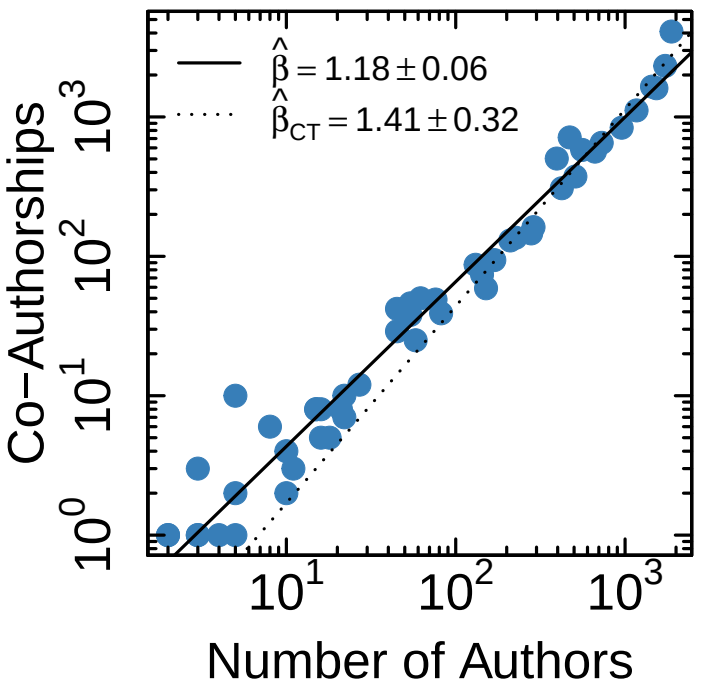

(c) Economic Sociology

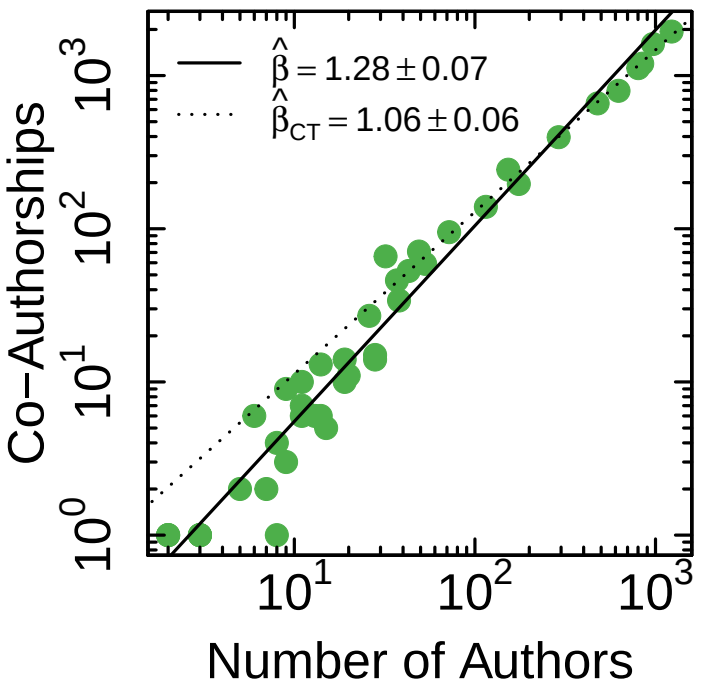

(d) Social Network Analysis

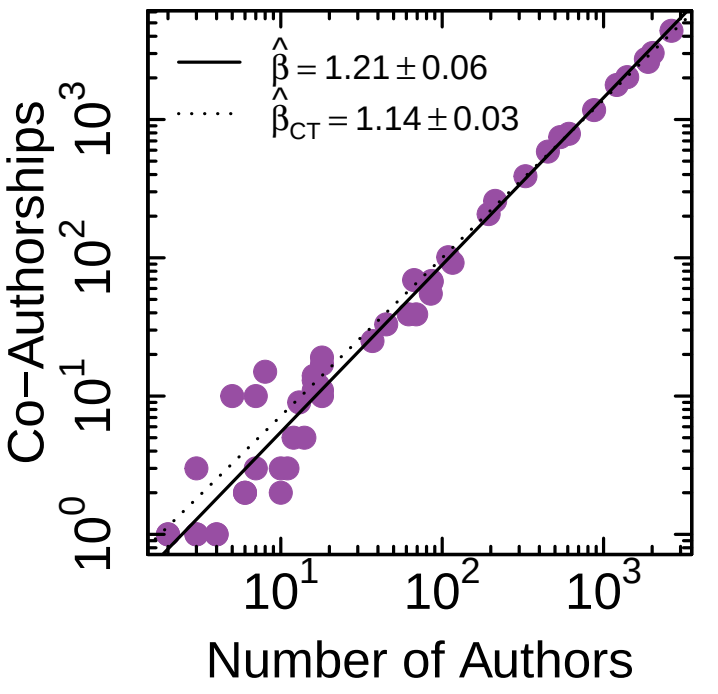

(e) Complexity Science

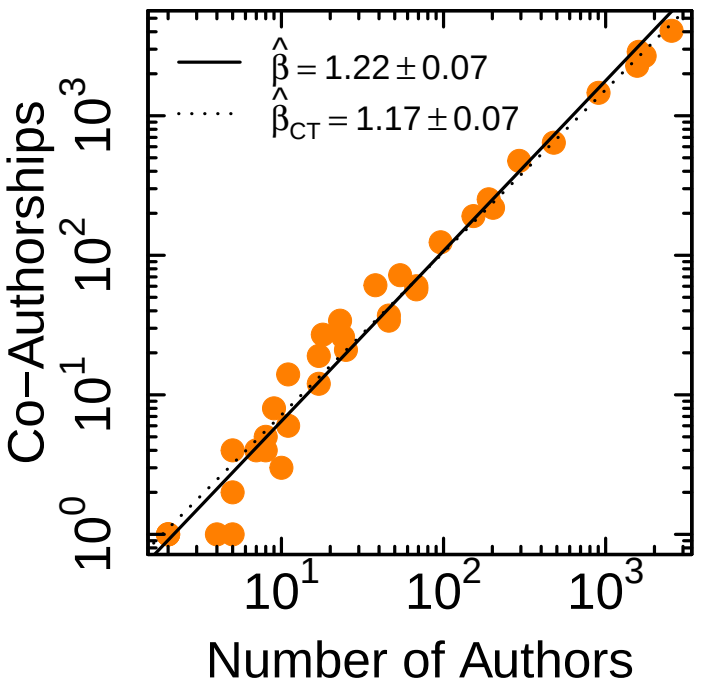

(f) Web Science

**Figure 3.14.: Co-authorship densification**

Exponents quantify the strength of emergence of a 4-identity's co-authorship structure. Scaling relationships are estimated using descriptive Standardizes-Major-Axis analysis. Each data point represents a year. All exponents exceed unity, i.e., these networks densify over time. Solid lines are fits to all years. For all plots, $r^2 \geq 0.97$, i.e., during the whole history of the (sub)domain(s), densification is self-similar with at least that quality. Different exponents are found when just the years after the Complexity Turn are fitted (dotted lines). Fits are better ($r^2 \geq 0.99$) except for Social Psychology ($r^2 = 0.97$) and Economic Sociology ($r^2 = 0.89$). The latter is the only subdomain that densifies more in the post-turn period than all throughout history.

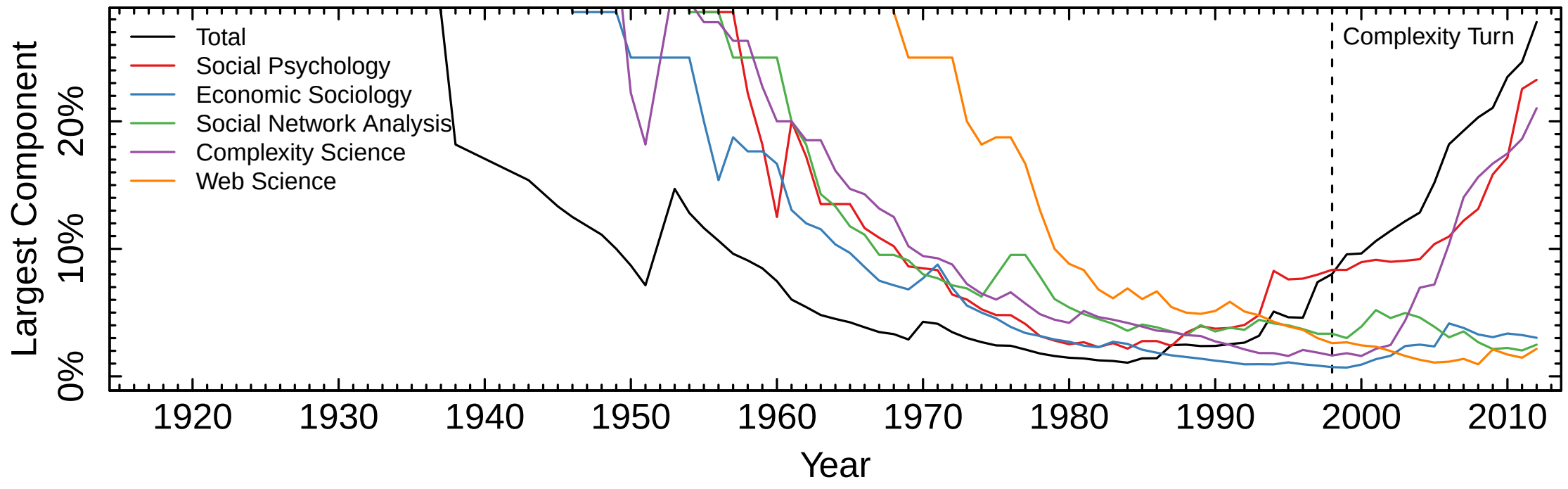

(a) Fraction of largest component

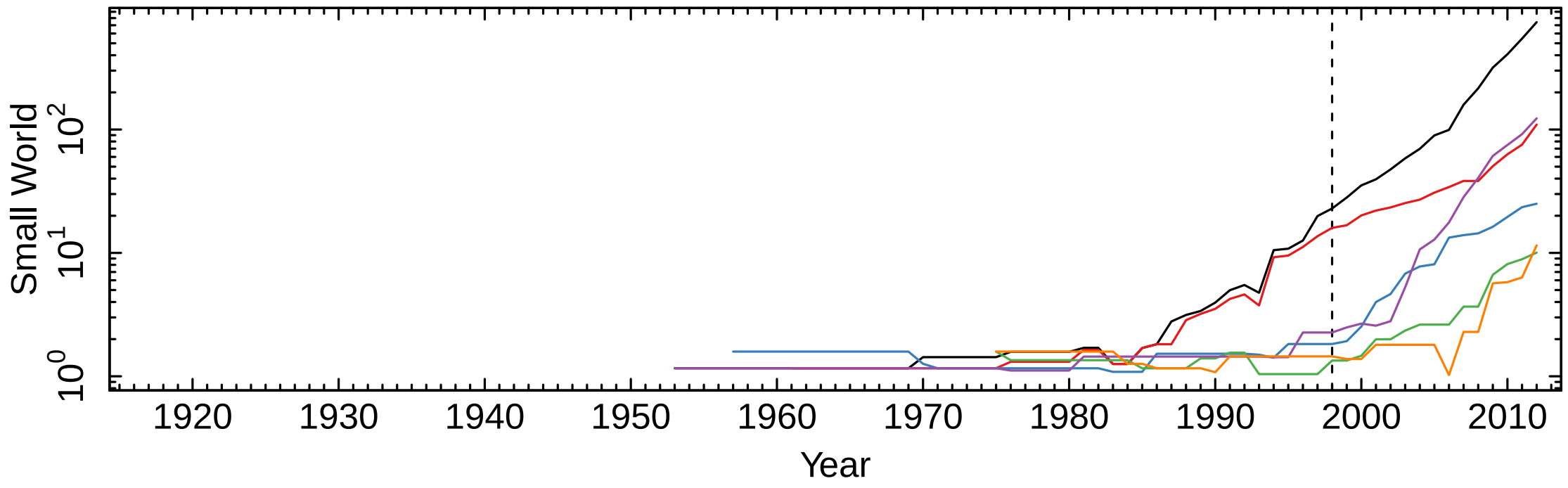

(b) Small World

**Figure 3.15.: Authorship emergence**
(a) The largest component in a network is the largest set in which all nodes are mutually reachable at least indirectly. Curves are for the co-authorship network cumulated over time. (b) The small-world coefficient is given for the largest component in the cumulative network. It is the ratio of the normalized average clustering coefficient and the normalized average path length: $sw = \frac{cc_{\text{obs}}/cc_{\text{exp}}}{l_{\text{obs}}/l_{\text{exp}}}$. In normalization, empirically observed clustering coefficients and lengths are divided by the values expected for a random network of similar size and density. The coefficient increases as observed networks have larger clustering coefficients than expected and approach the short average path length of the random network. We use the transitivity-based definition of the small-world coefficient (Humphries & Gurney, 2008).

in the research fronts, we turn towards the structural outcomes of the percolation process. The emergence of a giant component is an inevitable consequence of superlinear densification, and the size of the largest component tells at which stage percolation is. Social Psychology and Complexity Science exceed a connectivity of 20%, but despite superlinear densification, no giant component has emerged in Economic Sociology, Social Network Analysis, and Web Science (figure 3.15a). Essentially, the absence of a giant component signals a decentralized or multi-peaked collaboration landscape.

Autocatalysis is the missing piece in the puzzle of embeddedness. In figure 3.16 we

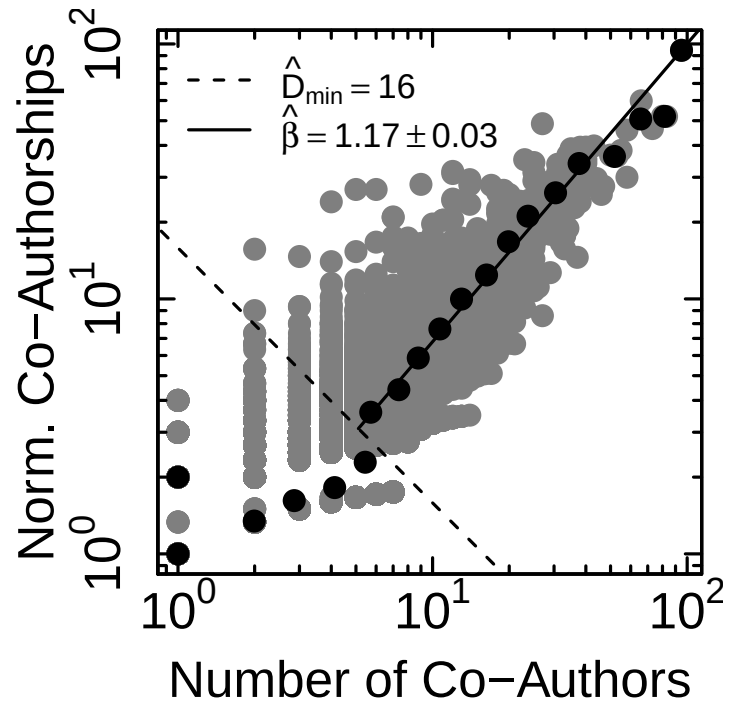

(a) Total

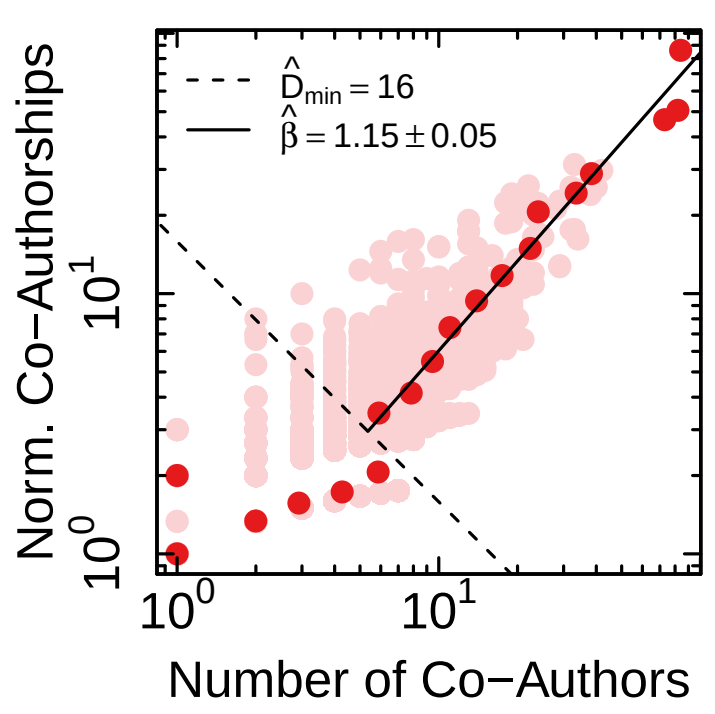

(b) Social Psychology

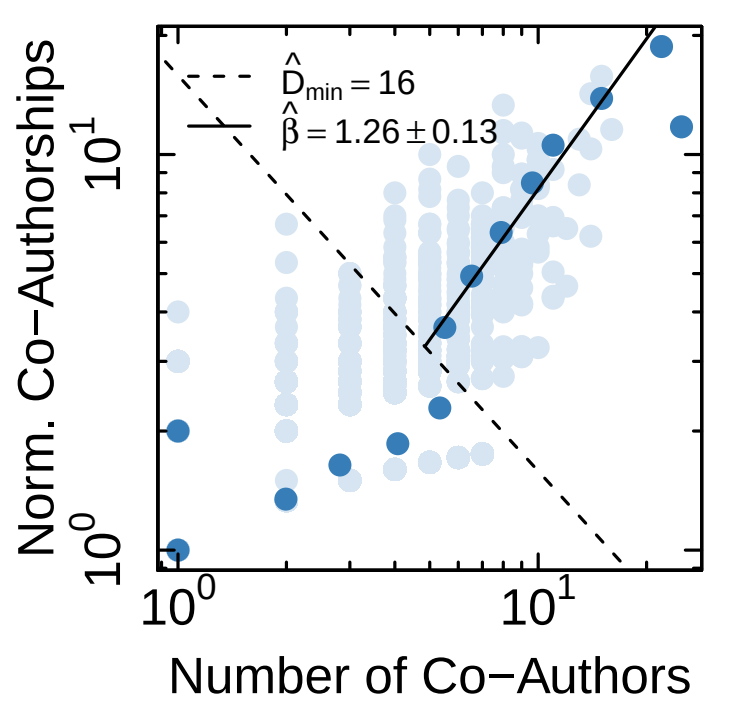

(c) Economic Sociology

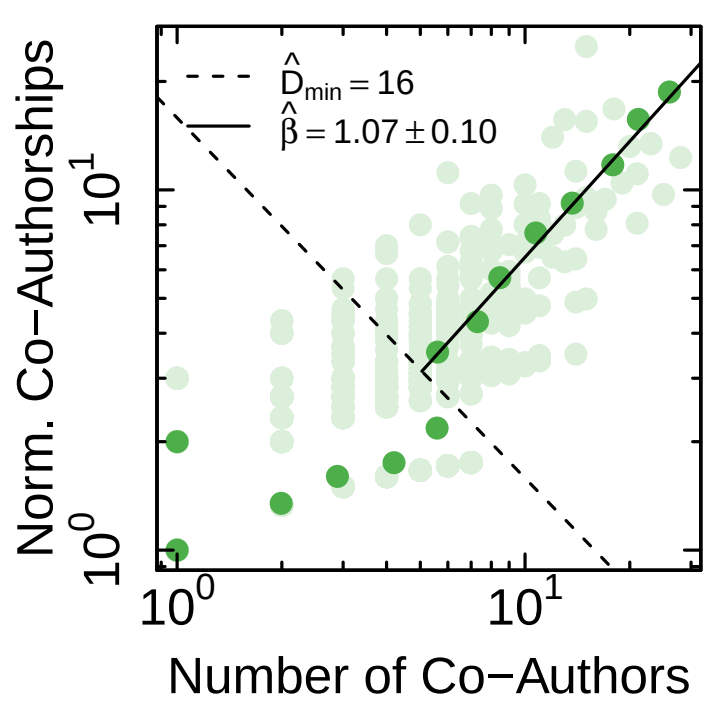

(d) Social Network Analysis

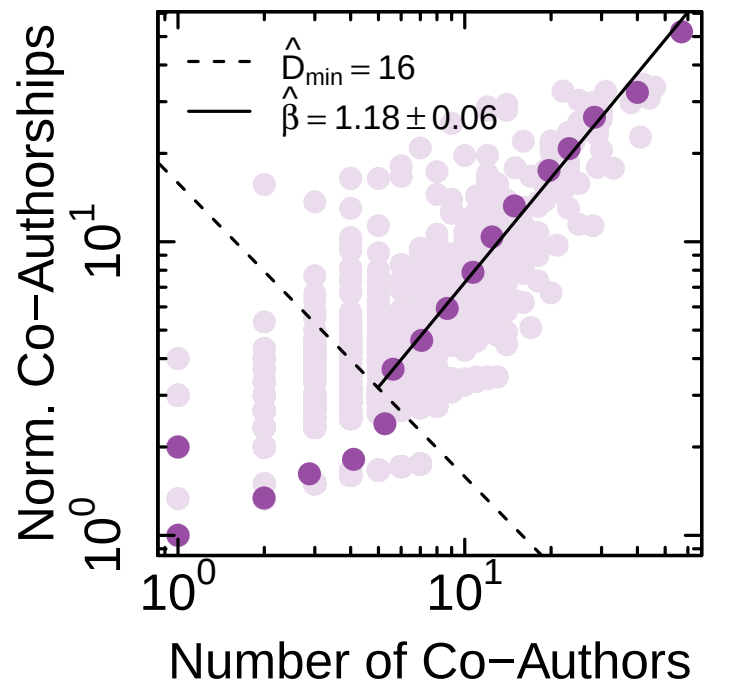

(e) Complexity Science

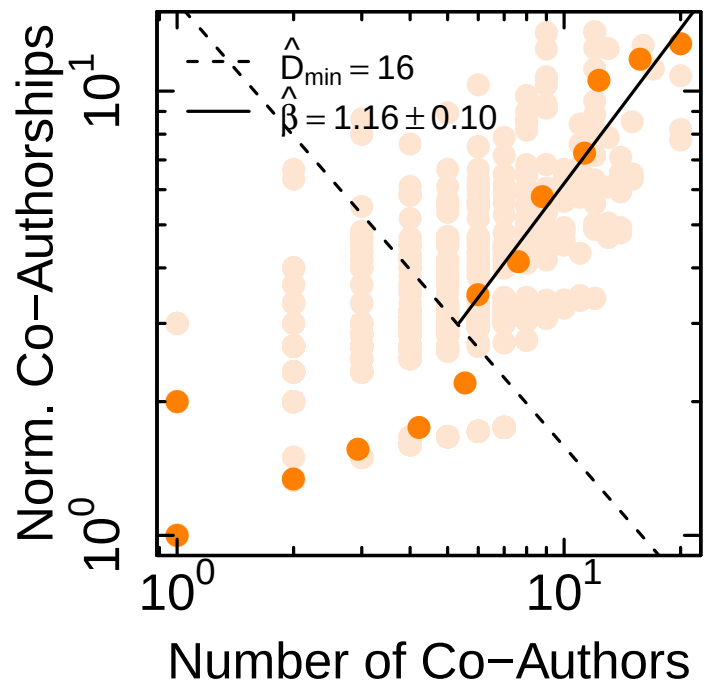

(f) Web Science

**Figure 3.16.: Co-authorship autocatalysis**

$\beta$ is the extent to which constitutive ties are increasingly reproduced towards the cores of co-authorship networks. Cores are represented by authors with many co-authors (large degrees). Each data point corresponds to an author. The scaling relationship of an author's unweighted and weighted degree (excluding self-loops) is estimated using descriptive Standardizes-Major-Axis analysis. Scaling was hidden in noise because papers with up to 84 authors created giant cliques. Therefore, from the 25,760 papers used to construct the co-authorship networks, 252 were removed because they were outliers with more than eight authors (mean plus three standard deviations). Fits are made to the shaded points above the dashed line which is a manually chosen cutoff resulting from noise reduction. Full points are a guide for the eye from double-logarithmic binning. Autocatalysis is present with increasing returns to degree ranging from 1.07 to 1.26.

study if strong constitutive ties emerge in the cores of collaboration networks or in the periphery. Now a scaling law is used to diagnose identities not in social time but in social space. After outlier papers with abnormally many authors have been removed, the domain and all subdomains exhibit increasing returns to degree regarding the average strength of a co-authorship tie. In other words, the exponent tells to which extent altruistic autocatalysis or team reproduction is stronger in the core than in the periphery. The goodness of fit is highest for Social Psychology and Complexity Science ($r^2 = 0.39$) but can be as low as 0.10 for Web Science.

This time, Economic Sociology stands out with the highest autocatalysis exponent (1.26). The subdomain that is least a team science, least contributes to the research fronts, and densifies weakest also reproduces its cores most strongly. In this case, strong superlinear autocatalysis means that reproduction occurs in many scattered clusters of authors, not in the core of a giant component. Such decentralization resembles high ambage or fuzz in social structure which contrasts the low ambiguity or cultural uncertainty associated with its highly cohesive co-citation core. This is how Economic Sociology keeps contingency critical. The low presence of Web Science in the invisible college paired with its high influence on the research fronts finds explanation in the subdomain's young age. It reproduces its cores comparable in strength to the average practice of the whole domain. Web Science is an emerging science which has not yet found direction through a coherent story set – there is no paradigmatic co-citation triad. It escapes the regime where identities die heat death through average densification and average autocatalysis.

### Small-World Emergence

The small-world coefficient is computed for the largest component in the cumulative network (see the caption of figure 3.15 for the maths of the coefficient). The largest component of the domain's co-authorship network (12,669 authors) has a small-world architecture with a coefficient of 741. Clustering ($cc_{\text{obs}} = 0.75$) is 1,447 times stronger than in a comparable random network. In the coefficient's denominator, the average path length ($l_{\text{obs}} = 9.81$) is about twice as long as in a comparable Erdős-Rényi graph. Compared to six degrees of separation, a typical value for co-authorship networks (Humphries & Gurney, 2008), 9.81 is a long distance. But, it gets much more small-worldly when one knows that it would be about 100 times longer if social life happened on a grid without any shortcuts. For comparison, the direct citation network has a small-world coefficient of 53. This is much smaller because, even though the average path length is almost as short as expected for randomness, clustering is only 68 times larger than in a random network.

Further investigation of the small-world effect reveals that Complexity Science and Social Psychology contribute most to the smallness of the world (figure 3.15b). From the mid 80s to the Complexity Turn, the effect was solely due to the dynamics in Social Psychology. After the turn, Complexity Science caught up within a decade. By 2012, their coefficients had reached 110 and 123, respectively. Other than the sizes of the largest component may indicate, Economic Sociology, Social Network Analysis, and Web Science do contribute to small-world connectivity. When the latter joined in the late 00s, small-world formation even accelerated. The meaning of emergence that the whole is more than the sum of its parts is expressed in the observation that, from 1998 onwards, the small-world coefficient of the domain is always larger than the sum of the subdomain coefficients. Global connectivity emerges from the overlap of different network domains with different story sets.

Short relational ties are evidence that collaborations are sought in fractal distinctions. However, though the world of Social Network Science is small, multiconnectivity is skewed towards, and mostly facilitated by, team-science subdomains with strong to average densification. According to our theory, superlinear densification resembles subcritical temperature that sooner or later leads to the divergence of paths. This is what we have actually observed for Complexity Science, the subdomain that practically represents the post-turn mainstream. In 2006, Evolutionary Game Theory split off from Complex Network Analysis. But the mapping from citation to authorship space is not neat. Besides community 3, the other research front matching Evolutionary Game Theory is community 7 with its focus on `AGENT_BASE_SIMULATION` (sic) of `SOCIAL_CONTAGION`.

Three authors lead the ranking of the most connected scholars: `LATKIN,_CARL` (129 co-authors), `BERKMAN,_LISA` (109), both in epidemiology (part of what we labeled Social Psychology), and `CARLEY,_KATHLEEN` (95). Like our idealtype, the co-authorship network is not scale-free, i.e., the degree distribution is not plausibly described by a power law. An average author in Social Network Science has four co-authors. Still, the network is self-similar in the sense of hierarchical modularity. An author's clustering coefficient is inversely proportional to its degree. This is another cross-sectional scaling law, see figure 3.17 for the statistics. For all subdomains but Economic Sociology, the result matches the exponent of the idealtype (figure 2.4d). Though they all differ in terms of densification and autocatalysis, they seek fractal architectures with a unit exponent. Economic Sociology deviates from this universal council architecture. Its exponent is much smaller, i.e., the hierarchy is flatter and the network is more decentralized, adding to a coherent picture of Economic Sociology being a fundamentally different science in terms of social structure.

### Towards Style

The scaling hypothesis says that identities at small and large scales are instantiations of an idealtypical identity that is not situated at a particular scale. According to our model, the five subdomains are styles that maintain a self-similar selection profile – in this case: authorship profile. As we have already seen, authorship is the most fluctuating practice of the three. Authors come and go as story sets stay. The size distribution which signals that the 5-identities as well as the context they embed into reside at the phase transition of normality and chaos is Lotka's Law (figure 3.8). Note that Lotka's Law is also manifest in social structure because it resembles the distribution of the weighted degrees in the normalized co-authorship network, i.e., the latter is scale-free under normalization.

Scaling laws have proven powerful tools to diagnose the macro level of meaning structures. To demonstrate that styles are scale-invariant, we now dissect networks at the level of individual scholars and their groups. Given the fuzzy boundaries of Social Net-

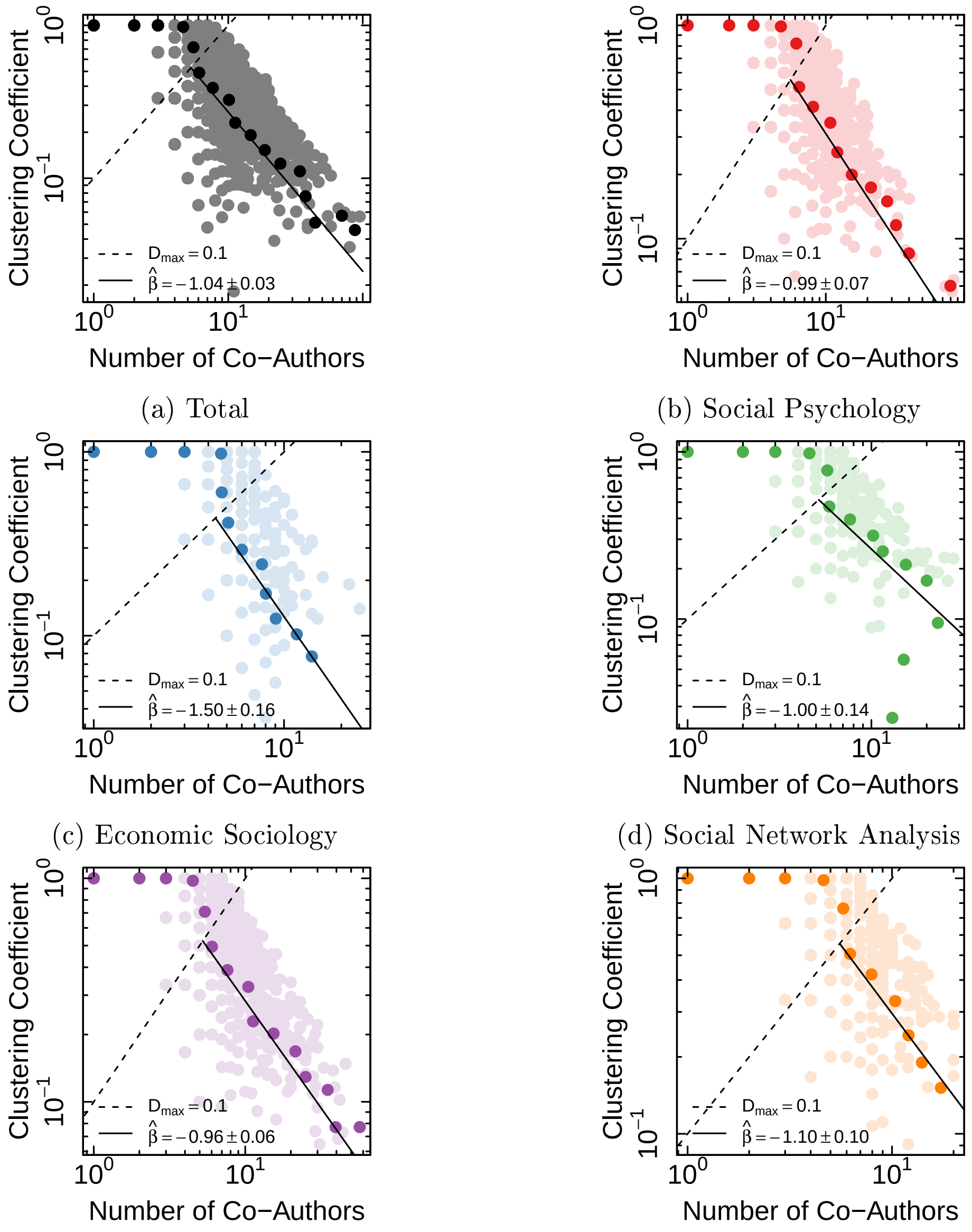

(a) Total (b) Social Psychology

(c) Economic Sociology (d) Social Network Analysis

(e) Complexity Science (f) Web Science

**Figure 3.17.: Hierarchal modularity in co-authorship**

Co-authorship networks are hierarchically modular because nodes' clustering coefficients are scaling functions of degree. Except for Economic Sociology, subdomain exponents are indistinguishable from $\beta = -1$ with 95% confidence. Scaling was hidden in noise, so noise was removed as described in figure 3.16. Functions are fitted to the raw data (shaded points) using Standardized-Major-Axis regression. Full points are a help for the eye from double-logarithmic binning. $D_{\max}$ is a manually chosen upper cutoff from noise reduction. Web Science required bootstrapping to not fit a positive exponent, i.e., $r^2 = 0.00$. The goodness of fit is best for Complexity Science ($r^2 = 0.10$).

**Figure 3.18.: Collaboration networks of H. C. White and A.-L. Barabási**
Co-authorship networks of two central scholars in Social Network Science. (a) WHITE,_HARRISON_C is credited for having envisioned the program that constituted the Harvard Breakthrough. *Wikipedia* names 17 students and mentees of him (en.wikipedia.org/wiki/Harrison_White, visited June 23rd, 2015). Twelve of them are in our dataset. All co-authors of WHITE,_HARRISON_C and his students and their relations are shown. The size of nodes depicts betweenness centrality. Two additional nodes are labeled that are essential for connectivity. All mentees' names are bold. (b) BARABASI,_ALBERT is credited for having initiated, alongside Duncan J. Watts, the Complexity Turn. In his c.v. from 2012 (www.barabasi.com/ALBarabasi_CV.pdf, visited June 21st, 2015), Barabási lists 20 scholars whose Ph.D. thesis he advised or whose postgraduate-scholarship he sponsored, 13 of which contributed to Social Network Science. As for White et al., the network has been expanded to include co-authors. Author names were further disambiguated. JEONG,_HAWOONG, e.g., had six synonyms because of his Korean name. The table gives the size and density of the networks, the modal year of the publications authored by the scholars, and how much they fractionally contribute to the five subdomains.

| | | | | Co-authorship share | | | | |
|---|---|---|---|---|---|---|---|---|
| **Group** | **Size** | **Density** | **Year** | **SP** | **ES** | **SNA** | **CS** | **WS** |
| White | 238 | 0.02 | 2008 | 29% | 18% | 17% | 28% | 8% |
| Barabási | 156 | 0.04 | 2007 | 4% | 3% | 8% | 80% | 6% |

work Science, we restrict ourselves to the styles of a social science and a natural science. The most social scientific subdomain is Economic Sociology. We have identified it in one corner of parameter space. Its co-authorship network is most decentralized as indicated by small teams, lowest densification, and a flat hierarchy. At the micro level, it will be represented by the sphere of influence of Harrison C. White.

Complexity Science is in the other corner. It is strongly a team science with high densification and an idealtypical centralization. In this subdomain, we observe the exact generative mechanism of the idealtype – the teaching of apprentices who then become sorcerers themselves. The scholar to represent Complexity Science is Albert-László Barabási because he is the most central figure in the largest research front. We have identified 13 of his former students in our dataset. Four of them are today professors themselves (ALBERT,_REKA, GOH,_KWANG_IL, GONZALEZ,_MARTA_C, WUCHTY,_STEFAN). The co-authorship network of BARABASI,_ALBERT and his students, greedily expanded by a depth of one, is depicted in figure 3.18b. The director of the Barabási Lab is most central, indicated by betweenness. ALBERT,_REKA has the second largest betweenness score, i.e., her research network is most decoupled. This is expected because she is Barabási's first Ph.D. student in the dataset. The largest (by degree) and second-most independent (by betweenness) subnetwork is JEONG,_HAWOONG's who also started publish-ing with Barabási in 1999. BIANCONI,_GINESTRA, GONZALEZ,_MARTA_C, GOH,_KWANG_IL, ARGOLLO_DE_MENEZES,_M, and SZABO,_GABOR have all established their own largely independent spaces with at least ten co-authors each. This kind of growth by duplication is reminiscent of the watchmaker Hora who builds watches by constructing modules which he combines into meta modules. We have also encountered this mechanism in the previous section: Physicists have a style of summing up knowledge in reviews. These are then cited as a milestone – the paradigm – that has been reached and from which further searches depart.

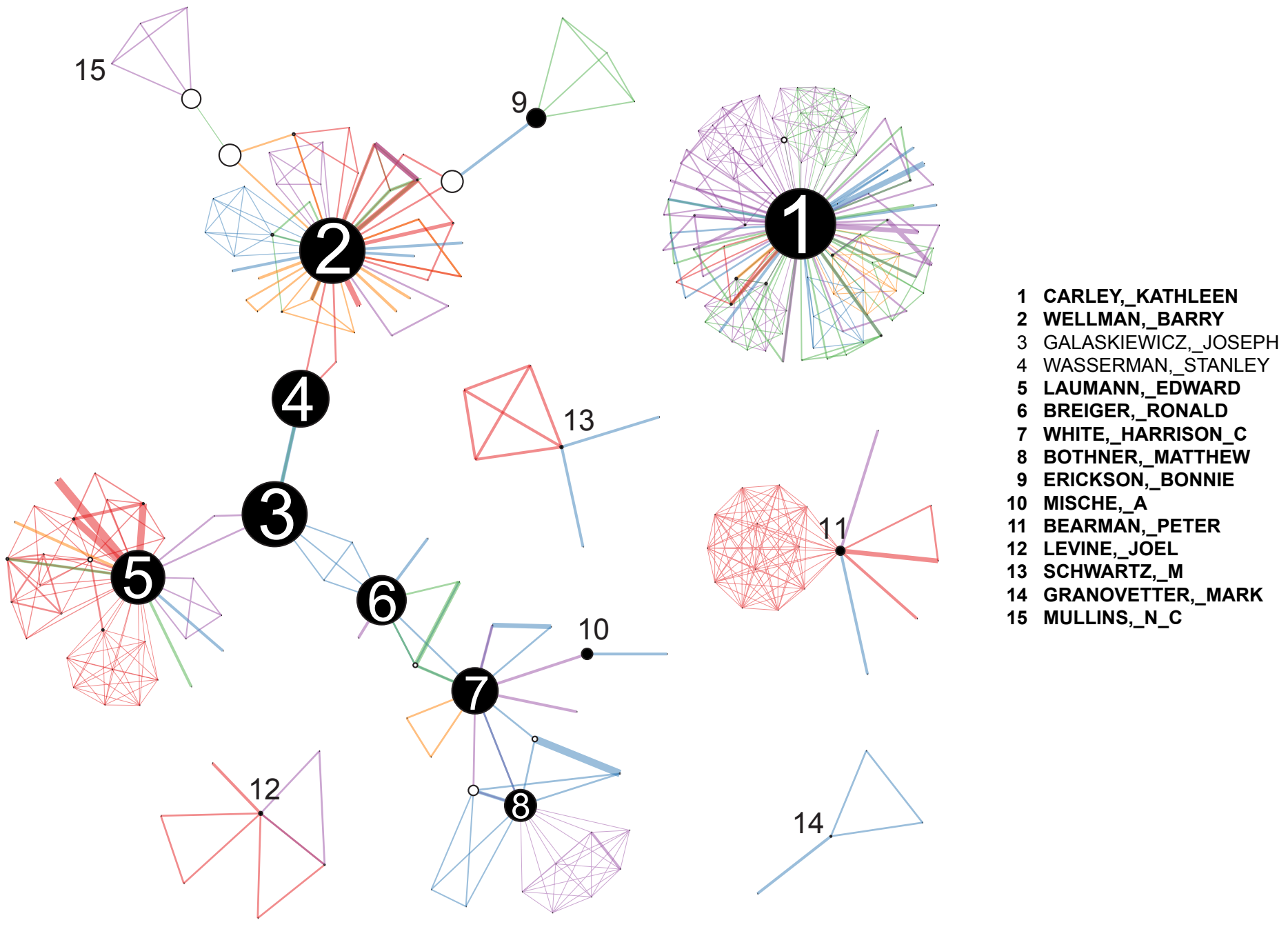

(a) Harrison C. White

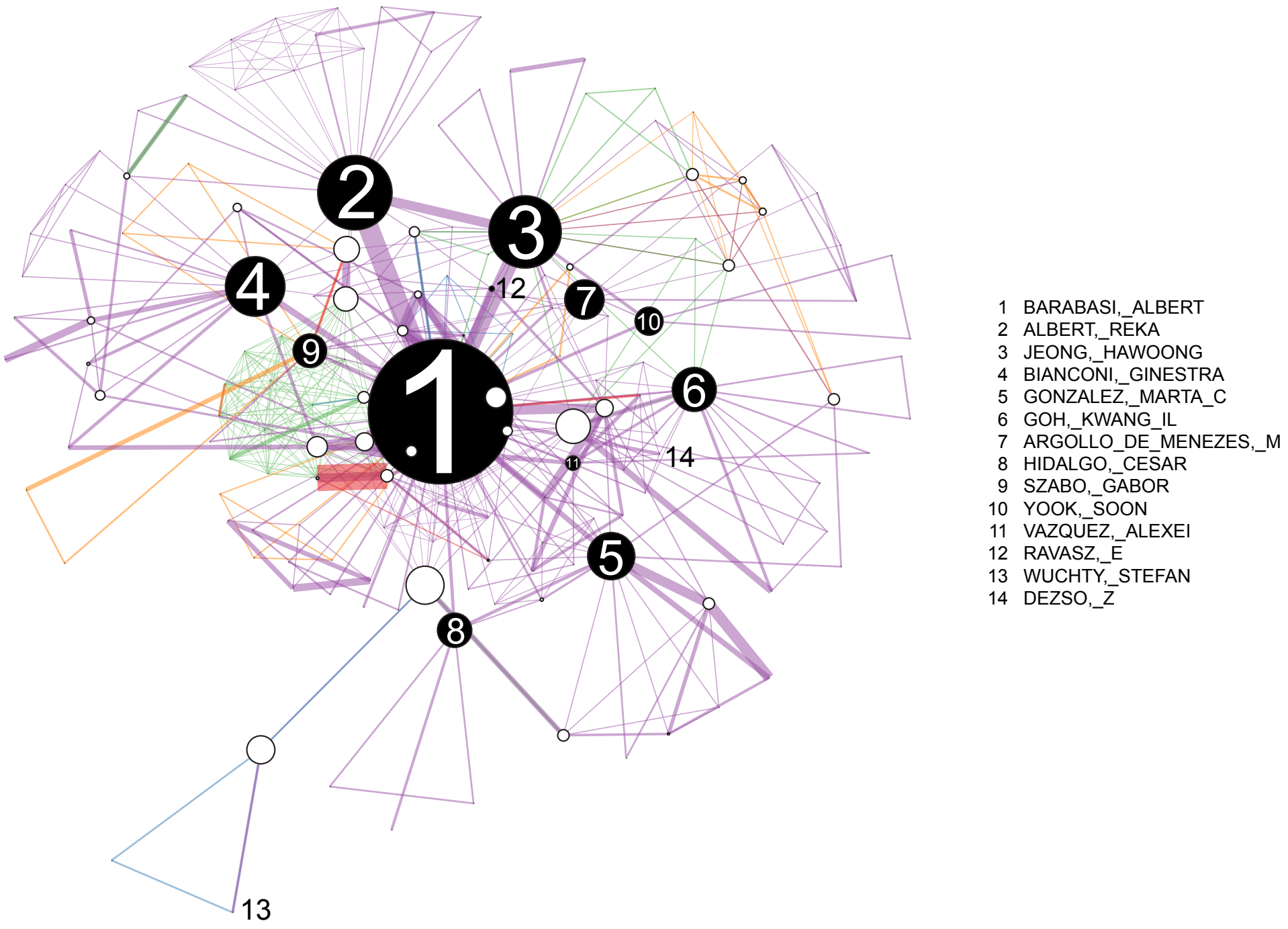

(b) Albert-László Barabási

**Figure 3.18.: Collaboration networks of H. C. White and A.-L. Barabási**

WHITE,_HARRISON_C has been chosen to represent the social sciences because his 1976 work on BLOCKMODELing was the most important paper in Social Network Science until 2001. He has contributed to all subdomains but Social Psychology. *Wikipedia* lists 17 prominent students of White, 15 of which are (or were) professors, 13 in sociology. The network of White et al. is fundamentally different than Barabási's. WHITE,_HARRISON_C is only seventh-most connected regarding betweenness and degree. The network is less coherent because in the social sciences, other than in disciplines like physics, the group leader is not usually listed as an author.[12] Of the mentees in our dataset, WHITE,_HARRISON_C has only co-authored with BREIGER,_RONALD, BOTHNER,_MATTHEW, and MISCHE,_A. WHITE,_HARRISON_C and GRANOVETTER,_MARK have never actually co-authored a paper.[13] Only three scholars are embedded in large groups. The strongest effort is by CARLEY,_KATHLEEN in the center on Computational Analysis of Social and Organizational Systems (CASOS). Being a professor in computer science, she organizes her group like in the natural sciences. The second-largest momentum is gained by WELLMAN,_BARRY who directs the NetLab devoted to studying the intersection of social, communication, and computer networks. As we have seen, papers by Wellman both attracted the main path (WELLMAN_1979_A_1201) and concluded (WELLMAN_1990_A_558) Social Support Studies. LAUMANN,_EDWARD is the hub in a large cluster on SOCIAL_NETWORK_ANALYSIS in the context of sexual networks. His paper with GALASKIEWICZ,_JOSEPH (LAUMANN_1978_A_455) is pivotal in both connecting the co-authorship network of the Harvard school and channeling the citation flow in Social Network Science.

While both White and Barabási have tremendous cultural influence on the domain, the latter's influence is much more manifest in social structure. Complexity Science's team style is mirrored in Barabási's extended ego network. The network from which the 5-identity BARABASI,_ALBERT emerges is twice as dense as WHITE,_HARRISON_C's. White's network is a small-scale version of Economic Sociology. It is anecdotal evidence for style as a longitudinally stable pattern of selection at the personal level that White kept his way of collaboration throughout the four decades he has left a trace in our dataset – he did not suddenly discover his physics roots and embed into a large group following the Complexity Turn.

The networks of the two sorcerers also differ in terms of functionality. While the Barabási Lab largely contributes to Complexity Science, the Harvard school is all over

[12] Though physicist NEWMAN,_M_E_J is most productive, he is not embedded in any 4- or 5-component because he does not run a group with a multitude of fractional authors.

[13] Confirmed by Mark Granovetter in personal communication, June 19th, 2015

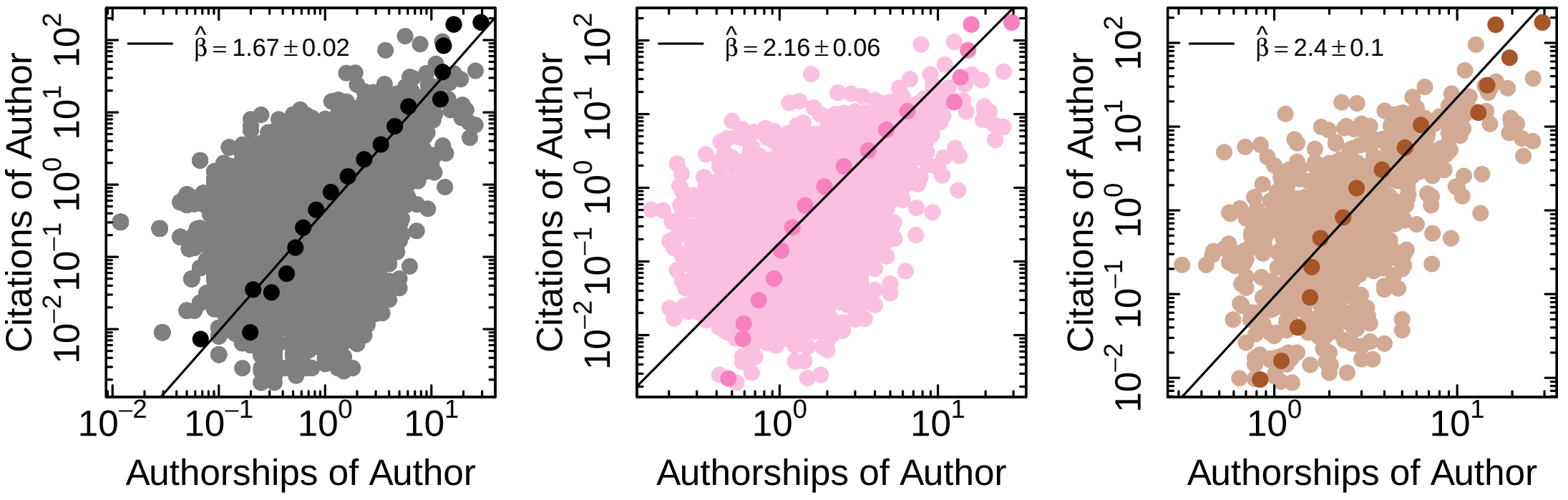

(a) Unifunctionality and more (b) Bifunctionality and more (c) Trifunctionality and more

**Figure 3.19.: Scaling of author impact**

The functionality of an author is the number of subdomains that he contributes to through at least one publication. Each point is a cited author, x gives the fractionally counted number of papers written by that author, and y the sum of the fractionally counted citations to all of that author's (wholly counted) papers. The scaling law is fitted using the Standardized-Major-Axis method on the shaded points. Full points are a help for the eye based on double-logarithmic binning. There are increasing returns of the average number of citations per publication to productivity. Returns increase as we move from (a) all 15,893 authors that are at least unifunctional ($r^2 = 0.09$), over (b) 3,690 authors that are at least bifunctional ($r^2 = 0.21$), to (c) 754 authors that are at least trifunctional ($r^2 = 0.33$). There is a cutoff for lowly productive authors in (b) and (c). Including these authors positively affects the goodnesses of fit because these data points are few. Our method of detecting the cutoff through a decrease of fitting quality is not applicable and, hence, the cutoff is ignored.

the place, though least in what is Web Science today. Especially `CARLEY,_KATHLEEN`'s and `WELLMAN,_BARRY`'s labs are large multifunctional research organizations. In section 3.2 we saw that the two contribute to all subdomains. These overlaps are interfaces through which knowledge spills over from one subdomain to another. Only 23 authors switch among all five and 139 among four subdomains. Switching is also the source of meaning. In this light, it is telling that Complexity Science, Economic Sociology, and Web Science are all structurally most equivalent to Social Network Analysis (see table in figure 3.10). We are interpreting the cosine similarity of subdomain authorship profiles as structural equivalence. The more two subdomains select the same authors, the more their story sets overlap. The three similarity scores signal the commitment of large parts of Social Network Science to the methods developed in Social Network Analysis. The only other pair similarity that matches those involving Social Network Analysis is the switching likelihood of Social Psychology and Economic Sociology, the oldest subdomains.

In section 3.2, we have already seen that there can be an advantage in teams in the sense that embeddedness translates to influence in coordinating the whole domain through the invisible college. Team sciences outreproduce others relationally. There is also an advantage in the diversity that teams can offer. A lone author will never be able to tap, combine, and draw innovation from multiple knowledge pools, but, in a team of five, each author can, in principle, be specialized in a different subdomain. In Social Network Science, we find such a creativity effect of multifunctionality at the author level. The creativity of an author is operationalized in terms of citations to his or her works. The underlying assumption is that creativity leads to quality which is rewarded through citation. We need to control for productivity which is a condition for multifunctionality and can explain citations through visibility. Therefore, we employ a type of scaling analysis the exponent of which is known in scientometrics as a scale-independent indicator (Katz, 2000). Figure 3.19a shows that there is a size effect related to impact: the more productive an author is, the higher his or her works are cited on average. In numbers, an author with 3 fractionally counted papers has an impact of about 0.9 normalized citations per paper, but the impact of an author with 30 papers is about 4.3. These increasing returns to scale are reflected in a scaling exponent of 1.67. As we restrict authors to at least bifunctional and trifunctional ones, the exponent increases to 2.16 (figure 3.19b) and 2.4 (figure 3.19c), respectively. An author that contributes to at least three subdomains and writes 30 fractionally counted papers has a normalized impact of about 10.2. Multifunctional authors are indeed more innovative than scholars who only dwell among similars, and, since larger teams can be more multifunctional, they can, at least in principle, be more reproductive. In other words, recombination or the mating of styles is a promising control strategy.

## Summary

The analysis of social structure has shed led light on the processes that have shaped Social Network Science throughout a century. The answer is more deep and more interesting than "yes, the world of Social Network Science is small and hierarchically modular." The emergence of the small-world architecture only began in the mid 80s and was, until the Complexity Turn, solely due to the dynamics of Social Psychology. As Complexity Science percolated in the 00s, no open elite or multifunctional network core emerged in the whole domain. Support for fractal distinctions only holds up to a structural-cohesion level of three. Above, a structural hole separates research fronts, specialized communities of practice with strongly different story sets and research directions. The most cohesive research fronts belong to Complexity Science (Complex Network Analysis) and Social Psychology (on public health), the two subdomains that most strongly qualify as team science. Densification exponents quantify the extent to which 4-identities build structure as a collective memory. Subdomains with large returns to scale are not only more likely to grow structurally more cohesive research fronts and relationally outreproduce less team-oriented subdomains. They also have an evolutionary advantage because embeddedness translates into collaborative influence and, through diversity, into heightened citation impact.

Socio-cultural temperature provides a model for the observation that the subdomain with the most decentralized network most strongly reproduces ties in the scattered cores, as revealed by an autocatalysis scaling law. Economic Sociology escapes lock-in by bal-ancing low cultural uncertainty through high social uncertainty. Web Science is still most atomized because it is very young. Despite differences of densification and autocatalysis, all subdomains find footing through stable intermediate forms and evolve into hierarchically modular networks governed by council discipline. Except for Economic Sociology, this architecture is universal. The social network of Social Network Science resides at a phase transition of stability and change described by Lotka's Law, and scaling laws (densification, autocatalysis, hierarchical modularity, and impact) describe the self-similar macro dynamics and structure at this critical point. The self-similarity of styles or sensibilities of selection was exemplified by showing that macro properties are mirrored in the extended ego networks of Harrison C. White and Albert-László Barabási.

### 3.3.3. Styles Self-Organize to the Normality-to-Chaos Transition

In the previous section we have shown that the recombination of different authorship styles is a promising strategy for reproduction through change. In this section, we want to know if authorship, citation, and word usage styles actually mate to change. Based on the quantitative analysis of the Matthew Effect we will determine the fitness of social facts and observe how they reproduce. When studying preferential attachment, an exponent $\beta < 1$ means that the Matthew Effect is muted and power laws do not emerge. When $\beta = 1$, autocatalytic reproduction is in full Markovian swing: facts that have been recognized as facilitators of collective control and consensus are memorized and are selected proportionally in the following time step. Bradford's and Zipf's Law result from styles just like Lotka's Law. Consequently, $\beta$, the Matthew exponent, is a measure for the strength of style. It quantifies the sensibility for selecting social facts from a distribution of facts selected in the immediate past. Before we start, we need to specify the memory $\Delta t$ as the time window for the (fractionally counted) number of selections used to predict the number of selections at the following time point. Golosovsky and Solomon (2013) use six years but give no reason for this particular window size. We use three years for the simple reason that most papers in Social Network Science have the highest citation rate in their third year. Price (1986 [1963], p. 72) had observed the same for physics.

Figure 3.20 demonstrates how exactly we determine the Matthew exponent. For 2012 and wholly-counted citations, references were subject to linear preferential attachment when they had reached an initial attractiveness of eight citations. In the remainder, we use fractional counting of selections. Figure 3.21 plots for the three practices how exponents evolve over time. For authorship, statistics are only available late in the life of the most productive subdomains Social Psychology and Complexity Science. Error bars are huge because the data points for fitting are few. In addition, what we are measuring as selection is not the number of co-authors but the number of authorships of authors, which is rather small compared to network degree. For where statistics are available, productivity dynamics can be explained by sublinear preferential attachment with $\hat{\beta} \approx 0.5$. For lack of statistics, our analysis of the authorship practice ends here.

Much better statistics are available for the more 'cultural' citation and word usage practices because those are more strongly concentrated, signaled by smaller exponents of the size distributions (cf. figure 3.8). These distributions are the outcomes of styles and resemble the actual sensibilities of identities. There is con*sensus* in Social Network Science to cite `WASSERMA_1994_SOCIAL` and use `COMMUNITY`. The Matthew Effect is how identities get to the phase transition. It is a marked result that preferential attachment converges to linearity in both practices. That means, we can interpret $\beta \in [0, 1]$ as a score how much an identity has fully found a style of selection and is at criticality. Being a 5-identity at the percolation threshold means that the 1-identities transact at the onset of global communication or at infinite correlation length.

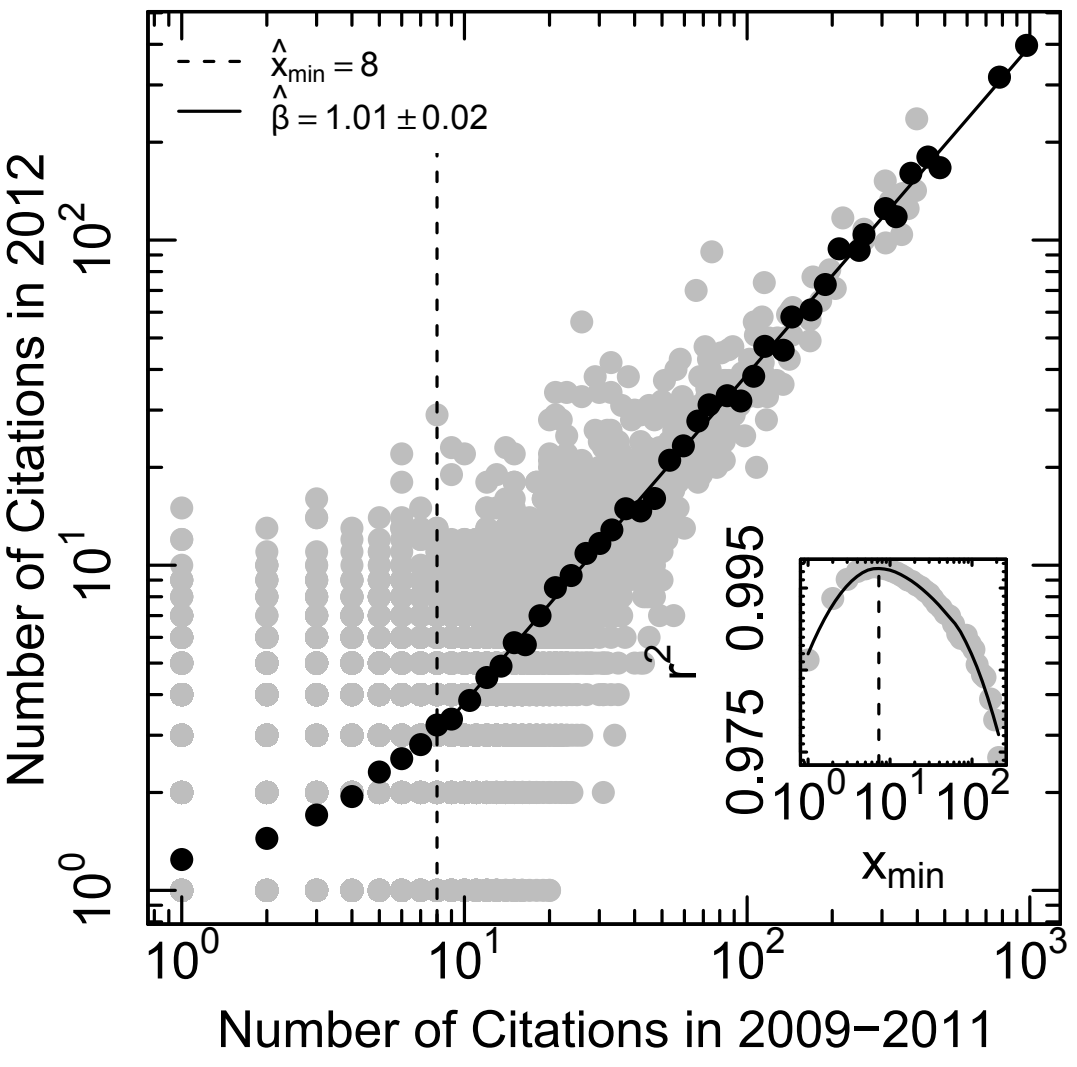

**Figure 3.20.: Preferential attachment in the citation practice (2012)**
The citation practice of Social Network Science in 2012 was governed by linear preferential attachment. The citations gathered by references during $\Delta t = 3$ years before 2012 are used to predict the number of citations in 2012. Each gray point is a reference. In this example, citations are counted fully for didactic reasons. The method is Ordinary-Least-Squares regression because this measurement serves a prediction. Fitting a scaling law to the raw data gives a nonsensical result because the statistical mass is in the low citations. Instead, the data averaged into 20 logarithmic bins per decade (black points) is fitted. Fitting is only attempted for ten and more binned data points. The curve reveals an initial attractiveness above of which preferential attachment sets in. To identify the exact cutoff, the goodness of fit $r^2$ is recorded as a decreasing number of points above $x_{\min}$ is fitted. The initial attractiveness is the number of citations where $r^2$ is maximized. The maximum is identified for the goodness-of-fit function that is smoothed using locally weighted polynomial regression, shown in the inset. In the example, linear preferential attachment sets in after references have accumulated eight citations. In our analysis, we do not report $r^2$ because the goodness-of-fit to binned data is misleading.

In citation, Social Psychology and Economic Sociology were the first to develop a style. Here, we find an explanation why Social Support Studies, not Structuralism, constituted the mainstream of Social Network Science from 1981 to 1993. In that time, exponents were relatively stationary at $\beta_{\text{SP}} = 0.61 \pm 0.07$ and $\beta_{\text{ES}} = 0.38 \pm 0.09$. Social Psychology's

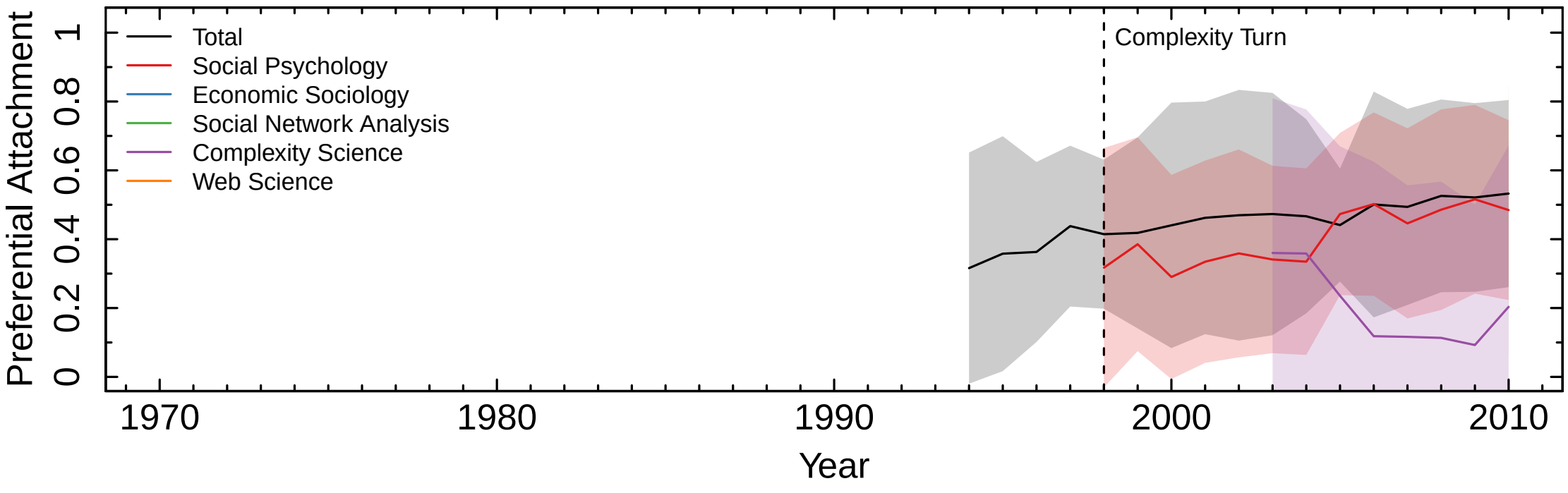

(a) Authorship

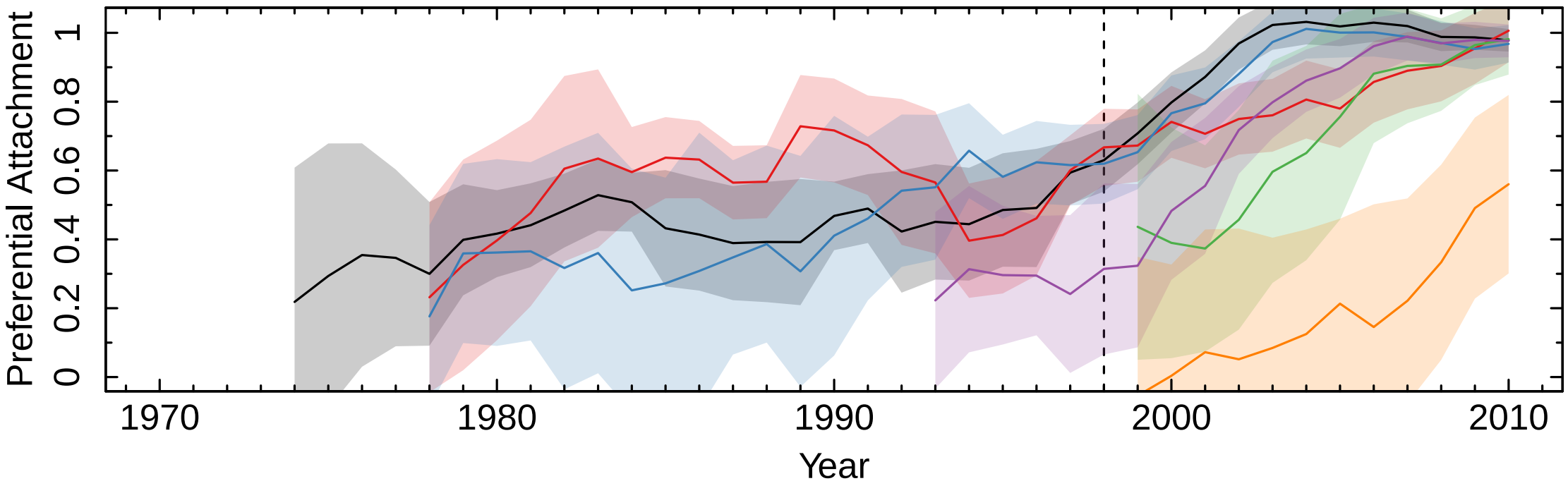

(b) Citation

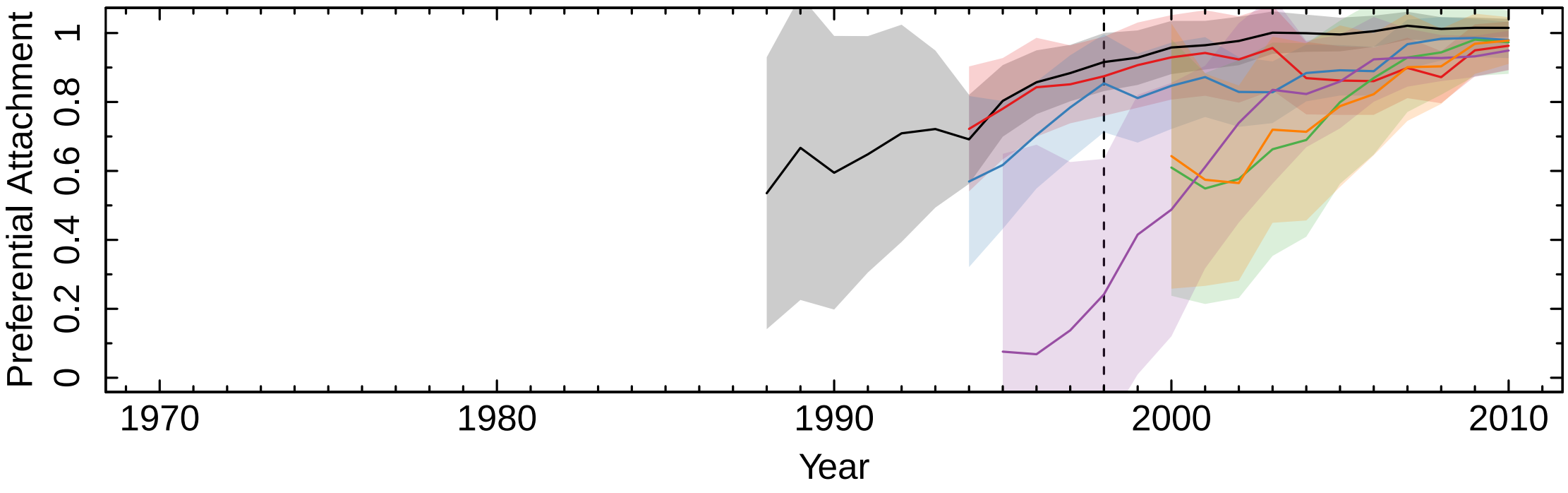

(c) Word usage

**Figure 3.21.: Emergence of linear preferential attachment**

Preferential attachment exponents are computed using fractional selection counting and three years for prediction. Error bars correspond to 95% confidence intervals. (a) Authorship is governed by sublinear preferential attachment but statistics are only available for two subdomains. Note that authorship is the selection of authors by agentic publications, not the selection of co-authors by authors. For citation (b) and word usage (c), preferential attachment converges to linearity following the Complexity Turn. Curves are smoothed using a 5-year moving average.

sensibility became commonsense because it was more visible. However, in 1993, the year when Structuralism took back the reins, Economic Sociology overtook Social Psychology. Economic Sociology permanently reached linear preferential attachment first in 2003 whereas Social Psychology was last in 2007. Social Network Analysis was second (2005) and Complexity Science third (2006). Like small-world emergence, criticality is a very recent phenomenon. By 2012, Web Science had not caught up which underlines its status as the new kid in the domain. The whole domain was globally correlated by 2002.

Differences in word usage are smaller but signals are also noisier. Because we have only title words until 1990, the signal can only start very late. Nevertheless, the fact that subdomain curves only begin in the mid 90s is evidence that the subdomains actually had not settled to a focused story set before. Again, Social Psychology and Economic Sociology were the first that reproduced a coherent narrative. Social Psychology was at criticality from 1999–2003 and is since 2009, Social Network Analysis since 2005, Economic Sociology and Web Science since 2007. Lexically, the latter is not newer than Social Network Analysis or Complexity Science. The subdomain that has brought the Complexity Turn has only occasionally fully reached linearity but is at over 95% since 2002. Linear preferential attachment, we conclude, is a universal property of the knowledge production system under observation.

The important message from these numbers is that, if the strength of the style gives the Kuhnian normality of science, then Social Network Science as a whole was *emerging science* before 2002 and normal science only starting that year. Requiring normal science to be critical, we have to withdraw our interpretation in section 3.3.1 that the Harvard renaissance from the mid 70s to the Complexity Turn is normal science. In the light of the Matthew Effect, Structuralism as represented by Economic Sociology was normal science to 62% by 1998. This is consequential for the remaining analyses. If Social Network Science is normal science only since 2002, then the Complexity Turn cannot have been a scientific revolution, and the styles that supposedly mate to change have not been fully developed until very recently.

### Fitness

While the Matthew exponent is a measure for the strength of a 5-identity which is influenced by a multitude of facts, fitness is a property of the individual facts, namely their influence on the style. The number of selections a social fact is probably going to have can be predicted because interface discipline creates stability through reproduction. Fitness is the error in predicting selection. A social fact is fit (unfit) if it is reproduced more (less) than an average fact with the same number of selections $\Delta t$ years before $t$. A fact's fitness is always relative to that of other facts. If one fact's fitness decreases ceteris paribus, another's fitness necessarily increases. Because fitness scores have finite mean and variance, we can average them. Figures E.1 and E.2 show the dynamics of individual facts and sets of facts in all of Social Network Science and its subdomains. For limits of interpretation and a detailed guide how to read these plots, please read the explanatory text in appendix E.

In general, the fitness of references is much more volatile than of words. This corrob-orates our earlier finding from the visual inspection of the corresponding static networks and the unconditional lifetime distributions that there is a broad topical consensus onto which perspective is modulated through the citation practice. Word usage is more inert than citation. We will, therefore, only discuss word fitness when signals are sufficiently distinguishable. The most striking result is that the paradigm of Complexity Science, which contains four references older than 1998, rose to prominence two years before the Complexity Turn and displaced the core of Social Network Analysis as the most fit paradigm in Social Network Science by 2000 (figure E.1d). From the Harvard breakthrough onward, except for three years in the mid 80s when Social Support Studies represented the mainstream, Social Network Analysis had been unrivaled in its extraordinary global fitness. In 1996, at the height of its reign and shortly before the end of Structuralism, Social Network Analysis's paradigm was 2.3 times as influential as expected. While Social Network Analysis plunged below expectation in the years 2000–2002, an observation that is obscured by the smoothing in figure E.1d, Complexity Science climbed to a maximum post-turn fitness of 1.9 in 2000. Social Network Analysis had been blessed with influence but was abruptly outshined by the powerfully emerging Complex Network Analysis.

To avoid the interpretational difficulties associated with the Kuhnian system of concepts, we first ask if the Complexity Turn constitutes systemic invention before we draw a conclusion whether or not it was a scientific revolution. Invention is systemic if the organizational innovation that created the situation lastingly alters the subdomain styles that constitute the whole. Since Social Psychology had lost its status of representing the mainstream of SNS in the early 90s, it had been turning towards modeling (Complexity Science), methodological (Social Network Analysis) and computational approaches (Web Science) (figures E.1e–E.1g). In 1999, the Social Psychology style had a strength of 67%. That year, this partially emerged 5-identity was strongly punctuated by the Complexity Science co-citation core which peaked with an average fitness of 4.2, the highest score ever reached in Social Psychology.[14] This organizational innovation turned into weak organizational invention as the Complexity Science core found footing in Social Psychology to some degree: from 2001 to 2004, it was the fittest core with a maximum of 1.4 in 2003.

Like Social Psychology, Economic Sociology was a similarly developed style in 1999 (65%) and was most strongly punctuated by Complexity Science that year (3.1). But differently, the Complexity Science paradigm remained the fittest core for a whole decade after the Complexity Turn. In 2000, Complexity Science core references were cited 1.8 times as much as expected (figure E.1j). Though organizational innovation was weaker, organizational invention was stronger than in Social Psychology. For Social Network Analysis, fitness curves are only available post-turn because this subdomain had only developed a style in 1997 which, however, dropped to 37% by 2001. From the mid 90s to the mid 00s, the subdomain was heavily and dominantly influenced by Complexity Science. In 2002, when knowledge spillover was largest, the import from Complexity Science was 3.8 times larger than expected (figure E.1m). The influence of Social Network

[14]This and other such peaks cannot be seen in the smoothed curves of figure E.1.

Analysis on Web Science has always been strong, both in terms of references (figure E.1s) and topics (figure E.2s). This corresponds to our earlier observation that the revisiting of its structuralist roots by Complex Network Analysis coincided with the convergence of Complexity Science and Web Science. Complementarily, again following organizational innovation in 1999 (3.1), the importance of Complexity Science for Web Science increased and reached an average fitness of 1.4 in 2010. Complexity Science itself was punctuated (3.2) at a time (1999) when it was abandoning Social Psychology and Economic Sociology and turning inward. Its own core reached a fitness maximum in 2001 (2.0) and stayed most influential until 2006, when Web Science took over (figure E.1p).

The meso level of the whole domain (figure E.1c) shows that the average fitness of Complexity Science references follows a circadian pattern with a wavelength of twelve years. During the Complexity Turn, the subdomain was recovering from an average fitness below expectation to where it had dropped by the mid 90s. This signal is mirrored for Complexity Science itself (figure E.1o), i.e., the breakthrough papers `WATTS_1998_N_440` and `BARABASI_1999_S_509` fell on fertile ground. In fact, these works caused the Complexity Turn. Of the Complexity Science core references shown in figure 3.9f, only Erdős & Rényi's "On random graphs I" (`ERDOS_1959_P_290`) and Milgram's popular science article on the small-world phenomenon (`MILGRAM_1967_P_60`) were published before 1998. But those works did not exceed an initial attractiveness until 2001 and 2005, respectively. Similarly, `SMALL_WORLD` did not show up again on the fitness landscape until 2002. This confirms the observation of a fractal distinction by Lazer, Mergel, and Friedman (2009) that the Complexity Turn revived research on the small-world problem.

Social Network Science as a whole underwent a major phase transition as it tipped from most strongly being attracted by the Social Network Analysis core to the Complexity Science core. Regarding its parts, to the extent that styles existed that could change, the Complexity Turn amounts to systemic invention if we exclude Social Psychology from the system. For Social Network Analysis and the three subdomains (Economic Sociology, Complexity Science, and Web Science) with which it is structurally most equivalent in terms of author overlap (figure 3.10), the Complexity Turn was a marked event. Economic Sociology lastingly altered its style as the Complexity Science paradigm became more influential year by year. Social Network Analysis had not been a style before the Complexity Turn but was strongly and lastingly given direction by the Complexity Science co-citation core. Web Science also emerged after the turn and increasingly leaned towards Complexity Science as it matured. The fact that Social Psychology was only weakly impacted by the Complexity Turn corresponds to its relative isolation in social structure. Also recall that Social Psychology has relatively weak personnel overlap with Social Network Analysis, Complexity Science, and Web Science. This suggests that structural holes are mirrored by "cultural holes" (Pachucki & Breiger, 2010). This, however, points at the importance of homophily as the additional factor how identities select social facts. Since paradigms can only shift where normal sciences exist, the Complexity Turn is not a scientific revolution, but a breakthrough. The two papers by Watts and Strogatz (1998) and Barabási and Albert (1999) shaped the emergence of Social Network Science as a normal science from 2002 onwards.

The Complexity Turn is the largest event during which styles mated to change, and the reconfiguration of cores can be observed all throughout history. Before 1996, Social Psychology mostly focused on its own paradigm but also tinkered with Web Science and Social Network Analysis. After the turn, it oscillated between Social Network Analysis and Complexity Science (figure E.1g). Economic Sociology has had varying sympathy for Web Science, Social Network Analysis, Complexity Science, and itself (figure E.1j). After Social Network Analysis and Complexity Science had been given direction by the Complexity Science core, Social Network Analysis and Web Science, respectively, became most influential. The Social Psychology core and also its meso and macro level references, with only few exceptions, have been continuously loosing influence. Since the turn, citation rates are at or below expectation, i.e., Social Psychology is loosing citation shares. The same can be observed for micro level word usage in the whole domain (figure E.2d). It seems that Social Psychology does not have much to offer to the other subdomains. A look at the fitness of individual facts in figures E.1a and E.2a shows that most paradigmatic facts only selected by Social Psychology (red curves) have effectively dropped below expectation, i.e., they are selected less and less.

The story of Social Network Science is a story of styles that are continuously transposed to other subdomains where they recontextualize the local meaning structure. Identities as sensibilities make such distinctions at all levels or length scales. In other words, styles make fractal distinctions. Through simultaneous reproduction and change, the parts of the whole mutually keep an identity from overspecializing or locking into uncreative paths. The only observation that suggests a division of labor is the socio-cultural hole between Social Psychology on the one hand and the other subdomains on the other hand.

### Summary

Social Network Science and its subdomains have self-organized to, and self-organize at, the percolation threshold between stability and change. Bradford's and Zipf's Law emerge from universal linear preferential attachment, our operationalization of Kuhnian normal science. The Matthew exponent is identified as a measure for the strength of a style. As identities reach a unit exponent, they have fully developed a Markovian sensibility for selecting social facts from a distribution of selections in the immediate past. Styles do not slavishly reproduce this predictable order but continuously reconfigure their cores by mating with other styles.

Social Network Science was not normal science until 2002, i.e., during the Harvard renaissance, it was still emerging. Consequently, the Complexity Turn initiated by the papers `WATTS_1998_N_440` and `BARABASI_1999_S_509`, though it changed the face of Social Network Science, does not constitute a scientific revolution, but a breakthrough. For Social Network Analysis, it ended a period of two decades during which its story set represented by a co-citation core was exceptionally fit and almost unrivaled. After the turn, Social Network Analysis embraced the new paradigm more than the other subdomains. For Economic Sociology, Complexity Science, and Web Science, all committed to structuralist methodology and strongly overlapping with Social Network Analysis in terms of authors, the turn meant either strong organizational invention or a boost to the emergence of style. The event fell on fertile ground because large parts of Social Network

Science had already turned towards knowledge from Complexity Science, it gave direction to further emergence, and it revived small-world research in another example of a fractal distinction in time. Only Social Psychology, the subdomain separated from Complexity Science through a structural hole, was not punctuated strongly by the Complexity Turn, which points at a complementary cultural hole and a partial division of labor.

# Discussion

There is no exact mapping of author co-authorship, reference co-citation, and word co-usage networks. Lifetime distributions show that the underlying practices crystallize at different time scales or with different speeds. Change in word usage is slow, language provides the story sets or topics that scientists work on. Change in citations is faster, through them cultural cores are renewed. Change in authorship is fastest, teams are quickly assembled and dissolved. "[I]nnovation in language usually lags, not leads, major organizational inventions" (Padgett, 2012a, p. 60). Accordingly, topical networks are much more coherent than collaboration structures. The word co-usage network basically has a multifunctional core and a partitioned periphery, the author co-authorship network is largely decentralized, and the reference co-citation network is between the extremes.

Persons are not the atoms of sociology, constant essentials from which higher levels of complexity are built by aggregation. Like groups, organizations, and objects observable at higher or lower levels of complexity, they are scale-free identities, variable dualities of transactions and stories, outcomes of social and cultural structures that co-constitute each other at all length scales of society. From the fruitful contact of diverse theories and a case study, the following model of identity as a sensibility of selection has emerged. An identity consists of parts in transaction (1-identities), and when it has found footing it exists as a whole (2-identity). Control is a collective and autocatalytic phenomenon that aims at constraining possible events. Collective control is the self-organization of 1-identities into blocks of structural equivalence, i.e., the parts of a whole share the same story set. Autocatalysis means that, from selections of social facts by 1-identities in transaction, a meaning structure of facts in co-selection emerges, while the emergent meaning structure influences which facts are selected by the 1-identities. The meaning structure provides a story set from which the content of transactions is drawn.

Since autocatalysis can only unfold in time, an identity is necessarily a process. Through the autocatalytic reproduction of patterns of selection, an identity becomes a path through a meaning structure in space and time (3-identity). The less an identity's parts are connected through constitutive ties, the more it is latent. But the higher the density of such ties, the more it is conscious about itself (4-identity). Realization of self usually occurs when a 3-identity as a flow of reproduction is interrupted by a punctuating event, like the emergence of a new fact. As a sensibility or style (5-identity), an identity ensures that, in the face of punctuations, an optimum of stability and change is maintained. It does so through a self-similar selection profile. Highly selected facts form the core of the meaning structure. Essentially, an identity self-organizes to the critical point because, there, conditions for adaptive reproduction are optimal in the evolutionary game of socio-cultural life.

The mechanism how identities attain the edge of chaos is the autocatalytic Matthew

Effect (Merton, 1973 [1968]). The Barabási/Albert model of preferential attachment is a simple rich-get-richer version of the mechanism. In the resulting equilibrium, a 3-identity predictably reproduces its meaning structure. Generalized preferential attachment (Bianconi & Barabási, 2001) is a fit-get-richer mechanism. We have made the mechanism more realistic by letting fitness emerge dynamically from collective selections. By allowing for variation of fitness, generalized preferential attachment becomes a far-from-equilibrium mechanism that can explain the reaction of autocatalytic flows of reproduction to punctuations. Those can be endogenous or exogenous. The emergence of a fit fact can cause previously fit facts to lose their fitness over time, no matter how highly selected they have been.

The Watts/Strogatz model is intimately connected to the Matthew Effect through search and homophily. Search corresponds to the random rewiring of the grid in the small-world model. Similarity breeds connectivity, i.e., 1-identities seek transactions with structurally equivalent 1-identities. Homophily formats social networks and ensures that culturally compatible parts are always proximate. As meaning structures they are memories of transactions among culturally compatible parts. 1-identities search facts in Lévy flights across transaction landscapes, i.e., the probability to find a fact at a given socio-cultural distance decays as a power law (Kleinberg, 2000; D. R. White et al., 2006). As styles, identities self-organize to criticality through regulating the search for facts. When cultural uncertainty of context is high, the core that gives autocatalysis direction is unstable. To counter the risk of drifting into chaos, styles reduce social uncertainty by decreasing searches at great distance. On the other hand, when cultural uncertainty is low, styles counter the risk of freezing in by increasing searches at great distance. Styles get to the phase transition of order and disorder through a biased random process, the Matthew Effect, because this is where they find the optimal mixture of autocatalysis and change that is required to stay in control.

In table 3.9, we sketch the identity model in pseudocode. It requires implementation in an agent-based simulation (Miller & Page, 2007) which is, however, beyond what we can do in our work here. The model generates two networks simultaneously: a social structure of agents – which take the role of 1-identities – in transaction and a meaning structure of social facts in co-selection. We expect the model to exhibit the following properties. First, practically starting from a random network, the social structure will transform into a small-world architecture. Because cultural similarity is transformed into structural proximity at the end of each time step, stratified Lévy-flight search (line 10 in pseudocode) mostly finds facts at close cultural distance. It will be interesting to see if clustering emerges all by itself or needs to be introduced by a triadic-closure rule. Because the Lévy mechanism occasionally draws distant actors, the short average paths of the random graph will be maintained. Second, due to generalized linear preferential attachment (line 13), a self-similar distribution of fact size will emerge. The identity develops a style of selection. When the fact co-selection network is constructed, the set of network hubs will form the core. Again, it will be interesting to what extent co-selection emerges.

Third, despite a style, autocatalytic equilibria will be punctuated. With greater-than-exponential probability, agents will create new facts with random fitness (lines 10–12),

**Table 3.9.: Pseudocode for the identity model**

In an identity or complex socio-cultural system, agents select social facts. Agents $a$ self-organize in social structures like collaboration networks. Facts $f$ self-organize in meaning structures. (1) Initially, agents are connected to their Moore neighborhood, i.e., they are arranged on a grid. (2) Each agent randomly selects $n$ facts $f_j$ with random fitness $\eta_j$ and size $k_j = 1$. $n$ can be thought of as the number of references or words cited or used by an agent. (3) Each time step $t$, agents' facts are ranked and a number of events takes place. (4–5) Facts are ranked such that those that had reached a high rank in the past and have dropped below that rank cannot reach a higher rank anymore. In other words, two sequences of increasing ranks, interrupted by a decrease, are disallowed. High ranks correspond to commonsense facts and low ranks provide texture to the high ranks. (6–13) In an event $e$, a randomly drawn active agent selects $n$ passive agents from which he samples facts for potential selection. Facts are sampled through stratified Lévy flights across the social structure, i.e., $n$ samples are drawn for rank 1 down to $n$ samples for rank $n$. Per rank, a fact is selected through generalized preferential attachment. Per rank, passive agents are drawn mostly at close distance, occasionally at large distance. Because $d$ can be 0, agents can reproduce their past selections. Novelty in the form of a new fact is created when no agent at distance $d$ exists. (14–17) After each time step, each fact's fitness and size is updated. Social networks are updated by connecting agents to the eight structurally most equivalent (in terms of selections) agents, i.e., structural equivalence is memorized through homophily. If less than eight agents have nonzero similarity, connections are completed randomly.

| | |
|---|---|
| 1 | initiate grid with moore neighborhood |
| 2 | for agent $a$ in 1 to $a_{\text{max}}$: select $n$ facts $f_j$ with $\eta_j = \text{rnd}$ and $k_j = 1$ |
| 3 | for time $t$ in 1 to $t_{\text{max}}$: |
| 4 |   for agent $a$ in 1 to $a_{\text{max}}$: |
| 5 |     order facts $f_j$ by $\eta_j k_j$ such that rank $\rho_j$, following an increase, only decr. |
| 6 |   for event $e$ in 1 to $e_{\text{max}}$: |
| 7 |     randomly draw an active agent |
| 8 |     for rank $\rho$ in 1 to $n$: |
| 9 |       repeat $n$ times: |
| 10 |         draw a passive agent $a$ at distance $d \in \mathbb{N}^0$ with prob. $(d+1)^{-\beta}$ |
| 11 |           if agent exists: memorize fact $f_j$ |
| 12 |           else: create and memorize new fact with $\eta_j = \text{rnd}$ and $k_j = 1$ |
| 13 |       from $n$ facts in memory select one $f_j$ with probability $\eta_j k_j / \sum_r \eta_r k_r$ |
| 14 |   update each fact's fitness $\eta_j$ and size $k_j$ |
| 15 |   if at least eight agents with nonzero similarity exist: |
| 16 |     connect to eight most similar ones |
| 17 |   else: complete connections by randomly drawn agents |

i.e., occasionally, the system will be perturbed by a social fact that has the right quality at the right time to become paradigmatic. Styles sense qualities, readable as high fitness, and the Matthew Effect propels fit facts into higher levels of selection exponentially with time: a fact that is twice as often selected as expected for three time steps in a row will be selected $2^3 = 8$ times more than at the beginning. At higher levels, facts increasingly attain the status of core facts. But the same mechanism that propels a fact into the core also works the other way. A fact that is half as often selected as expected for three time steps in a row will plunge into nothingness as quickly as it rose to stardom (Sterman & Wittenberg, 1999). Such exponential dynamics are characteristic of the rapid phase transitions during which new meaning emerges or old meaning is forgotten. These feedback loops are supposed to be enforced by preventing facts from rising to old fame they had once lost (line 5, cf. Bornholdt et al., 2011).

### The New Story of Social Network Science

The identity model was evaluated in a case study of Social Network Science. In that course, we have added, confirmed, and revised empirical findings about this research domain. Throughout the study, we have treated as social facts or identities the whole research domain, its five subdomains, five research paths, research fronts, and individual facts like authors, cited references, and words. Great benefit was reaped from Abbott's (2001a) heuristic of fractal distinctions, one of sociology's most developed treatments of scale invariance. Around this idea, three specific research questions have been posed that formulate expectations used for calibrating our model.

Overall, we have found that Social Network Science and its parts are not stable essentials but emergent formations that position themselves between order and disorder where their normal flow of reproduction is occasionally subject to large change. In the 20th century, Social Network Science is basically the history of Social Psychology and Economic Sociology. In the mid 50s, Social Support Studies with roots in anthropological network analyses were born within Social Psychology as a branch of Social Network Science parallel to the sociological mainstream of Structuralism. In the 70s, Social Network Science witnessed the Harvard breakthrough, brought forth by Economic Sociology which strongly represented Structuralism. For a quarter of a century, the blockmodels of H. C. White et al. (1976) had been the starting point for modeling communities, the most central topic throughout Social Network Science. Despite this innovation, Social Psychology had become a stronger citation style than Economic Sociology by the early 80s, and Social Network Science as a citation path or 3-identity switched from Structuralism to the more focused Social Support Studies. Social Psychology is the only subdomain whose paradigm consists of two co-citation cores concerned with social support. Until the early 90s, Social Support Studies rediscovered its structuralist roots, measurable because the newer core which overlaps with that of Economic Sociology became increasingly attractive. In 1993, when the paradigm of Economic Sociology had become fitter than Social Psychology's, the paths of Structuralism and Social Support Studies converged. This resembles the first major fractal distinction because Social Psychology returned to the structuralist roots it had once abandoned after its birth. This account slightly revises, and adds detail to, the review of three intellectual lineages by Scott (2012).

Starting in 1998, Social Network Science was most strongly punctuated by the publications of Watts and Strogatz (1998), Barabási and Albert (1999), and following publications belonging to Complexity Science. During the Complexity Turn, Structuralism as a 3-identity was displaced by Complex Network Analysis. In this event, which we have called a Bayesian fork because it created a singularity of path that cast the future of Social Network Science wide open, much knowledge got lost by the mainstream. In retrospective, the reference core of the domain mainstream was completely renewed by mostly physics references. The former breadth of community detection methods, from various role models to cohesion models, gave way to modularity maximization methods, as observed by Freeman (2011). The main pre/post-turn continuity is the small-world problem which was significantly revived following Watts and Strogatz (cf. Garfield, 2004; Lazer, Mergel, & Friedman, 2009).

The Complexity Turn was not a Kuhnian scientific revolution because Social Network Science had not been normal science before. In our definition, science is normal when the Matthew Effect is fully autocatalytic or when preferential attachment is linear. In 1998, the Matthew exponent was slightly larger than 0.6, i.e., the domain was not at criticality where normal science resides. The Complexity Turn occurred while the domain was still emerging science. Four of its five subdomains attained criticality between 2003 and 2007 while Web Science, the youngest subdomain, was still getting there in 2012. This is a strict definition of normal science, but it is one that is amenable to computation. In contrast, Hummon and Carley (1993) had subjectively concluded that Structuralism was normal science because it incrementally reproduced along main paths with "a single identifiable coherent substantive concern". Our interpretation is supported by the observation that, until the Complexity Turn, the emergence of a small-world collaboration network was solely due to the densification dynamics of Social Psychology. All other subdomains only helped building the small world after 1998.

The Complexity Turn was a scientific breakthrough that indeed created a "new" science of social networks – not because of the "excitement" and the "unprecedented degree of synthesis" described by Watts in the introduction, but because an innovation originating in Complexity Science spilled over into three other subdomains and altered the local styles. The way Social Network Science changed resembles the emergence of partnership systems in Renaissance Florence and of Dedicated Biotechnology Firms in the 90s (cf. Padgett & Powell, 2012a). The Watts/Strogatz and Barabási/Albert models were made possible by an innovation in the technology realm. Bornholdt and Schuster wrote in 2003: "Triggered by recently available data on large real world networks ... combined with fast computer power on the scientist's desktop, an avalanche of quantitative research on network structure and dynamics currently stimulates diverse scientific fields" (p. V). The papers on small-world and scale-free networks are among the first that rely on such data (like the International Movie Database) and computational methods. The opportunities were most quickly seized by physicists and, according to Watts (2003), "no one descends with such fury and in so great a number as a pack of hungry physicists, adrenalized by the scent of a new problem" (p. 62).

The post-turn sentiment of being "invaded" likely diffused especially in the subdomain of Social Network Analysis because it had been the methodological powerhouse of Structuralism. Since the Harvard breakthrough, the Social Network Analysis paradigm had most strongly given direction to Social Network Science. But after the turn, not only did the physicists not cite this work, the Social Network Analysis paradigm was only second or third to the Complexity Science core in most subdomains. Social Network Analysis itself totally embraced the complexity paradigm. Economic Sociology was lastingly and Web Science lately influenced. Following the transposition of a technological innovation into the domain, multiple subdomains organizationally invented their styles. In fact, the complex network innovation fell on fertile ground as Social Network Science had been turning towards references of Complexity Science since the mid 90s. Until 2012, the transformation of the domain was no systemic invention for the only reason that Social Psychology was hardly impacted by the new paradigm. Social Psychology's own paradigm established as the least fit, and the subdomain significantly lost publication shares. Sentiment was voiced in Social Network Analysis because their reign had ended and contribution was not acknowledged.

Through systematic analyses of fitness, we have shown that the turning of a style towards the core of another style in order to get fresh action did not only occur after the Complexity Turn but all throughout the Social Network Science narrative. The mating of styles in order to change is the rule, not an exception. The merger of Structuralism and Social Support Studies was the first major example identified through community detection in the direct citation network. Similarly, Complex Network Analysis changed in a fractal distinction when, in 2010, it took up the concerns of Structuralism and partially changed its face or 2-identity.

Fractal distinctions are the engine of history and find a counterpart in power-law size distributions that quantify the stochastic regularity of an identity at the percolation transition. Citation and language styles result in Bradford's and Zipf's Law, signatures of fractal order in space. Highly selected social facts have long lifetimes, i.e., few stories like `COMMUNITY`, `SOCIAL_SUPPORT`, `SOCIAL_CAPITAL`, `SOCIAL_NETWORK_ANALYSIS`, `COMPLEX_NETWORK`, or `USER` offer long-term stability. Facts in the lower tail offer short-term stability and change. Only in the rare cases that they displace paradigms from their core position will change be large. One must not think of these power laws as causal laws such that b follows from a, no matter what. Sociological laws are not like gravity. When an identity is in the domain of attraction of a core in a meaning structure, then it does not inescapably select facts of that attractor. The laws of sociology are like phase transitions in water. When the temperature decreases, the core will freeze in. Sociological laws are observational laws of patterns and mechanisms. Their existence is empirically demonstrated through universality. Linear preferential attachment is universal because all subdomains are governed by it.

Universality is not necessarily found in the form of power-law exponents. The memory for authors, references, and words that is encoded in the transaction network of 1-identities is universal. Web Science, e.g., is a much younger subdomain than Social Psychology and, consequently, facts that influence the latter can have much larger lifetimes. But when their lifetime distributions are rescaled, they collapse onto a universal curve characteristic for all identities in the domain. The underlying law states that all subdomains in the universality class called Social Network Science emerge in self-similar ways through the same mechanism. As long as we are only interested in a stochastic quantity, not in qualities like content, we can study one subdomain and still learn about all of them.

Four metrics in concert allow classifying the collaboration structure of subdomains: the rate of the average number of authors per paper and the scaling exponents for densification, macro autocatalysis, and hierarchical modularity. Social Psychology, Social Network Analysis, Complexity Science, and Web Science are Priceian big sciences or team sciences, Economic Sociology is little science. This is an important result because it means that the heuristic of fractal distinctions, originally developed for the social sciences, also works for a mix of little and big science. All subdomains build small-world networks like Hora in Simon's (1962) parable builds watches: authors cluster in overlapping groups which cluster in overlapping groups and so on, i.e., collective control evolves through stable intermediate forms. The co-authorship networks of team sciences are even universal with a unit exponent.

While, for a start, big science is just another control strategy than small science, one that is associated with the natural sciences, there is also a structural advantage in big science. First, big science outreproduces little science relationally. The most cohesive research fronts in Social Network Science practically belong to the biggest sciences Complexity Science and Social Psychology. Second, the strong embeddedness of big science translates into social influence. The two subdomains are structurally more efficient in coordinating the whole domain through the invisible college. Third, the bigger a science, the more fractional authors there will be. Teams can tap multiple knowledge pools, and multifunctional authorship is higher cited on average. In a fixed time window, big science can generate greater complexity and, therefore, evolvability. Its rhythm is faster.

Whether or not this advantage plays out depends on degrees of homophily. Despite the ubiquity of fractal distinctions and the emergence of a hierarchically modular small-world network, authors in Social Network Science are coordinated by an open elite only to a minor extent. Instead of a permeable boundary that facilitates the open exchange of ideas across all subdomains, a structural hole walls off Social Psychology from a multifunctional cluster that is, in turn, strongly dominated by Complexity Science. Social psychologists also appear in packs, but only the physicists occupy the main core of the domain's collaboration landscape. The structural hole is mirrored by the cultural hole that decouples the larger and older of Social Psychology's two reference cores from the whole's paradigm. The subdomain does not fit into the otherwise coherent image of emergence through fractal distinctions. The observations are diagnostic of a Durkheimian division of labor in the domain. This makes sense because, unlike any other subdomain, Social Psychology is largely concerned with public health issues. Research fronts like those of `FRIEDMAN,_SAMUEL_R` and `LATKIN,_CARL` on the epidemiology of infectious diseases differ from all others in that they are invasive, i.e., they aim at improving the state of the observables.

The structure and dynamics of Social Network Science can be modeled on socio-cultural temperature as a trade-off among ambage (social uncertainty), ambiguity (cultural uncertainty), and contingency. Styles strive to maintain the critical temperature, i.e., they try to avoid strong/static cores that lock them into a frozen path and weak/unsettled cores that offer no escape from chaos. Anecdotal evidence is sufficient for three subdomains (cf. figure 1.7). First, following the Complexity Turn, the complexity paradigm was very attractive, Complexity Science significantly increased its publication share, and quickly percolated into a small-world collaboration network. The combination of low ambage and low ambiguity was not sustainable. Lock-in was avoided by two events that increased the two types of uncertainty. In 2006, Complex Network Analysis specialized by splitting off a path on Evolutionary Game Theory, and in 2010, it innovated by turning towards the topics of Web Science in the course of which Structuralism was revisited.

Second, as the only small science, Economic Sociology counters the high ambage of decentralized but strongly reproducing groups through the low ambiguity of a very coherent cultural core that consists of six references in complete co-citation. In addition, the sociologists lastingly renewed their style after the turn. Third, the interest of the domain for the story of Social Psychology was at an all-time low by 2012. The last time it managed to attract selections above expectation was in the mid 90s. Social Psychology is the only subdomain that was hardly affected by the Complexity Turn. It is the part of Social Network Science that prevents this turning point from being a systemic invention. The style hesitates to fundamentally change by mating with others. This low ambiguity of sitting in a contingent niche combined with low ambage in terms of strong densification is a call for adaption to avoid lock-in and a further loss of reproductive fitness.

Finally, like in any complex narrative, single and initially small social facts can have large effects as they attract more and more selections. Social Network Science can be seen as a story of two men with completely different styles. The first man is Harrison C. White, the leading figure of the Harvard renaissance. He is like Otto von Bismarck, the first chancellor of Germany. Bismarck formed Germany through a combination of two mechanisms. First, by dually including the contradictory principles of autocracy and democracy, he accelerated the emergence of new political actors. Second, by attacking Austria, he split third parties and created ties to some of the resulting parts (Obert & Padgett, 2012). Similarly, White formed Social Network Science by mating sociology and physics. Blockmodeling is inspired by the renormalization procedure of statistical mechanics and most strongly gave direction to the domain until the Complexity Turn. By attacking his enemy Parsons, he cast doubt in other sociologists' heads and pulled some of them over into the relational camp. None of this influence can be seen by looking at the collaboration network White has embedded into. Being a sociological style, his network is sparse.

The other man is Albert-Lászlό Barabási. Unlike Watts who turned into a sociologist or Newman who visited the social science literature, Barabási has been forging a new science early on. He is like Cosimo de Medici, the ruler of Renaissance Florence, who strategically arranged marriages and market ties to consolidate the organization that he was the patron of (Padgett & Ansell, 1993). We do not want to say that Barabási is interested in personal power or chooses collaborations other than scientifically, but, like Medici, he is the ultimate broker in the organization he has built.[15]

[15] I am grateful to John Padgett who pointed out these analogies.

## Methodological Issues and Future Work

Our work rests on three rules of method. The first rule is that, in order to explain stability and change, we need to model identities using a relational, not reductionist, ontology. Identity and selection were operationalized through a bipartite graph of 1-identities, social facts, and selections as second-level observations (Batagelj & Cerinšek, 2013). Meaning structures were modeled as autocatalytic networks where arcs are catalyses. Bibliographic data from the *Web of Science* favors journal science, i.e., our results must be taken with a grain of salt. But it proved to be extremely useful to study the more social dimension of identities through the practice of authorship, the more cultural dimension through citation and word usage, and interface discipline through direct citation networks.

We hope that our data model can be a general framework for other socio-cultural phenomena and contexts as well. Since identities derive meaning from being distinct, modeling multiple network domains and their overlap is of utmost importance. For an identity representing a domain, we have developed a graph theoretical method that detects subdomains or story sets in networks of structurally equivalent 1-identities or publications. Then, meaning structures are link communities of social facts with potentially as many types of tie as there are story sets. Delineating subdomains through citations and word usage allowed us to study the emergence of collective consciousness or memory – the transformation of 0-blocks into 1-blocks – in co-authorship networks.

Using publications as agentic 1-identities seems counterintuitive for authorship because publications 'select' authors, but it allows modeling all practices alike. However, the unified approach has one trying to explain Lotka's Law by linear preferential attachment of papers to authors, which is actually unrealistic. Authors are not highly productive today because they have written many publications yesterday. The literature does report preferential attachment for the growth of co-authorship networks, though. Were we to repeat our analysis, we would, therefore, treat the authorship practice differently and use authors as both agents and facts.

Cohesion models are the clustering method of choice to study fact networks, like we did for co-authorship. In future work, we want to extend the analysis of structural cohesion to reference co-citation and word co-usage networks which, however, are very dense and require edge filtering. An algorithm for weighted networks may be a useful innovation. Fractional selection counting ensures that the different types of tie are actually comparable (Leydesdorff & Opthof, 2010). Information science was also very profound in dealing with the boundary problem. Bibliometrically enhanced information retrieval is deeply sociological by taking advantage of the selection mechanism. A set of publications is delineated by identifying the core facts that these publications select (Zitt, 2015). We have improved this procedure by applying it on the level of subdomains.

The second rule of method is to take a narrative approach "based on stories." Stories are the symbolic communications through which a complexity emerges that is not present in non-socio-cultural systems. Events are unperturbed contingencies of meaning with some duration. Narratives relate stories and events in time. These are working definitions that will benefit from theoretical and empirical polishing. Interface networks in the form of direct citations are historical reconstructions of chains of events. Search path counting (Batagelj, 2003) in combination with betweenness-based community detection (Girvan & Newman, 2002) is capable of drawing an improved picture of fractal stability and change in science. Applied to a long-established research domain or discipline, further insights about *The Structure of Scientific Revolutions* (Kuhn, 2012 [1962]) should be possible. In such a study it would be desirable to not filter arcs as we did, because that restricted our analysis to the mainstream of the research path.

In non-linear modeling, causality has meaning as part of a mechanism like autocatalysis or the Matthew Effect (Hedström & Bearman, 2009b). In linear, variable-centered approaches to causality, ordinary-least-squares fitting is routinely used which assumes that independent variable can be predicted by the independent variable(s). We have benefited from bivariate line-fitting methods developed in biology, particularly standardized-major-axis fitting, which does not assume predictability (Warton, Wright, Falster, & Westoby, 2006). We did use prediction (ordinary-least-squares regression) for fitting the Matthew exponent because the Matthew Effect is an univariate feedback process. But in none of the bivariate scaling analyses did we find it plausible that one variable is the cause and the other the effect. For example, in hierarchical modularity, why should a high degree explain a low clustering coefficient or vice versa? The two author properties co-evolve. Had we used the standard method of linear modeling we often would have gotten markedly different results.

For bivariate analyses, we used the coefficient of determination as a goodness-of-fit measure (Warton, Duursma, Falster, & Taskinen, 2012) and often obtained $r^2 < 0.1$ for a scaling function that is obviously right. We did not present $p$-values because these would have only told us that we can still trust the fit. Significance tests are rather useless for analyses of unsampled process data with large $n$. Next steps should, therefore, involve the bootstrapping methods of univariate scaling analysis (Clauset et al., 2009).

The third rule of method to account for different levels of observation was central in research design and sharpened our analytical eye. When designing research, we decided to take, in several respects, a static third-level observer position at 2012, i.e., we have observed Social Network Science from the perspective of what it was that year and how it got there. Put differently, each event got its meaning from being embedded in the full history of the domain. The paradox that a subdomain is called Web Science but dates back to 1916 when the Web did not exist is due to this approach. 75% of all Social Network Science papers are from 2002 and later. We chose to detect subdomains through time-integrated structural equivalence because it delineates enduring story sets. In fact, the whole dataset is built that way.

Regarding citation flow, the static observer position is unproblematic because the only information that gets lost is how a reference had become what it was at $t_{\max}$. The research fronts identified in the integrated co-authorship network are also meaningful when they are described in their temporal characteristics ex post. Readers may object that we held constant a subdomain's paradigm because it kept us from measuring the fitness of that subdoman's paradigm at shifting presents. This is also something we would do differently, were we to repeat our analysis.

Different levels of observation have proven to be extremely valuable sharpening the identity model. The import from Luhmannian systems theory helped resolve a paradox of socio-cultural memory. On the one hand, an identity memorizes few social facts for a very long time while most are quickly forgotten, but on the other hand, the decision which facts to select is based on how many times they have been selected in a fixed Markovian memory window $\Delta t$ before the present. A selection is a second-level observation by a 1-identity, but reasoning about lifetimes is a third-level observation of Social Network Science. The latter is what we do as sociologists of Social Network Science. But identities, too, can observe themselves at a third level. When they do, they are 4-identities. Usually self-consciousness about lives and lifetimes emerges at Bayesian Forks. When the heat is turned on by a punctuation, identities become conscious of the meanings encoded in the ordering power laws and the opportunity arises to strike a new path.

## Conclusion

By studying the three scholarly practices of authorship, citation, and word usage in a research domain, we have demonstrated that identities are not dichotomies but dualities of social network and cultural domain, micro and macro phenomena, as well as stability and change. First, identities are, embed into, and co-constitute dualities of social networks (transaction structures) and cultural domains (meaning structures). The latter story sets fluctuate less, are less distinctive and more inert, than the individuals that work on these topics. Words have longer average lifetimes than authors, and word co-usage meaning structures are more centralized than co-authorship networks. Second, identities are scale-free. Not only are persons, groups, organizations, etc. manifestations of an idealized identity at different levels of socio-cultural complexity. Concrete identities also extend over multiple such levels.

To explain the stability and change of identities, we found useful tools in Relational Sociology. Five senses are diagnostic of different aspects of identity, and when they come together in socio-cultural process, a complex socio-cultural system comes into existence. Models of H. C. White (2008) and Padgett and Powell (2012a) were successfully combined with the scaling hypothesis of percolation theory according to which the convergence to linear preferential attachment indicates that stability and change co-exist at a fractal phase transition.

Fractal symbolic communication is *what* an identity does. Its building blocks are globally correlated through a story set. The Matthew Effect is *how* story sets become fractal. Fit social facts become richer. Optimization is *why* story sets become fractal. Identities that do not balance stability and change are, in the long run, outreproduced in socio-cultural life. Styles are *who* self-organize to become fractal network domains. Identities are stochastic flows through ragged fitness landscapes. It is in this percolation model that the importance of the works that initiated the Complexity Turn becomes apparent: Barabási and Albert (1999) provide a network model of the Matthew Effect and Watts and Strogatz (1998) model the search for social facts at the phase transition. In the resulting *historical and analytical sociology of complexity and process*, an old dispute between Simon and Mandelbrot is reconciled. The Matthew Effect is based on *both* biased randomness and optimization. It operates because it provides optimal conditions of stability and change, of reproductive normality and creative chaos.

This identity model was evaluated in a case study of Social Network Science. As expected from percolation theory, the fractal state can be described by a number of scaling laws. Self-organization to criticality is expressed through hierarchically modular small-world social structures and self-similar meaning structures and dynamics. The evolution of the domain is convergent, i.e., it does not progress in a division of labor but in distinctions that repeat in themselves. Social Psychology is an exception because its story set is too different. All other subdomains continuously change through mating with other styles. The Complexity Turn of 1998 was not a scientific revolution because Social Network Science was not normal science until 2002. It was a scientific breakthrough that caused all subdomains but Social Psychology to markedly innovate.

To the social scientists, the influx of physicists into the research domain had the connotation and – as we showed – the reality of an "invasion." Because natural science is team science and team science has an evolutionary advantage through greater embeddedness, the scientific mainstream moved into the physicists' domain. During this transition, much knowledge got lost, most notably the idea of structural equivalence. This is unfortunate because the homeostasis we have uncovered, the stability of a system's meaning when parts change, points at the importance of positions. But from the perspective of a third-level observer, occupants of positions in one story set cannot be expected to know and cite another story set. Subdomain paradigms are necessarily incommensurable to some extent (S. Fuchs, 2001, p. 55). In the case of the subdomain of Social Psychology that is concerned with actively improving the health of networked persons, a story set that has been stable for generations has kept many scientists from switching into the other subdomains, particularly into Complexity Science.

Despite differences, our analysis has uncovered universality in Social Network Science, i.e., different subdomains are generated through the exact same mechanism. The most notable universal property is a convergence to linear preferential attachment of citations and word usages. A second universality is the institutionalization of social facts. In addition, all team sciences progress through stable intermediate forms governed by discrete scale invariance. In time, these building blocks evolve along paths of fractal distinctions. Just like the Koch snowflake reveals the same geometry when we zoom into an edge, phases of reproduction separated by turning points repeat in themselves. The Complexity Turn seemingly divided Social Network Science into two enduring events or periods during each of which knowledge was produced in equilibrium. But each event is again split into shorter periods in stronger equilibrium. Science is only "nearly decomposable" to equilibria (cf. Simon, 1962).

Scaling laws confirm this observation in a statistical way and have prompted criticism of Kuhn's theory of change in science. The argument is that power laws show that there are no typical sizes of scientific revolutions or lifetimes of periods of normal science, just self-similar phenomena at all length scales. "Science is always revolutionary, but by the typical statistics of complex systems, there are mostly smaller and only rarely big breakthroughs" (Van Raan, 2000, p. 360).

There are two answers. One is by Kuhn (2012 [1962]) himself who wrote in the postscript of 1969:

> A revolution is for me a special sort of change involving a certain sort of reconstruction of group commitments. But it need not be a large change, nor need it seem revolutionary to those outside a single community, consisting perhaps of fewer than twenty-five people. It is just because this type of change, little recognized or discussed in the literature of the philosophy of science, occurs so regularly on this smaller scale that revolutionary, as against cumulative, change so badly needs to be understood. (pp. 180)

Kuhn actually wanted to raise awareness for changes of all sizes. Like Durkheim, Kuhn may have been misread. The second answer is already implied by Kuhn and is related to the fact that, though the phenomena are similar, meanings are not. Transitions of styles at macro levels of whole sciences have completely different meanings than when a research group or a specialty changes direction. Below some threshold, punctuations will not be scientific revolutions but minor recontextualizations of cores.

We have evaluated this model for identities in a particular realm of science. Originally, Abbott had introduced fractal distinctions as a heuristic for studying the social sciences, and Kuhn had modeled phases of the natural sciences. But discussing whether or not models developed for one or the other science can be applied to an interdisciplinary domain like Social Network Science would distract from our central result that small and big science did evolve in fractal distinctions almost up to a systemic invention. Fractal distinctions also govern mixtures of social and natural sciences. The point is that, when different sciences co-evolve, big science has an evolutionary advantage due to a greater speed to complexity.

Constructivism's strength is particularization, the description of unique identities and what makes them distinct. Realism's strength is constructivism's weakness, the ability to abstract from the particular and identify the universality among unique identities and to make them predictable. In a scientific revolution, the set of rules for compounding distinctions with one another changes. Kuhn losses like the extinction of blockmodeling in the new mainstream of Social Network Science can be explained when distinctions are affiliated. In Abbott's (2001a) words, edited to describe our situation:

> [If realists are always opposing constructivists, and realists always seek universalistic explanations and constructivists particularistic ones, then even the fact that realists are themselves internally split into realists and constructivists, as are the constructivists as well, and so on and on, even that fact will not allow us to explore very many of the possible knowledges of society.] (p. 30)

The emergentism we advocate is a style that mates the realist constructivism of Relational Sociology and the constructivist realism of Complexity Science. Constructivism and realism are complementary. The common philosophical ground is the relational ontology and the narrative approach that does not know neutral observers but identities particular in content and universal in pattern.

Shortly before the Complexity Turn, the Gulbenkian Commission on the Restructuring of the Social Sciences (Wallerstein, 1996) called for such a *pluralistic universalism*. The practical inseparability of research on society, polity, and economy should be done justice by overcoming anachronistic disciplinary boundaries that separated sociology, political science, and economics. The Complexity Turn is something that the commission had actually wanted to avoid. It brought organizational invention from outside as the social sciences realized that, if they do not embrace the new science of networks, then they simply will not be a part of it.

The Complexity Turn is way more than a change of Social Network Science's face. It is part of an ongoing transformation that introduces, in all of science, a new era of computational research driven by relational data. In the last decade, the problem arose how "to handle an avalanche of traces whose magnitude and diversity is unprecedented for the social sciences. For the first time in their history, social scientists have continuous information about their objects" (Venturini, Jensen, & Latour, 2015). As we have reconstructed, the reaction of the science system has been interdisciplinary. For a while now, a research front is called Computational Social Science (Lazer, Pentland, et al., 2009). In our dataset, we have identified the part of it that develops the community detection methods for this domain. This part emerged when Complex Network Analysis embraced Web Science and revisited some structuralist concepts.

A pressing question in Computational Social Science is how we can study social media and learn something about society rather than about social media use (Rogers, 2013). Facing new data, much current work is understandably concerned with description rather than modeling. Venturini et al. (2015) propose that "[t]he time is now to develop the formal techniques necessary to unfold the origami of collective existence and this should be the aim of the renewed alliance between the social and natural sciences. For the next few years, at least, efforts should be shifted from simulating to mapping, from simple explanations to complex observations."

But there are urgent questions that do require modeling and simulation. The Gulbenkian Commission recommended reform because realms like society, polity, and economy are overlapping, but today it has become clear that they are also coupled to science, technology, agriculture, the financial and other systems on a global scale. Perturbations cascade through all these realms like the data/computation innovation was transposed from technology to science and diffused through the subdomains of Social Network Science. How can we handle the risks associated with these globally densifying networks to arrive at sustainable institutions (Bettencourt et al., 2007; Helbing, 2013)? How can we nurture breakthroughs from understanding the structure and dynamics of complex systems (Sornette, 2008)?

Relational Sociology informed by Complexity Science has much to contribute to the great endeavor Computational Social Science is facing. H. C. White's work (1992, 2008) is fit for a lead role because, through his training in sociology and physics, complexity is already accounted for. Accounting for complexity means allowing some degree of realism – not the realism of the functionalist but of the structuralist Merton. *Identity and Control* is "a coherent, logical, necessary system of general ideas in terms of which every element of our experience can be interpreted" (Whitehead, 1978, p. 3). Through sometimes diffuse abstraction, it is capable of describing the self-similarity and universality of particular processes. For example, the change of styles in music is modeled just like in science

(H. C. White, 2008, pp. 331). Like systems theory, *Identity and Control* can explain the observation of itself. In fact, it is more general than a prototypical theory of the middle range. Because persons all the way up to large-scale network domains are versions of an idealized identity, it is a theory of complex socio-cultural systems, though not in the sense of Luhmann's essentialist individualism (Padgett, 2012a, p. 56).

Order emerges spontaneously, but self-organization does not mean that we resort to a sociological naturalism. On the contrary, sociology puts meaning and culture into complexity. Order is "for free" (Kauffman, 1993) because the optimization mechanism sorts out identities that do not reproduce through the Matthew Effect. Without a corresponding meaning structure, any percolation in transaction networks must seem like it is caused by a mysterious negentropy. Through the feedback loop of emergence and downward causation, percolation finds explanation in a simple process of biased randomness in the meaning structure. A public can be thought of as a *sixth sense* of identity. It is a highly sparse array of all facts and cores that are at least minimally relevant for an identity and which the identity can select. The sixth sense is not what an identity is but all that it *could be*. Publics facilitate netdom switchings and the emergence of meaning through distinctions. Context is what larger sets of identities make of the public.

Sociological explanation means making a prediction about the behavior of an identity in the future or, cross-sectionally, about the behavior of identities in the same universality class (Watts, 2014). No model can predict what empirical facts arise from non-linear dynamics in complex networks, but post-emergence dynamics or stylized facts (patterns) are predictable (Hedström & Bearman, 2009b). We have encountered the Matthew Effect as a mechanism that allows *predicting stability* from reproduction through interface discipline. D. Wang et al. (2013) have been able to predict an average reference's citation career by keeping fitness constant. Identities modeled on socio-cultural temperature are a first and careful step towards an uncertainty calculus that mathematically describes the trade-off among social and cultural uncertainty and contingency. The test of such a calculus is *predicting change* (H. C. White, 2008, p. 141). Our idea of the uncertainty calculus also involves the fitness of social facts but aims at predicting identities as bundles of social facts: Identities that stray from criticality and become either too frozen or too liquid will sooner or later return to criticality by adjusting social or cultural uncertainty or both. When Complexity Science rapidly became structurally cohesive, a divergent event like the emergence of Evolutionary Game Theory or a change of face was expected and occurred because it increased cultural uncertainty. To be a calculus, the dimensions in figure 1.7 will have to be represented by carefully selected parameters.

With the availability of digital traces it has become possible to map publics, observe them through macroscopes, and monitor the emergence and decline of identities (Watts, 2011). Knowing what styles exist is the condition for identifying potential novelty from recombinations of styles. Social, political, and scientific implications are that institutions can be monitored if they are at risk of freezing or melting, warning signs of crises can be early-detected, and breakthroughs can be nurtured. An uncertainty calculus as sketched bears a resemblance to methods for detecting bubbles in the financial system (Sornette, 2008; Garcia, Tessone, Mavrodiev, & Perony, 2014) which is worth developing.

# Glossary

### Agent

Identity in the first sense with a capacity to act; persons, groups, or aggregates.

### Ambage

Social uncertainty; uncertainty in a meaning structure of agents (H. C. White, 2008, p. 58). Operationalization: Interpretation of densification exponent (Leskovec et al., 2005), autocatalysis exponent, and small-world coefficient (Humphries & Gurney, 2008) in author co-authorship networks.

### Ambiguity

Cultural uncertainty; uncertainty in a meaning structure of social facts that cannot make selections (H. C. White, 2008, p. 58). Operationalization: Interpretation of structural cohesion (D. R. White & Harary, 2001) and fitness of reference co-citation cores.

### Arena discipline

Discipline that creates boundaries with attachment valued by purity and governed by homophily (H. C. White, 2008, p. 95–104). Operationalization: Screening of mixture of types of tie in core.

### Autocatalysis

(Co-)reproduction of social fact(s) due to repeated selection; requires memory (Padgett, 2012a). Operationalization at micro level: Additive directed tie in fact co-selection matrix (Batagelj & Cerinšek, 2013, sec. 3.3). Operationalization at macro level: Scaling law of degree and weighted degree.

### Autocatalysis exponent

Macro operationalization of autocatalysis; extent to which ties are reproduced in cores where degrees are large; exponent in scaling relationship of a social fact's unweighted and weighted degree.

### Barabási/Albert model

Model of scale-free random networks with a power-law degree distribution from growth and preferential attachment (Barabási & Albert, 1999).

### Bayesian fork

Punctuation that erases structural memory and creates an opportunity for a new identity in the third sense (H. C. White, 1995). Operationalization: Termination of path in direct citation network (Girvan & Newman, 2002).

### Betweenness

Brokerage measure, number of shortest paths that traverse a network node (Freeman, 1977).

### Blockmodeling

Detection of blocks or positions consisting of structurally equivalent nodes in unipartite networks (H. C. White et al., 1976).

### Boundary

Fuzzy region in the periphery of a core that results from inside/outside distinctions at multiple levels (Abbott, 2001a). Operationalization: Detection of cores through bibliometrically enhanced publication retrieval (Zitt & Bassecoulard, 2006; Mogoutov & Kahane, 2007).

### Boundary precision

Minimum quality of all subdomain boundaries; parameter in delineation procedure.

### Change → Chaos

### Chaos

Extreme temporal sensitivity of a system to initial conditions and punctuations (Kauffman, 1993, ch. 5); causally related to spatial order; other side of normality in a phase transition.

### Co-selection

Selection of two social facts by 1-identities; additive. Operationalization: Multiplication of normalized selection matrix (Batagelj & Cerinšek, 2013).

### Complex socio-cultural system

Identity that exists in all six senses.

### Conditional lifetime

Number of time units that a social fact or event endures; depends on level of analysis. Operationalization: Number of years in which a fact (author/reference/word) has a selection (authorship/citation/usage) fraction not below a given threshold.

### Constitutive tie

Transaction inside the boundary of identity; devoted to reproduction (Padgett, 2012b, p. 100–8). Operationalization: Maximized through modularity maximization (Blondel et al., 2008).

### Context

What identity embeds into; meaning structure of higher-level identity. Operationalization as meaning structure: Fact co-selection matrix. Operationalization as distribution: Size distribution of social facts (Clauset et al., 2009).

### Contingency

Possibility of path (Mahoney & Schensul, 2006, p. 461). Operationalization of strength: Socio-cultural temperature.

### Control

Footing in otherwise stochastic networks; attained in collective self-organization. Operationalization: Second to fifth sense of identity; observable pattern in meaning structure.

### Core

Central position in a meaning structure (S. Fuchs, 2001; D. R. White & Harary, 2001). Operationalization: Social facts in the fat tail of size distribution; structural cohesion (D. R. White & Harary, 2001).

### Council discipline

Discipline that organizes diversity with attachment valued by prestige and governed by triadic closure (H. C. White, 2008, p. 86–95). Operationalization: Hierarchical modularity of meaning structure (Ravasz & Barabási, 2003); screening of mixture of types of tie.

### Degree

Number of edges of a network node.

### Densification exponent

Longitudinal operationalization of 4-identity; extent to which 4-identity becomes cohesive or conscious about itself; exponent in scaling relationship of a meaning structure's number of nodes and edges over time (Leskovec et al., 2005).

### Discipline

Meta mechanism that structures paths through different attachment mechanisms; structural blueprint that reproduces a network of positions; exists in arena, council, and interface types (H. C. White, 2008, ch. 3).

### Distinction

Interpretation by identities in the first sense whether a social fact in the public is allowed into the meaning structure (Abbott, 2001b, ch. 9). Operationalization: Detection of positions or subdomains in a publication network (Blondel et al., 2008).

### Emergence

Bottom-up process whereby a meaning structure results from transactions among identities in the first sense (Anderson, 1972). Operationalization: Non-trivial meaning structure.

### Emerging science

Research domain without a coherent cultural core; stage before normal science (cf. Bettencourt et al., 2009). Operationalization: Sublinear preferential attachment of citation or word usage (Golosovsky & Solomon, 2013).

### Equilibrium

Identity's state of being at rest due to isolation from context.

### Event

Story realized in time; contingency of meaning; nearly decomposable to an equilibrium, i.e., endures as long as it is not punctuated; social fact when it punctuates (H. C. White, 2008, p. 186). Operationalization: Community detection in direct citation networks (Girvan & Newman, 2002).

### Feedback

Circular causality; traced back to Durkheim (1982 [1895]). Operationalization: Central mechanism of the identity model (figure 1.2).

### Feedback network

Network model that integrates preferential attachment and Lévy flight search; creates structural cohesion (D. R. White et al., 2006).

### Fitness

Ability of social fact ('trait') to outreproduce others and gain selection shares (cf. Darwin, 2003 [1859]). Operationalization: Observed number of selections over number of selection expected from preferential attachment; error in predicting selection success.

### Fractal

Infinite scale-invariant mathematical set with non-integer dimensionality (Mandelbrot, 1982, p. 15).

### Fractal distinction

Distinction that repeats in itself (Abbott, 2001a). Operationalization in time: Story that is selected in a distinction at a present but was not selected in a preceding distinction.

### Homeostasis

Stability of meaning when positions in meaning structure are filled differently.

### Homophily

Principle that cultural similarity breeds social connectivity (Kossinets & Watts, 2009). Operationalization: Screening of mixture of types of tie in core.

### Hub

Strongly connected network node.

### Identity

Object of observation; socio-cultural building block of social life; self-similar concept because not situated at particular socio-cultural length scale; exists in six senses (H. C. White, 2008).

### Identity in the first sense (1-identity)

Fundamental building block (distinction, event, person, group, organization) that identity consists of but to which it is not reducible; engage with each other through transactions. Operationalization: Publication in bipartite selection matrix (figure 2.1).

### Identity in the second sense (2-identity)

Collection of identities in the first sense that have found footing through collective control. Operationalization: Existence of cohesive core in meaning structure.

### Identity in the third sense (3-identity)

Identity as path through socio-cultural space-time; narrative. Operationalization: Search path counting in direct citation networks (Batagelj, 2003).

### Identity in the fourth sense (4-identity)

Identity that is conscious about its identity and path; percolation strength. Operationalization in space: Small-worldliness of co-authorship network (Humphries & Gurney, 2008) Operationalization in time: Densification of author co-authorship network (Leskovec et al., 2005).

### Identity in the fifth sense (5-identity)

Sensibility of making distinctions; trade-off among social integration and cultural regulation; results in self-similar size distributions; activated by punctuations (H. C. White, 2008, ch. 4). Operationalization of strength: Exponent of attachment scaling law (Golosovsky & Solomon, 2013). Operationalization of fluctuations of fact selections: Variable fitness of social facts.

### Identity in the sixth sense (6-identity)

Highly sparse network of overlapping domains in which switchings occur and from which punctuations originate (Mische & White, 1998); array of all the ways identity could be. Operationalization: (Story set of) publications that use `SOCIAL` and `NETWORK*` as words.

### Institution

Social fact that is selected for a certain duration (Durkheim, 1982 [1895], ch. 1; H. C. White, 2008, ch. 5). Operationalization: Conditional lifetime; number of time units that a social fact endures.

### Interface discipline

Discipline that organizes production flows with attachment valued by quality and governed by the Matthew Effect (H. C. White, 2008, p. 80–6). Operationalization: Structure of direct citation network.

### Invisible college

Set of prestigious scholars that coordinates a research domain (Price, 1986 [1963], p. 119). Operationalization: Co-authorship network of authors at the micro level.

### Level

Layer of emergence; higher levels provide context for lower levels. Operationalization of levels of analysis: Meaning structure with all social facts (except citation practice) on the macro level, moderately selected facts on the meso level, and core (paradigmatic) facts on the micro level.

### Level of analysis → Level

### Lévy flight

Biased random walk where the search distance decays with a power law tail (Kleinberg, 2000).

### Lifetime → Conditional lifetime

### Mainstream

Main narrative. Operationalization: Top 1% of the most traversed citations (Batagelj, 2003).

### Matthew Effect

Positive feedback effect according to which the present number of selections is proportional to the past number of selections (rich get richer and poor get poorer) (Merton, 1973 [1968]). Operationalization of rich-get-richer part: Generalized preferential attachment (Bianconi & Barabási, 2001).

### Matthew exponent

Operationalization of strength of identity in the fifth sense; exponent in scaling relationship of past and present selections of social facts (Golosovsky & Solomon, 2013).

### Meaning structure

Network of social facts; emerges from, and influences, selections made in transaction structure; memory of identity; upper network in identity model (figure 1.2). Operationalization: fact co-selection matrix like author co-authorship network, reference co-citation network, and word co-usage network (figure 2.1).

### Memory

Function of meaning structure for identities in the first sense to recall successful transactions; condition for autocatalysis (Padgett, 2012b, p. 93–100). Operationalization: Fact co-selection matrix.

### Narrative

Chain of events; story set in time; path through socio-cultural space-time. Operationalization: Community detection in direct citation networks (Girvan & Newman, 2002).

### Netdom → Network domain

### Network domain

Duality of transaction structure and meaning structure; consists of a set of identities in the first sense and a set of social facts (Mische & White, 1998). Operationalization: Set of publications.

### Normal science

Phase of unpunctuated knowedge production; terminated by a scientific revolution (Kuhn, 2012 [1962], ch. 3). Operationalization: Non-sublinear preferential attachment of citation or word usage (Golosovsky & Solomon, 2013). Operationalization of beginning and end: betweenness-based community detection (Girvan & Newman, 2002) in direct citation networks.

### Normality

Extreme temporal insensitivity of a system to initial conditions and punctuations; causally related to spatial disorder; other side of chaos in a phase transition.

### Open elite

Multifunctional core (D. R. White et al., 2004). Operationalization: Structural cohesion (D. R. White & Harary, 2001) and mixture of types of tie.

### Optimization

Improvement of control; reason for the Matthew Effect.

### Organizational innovation

Punctuation of style (Padgett & Powell, 2012a, p. 7–11).

### Organizational invention

Change of style due to organizational innovation (Padgett & Powell, 2012a, p. 7–11). Operationalization of strength: Cosine similarity of cores before and after invention.

### Paradigm

Story set or set of rules that is exemplary for how puzzles are solved in normal science (Hoyningen-Huene, 1993, pp. 135). Operationalization: Co-selection core of strongly selected social facts; micro level of analysis.

### Paradigm shift → Scientific revolution

### Percolation

Emergence of a giant component in a transaction or meaning structure. Operationalization of percolation strength: Network densification (Leskovec et al., 2005).

### Phase transition

Co-existence of order and disorder in a meaning structure; ground state of identities according to the scaling hypothesis. Operationalization: Power-law size distributions like Lotka's Law (1926), Bradford's Law (1985 [1934]), or Zipf's Law (2012 [1949]).

### Position

Set of nodes in a blockmodel; positions can be filled by concrete social facts; e.g., core, periphery (H. C. White et al., 1976).

### Practice

Scientific procedure. Operationalization: Authorship, citation, word usage.

### Preferential attachment

Operationalization of the Matthew Effect (Bianconi & Barabási, 2001).

### Public → Identity in the sixth sense

### Punctuation

Interruption of a story set's reproduction by an event that is felt to matter; necessitates a distinction (Mische & White, 1998, p. 711).

### Relational tie

Transaction outside the boundary of identity; devoted to production (Padgett, 2012b, p. 93–100). Operationalization: Minimized through modularity maximization (Blondel et al., 2008).

### Reproduction → Autocatalysis

### Research front

Core of active and current research in a domain (Price & Beaver, 1986 [1966]). Operationalization: Structurally cohesive author co-authorship networks with $k \leq 4$ (D. R. White & Harary, 2001).

### Scale invariance

Property of a system to be similar at different scales of observation; mathematical formulation of self-similarity. Operationalization: Power laws (Clauset et al., 2009).

### Scaling → Scale invariance

### Scaling law

Bivariate power laws that describe the phase transition. Operationalization: Scaling functions (appendix A).

### Scientific revolution

Punctuation of normal science that greatly changes the composition of its core (Kuhn, 2012 [1962], ch. 9). Operationalization of strength: Cosine similarity of preceding and following subdomain paradigms.

### Selection

Exertion of influence by a social fact on an identity in the first sense. Operationalization: Authorship by author of publication, citation of reference by publication, or usage of word by publication; data model (figure 2.1).

### Self-organization

Spontaneous emergence of control (H. C. White, 1973).

### Self-similarity → Scale invariance

### Social fact

Story that (a) emerges from transactions among identities in the first sense and (b) exerts an influence on at least one identity; identity in second-level observation (Durkheim, 1982 [1895], p. 59–62). Operationalization: author, cited reference, or word.

#### Stability → Normality

#### Story

Unit of analysis; as network tie a transaction; as network node a social fact (H. C. White, 2008, p. 27–30). Operationalization: social fact as node in meaning structure.

#### Story set

Pattern of social facts coupled through transactions or co-selections (Godart & White, 2010). Operationalization: Fact co-selection matrix.

#### Structural cohesion

Hierarchical embeddedness of increasingly denser but smaller subnetworks. Operationalization: Detection of $k$-components (D. R. White & Harary, 2001).

#### Structural equivalence

Property of network nodes in unipartite matrices to have identical relational profiles (Lorrain & White, 1971). Operationalization: Generalized blockmodeling of bipartite matrices (Doreian et al., 2004).

#### Style → Identity in the fifth sense

#### Subdomain

Substructure of a networm domain. Operationalization: Detected through distinctions.

#### Switching

Transition from observing one to another meaning structure (Godart & White, 2010).

#### Systemic invention

Organizational invention of all subdomains in a domain (Padgett & Powell, 2012a, p. 7–11). Operationalization of strength: Cosine similarity of all subdomains' cores before and after invention.

#### Talk → Transaction

#### Temperature

Operationalization of contingency; extent to which the state of percolation allows punctuating social facts to take a core position.

### Transaction

Relation among identities in the first sense; all transactions are stories, but only stories as ties are transactions.

### Transaction structure

Network of identities in the first sense; origin of selections; gives rise to, and is formatted by, meaning structure; lower network in identity model (figure 1.2). Operationalization: Direct citation network (Batagelj, 2003); author co-authorship are modeled as meaning structures.

### Triad

Fully connected network of three nodes.

### Type of tie

Tie that belongs to one particular story set. Operationalization: One type per subdomain.

### Universality

In physics: Quality of an entity that is consistent in the universe (Stanley, 1999). In sociology: Quality of a social fact that is consistent at a middle range of observation (Hedström & Udéhn, 2009). Operationalization: Invariance of scaling law exponents; data collapse (Stanley, 1999).

### Watts/Strogatz model

Small-world network model at the transition from order to disorder from randomly rewiring a grid (Watts & Strogatz, 1998).

# A. Statistical Properties and Measurement of Power Laws

The *linear model* is at the core of sociological methodology. It has its root in survey research where populations are necessarily sampled. The philosophical assumption is that the social world is populated with static identities that have attributes which are measured as variables. The linear model relies on an essentialist ontology (Schelling, 2006 [1978], ch. 2; Abbott, 2001b, ch. 1).

To be able to make statements about populations "on average," the variables that describe the attribute values are assumed to be independent and identically distributed. Independent means that observations do not influence each other, identically distributed that they are drawn from the same underlying distribution. A third and essential assumption is that the arithmetic mean and variance of the underlying distribution are finite. The Central Limit Theorem states that, when a population that fulfills these assumptions is sampled repeatedly with sufficiently large samples, the means of the samples will be normally distributed, the mean of the sample means will converge to the mean of the population, and the standard deviation of the sample means will converge to the standard error of the population. The Central Limit Theorem assures that sampling methods actually yield valid findings about the population. The normal distribution is at the heart of the linear model because it conforms to the Central Limit Theorem. The normal distribution is described by the mean and standard deviation, and these parameters set the scale of the distribution. The tails of the distribution decay exponentially, i.e., extreme values have very low probabilities.

The most common application resting on these assumptions is causal explanation. In such models, e.g., in linear regression, linearity means that the independent variable determines the dependent one up to an error term. As a consequence, big effects must have big causes. A mechanism of the linear model that generates independent and identically distributed variables is the uncorrelated Gaussian process (Mandelbrot & Van Ness, 1968). As a purely random process it generates normal distributions. In sum, the linear model assumes identically and independent distributed variables, it considers social dynamics as random and contextless processes, thereby gains explanatory power, but does so at the cost of losing applicability where the parametric conditions are not fulfilled (Abbott, 2001b, ch. 6).

For relational analyses, sampling is not necessary because networks typically represent complete populations. Network nodes are not the attributes of, e.g., persons but the persons themselves. The linear model as discussed is not applicable because nodes influence each other through relations. The generating processes of complex networks are not purely random but best described by feedback dynamics that potentially produce non-normal distributions (D. R. White et al., 2006). In the *non-linear models* used in this work, variables are interdependent, identities are embedded in context, and small events can have big consequences.

## Univariate Scaling

A variable $x$ exhibits power-law behavior if the distribution has the function

$$p(x) = Cx^{-\alpha} \tag{A.1}$$

where $\alpha > 1$ is the exponent, $x > 0$, and $C$ is a normalization constant. Power laws show as straight lines on double-logarithmic axes, as can be shown by taking the logarithm of both sides of equation A.1:

$$\log(p(x)) = \log(Cx^{-\alpha}) \propto -\alpha \log(x). \tag{A.2}$$

Because the fat (upper) tail of the distribution decays much slower than that of a normal distribution, extreme values have a much greater probability of occurring.

A defining difference between the power law and the normal distribution is their moments. In general, the moment $m$ describes the shape of a distribution. In statistics, the first moment is the arithmetic mean and the second moment is related to the standard deviation. For a normal distribution, both moments are finite. For a power law, the moments are only finite for $m < \alpha - 1$. If $\alpha \leq 3$, the standard deviation is infinite. If $\alpha \leq 2$, even the mean is infinite. For $\alpha \leq 3$ the distribution can not be meaningfully described using parametric statistics – it is *scale-free* because no mean and standard deviation describe the intrinsic scale of the population. Mean and standard deviation can be calculated but they are meaningless and introduce an artificial scale. The Central Limit Theorem does not apply (Barabási, 2015, ch. 4).

Saying that a distribution is scale-free is another way of saying that it is scale-invariant, i.e., multiplying $x$ with a constant $a$ results in a proportionate scaling of the function itself:

$$p(ax) = C(ax)^{-\alpha} = a^{-\alpha}p(x) \propto p(x). \tag{A.3}$$

Put differently, rescaling a power law does not change the shape of the distribution (Newman, 2005). A normal distribution is not scale-invariant, because rescaling $x$ results in a different parameter set.

The following parameter estimations and statistical tests for the power law are taken from Clauset et al. (2009), as implemented by Gillespie (2014). The exponent is estimated using the method of maximum likelihood. For an integer variable $x$ with observed values $i, ..., n$,

$$\hat{\alpha} \simeq 1 + n \left[ \sum_{i=1}^{n} \ln \frac{x_i}{x_{\min} - \frac{1}{2}} \right]^{-1} \tag{A.4}$$

Here, $x_{\min}$ is a lower bound. If the full distribution obeys a power law, $x_{\min} = 1$, but often power-law behavior is only found in the upper tail of the distribution. Before $\alpha$, $x_{\min}$ is estimated by making the probability distribution of the empirical data and

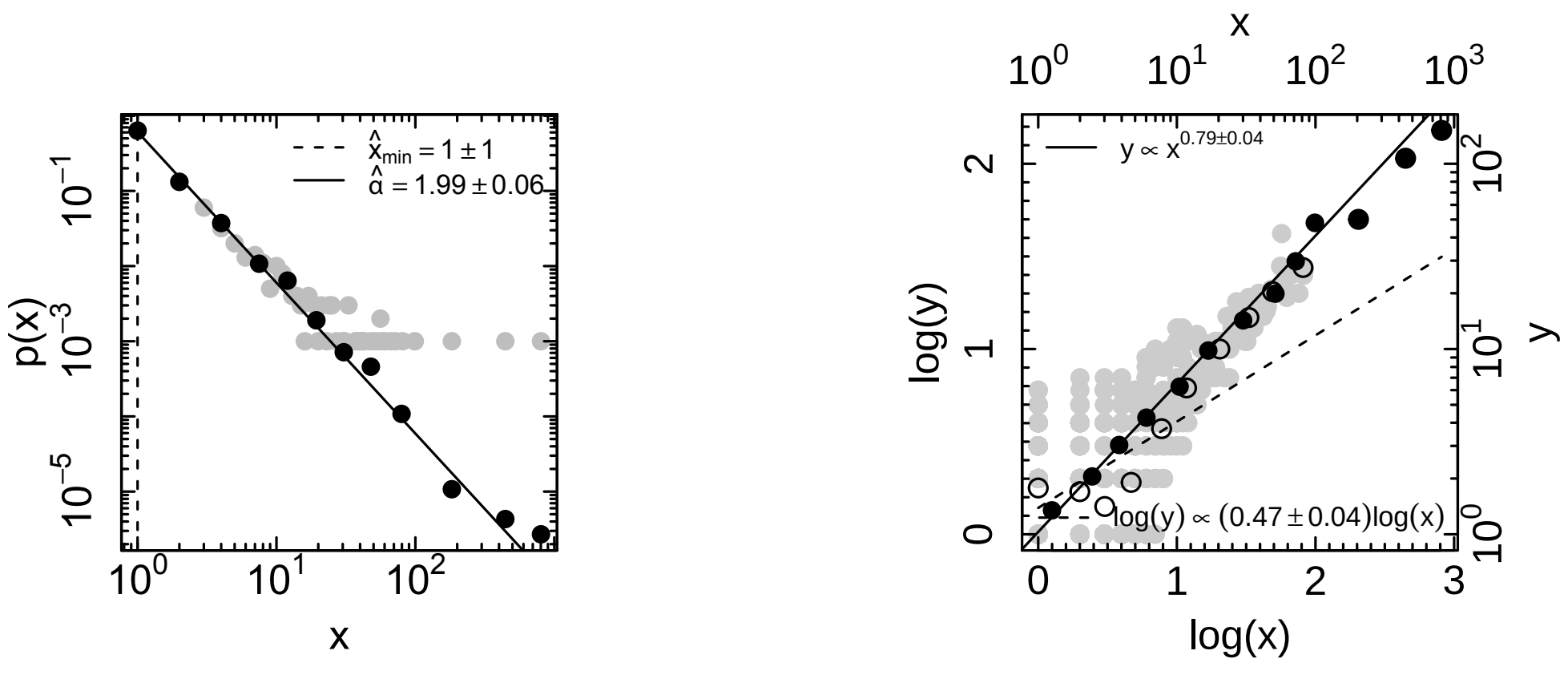

(a) Univariate

(b) Bivariate

**Figure A.1.: Scaling laws**

(a) Synthetic power-law distribution of variable $x$ generated from $\alpha = 2$ and $x_{\min} = 1$. The synthesizing parameters are well estimated and a power law is a plausible fit ($p = 0.2$). (b) Scaling relationship synthesized from $y \propto x^{0.8}$ with noise added in both variables. The dashed line is obtained by Ordinary-Least-Squares (OLS) regression of log-transformed variables in a linear model and yields the wrong estimate $\hat{\beta}_{\text{OLS}} = 0.47 \pm 0.04$ because determination of $y$ by $x$ is assumed. A scaling law, estimated using Standardized-Major-Axis (SMA) regression, yields the right estimate $\hat{\beta}_{\text{SMA}} = 0.79 \pm 0.04$. The goodness-of-fit is the same for both methods ($r^2 = 0.96$). Unfilled points in (b) result from logarithmic binning, representing each logarithmic decade on the x-axis by five equally-spaced data points. The filled black points are obtained through double-logarithmic binning, representing two logarithmic decades on axes x and y by five data points. These points reveal the logic of OLS and SMA regression, respectively.

the fitted power-law function as similar as possible. This is achieved by minimizing the Kolmogorov-Smirnov (KS) distance – the smaller it is, the better the fit is. As with the estimate for $\alpha$, the lower bound $\hat{x}_{\min}$ and distance $\hat{KS}$ that best fit the observed data are marked with a hat.

The uncertainty in these estimates is measured by sampling $n$ values with replacement (bootstrapping) from the original dataset and estimating $x_{\min}$ and $\alpha$ using the KS statistic. This is done 1,000 times. Note that power-law parameters are sampled, not means, so the CLT does apply. The means ($\bar{\alpha}$, $\bar{x}_{\min}$, $\bar{KS}$) and standard deviations ($\sigma_\alpha$, $\sigma_{x_{\min}}$, $\sigma_{KS}$) of the sampling distributions then give the uncertainty in the estimates.

Finally, to determine how good the best fit fits the empirical data, a goodness-of-fit test is performed. 1,000 synthetic power laws are generated using $\hat{\alpha}$ and $\hat{x}_{\min}$. For each synthetic data set, the KS distance to the best-fit power law is calculated. The goodness-of-fit $p$ is then the fraction of the 1,000 KS distances that is larger than $\hat{KS}$. As a rule of thumb, power-law behavior is ruled out if $p \leq 0.1$. In other words, for the power-law fit to be plausible, in more than 10% of the 1,000 synthetic data sets the KS distance to the best fit should be larger than the distance of the best fit to the empirical data.

Figure A.1a shows a synthetic power law generated from $\alpha = 2$ and $x_{\min} = 1$. Gray points show the original data. Black points show logarithmically binned data, but only as a help for the eye (Alstott, Bullmore, & Plenz, 2014). Binning denoises the fat tail and shows that the whole distribution is indeed a straight line. The parameters are estimated for the unbinned data. The exponent and lower bound given in the plot are the mean and standard deviations of the bootstrapping procedure. Most often throughout this study, the best fit estimates correspond closely to the mean. The solid and dashed lines display the estimates. Estimates are only shown if the data is plausibly fit by a power law. In this case, the fit is plausible with $p = 0.20$.

### Bivariate Scaling

When insights about two variables' association or mathematical relationship are sought, the standard tools of the linear model are Pearson correlation and linear regression. As we will shortly see, non-linear models are similar in the mathematics but different in the interpretation. We continue the discussion for the case that two variables are not linearly related.

One typical example of a non-linear relationship between two variables $x$ and $y$ is the scaling relationship,

$$y(t) = D(t)x(t)^{\beta}, \tag{A.5}$$

where $\beta$ is the scaling exponent, $D$ is a normalization constant, and $t$ is time (irrelevant for now). As this function is of the same form as equation A.1, it is also scale-invariant and shows as a straight line on a log-log plot. Consider a variable $y = x^{0.8}$ and random noise added such that $\beta = 0.8$ is preserved. The scatterplot is depicted gray in figure A.1b. Like the original synthetic variable, the new $x$ and $y$ have infinite variance.

In linear modeling, if a statement about association is intended, the Pearson correlation coefficient $r$ gives the linear dependence of $x$ and $y$. $r$ is scale-invariant because the correlation does not change if the variables are rescaled. However, since $r$ is based on the means and standard deviations of $x$ and $y$ but both variables have infinite variance, we must not use the coefficient in our toy example. In such situations, textbooks propose data transformations (Baguley, 2012, pp. 342). Taking the logarithm[1] of $x$ and $y$ effectively renders their means and standard deviations finite and results in a correlation of $r = 0.58$. The linear model was tailored to a non-linear situation.

If a statement about a mathematical relationship between the two variables is intended, the first thing the linear model offers is linear regression with OLS fitting. In the most simple form, the model shows how a dependent variable $y$ minus an error term $\epsilon$ can be predicted from an independent variable $x$ by fitting the linear function

$$y = \beta x + \epsilon. \tag{A.6}$$

The slope $\beta$ then gives the average increase in $y$ for a one-unit change in $x$. The original model is not scale-invariant because it is additive and multiplying $x$ or $y$ with a constant $b$ will cause the slope to be divided or multiplied by $b$, respectively. But rescaling may be desirable, e.g., if one variable is counted in thousands and the other in millions. Taking the logarithm of both variables not only renders their variance finite, it also makes the model scale-invariant. Because of equation A.2, the slope of the linear regression line is the exponent in the scaling relationship.

[1] Throughout this work, the common logarithm with base 10 is used.

OLS regression of $\log(y) \propto \log(x)$ gives $\hat{\beta}_{\mathrm{OLS}} = 0.47 \pm 0.04$ which is far from the real value of 0.8. The 95% confidence interval $[0.43, 0.51]$ for the slope gives the uncertainty in the measurement, and the coefficient of determination $r^2 = 0.34$ serves as the goodness-of-fit statistic. The result is shown in figure A.1b. The reason that the fit is so far off is that the OLS method measures residuals vertically because $x$ is implicitly assumed to determine $y$. The unfilled points in the figure are obtained by logarithmic binning. These bin averages reveal the vertically inferential logic of OLS regression.

The way we have created the synthetic data, $x$ and $y$ are non-linearly associated. In such a case, the correct way is to use Standardized-Major-Axis (SMA) regression. The SMA method is described by Warton et al. (2006) and implemented by Warton et al. (2012). The direction that residuals are measured is the fitted line reflected about the $y$ axis. It results in a realistic fit $\hat{\beta}_{\mathrm{SMA}} = 0.79 \pm 0.04$. The corresponding filled black points are obtained from double logarithmic binning, i.e., data is averaged for both axes simultaneously. Essentially, the data points from binning are only a help for the eye. Regression is performed on the full data.

SMA can be applied in linear and non-linear modeling. In both, $\beta$ is estimated by logarithmically transforming the variables and fitting a regression line. Mathematically, $\beta$ is interpreted as a slope in the linear model and as an exponent in the non-linear model. In terms of meaning, the interpretation is similar in both models, namely, that the average $y$ per $x$ exhibits decreasing returns to scale if $\beta < 1$ and increasing returns if $\beta > 1$. In the linear model, the data is transformed into the linear realm and there it is interpreted – the transformation is mathematical and theoretical. In the non-linear model, the transformation is only mathematical. The interpretation takes place in the non-linear realm. Linear regression with log-transformed variables is a non-linear model expressed in linear terms.

# B. Data Pre-Processing

### Author Name Disambiguation

Author names are already preprocessed in the *Web of Science* database. Accents on vowels have been removed and language-specific letters have been replaced. In addition, we have whitespaced hyphens and removed call names in brackets. Unfortunately, we cannot use the information contained in lower- and uppercase because some names are all uppercase. All disambiguation proceeds on standardized uppercase names. In the database, author names are attributed to publications and the goal of author name disambiguation is to identify author identities from author names. Author names are provided fully as supplied by the authors of the manuscripts (full names) and given names are reduced to initials (abbreviated names). Note that full names not necessarily contain spelled-out given names. It just means that they are not reduced to initials if they are spelled out.

If one of these codifications is used as identities, errors result. If full names are used to represent author identities, false negatives occur. For example, in all his authorships, Luís A. Nunes Amaral has provided his name as `AMARAL,_LUIS_A_NUNES`, `AMARAL,_LUIS_A_N`, and `AMARAL,_L_A_N`.[1] The disambiguation task is to unify these synonyms and reduce them to the one identity. If abbreviated names are used, false positives result. For example, `ZHANG,_Y` has 50 authorships but subsumes nine different spelled-out given names from `YAN` to `YUAN`. Here the disambiguation task is to split homonyms. This is complicated by real-life homonyms that do not result from data processing. This is the case when authors share exactly the same name.

We apply a naive semi-automatic disambiguation procedure that splits homonyms and unifies synonyms. For homonym splitting, we apply the disambiguation algorithm by Soler (2007). The implementation of U. Sandström and Sandström (2010) uses all available meta data. Because many given names are supplied by the *Web of Science* only as initials, only first initials are used.

Spelled-out given names are used to subsequently and manually unify synonyms. The rationale of the procedure is to use full names but coherently simplify them as long as the changes do not introduce an error. In pseudo-algorithmic terms, for all authorships involving full names with similar family name, similar initials, $n$ given names – in our case $n_{max} = 4$ –, and at least one spelled-out given name in these authorships: check for each given name position if the spelled-out given name is unique. If it is, eliminate the last given name. Repeat the procedure just described until only the first given name is left. The coded name is then the most-reduced name where the initials are replaced by the unique spelled-out given names. For example, Luis A. Nunes Amaral's coded name

[1] All exemples are from the solution dataset.

is AMARAL,_LUIS because for the initials L, A, and N there are no LUIGI, ANDREA, or NICK.

This way, two thirds out of 75,714 authorships get disambiguated. The rest consists of names where multiple spelled-out given names for one initial exist. These names require special treatment. Again in pseudo-algorithmic terms, for all authorships involving full names with similar family name and initials where at least one spelled-out given name in these authorships is not unique: check for each organization and given-name-position if the spelled-out given name is unique. If it is, the coded name is a string of the full name, where initials are replaced by unique spelled-out given names, and the organization's name. If it is not unique: add the name of the subdomain to the final string. For example, not using the organizational context, it would have been impossible to replace the initials in the abbreviated name ZHANG,_Y_C as there is a ZHANG,_YAN_CHAO as well as a ZHANG,_YI_CHENG. After disambiguation there are six identities of ZHANG,_YI_CHENG, each at a different organization, and ZHANG,_Y_C has been merged with one of them. Of course, it is possible that some of these six identities are synonymous.

At this point, 17,119 authorships remain that involve abbreviated names whose initials were not replaced by spelled-out given names either because it was not successful (3% of the remaining authorships) or because it was not possible in the first place – no spelled-out given names existed to replace (97%). The former are mostly Chinese and Korean names and are subjected to hand coding. The latter do not involve many Chinese and Korean names but often belong to authors who have consistently used only their (multiple) initials in their authorships. Knowing that false positives from homonyms create erroneous connectivity, we use the original abbreviated name but subject them to hightened attention in the final sanity check. All in all, despite the roughness and naivety of this procedure, it provides sufficiently accurate results as our conservative goal was to prevent artifactual connectivity (increased density in co-authorship networks). For this reason we did not invest more time into the disambiguation.

### Reference Matching

To match cited references to citing publications, we have transformed both into matchkeys. Most sophisticated approaches developed to only match journal articles proceed in an iterated, rule-based way (Olensky, Schmidt, & van Eck, 2015). Beyond matching journal articles, we must also assign matchkeys to books and book chapters to identify the unique social facts that multiple publications select. Other than for journal articles, for which usually the first page is cited, there is much more variation in the page numbers of cited books. Therefore, we use a rule-based key generation procedure that discriminates different reference types and compare it to an indiscriminating one. Before matchkeys are generated, all text strings are transformed into uppercase. Author names are reduced to the first eight letters of the family name and cut off after commas or periods. Source names (titles of journals, proceedings, or books) are cut off after hyphens, colons, and brackets. To improve the matching especially of proceedings articles, serial numbers between 1 and 99, either standing alone or followed by ST, ND, RD, or TH ("13th Conference on ...'), and years from 1950 to 2012 ("2012 Conference on ...") are removed when they lead the name of the cited source (exception: 1 MONDAY).

The indiscriminating matchkey consists of the author name, the publication year, the first letter of the source name, and the first or cited page. The role of a page can be taken by a DOI. This is the standard method for all citing publications in the seed and the solution sets. The discriminating method is applied to the cited references and proceeds as follows. If a page or DOI is cited, then check if a volume is cited. If yes, then the reference is coded as an *article* and the matchkey is constructed the standard way. If no volume is cited, then the reference is coded as a *chapter* and the matchkey is constructed from the author name, the publication year, all letters of the source name up to the first whitespace, and the first or cited page. If no page or DOI is cited, then the reference is coded as a *book* and the matchkey is constructed from the author name, the publication year, and all letters of the source name up to the first whitespace. Table B.1 gives examples for the three types. Chapters can include proceedings articles when those do not have a volume. This method produces errors when journals do not number their volumes or when books have volumes, but is considered the best procedure to separate articles from non-articles (Moed, 2005, pp. 122). Table B.2 shows that the discriminating matchkey is more conservative because it creates more distinct matchkeys for cited references. On the other hand, it identifies $171,303$ cited books. In total, $7,037$ or 29.9% of $23,529$ distinct seed matchkeys are cited by the seed.

### Natural Language Processing

The processing of natural language and keyword usage is necessary for a valid analysis of lexical concepts. Typically, titles, abstracts, and author keywords are used as data sources for lexical concept analysis. In the following, words are meant to include sequences of $n$ words ($n$-grams) such as `COMPLEX_NETWORK` and `SOCIAL_NETWORK_ANALYSIS`, i.e., lexical

**Table B.1.: Examples of matchkey types**

Articles have a page or DOI and a volume. Chapters have a page or DOI but no volume. Books have neither page or DOI nor volume.

| | Article | | Chapter | Book |
|---|---|---|---|---|
| **Author** | GRANOVETTER MS | NEWMAN MEJ | BOURDIEU P | WASSERMAN S |
| **Year** | 1973 | 2004 | 1986 | 1994 |
| **Source** | AM J SOCIOL | PHYS REV E | HDB THEORY RES SOCIO | SOCIAL NETWORK ANAL |
| **Volume** | 78 | 69 | N/A | N/A |
| **Page** | 1360 | N/A | 241 | N/A |
| **DOI** | N/A | 10.1103/ PhysRevE.69. 026113 | N/A | N/A |
| **Matchkey** | GRANOVET_1973_ A_1360 | NEWMAN_2004_ P_10.1103/ PhysRevE.69. 026113 | BOURDIEU_1986_ HDB_241 | WASSERMA_1994_ SOCIAL |

concepts consisting of two and three words, respectively. Using all words that occur in titles, abstracts, and author keywords would require the elimination of words that belong to general language use and do not contribute to distinguishing publications. To skip the requirement of removing such stop words, a crowd-sourcing approach is followed. The assumption is that all relevant words are used as keywords by the author collective of a domain and that irrelevant words, such as those commonly used as stop words, are not used as keywords (cf. Glänzel & Thijs, 2011).

The publication set used as a seed for the delineation procedure contains 27,139 unique author keywords. The problem with author keywords is that they are only available for publications from 1990 onwards. This meta data was historically not kept available because of data storage and processing limitations. Therefore, the approach is to use author keywords as the vocabulary and identify these words also in titles and abstracts.

**Table B.2.: Performance of different citation matching methods**

To match cited references to publications, they are transformed into text strings. The indiscriminating method attributes an identically constructed standard matchkey to cited articles, books, and chapters. In the discriminating method we introduce, the standard matchkey is only used for articles but different constructions are used for books and chapters. The 23,568 articles in the seed cite 512,036 different reference strings, represented by IDs, of which given numbers in the first column are articles, books, or chapters according to the definition given in the text. These references are translated into 481,128 unique matchkeys cited by the seed using the standard method and 492,921 using the discriminating method. Compression rates tell how strongly different reference strings are reduced to unique matchkeys. Compression is higher for the standard method (6.0% vs. 3.7%) because the number of unique matchkeys is lower even though, in the discriminating method, books are compressed more strongly than the total. The seed, which by definition only contains articles, is represented by 23,529 unique standard matchkeys. These are matched by 7,552 references using the standard method and by 7,037 references using the discriminating method, resulting in different matching rates. The matching rate for the standard method is higher (32.1% vs. 29.9%), i.e., our method is more conservative. This is because some reference strings which our method treats as books are matched to articles in the standard method. Note that the sum for the three publication types in the second column is larger than the corresponding total because some chapters' first title word has length one and may therefore have the same matchkey as an article.

| | **IDs cited by seed** | **Matchkeys cited by seed** | **compression rate** | **Seed matchkeys matched** | **Matching rate** |
|---|---|---|---|---|---|
| **Standard** | 512,036 | 481,128 | 6.0% | 7,552 | 32.1% |
| **Article** | 260,740 | 258,387 | 0.1% | 7,012 | 29.8% |
| **Book** | 187,486 | 171,303 | 8.6% | 0 | 0% |
| **Chapter** | 63,810 | 63,492 | 0.5% | 42 | 0.2% |
| **Total** | 512,036 | 492,921 | 3.7% | 7,037 | 29.9% |

The assumption is that words which had been relevant until 1990 were used as keywords after 1990. Technically, we construct the word usage matrix $G_{\text{wrd}}$ by using unique author keywords as cultural facts $F_{\text{wrd}}$ and creating an edge to a publication if the latter uses the author keyword in either its title, abstract, or author keywords list. As with all natural language, author keywords are used very inconsistently – in singular/plural, as lowercase/uppercase, with spelling mistakes, etc. Hence, titles, abstracts, and author keywords were pre-processed.[2] Text pre-processing reduced the list of unique author keywords to 26,689.

It turned out that stop words do show up in keyword lists. For example, `THE` is used rarely as a stand-alone author keyword, very likely as a consequence of automated text processing at the side of the database producer. Occasionally, the article leads meaningful keywords. Hence, typical stop words were removed if they stood alone or led the keyword string. For a list of the stop words, the extensive list of AutoMap 3.0.2 was used, reproduced in table B.3. Stand-alone ISO 3166-1 country names were also removed. Only keywords with three or more letters and five or less words were kept to become facts.

[2] All numbers, whitespaces, symbols (including `-`, excluding `&` and `@`), and brackets were replaced by underscores. Exceptions for number replacement were selected manually: `1/F_NOISE`, `2-MODE`, industry standard `802.11`, `PEER_2_PEER`, `PEER-2-PEER`, `P2P`, `1.0`, `2.0`, `3.0`, `B2B`, and `B2C`. Words were stemmed. Morphological stemming was chosen over Porter stemming because the former tends to create less overstemming. Not stemming uppercase is not useful because the *Web of Science* uses uppercase for old text. All letters were transformed to uppercase. AutoMap versions 3.0.2 and 3.0.10 were used for all text processing.

**Table B.3.: Stop words removed from titles, abstracts, and author keywords**
General stop words had to be removed because such unexpectedly showed up in author keyword lists. General science stop words had to be removed because they represent general scientific language. The left list is a standard, the right one was hand-crafted.

| **General stop words** | | **General science stop words** | | |
|---|---|---|---|---|
| a | she | address | feature | per |
| an | so | affect | framework | potential |
| and | that | analyse | general | present |
| as | the | approach | give | problem |
| at | their | area | high | propose |
| but | theirs | argue | home | relate |
| for | them | article | importance | report |
| he | they | base | index | research |
| her | to | be_use | invention | right |
| her | us | bee | issue | run |
| hers | we | case | lead | see |
| him | who | case_study | let | self |
| his | whoever | concept | level | study |
| i | whom | concern | like | term |
| it | whomever | conduct | live | tip |
| its | will | data | mean | type |
| me | would | date | name | use |
| mine | you | design | need | usefulness |
| my | your | difference | new | value |
| nor | yours | effect | non | view |
| of | yourself | etc | number | well |
| or | | evidence | other | world |
| our | | example | paper | |

The crowd-sourcing approach guarantees that general language is largely excluded from the concept lists generated so far. Nevertheless, there are still some concepts that are specific to science in general. To remove these, a general science stop word list was handcrafted (see right part of table B.3). The most frequent concept `SOCIAL_NETWORK` that was used to retrieve the original publication set (17,767 out of 23,568 publications use it as a stand-alone concept, the remaining 5,801 publications use it in combination with other words, e.g., `SOCIAL_NETWORK_ANALYSIS`) as well as its parts `SOCIAL` (2,964 stand-alone usages) and `NETWORK` (6,329 stand-alone usages) were removed. The resulting lexical concept set $\mathcal{F}_{\text{wrd}}$ contains 22,754 words (incl. $n$-grams, $n \leq 5$).

To establish edges between these words and publications, the former were sorted descendingly, i.e., 5-grams are linked first and removed from the pre-processed titles, abstracts, and keyword lists, then 4-grams are linked and removed, etc. For example, if a publication uses the concept `ACTOR_NETWORK_THEORY`, this publication and word are linked by a usage edge. An edge to `ACTOR_NETWORK` is only set if this word also occurs not followed by `THEORY`, to `ACTOR` only if it also occurs not followed by `NETWORK_THEORY`, and so on. $(n-1)$-grams embedded in $n$-grams were not linked to publications because $n$ is inversely proportional to word usage, resulting in sparser networks and hopefully clearer-cut subfield boundaries. The resulting network is the bipartite matrix $G_{\text{wrd}}$ of 23,568 publications and 22,754 words, 14,956 of which were used at least twice. Note that these numbers are for the seed described in section 3.1.

# C. Matrices for the Idealtype of Knowledge Production

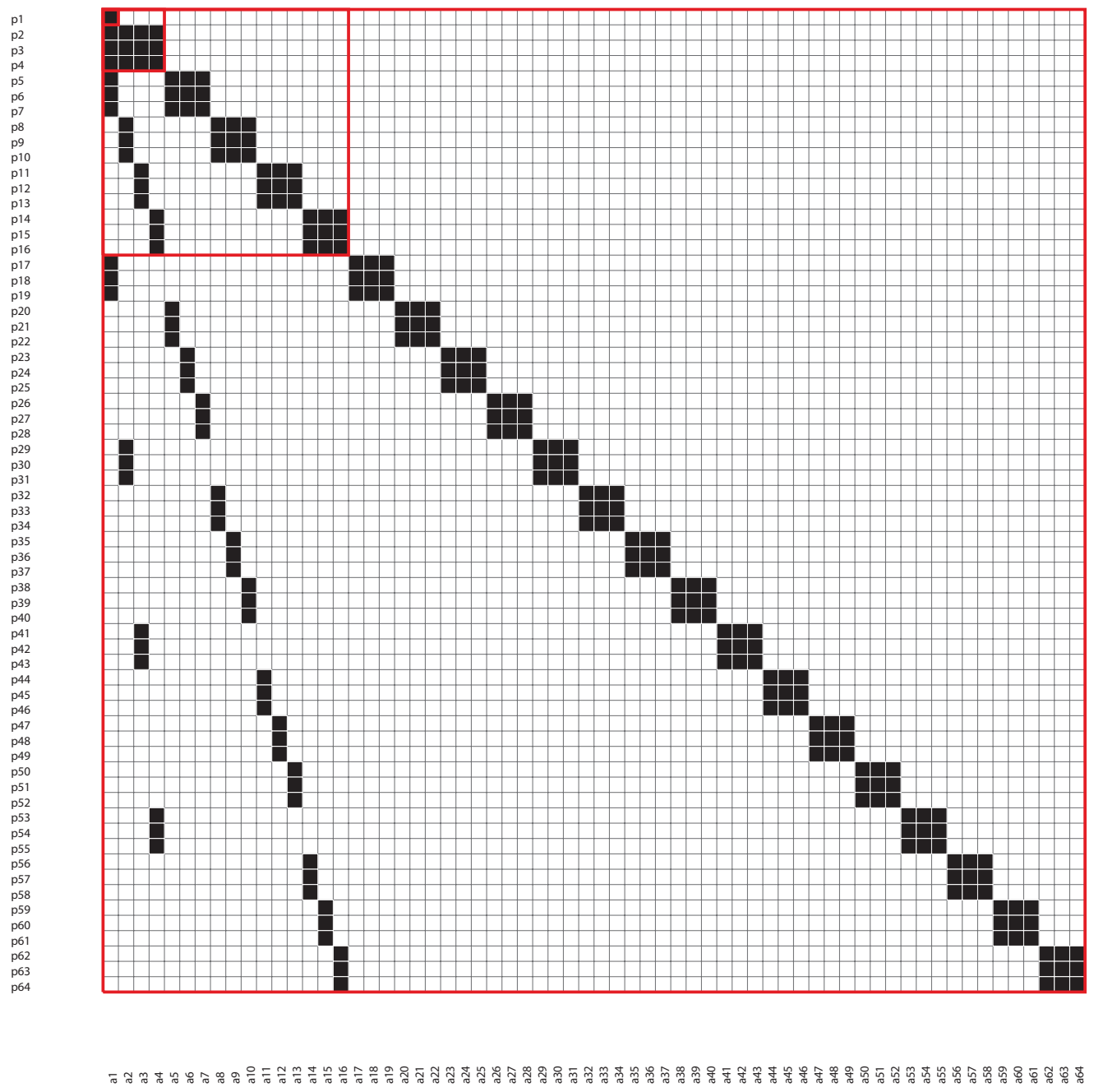

(a) Author selection

**Figure C.1.: Matrices for the idealtype of knowledge production**
Matrices show publications (rows) selecting (authored by/citing/using) social facts (authors/references/words). The knowledge production process represented by these matrices is described in figure 2.2.

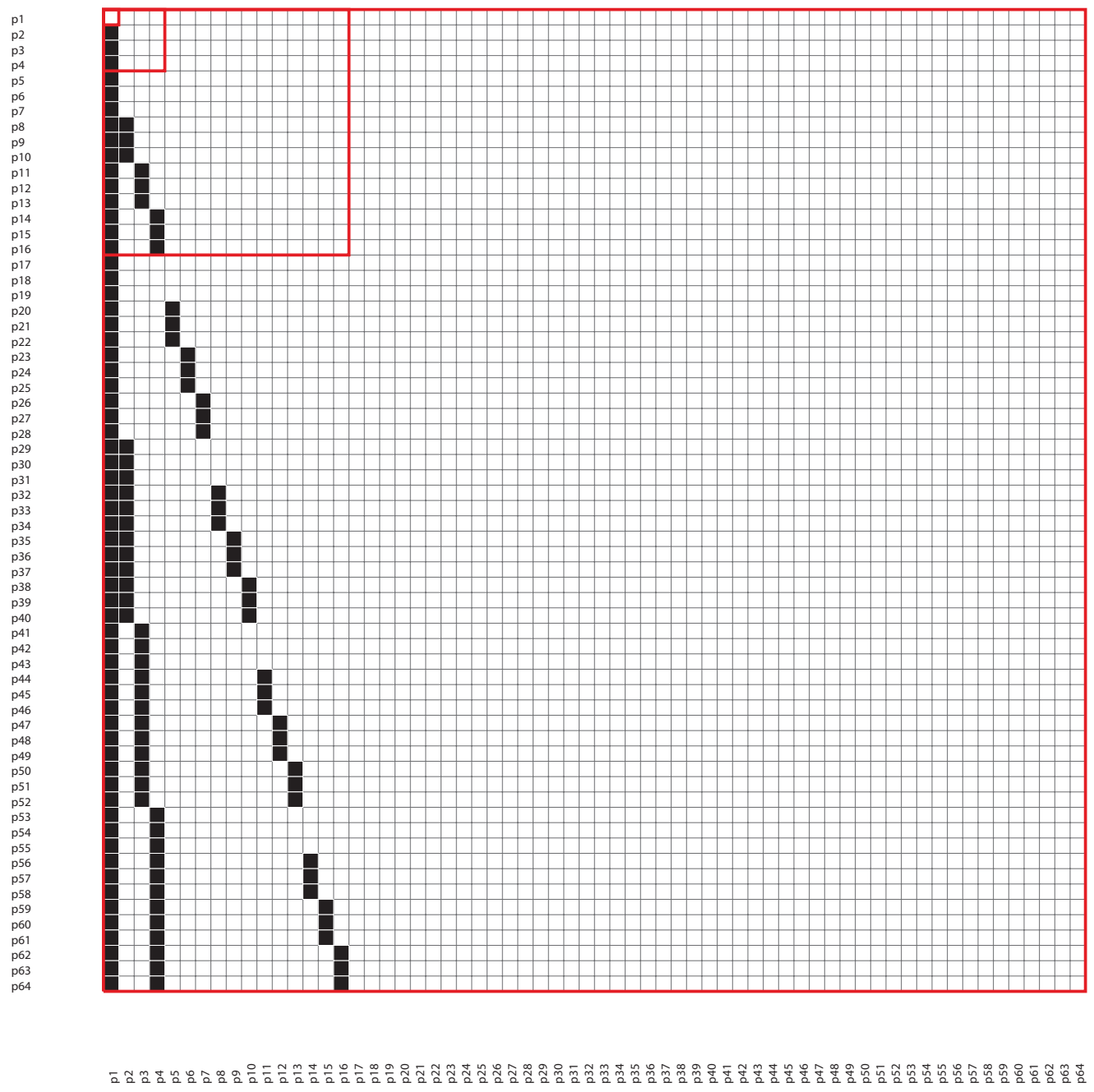

(b) Reference selection

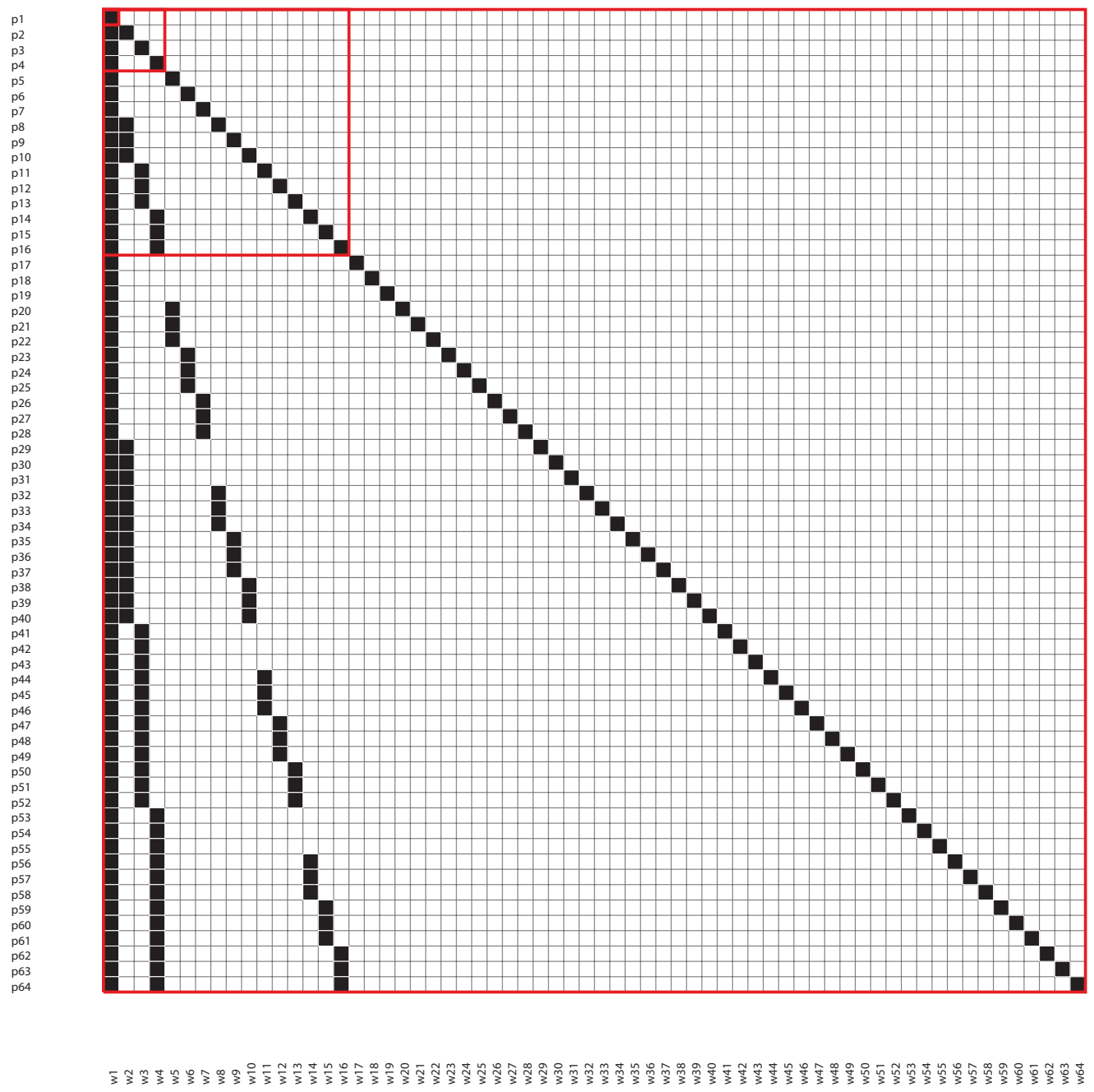

(c) Word selection

**Figure C.1.: Matrices for the idealtype of knowledge production**

# D. Boundary Papers

We define Social Network Science as the research domain that results from the union of Social Network Analysis (Freeman, 2004, 2011; Scott, 2012) and Network Science (Brandes et al., 2013; Barabási, 2015). This domain is a multidisciplinary science of social networks, not a sociological network science. By definition of our delineation procedure, the papers listed in table 3.2 belong to the domain and serve as exemplars for Social Network Science. In the following, problematic papers are listed that exist in the boundary of Social Network Analysis. 15 publications that are ruled inside or outside illustrate our delineation. The criterion to be inside and outside is a relational and metaphorical use of the `NETWORK` concept, respectively.

### Papers Ruled Inside

Adam, F. & Roncevic, B. (2003). Social capital: Recent debates and research trends. *Social Science Information*, *42*(2), 155–183. doi:10.1177/0539018403042002001

> The aim of this article is not only to provide an overview of the state of recent discussion about the concept of social capital, it is also an attempt at critical reflection on theoretical and empirical research efforts. The question is whether the concept of social capital is a fashionable (and short-lived) term proposed as a cure-all for the maladies affecting contemporary communities, organizations and societies as a whole or whether it has more long-term strategic – theoretical as well as applicable – meaning for sociology and other social-science disciplines. Despite the deficiencies of the recent research findings, we argue that the latter is true. The concept represents a very important conceptual innovation which can facilitate the theoretical integration within sociology and the inter- and trans-disciplinary collaboration of sociology and other disciplines, especially economics. The article emphasizes the problems of reception, definition and operationalization, and the developmental role of social capital.

Ruled inside because: A core concept is reviewed and argued to be substantially relational.

---

Baptista, A., Comes, M. C., & Paulin, H. (2012). Session-based dynamic interaction models for stateful web services. *Exploring Services Science*, *103*, 29–43.

> The prevalence of the service paradigm spans diverse domains like commercial, social, technological or scientific. Due to its simplicity and familiar semantics, it provides a powerful general abstraction for system programming,

> interaction, and integration. Several standardisation efforts have further contributed to the popularity of the service concept and its usage, since this provides a uniform access to and aggregation of entities with different characteristics and at different levels of the cyberinfrastructure. The perceived current trend on making everything accessible as a service (XaaS) builds on such service characteristics, and examples range from Web-enabled Wireless Sensor Networks, the Internet of Things and Web of Things, to Cloud computing, and the Internet of Services. Upon the acknowledgement of such high heterogeneity and of the extreme large scale of emerging service systems, Service Science presents a novel and overarching view on analysing and developing further the service paradigm. The high complexity of current and future service systems in this domain, require innovative solutions to be developed in order to improve service productivity and quality. To this extent, this work concentrates on service engineering Web services and proposes a solution based on the Session concept contributing to solve open problems on service system interaction and adaptation. The focus is on the interactions between Web services interfacing stateful resources and its clients, in particular. The session abstraction is used to: a) capture the service/users interaction context, b) support dynamic interaction models within, and c) contextualize on demand and automatic dynamic adaptations. The major goal is to capture Web service/users interactions modelled as Sessions, in order to simplify their re-use and adaptation in the context of the cited Services Sciences' complex systems.

Ruled inside because: Networked socio-technical services are designed, and the paper is not purely about implementation.

---

Cabezas-Clavijo, A., Torres-Salinas, D., & Delgado-Lopez-Cozar, E. (2009). Science 2.0: Tools catalogue and consequences for scientific activity. *Profesional de la Informacion, 18*(1), 72–79. doi:10.3145/epi.2009.ene.10

> The concept of Science 2.0 is introduced and analysed based on its principal characteristics: user participation and collaboration, as well as free information exchange by means of web applications. A categorisation of tools for main web 2.0 functionalities for scientists is detailed: blog networks, journals with 2.0 tools, online reference managers and social tagging, open data and information reutilisation, social networks, and audio and video-science. Main factors influencing the use of these tools are presented. Finally, the consequences for scientific activity of general adoption of these set-vices and applications are discussed.

Ruled inside because: The discussed social service is collaborative.

---

Campbell, D. (2011). Reconsidering the implementation strategy in faith-based policy initiatives. *Nonprofit and Voluntary Sector Quarterly, 40*(1), 130–148. doi:10.1177/-0899764009349847

Despite their hypothesized potential to change lives and promote social policy goals, the track record of faith-based public policies is less than stellar. Using survey, administrative, and organizational case study data from a California faith-based initiative, we find that the commonly used implementation strategy-featuring a top-down, short-term approach that funds isolated local organizations demonstrates little power to nurture enduring programs and partnerships at the community level. Our analysis of why this is so suggests the possible viability of an alternative implementation strategy that takes community social service networks as the unit of action and analysis. A next step for policy and research might be carefully monitored local demonstrations that foster government partnerships with a broad cross-section of community and faith-related organizations within a single locale, seeking to develop a coordinated continuum of community care by patiently building network capacity for the long haul.

Ruled inside because: Monitoring and analysis of social networks are discussed as means to inform policy decisions.

---

Cheshire, L., Wickes, R., & White, G. (2013). New suburbs in the making? Locating master planned estates in a comparative analysis of suburbs in South-East Queensland. *Urban Policy and Research*, *31*(3), 281–299. doi:10.1080/08111146.2013.787357

In urban scholarship Master Planned Estates (MPEs) are viewed as illustrative of broader changes to the urban environment and characterised as homogenous, affluent enclaves where community life is largely orchestrated by the developer. Yet no study has fully considered if, and to what extent, MPEs can be distinguished from other suburb types in terms of their residential composition and their levels of sociability and community attachment. In this article, we empirically test if MPEs are different from conventional' suburbs by examining them structurally in terms of their demographic and socio-economic characteristics, as well as in terms of their key community social processes. Using data from a 2008 study of 148 suburbs across Brisbane, Australia (which includes data from two MPEs), we undertake a comparative analysis of suburbs and examine the density of neighbour networks, residents' reports of place attachment and cohesion and neighbourly contact in MPEs compared to other residential suburbs. Our findings suggest that MPEs are not distinct in terms of their degree of homogeneity and socio-economic characteristics, but that connections among residents are lower than other suburbs despite or perhaps because of the active interventions of the developer.

Ruled inside because: Social Network Analysis is employed in urban planning.

---

Huang, L. (2009). A P2P framework of virtual society based on virtual hierarchical tree grid organizations (VIRGO). *2009 International Conference on New Trends in Information and Service Sciences (NISS 2009)*, *1–2*, 1393–1396. doi:10.1109/NISS.2009.189

This paper presents a P2P framework for Virtual Society based on virtual hierarchical tree Grid organizations (VIRGO). The virtual social persons, which have human's actions and responses modelled by computer algorithms, join VIRGO network to construct virtual society, which simulates real society. The nodes hosting virtual social persons construct n-tuple overlay virtual hierarchical overlay network according to their joining to virtual organizations. Because of the cached addresses of nodes, the overload of traffic in tree structure can be avoided. Every virtual social person can find each other effectively. We can use this framework to simulate social phenomena. We also give out a primary design of virtual society simulation.

Ruled inside because: A virtual social system is simulated using networks.

---

Klein, R. (2004). Sickening relationships: Gender-based violence, women's health, and the role of informal third parties. *Journal of Social and Personal Relationships*, *21*(1), 149–165. doi:10.1177/0265407504039842

Over the last two decades, two trends have emerged that highlight the relational and communal dimensions of health and well-being. The Ottawa Charter of 1986 proposed a model of health that encompasses personal, social, and environmental well-being and emphasizes the role of communities in creating healthy living conditions. In a parallel development, battered women's advocates identified violence against women as a major health issue, and situated effective interventions within coordinated community response strategies. Both trends point to the significance of social networks and a wide range of social and personal relationships in the promotion of women's health and well-being.

Ruled inside because: The role of social relations in individual well-being is discussed.

---

Longan, M. W. (2005). Visions of community and mobility: The community networking movement in the USA. *Social & Cultural Geography*, *6*(6), 849–864. doi:10.1080/-14649360500353178

Different visions of community and mobility may influence the ability of community computer network organizations to promote social change. Community networks successfully generate instrumental mobility, the literal movement of information across space, but have difficulty creating successful online spaces that promote communicative mobility, the metaphorical movement of people towards common understandings of a shared situation. Interviews with community networking activists explore the ways that community networks generate instrumental mobility online as well as barriers that community networks face in creating online spaces for communicative mobility. Ironically, given their technological focus, community networks have little difficulty generating communicative mobility in face-to-face situations. Differentiating between instrumental and communicative mobility allows this research to move beyond a simple discussion of the geography of the conduits of communication to consider the geography of communication itself. It therefore contributes a more detailed understanding of the role of electronic communication in social and political change.

Ruled inside because: Socio-cultural consensus and change are discussed from the perspective of online social networks.

---

Mayer, A. (2009). Online social networks in economics. *Decision Support Systems*, *47*(3), 169–184. doi:10.1016/j.dss.2009.02.009

> This paper describes how economists study social networks. While economists borrow from other fields like sociology or computer science, their approach of modeling of social networks is distinguished by the emphasis on the role of choices under constraints. Economists investigate how socioeconomic background and economic incentives affect the structure and composition of social networks. The characteristics of social networks are important for economic outcomes like the matching of workers to jobs and educational attainment. I review the theoretical and empirical literature that investigates these relationships and discuss possible implications of new, Internet based, forms of social interactions.

Ruled inside because: Social Network Analysis in economics is described.

---

Neufeld, A. & Kushner, K. E. (2009). Men family caregivers' experience of nonsupportive interactions context and expectations. *Journal of Family Nursing*, *15*(2), 171–197. doi:10.1177/1074840709331643

> Men's involvement as family caregivers has grown as the prevalence of dementia has increased. Men rely on support from others for caregiving but also experience nonsupportive interactions. The purpose of this ethnographic study of 34 men (24 spouses and 10 sons) caring for a relative with dementia, 5 assisting caregivers, and 15 professionals was to identify primary caregivers' perceptions of nonsupportive and supportive interactions in relationships with kin and friends as well as professionals. Thematic analysis of transcribed data generated from interviews, diaries, and focus group discussions revealed the nature of men's caregiving journeys, the characteristics of their social networks, and their expectations of supportive interactions. The nonsupportive interactions men caregivers experienced included a lack of orientation to the caregiving situation, an unsatisfactory linkage to support sources, insufficient support, and hurtful interactions. Information about nonsupportive interactions can sensitize kin and friends as well as professionals to the complexity of men's experience and potentially avoid unintended negative consequences of support efforts.

Ruled inside because: Social support is studied through the ethnographic observation of relationships.

---

Park, J. & Van Der Schaar, M. (2010). A game theoretic analysis of incentives in content production and sharing over peer-to-peer networks. *IEEE Journal of Selected Topics in Signal Processing*, *4*(4), 704–717. doi:10.1109/JSTSP.2010.2048609

> Peer-to-peer (P2P) networks can be easily deployed to distribute user-generated content at a low cost, but the free-rider problem hinders the efficient utilization of P2P networks. Using game theory, we investigate incentive schemes to overcome the free-rider problem in content production and sharing. We build a basic model and obtain two benchmark outcomes: 1) the non-cooperative outcome without any incentive scheme and 2) the cooperative outcome. We then propose and examine three incentive schemes based on pricing, reciprocation, and intervention. We also study a brute-force scheme that enforces full sharing of produced content. We find that 1) cooperative peers share all produced content while non-cooperative peers do not share at all without an incentive scheme; 2) by utilizing the P2P network efficiently, the cooperative outcome achieves higher social welfare than the non-cooperative outcome does; 3) a cooperative outcome can be achieved among non-cooperative peers by introducing an incentive scheme based on pricing, reciprocation, or intervention; and 4) enforced full sharing has ambiguous welfare effects on peers. In addition to describing the solutions of different formulations, we discuss enforcement and informational requirements to implement each solution, aiming to offer a guideline for protocol design for P2P networks.

Ruled inside because: Emergent effects of collaboration are simulated.

---

Seppa, P., Johansson, H., Gyllenstrand, N., Palsson, S., & Pamilo, P. (2012). Mosaic structure of native ant supercolonies. *Molecular Ecology*, *21*(23), 5880–5891. doi:10.1111/-mec.12070

> According to the inclusive fitness theory, some degree of positive relatedness is required for the evolution and maintenance of altruism. However, ant colonies are sometimes large interconnected networks of nests, which are genetically homogenous entities, causing a putative problem for the theory. We studied spatial structure and genetic relatedness in two supercolonies of the ant Formica exsecta, using nuclear and mitochondrial markers. We show that there may be multiple pathways to supercolonial social organization leading to different spatial genetic structures. One supercolony formed a genetically homogenous population dominated by a single mtDNA haplotype, as expected if founded by a small number of colonizers, followed by nest propagation by budding and domination of the habitat patch. The other supercolony had several haplotypes, and the spatial genetic structure was a mosaic of nuclear and mitochondrial clusters. Genetic diversity probably originated from long-range dispersal, and the mosaic population structure is likely a result of stochastic short-range dispersal of individuals. Such a mosaic spatial structure is apparently discordant with the current knowledge about the integrity of ant colonies. Relatedness was low in both populations when estimated among nestmates, but increased significantly when estimated among individuals sharing the same genetic cluster or haplogroup. The latter association indicates the important historical role of queen dispersal in the determination of the spatial genetic structure.

Ruled inside because: The analysis of animal social networks is relevant for evolutionary network theory.

---

Trewern, A. & Lai, K. W. (1999). Developing an interactive learning environment for teachers in the New Zealand context. *Advanced Research in Computers and Communication in Education, Vol 1: New Human Abilities for the Networked Society*, *55*, 950–953.

Online interactive learning environments are increasingly being used by teachers and educators to accessing and sharing professional resources and for communicative purposes. In this paper the design principles used in setting up an online professional learning network, called the New Zealand Learning Network, for New Zealand primary and secondary school Leachers are documented. As well, some preliminary observations of the success factors involved in constructing such a network are also briefly described in this paper.

Ruled inside because: An engineered social system is evaluated through observation.

---

Van Tilburg, T. (1998). Losing and gaining in old age: Changes in personal network size and social support in a four-year longitudinal study. *Journals of Gerontology Series B – Psychological Sciences and Social Sciences*, *53*(6).

Objectives. Previous studies have shown that most older people have a significant number of relationships. However, the question of whether the aging of old people produces losses in their personal networks remains open for discussion. This study models the individual variability of the changes affecting multiple personal network characteristics. Methods. Personal interviews were conducted with 2,903 older Dutch adults (aged 55–85) in three waves of a four-year longitudinal study. Results. A stable total network size was observed, with an increasing number of close relatives and a decreasing number of friends. Contact frequency decreased in relationships, and the instrumental support received and emotional support given increased. Age moderated

> the effect of time for some of the network characteristics, and for many of them, effects of regression towards the mean were detected. Furthermore, major variations in the direction and the speed of the changes were detected among individual respondents, and nonlinear trends were observed. Discussion. The widely varying patterns of losses and gains among the respondents squares with the focus on the heterogeneity of developments among aging people. The instability of the network composition might reflect the natural circulation in the membership of networks.

Ruled inside because: The social support study is truly structural.

---

Wright, K. B. & Bell, S. B. (2003). Health-related support groups on the Internet: Linking empirical findings to social support and computer-mediated communication theory. Journal of Health Psychology, 8 (1), 39–54. doi:10.1177/1359105303008001429

> This literature review of research on health-related computer-mediated support groups links features of these groups to existing theory from the areas of social support and computer-mediated communication research. The article exams computer-mediated support groups as weak tie networks, focuses on how these support groups facilitate participant similarity and empathic support and identifies changes in supportive communication due to characteristics of the medium.

Ruled inside because: Network structural methods are employed.

#### Papers Ruled Outside

Brown, J. P. (2011). “Touch in transit”: Manifestation/manifestacion in Cecilia Vicuna’s cloud-net. *Contemporary Women’s Writing*, *5*(3), 208–231. doi:10.1093/cww/vpr012

> With poems typeset in string-like arrangements, Cecilia Vicuna’s cloud-net connects the poetic line with large-scale gallery installations of white woolen skeins. Together the poems and installations weave a textual and material web, incorporating not only different media, but also languages, feminine and domestic craft, history and fate, the World Wide Web, social networking, globalization, memory, and a mindfulness for the present. Vicuna’s merging of verse, lyric, song, dance, sculpture, performance, and installation, comes in response to a world overwhelmed by the effects of globalization and ecological deterioration. As Vicuna and her readers touch and move their bodies among cloud-net’s webs, they open a shared transitive space, reclaiming the original energy of universal genesis, and its potential to heal the destruction wrought since then.

Ruled outside because: This is about art.

---

Burgers, J. & Musterd, S. (2002). Understanding urban inequality: A model based on existing theories and an empirical illustration. *International Journal of Urban and Regional Research*, *26*(2), 403. doi:10.1111/1468-2427.00387

> In the debate on urban inequality, Sassen's theory on social polarization and Wilson's theory on spatial mismatch have received much attention. Where Sassen highlights the decline of the middle classes, Wilson focuses on the upgrading of urban labour markets. In this article we argue that both theories may be valid, but that they have to be put in a more extended theoretical framework. Of central importance are national institutional arrangements, membership of different ethnic groups and networks, and place-specific characteristics rooted in local socio-economic histories. As a first empirical illustration of our model, we use data on the labour markets of Amsterdam and Rotterdam and show that different forms of inequality can be found both in economic sectors and within ethnic groups. The model we present could be used both to reinterpret existing data and as an analytical framework for the analysis of different forms of urban inequality.

Ruled outside because: No indication that networks are used analytically.

---

Dalsgaard, C. & Paulsen, M. F. (2009). Transparency in cooperative online education. *International Review of Research in Open and Distance Learning*, *10*(3).

> The purpose of this article is to discuss the following question: What is the potential of social networking within cooperative online education? Social networking does not necessarily involve communication, dialogue, or collaboration. Instead, the authors argue that transparency is a unique feature of social networking services. Transparency gives students insight into each other's actions. Cooperative learning seeks to develop virtual learning environments that allow students to have optimal individual freedom within online learning communities. This article demonstrates how cooperative learning can be supported by transparency. To illustrate this with current examples, the article presents NKI Distance Education's surveys and experiences with cooperative learning. The article discusses by which means social networking and transparency may be utilized within cooperative online education. In conclusion, the article argues that the pedagogical potential of social networking lies within transparency and the ability to create awareness among students.

Ruled outside because: Social networking only is an object of study.

---

Del Valle Barrera, M., Koch, T., & Aguirre, B. E. (2013). Commemorating Chile's coup: The dynamics of collective behavior. *Latin American Politics and Society*, *55*(2), 106–132. doi:10.1111/j.1548-2456.2013.00195.x

> This article examines the dynamics of collective behavior in Santiago, Chile every September 11, the date of the 1973 coup that brought General Augusto Pinochet to power. It uses a multiple-method strategy that includes participant observation, personal interviews, and content analysis of three major newspapers during the period 2003–8. The theoretical approach emphasizes time and space coordinates of specified social actors, sociocultural emergence, a limited range of dominant emotions, and dramaturgy to describe the complexity of ritualized commemorations. It shows that incidents occurring on this date are not primarily caused by the actions of social movement organizations. Moreover, the dichotomy of "day and night" used to understand the peaceful and violent commemorations is an oversimplification of a complex network of events, actors, and scenarios that has the effect of denying any legitimacy to actions that fall outside the state-approved practices.

Ruled outside because: "Complex network" is used to tell a story.

---

Harrison, M. (2011). Forging success: Soviet managers and accounting fraud, 1943–1962. *Journal of Comparative Economics*, *39*(1), 43–64. doi:10.1016/j.jce.2010.12.002

> Attempting to satisfy their political masters in a target-driven culture, Soviet managers had to optimize on many margins simultaneously. One of these was the margin of truthfulness. False accounting for the value of production was apparently widespread in some branches of the economy and at some periods of time. A feature of accounting fraud was that cases commonly involved the aggravating element of conspiracy. The paper provides new evidence on the nature and extent of accounting fraud: the scale and optimal size of conspiratorial networks: the authorities' willingness to penalize it and the political and social factors that secured leniency: and inefficiency in the socialist market where managers competed for political credit.

Ruled outside because: No indication that networks are used analytically.

---

Holzhaider, J. C., Sibley, M. D., Taylor, A. H., Singh, P. J., Gray, R. D., & Hunt, G. R. (2011). The social structure of New Caledonian crows. *Animal Behaviour*, *81*(1), 83–92. doi:10.1016/j.anbehav.2010.09.015

> New Caledonian (NC) crows, Corvus moneduloides, have impressive tool-manufacturing and tool-using skills in the wild, and captive birds have displayed exceptional cognitive abilities in experimental situations. However, their social system is largely unknown. In this study we investigated whether the social structure of NC crows might have had a role in the development of their cognitive skills. We observed crows in their natural habitat on the island of Mare, New Caledonia, and estimated their social network size based on tolerance to family and nonfamily crows at feeding tables. Our findings suggest that NC crows are not a highly social corvid species. Their core unit was the immediate family consisting of a pair and juveniles from up to two consecutive breeding years. Pairs stayed together year round, and were closely accompanied by juveniles during their first year of life. Parents were highly tolerant of juveniles and sometimes continued to feed them well into their second year. NC crows predominantly shared feeding tables with immediate family. Of the nonfamily crows tolerated, juveniles were overrepresented. The main mechanism for any social transmission of foraging skills is likely to be vertical (from parents to offspring), with only limited opportunity for horizontal transmission. The social organization we found on Mare is consistent with the idea that NC crows' multiple pandanus tool designs on mainland Grande Terre are an example of cumulative technological evolution.

Ruled outside because: Relevance for socio-cultural systems is not stated.

---

Horgan, A. & Sweeney, John. (2012). University students' online habits and their use of the internet for health information. *Computers Informatics Nursing*, *30*(8), 402–408. doi:10.1097/NXN.0b013e3182510703

Studies have explored the use of the Internet for health information, but few have focused on the young adult population, a population that is known to have difficulties in accessing mainstream health services. It has been acknowledged that young people are active users of the Internet, and this mode of health service delivery warrants further exploration. This study aimed to determine university students' online habits and their use of the Internet for health information using a quantitative descriptive design. Data were collected from 922 university students in Ireland, aged between 18 and 24 years. The findings indicated that university students are active users of the Internet and of social networking sites, particularly for communication purposes. It was also found that 66.1% of participants had used the Internet to search for health information, for a variety of reasons, including information on specific illnesses, sexual health, and fitness and nutrition. It is concluded that the use of the Internet to communicate with young people in relation to their health needs to be explored.

Ruled outside because: Social-networking-service usage is observed, not the traces left by usage (which the authors recommend to do).

---

Jimenez, R., Magrinya, F., & Almandoz, J. (2010). The role of urban water distribution networks in the process of sustainable urbanisation in developing countries case study: Wukro water supply, Wukro Town, Ethiopia. *Proceedings of the First International Conference on Sustainable Urbanization* (ICSU 2010), 1262–1267.

The development of urban distribution networks in the South is often carried out by agents of the North and analyzed under the financial and operational indicators used also in the North. Using the results of a case study in a small urban centre in Ethiopia, we propose a different evaluation model to measure the efficiency of such implementations across the concepts of Sustainable Urbanization. GIS maps are used to analyze the evolution of urban basic services, the dynamics of population movements and economic changes. With those results we can address the social dimension of inequality and exclusion that should be considered in the design and planning of water networks. The aim is to reduce the risk of slums formation and to promote inclusive growth in services.

Ruled outside because: Paper is about social dimension of infrastructure networks.

---

Kouijzer, M. E. J., De Moor, J. M. H., Gerrits, B. J. L., Buitelaar, J. K., & Van Schie, H. T. (2009). Long-term effects of neurofeedback treatment in autism. *Research in Autism Spectrum Disorders*, *3*(2), 496–501. doi:10.1016/j.rasd.2008.10.003

Previously we demonstrated significant improvement of executive functions and social behavior in children with autism spectrum disorders (ASD) treated with 40 sessions of EEG neurofeedback in a nonrandomized waiting list control group design. In this paper we extend these findings by reporting the long-term results of neurofeedback treatment in the same group of children with ASD after 12 months. The present study indicates maintenance of improvement of executive functions and social behavior after 12 months in comparison with the immediate outcomes. Neurofeedback mediated suppression of theta power is supposed to promote more flexible functioning of the brain by enhancing activation in the medial prefrontal cortex and improving flexibility of activation in the default mode network supporting the improvement of executive functions and theory of mind in ASD.

Ruled outside because: Paper is not about the social dimension of autism.

---

Laroche, M., Habibi, M. R., Richard, M.-O., & Sakaranarayanan, R. (2012). The effects of social media based brand communities on brand community markers, value creation practices, brand trust and brand loyalty. *Computers in Human Behavior*, *28*(5), 1755–1767. doi:10.1016/j.chb.2012.04.016

Social media based brand communities are communities initiated on the platform of social media. In this article, we explore whether brand communities based on social media (a special type of online brand communities) have positive effects on the main community elements and value creation practices in the communities as well as on brand trust and brand loyalty. A survey based empirical study with 441 respondents was conducted. The results of structural equation modeling show that brand communities established on social media have positive effects on community markers (i.e., shared consciousness, shared rituals and traditions, and obligations to society), which have positive effects on value creation practices (i.e., social networking, community engagement, impressions management, and brand use). Such communities could enhance brand loyalty through brand use and impression management practices. We show that brand trust has a full mediating role in converting value creation practices into brand loyalty. Implications for practice and future research opportunities are discussed.

Ruled outside because: A non-relational approach is taken where interview data is used in structural equation modeling.

---

McCosker, A. & Darcy, R. (2013). Living with cancer: Affective labour, self-expression and the utility of blogs. *Information Communication & Society*, *16*(8), 1266–1285. doi:10.1080/1369118X.2012.758303

Blogs have been used extensively to self-document the intimate and often intensive experiences of living with serious illness, charting their author's health and treatment often over many years, connecting with others and drawing attention and concern along the way. This paper analyses a selection of typical cancer blogs with the aim of understanding the kinds of personal investment, or labour, involved in the process of forming and maintaining them over a sustained period. Where previous research has investigated what is often seen as the empowering role of these blogs, we attempt to qualify these claims. Our qualitative case study analysis draws on and expands the theory and debates around the nature of immaterial and affective labour. We highlight the value of cancer blogging as personal, in the form of identity and affect management, network-enabling in generating online spaces for shared experience and support, and social in what is recouped in the forms of non-institutional management of serious illness. In addition, this labour helps to shape the broader social understanding of cancer, its experience and personal affects.

Ruled outside because: Blogs are found to facilitate networking; no relational analysis.

---

Melchiorre, M. G. and Chiatti, C., Lamura, G., Torres-Gonzales, F., Stankunas, M., Lindert, J., Ioannidi-Kapolou, E., Barros, H., Macassa, G., & Soares, J. F. J. (2013). Social support, socio-economic status, health and abuse among older people in seven European countries. *PLoS ONE*, *8*(1). doi:10.1371/journal.pone.0054856

Background: Social support has a strong impact on individuals, not least on older individuals with health problems. A lack of support network and poor family or social relations may be crucial in later life, and represent risk factors for elder abuse. This study focused on the associations between social support, demographics/socio-economics, health variables and elder mistreatment. Methods: The cross-sectional data was collected by means of interviews or interviews/self-response during January-July 2009, among a sample of 4,467 not demented individuals aged 60-84 years living in seven European countries (Germany, Greece, Italy, Lithuania, Portugal, Spain, and Sweden). Results: Multivariate analyses showed that women and persons living in large households and with a spouse/partner or other persons were more likely to experience high levels of social support. Moreover, frequent use of health care services and low scores on depression or discomfort due to physical complaints were indicators of high social support. Low levels of social support were related to older age and abuse, particularly psychological abuse. Conclusions: High levels of social support may represent a protective factor in reducing both the vulnerability of older people and risk of elder mistreatment. On the basis of these results, policy makers, clinicians and researchers could act by developing intervention programmes that facilitate friendships and social activities in old age.

Ruled outside because: Data collection is reductionist.

---

Nan, L. C., Xing, L., & Li, L. (2006). Performance measurement for investment decision-making based on wavelet network model. *Wavelet Active Media Technology and Information Processing, 1–2*, 972–978.

Follow the trend of the combination of financial and non-financial measures of strategic performance measurement system. A performance measurement index system for investment decision-making was designed which consisted of six-measured, including financial condition, theological innovation, customers, business process, latent development capability, and social and environment protection. This index system is comprehensive flirtation combined by Benchmarking and Principal Component analysis. Wavelet network being better than neural network for solving nonlinear problems overcoming shortcomings of neural network's constringency with slow velocity and easily entering into local minimum, the performance evaluation model for investment decision-making based on wavelet network was devised. An empirical research of this model was done, based on the date from 10 projects in cooking industry, in 2000. The results showed the wavelet network training up to 7933 times with convergence precision of 0.001, taking 3.31 seconds.

Ruled outside because: Relations only connect variables in a neural network model.

---

Walker, R. K. (2012). The right to be forgotten. *Hastings Law Journal, 64*(1), 257–286.

Information posted to the Internet is never truly forgotten. While permanently available data offers significant social benefits, it also carries substantial risks to a data subject if personal information is used out of context or in ways that are harmful to the subject's reputation. The potential for harm is especially dire when personal information is disclosed without a subject's consent. In response to these risks, European policymakers have proposed legislation recognizing a "right to be forgotten." This right would provide persons in European Union countries with a legal mechanism to compel the removal of their personal data from online databases. However, only a limited form of the right to be forgotten a right to delete data that a user has personally submitted would be compatible with U.S. constitutional law. By itself this limited right is insufficient to address the myriad privacy issues raised by networked technologies, but it is nevertheless an essential component of a properly balanced regulatory portfolio as existing privacy tort law is inadequate in this context. As such, this Note argues that Congress should recognize this limited right through adoption of a default contract rule where an implied covenant to delete user-submitted data upon request is read into website terms of service contracts.

Ruled outside because: Paper is about a legal issue.

---

Zea, M. C., Reisen, C. A., Poppen, P. J., Bianchi, F. T., & Echeverry, J. J. (2005). Disclosure of HIV status and psychological well-being among Latino gay and bisexual men. *AIDS and Behavior*, *9*(1), 15–26. doi:10.1007/s10461-005-1678-z

This study examined disclosure of HIV-positive serostatus by 301 Latino gay and bisexual men to members of their social networks and the mental health consequences of such disclosure. The sample was recruited from clinics, hospitals, and community agencies in New York City, Washington, DC, and Boston. Proportions disclosing differed depending on the target, with 85% having disclosed to closest friend, 78% to male main partner, 37% to mother, and 23% to father. Although there were differences depending on the target, disclosure was related to greater quality of social support, greater self-esteem, and lower levels of depression. Moreover, findings indicated that social support mediated the relationship between disclosure of serostatus and both self-esteem and depression. Thus, disclosure resulted in greater social support, which in turn had positive effects on psychological well-being. Findings demonstrate that generally Latino gay men are selective in choosing people to whom they disclose their serostatus and that disclosure tends to be associated with positive outcomes.

Ruled outside because: Paper takes a variable-centered, not relational, approach.

# E. Fitness of Social Facts

The fitness $\eta$ of a social fact is its ability to influence an identity. It is computed annually as the deviation from the scaling law describing the Matthew Effect, described in figure 3.20. For the plots here, fitness scores are transformed by $\eta^{\mathrm{T}} = (\eta - 1)/(\eta + 1)$, i.e., a fact with a transformed fitness of 0.3 (0, -0.3) is about 1.9 (1, 0.54) times as fit as an average fact with the same normalized number of selections $\Delta t$ years before $t$. All curves are smoothed by a 5-year moving average. Shaded regions depict 95% confidence intervals for estimates of preferential-attachment exponents. Figure E.1 is for cited references. Word usage shown in figure E.2 is less dynamic than citation.

The following is a how-to-read guide for figure E.1. (a) Fitness of individual paradigmatic references shown in figure 3.9f in Social Network Science. Axis labels correspond to those in (b). For example, `BARABASI_1999_S_509` is first attributed a fitness in 2003, four years after publication, because it had reached an initial attractiveness in 2001 and two years are removed due to smoothing. The paper is fitter than expected in each year since publication. Its error bar is shaded gray because the fact is paradigmatic (belongs to the micro-level set) in multiple subdomains. If a fact is not multifunctional, it is given the color of the subdomain it is paradigmatic for.

(b–d) Average fitness of references in Social Network Science conditional on subdomain levels. Levels are defined in figure 3.8. At micro levels, facts are paradigms. Curves can all be below or above expectation because the fitness of facts is averaged in time but facts are selected following integration over time. For example, the purple curve in (b) shows that, in Social Network Science, the average fitness of references cited by Complexity Science on the macro level was below expectation since 1980 and above expectation from 1998–2005. (c) The meso-level references of Complexity Science exhibit a circadian fitness pattern. At the peaks, in 1977, 1989, and 2001, they had the largest influence on the citation practice of the whole Social Network Science domain. Colored ranges indicate where results are biased towards old references. References of Complexity Science conditional on the meso level have an average lifetime of 11 years (cf. figure 3.8b). Since references considered here exceed this mean by definition, no reference at these levels is younger than 11 years. Only Web Science has an shorter conditional lifetime. This effect explains why most curves converge to zero. (d) From 1999–2005, the paradigm of Complexity Science was most influential in Social Network Science. Subfigures (e–g) depict the average fitness of references in Social Psychology conditional on subdomain levels, subfigures (h–j) the average fitness of references in Economic Sociology, and so on. For example, (p) in Complexity Science, its own paradigm was the fittest of all subdomains' paradigms from the Complexity Turn until 2005, when Web Science took over.

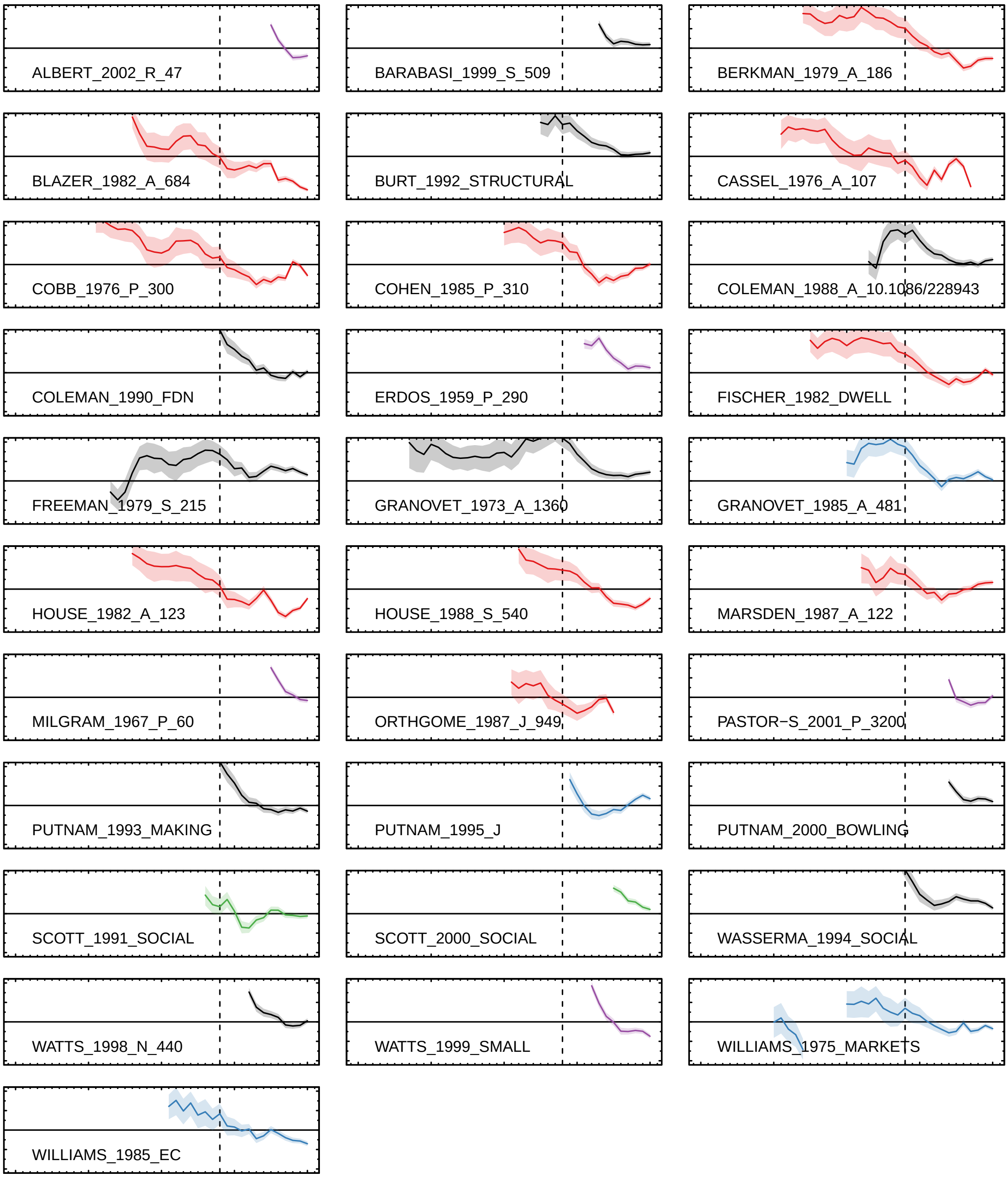

(a) Micro level citation in the whole domain

**Figure E.1.: Fitness of cited references**

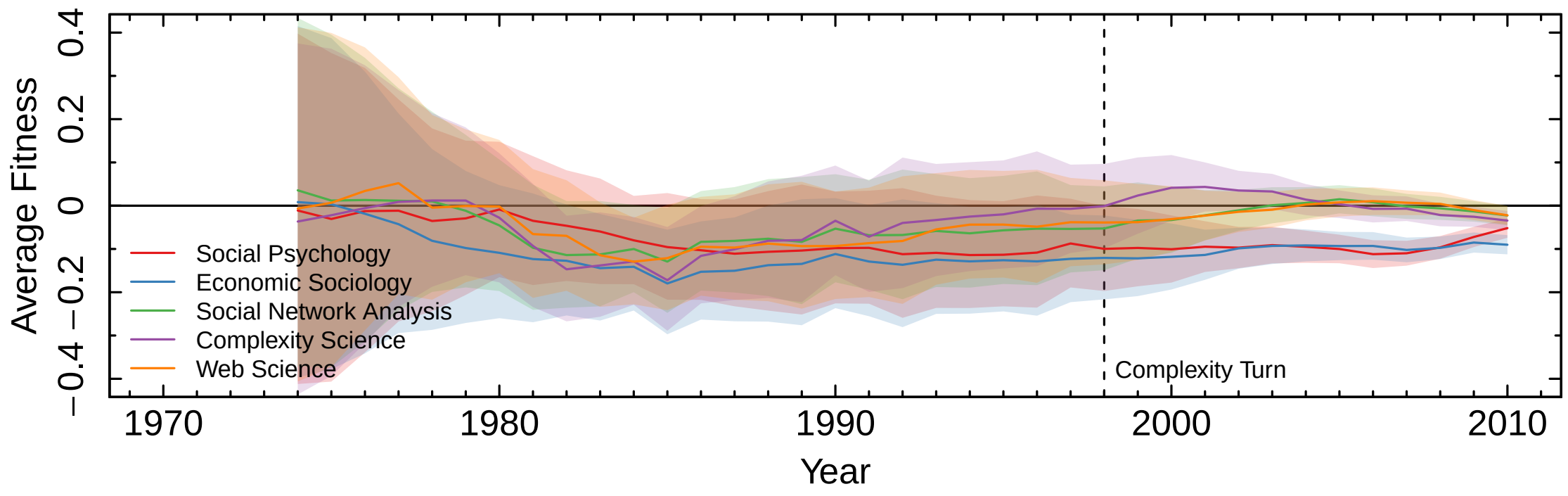

(b) Macro level citation in the whole domain

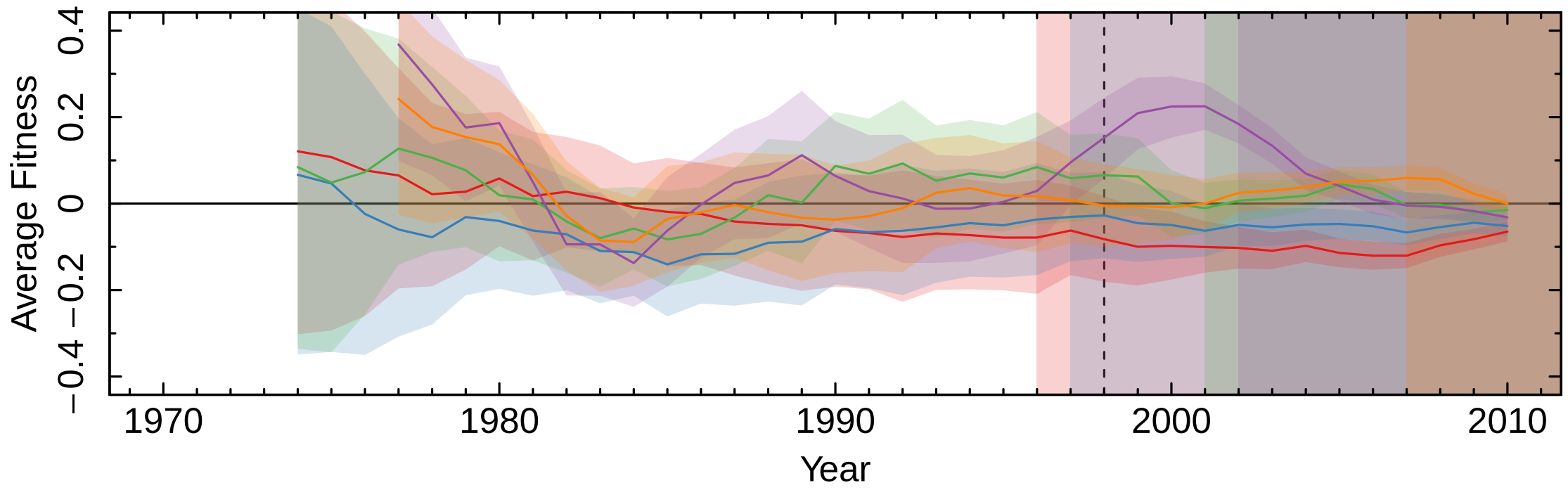

(c) Meso level citation in the whole domain

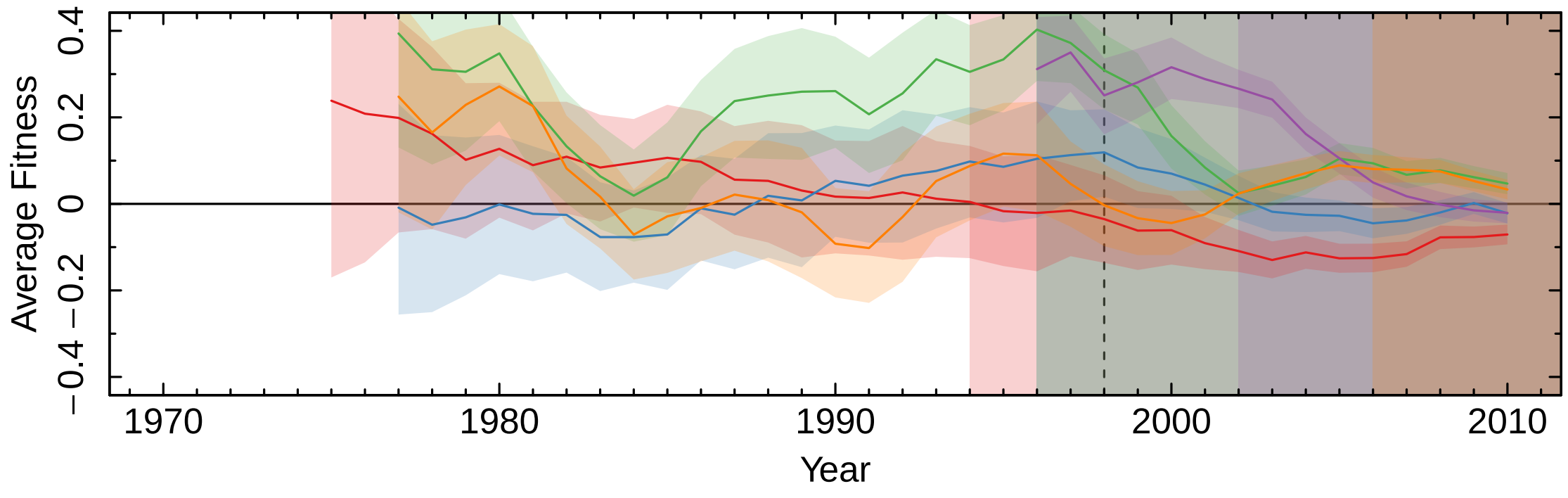

(d) Micro level citation in the whole domain

**Figure E.1.: Fitness of cited references**

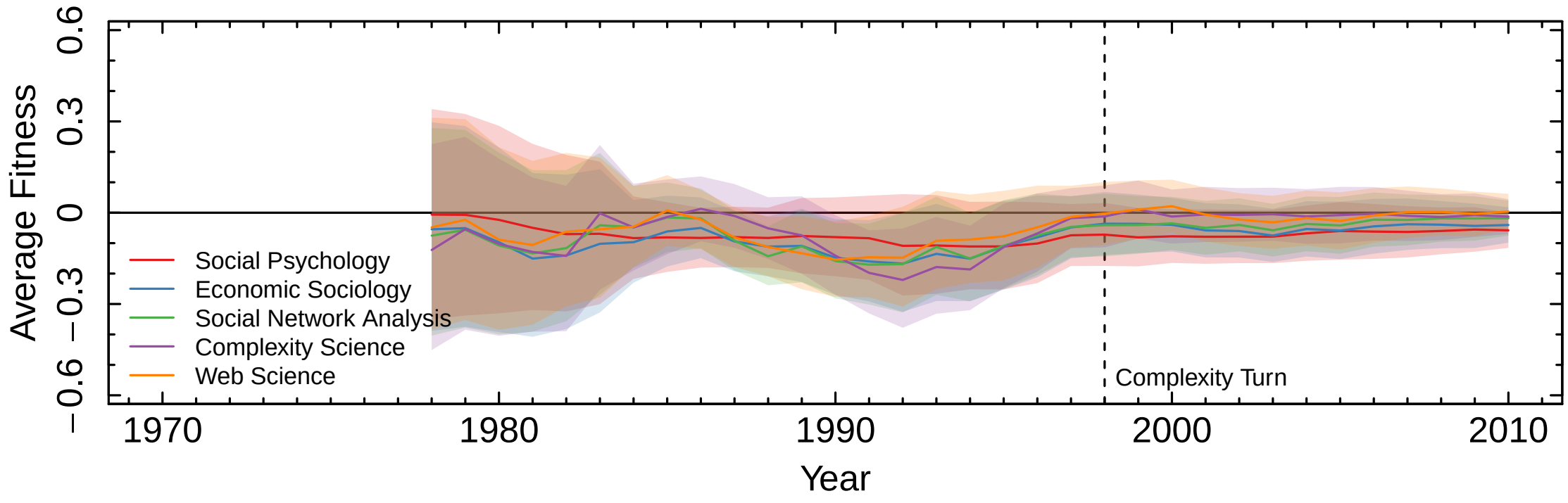

(e) Macro level citation in Social Psychology

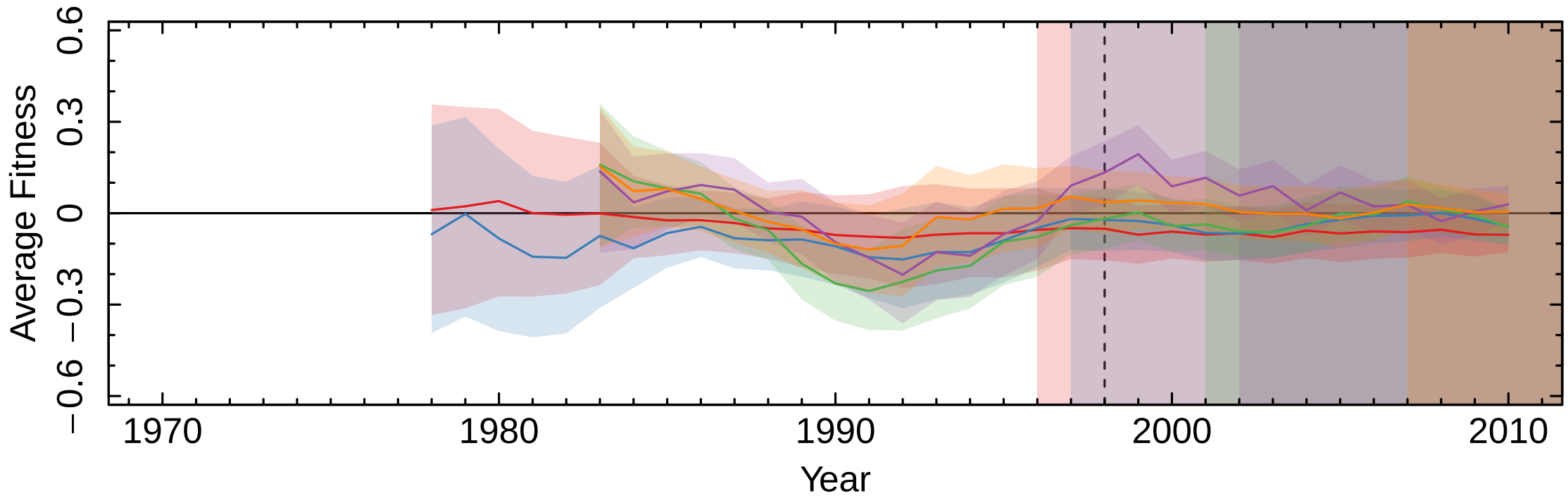

(f) Meso level citation in Social Psychology

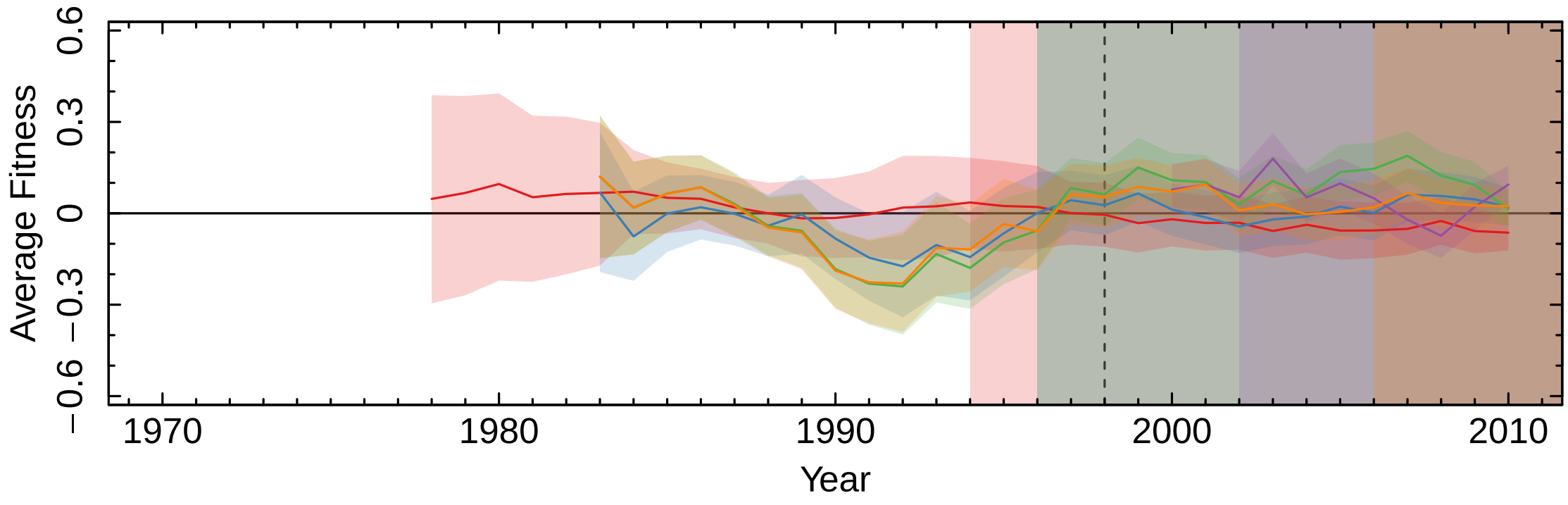

(g) Micro level citation in Social Psychology

**Figure E.1.: Fitness of cited references**

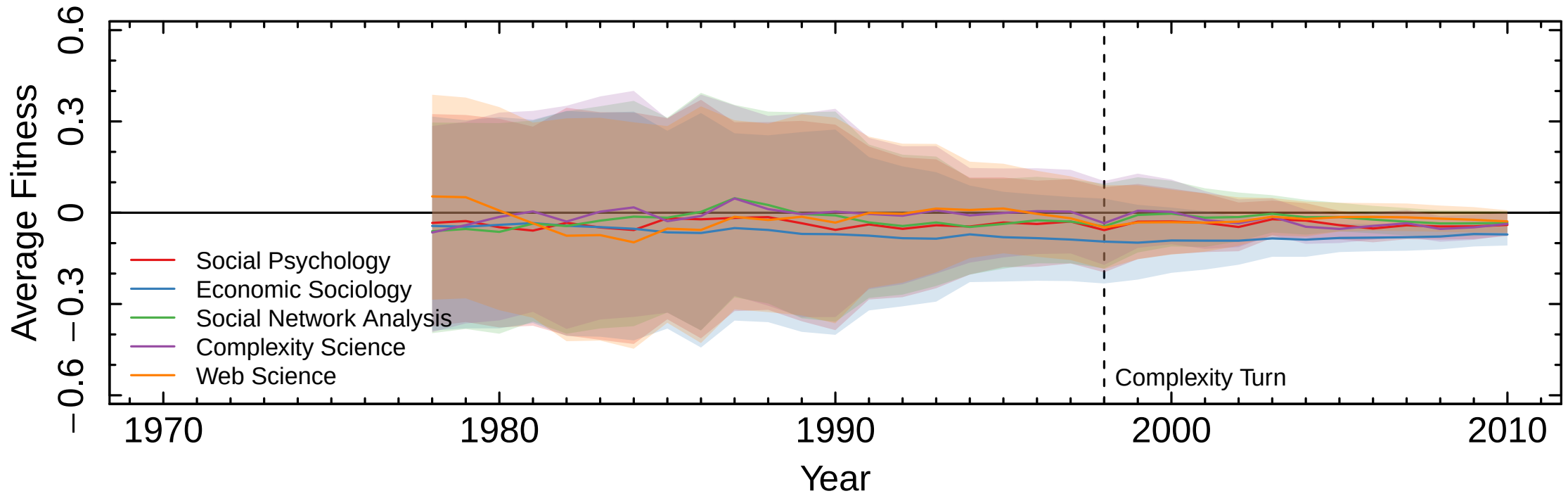

(h) Macro level citation in Economic Sociology

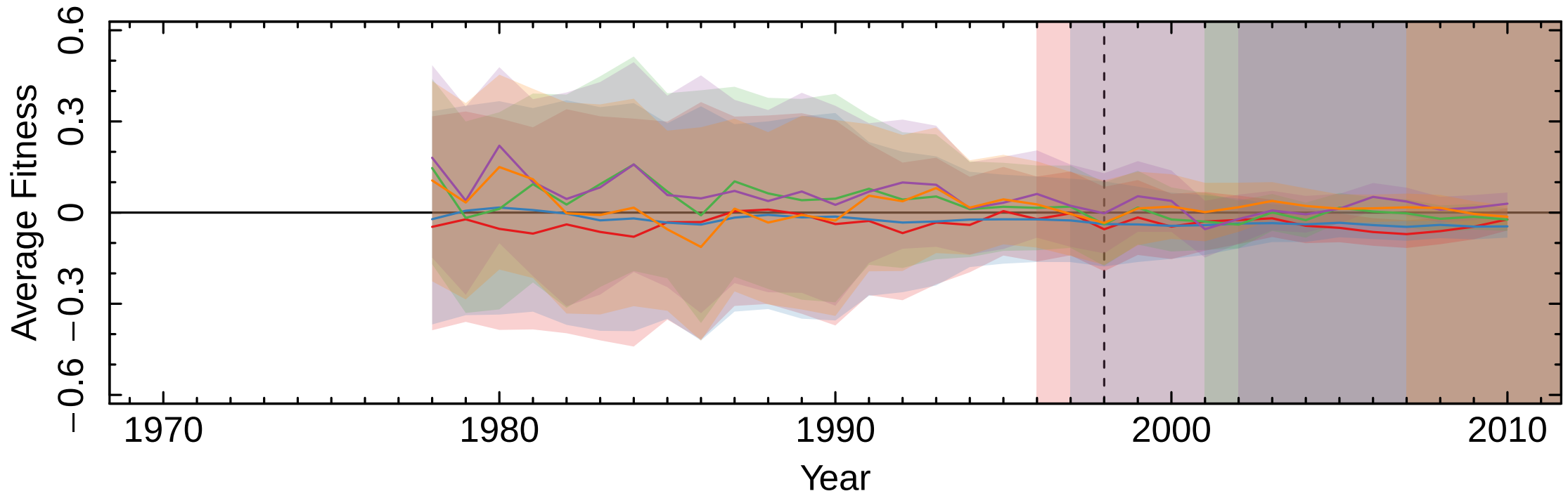

(i) Meso level citation in Economic Sociology

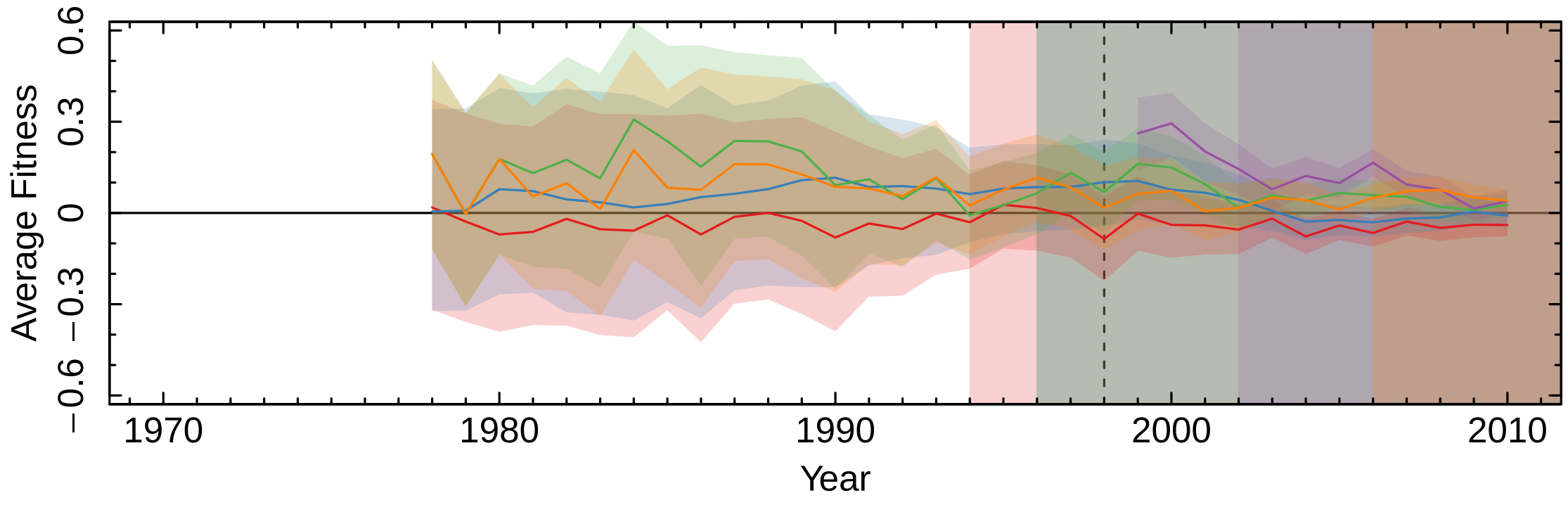

(j) Micro level citation in Economic Sociology

**Figure E.1.: Fitness of cited references**

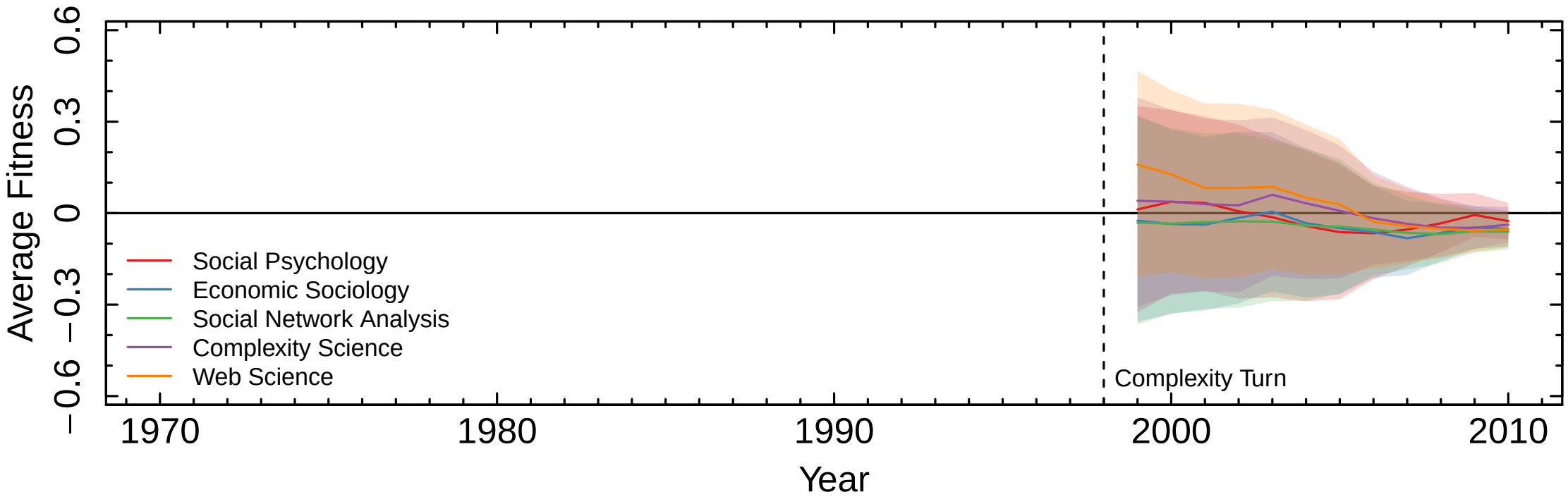

(k) Macro level citation in Social Network Analysis

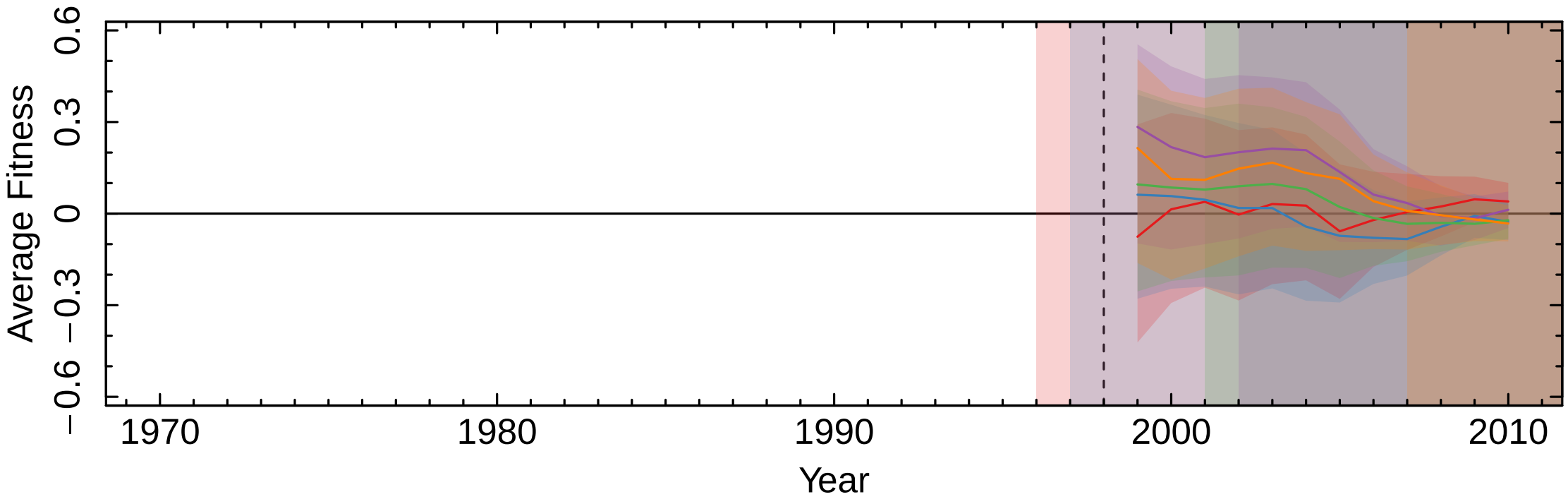

(l) Meso level citation in Social Network Analysis

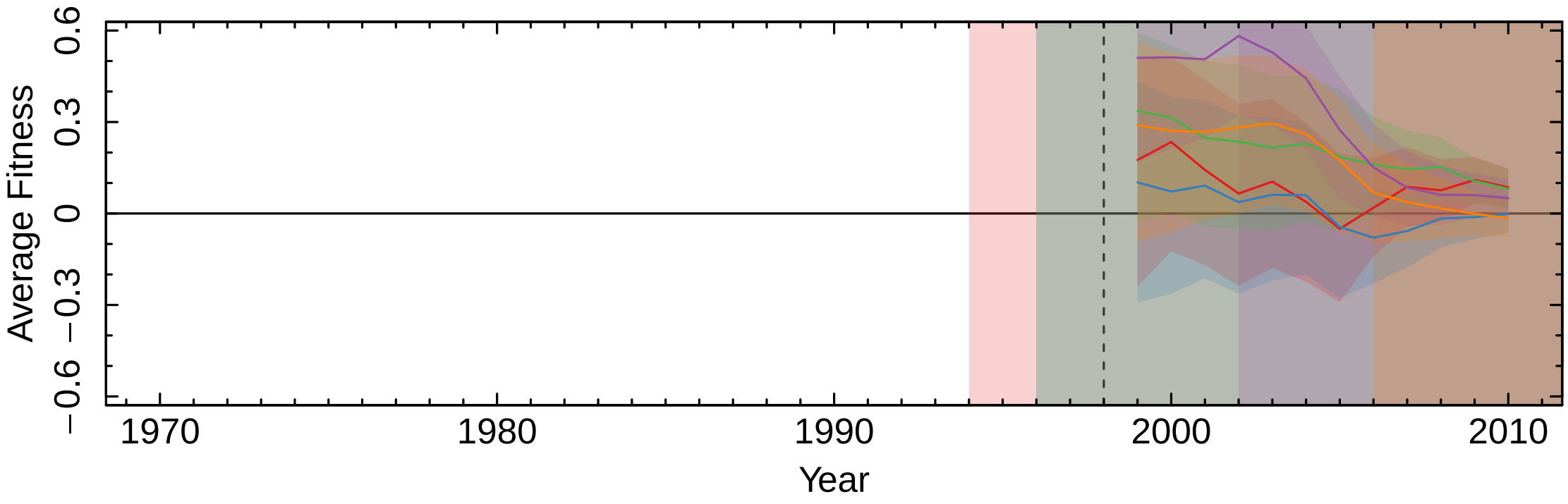

(m) Micro level citation in Social Network Analysis

**Figure E.1.: Fitness of cited references**

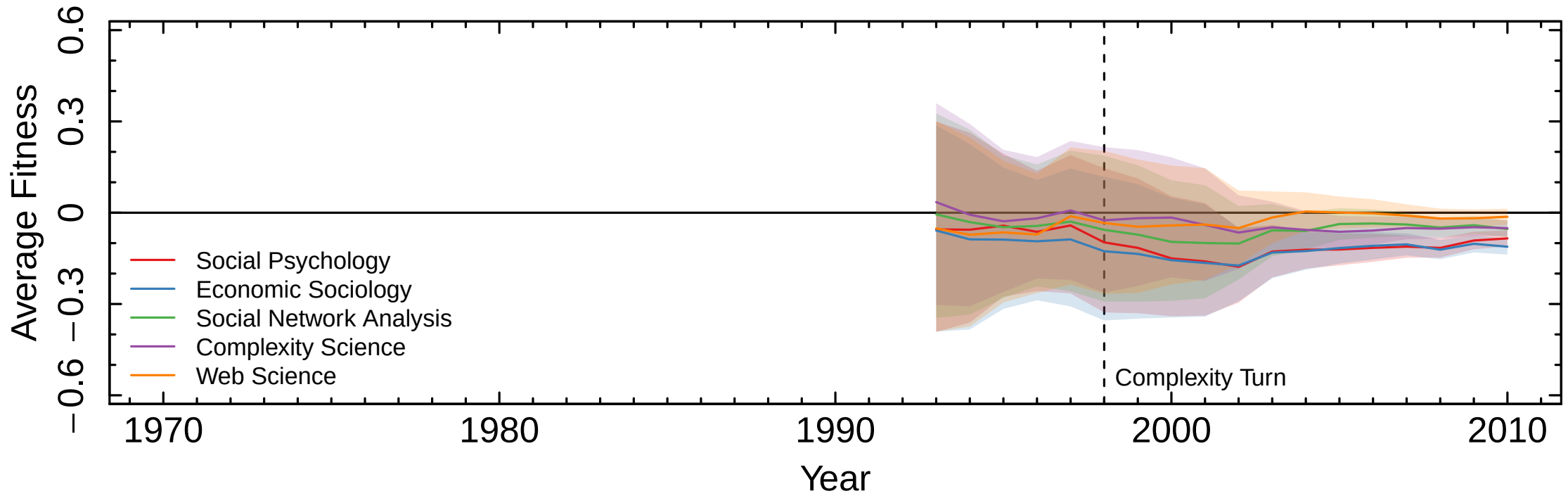

(n) Macro level citation in Complexity Science

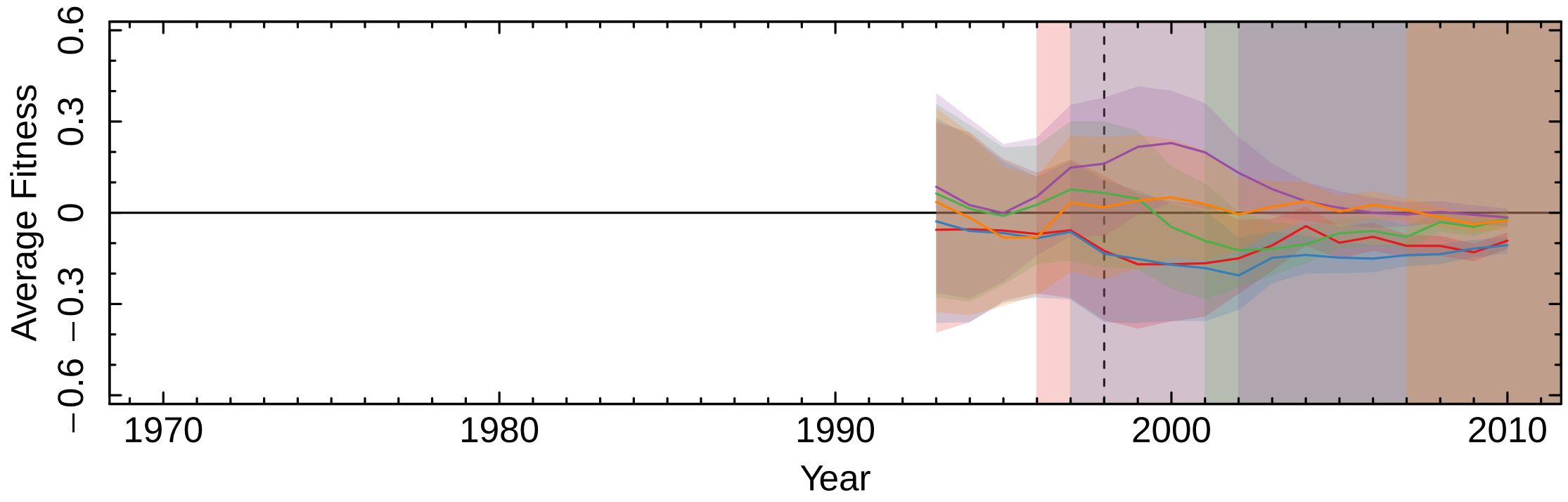

(o) Meso level citation in Complexity Science

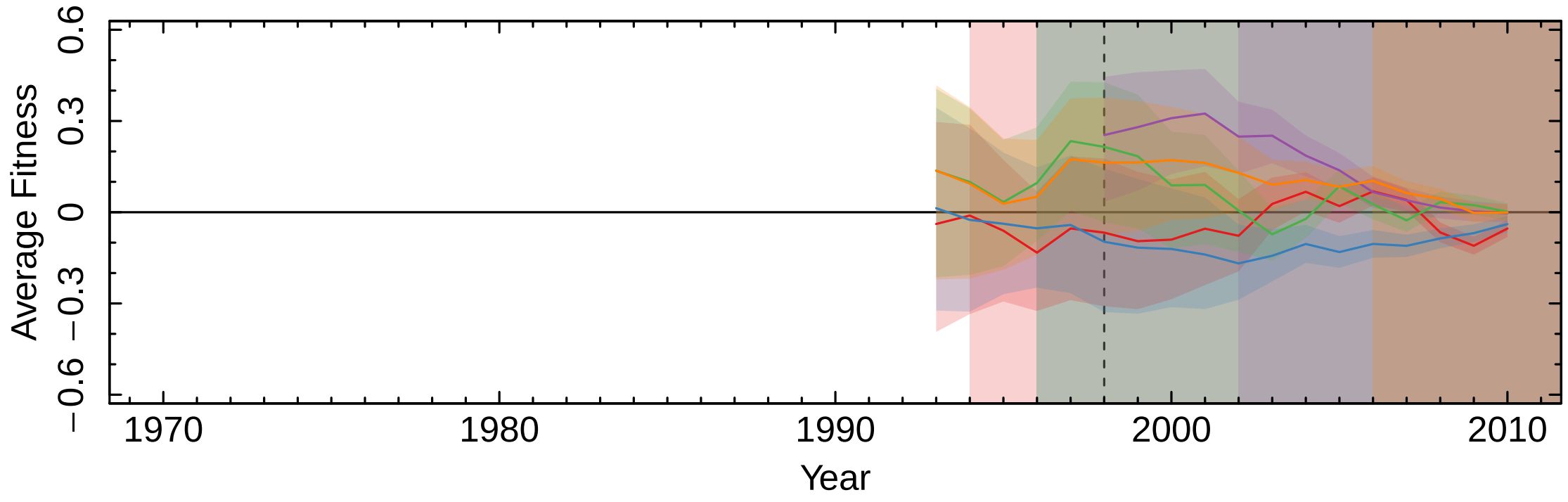

(p) Micro level citation in Complexity Science

**Figure E.1.: Fitness of cited references**

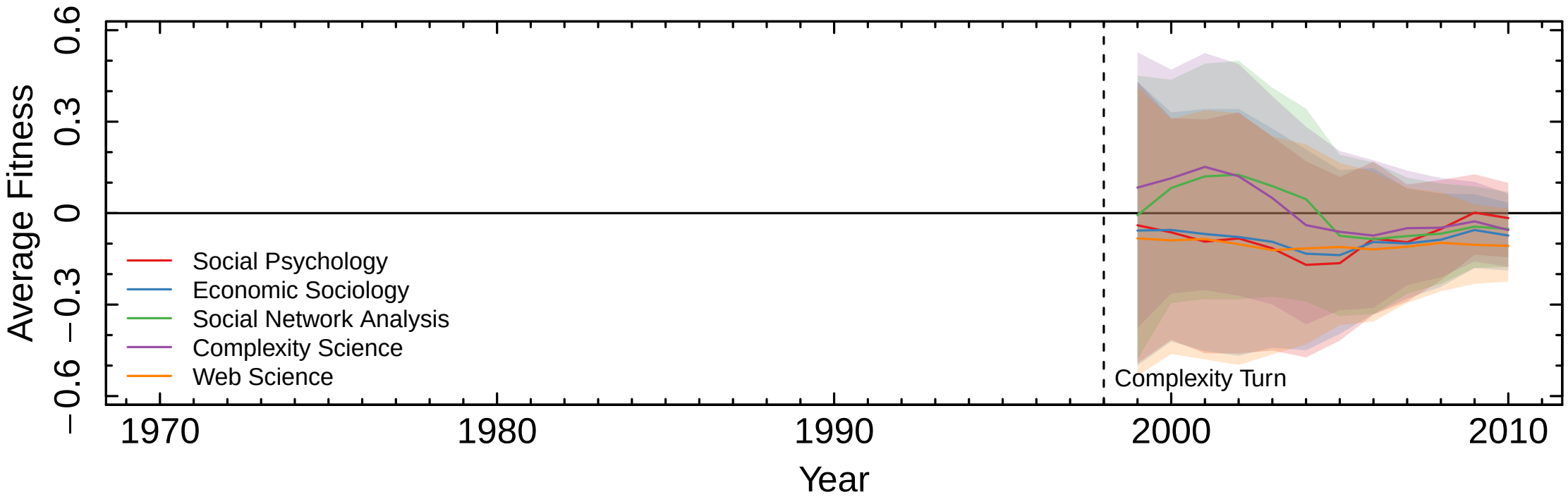

(q) Macro level citation in Web Science

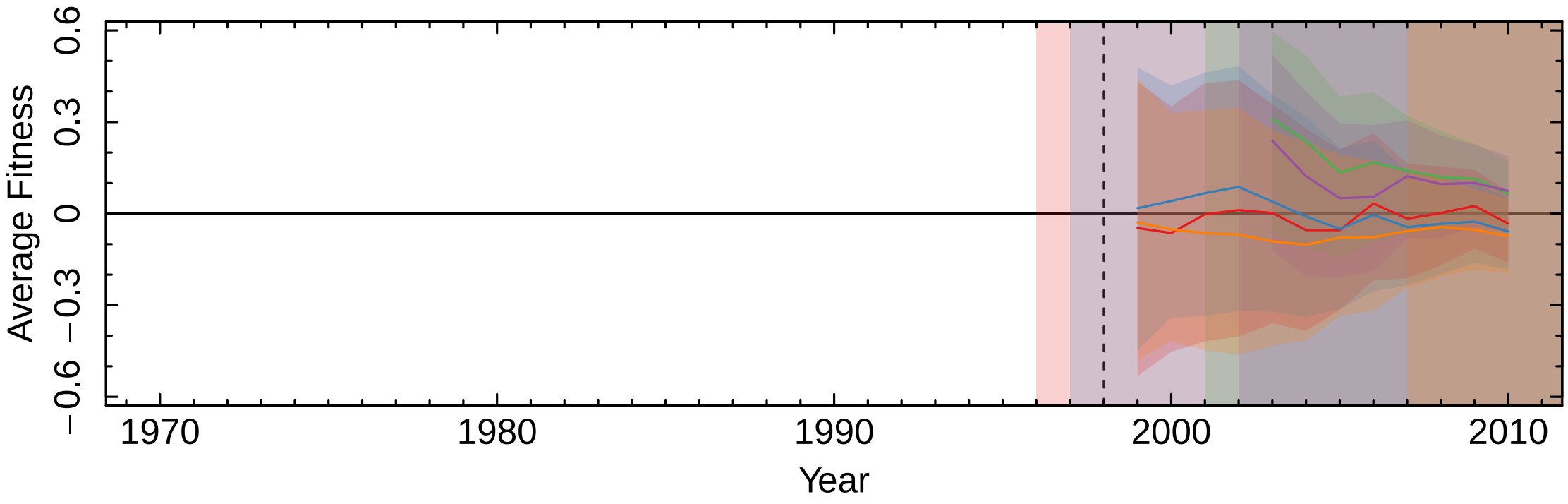

(r) Meso level citation in Web Science

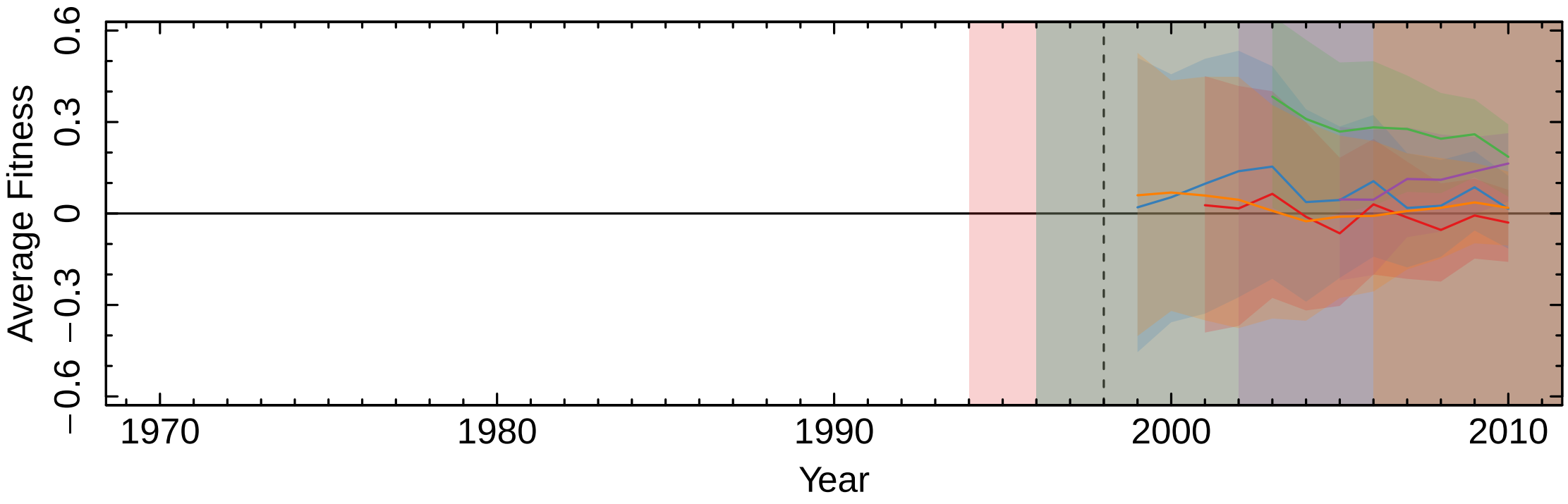

(s) Micro level citation in Web Science

**Figure E.1.: Fitness of cited references**

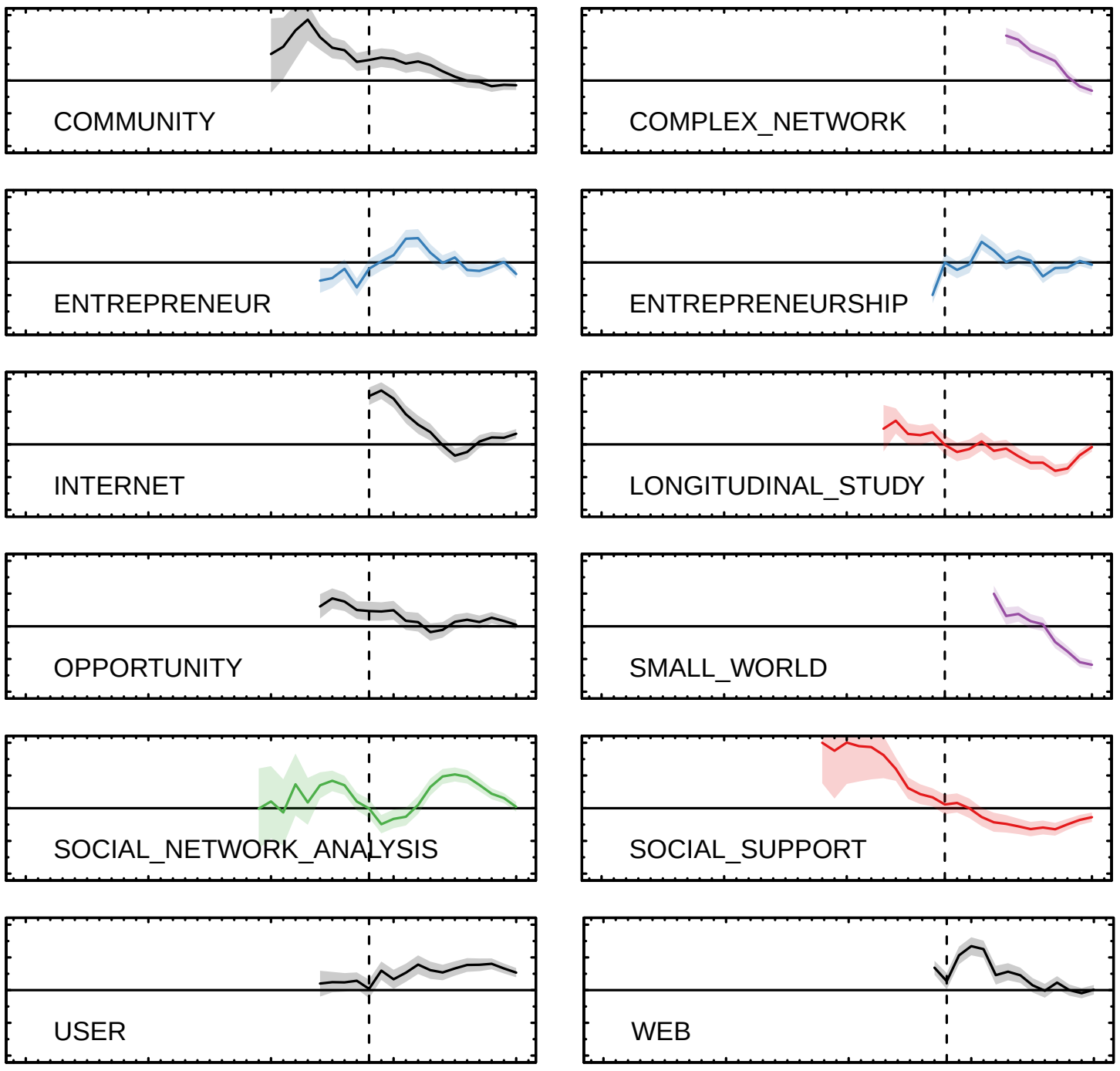

(a) Micro level word usage in the whole domain

**Figure E.2.: Fitness of words**

Word usage is less dynamic than citation. Descriptions as in figure E.1.

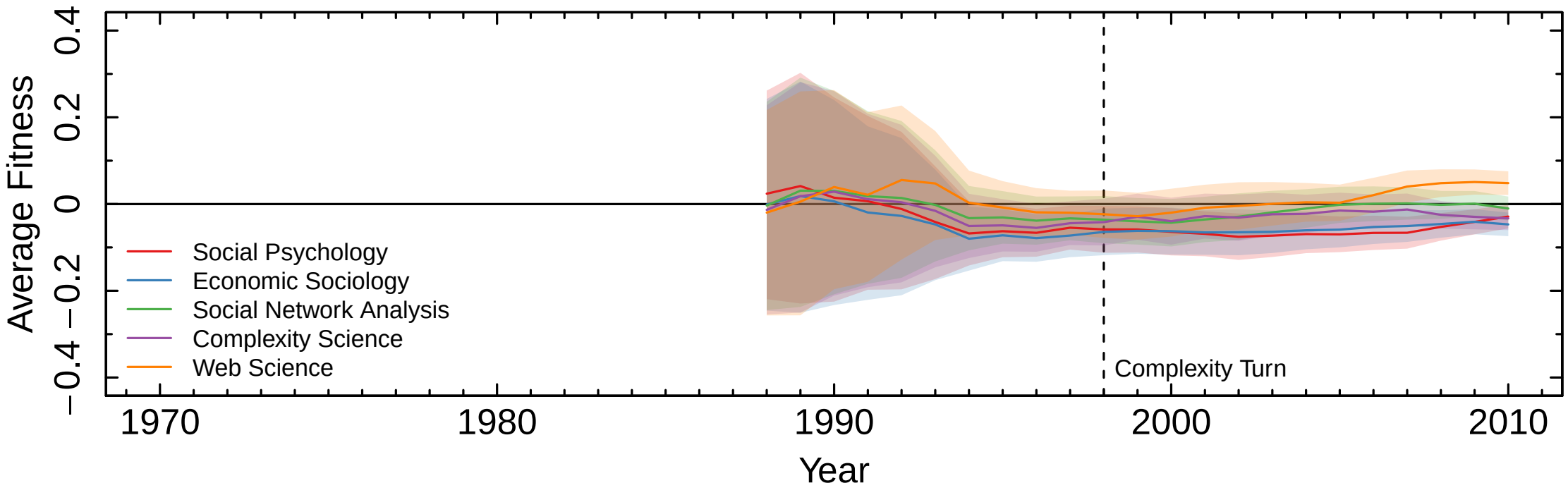

(b) Macro level word usage in the whole domain

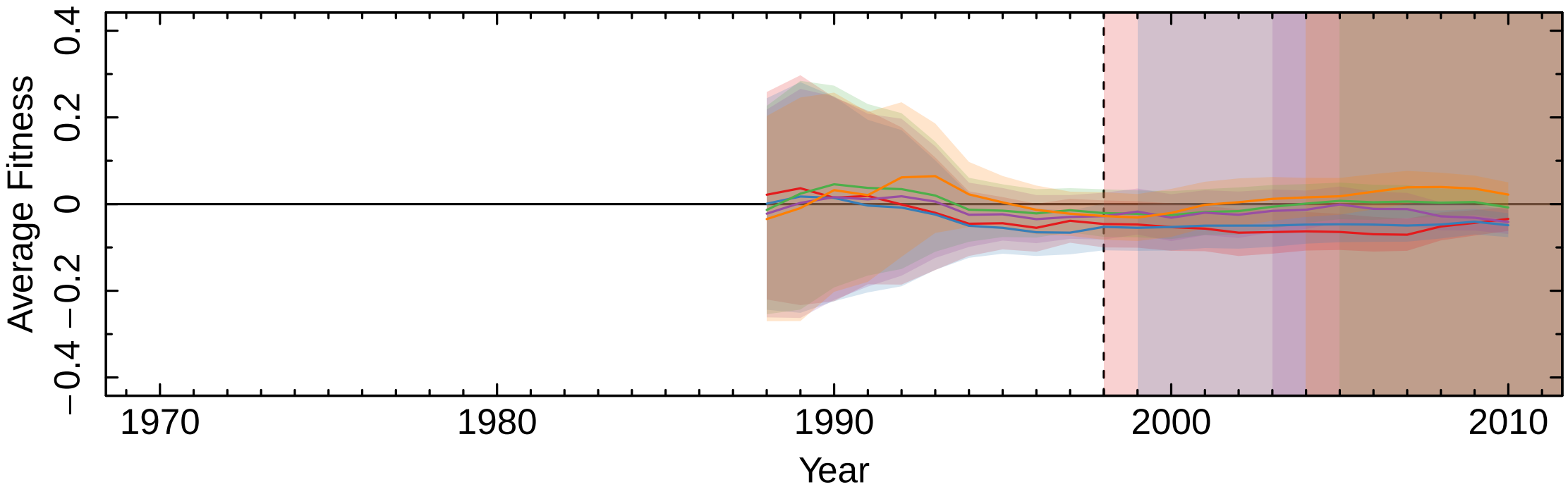

(c) Meso level word usage in the whole domain

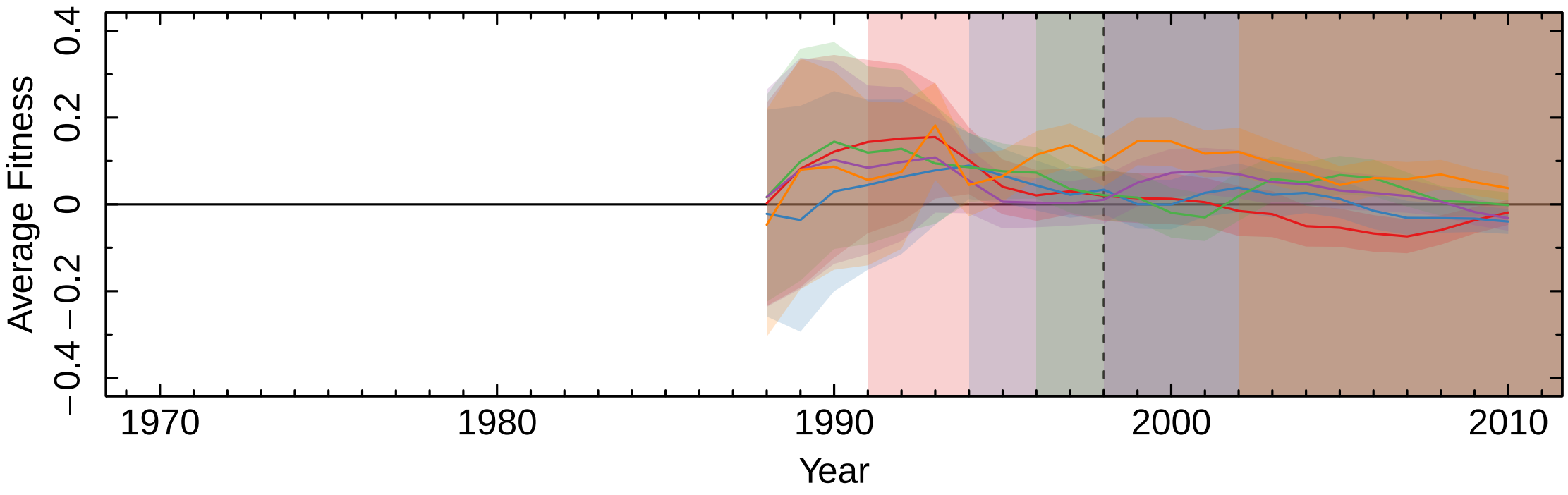

(d) Micro level word usage in the whole domain

**Figure E.2.: Fitness of words**

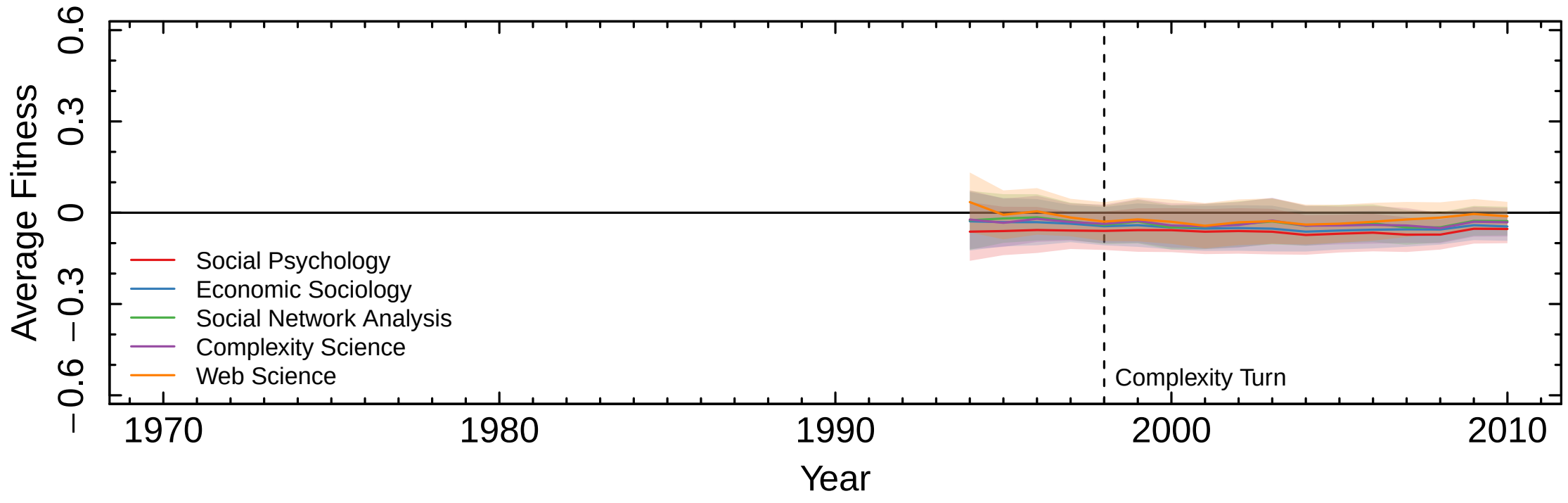

(e) Macro level word usage in Social Psychology

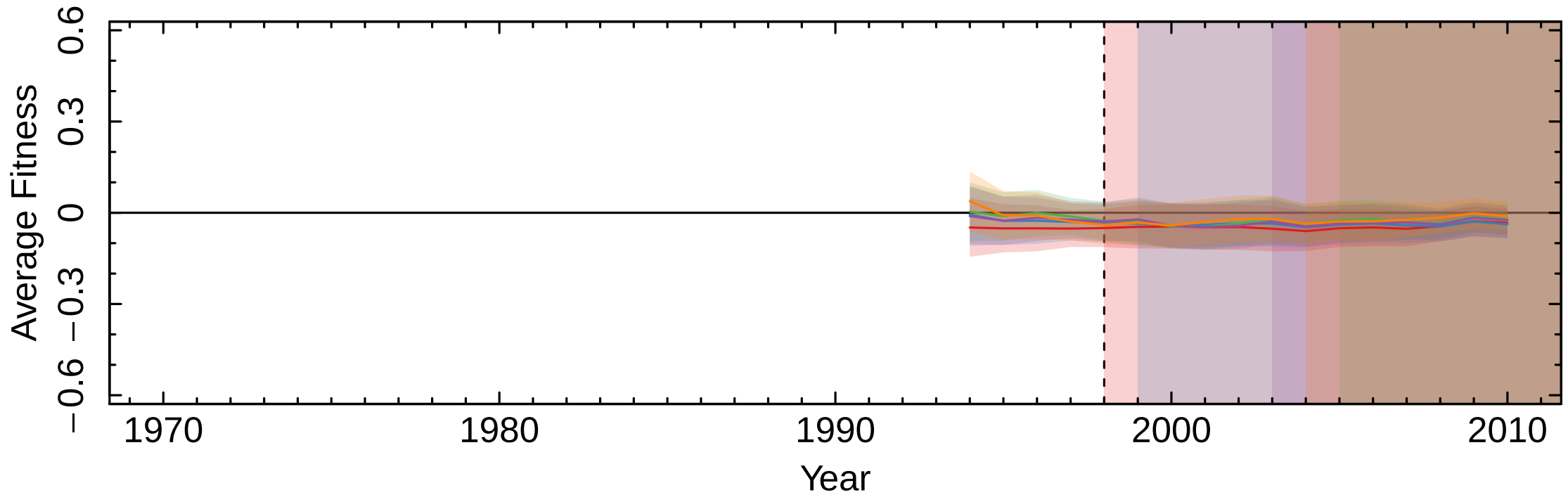

(f) Meso level word usage in Social Psychology

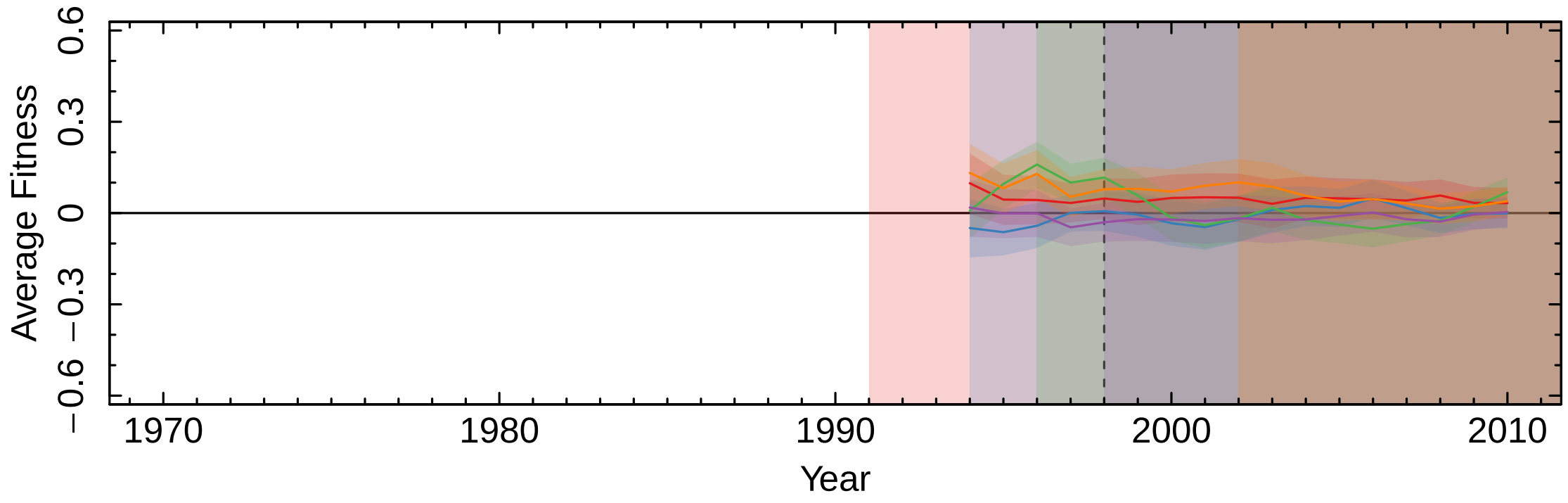

(g) Micro level word usage in Social Psychology

**Figure E.2.: Fitness of words**

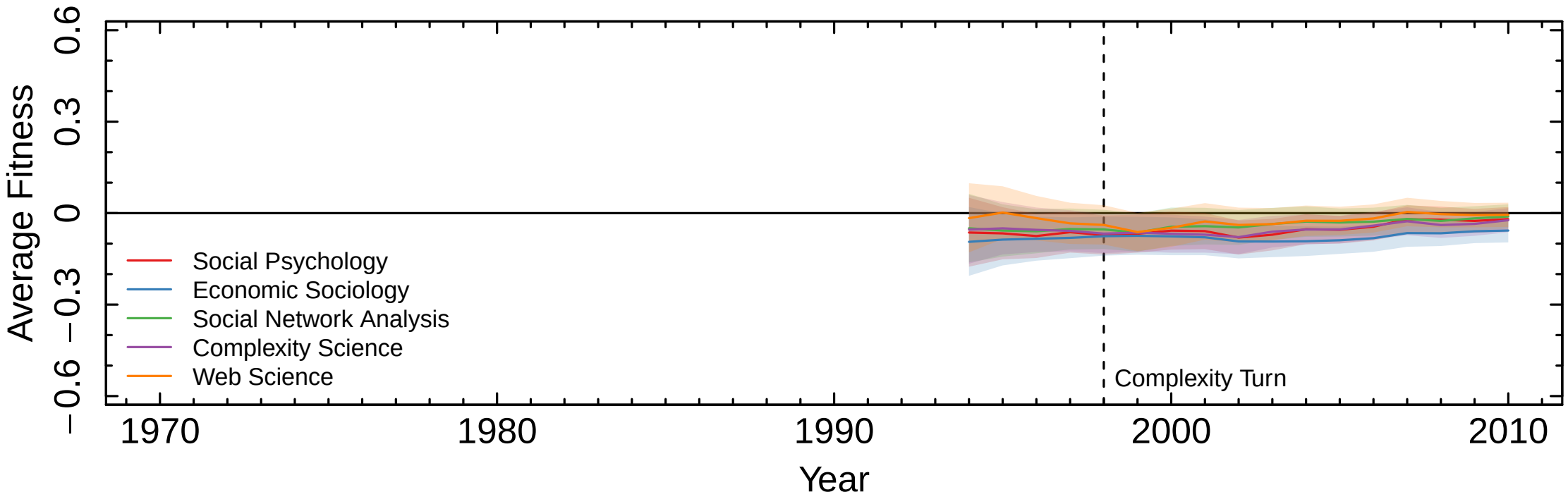

(h) Macro level word usage in Economic Sociology

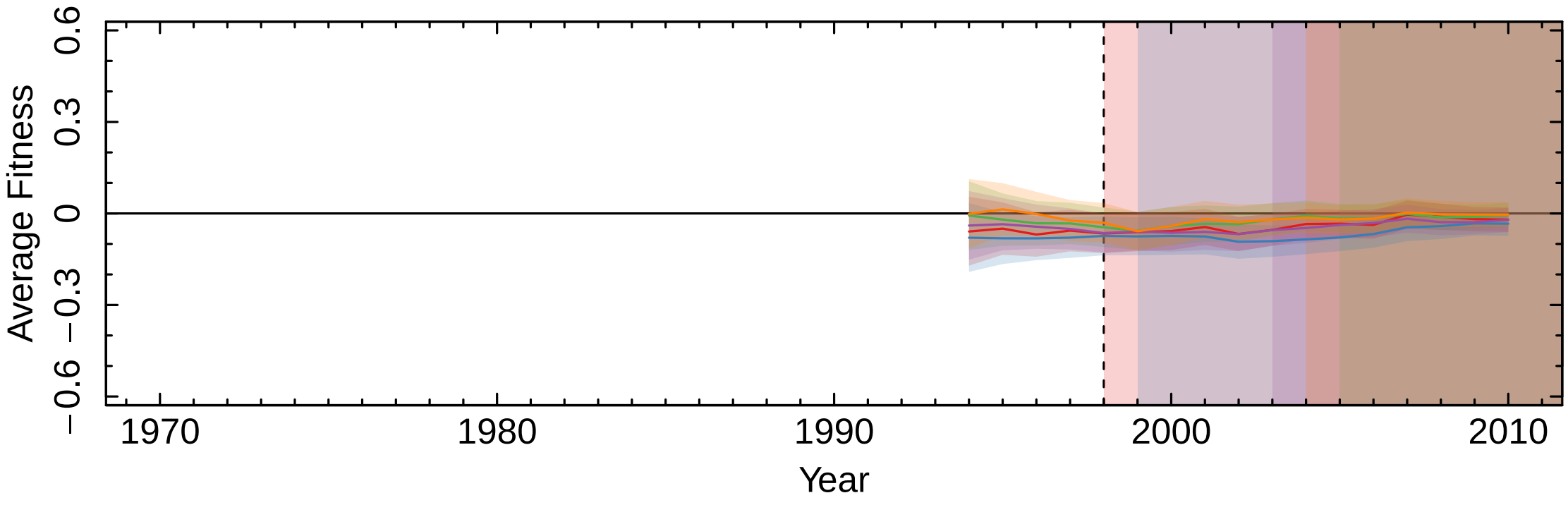

(i) Meso level word usage in Economic Sociology

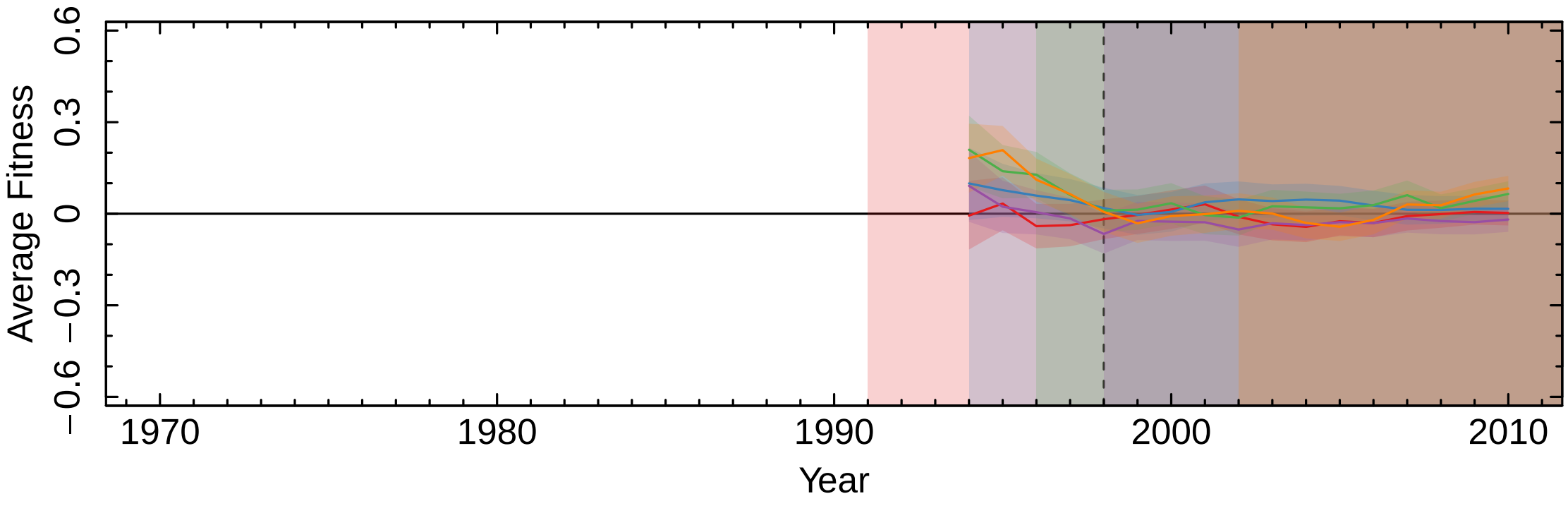

(j) Micro level word usage in Economic Sociology

**Figure E.2.: Fitness of words**

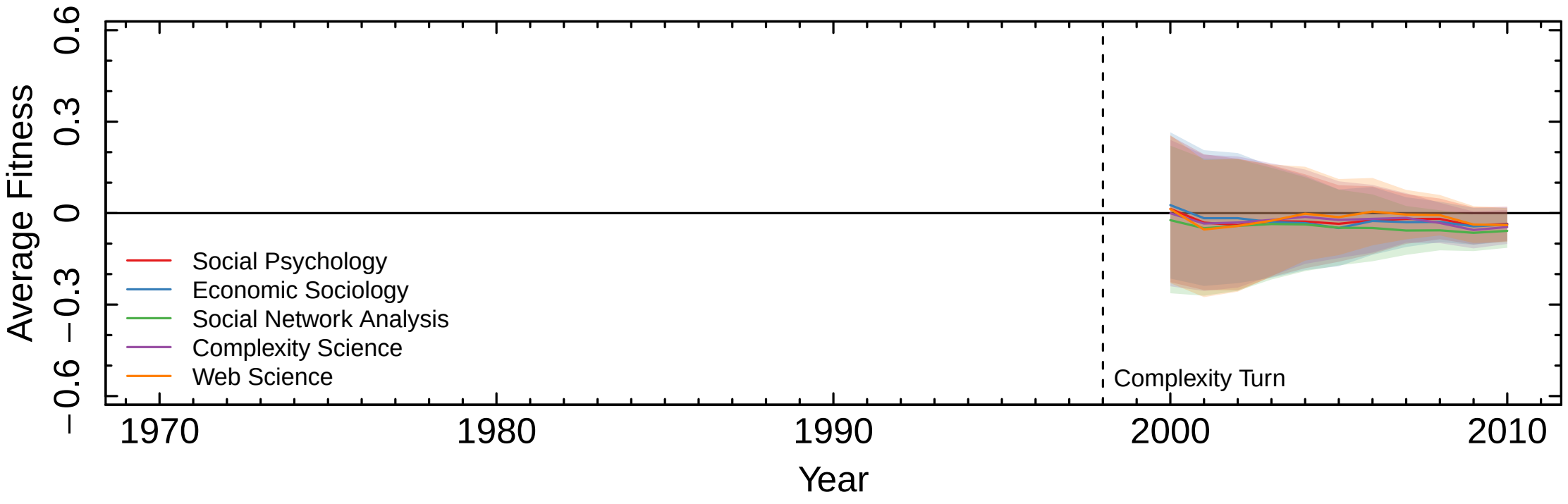

(k) Macro level word usage in Social Network Analysis

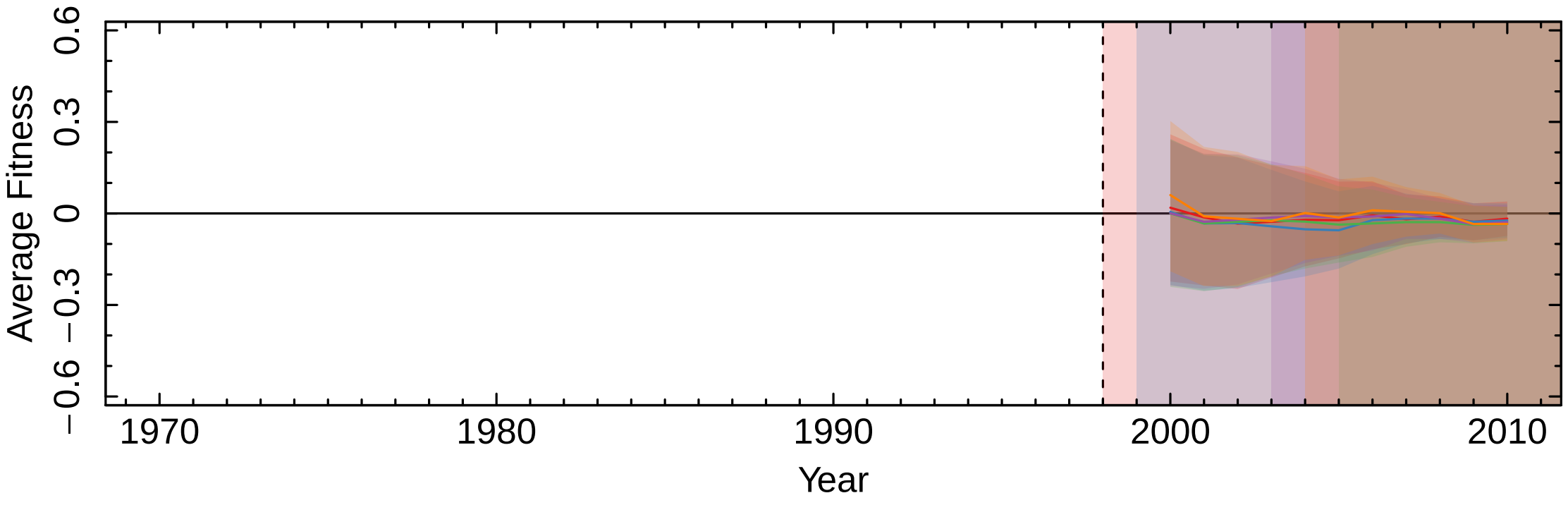

(l) Meso level word usage in Social Network Analysis

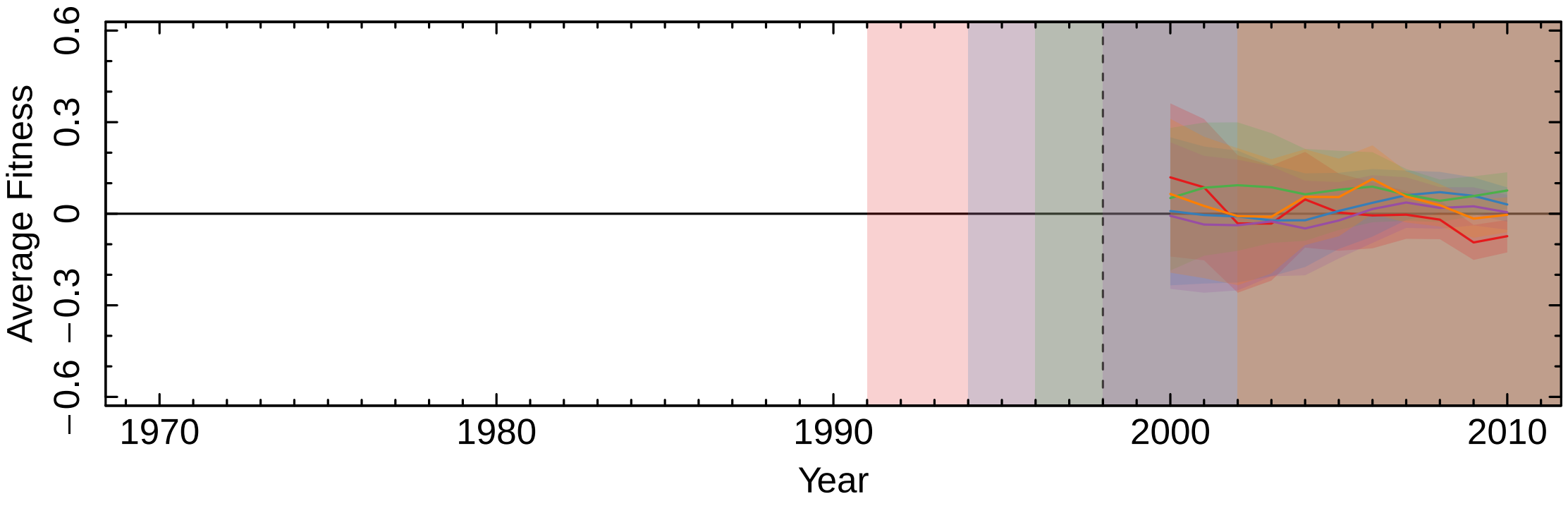

(m) Micro level word usage in Social Network Analysis

**Figure E.2.: Fitness of words**

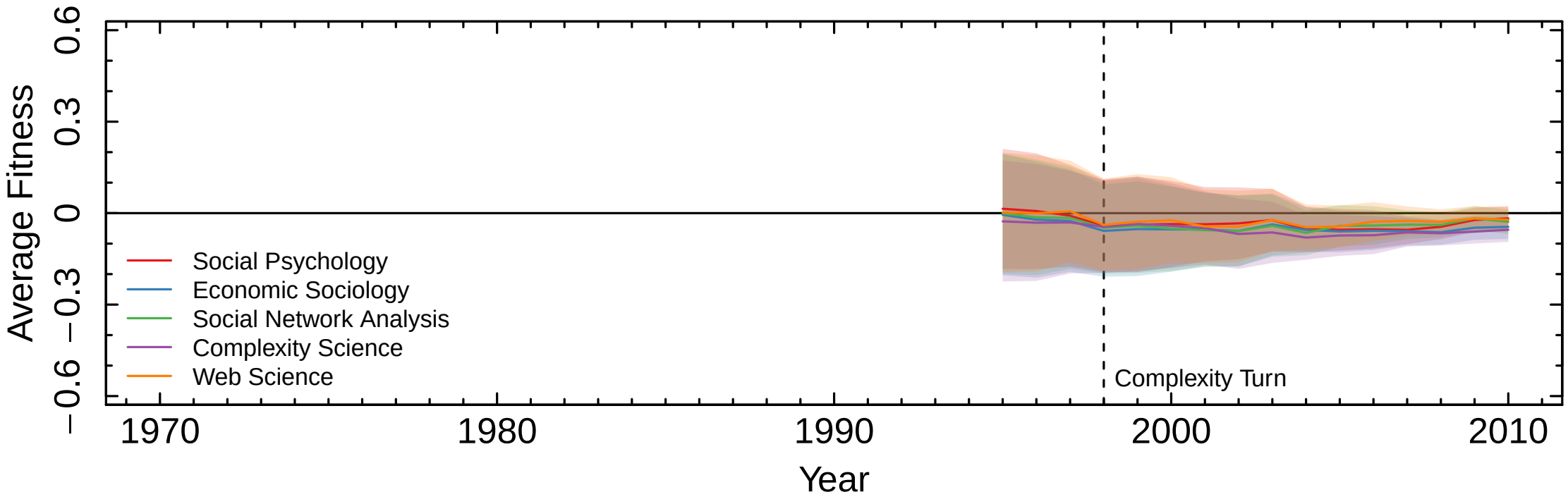

(n) Macro level word usage in Complexity Science

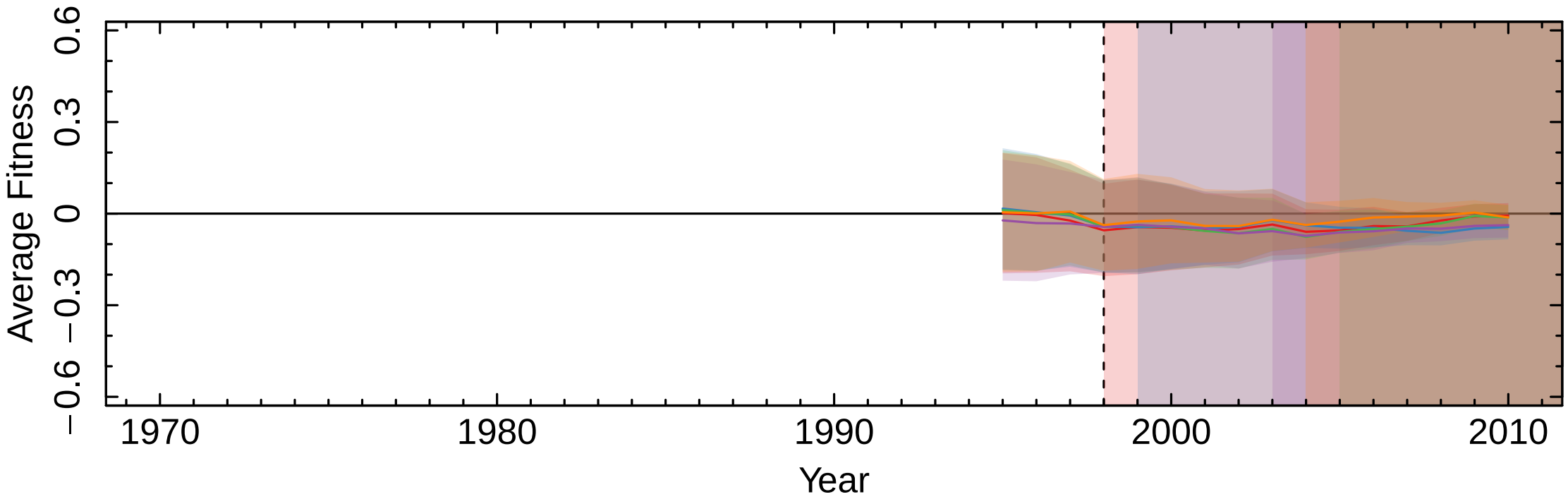

(o) Meso level word usage in Complexity Science

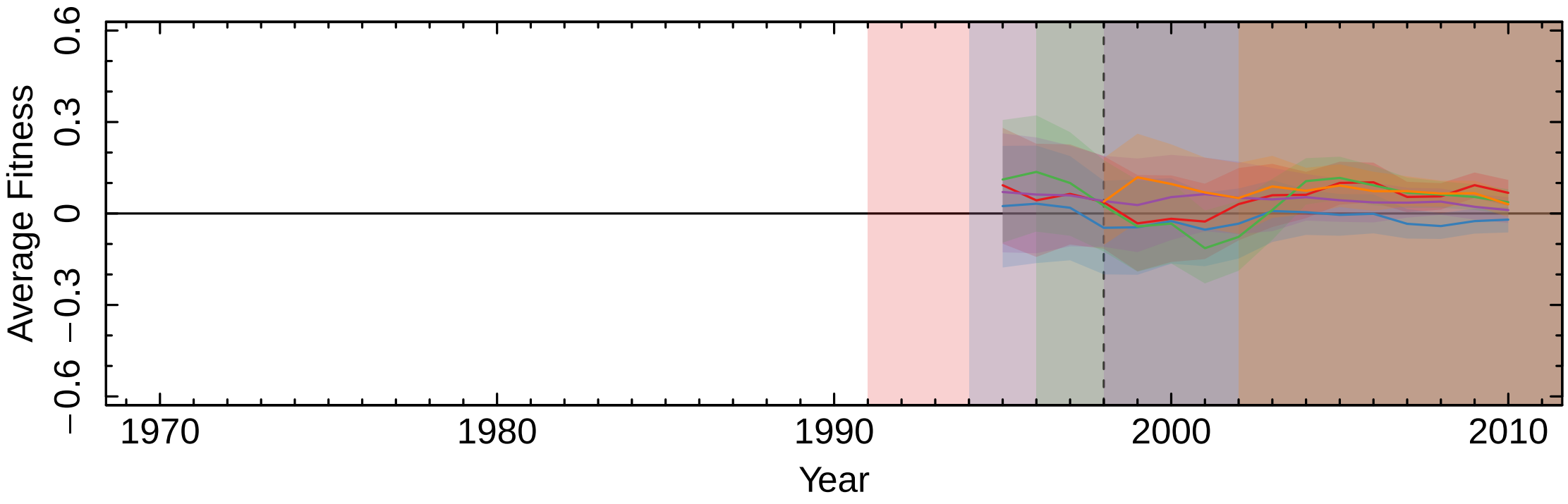

(p) Micro level word usage in Complexity Science

**Figure E.2.: Fitness of words**

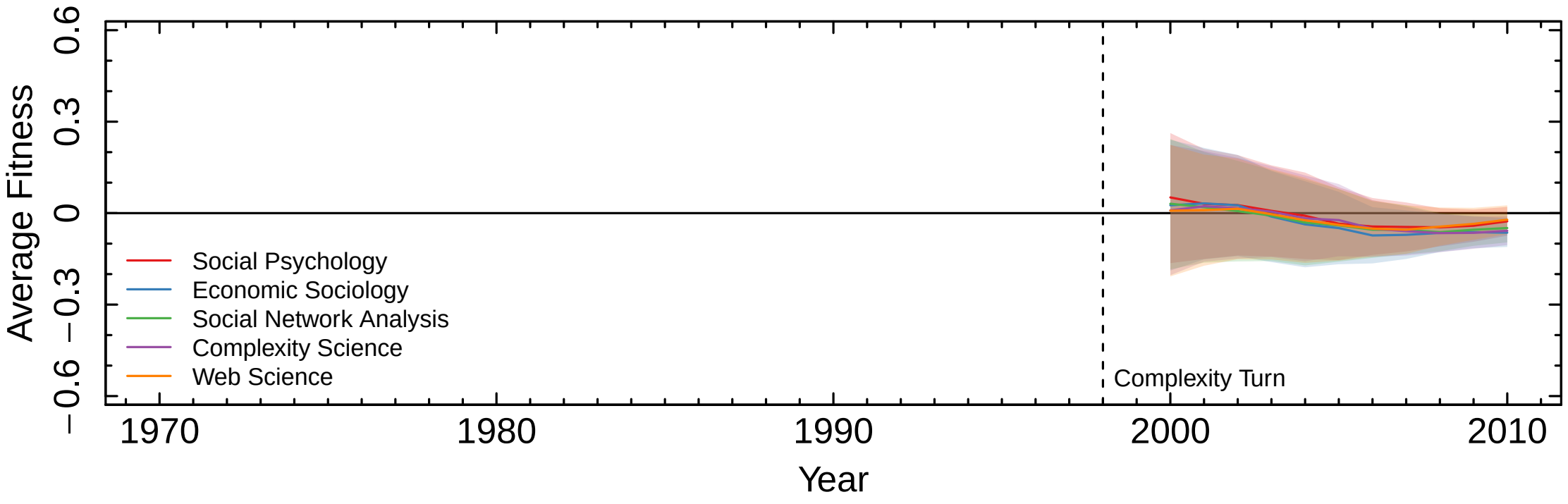

(q) Macro level word usage in Web Science

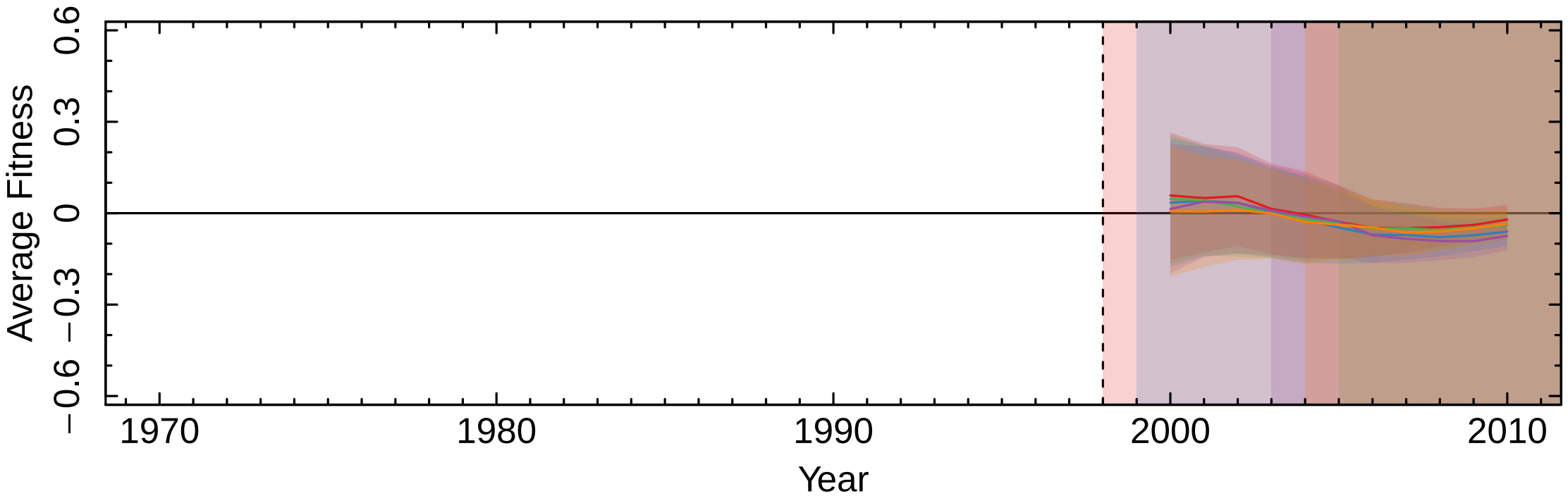

(r) Meso level word usage in Web Science

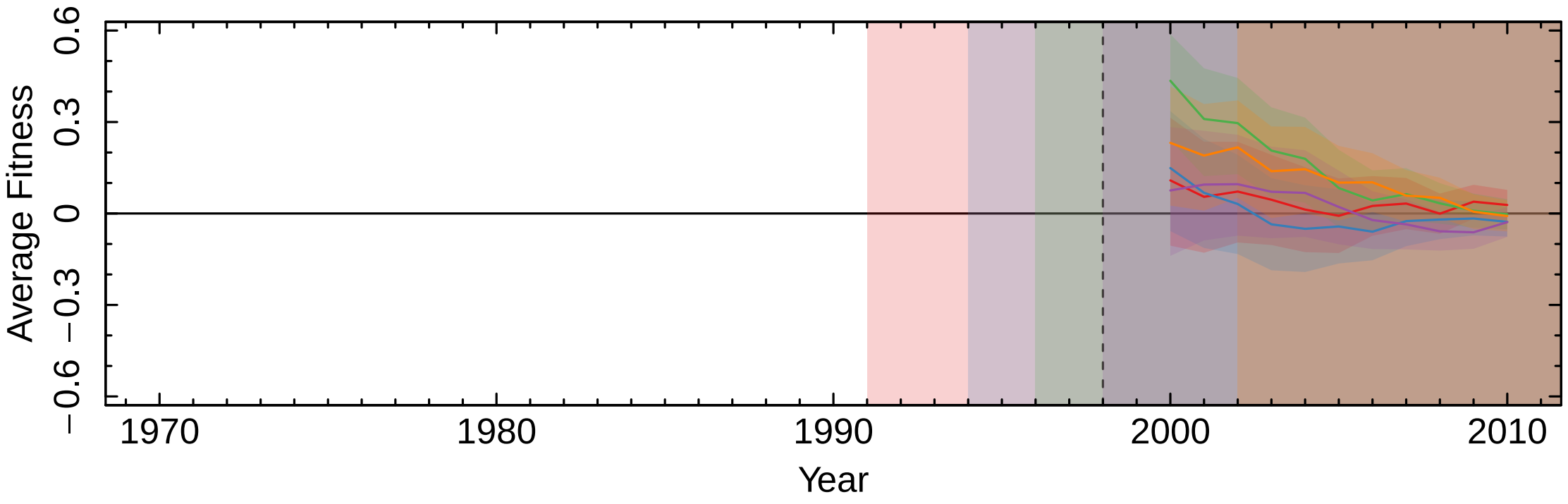

(s) Micro level word usage in Web Science

**Figure E.2.: Fitness of words**